\documentclass[apj,twocolumn]{emulateapj_mod}
\usepackage{epsfig,apjfonts,mathptmx}
\usepackage{graphicx}
\usepackage{float}
\usepackage{placeins} 
\usepackage{amsfonts,amsmath,amssymb}
\usepackage{natbib}
\usepackage{url}
\usepackage[colorlinks,citecolor=blue]{hyperref} 
\usepackage{xcolor}
\usepackage{changes}
\usepackage[compatibility=false]{caption} 
\usepackage{subcaption} 
\usepackage{soul}

\newcommand{\angstrom}{\text{\normalfont\AA}}

\newcommand{\REVISED}[2][]{%
\ifthenelse{\isempty{#1}}%
{\textbf{\textcolor{blue}{#2}}}%
{\textbf{\textcolor{blue}{#2}}}%
} 
\renewcommand{\REVISED}[2][]{#2}

\newcommand{\REREVISED}[2][]{%
\ifthenelse{\isempty{#1}}%
{\textbf{\textcolor{blue}{#2}}}%
{\textbf{\textcolor{gray!50!white}{\sout{#1}}}\textbf{\textcolor{blue}{\ul{#2}}}}%
} 

\renewcommand{\REREVISED}[2][]{%
\ifthenelse{\isempty{#1}}%
{\textbf{\textcolor{blue}{#2}}}%
{\textbf{\textcolor{blue}{#2}}}%
} 

\newcommand{\MINORREREVISED}[2][]{%
\ifthenelse{\isempty{#1}}%
{\textbf{\textcolor{blue}{#2}}}%
{\textbf{\textcolor{blue}{#2}}}%
} 

\renewcommand{\MINORREREVISED}[2][]{%
\ifthenelse{\isempty{#1}}%
{#2}%
{#2}%
} 

\renewcommand{\REREVISED}[2][]{%
\ifthenelse{\isempty{#1}}%
{#2}%
{#2}%
} 

\newcommand{\Superdb}{{Super-deblended}}
\newcommand{\superdb}{{super-deblended}}
\newcommand{\crowdedness}{\textit{crowdedness}}
\newcommand{\goodArea}{\textit{goodArea}}
\newcommand{\galfit}{\textit{galfit}}

\def\gtsima{$\; \buildrel > \over \sim \;$}
\def\ltsima{$\; \buildrel < \over \sim \;$}
\def\prosima{$\; \buildrel \propto \over \sim \;$}
\def\gsim{\lower.5ex\hbox{\gtsima}}
\def\lsim{\lower.5ex\hbox{\ltsima}}
\def\simgt{\lower.5ex\hbox{\gtsima}}
\def\simlt{\lower.5ex\hbox{\ltsima}}
\def\simpr{\lower.5ex\hbox{\prosima}}

\def\h1{$h^{-1}$}
\def\beq{\begin{equation}}
\def\eeq{\end{equation}}
\def\8mu{8\,$\mu{\rm m}$}
\def\16mu{16\,$\mu{\rm m}$}
\def\24mu{24\,$\mu{\rm m}$}
\def\70mu{70\,$\mu{\rm m}$}

\shorttitle{Super-deblended IR photometry of star forming galaxies}
\shortauthors{D. Liu et al.}

\begin{document}

\title{``\Superdb'' Dust Emission in Galaxies: I. The GOODS-North Catalog\\ and the Cosmic Star Formation Rate Density out to Redshift 6}

\author{%
    Daizhong Liu\altaffilmark{1,2,3,4}, 
    Emanuele Daddi\altaffilmark{2}, 
    Mark Dickinson\altaffilmark{5}, 
    Frazer Owen\altaffilmark{6}, 
    Maurilio Pannella\altaffilmark{7}, 
    Mark Sargent\altaffilmark{8}, 
    Matthieu B{\'{e}}thermin\altaffilmark{9,10}, 
    Georgios Magdis\altaffilmark{11}, 
    Yu Gao\altaffilmark{1}, 
    Xinwen Shu\altaffilmark{2,12}, 
    Tao Wang\altaffilmark{2}, 
    Shuowen Jin\altaffilmark{2},
    Hanae Inami\altaffilmark{13}
}
\altaffiltext{1}{Purple Mountain Observatory/Key Laboratory for Radio Astronomy, Chinese Academy of Sciences, No. 2 West Bejing Road, Nanjing, China
(dzliu@mpia.de)
}
\altaffiltext{2}{CEA, IRFU, DAp, AIM, Universit\'{e} Paris-Saclay, Universit\'{e} Paris Diderot, Sorbonne Paris Cit\'{e}, CNRS, F-91191 Gif-sur-Yvette, France
}
\altaffiltext{3}{University of Chinese Academy of Sciences, 19A Yuquan Road, Shijingshan District, 10049, Beijing, China}
\altaffiltext{4}{Max Planck Institute for Astronomy, K\"{o}nigstuhl 17, D-69117 Heidelberg, Germany}
\altaffiltext{5}{National Optical Astronomy Observatory, Tucson, Arizona 85719, USA}
\altaffiltext{6}{National Radio Astronomy Observatory, Socorro, NM 87801, USA}
\altaffiltext{7}{%
Faculty of Physics, Ludwig-Maximilians Universit\"at, Scheinerstr.\ 1, 81679 Munich, Germany}
\altaffiltext{8}{Astronomy Centre, Department of Physics and Astronomy, University of Sussex, Brighton, BN1 9QH, UK}
\altaffiltext{9}{European Southern Observatory, Karl Schwarzschild Strasse 2, 85748, Garching, Germany}
\altaffiltext{10}{Aix Marseille Univ, CNRS, LAM, Laboratoire d'Astrophysique de Marseille, Marseille, France}
\altaffiltext{11}{Dark Cosmology Centre, Niels Bohr Institute, University of Copenhagen, Juliane Mariesvej 30, DK-2100 Copenhagen, Denmark}
\altaffiltext{12}{Anhui Normal University, Wuhu, Anhui, China}
\altaffiltext{13}{Observatoire de Lyon, 9, avenue Charles Andr\'{e}, 69561 Saint Genis Laval, France}

\begin{abstract}
We present a new technique to measure multi-wavelength ``\superdb{}'' photometry from highly confused images, which we apply to \textit{Herschel} and ground-based far-infrared (FIR) and (sub-)millimeter (mm) data in the northern field of the Great Observatories Origins Deep Survey (GOODS). There are two key novelties. First, starting with a large database of deep \textit{Spitzer} 24~$\mu$m and VLA 20~cm detections that are used to define prior positions for fitting the FIR/submm data, we perform an \textit{active} selection of \textit{useful} priors independently at each frequency band, moving from less to more confused bands. Exploiting knowledge of redshift and all available photometry, we identify \textit{hopelessly faint} priors that we remove from the fitting pool. This approach significantly reduces blending degeneracies and allows reliable photometry to be obtained for galaxies in FIR+mm bands. Second, we obtain well-behaved, nearly Gaussian flux density uncertainties, individually tailored to all fitted priors for each band. This is done by exploiting extensive simulations that allow us to calibrate the conversion of formal fitting uncertainties to realistic uncertainties depending on directly measurable quantities. We achieve deeper detection limits with high fidelity measurements and uncertainties at FIR+mm bands.  As an illustration of the utility of these measurements, we identify 
70 
galaxies with $z\ge3$ and reliable FIR+mm detections.
We present new constraints on the cosmic star formation rate density at $3<z<6$, finding a significant contribution from $z\ge3$ dusty galaxies that are missed by optical-to-near-infrared color selection. Photometric measurements for 3306 priors, including over 1000 FIR+mm detections are released publicly with our catalog.
\end{abstract}


\section{Introduction}

\label{Section_Introduction}


A wealth of infrared (IR) to millimeter (mm) deep survey observations has accumulated in the last decade, with data obtained by the \textit{Spitzer Space Telescope} \citep[hereafter \textit{Spitzer};][]{Werner2004}, \textit{Herschel Space Observatory} \citep[hereafter \textit{Herschel};][]{Pilbratt2010} and many ground-based single-dish telescopes (e.g., the IRAM 30m and JCMT 15m telescopes). 
These observations are indispensable for understanding the evolution of star formation and of the interstellar medium of galaxies from early cosmic epochs to the present. Dust grains are produced on short time scales during the evolution of massive stars, and hence are a direct product of star formation activity in young star-forming galaxies\footnote{Dust grains can also be produced by low- and intermediate- mass stars during their Asymptotic Giant Branch (AGB) stage, on longer time scales, e.g., \citealt{Draine2003ARAA,Gail2009,Zhukovska2013}.}.  Dust absorbs starlight at ultra-violet (UV) to optical wavelengths and re-emits the energy as longer-wavelength IR photons. 
Large amounts of dust can exist in galaxies with intense star formation activity, strongly attenuating their rest-frame UV-to-optical light. 
Hence, with UV to optical observations alone, one can directly measure only relatively unobscured star formation in galaxies. 
The dust attenuation can be estimated from the UV to optical photometry, for example from the UV continuum slope or the UV-to-optical spectral energy distribution (SED), but these estimates are indirect, they can have large uncertainties, and they can entirely fail to detect large amounts of star formation hiding behind optically thick dust. Far-IR (FIR) observations have thus become an essential tool for directly measuring  obscured star formation rates (SFRs) in galaxies across cosmic time
(e.g., \citealp{Draine2007SINGS}; \citealp{Ciesla2014}; \citealp{Magdis2012SED}; \citealp{Tan2014}). 

\textit{Spitzer} was the first IR space telescope efficient and sensitive enough to survey large areas with adequate sensitivity to  detect galaxies at cosmological redshifts (e.g., \citealp{LeFloch2005}; \citealp{LeFloch2009}; \citealp{Frayer2006}; \citealp{Frayer2009}; \citealp{Magnelli2009}; \citealp{Magnelli2011}; \citealp{Ashby2013,Ashby2015}). 
\textit{Spitzer} observations are very sensitive at $3.6$--$8.0\,{\mu}\mathrm{m}$ \citep[the Infrared Array Camera, IRAC;][]{Fazio_2004_IRAC_Instrument} and at 24~$\mu$m \citep[the Multiband Imaging Photometer, MIPS;][]{Rieke_2004_MIPS_Instrument}, but is much less sensitive at FIR wavelengths (70 and 160~$\mu$m) than the later IR space telescope \textit{Herschel}. 

\textit{Herschel} observed at FIR wavelengths from 70~$\mu$m to 500~$\mu$m.  Its Photoconductor Array Camera and Spectrometer (PACS; \citealt{Poglitsch2010}) provided more than 4 times better angular resolution than \textit{Spitzer} at 70 to 160~$\mu$m, while the Spectral and Photometric Imaging Receiver (SPIRE; \citealt{Griffin2010}) covered longer wavelengths (250, 350 and 500~$\mu$m) at very efficient mapping speeds.  \textit{Herschel} enabled direct measurement\MINORREREVISED[]{s} of FIR emission for a large number of galaxies at cosmological redshifts (e.g., \citealp{Magnelli2013}; \citealp{Gruppioni2013}; \citealp{Lee2013}; \citealp{Bethermin2015}; \citealp{Schreiber2015}; \citealp{Shu2016}; \citealp{WangTao2016}). 


However, source confusion problems can introduce substantial biases in photometric works. 
\textit{Herschel} SPIRE images have point spread functions (PSFs) that are several times larger (\REVISED[20--38]{17.6--35.2}$''$) than those of PACS images \REVISED{(7--12$''$)}. 
The fluxes of individual galaxies are often difficult to measure with SPIRE because of blending from close neighbors.  
If one considers all possible star forming galaxies that might contribute to the SPIRE signal, and try to simultaneously fit for all of them, no individual measurement can be obtained because of degeneracies.  In sufficiently deep data, the average number of sources that {\it potentially} contribute within each SPIRE beam is always $\gg 1$.
On the other hand, measured fluxes will be biased if we simply ignore a fraction of the potential contributing sources in the \textit{prior-extraction} method \footnote{The prior-extraction method adopts a list of prior source positions as the input for the source fitting with PSF or other models to the image data.}, or if we treat several blended sources as one source (e.g., the \textit{blind-extraction} method~\footnote{%
The blind-extraction method uses automatic searching algorithms to identify \MINORREREVISED[sources above the sky background in the image data]{groups of pixels with values significantly greater than the sky background}.
}). 
\MINORREREVISED[Therefore, a careful approach that considers both the preselection of prior sources (in such a way that the final density of sources corresponds in all bands to $\simlt1$ sources per beam) and the contributions of faint sources is needed in order to obtain high quality photometry and scientific results.]{%
Therefore, a careful approach is needed to preselect the sources used as fitting priors so that the actual number of fitted sources is $\lesssim 1$ per beam for all bands.  In addition, the flux contribution of faint sources that are {\em not} included in the fitting should be taken into account.}

Several solutions have been explored for prior-extraction of galaxy fluxes in FIR/mm images with potentially significant source confusion. 
\citet{Bethermin_2010_FASTPHOT} started from an input catalog of MIPS 24~$\mu$m source positions \citep{Bethermin_2010_SWIRE_MIPS_24} and applied an IDL-based routine with a linear inversion algorithm (FASTPHOT) to perform prior-extraction photometry for BLAST data in the Extended Chandra Deep Field South. 
\citet{Roseboom2010} used MIPS 24~$\mu$m-detected sources ($\mathrm{S/N}\ge5$) as priors for SPIRE photometry 
with their cross-identification (XID) tool~\footnote{%
\REVISED{Similar to FASTPHOT, or other tools like T-PHOT \citep{Merlin2015}, XID (also named DESPHOT) also uses a linear inversion algorithm as its core fitting function. It additionally adopts a ``top-down'' approach to iteratively jackknife the actual prior list at each band so as to achieve a best model fit.}
}%
.
\citet{Elbaz2011}~%
\footnote{Their catalogs are published as \citet{Elbaz2013}.} also use MIPS 24~$\mu$m-detected ($\mathrm{S/N}\ge5$)
sources for PACS 160~$\mu$m and SPIRE 250~$\mu$m photometry in GOODS-North and GOODS-South fields, adopting SPIRE 250~$\mu$m $\mathrm{S/N}>2$ sources as the prior list for SPIRE 350 and 500~$\mu$m photometry. 
\citet{Bethermin2012,Bethermin2015} used MIPS 24~$\mu$m sources as priors as well, but have done more detailed selections. They first perform stacking to derive an average relationship between redshift, stacked flux and SPIRE-to-24~$\mu$m color (flux ratio), based on which they predict the SPIRE flux for each prior source. A 24~$\mu$m source is selected as a prior only if it has the highest predicted SPIRE flux within $0.5\times\mathrm{PSF\;{FWHM}}$, while the remaining fainter sources are ignored. 
\citet{Lee2013} make use not only of MIPS 24~$\mu$m but also VLA $1.4\,\mathrm{GHz}$ sources as priors for \MINORREREVISED[COSMOS]{the} SPIRE photometry \MINORREREVISED[]{in the COSMOS field}. They use the XID algorithm and the iterative jackknife approach for the photometry as also done by \citet{Roseboom2010,Roseboom2012}. 
\citet{YanHaojing2014} presented \REVISED{a} careful deblending work toward a few\MINORREREVISED[,]{} specific sources. Their approach uses $H$ band prior sources to form different decomposition schemes, and runs iteratively to determine a best fitting scheme. 

\citet{Safarzadeh2015} developed a novel technique based on \MINORREREVISED[Monte Carlo Markov Chain]{Markov chain Monte Carlo} (MCMC) sampling for prior-extraction photometry. Their approach uses $H$-band sources as position priors, and \MINORREREVISED[adopts]{fits} UV-to-optical SEDs to predict FIR flux densities (accurate within $\pm$~1 dex; \MINORREREVISED[]{these serve as the initial guesses for MCMC fitting}). 
\MINORREREVISED[
	Their approach also considers FIR flux from other catalogs for sources brighter than 3 times the confusion limit. Then, all the prior sources are divided into disjoint confusion-groups based on their FIR fluxes and positions (so as to save computational resources), and the pixels belonging to each confusion-group (usually irregular 2-dimensional shapes) are fit in a MCMC way so as to obtain the Bayesian probability distribution function (PDF) of each prior source flux.%
]{%
    Then, they separate all priors into blended groups (so as to save computational resources) and fit each blended group in a MCMC way to obtain the posterior probability distribution function (PDF).
} 
They tested their technique with a simulated PACS 160~$\mu$m image \MINORREREVISED[]{and demonstrated such an approach can obtain reliable fluxes below the nominal confusion limit.}. 
%
\citet{Hurley2016} developed a similar MCMC-based prior-extraction tool XID$^+$, which can fit all prior sources to obtain the \MINORREREVISED[]{posterior} PDF of the flux of each prior source. Then they measured SPIRE fluxes for COSMOS 24~$\mu$m prior sources, tiling the full image into diamond-shaped regions to save computation\MINORREREVISED[]{al} resource\MINORREREVISED[]{s}. This algorithm is efficient in obtaining Bayesian PDFs for all prior sources, and thus flux uncertainties can be estimated. However, XID$^+$ does not use prior information on the FIR flux, and thus might still suffer from some limitations, for example, in the case that there are high-$z$ sources that are missing from the prior list but which contribute substantial flux in SPIRE images, or for sources that are close together on the sky but which have very different redshifts (the low redshift ones often will not contribute substantially at SPIRE wavelengths).
%
%
\REVISED{%
    As an extension to XID$^+$ photometry, \cite{Pearson2017} developed a method of incorporating galaxy SEDs as ``informed priors'', finding improvements in the detection of faint sources.
}
%
%
\REVISED{%
    More aggressively, there are also methods that fit multi-band images simultaneously by fixing the shape of galaxy SEDs \citep[e.g.,][]{MacKenzie2016} in an approach inspired by Bayesian techniques \citep[][e.g., their Eq.~19]{Budavari2008}. The results from these approaches can be substantially affected by the assumptions about galaxy SED shapes.
}


From this brief review of earlier photometric methods, it appears that one of the most important steps is choosing the most appropriate prior source list for fitting at each PACS, SPIRE and (sub)mm band. The prior source list should not be redundant, i.e., it should have about or less than one source per beam, see also \citealt{Dole2003,Magnelli2013}). At the same time, it should be as complete as possible, to avoid ignoring any sources that make non-negligible contribution to the total, observed flux. 


In this paper we develop a highly optimized method for the extraction of sources in highly confused images, and apply it to \textit{Herschel} PACS and SPIRE and ground-based single-dish (sub)mm images in the GOODS-North field (hereafter GOODS-N). Our method includes choosing an appropriate prior source list for photometry at each band with the assistance of state-of-the-art SED fitting over all \MINORREREVISED[mid-infrared (mid-IR) to radio]{mid-IR to radio} bands. Benefiting from the latest understanding of the evolution of IR galaxy SEDs with redshift, and from the availability of high-quality catalogs of \MINORREREVISED[]{optical/near-IR} photometric and spectroscopic redshifts, we demonstrate that it is possible to reliably predict the fluxes of galaxies in highly blended FIR bands once flux densities in less blended images have been measured.  

These flux density predictions are used exclusively to determine which sources are hopelessly faint and should therefore be excluded (by subtracting their very weak \MINORREREVISED[expected]{SED-predicted} flux densities from the data before further measurements), and which ones are to be kept for the prior-extraction photometry.  These flux density predictions do not constrain measurements for the remaining priors; their fluxes are extracted directly from the observed data after the hopelessly faint sources have been flagged and removed. 
We describe the SED fitting in Section~\ref{Section_SED_Fitting}, and the selection of fitted/excluded prior sources in Section~\ref{Section_Choosing_Prior_Source_List}.  The prior-extraction photometry is presented in Section~\ref{Section_Prior_Extraction_Photometry}, and a few examples are illustrated in detail in Section~\ref{Section_Example_Superdeblending}. 

In addition, we blindly extract sources from the residual image after each photometry step for each band. These additional sources (when detected) have a high likelihood of being high redshift galaxies that are too faint at near-IR and mid-IR (24~$\mu$m) wavelengths, and thus are not included in our initial prior catalog (although they might also be blends of several low-luminosity lower-redshift sources, or spurious noise peaks). After extracting the additional sources, we re-run the prior-extraction photometry including them\MINORREREVISED[, at each band]{}. The details of this procedure are given in Section~\ref{Section_Additional_Sources_In_Residual}. 

Meanwhile, to verify the performance of the photometry and to provide statistically sound estimates of uncertainties, we run Monte-Carlo simulations for each band from 24~$\mu$m all the way to 1.16~mm and 20~cm. We propose a recipe for correcting formal flux density uncertainties for each individual source for each band, based on a number of directly measurable parameters, calibrated by simulations. The details of the relevant procedures are given in Section~\ref{Section_Simulation}. 

We apply these SED fitting, photometry, simulation and correction steps to GOODS-N, which is a survey field with some of the deepest and richest multi-wavelength data currently available. 
\MINORREREVISED[]{The overall flow-chart is summarized in} Section~\ref{Section_Flow_Chart}.
\MINORREREVISED[We]{We present detailed quality checks in  Section}~\ref{Section_Quality_checks}\MINORREREVISED[]{, where we} compare our final GOODS-N catalog with several catalogs from the recent literature in Section~\ref{Section_Compare_Measurements}, finding that the flux density measurements are generally consistent, while often we have achieved better deblending performance for individual cases. This work also leads to somewhat deeper detection limits in the SPIRE bands. We emphasize, however, that the main advantage of our work is that uncertainties are well-behaved (quasi-Gaussian) and the measurement strategy is, in our opinion, nearly ideal, while still manageable and reproducible, and highly optimized. We believe that the approach described in this paper should be portable to other fields, of course while taking into account the relative depths of the data sets available for prior selection and FIR measurements. 
A future work (S. Jin et al., 2017, in preparation) will present the application of this technique to the 2-square degree COSMOS field.

We present the final sample of GOODS-N FIR-to-mm (FIR+mm) detected galaxies \MINORREREVISED[at $z\ge2.5$]{} in Section~\ref{Section_Final_Catalog_Highz}, including photometric measurements, uncertainties, and IR SED photometric redshifts. 
We derive the dust and star formation properties from the SEDs, and constrain the cosmic star formation rate density (CSFRD) up to redshift 6 in Section~\ref{Section_CSFRD} from directly-detected galaxies. Using stellar mass functions (SMFs) from the literature, and assuming the star-forming main-sequence (MS; e.g., \citealp{Noeske2007}; \citealp{Elbaz2007}; \citealp{Daddi2007}; \citealp{Pannella2009}; \citealp{Rodighiero2010}; \citealp{Karim2011}) correlations at each redshift, 
we estimate completeness correction to the SFR densities. Although the FIR+mm high redshift ($z>3$) samples have a high incompleteness in stellar mass, they appear to contribute a substantial fraction to the total SFR, 
\REVISED[sometimes even exceeding the predicted empirical SFR when binning by stellar mass.]{and sometimes the number of most massive galaxies at high-$z$ even exceeds what is predicted by empirical stellar mass functions.}
We further compare rest-frame UV color selections with our sample and discuss how optical to near-infrared studies could miss the dusty galaxies.

Our imaging data sets come from several surveys.  MIPS 24~$\mu$m data come from the GOODS-\textit{Spitzer} program (PI: M. Dickinson). We use coadded PACS images \citep{Magnelli2013} combining data from PEP (PI: D. Lutz) and GOODS-\textit{Herschel} (PI: D. Elbaz), and SPIRE data from GOODS-\textit{Herschel}. 
The SCUBA-2 data are from the S2CLS program 
\citep{Geach_2016_SCUBA2}%
, and the AzTEC+MAMBO coadded data are from \citet{Penner2011}. 
\REREVISED[]{%
We used the original SCUBA-2 maps rather than the matched-filtered versions (which are also provided by \citealt{Geach_2016_SCUBA2}) where the convolution with PSF reduces the effective angular resolution~\footnote{\REREVISED[]{%
As a general comment, there are matched-filtering approaches that do not necessarily imply an increase of the PSF, e.g., by using functions with negative rings like in the method discussed by \cite{Chapin_2011_SCUBA2,Chapin_2013_SCUBA2}.
}}%
~\footnote{\REREVISED[]{%
We have also carried out our full photometry using the matched-filter version of the SCUBA-2 maps, finding a similar detection depth and number of detections as for the unfiltered version.
For the matched-filter map, we constructed the PSF image with double-Gaussian function according to the Equation (1) of \citet{Geach_2016_SCUBA2}.
}}.%
}
\MINORREREVISED[Our]{We used the} deep radio imaging \MINORREREVISED[is]{}from Owen~(2017), \MINORREREVISED[]{and the shallower radio imaging from \citealt{Morrison2010}}. 

We adopt $H_{0}=73$, $\Omega_{M}=0.27$, $\Lambda_{0}=0.73$, and a Chabrier IMF \citep{Chabrier2003} unless specified in the text for specific comparisons to other works. 
Where necessary, we add -0.04 dex to logarithmic quantities to convert literature measurements from a Kroupa IMF \citep{Kroupa2002} to a Chabrier IMF, and -0.238 dex to convert from a Salpeter IMF \citep{Salpeter1955}. 

\begin{figure}
\centering
\includegraphics[width=0.470\textwidth, trim=4.2mm 17mm 0 22mm]{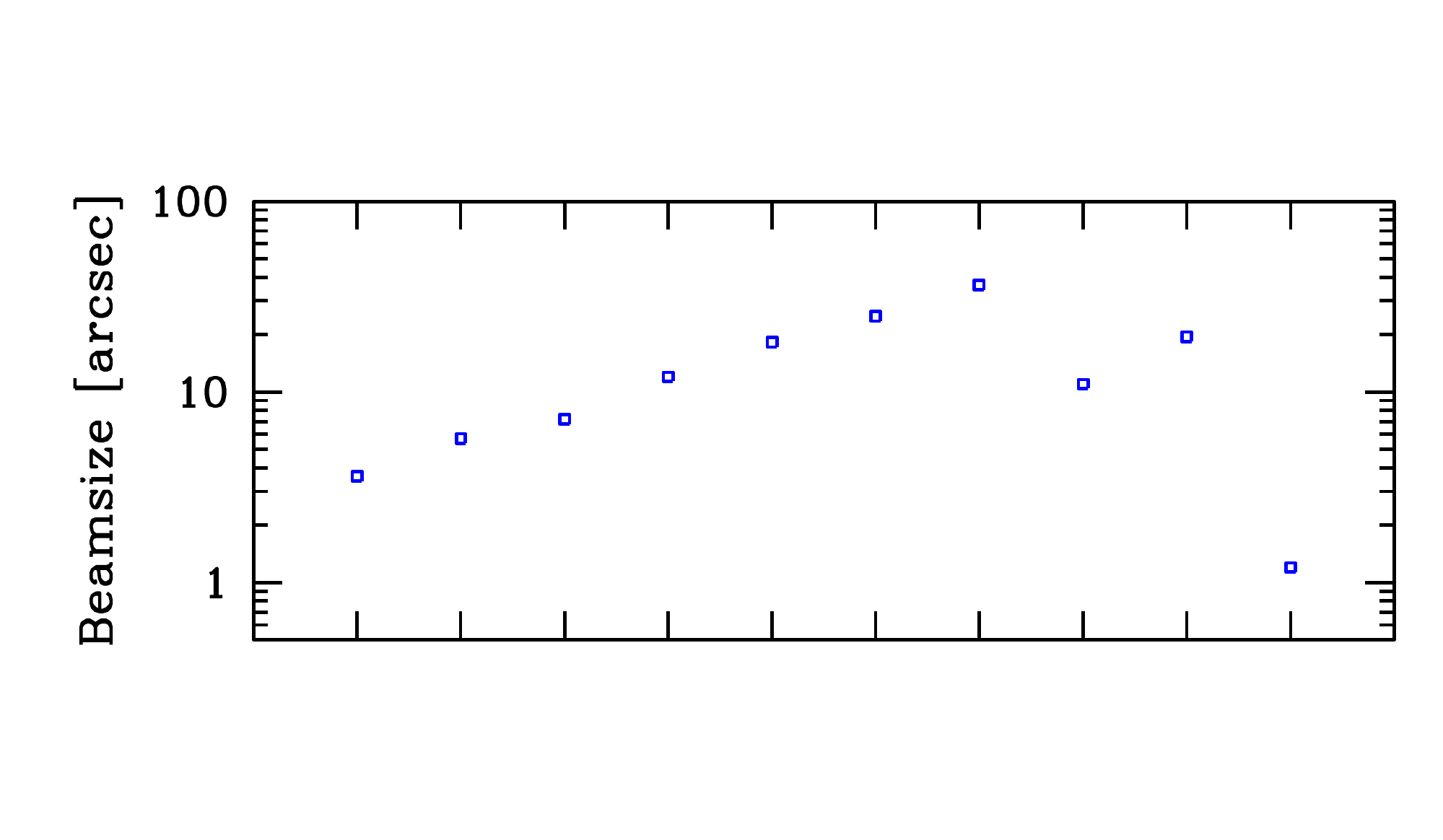}
\includegraphics[width=0.465\textwidth, trim=0 0 0 17mm]{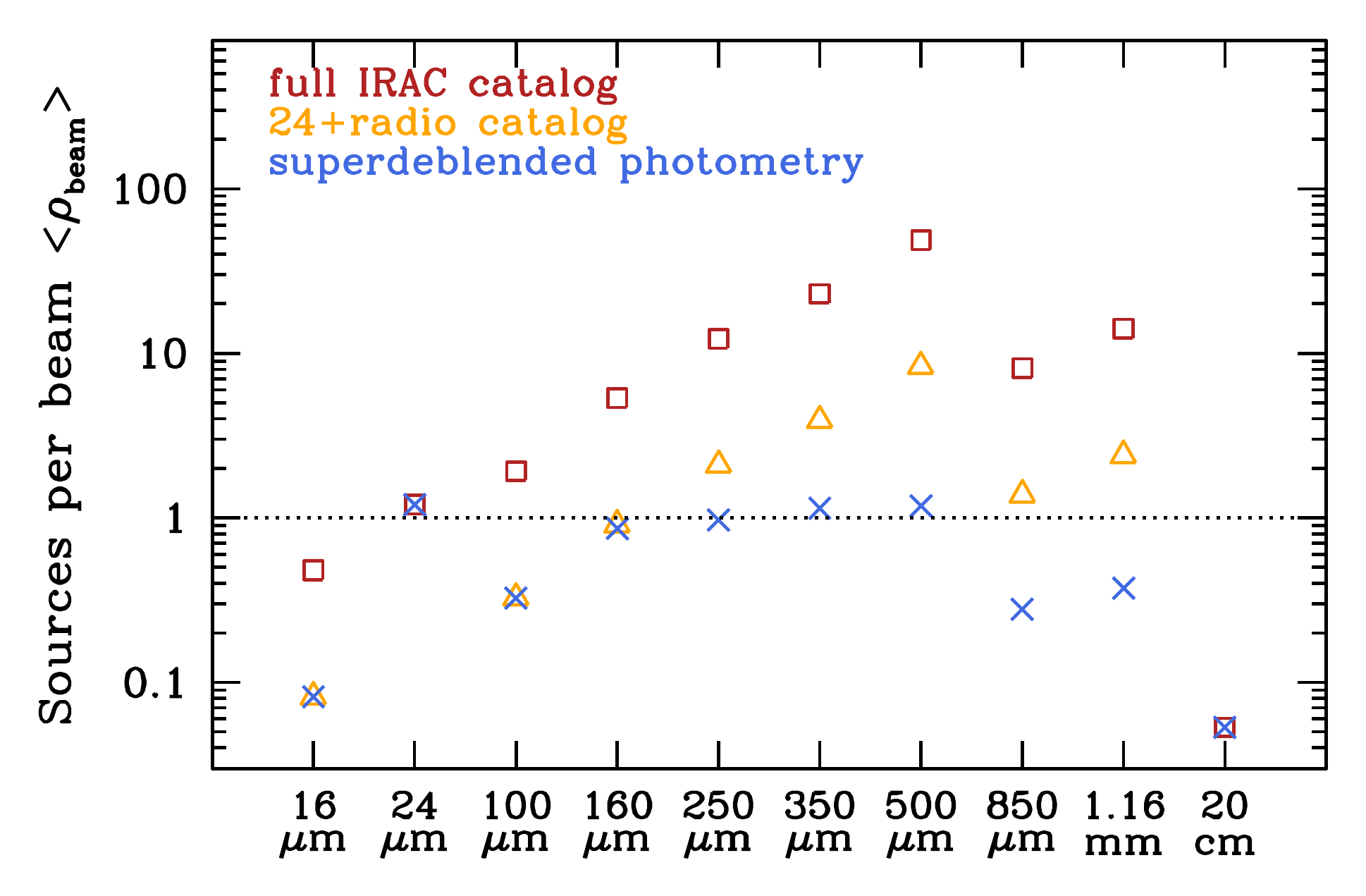}
\caption{
	The upper panel shows the beamsizes (FWHM) of the GOODS-N images in our analysis. 
    The lower panel shows how our prior source selection as described in Section~\ref{Section_Superdeblending} can reduce the number of sources per beam $\left<\rho_{\mathrm{beam}}\right>$ to about 1 at FIR and mm bands, yielding a balance between source confusion and detection depth. The source densities of the full IRAC catalog are shown as the red squares, the 24~$\mu$m- and radio-detected sources are orange triangles, while the ``\superdb{}'' prior sources that we selected and actually fitted are shown as blue crosses.
    We compute the source density by $\left<\rho_{\mathrm{beam}}\right> \equiv N_{prior} / A_{\mathrm{GOODS-N}} \times \left( {\pi}/({4\ln2}) \times \theta^{\;\,2}_{\mathrm{beam}} \right)$, where $N_{\mathrm{prior}}$ is the number of prior sources in GOODS-N, $A_{\mathrm{GOODS-N}}$ is the GOODS-N area ($10.0\times16.5\,\mathrm{arcmin}^{2}$) and ${\theta_{\mathrm{beam}}}$ is the beamsize (FWHM). 
    The numbers are also listed in Table~\ref{Table_1}. 
    \label{Fig_Galsed_Plot_Number_per_Beam}%
%
}
\end{figure}


\section{Setting the general prior source list}
\label{Section_Initial_IRAC_Catalog}

In this section we describe how we set up a prior source list that has enough data points to include, in principle, all possible contributors to the flux densities in FIR to (sub)mm images that we aim to analyze further. 
\REVISED{As far as we know, there are no distant star forming galaxies reliably detected in current FIR/submm data that do not also have a counterpart in GOODS-depth IRAC images.} 
This is the natural outcome of the close connection between stellar mass and star formation, i.e., the Main Sequence paradigm, coupled with the great sensitivity of \textit{Spitzer} IRAC imaging in GOODS. We therefore start with a list of IRAC-detected galaxies. Most of these galaxies are nevertheless going to be very faint in FIR+mm bands, and we define our general prior list as the subset of them that are detected in either 24~$\mu$m or 20~cm images, as described below.

%
%

%
%
We begin with \MINORREREVISED[an]{the} IRAC source catalog from the GOODS-\textit{Spitzer} project (PI: M. Dickinson), generated using \textit{SExtractor} \citep{Bertin1996} to detect objects in a coadded IRAC 3.6$\,\mu$m+4.5$\,\mu$m image.  The same catalog has also been used as input priors for several previous {\it Spitzer} and {\it Herschel} mid- to far-IR catalogs \citep[e.g.,][]{Magnelli2011,Elbaz2011,Elbaz2013,Magnelli2013}, and contains 19437 IRAC sources.  
\MINORREREVISED[]{%
We cross-matched the IRAC source positions with near-IR/optical photometric catalogs using a 1$''$ separation limit.  Primarily, we used the 3D-HST catalog \citep{Skelton2014}, which detected sources in \textit{Hubble Space Telescope} (HST) WFC3 near-IR data from CANDELS (\citealt{Grogin_2011_CANDELS}, \citealt{Koekemoer_2011_CANDELS}).  If no counterpart was found in the 3D-HST catalog, we used the catalog of \cite{Pannella2015}, which detected sources in CFHT/WIRCAM data from \cite{Lin2012}.  In this way, we associate near-IR/optical fluxes, photometric redshifts, and stellar masses with each IRAC source.}
We find overall very good agreement between the two catalogs.
16862 of the IRAC sources have photometric redshifts and stellar masses. 

We have also cross-matched with spectroscopic redshift catalogs from the literature: 
\cite{Lowenthal1997}; 
\cite{Phillips1997}; 
\cite{Cohen1996,Cohen2000}; \cite{Cohen2001}; 
\citeauthor{Steidel1996b} (\citeyear{Steidel1996b}, \citeyear{Steidel1999}, \citeyear{Steidel2003});  
\cite{Dawson2001}; 
\cite{Barger2002,Barger2008}; 
\cite{Wirth2004}; 
\citeauthor{Chapman2004} (\citeyear{Chapman2004}, \citeyear{Chapman2005});
\cite{Strolger2004,Strolger2005}; 
\cite{Treu2005};
\cite{Reddy2006}; 
\cite{Kakazu_Cowie_2007}; 
\cite{Pope2008}; 
\citeauthor{Daddi2008BzK} (\citeyear{Daddi2008BzK}, \citeyear{Daddi2009GN10}, \citeyear{Daddi2009GN20}, \citeyear{Daddi2010BzK}); 
\cite{Murphy2009}; 
\cite{Hu_Cowie_2010}; 
\cite{Yoshikawa2010}; 
\cite{Cooper2011}; 
\citet[][MODS survey]{Kajisawa2011}; 
\cite{Stark2011}; 
\cite{Riechers_2011_HDF242}; 
\cite{Casey2012}; 
\cite{Kriek2015}; 
and \citet[][MOSDEF Survey]{Wirth2015}. 
Besides, we also use some previously unpublished redshifts from our team (H. Inami et al., in preparation.; D. Stern et al., in preparation). \REVISED{The references are also numbered in our released catalog.}
%
%
Finally, 16.2\% of the IRAC sources have secure spectroscopic redshifts. 

\begin{figure}
\centering
\includegraphics[width=0.48\textwidth]{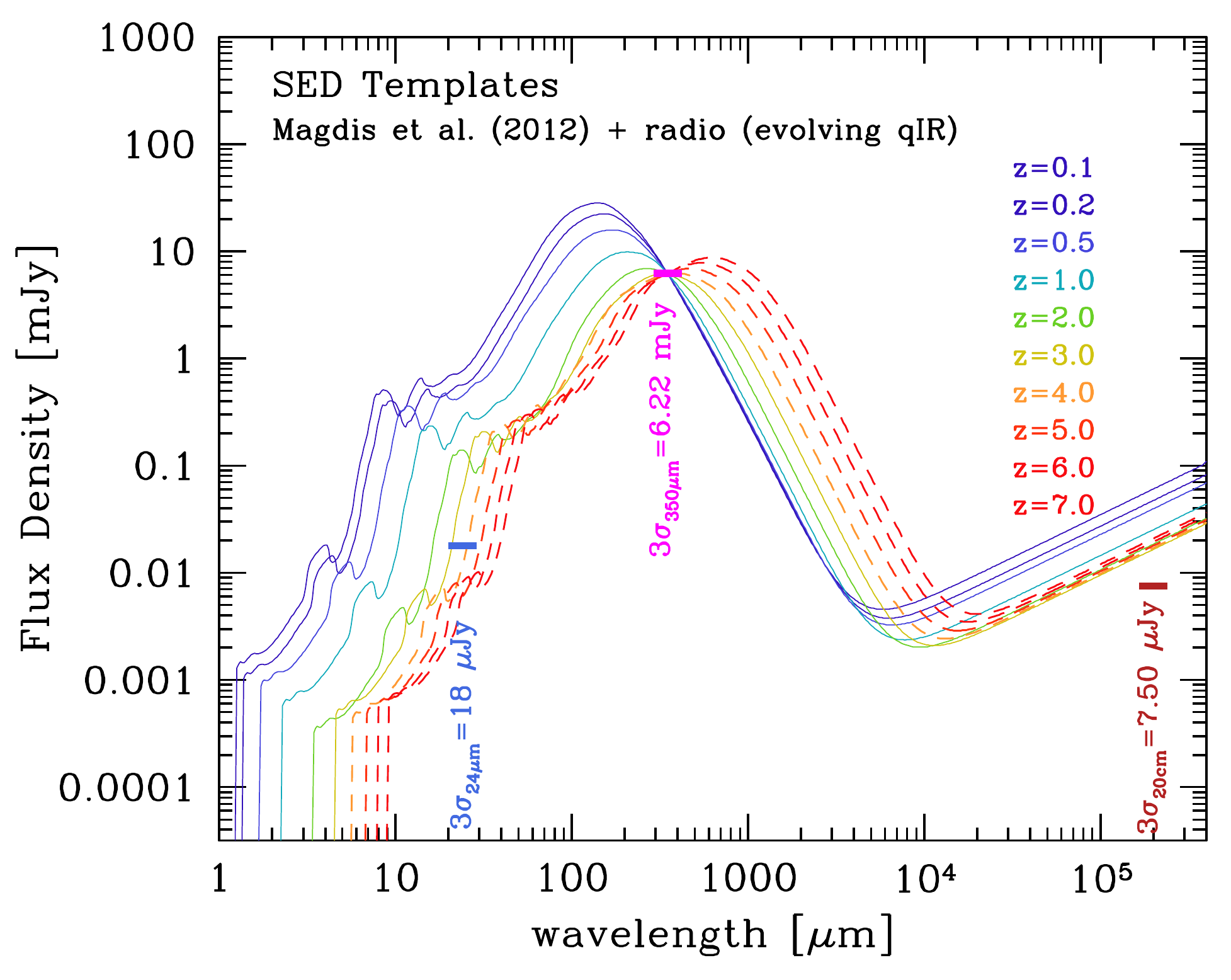}
\caption{%
    In order to illustrate prior completeness and depth issues, we show SED templates from the \citet{Magdis2012SED} library
    (see Section~\ref{Section_SED_Fitting}; color coded by their redshifts), which have been normalized to the $3\,\sigma$ detection limit at $350\,{\mu}$m that we derive in our analysis: $S_{350\,{\mu}\mathrm{m}}=6.22$~mJy (see Table~\ref{Table_1}). 
    The $3\,\sigma$ detection limits at 24~$\mu$m, 350~$\mu$m and 20~cm are indicated by short solid bars and vertical text labels.
    The 24$\,\,u$m and 20~cm data should be sufficiently deep to provide a complete list of fitting priors for all galaxies with measurable 350$\,\mu$m fluxes at $z < 2$ to 3 (solid line SED templates), largely independently of their detailed SED shapes.
    \REVISED{At higher redshifts (dashed templates), sources with measurable flux at 350$\,\mu$m may be undetected at 24$\,\mu$m, but the 20~cm radio limit is sensitive enough to detect all priors needed to fit the Herschel data. 
    Though valid in general, in practice this may not be true in absolutely all cases, as discussed in Section~\ref{Section_Additional_Sources_In_Residual}. 
    }
\label{Fig_SED_Templates}%
%
%
}
\end{figure}



\begin{figure}[ht]
\centering
\includegraphics[width=0.48\textwidth]{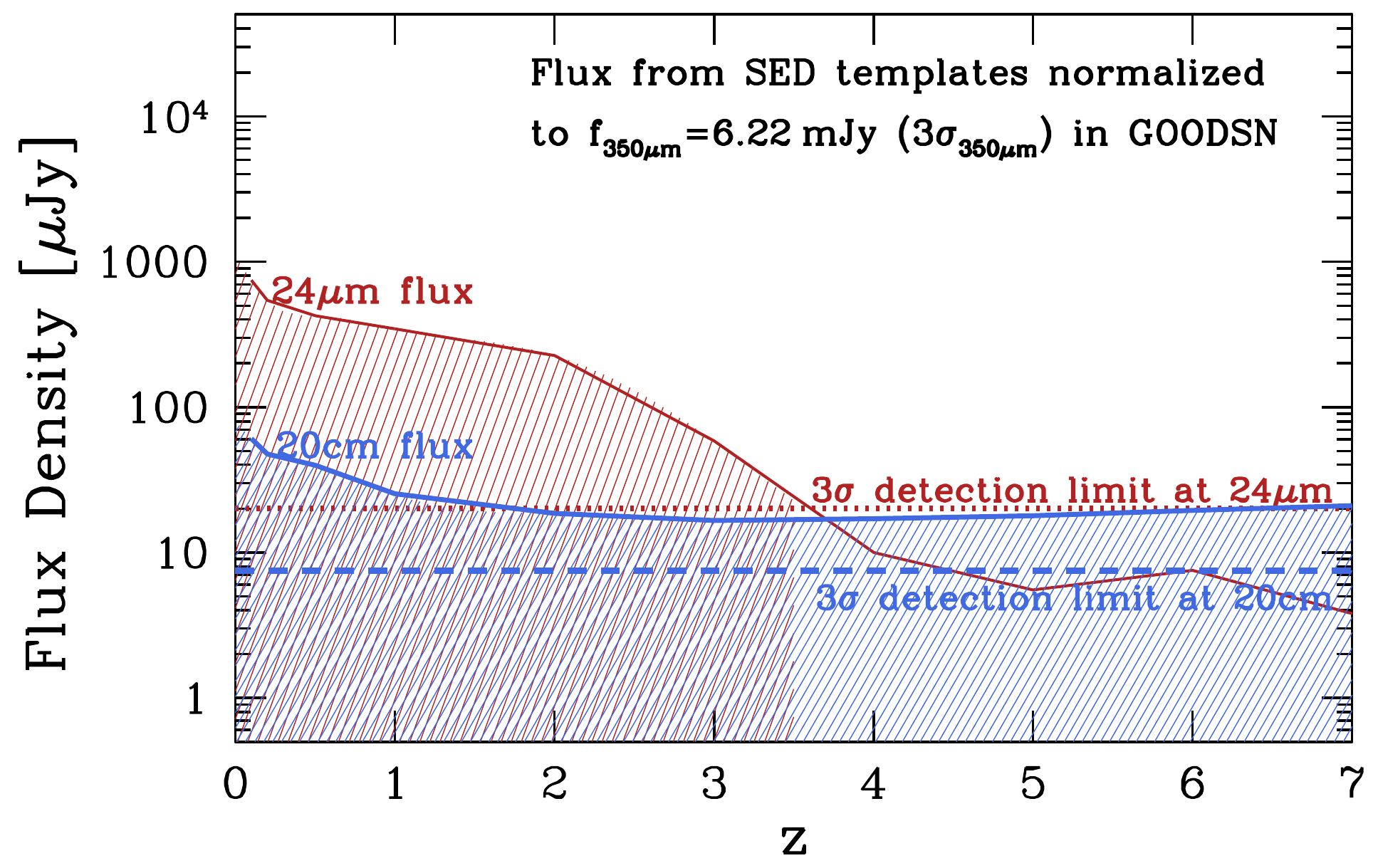}
\caption{%
    Similar to Fig.~\ref{Fig_SED_Templates}, showing the predicted flux densities at 24~$\mu$m (red) and 20~cm (blue) \REVISED{for the \citet{Magdis2012SED} SED templates as a function of redshift, normalized to the 3$\sigma$ detection limit at 350$\,\mu$m.}
    The shaded regions indicate the redshift ranges over which 24~$\mu$m or 20~cm emission can be detected, i.e., the predicted flux densities are above the detection limits.
    The flux density at 24~$\mu$m fluctuates somewhat with redshift as rest-frame mid-infrared features (e.g., the Polycyclic Aromatic Hydrocarbons; PAHs) move through the passband, then falls below the detection threshold at $z>3.5$. 
    The flux density at 20~cm slowly decreases with redshift and then flattens to a nearly constant value, and is in principle always detectable for these SPIRE-normalized SED templates that assume the FIR-radio correlation. 
\label{Fig_SED_expected_flux_against_z}%
}
\end{figure}

In Fig.~\ref{Fig_Galsed_Plot_Number_per_Beam}, we present the mean number of prior sources per PSF beam area $\left<\rho_{\mathrm{beam}}\right>$ for the catalogs  used in this work. The IRAC catalog, represented by \MINORREREVISED[]{red} squares, is clearly too crowded to be used in its entirety for PSF fitting at PACS and SPIRE bands, resulting in
$\left<\rho_{\mathrm{beam}}\right> > 10$ at 250$\,\mu$m and approaching 100 at 500$\,\mu$m, making deblending virtually impossible. We note in passing that the drop in prior density (which is larger than the drop in beam size) when going from SPIRE to the (sub)mm bands is due to the fact that the 850~$\mu$m and 1.16~mm data are effectively shallower than the SPIRE images for sources at the typical redshifts of the priors. 
However, 
$\left<\rho_{\mathrm{beam}}\right>$
is low enough ($\simlt1$) for 16~$\mu$m, 24~$\mu$m and 20~cm, so that we can proceed directly to fitting these bands as described below.

\REVISED{We do not remove galaxies classified as passive  (e.g., using $UVJ$ colors) from the prior pool. In practice, passive sources that might be discarded in this way amount only to a small fraction (of order 10-15\%) of the total. This classification as passive is not always fully reliable, and in some cases galaxies classified as such might actually have positive IR emission (\citealt{Man2016}; \citealt{Gobat2017b}). By discarding passive galaxy candidates, we could erroneously miss some genuine FIR-emitting galaxies in trade for a fairly small benefit.}

\subsection{PSF fitting methodology}

In our analysis, we fit source fluxes in the images using \galfit{} \citep{Peng2002,Peng2010}.  We use PSF fitting, i.e., we treat the sources as unresolved. The IR data with the highest angular resolution\footnote{The PSF FWHM of the VLA data is 2$''$, and whenever possible we use flux measurements from Owen (2018, in prep.) that carefully account for possible extended emission, see next sections.} are the {\it Spitzer} 16$\,\mu$m and 24$\,\mu$m images, with FWHM 4--6$''$, a scale at which the intrinsic sizes of most distant galaxies can be neglected.  This might not be true for some low redshift galaxies in the images, but these are not the main focus of our efforts.

Since it is not practical to use \galfit{} to simultaneously fit a very large number of prior sources in a large image, our code will first divide the image under examination into small regions (boxes). When fitting prior sources in each box we also consider prior sources from a buffer region of surrounding boxes to avoid edge effects.  The buffer size is at least 2--3$\times$ the PSF FHWM of the image being analyzed. 
%
\REVISED{%
We performed tests to verify that, as long as the box size and buffer size are several times larger than the PSF FWHM, the \galfit{} results are unaffected by the exact choices; only the computing speed is affected.}
We run \galfit{} PSF fitting in each box and then combine all boxes to make the full source model images and residual images. All our PSF fitting is performed at fixed RA-DEC positions as determined from the IRAC catalog, after checking and correcting astrometric differences with each data set under consideration. However, in order to improve the fitting for bright sources, we perform a second-pass fit (using first-pass results as first guesses) allowing high $\mathrm{S/N}$ (e.g., $\gtrsim10$) sources to vary their positions. This second-pass exercise also allows us to evaluate the extra photometric noise (and possible flux biases) introduced by residual (uncorrected) astrometric distortions between IRAC and each IR dataset. These terms are generally small but we correct for them in the analysis.


\subsection{Photometry at 24~$\mu$m}
\label{Section_Photometry_24}

%
%

We fit simple PSFs at 24~$\mu$m rather than extended source models because the PSF FWHM at this band is about 5.7$''$, much larger than the typical sizes of $z \ge 0.2$ star-forming galaxies, which are $\sim$8~kpc or 2.5$''$ at $z=0.2$ \citep[e.g.,][]{Conselice_2014_ARAA}. For lower redshift galaxies in the image, we caution that this approach will lead to underestimated flux measurements, for example, for ID~11828, which is the largest spiral galaxy at $z_{spec}=0.106$ in GOODS-N%
, and for 30 more $z<0.2$ FIR-to-mm detected galaxies in our final catalog. 
To properly remove the background%
\footnote{%
\REVISED{Here and in the rest of the paper we define {\it background} to be any pedestal level above which sources are emitting, regardless of its origin.}%
}
in the 24~$\mu$m image, we run a first pass of PSF fitting, then median filter the residual image on a scale of 30$\times$30 pixels, then subtract the smoothed background image from the original image. We run a second pass of PSF fitting with fixed prior source positions, and a third pass with varied prior source positions for the highest $\mathrm{S/N}$ sources from the second-pass fitting. Then we obtain the final 24~$\mu$m fluxes and formal \galfit{} errors.


\subsection{Photometry at VLA 20cm}
\label{Section_Photometry_20cm}

The GOODS-N field has very deep VLA $\sim$20~cm observations from Owen (2018) and shallower observations from \cite{Morrison2010}.\footnote{The VLA data from Owen (2018) were obtained with an average frequency of 1.525~GHz, higher than the 1.4~GHz of \cite{Morrison2010}. We convert to 1.4~GHz assuming a canonical radio slope of -0.8 for the comparison and analysis. However, we also note that a radio slope of -0.8 is not applicable to all sources (e.g., \citealt{Kimball2008,Marvil2015})} 
The \cite{Morrison2010} catalog covers a wider region (diameter of $\sim$15') than the circular area covered by the Owen (2018) catalog (diameter of $\sim$9'). 
In this work, we use the radio data for two main purposes:  to complete the list of prior sources at high redshifts, where galaxies might be too faint to be detected at 24~$\mu$m, and to help constrain the overall IR luminosity of each prior source based on the FIR-radio correlation.

The importance of including radio prior sources is illustrated in Fig.~\ref{Fig_SED_Templates}, where we present a series of redshifted SEDs based on templates that we used in this work {(see Section~\ref{Section_SED_Fitting})}. 
All SEDs in the figure are normalized to a common $S_{350\,{\mu}\mathrm{m}}=6.22$~mJy, which is the 3$\,\sigma_{350\,{\mu}\mathrm{m}}$ detection limit derived from our simulation (see the following sections). Normalizing to other SPIRE bands will lead to a similar conclusion: the expected $S_{20\,\mathrm{cm}}$ flux is always somewhat brighter than the empirical detection limit at 20~cm (3$\,\sigma_{20\,\mathrm{cm}}\approx7.5\,{\mu}\mathrm{Jy}$) whereas we do not expect to detect the faintest SPIRE sources with $z \simgt 3$ at 24~$\mu$m. This is also illustrated in Fig.~\ref{Fig_SED_expected_flux_against_z}, where we plot the predicted flux density at 24~$\mu$m and 20~cm as a function of redshift, for the same normalization at 350~$\mu$m. The radio data are effectively deeper than the MIPS 24~$\mu$m data for selecting sources at $z\gtrsim3$, by roughly a factor of $\simgt4$. The predicted radio flux densities are $\sim2\times$ brighter than our 20~cm detection limit at any redshift, while the predicted 24~$\mu$m flux densities at $z \simgt 4$ are $\sim2\times$ fainter than the detection limit.
Therefore, by using the VLA 20~cm data, we can have a more complete prior source catalog especially for $z\gtrsim3$ sources, improving our completeness for fitting PACS, SPIRE and (sub)mm photometry.

We use the radio catalog from Owen (2018) as the main source of 20cm photometry in this study, with detections down to a 5$\,\sigma$ level of significance. Owen (2018) carefully derived best estimates for the flux, accounting for possible spatial extension of the sources by comparing fluxes in radio images of different resolution from 1$''$ to 3$''$. In order to match our prior catalog with  IRAC-based positions to the radio we adopt the same matching scheme of Owen 2018 (in prep.; their Eq.~1), with a tolerance depending on the radio $\mathrm{S/N}$ and the sizes of the sources, and including a 0.2$''$ term in quadrature to account for typical IRAC astrometric accuracy.

In order to provide a radio measurement for sources weaker than the Owen 2018 5$\,\sigma$ detection limit (and/or upper limit), we also performed prior-extraction photometry and Monte-Carlo simulations independently on the radio imaging data of both \cite{Morrison2010} and Owen (2018). 
Small astrometric misalignments 
\REVISED{(including non-linear distortions)} 
between the images under exam and the IRAC catalog have been corrected in our photometry by adjusting the prior source positions, at radio and in all other bands discussed elsewhere in this paper.
\REVISED{(Note that this is only important for 24~$\mu$m, 16~$\mu$m, 20~cm and PACS image data, where the PSFs are still relatively small.)} 
We first run \galfit{} to fit PSF models at fixed positions, then we run \galfit{} again allowing high $\mathrm{S/N}$ source positions to vary by a small amount (maximum offset less than 1 pixel).  
We fit on the radio images without primary beam correction, then calculate the primary beam correction according to each prior source position and apply it to each photometry measurement \footnote{We used the primary beam correction equations on \url{https://www.cv.nrao.edu/vla/hhg2vla/node41.html}. Most of our correction factors are smaller than 1.5.}.

Note that the PSF of Owen's radio image is about $\sim$2$''$ (provided by F.\ Owen especially for this work), and $\sim$1.7$''$ for Morrison's radio image.%
\footnote{%
\REVISED{The VLA, with its wide bandwidth and multi-frequency synthesis produces a very clean beam, so the effective beam is very close to a Gaussian.}}
Some sources might be resolved in the radio data, typically at low redshift. In order to verify the performance of our photometry we compare the measurements derived with our procedure to those from the Owen (2018) catalog. We find that there is a good agreement in general: the median of the ratio between the different measurements is very close to 1, while the semi-interquartile range is  about 6\% for bright sources ($> 80\,\mu$Jy), and reaching 10\% for fainter fluxes down to the 5$\,\sigma$ limit. This includes effects of flux losses due to over-resolution of the sources, and is accurate enough for a statistical use of the fluxes from our procedure below the 5$\,\sigma$ limit, 
despite its simplicity in assuming unresolved emission and fitting at fixed positions.

\begin{figure}
\centering
\includegraphics[width=0.48\textwidth]{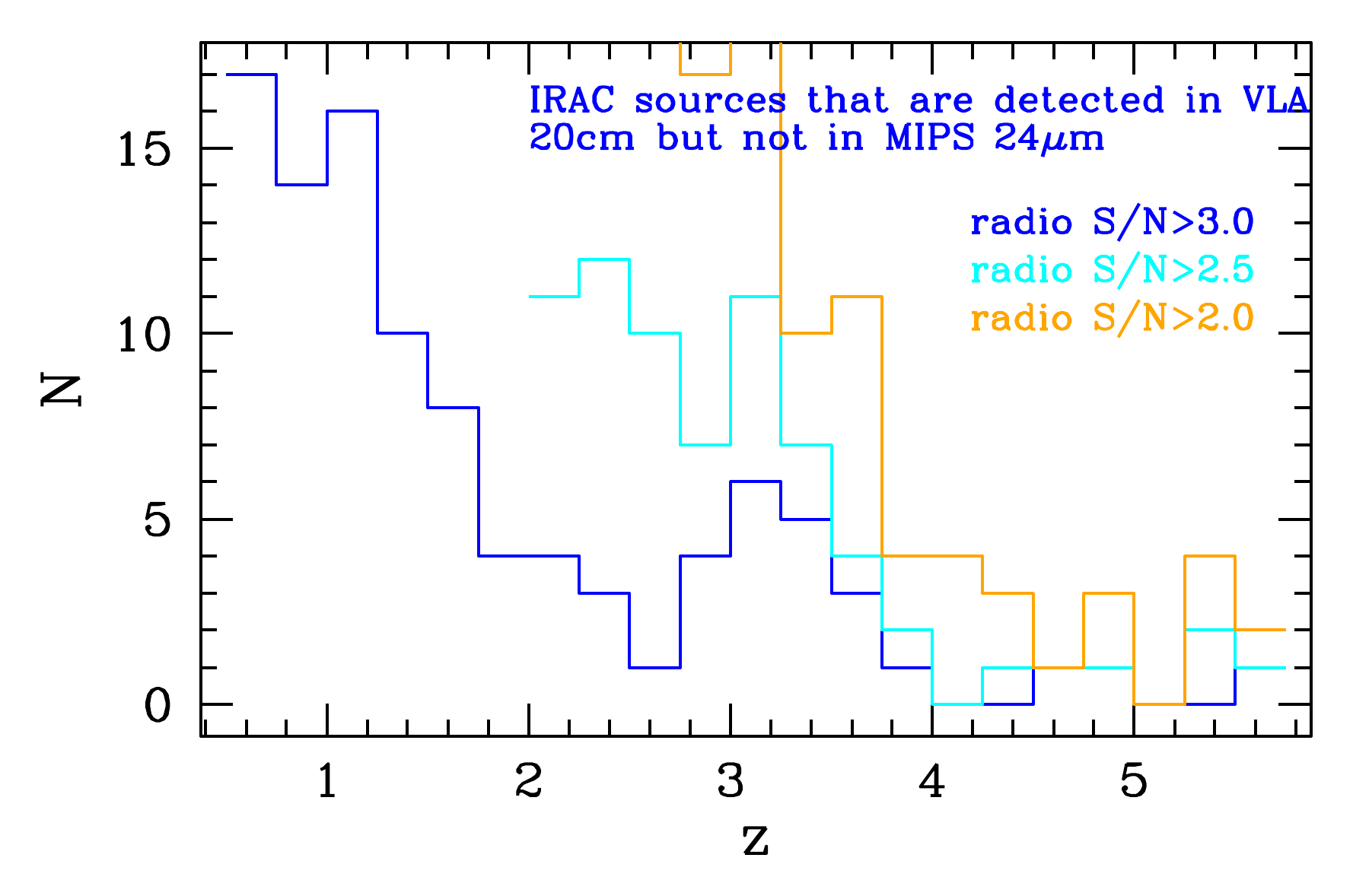}
\caption{%
    Histograms of optical/near-infrared photometric redshifts for the IRAC catalog sources that are detected at 20~cm but not at 24~$\mu$m (with 24~$\mu$m $\mathrm{S/N}<3$). These sources could be either \REVISED[lower-$z$]{} radio-loud AGNs, or high redshift ($z\sim3$) dusty star-forming galaxies (See Section~\ref{Section_Photometry_20cm}), \REVISED {or in some cases spurious detections}. 
    The blue histogram shows sources with 20~cm $\mathrm{S/N}$ $>3$, while the cyan and yellow histograms show the effects of lowering the 20~cm $\mathrm{S/N}$ threshold to 2.5 and 2.0, respectively, for sources at $z>2$. 
\label{Figure_Radio_Detection_Histogram}%
}
\end{figure}

As a result of these measurements, we find 1334 IRAC sources with $\mathrm{S/N}_{20\,\mathrm{cm}}>3$ (554 from the Owen 2018 catalog), of which 112 are not detected with $\mathrm{S/N}$ of 3 at MIPS 24~$\mu$m.  
The 24~$\mu$m-undetected radio sources could either be radio-loud Active Galactic Nuclei (AGNs)\REVISED[ at lower redshifts]{}%
, \MINORREREVISED[moderate]{intermediate} redshift (e.g., $z\sim 1.5$) galaxies with strong rest-frame 9.7$\,{\mu}\mathrm{m}$ Silicate absorption \citep[e.g.,][]{Magdis2011Dropout24}, 
higher redshift star-forming galaxies, 
\REVISED{although in some cases they could be spurious detections boosted by the noise in the radio data.}
In Fig.~\ref{Figure_Radio_Detection_Histogram}, we plot the histogram of photometric redshifts for these sources. Indeed two peaks can be found, and the $z\sim3$ peak likely represents the high-$z$ star-forming galaxies missed by the MIPS 24~$\mu$m selection. 
We also note that reducing the radio $\mathrm{S/N}$ threshold to 2.5 for the high-$z$ (e.g., $z>2$ in the figure) IRAC sources will almost double their number, while making the gap at $z\sim2$ shallower (indicated by the cyan histogram in Fig.~\ref{Figure_Radio_Detection_Histogram}). Thus in the next section we use 2.5 as the radio $\mathrm{S/N}$ threshold for $z>2$ sources.

\subsection{The 24+radio prior source catalog}
\label{Section_The_24_Radio_Catalog}

By combining detections at 24~$\mu$m and 20~cm we produce our 24+radio prior source catalog, which will be used for the analysis in the next steps:  SED fitting (Section~\ref{Section_SED_Fitting}) and photometry over the FIR-mm bands \MINORREREVISED[]{(Sections}~\ref{Section_Choosing_Prior_Source_List}~\MINORREREVISED[]{to}~\ref{Section_Prior_Extraction_Photometry}\MINORREREVISED[]{)}. 
There are 2983 sources in the IRAC catalog with $\mathrm{S/N}_{24\,{\mu}\mathrm{m}} \ge 3$, plus 112 sources with $\mathrm{S/N}_{24\,{\mu}\mathrm{m}} < 3$ but with $\mathrm{S/N}_{20\,\mathrm{cm}} \ge 3$.
%
%

In order to improve completeness for potential high-$z$ prior sources which might be detectable in the FIR and (sub)mm data due \MINORREREVISED[]{to} their redshifted SED peaks, we also include lower $\mathrm{S/N}$ sources (down to S/N$=2.5$) with photometric redshift above 2.0 in the 24+radio prior source list. 
In this way, we selected a total of 3306 sources in the 24+radio prior source catalog, with these criteria: $\mathrm{S/N}_{24\,{\mu}\mathrm{m}} \ge 3$, or $\mathrm{S/N}_{20\,\mathrm{cm}} \ge 3$, or $\mathrm{S/N}_{24\,{\mu}\mathrm{m}} \ge 2.5$ when $z_{\mathrm{phot}}>2.0$, or $\mathrm{S/N}_{20\,\mathrm{cm}} \ge 2.5$ when $z_{\mathrm{phot}}>2.0$. 
%
{%
47.9\% of sources in this full 24+radio prior catalog have spectroscopic redshifts, but only 4.66\% for sources with photometric redshifts $z_{\mathrm{phot}}>2$.
}

Although the flux error measurements at 24~$\mu$m and 20~cm are fairly close to Gaussian (see \MINORREREVISED[next Sections]{Section}~\ref{Section_Simulation_Correction_df_corr} and the Appendixes), starting from a large number of 19437 IRAC sources would inevitably result in a number of noise-dominated spurious detections \REVISED{at these radio/mid-IR bands even assuming simple Gaussian statistics}. 
We expect a total of $\sim50$ spurious detections above $3\,\sigma$ combining the two bands (24~$\mu$m and 20~cm). 
Given that only $\sim4000$ IRAC priors have $z_{\mathrm{phot}}>2$, they would also produce some additional $\sim50$  sources with spurious radio/mid-IR detection at $z_{\mathrm{phot}}>2$ and \MINORREREVISED[above $2.5\,\sigma$]{S/N between 2.5 and 3}. 
These $\approx100$ objects would get included among our 3306 priors spuriously, lacking any actual 24~$\mu$m or radio detections. 
This number is small enough that it should not adversely affect the detectability of IR galaxies, \MINORREREVISED[as the resulting increase of the surface density of priors is small at all bands]{as it only marginally increases the number density of priors fitted at all bands}. In most cases we expect these spurious priors to remain undetected in the FIR/mm bands following the super-deblending analysis.

For the remaining 16131 IRAC sources that were not included in our prior list, we assume that their flux contributions to the PACS, SPIRE and mm image data are negligible and we do not consider them further in the rest of this work. Even though their fluxes will not be exactly zero, their presence will act as a background whose average level will be consistently taken into account by our procedure. Their possibly inhomogeneous distribution will act as additional confusion noise, which our procedure will also quantify.


\REVISED{Figs.~\ref{Fig_SED_Templates}~and~\ref{Fig_SED_expected_flux_against_z} suggest that our prior source list is fairly complete and will include, in principle, any FIR-mm detectable star-forming galaxy, even at the highest redshifts thanks to our very deep 20~cm detection limit. This of course is no longer entirely true when scatter in the FIR-radio correlation and SED shape variations are considered instead of assuming galaxies to be perfectly represented by \cite{Magdis2012SED} templates, as in Figs.~\ref{Fig_SED_Templates}~and~\ref{Fig_SED_expected_flux_against_z}. Also, the presence of noise in the measurements might affect \MINORREREVISED[]{the} detectability of sources close to the detection limits. Therefore,}
despite using the deepest 24~$\mu$m and radio catalogs, we do expect some incompleteness in our prior list for $z>3$ galaxies at the faintest (but detectable) flux levels, especially over the SPIRE bands and in the (sub)mm. This motivates {\em a posteriori} addition of sources found in residual map (mainly) in these bands, as will be discussed in Section~\ref{Section_Additional_Sources_In_Residual}. 

The combination of 24~$\mu$m and radio priors is also useful to overcome \MINORREREVISED[]{the} limitations at 24~$\mu$m where we fit all IRAC priors with a density of about 1 source per beam. This will imply that the effective depth of our catalog at 24~$\mu$m will depend on local crowding (see further discussions about \crowdedness{} in this paper). However, the beam at 20~cm is small enough that crowding is irrelevant. Using the radio data should reduce the possibility that we are missing useful priors in regions where 24~$\mu$m is most crowded.

A more sophisticated way to complete the prior sample at high redshifts would be to cull more sources from the IRAC catalog and to use the correlation between stellar mass and SFR to single out appropriate additional high redshift priors. We have carried out a preliminarily investigation of this idea in GOODS-N, and find that this does not provide an obvious improvement to what has been achieved in the current work. The stellar mass selection will be explored in more detail in future work in COSMOS (S. Jin et al., 2017, in preparation) where the prior catalogs at 24~$\mu$m and radio wavelengths are shallower than in GOODS-N.



\subsection{Photometry at 16~$\mu$m}
\label{Section_Photometry_16um}

Despite having a fairly small PSF (FWHM~$\approx 4.5''$; hence a small number of IRAC sources per beam), the GOODS-N 16~$\mu$m imaging data observed by the \textit{Spitzer} IRS Peak-Up Imaging (PUI) 
\REVISED{are a factor of $\simeq 2$ shallower (for the same SFR) than the MIPS 24~$\mu$m image data ($\sim$7.5$\mu$Jy vs. $\sim$5$\mu$Jy sensitivities, see Table~\ref{Table_1}).} Therefore in this work we directly fit the 16~$\mu$m image data at all 24+radio prior positions to measure 16~$\mu$m fluxes. 
Only 40\% of our 3306 24+radio prior sources have $\mathrm{S/N}_{16\,{\mu}\mathrm{m}} \ge 3$.

\vspace{1truecm}


\section{\Superdb{} photometry for blended FIR/mm images}
\label{Section_Superdeblending}

In this section we describe the core of our photometric method to obtain \superdb{} photometry in confused FIR/mm images.

We proceed one band at a time, \REVISED{working toward longer wavelengths and (generally) larger beam sizes}. For example, at this stage, we have obtained 16~$\mu$m, 24~$\mu$m and 20~cm \MINORREREVISED[coverage]{photometry} for the 3306 24+radio sources in our full prior list derived from a parent IRAC catalog. We use this information to optimize photometry in the next band, PACS 100~$\mu$m, and continue similarly for other bands.

We use SED fitting of all available photometry at each stage to predict the  flux at the next FIR/mm band, with the aim of optimizing the pool of prior sources to be fitted at each FIR/mm band. Once we have predicted fluxes \REVISED{and uncertainties} for all sources in a given band, we determine a criterion to distinguish faint sources (whose fluxes are then subtracted from the original images), from those that are retained and eventually used as prior positions for flux measurement by PSF fitting. 

\REVISED{%
    We note that in this way, the ``earlier'' bands are simply used to decide which priors to use for \galfit{} analysis of the data.  PSF fitting with \galfit{} is carried out one band at a time, i.e., we are not making photometric measurements simultaneously in multiple bands.
    Such an approach of simultaneous multi-band fitting could have benefits, but also has disadvantages, notably the fact that complex inter-band dependencies would be added by the need to use some kind of SED\MINORREREVISED[]{s} to connect information across bands, making it much harder to reliably derive photometric uncertainties for each band.
}

\REVISED{%
    We emphasize that if a source is discarded for fitting at any given band, e.g., say at 100~$\mu$m because it is predicted to be too faint there, this does not exclude the possibility that the same source might be fitted in further steps at longer wavelengths, e.g., in SPIRE or SCUBA2. Indeed, this is quite likely to happen for prior sources with a high redshift.
}

Finally, we inspect the residual images to perform blind detection and extraction of sources that might still be present after all sources fit with priors have been removed. In later sections, we provide a detailed example of the procedure for galaxies in the area of GN20 \citep{Daddi2009GN20}, and some comparison of our measurements to published catalogs.

We have applied a factor of 1.12$\times$ to the final PACS flux densities and uncertainties (used in this paper and released in our catalog) to account for the flux losses from the high-pass filtering processing of PACS images (e.g., \citealp{Popesso2012}; \citealp{Magnelli2013}). 


\subsection{SED Fitting Algorithm}
\label{Section_SED_Fitting}

We consider four distinct SED components in fitting procedure.  From shorter to longer wavelengths, these are:
1) a stellar component from \citet[][hereafter BC03]{BC03} with a Small Magellanic Cloud (SMC) attenuation law; 
2) a mid-infrared AGN torus component from \citet{Mullaney2011}; 
3) dust continuum emission as modeled by the \citet{Magdis2012SED} library; 
and 4) a power-law radio component (see text below). 

The \citet{Magdis2012SED} library is based on  \citet[][hereafter DL07]{Draine2007SED} templates fitted to the average SEDs of main sequence (MS) and starburst (SB) galaxies as a function of redshift. DL07 templates are parametrized by {a number of} physical properties: the minimum interstellar radiation field (ISRF) intensity $U_{\mathrm{min}}$, the maximum ISRF intensity $U_{\mathrm{max}}$, the dust mass $M_{\mathrm{dust}}$, the mass fraction of PAH to total dust mass $q_{\mathrm{PAH}}$, and the mass fraction of dust grains located in Photo-Dissociation Regions (PDR) $\gamma$, etc. Hence they provide quite a wide range of SED shapes. However, \citet{Magdis2012SED} simplified the parametrization of DL07 templates for galaxies at various redshifts with only two parameters: the IR luminosity per dust mass $L_{\mathrm{IR}}/M_{\mathrm{dust}}$ or the mean ISRF intensity $\left<U\right>$~\footnote{$\left<U\right>$, by its definition, is proportional to $L_{\mathrm{IR}}/M_{\mathrm{dust}}$ (see e.g., \citealt{Magdis2012SED}).}, and whether the source is on the MS or not. 
\citet{Magdis2012SED} find that the shapes of the dust SEDs of MS galaxies at a given redshift, as traced by $L_{\mathrm{IR}}/M_{\mathrm{dust}}$, is not expected to vary significantly with increasing $L_{\mathrm{IR}}$ or SFR. 
Meanwhile $L_{\mathrm{IR}}/M_{\mathrm{dust}}$ changes as a function of redshift, and \MINORREREVISED[]{the} variation of SED shape among MS galaxies is expected to depend only on $L_{\mathrm{IR}}/M_{\mathrm{dust}}$, or \MINORREREVISED[]{equivalently} $\left<U\right>$. 
\citet{Magdis2012SED} constructed a series of dust SED templates for MS galaxies parametrized only by $\left<U\right>$, as shown in their Fig.~16. These templates assumed an evolution of $\left<U\right> \propto (1+z)^{1.15}$, no evolution in PDR fraction, and a small evolution of $q_{\mathrm{PAH}}$ beyond $z=2$, as indicated by their data. \citet{Bethermin2015} updated the evolution of $\left<U\right> \propto (1+z)^{1.8}$ {based on the latest data in the COSMOS field}. Therefore, in this work, we use the \citet{Magdis2012SED} dust SED templates (which depend on $\left<U\right>$) with redshift evolution from \citet{Bethermin2015} (which determines $\left<U\right>$ at each redshift) to fit galaxy SEDs and predict photometric redshift and FIR/mm fluxes. Examples of our SED templates are shown in Fig.~\ref{Fig_SED_Templates}. 

Meanwhile, at all redshifts, there is a small fraction of SB galaxies which have SFRs higher than those of MS galaxies with similar stellar masses. Their $L_{\mathrm{IR}}/M_{\mathrm{dust}}$ ratio\MINORREREVISED[]{s} or $\left<U\right>$ \MINORREREVISED[is]{are} found to be higher than \MINORREREVISED[that]{those} for MS galaxies (at least at $z<1.5$) and likely \MINORREREVISED[does]{do} not vary with redshift (\citealp{Magdis2012SED}; \citealp{Tan2014}; \citealp{Bethermin2015}). So their SEDs can be also parametrized in the same way, by a constant $\left<U\right>$. 
When fitting MS galaxies at a given redshift we allow for a range of $\left<U\right>$ values spanning $\pm0.2$~dex around the expected value, to allow for the observed scatter of dust temperature among MS galaxies (defined by $z=0$ results, see \citealt{Magdis2012SED}). 
%


A power-law radio SED has been added to describe the radio continuum as ($S_{\nu}\propto\nu^{-0.8}$)
\footnote{We choose a slope of -0.8 in this work, e.g., \citealt{Kellermann1988,Kimball2008}. But we also note that a single slope may not apply to all galaxies, e.g., \cite{Marvil2015}. Since we use radio data mainly for optimizing the fitted prior sources at FIR/mm band, and then focus our attention on the FIR/mm properties of the sources, we do not implement a galaxy-dependent radio slope for the SED fitting.}
and tied to $L_{\mathrm{IR}}$ by the FIR-radio logarithmic ratio 
$q_{\mathrm{IR}} \equiv \log_{10} (S_{\,\mathrm{IR},\,8-1000\,{\mu}\mathrm{m}}/3.75\times10^{12}\,\mathrm{W\,m^{-2}})
- \log_{10} (S_{\,1.4\,\mathrm{GHz}}/\mathrm{W\,m^{-2}\,Hz^{-1}})$
(e.g., \citealp{Helou1985}; \citealp{Condon1992}; \citealp{Ivison2010}; \citealp{Sargent2010RadioIR}; \citealp{Tan2014}), so that the radio data point can be used in the fit together with FIR/mm to constrain $L_{\mathrm{IR}}$. Here we adopt the latest results that suggest a slowly evolving $q_{\mathrm{IR}} = 2.35 \times (1+z)^{-0.12} + \log(1.91)$, following \cite{Magnelli2015} and \cite{Delhaize2017}.\footnote{$\log(1.91)$ accounts for the conversion between $L_{\mathrm{FIR}}$ (rest-frame 42 to 122~$\mu$m) and $L_{\mathrm{IR}}$ (rest-frame 8 to 1000~$\mu$m) according to \cite{Magnelli2015}.} We find that such \MINORREREVISED[]{a} small evolution is also generally warranted by our global SEDs. We verified that adopting a constant $q_{\mathrm{IR}}$ would have negligible impact on the photometric analysis carried out in this paper.

\subsubsection{SED fitting parameters}

In the case of radio-excess sources or radio-loud AGNs, which are outliers of the FIR-radio correlation, the radio emission does not predominantly arise from star formation. In these cases the radio photometry should not be used in SED fitting. \REVISED{Also, we find that mid-IR rest frame emission is at times hard to fully reproduce with models that fit the FIR emission. }
For this and other analogous situations in the following we introduce three parameters which summarize the SED fitting approach and which are included in our final catalog: 1) a flag to distinguish between MS and SB galaxies; 2) a flag to identify radio-excess; and 3) a flag for high-quality FIR photometry, \REVISED{(in which case radio and 24~$\mu$m fluxes are not used in the SED fitting and flux prediction; they are un-necessary and can possibly be misleading if they are affected by a radio-loud or mid-IR AGN).}

The first parameter is the MS/SB classification (noted as ``\textit{Type\_SED}''), corresponding to the use of MS/SB types of SED templates. This is determined by carrying out the SED fitting twice, as we need a first estimate of SFR to decide if the galaxy is MS or SB. In the first pass, we fit all MS and SB SED templates for each source and derive the best fitting values and uncertainties for $z_{\mathrm{phot}}$, $\left<U\right>$, $M_{\mathrm{dust}}$ and $L_{\mathrm{IR}}$, based on the equations in 
\REVISED[Avni (1976); Avni et al. (1980); Lampton et al. (1976);]{\citet{Press1992}}.
%
SFRs are computed from the integrated $L_{\mathrm{IR}}$ assuming a Chabrier IMF \citep{Chabrier2003} and the relation $\mathrm{SFR}=L_{\mathrm{IR}}/\left(1\times10^{10}\,L_{\odot}\right)\,\mathrm{M}_{\odot}\,\mathrm{yr}^{-1}$ \citep{Daddi2010SFL}. 
We then compare the $\mathrm{SFR}$ and its uncertainty ${\sigma}_{\mathrm{SFR}}$ with the MS-based SFR ($\mathrm{SFR_{MS}}$) according to Eq.~(A1) of \citet{Sargent2014}: 
$\mathrm{SFR_{MS}} = M_{*} \times N(M_{*}) \times \exp \left( \frac{A \cdot z}{1 + B \cdot z^{C}} \right)$, 
where the redshift $z$ comes from the SED fitting and the stellar mass $M_{*}$ is from \REVISED{optical/near-IR catalogs} as described in Section~\ref{Section_Initial_IRAC_Catalog} when available. 
%
\REVISED{The parameter values $N_{M_*}$, $A$, $B$ and $C$ are taken from \citet[][Appendix]{Sargent2014} and are described therein.}
%
%
Sources are set to pure SB (\textit{Type\_SED} $=1$) if 
$\log (\mathrm{SFR} / \mathrm{SFR_{MS}}) > 0.6 \ \mathrm{dex}$
and $\mathrm{SFR}/{\sigma}_{\mathrm{SFR}}>3$, 
and to pure MS (\textit{Type\_SED} $=-1$) if
$\log (\mathrm{SFR} / \mathrm{SFR_{MS}}) < 0.4 \ \mathrm{dex}$
and $\mathrm{SFR}/{\sigma}_{\mathrm{SFR}}>3$.
For all other sources, e.g., those without $M_*$ estimate or those for which the SFR does not meet the criteria above, 
\REVISED{we assume that we cannot conclusively decide whether they are MS or SB galaxies, and we set \textit{Type\_SED} $=0$, using all MS and SB templates to search for the best fit to their SEDs.}


The second parameter is the classification of radio-excess sources or radio-loud AGNs (noted as ``\textit{Type\_20cm}''). This is also  determined using a two-pass SED fitting approach. In the first pass, we do not use the observed radio flux density ($S_{\mathrm{OBS,radio}}$). Since the radio SED component is tied to the $L_{\mathrm{IR}}$ by the FIR-radio correlation,  
we can derive an expected radio flux \MINORREREVISED[]{density ($S_{\mathrm{SED,radio}}$)} and associated uncertainty from the IR SED fit. Then, the (dis)agreement between \MINORREREVISED[the observed radio flux and the flux predicted from the best-fit IR SED]{$S_{\mathrm{OBS,radio}}$ and $S_{\mathrm{SED,radio}}$} determines our choice of the parameter 
\textit{Type\_20cm}: if $(S_{\mathrm{OBS,radio}}-S_{\mathrm{SED,radio}})/\sqrt{{\sigma}_{{\mathrm{OBS,radio}}}^{\ 2}+{\sigma}_{{\mathrm{SED,radio}}}^{\ 2}}>3$ and $S_{\mathrm{OBS,radio}}>2{\times}S_{\mathrm{SED,radio}}$, then we set \textit{Type\_20cm} $=1$, otherwise \textit{Type\_20cm} $=0$. 
Then, in the second pass, if a source has \textit{Type\_20cm}~$=1$, we do not use its radio data, while for the remaining sources we use the radio data to provide tighter constraints on the physical properties. Overall, we have flagged 10.3\% of the 24+radio source as radio-excess/radio-loud candidates. Our approach is very conservative, flagging a source as a possible radio AGN when both the bolometric $L_{\mathrm{IR}}$ and radio are well measured and when the radio flux exceeds the value predicted by the bolometric  $L_{\mathrm{IR}}$ by a factor of two or more, which is comparable to the scatter of the radio-IR correlation. Some of these sources will not actually be AGNs but we refrain from using their radio fluxes for SED fitting when we have a well-determined IR SED.

The third parameter is the combined $\mathrm{S/N}$ computed over the FIR/mm bands (noted as ``\textit{Type\_FIR}''). 
For sources with FIR/mm combined $\mathrm{S/N}$ $\ge5$ \footnote{For example, at 350~$\mu$m we calculate the combined $\mathrm{S/N}$ from 100~$\mu$m to 250~$\mu$m as $\sqrt{\mathrm{S/N}_{100\,{\mu}\mathrm{m}}^{\ 2} + \mathrm{S/N}_{160\,{\mu}\mathrm{m}}^{\ 2} + \mathrm{S/N}_{250\,{\mu}\mathrm{m}}^{\ 2}}$, and similarly at other bands.}, we set \textit{Type\_FIR} $=1$ and do not fit the 24~$\mu$m and radio data points anymore. In this way our SED fitting procedure will not be affected by the scatter of the FIR-radio relation, or by galaxy-galaxy variations of the mid-IR dust SED features (e.g., variations of the IR to rest-frame 8~$\mu$m  ratio, IR8, \citealt{Elbaz2011}). This optimization improves the FIR/mm part SED fitting of individual sources, but does not lead to obvious difference in the statistical results in Section~\ref{Section_Final_Catalog_Highz}.

\vspace{1truecm}

\begin{figure}
\centering
\includegraphics[width=0.45\textwidth]{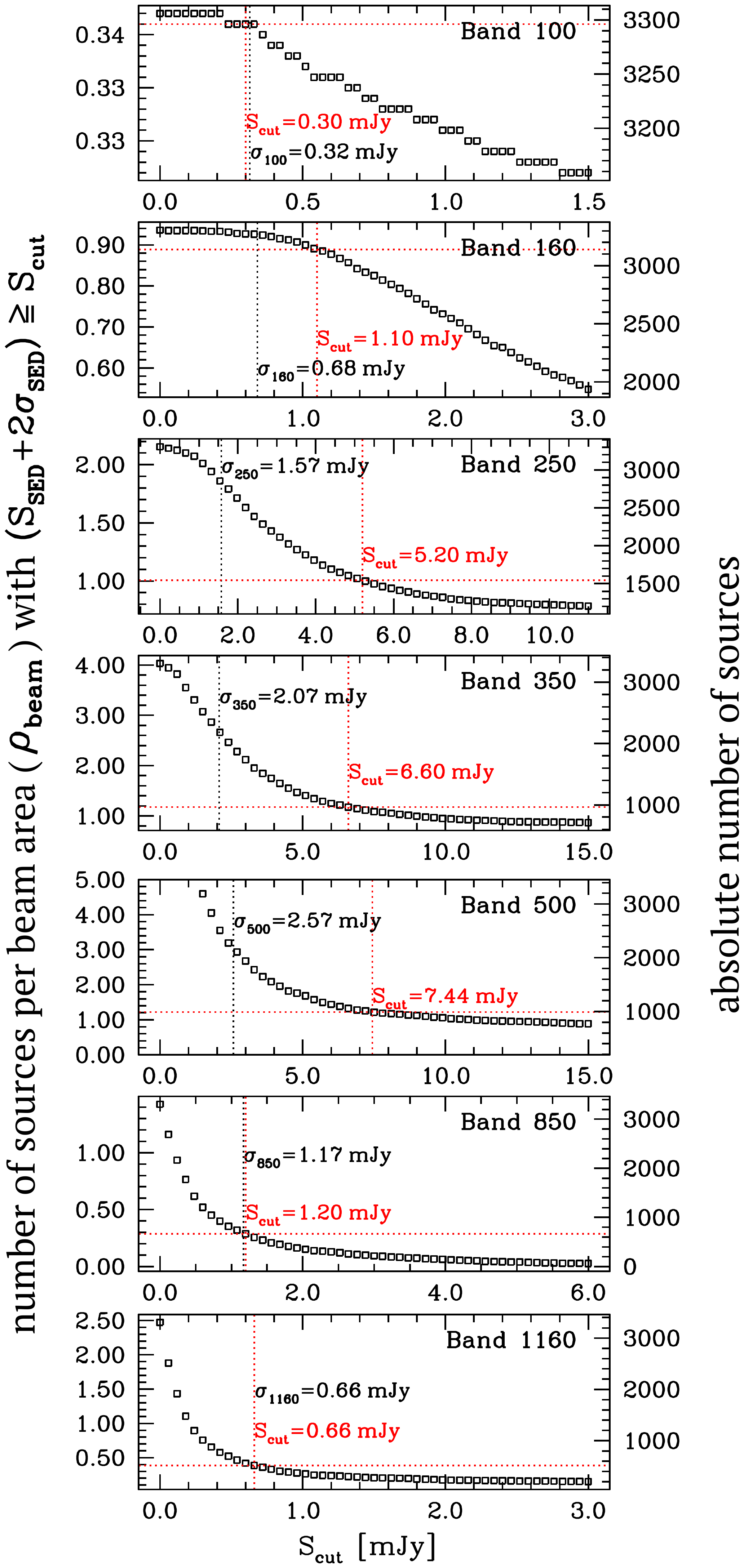}
\caption{%
    The cumulative number density $\rho_{\mathrm{beam}}$ of prior sources versus \REVISED{the threshold (cut) value applied to} their expected flux densities, conservatively increased by twice the flux density uncertainty, as discussed in the text. 
    \REVISED{This is conservative because the large majority of galaxies will have actual flux densities lower than this value.} We select sources for PSF fitting with \galfit{} PSF when their flux densities are larger than the cut value: $S_{\mathrm{SED}}+2\,{\sigma}_{{\mathrm{SED}}} \ge S_{\mathrm{cut}}$, where $S_{\mathrm{SED}}$ is the SED best fitting flux density and ${\sigma}_{{\mathrm{SED}}}$ is the SED flux density uncertainty derived from $\chi^2$ distribution statistics (Section~\ref{Section_SED_Fitting}). The left Y axis $\rho_{\mathrm{beam}}$ has the same definition as in Fig.~\ref{Fig_Galsed_Plot_Number_per_Beam}. The right Y axis indicates the total number of sources for fitting at each band. The posterior 1${\sigma}$ detection limit for each band is labelled with black text in each panel. The flux density cut value $S_{\mathrm{cut}}$ is labelled in red. 
    \label{Fig_Galsed_cumulative_number_function}%
}
\end{figure}

\begin{figure*}
\begin{center}
\includegraphics[width=\textwidth]{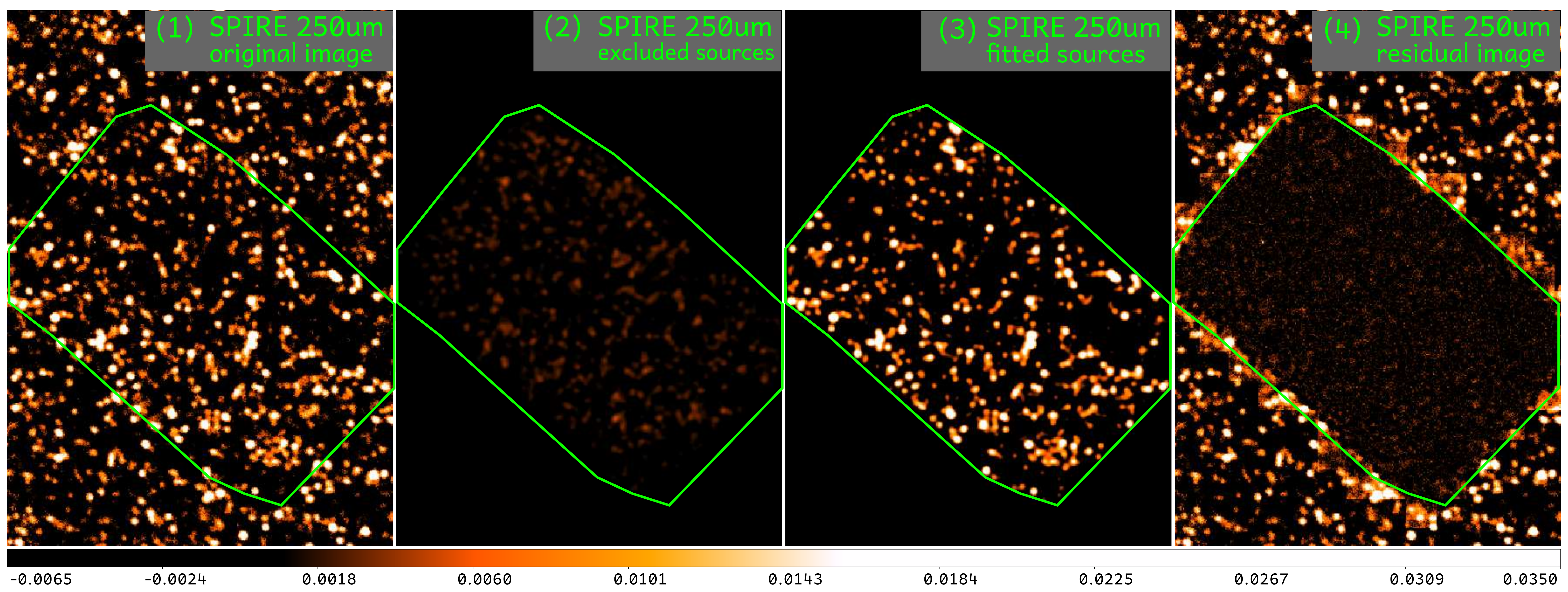}
\caption{%
    SPIRE 250~$\mu$m images, shown as an example of our photometry method. Panel (1) is the original SPIRE 250~$\mu$m image. The green boundary indicates the sky coverage of our 24+radio catalog (Section~\ref{Section_The_24_Radio_Catalog}). Panel (2) is the modeled image of the sources that were excluded from the \superdb{} prior list (Section~\ref{Section_Choosing_Prior_Source_List}) based on their fluxes predicted by SED fitting. These faint sources were not used for PSF fitting, while all other sources are fitted as priors because of their higher likelihood of being detected at 250~$\mu$m. Panel (3) shows the best fitting model image of these prior sources after running \galfit{} photometry (e.g., Section~\ref{Section_Photometry_24}), and panel (4) is the resulting residual image. The image in panel (1) is the sum of panels (2),  (3) and (4). The intensity scales are the same for all panels as indicated by the bottom color bar. The flux density unit is $\mathrm{Jy/beam}$. 
    \label{Fig_Photometry_Images}%
}
\end{center}
\end{figure*}

\begin{figure*}
\begin{center}
\includegraphics[width=\textwidth]{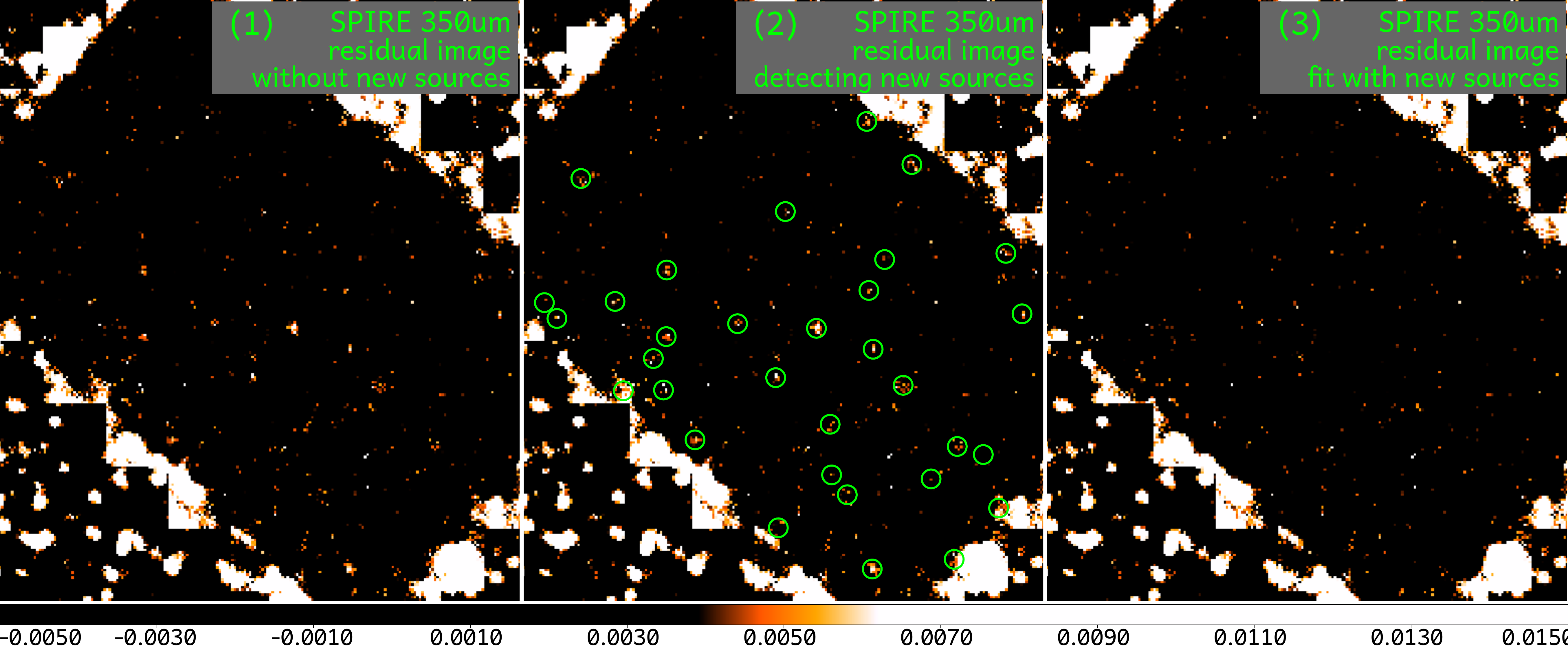}
\caption{%
    Here we show the extraction of additional sources from the residual image. Panels (1) and (2) show the residual image after fitting and subtracting sources from the catalog of priors. The two images are identical, but panel (2) shows the new sources detected by SExtractor. Panel (3) is the new residual image after re-running the photometric fitting including these additional sources. All panels have the same intensity scale and the flux density unit is $\mathrm{Jy/beam}$. 
    \label{Plot_Spdb_350_Images_Demonstration}%
}
\end{center}
\end{figure*}


As has already been briefly mentioned, the SED fitting is (re)run for each FIR/mm band prior to extracting photometry for that band. For example, before we measure PACS 100~$\mu$m fluxes, we run SED fitting to the $K_s$, IRAC, 16~$\mu$m, 24~$\mu$m, and 20~cm (if not radio-excess/radio-loud) data points for each 24+radio prior source. Similarly, before extracting SPIRE 350~$\mu$m photometry, we run SED fitting with the aforementioned bands plus the newly measured 100~$\mu$m, 160~$\mu$m and 250~$\mu$m data points. The purpose of this SED fitting is to provide the best possible prediction for choosing the final prior sources that are used for fitting.  This is a crucial step towards obtaining \superdb{} photometry. 

\MINORREREVISED[]{For sources which have reliable spectroscopic redshifts we fix the redshifts to those values for the SED fitting.  For those with optical/near-IR photometric redshifts, we perform SED fitting allowing the redshift values to span a range}
of $\pm$10\% in $(1+z)$, corresponding to an uncertainty range of about $\pm2\sigma$ 
\citep{Pannella2015,Skelton2014}.  We do not allow redshifts to vary beyond this range even if the minimum $\chi^2$ were to be found at the edge of the interval, to avoid getting solutions that are too inconsistent with the optical constraints. For IRAC sources without any available optical/near-IR photometric redshift ($\sim10$\% of the total IRAC catalog,  but only 2.3\% of the 3306 24+radio priors %
\footnote{These could be real extremely red galaxies, or perhaps blends with displaced centroids that could affect cross-matching to the optical data. 
Further exploration of this is beyond the scope of this paper.}) we allow the full range $0<z<8$ for the SED fitting.
\REREVISED[]{The IR-to-radio SED redshifts are presented as $z_{\mathrm{IR}}$ in our catalog (e.g., Table~\ref{Table_2}).}

For each band, the SED fitting procedure returns a most likely expected flux \MINORREREVISED[]{density} ($S_{\mathrm{SED}}$), as well as its uncertainty (${\sigma}_{{\mathrm{SED}}}$) based on the $\chi^2$ statistics.
This procedure requires that we can associate a well-behaved, (quasi-)Gaussian uncertainty to the fluxes measured in all shorter-wavelength bands. The description of the derivation of reliable flux uncertainties will be presented in 
Section~\ref{Section_Simulation}.



\subsection{Optimized prior source lists for each FIR/mm band}
\label{Section_Choosing_Prior_Source_List}


Fig.~\ref{Fig_Galsed_cumulative_number_function} shows the number of prior sources per beam \MINORREREVISED[plotted as a function of]{with} $S_{\mathrm{SED}}+2\,{\sigma}_{{\mathrm{SED}}}$ \MINORREREVISED[]{above threshold values indicated by the X axis}. We add twice the flux density uncertainty to the best-fit flux density in order to conservatively estimate a plausibly maximum flux density for each source. For each band we show a black vertical line that indicates the median 1$\,\sigma$ flux density uncertainty that we finally obtain from our procedure (see Table~\ref{Table_1}). 

\REVISED{Using these figures, we can optimally determine the threshold value $S_{\mathrm{cut}}$ for our analysis for each band. We recall that sources with $S_{\mathrm{SED}}+2\,{\sigma}_{{\mathrm{SED}}}$ larger than $S_{\mathrm{cut}}$ are included in the list of prior sources for \galfit{} photometric measurements in a given band (henceforth, \textit{selected} sources, for that band). Sources with $S_{\mathrm{SED}}+2\,{\sigma}_{{\mathrm{SED}}}$ fainter than $S_{\mathrm{cut}}$ in a given band are excluded from \galfit{} fitting in that band (henceforth, \textit{excluded} sources), and their flux \MINORREREVISED[]{densities}, as predicted from the SED fitting, will be subtracted from the image, as discussed in the next section. 
A larger value of $S_{\mathrm{cut}}$ will lead to less confusion in \galfit{} fitting (i.e., a smaller number density of fitted sources, $\rho_{\mathrm{beam}}$) but a larger number of fainter sources to be subtracted, and vice versa. Also, setting too large a value for $S_{\mathrm{cut}}$ might prevent us from obtaining measurements for galaxies with detectable FIR/mm emission. We therefore set $S_{\mathrm{cut}}$ by requiring both $\rho_{\mathrm{beam}} \lesssim 1$ and also satisfying $1\,{\sigma} \lesssim S_{\mathrm{cut}} \lesssim 3\,{\sigma}$ at each band, where ${\sigma}$ is the empirical median noise per beam at this band calculated from simulations as described below.
The lower boundary is set to $1\,{\sigma}$ to avoid fitting sources that are too faint for the data in hand.
The  upper boundary of $3\,{\sigma}$ is still faint enough 
(particularly considering that we have added $2\,{\sigma}_{{\mathrm{SED}}}$ to the SED flux densities when evaluating the cutoff threshold) that we do not run much risk of discarding priors for any sources that are individually detectable in the data.}

Regardless of these criteria, we also always keep as a prior any source that was significantly detected in the previous (shorter wavelength) band with $\mathrm{S/N}>3$.
In this way, for sources that have high $\mathrm{S/N}$ in the preceding band but which are predicted to be too faint in band currently under analysis, we can still obtain an upper limit which is useful in SED fitting.

Fig.~\ref{Fig_Galsed_cumulative_number_function}, shows the effect of imposing $S_{\mathrm{cut}}$ thresholds on each band. 
The number of the \textit{selected} sources is reduced to only $\sim$30\% of the total number (3306) at 350~$\mu$m and 500~$\mu$m bands (overcoming blending), and to $\sim$15\% at 1.16~mm (which avoids fitting overly-faint sources). 



\subsection{Subtract excluded sources and fit selected sources}
\label{Section_Prior_Extraction_Photometry}


Sources that are excluded from \galfit{} PSF fitting through the previous section analysis 
do still contribute some weak flux in the observed images, 
and could thus boost the photometry for sources that are retained for fitting. 
Therefore we first make a model image of those excluded sources by fixing their fluxes to the predicted values, then subtract the faint-source model image from the observed original image and obtain a faint-source-subtracted image. 
The \galfit{} photometry is then performed on the faint-source-subtracted image for the selected prior sources in a similar way as described in Section~\ref{Section_Photometry_24}. 
In Fig.~\ref{Fig_Photometry_Images}, we present the SPIRE 250~$\mu$m images in each stage of this procedure. 
{
In the Appendix we show all of the images produced during these steps for all of the bands analyzed in this paper. It is interesting to compare the total flux subtracted from the original images (for faint excluded sources) with that actually extracted from the fitted sources. This reaches up to 30\% of the total flux in the worst cases for the
longer wavelength
SPIRE bands, where only 15-20\% of the original priors are actually fitted, and is generally much smaller for other bands. 
}

{
The list of prior sources that are selected for fitting extends to sufficiently faint (i.e., $<3\sigma$) flux densities for each band that we do not expect all sources used for the PSF fitting to actually be detected with S/N$>3$ in the extraction process.  The number of S/N$>3$ detections ($N_{\mathrm{S/N}>3}$) is generally of order of $1/3$ of all fitted sources ($N_{\mathrm{fit}}$). 
$N_{\mathrm{fit}}$ and $N_{\mathrm{S/N}>3}$ of all bands are listed in Table~\ref{Table_1}. 
}




\subsection{Additional sources in the residual image}
\label{Section_Additional_Sources_In_Residual}


Although the 24+radio combination provides a very deep prior source catalog for fitting the FIR/mm bands, some less massive, radio-quiet high-$z$ galaxies could still be missed. For example, in Fig.~\ref{Fig_SED_Templates}~and~\ref{Fig_SED_expected_flux_against_z}, the $z\sim$3--5 SEDs have 24~$\mu$m and 20~cm fluxes that are close to or even below the detection limits. Due to the diversity of galaxy properties, i.e., PAH fraction or FIR-radio correlation scatter, some of these sources will be missed in our 24+radio prior list, but might have detectable SPIRE or (sub)mm fluxes. 

We carry out a ``blind'' search (i.e., unguided by priors) for additional sources by running SExtractor \citep{Bertin1996} on the residual images.   We combine any new sources detected this way with the previous list of priors, and
then re-run \galfit{} to see if the new, blindly-selected sources can be detected. 
%
%

We limit our blind-extraction region to fall within the 24+radio catalog coverage. This is done by measuring the perimeter of the 24+radio sources and setting pixels outside of the perimeter to NaN (Not-a-Number) values. Then we run SExtractor on the masked residual image with a detection threshold of about 
\REVISED[3.0]{2.5--3.0 (depending on the band, with a GLOBAL background) to detect all possible residual sources.  We subject the resulting SExtractor catalogs to careful visual inspection.  
We run \galfit{} again, fitting over all previously selected prior sources in that band plus the new sources from blind detection that pass the visual inspection. The new residual source candidates are finally kept and included in our  catalog if their final $\mathrm{S/N}$ is above 3, typically only 1/4 of 
the initial SExtractor candidates.
}


We illustrate this blind detection in Fig.~\ref{Plot_Spdb_350_Images_Demonstration}, using SPIRE 350~$\mu$m data. Panel (1) and panel (2) show the same residual image before extracting additional sources, and the green circles in panel (2) highlight the additional sources which are extracted by SExtractor then measured with \galfit{} to have $\mathrm{S/N}$ $>3$. 
%
%
Panel (3) shows the new residual image after we run the final \galfit{} fitting including prior sources and the additional sources that were retained. We list the number of final detected additional sources at each band in Table~\ref{Table_1}. 


\REVISED{%
We have decided not to include these additional sources in the scientific analysis presented in the final sections of this paper, for a number of reasons.  Their positions are uncertain and in most cases it is not trivial to associate them to optical/near-IR counterparts. 
Also, a certain fraction of these blindly-detected sources are probably spurious, often a superposition of a number of fainter sources simulating a brighter object. 
\MINORREREVISED[Instead, it is unlikely that they are due to poor source fits creating residuals around bright sources that almost never happen in the \textit{Herschel}/(sub)mm bands.]{It is unlikely that these detections are residuals from poor fits to brighter sources, as this is rarely seen in the FIR/mm residual images.}
Nevertheless, there are also likely genuine $z>3$ sources among these objects,
for example, the $z=5.3$ submillimeter galaxy HDF850.1 \citep{Walter2012} is the additional source ID~850011 in our additional source catalog. 
}

\vspace{1truecm}

\begin{figure}
\begin{center}
\includegraphics[width=0.40\textwidth]{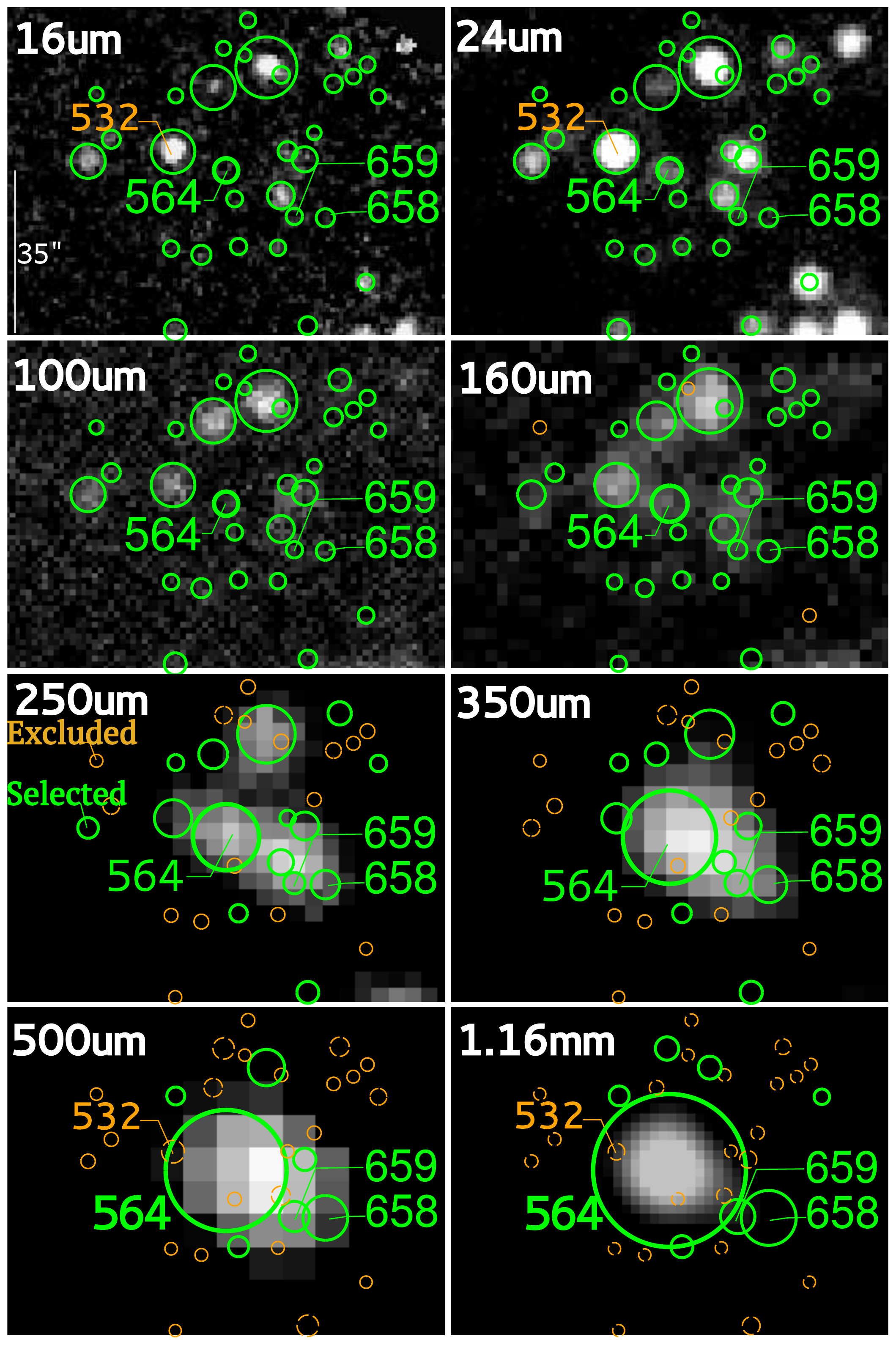}
\caption{%
    Multi-wavelength cutouts around the bright submillimeter galaxy GN20 \REVISED{(\citealt{Pope2006}; \citealt{Daddi2009GN20})}\REVISED{, which is} ID~564 in this work. Each circle represents a 24+radio prior source (Section~\ref{Section_Initial_IRAC_Catalog}). Green circles are selected for fitting (marked ``Selected'' in the 250~$\mu$m panel) while orange circles are excluded (marked ``Excluded''). 
    The circles have sizes proportional to the source flux densities for each band (i.e., the measured fluxes for fitted sources, or the SED-predicted fluxes for excluded sources, but with a minimum size of 0.5$''$ and a maximum size equal to the PSF FWHM). 
    The SEDs of the four marked sources, ID~564 (GN20), 658 (GN20.2a), 659 (GN20.2b) and 532, are shown in Fig.~\ref{Plot_SED_GN20}. The SED-predicted flux density for ID~532 (plus twice the uncertainty) is too faint at 500~$\mu$m (as indicated by the vertical dashed line in the SED figure) and hence is excluded from the fitting list in this band. 
    \label{Figure_Cutouts_GN20}%
}
\end{center}
\end{figure}

\begin{figure*}
\begin{center}
\includegraphics[width=0.45\textwidth, trim=0 0.0cm 0 0.3cm, clip]{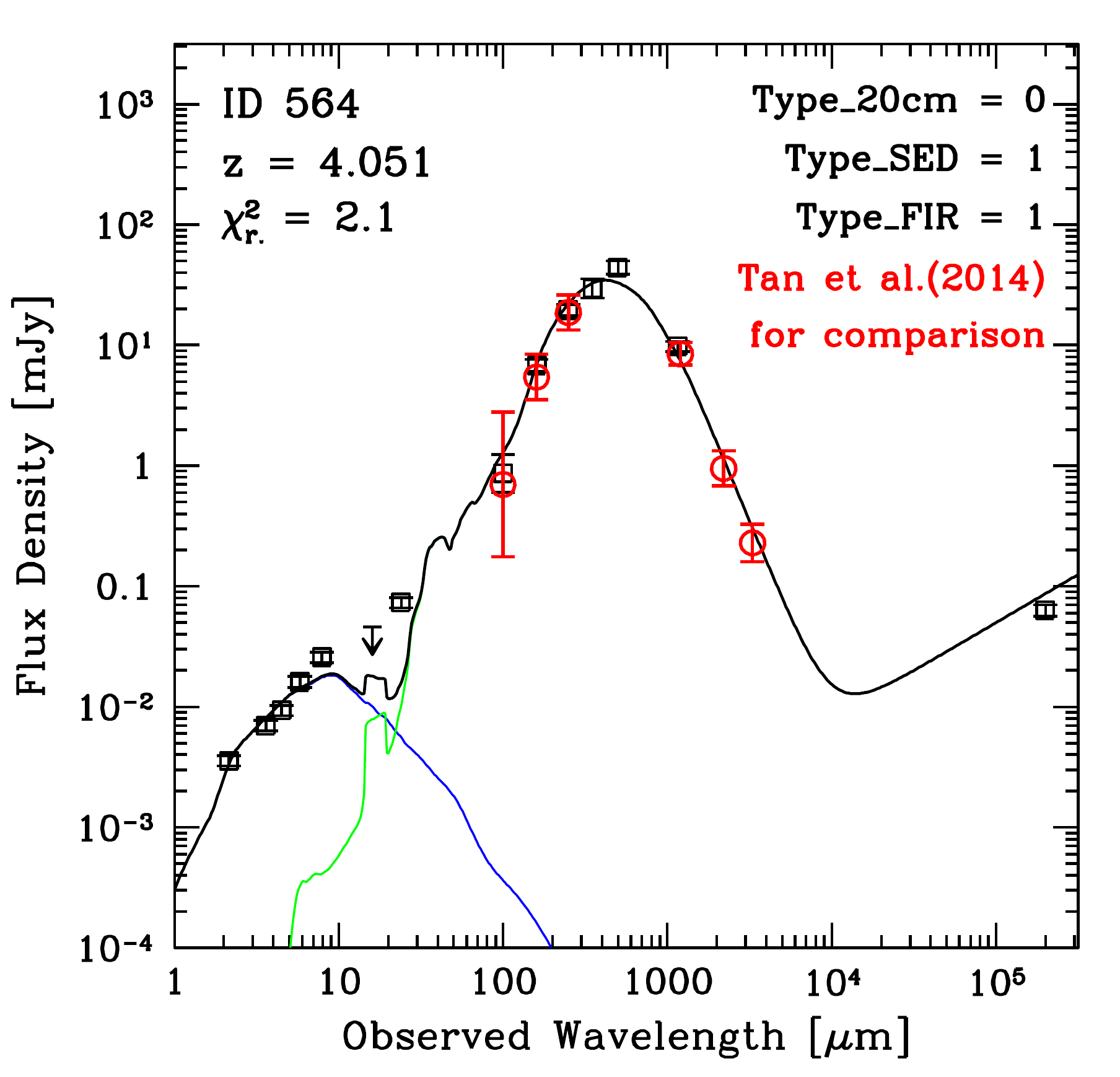}
\includegraphics[width=0.45\textwidth, trim=0 0.0cm 0 0.3cm, clip]{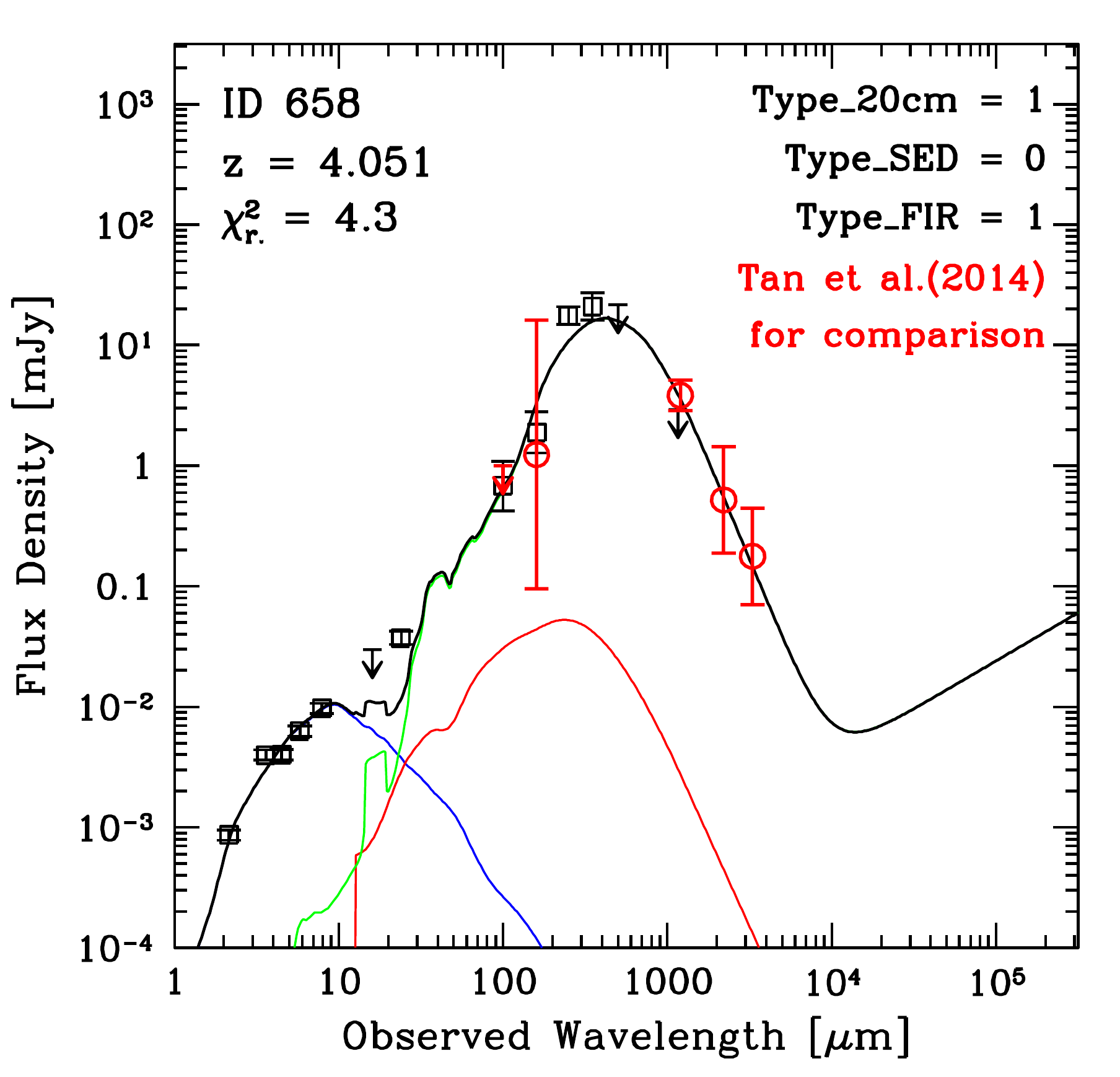}
\includegraphics[width=0.45\textwidth, trim=0 0.0cm 0 0.3cm, clip]{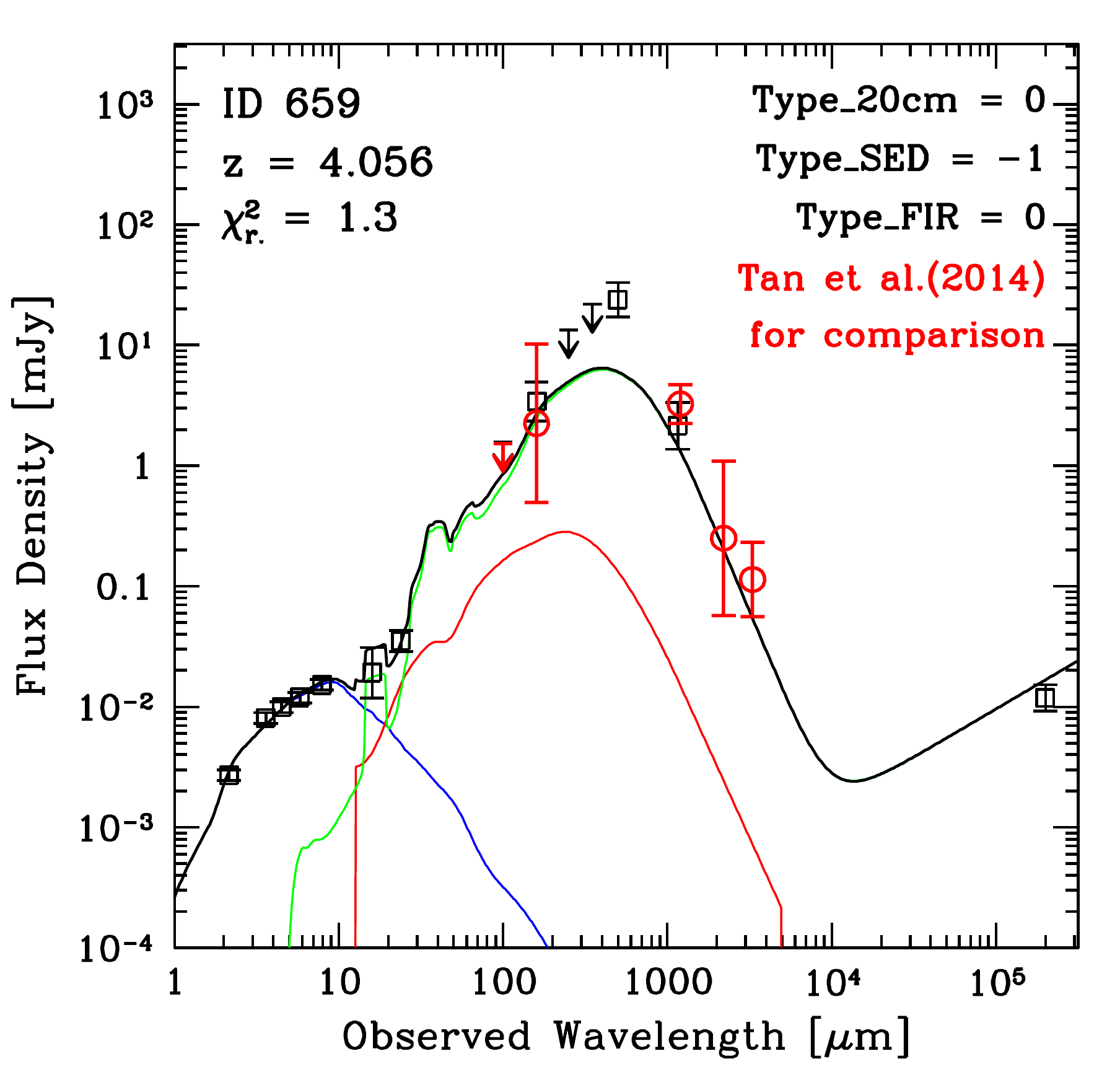}
\includegraphics[width=0.45\textwidth, trim=0 0.0cm 0 0.3cm, clip]{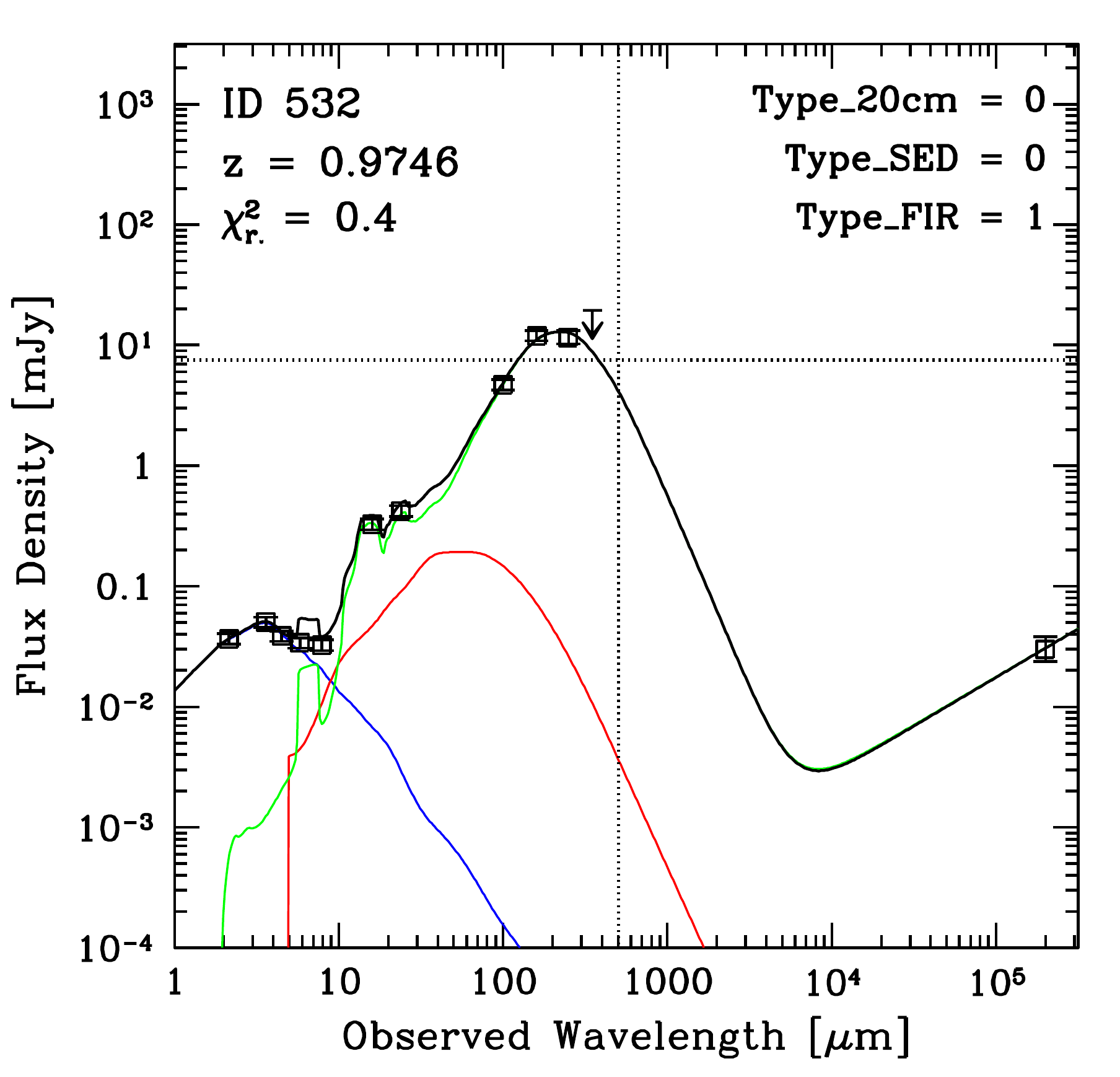}
\caption{%
    The SEDs of the four sources marked in Fig.~\ref{Figure_Cutouts_GN20}.
    In the panels for ID~564 (GN20), 658 (GN20.2a) and 659 (GN20.2b), we show the red data points from \citet{Tan2014} for comparison. These are not used in the SED fitting, but show very good consistency with our measurements.
    For ID~532, the 500~$\mu$m SED-predicted flux density (indicated by the vertical dashed line) becomes fainter cutoff threshold (the horizontal dashed line), even when increased by twice the uncertainty, and hence this source is excluded from fitting at this band. 
    \REVISED{In the upper-left corners we show the redshift and best fitting SED reduced $\chi^2$ ($\chi_r^2$) which is computed over the wavelength range 100--1100$\mu$m.}
    The three SED tuning parameters are shown in the upper right corners. See Section~\ref{Section_SED_Fitting} for details. 
\label{Plot_SED_GN20}%
}
\end{center}
\end{figure*}

\begin{figure*}
	\begin{center}
		\includegraphics[height=4cm, trim=0 0 -1.0cm 0]{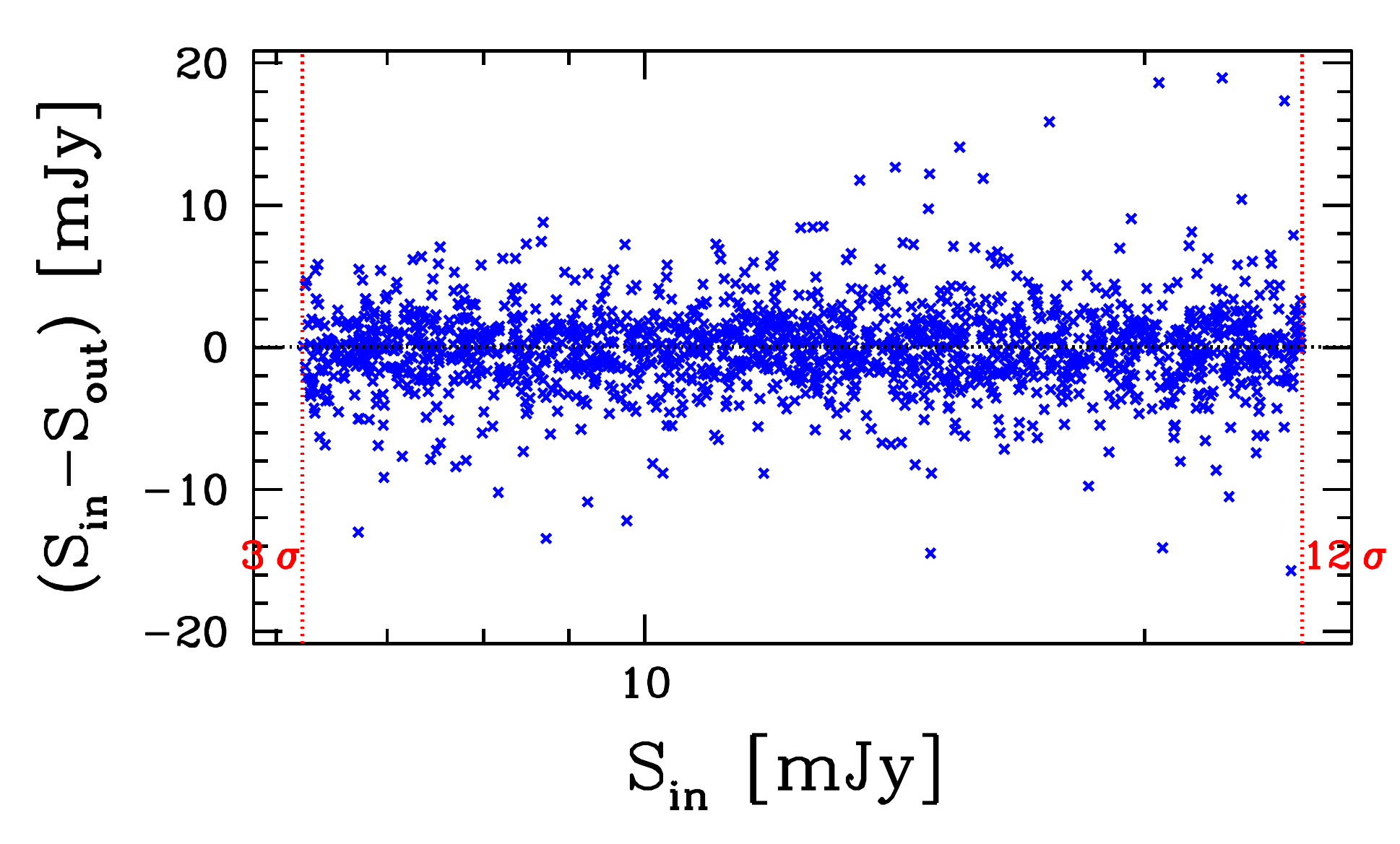}
		\includegraphics[height=4cm, trim=-1.5cm 0 0 0]{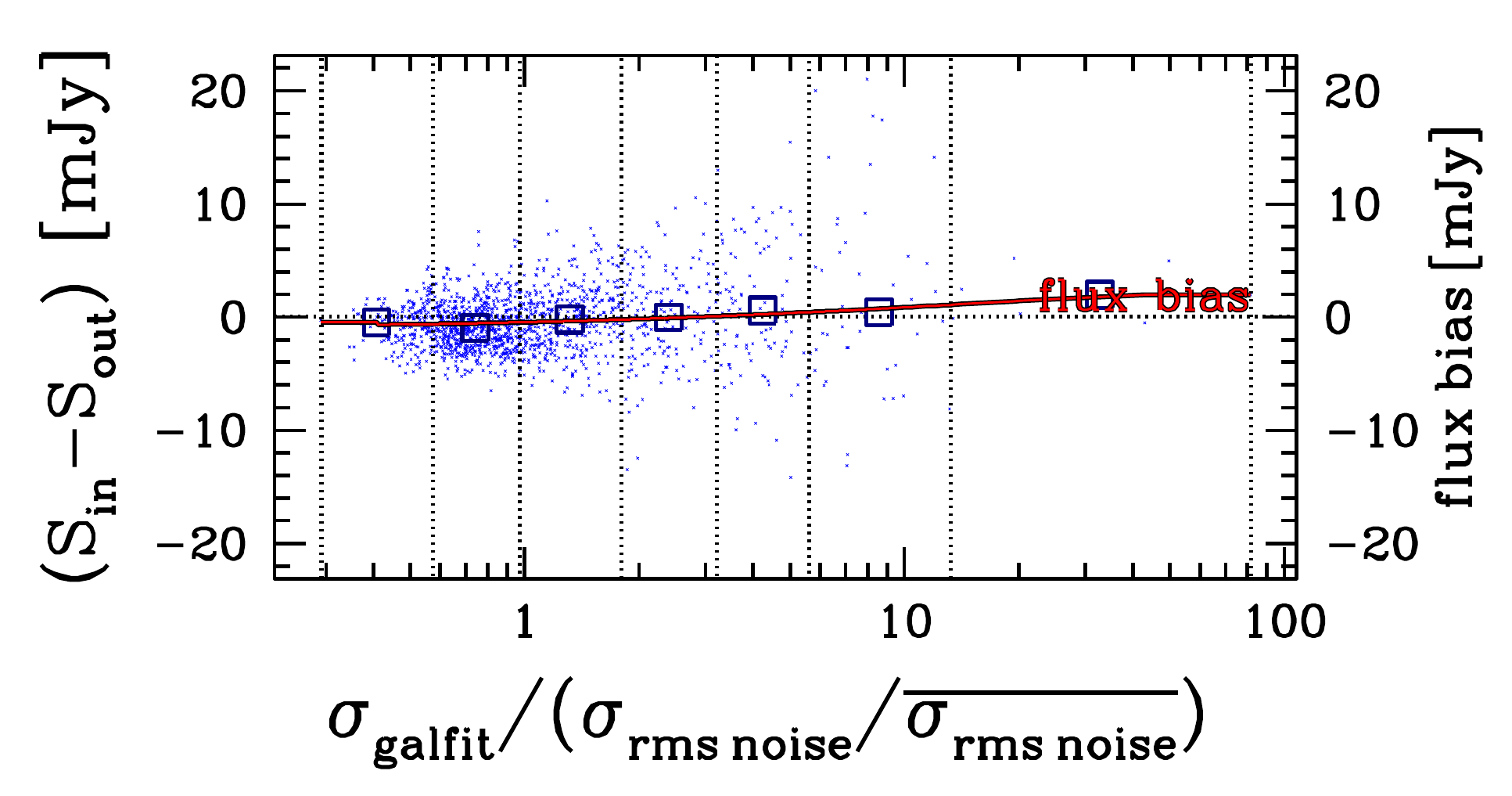}
		\includegraphics[height=4cm]{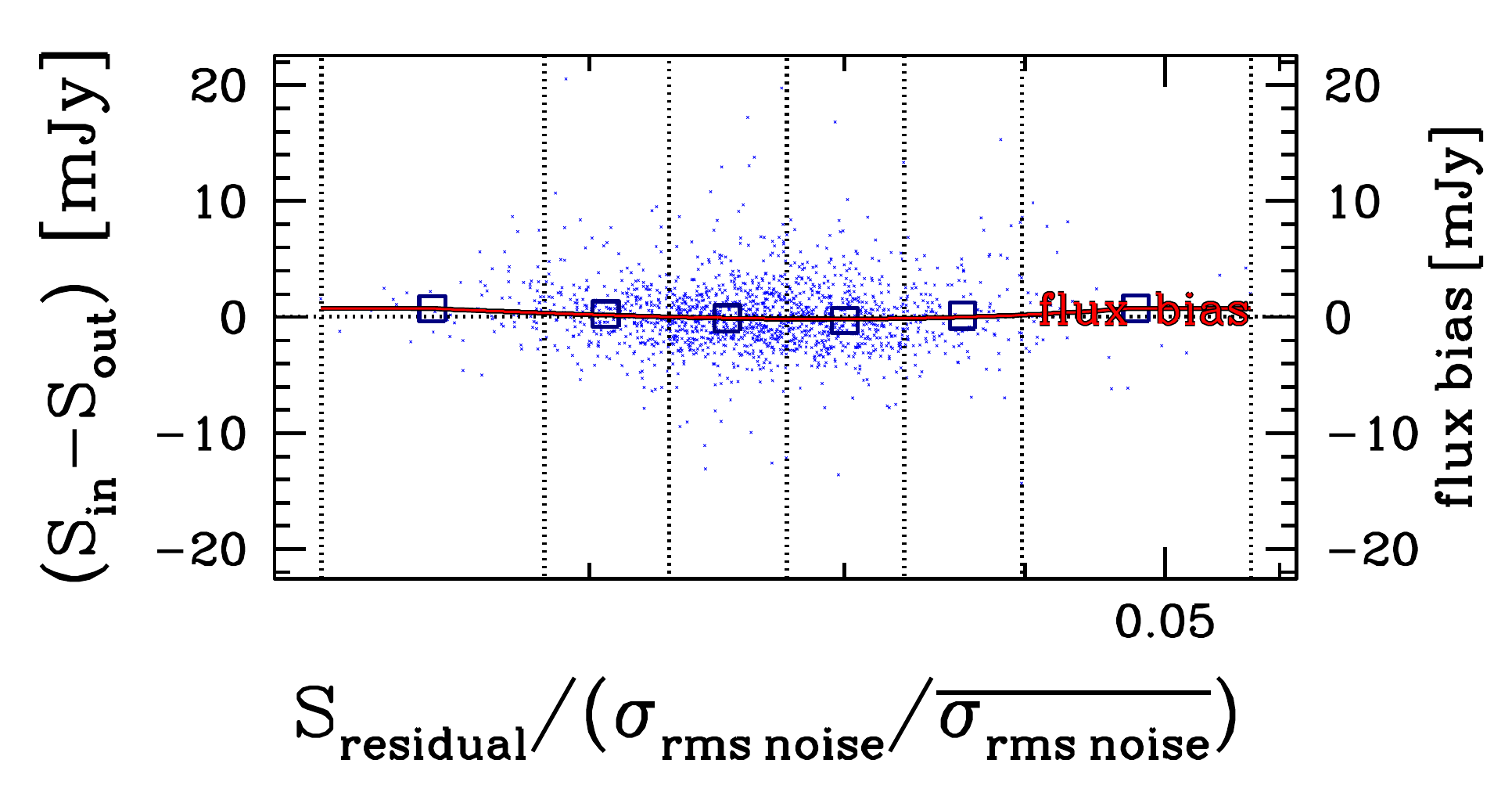}
		\includegraphics[height=4cm]{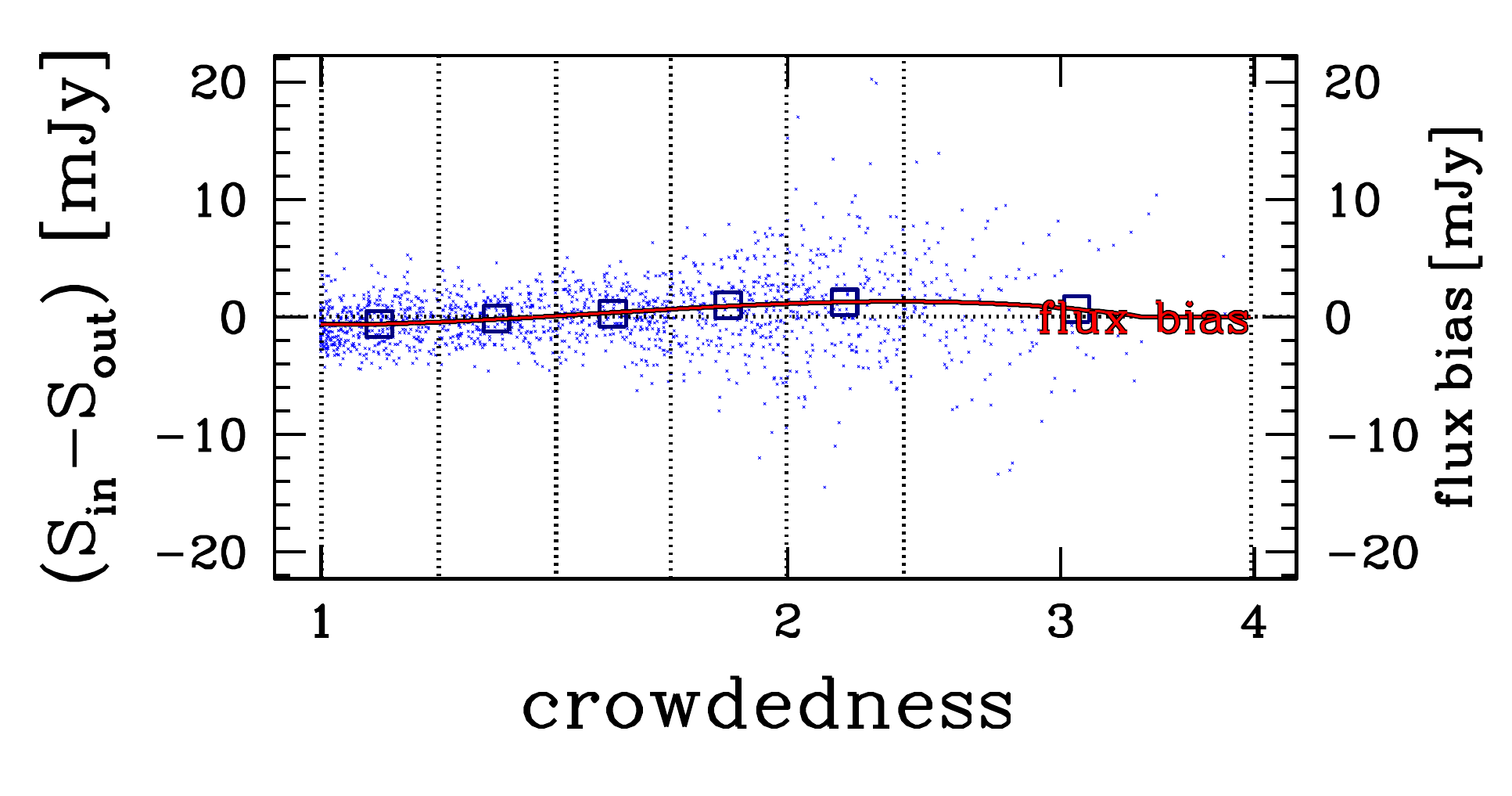}
		\caption{%
			Panel (1) shows the difference between input and output flux densities of simulated sources ($S_{\mathrm{in}}-S_{\mathrm{out}}$) versus the input (i.e., true) flux density \REVISED{for SPIRE 350~$\mu$m Monte-Carlo simulation}. In panel (2) to (4), we bin simulated sources by three measurable parameters: the \galfit{} flux density uncertainty normalized by the local rms noise, the flux density on residual image also normalized by the local rms noise, and the \crowdedness{} parameter which measures the blending situation at each source position (see text in Section~\ref{Section_Simulation_Correction_fbias}). Dashed vertical lines indicate the bins. We measure the mean of $S_{\mathrm{in}}-S_{\mathrm{out}}$ (blue square data points), which is just the flux bias, then fit a polynomial function to link the flux bias to each parameter (red curves). In this way, bias can be better corrected for different sources. Similar figures for all other bands are shown in Appendix. 
			\label{Fig_Galsim_flux_bias_SPIRE}%
		}
	\end{center}
\end{figure*}

\begin{figure*}
	\begin{center}
		\includegraphics[height=3.5cm]{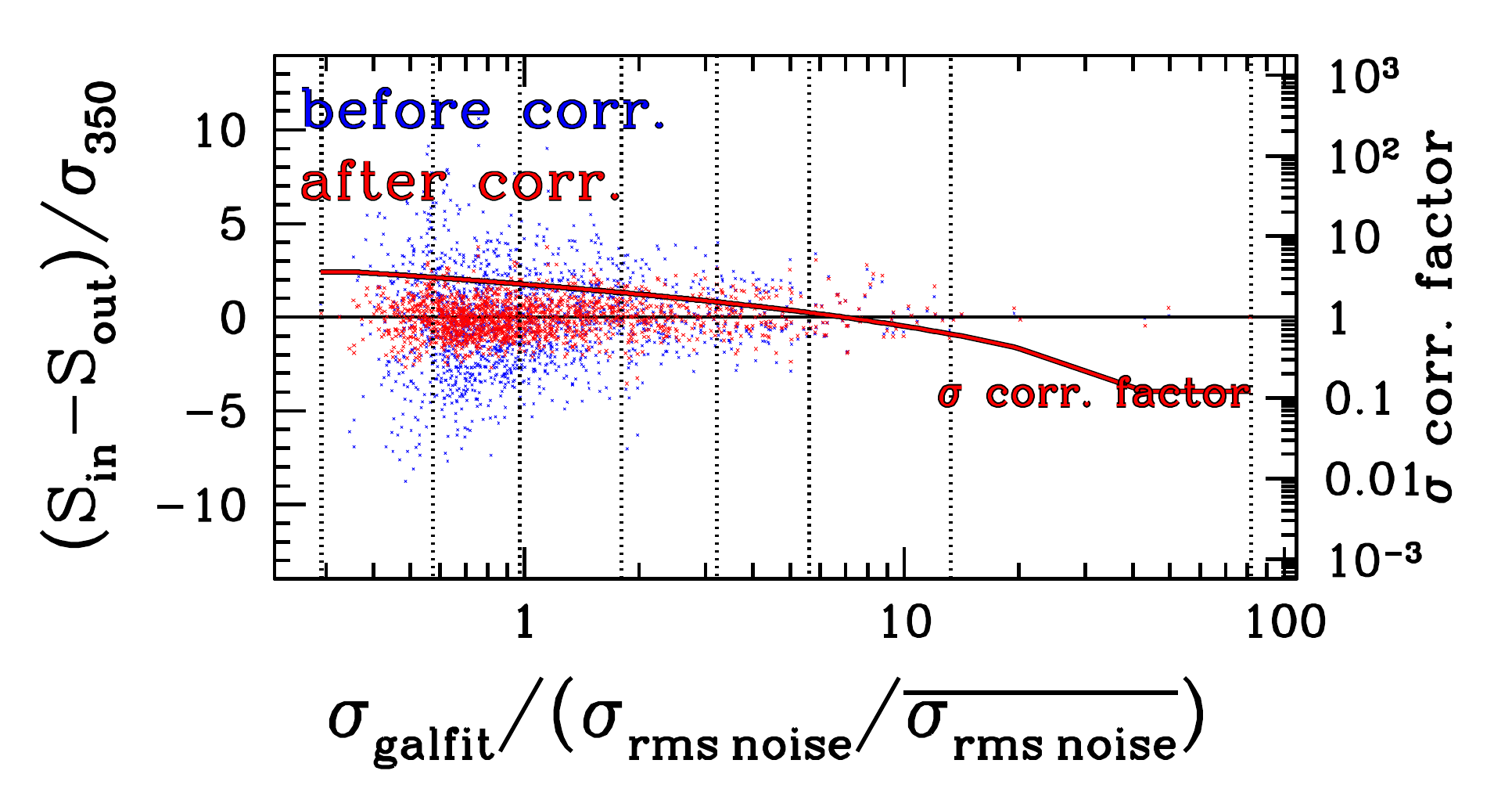}
		\includegraphics[height=3.5cm]{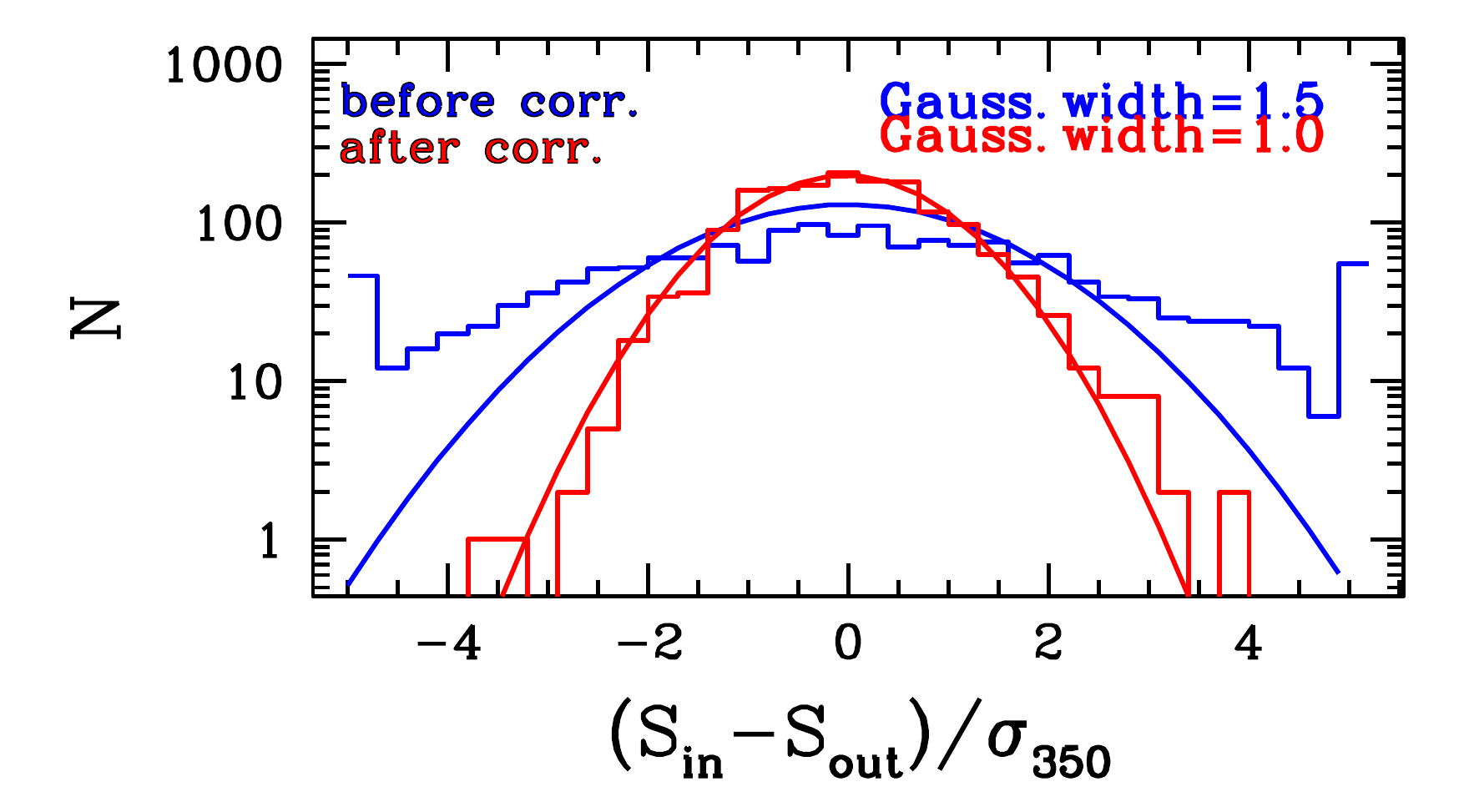}
		\includegraphics[height=3.5cm]{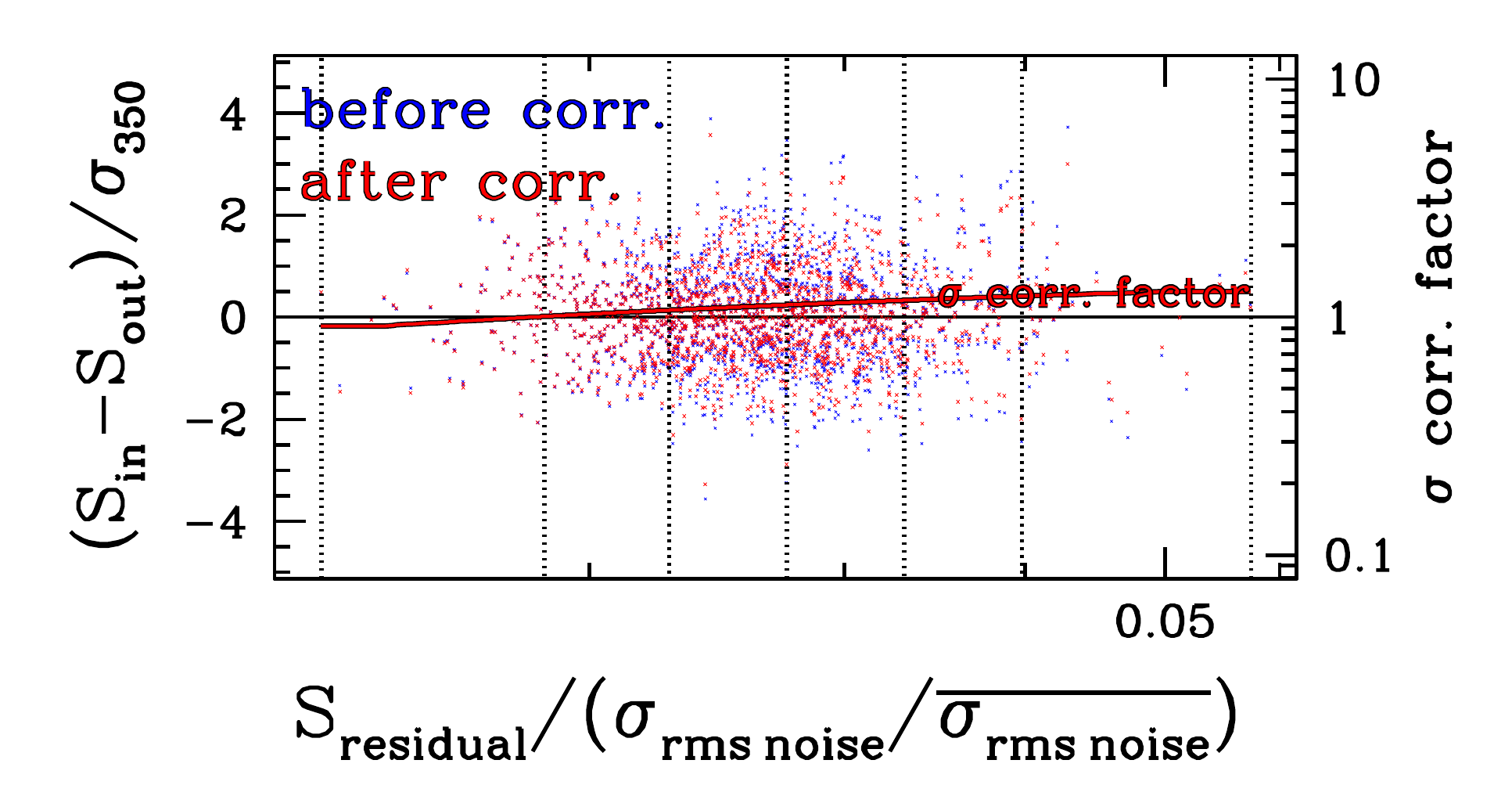}
		\includegraphics[height=3.5cm]{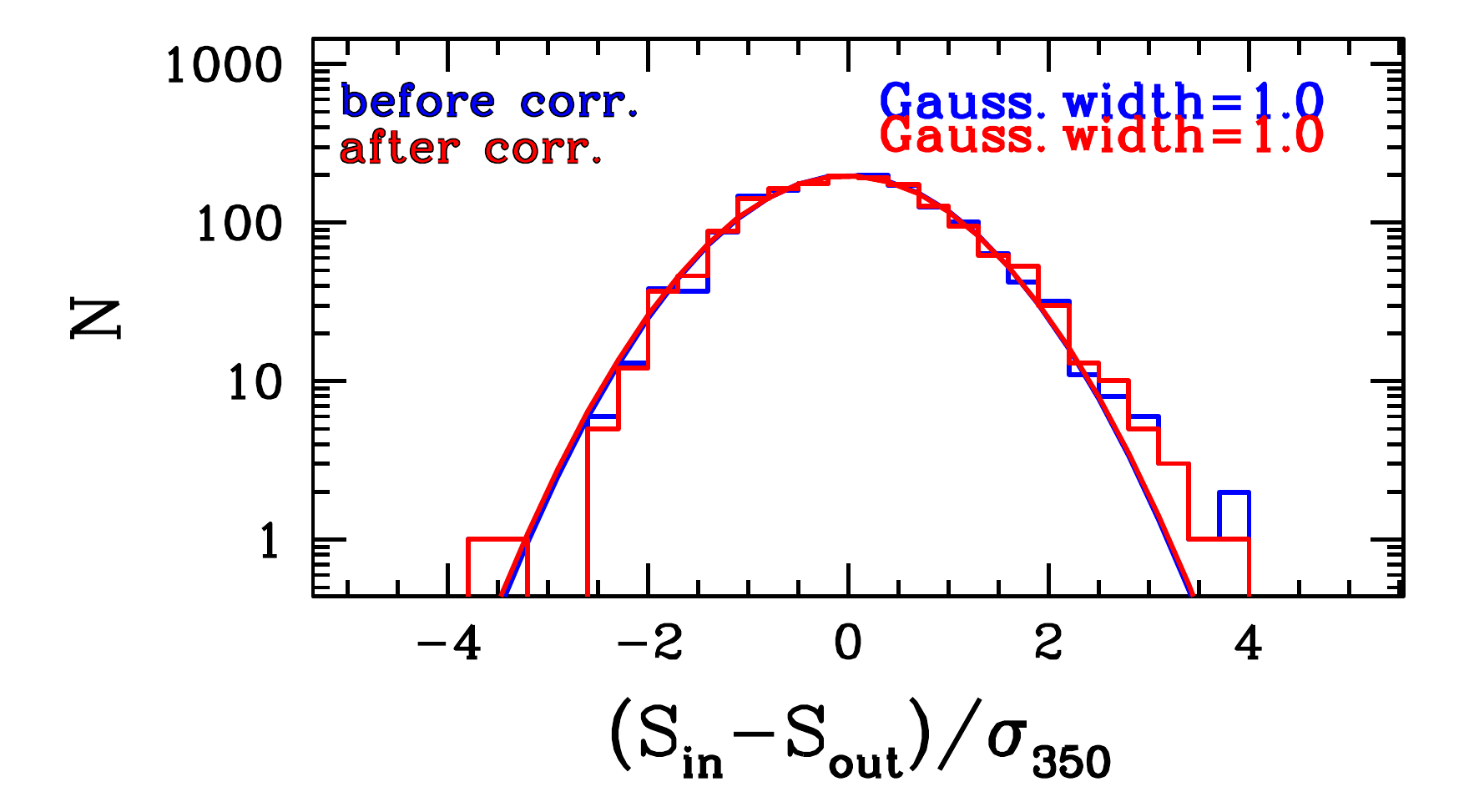}
		\includegraphics[height=3.5cm]{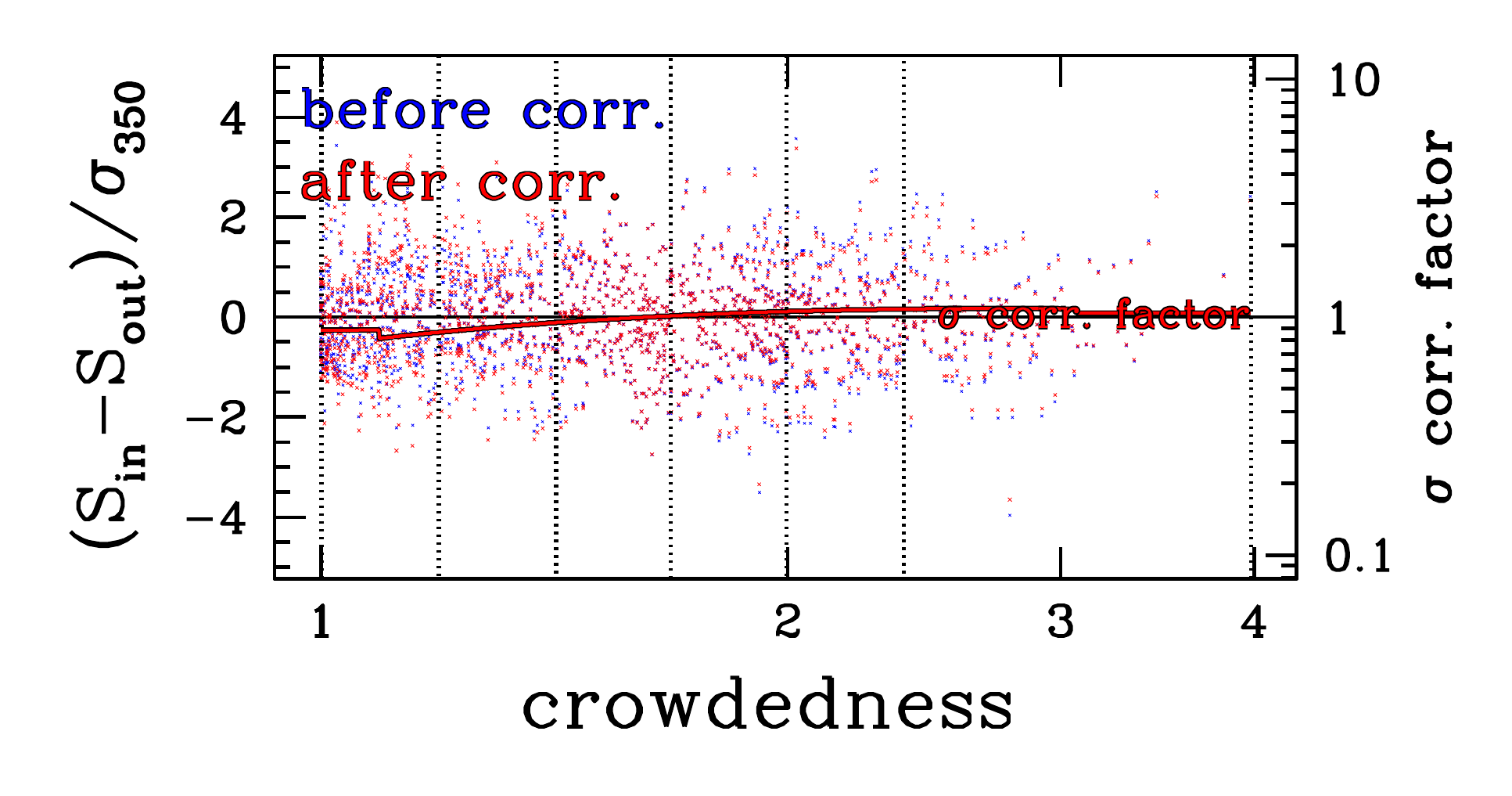}
		\includegraphics[height=3.5cm]{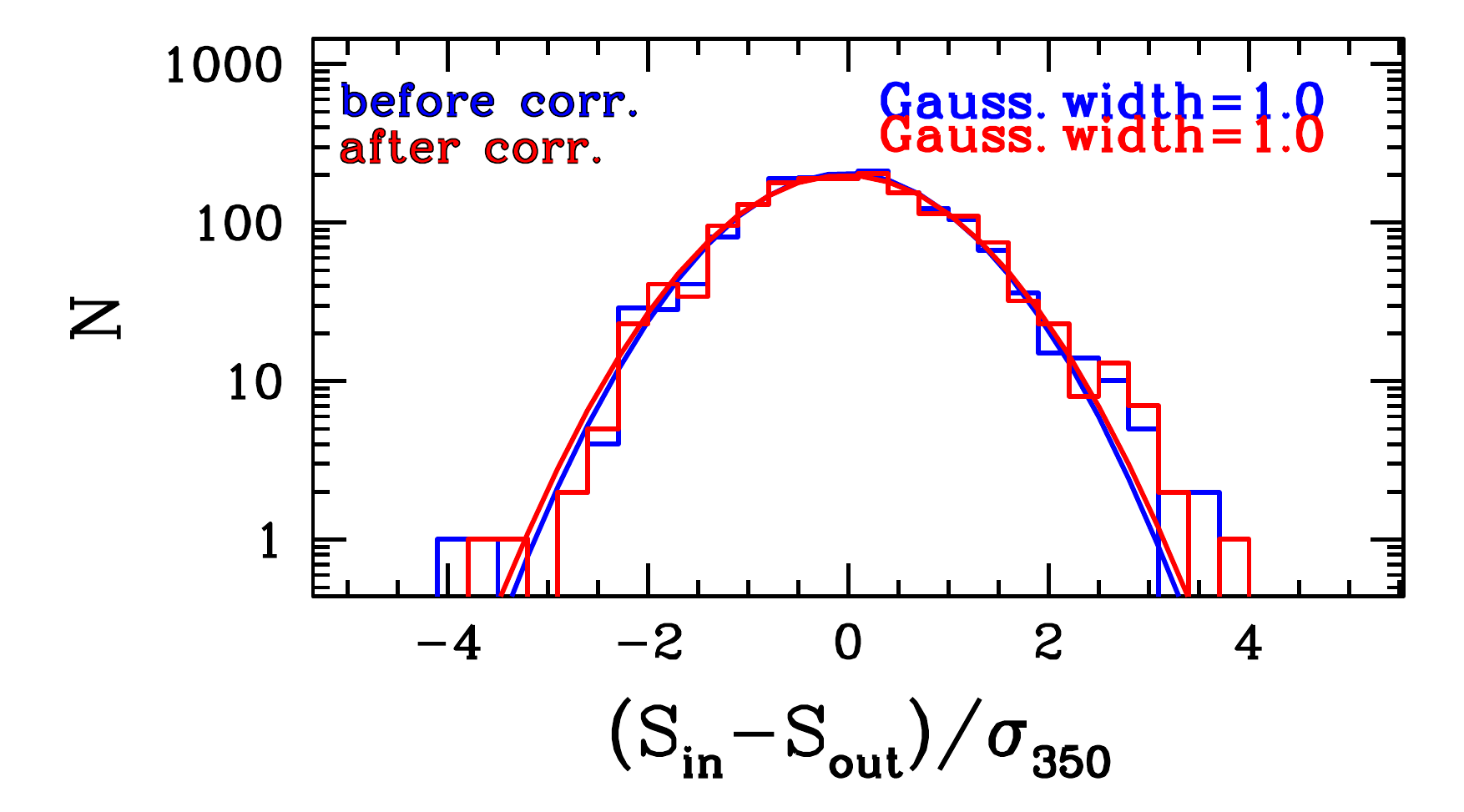}
		\caption{%
			\textbf{(Left panels)} For simulated objects (at 350~$\mu$m as an example), we show the difference between input (real) and output (measured) flux divided by flux uncertainty in Y axis, i.e., $(S_{\mathrm{in}}-S_{\mathrm{out}})/{\sigma}$. From top to bottom, we analyze this quantity against \textcolor{green!50!black}{three} parameters as indicated by the X axis label. We bin objects as indicated by the vertical dashed lines and compute the rms in each bin for deriving the correction factors as described in Section~\ref{Section_Simulation_Correction_df_corr} (i.e., imposing that, in bins defined for quantities in the X axes, the rms dispersion of $(S_{\mathrm{in}}-S_{\mathrm{out}})/{\sigma}$ is equal to 1.0). 
			We show the data points before and after correction (i.e., $(S_{\mathrm{in}}-S_{\mathrm{out}})/({\sigma,\,\mathrm{uncorrected}})$ and $(S_{\mathrm{in}}-S_{\mathrm{out}})/({\sigma,\,\mathrm{corrected}})$) in blue and red respectively. 
			\textbf{(Right panels)} The histograms of the $(S_{\mathrm{in}}-S_{\mathrm{out}})/{\sigma}$ in logarithm space, and with a solid line representing a best fitting Gaussian profile to the inner part of each histogram. 
			After the \textcolor{green!50!black}{three}-step corrections the histograms are well-behaved and generally well consistent with a Gaussian distribution. 
			Similar figures for all other bands are shown in Appendix~\ref{Section_Simulation_Performance}.
			\label{Fig_Galsim_df_corr_SPIRE}%
		}
	\end{center}
\end{figure*}

%
%
\section{An example: optimizing the prior source list around GN20}
\label{Section_Example_Superdeblending}

Fig.~\ref{Figure_Cutouts_GN20} illustrates the procedure of choosing 
prior sources for fitting based on their SED predicted fluxes for the fairly crowded field 
around the luminous $z=4.055$ starburst GN20 \REVISED{(\citealt{Pope2006}; \citealt{Daddi2009GN20})}, or ID~564 in our catalog. The circles in the figure represent all 24+radio prior sources (Section~\ref{Section_Initial_IRAC_Catalog}), including those \textit{selected} (green) and \textit{excluded} (orange) from fitting. 

As described in the previous sections, for each band we first construct the SED for each 24+radio prior source using the photometry available so far, then choose which objects to fit in the current image and which to exclude. Here we consider 4 marked sources in Fig.~\ref{Figure_Cutouts_GN20} as examples: ID~532, 564, 658 and 659. ID~532 is one of the brightest sources at 16~$\mu$m and 24~$\mu$m bands, but becomes relatively fainter in the SPIRE bands. It has a spectroscopic redshift of 0.9746, and thus the redder SPIRE wavelengths fall longward of the FIR \MINORREREVISED[]{SED} peak. The extracted SED is shown in the last panel of Fig.~\ref{Plot_SED_GN20}. Based on SED fitting at wavelengths shorter than 500~$\mu$m, the SED-predicted 500~$\mu$m flux density of ID~532, augmented by twice the uncertainty, falls below the cut-off threshold: $S_{\mathrm{SED}}+2\,{\sigma}_{{\mathrm{SED}}} < S_{\mathrm{cut},\,500\,\mu\mathrm{m}}$, where the $S_{\mathrm{cut},\,500\,\mu\mathrm{m}}$ is the critical value we chose as shown in the 5th panel of Fig.~\ref{Fig_Galsed_cumulative_number_function}. Thus we do not fit ID~532 at 500~$\mu$m. Meanwhile, its predicted flux contribution to the observed 500~$\mu$m image is subtracted as described in Section~\ref{Section_Prior_Extraction_Photometry}, so that we can measure the real, deboosted flux for the remaining prior sources. 

ID~658 and ID~659 are two fainter sources near GN20 that have the same spectroscopic redshift $z = 4.055$ (GN20.2a, GN20.2b respectively; they are part of a proto-cluster at that redshift). \citet{Tan2014} presented photometry and SEDs for these objects, as well as mm continuum and spectral line observations. The mm photometry provides tight constraints on the Rayleigh-Jeans tail of the dust SED. In Fig.~\ref{Plot_SED_GN20}, we compare our deblended photometry and SEDs with the photometry of \citet{Tan2014}. 
They did not measure the SPIRE fluxes for GN20, and their flux uncertainties at PACS 100~$\mu$m and 160~$\mu$m are larger than ours. Their mm photometry agrees very well with our SEDs, noting that the SEDs are fitted only with the black data points in this work. For the fainter GN20 proto-cluster members, ID~658 and ID~659, \citet{Tan2014} provided only PACS fluxes, which all have $\mathrm{S/N} < 3$. Our photometry provides more data points, and our SEDs are in very good consistency with the previous mm continuum observations.

\vspace{1truecm}

%
%
\section{Simulation-based flux and uncertainty corrections}
\label{Section_Simulation}


We use extensive Monte Carlo simulations in our \superdb{} photometry for two main purposes. Firstly, simulations can verify potential flux biases, so that we can correct for any imperfect sky background or other subtle systematic effects. Secondly, they help us to understand and calibrate the uncertainties of photometric measurements.

In fact, the flux uncertainties estimated by \galfit{} are based on diagonalizing and projecting the covariance matrix, and hence may not properly reflect real errors \citep{Peng2002}. In simulations, we have not only the \galfit{} output but also the simulated input information. Therefore we can link the real uncertainty to the \galfit{} output with measurable source properties. We will present the Monte Carlo simulation method in Section~\ref{Section_Simulation_Method}, the flux bias analysis in Section~\ref{Section_Simulation_Correction_fbias}, and the flux uncertainty analysis in Section~\ref{Section_Simulation_Correction_df_corr}.


\subsection{Monte Carlo simulation}
\label{Section_Simulation_Method}

We simulate one source at a time in the faint-source-subtracted image.\footnote{The faint-source-subtracted image is the image with {excluded} source models subtracted from the original image; see Section~\ref{Section_Prior_Extraction_Photometry}.}
The simulated source positions are generated randomly within the survey area, and do not avoid other real, bright sources. Their input flux densities, $S_{\mathrm{in}}$, are uniformly distributed in log space within the range from $\sim1\sigma$ to $\sim30\sigma$, where $\sigma$ is the {mean flux density uncertainty} at each band (see Table~\ref{Table_1}). 

We add each extra simulated source to the 24+radio prior list which has already been filtered for that band (i.e., with all selected sources, \MINORREREVISED[]{see Section}~\ref{Section_Choosing_Prior_Source_List}, as well as additional sources detected in the residual images, see Section~\ref{Section_Prior_Extraction_Photometry}), and then perform the PSF fitting photometry including all active priors and an extra prior at the position of the simulated source. 
We also perform two-pass \galfit{} fitting as described in Section~\ref{Section_Photometry_24} to allow variation in the positions of priors in the second pass if they have high $\mathrm{S/N}$ in the results of the first pass. 
At the end of each pass, we run the corrections described in the following Sections~\ref{Section_Simulation_Correction_fbias}~and~\ref{Section_Simulation_Correction_df_corr}. 
The \galfit{} output flux density of the simulated source is $S_{\mathrm{out}}$, and the \galfit{} flux density uncertainty is ${\sigma}_{\mathit{galfit}}$. 

Finally, this process is repeated $\sim$3000 times. Therefore, for each band, we have $\sim$3000 values for $S_{\mathrm{in}}$, $S_{\mathrm{out}}$ and ${\sigma}_{\mathit{galfit}}$. In addition, we measured several additional properties for each simulated source: for example the local rms noise value ${\sigma}_\mathrm{rms\,noise}$ for each source, measured in the instrument noise image data;  the local absolute flux \MINORREREVISED[]{density} in the residual image (hereafter $S_{\mathrm{residual}}$), measured by computing the total absolute pixel values 
in the PSF aperture for each fitted prior source; the local scatter in the residual image, measured also in the PSF circle; and the \crowdedness{} parameter, computed by summing up the Gaussian weighting of all sources at current source $i$ position: 
\begin{equation}
\sum\limits_{j=1}^{N} e^{\left(-d_{j,i}^2/d_{\mathrm{PSF}}^2\right)}
\end{equation}
where $d_{j,i}$ is the angular distance in arcsec from source $j$ to current source $i$ and $d_{\mathrm{PSF}}$ is the FWHM in arcsec of the PSF for the band being analyzed. Defined in this way, the \crowdedness{} is a weighted measure of the number of sources present within the beam, including the specific source under consideration.

These measurable parameters are designed to provide key information about the quality of fitting and the local crowding (hence blending) of prior sources. We will make use of them with the simulation results in order to calibrate the best possible flux bias and flux uncertainty corrections. 

In addition to the uncertainties estimated in this way, we add in quadrature to each object's flux uncertainty the appropriate contribution (when relevant) to account for astrometric dispersion when fitting at fixed spatial positions. This term is estimated by computing the dispersion in measured fluxes for bright sources at fitted position versus those at the positions optimized in the fit.


\subsection{Flux bias correction}
\label{Section_Simulation_Correction_fbias}

The statistics of the differences between input and output fluxes, $S_{\mathrm{in}}-S_{\mathrm{out}}$, can be used to verify and correct the bias of the \galfit{} flux measurements. 
In Fig.~\ref{Fig_Galsim_flux_bias_SPIRE} we present the \MINORREREVISED[analyses]{analysis} of $S_{\mathrm{in}}-S_{\mathrm{out}}$ based on simulation data for SPIRE 350~$\mu$m as an example.  
In panel (1), $S_{\mathrm{in}}-S_{\mathrm{out}}$ is plotted against the simulation input flux density $S_{\mathrm{in}}$, which covers the range $3-12\times$ in terms of the detection limit ${\sigma}_{350\,{\mu}\mathrm{m}}$. The overall median of $S_{\mathrm{in}}-S_{\mathrm{out}}$ is generally small, indicating that the (constant) background levels used in the \galfit{} work are quite accurate. 

%
%
For all bands we explore the dependence of $S_{\mathrm{in}}-S_{\mathrm{out}}$ on the following three key  properties: (1) the flux uncertainty ${\sigma}_{\mathit{galfit}}$, which is a direct output of \galfit{} fitting; (2) the residual flux $S_{\mathrm{residual}}$, measured in the output residual image; and (3) the \crowdedness{}, which indicates prior source blending as described in the previous section. When testing parameters (1) and (2), we further use the rms noise ${\sigma}_\mathrm{rms\,noise}$ to normalize these two parameter values in order to account for the local rms variation in case of non-uniform depth of the data over the whole GOODS-N region. For example, if sources at the map center have smaller rms noise than sources near the edges, the outer sources will likely have higher ${\sigma}_{\mathit{galfit}}$ and $S_{\mathrm{residual}}$ simply because of their locations. 


{%
As shown in panels (2) to (4) of Fig.~\ref{Fig_Galsim_flux_bias_SPIRE}, we bin simulated sources by \MINORREREVISED[source properties]{each of the key properties}. In each bin we compute the median of $S_{\mathrm{in}}-S_{\mathrm{out}}$, which is just the flux bias $S_\mathrm{bias}$. Then we fit $S_\mathrm{bias}$ in each bin by a 3-order polynomial function of the X axis parameter. The fitted polynomials are shown as the red curves in panels (2) to (4). Note that the quantities along the X axes are measurable for each real fitted source, thus these functions are also applicable to each real source. 
}

\begin{figure*}
	\centering
	\includegraphics[width=0.29\textwidth]{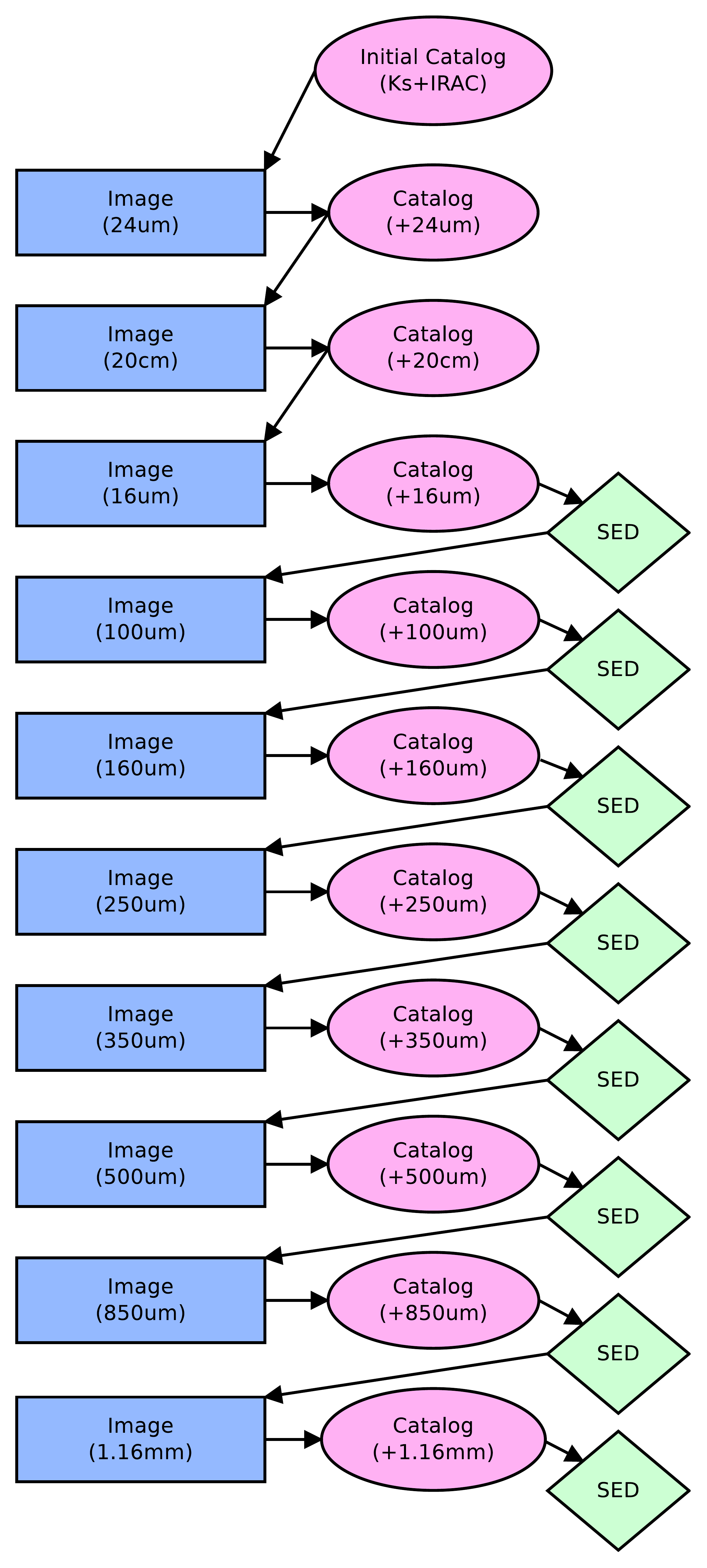}
	\includegraphics[width=0.70\textwidth]{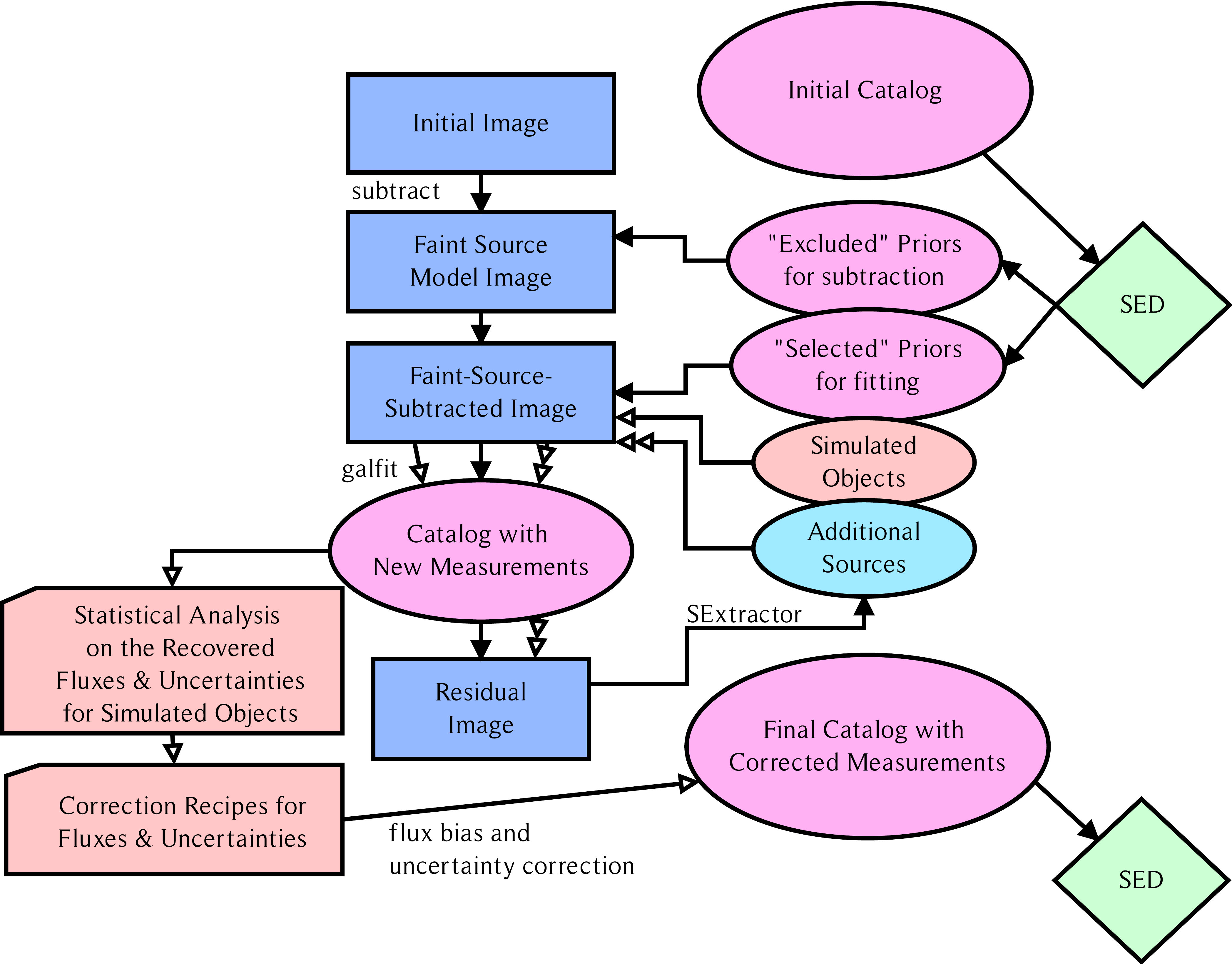}
	\caption{%
		\textbf{(Left:)} The flow chart of our photometry procedures as applied to all bands. \REVISED[We]{Catalogs are indicated by purple ellipses, SED fitting processes by green diamonds, and images by blue rectangles. From top to bottom, following the arrows, we} start with the IRAC catalog and perform photometry at 24~$\mu$m, 20~cm and 16~$\mu$m as described in Section~\ref{Section_Initial_IRAC_Catalog}. Then SED fitting 
		is used to help determine an updated source prior list for each FIR/mm band, as discussed in Section~\ref{Section_Superdeblending}. 
		\textbf{(Right:)} The \superdb{} procedures at each FIR/mm band. 
		For each source in the initial catalog, we run SED fitting and predict the flux and uncertainty in the current band (Section~\ref{Section_SED_Fitting}). Then we split prior sources into two lists: ``priors to subtract'' and ``priors to fit'' (Section~\ref{Section_Choosing_Prior_Source_List}). 
		We make a source-model-image for sources in the former list and subtract it from the initial observed image, resulting in a faint-source-subtracted image on which the \galfit{} photometry is then performed for sources in the latter list (Section~\ref{Section_Prior_Extraction_Photometry}). 
		We further blindly extract additional sources in the residual image and re-run the \galfit{} photometry with ``priors in residual'' (Section~\ref{Section_Additional_Sources_In_Residual}), which results in the ``catalog with new measurements''. 
		Meanwhile, we generate simulated objects in the faint-source-subtracted-image, then recover fluxes and analyze and calibrate uncertainties as described in Section~\ref{Section_Simulation}. 
		We derive ``correction recipes'' that make use of measurable parameters; thus for each fitted source, we can measure these parameters and apply the corrections to real measurements. 
		Finally, we obtain the ``catalog with corrected measurements'' for each band and fit the SED again, which will be used in the next band or as the final output after the last FIR/mm band has been analyzed.
	}
	\label{Figure_FlowChartDiagram}
\end{figure*}


\subsection{Flux uncertainty correction}
\label{Section_Simulation_Correction_df_corr}

The Monte Carlo simulations also provide a path towards obtaining reliable estimates of flux uncertainties. The flux error returned by \galfit{} (${\sigma}_{\galfit{}}$) only reflects the fitting uncertainty in a formal way \citep[e.g.,][]{Peng2002}. In fact, the \galfit{} formalism is optimized for optical/near-IR images and assumes uncorrelated noise (all pixel are independent) in the data. Thus it cannot fully account for real uncertainties, especially in the presence of correlated noise and confusion noise. Imperfect flux uncertainties will strongly affect the validity of SED fitting for individual sources and the assessment of detections or lack thereof. 

We analyze the statistics of $(S_{\mathrm{in}}-S_{\mathrm{out}})/{\sigma}_{\mathit{galfit}}$ with simulation data, which should have a rms dispersion of 1.0 by definition if uncertainties are well defined and Gaussian-like. 
%
%
%
Taking the 350~$\mu$m simulation as an example, in the left upper panel of Fig.~\ref{Fig_Galsim_df_corr_SPIRE} we bin simulated objects by the measurable parameter (${\sigma}_{\mathit{galfit}}/{\sigma}_{\mathrm{rms\,noise}}$) as in previous section, and in the middle and right upper panel we show the histograms of $(S_{\mathrm{in}}-S_{\mathrm{out}})/{\sigma}_{\mathit{galfit}}$ (in blue) and $(S_{\mathrm{in}}-S_{\mathrm{out}})/{\sigma}_{1}$ (in red), where ${\sigma}_{1}$ is the first-step corrected flux uncertainty. 
The width of the blue histogram is clearly much broader than 1.0, indicating that the values ${\sigma}_{\mathit{galfit}}$ are substantially underestimated. 

%
We determine correction factors as a function of the X axis parameter, requiring that, in bins defined for quantities in the X axis, the rms dispersion of $(S_{\mathrm{in}}-S_{\mathrm{out}})/{\sigma}$ is equal to 1.0. 
In each bin, we compute the rms of $(S_{\mathrm{in}}-S_{\mathrm{out}})$ and the median of ${\sigma}_{\mathit{galfit}}$, then calculate their ratio, which is exactly the factor that we need for correcting ${\sigma}_{\mathit{galfit}}$ to the real uncertainty of $(S_{\mathrm{in}}-S_{\mathrm{out}})$. As in the previous section, we fit the correction factor as a 3-order polynomial function against the X axis parameter. The fitted polynomial functions are shown as the red curves in the figure. In this way, we obtain the first-step corrected ${\sigma}_{1}$, whose histogram is shown in red, and is much closer to a width of 1.0.

%
The first-step corrected uncertainty histograms can sometimes have asymmetric shapes, which indicate that there are still some systematic biases. Thus we take the first-step corrected uncertainty ${\sigma}_{1}$ as input and perform the second and the third steps. In the second step, we first perform the flux bias correction with parameter $S_{\mathrm{residual}}/{\sigma}_{\mathrm{rms\,noise}}$, then perform the uncertainty correction with the same parameter as illustrated by the middle row panels of Fig.~\ref{Fig_Galsim_df_corr_SPIRE}. Then, similarly, in the third-step we apply the flux bias correction, followed by the flux uncertainty correction with the parameter \crowdedness{}. Therefore, with these correction recipes, we are able to apply the correction to real fitted sources. 

We have also tried multiple iterations or different combinations of the three steps. For example after the three-step processing, we use the output corrected flux and uncertainty as the input for several more iterations of the three-step processing. However, we find that the corrected values and histograms are stable and cannot be further improved in this way. 

Similar figures for all examined bands are reported in the Appendix~\ref{Section_Simulation_Performance} (Figures~\ref{Figure_galsim_160_bin}~to~\ref{Figure_galsim_24um_bin}). 
\REVISED{Generally, for objects whose \galfit{} uncertainties are around or below the median for that band, we find that the flux uncertainties are often quite underestimated, while for objects with the largest uncertainties (typically those affected by substantial deblending uncertainties) the values are often overestimated (i.e., the real errors are smaller as judged from our simulations).}
An analogous effect is seen for the \crowdedness{} parameter. We have analyzed a number of other parameters (including flux itself, spatial positions, redshift, different kind of \crowdedness{} and residual definitions, etc.)  and we found no other significant dependency to further correct flux uncertainties.

%
%
%

\begin{table*}

\begin{center}

\caption{ %
    GOODS-N {\em ``Super-deblended''} Photometry Results %
    \label{Table_1} %
}

\begin{tabular*}{0.85\textwidth}{ @{\extracolsep{\fill}} c c c c c c c c c }

        \hline
        Band         &       Instrument  & Beam FWHM \tablenotemark{a} %
                                         & $\rho_{\mathrm{fit}}$ \tablenotemark{b} %
                                         & $N_{\mathrm{fit}}$ \tablenotemark{c} %
                                         & $N_{\mathrm{excl.}}$ \tablenotemark{d} %
                                         & $N_{\mathrm{S/N}{\ge}3}$ \tablenotemark{e} %
                                         & $N_{\mathrm{add.}}$ \tablenotemark{f} %
                                         & $1\,\bar{{\sigma}}$ \tablenotemark{g} \\
                     &                   & arcsec %
                                         & beam$^{-1}$ %
                                         & %
                                         & %
                                         & %
                                         & %
                                         & mJy \\
        \hline
        $24{\mu}m$   &        Spitzer/MIPS  &                        5.7 &           1.205 &         19437 &           0 &             3056 &                      0 &        5.165$\times 10^{-3}$ \\
        $20{c}m$     &                 VLA  &  1.7/2.0 \tablenotemark{h} &           0.107 &         19437 &           0 &             1328 &                      0 &        2.744$\times 10^{-3}$ \\
        $16{\mu}m$   &     Spitzer/IRS/PUI  &                        3.6 &           0.082 &          3306 &           0 &             1335 &                      0 &        7.681$\times 10^{-3}$ \\
        $100{\mu}m$  &       Herschel/PACS  &                        7.2 &           0.326 &          3294 &          12 &             1178 &                      0 &                 0.315 \\
        $160{\mu}m$  &       Herschel/PACS  &                       12.0 &           0.862 &          3137 &         169 &             1153 &                     18 &                 0.681 \\
        $250{\mu}m$  &      Herschel/SPIRE  &                       18.2 &           0.973 &          1540 &        1766 &              668 &                     13 &                 1.571 \\
        $350{\mu}m$  &      Herschel/SPIRE  &                       24.9 &           1.142 &           966 &        2340 &              292 &                     10 &                 2.072 \\
        $500{\mu}m$  &      Herschel/SPIRE  &                       36.3 &           1.181 &           470 &        2836 &              125 &                     17 &                 2.570 \\
        $850{\mu}m$  &         JCMT/SCUBA2  &     11.0 \tablenotemark{i} &           0.154 &           668 &        2638 &               32 &                     15 &                 1.249 \\
        $1160{\mu}m$ &         AzTEC+MAMBO  &                       19.5 &           0.374 &           515 &        2791 &               43 &                     11 &                 0.661 \\
        \hline
        
\end{tabular*}

\begin{minipage}{0.82\textwidth}
    
\flushleft

$^\mathrm{a}$ 
    Beam FWHM is the full width at half maximum of the circular-Gaussian-approximation point spread function of each image data. %

$^\mathrm{b}$ 
    $\rho_{\mathrm{fit}}$ is the number density of prior sources fitted at each band, normalized by the Gaussian-approximation beam area. %

$^\mathrm{c}$ 
    $N_{\mathrm{fit}}$ is the number of prior sources fitted at each band. %

$^\mathrm{d}$ 
    $N_{\mathrm{excl.}}$ is the number of prior sources excluded from fitting at each band. These sources are subtracted from original image with their spectral energy distribution predicted flux density at each band. %

$^\mathrm{e}$ 
    $N_{\mathrm{S/N}{\ge}3}$ is the number of prior sources with $\mathrm{S/N}\ge3$ (i.e. detected) at each band. %

$^\mathrm{f}$ 
    $N_{\mathrm{add.}}$ is the number of $\mathrm{S/N}\ge3$ additional sources that are not in the prior source catalog but blindly-extracted from the intermediate residual image product at each band (see Section~\ref{Section_Additional_Sources_In_Residual}). %

$^\mathrm{g}$ 
    $1\,\bar{{\sigma}}$ is the median of the flux density uncertainties of all sources detected with S/N${\ge}3$ in each band. 
    %

$^\mathrm{h}$ 
    The shallower VLA image data from \cite{Morrison2010} have a beam FWHM of 1.7$''$, and the deeper VLA image data from Owen (2017) are especially produced with a beam FWHM of 2.0$''$. 

$^\mathrm{i}$ 
    We use the unfiltered image, which has a narrower PSF FWHM than that of the matched-filter convolved image. We measured FWHM~$\approx 11''$ in our PSF reconstruction for the unfiltered image, which is roughly consistent with Fig.~14 of \cite{Chapin_2013_SCUBA2}, and compares with 14.8$''$ for the matched-filter image.

\end{minipage}

\end{center}

\end{table*}



%
%

\vspace{1truecm}

%
%
\section{\Superdb{} photometry --- overall work flow}
\label{Section_Flow_Chart}

\begin{figure}
\centering
\includegraphics[width=0.48\textwidth]{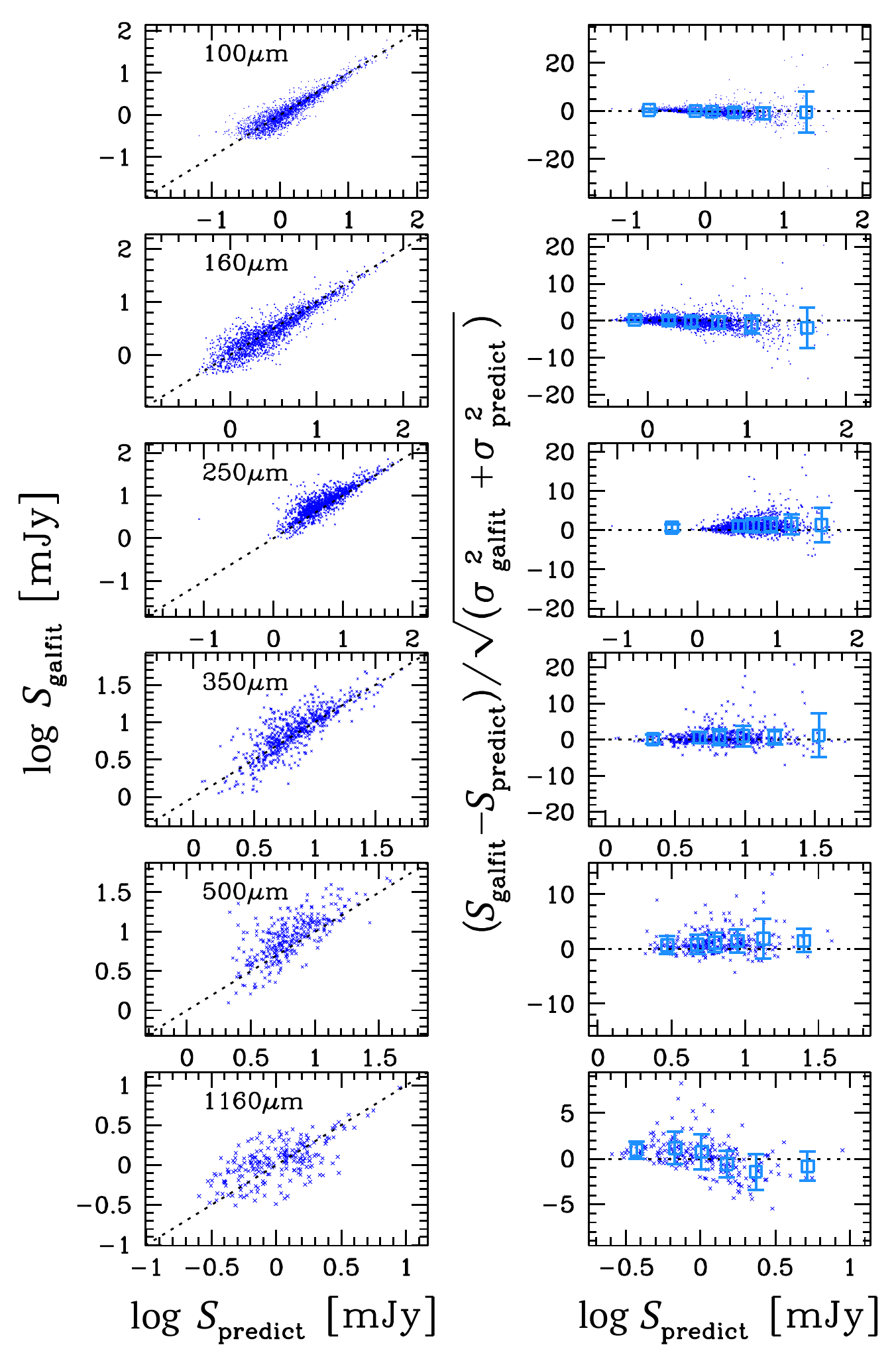}
\caption{%
    Comparison between SED-predicted flux densities ($S_{\mathrm{predict}}$) before the \galfit{} photometry step, and the newly measured flux densities ($S_{\mathit{galfit}}$) after the \galfit{} photometry step for each band, as a demonstration of the performance of our SED prediction. The left panels show the flux-flux comparison in log space, while the right panels show flux versus flux difference weighted by the combined error in linear space, for the same bands as the left panels. 
    Because the SED flux and \MINORREREVISED[error]{uncertainty} ($S_{\mathrm{predict}}$ and ${\sigma}_{{\mathrm{predict}}}$) values are used \textbf{(a)} to determine which sources are too faint to be detectable, and \textbf{(b)} to subtract their flux contributions from the observed data (see Section~\ref{Section_Prior_Extraction_Photometry}), it is essential to ascertain that SED predictions are not subject to strong biases.
    In this figure, $S_{\mathrm{predict}}$ and $S_{\mathit{galfit}}$ are generally consistent in all bands, although with larger dispersion (indicated by the error bars in the right panels) at longer wavelengths. The small skewness in the results be explained by the fact that only significant ($>3\,\sigma$) detections are shown, which produces a bias toward sources with higher $S_{\mathit{galfit}}$/$S_{\mathrm{predict}}$ ratios.
    \label{Fig_Galsed_SED_flux_vs_Galfit_flux}%
}
\end{figure}


Based on the simulation analyses, we apply the correction recipe to real fitted sources. For each fitted source, we begin with the flux uncertainty given by \galfit{} fitting (${\sigma}_{\mathit{galfit}}$), 
then measure the local noise on the rms image (${\sigma}_\mathrm{rms\,noise}$), the local residual flux on the residual image ($S_{\mathrm{residual}}$), and the \crowdedness{} parameter using all fitted prior source coordinates at each band. Therefore the aforementioned three measurable parameters can be obtained for each fitted source. We compute the correction factor from each parameter and correct flux bias and uncertainty sequentially in the steps described in Section~\ref{Section_Simulation_Correction_fbias}~and~\ref{Section_Simulation_Correction_df_corr}. Finally, we obtain the \superdb{} GOODS-N FIR+mm photometry results.

Our \superdb{} method of optimizing prior source lists with the help of the optimized SED fitting ensures that we are fitting all the most likely detectable prior sources for each band. 
%
The flux and uncertainty corrections we derive on a source-by-source basis make our detections and SED fitting more reliable.
We list the final \REVISED[detection limit ($1{\sigma}^{det.}$)]{mean flux density uncertainty ($1{\bar{\sigma}}$)} 
and the number of $\mathrm{S/N}\ge3$ sources ($N_{\mathrm{S/N}}\ge3$, i.e., single band detected catalog sources) at each band in Table~\ref{Table_1}. 

\begin{figure*}
\centering
\includegraphics[width=0.45\textwidth, trim=0 10mm 0 8mm]{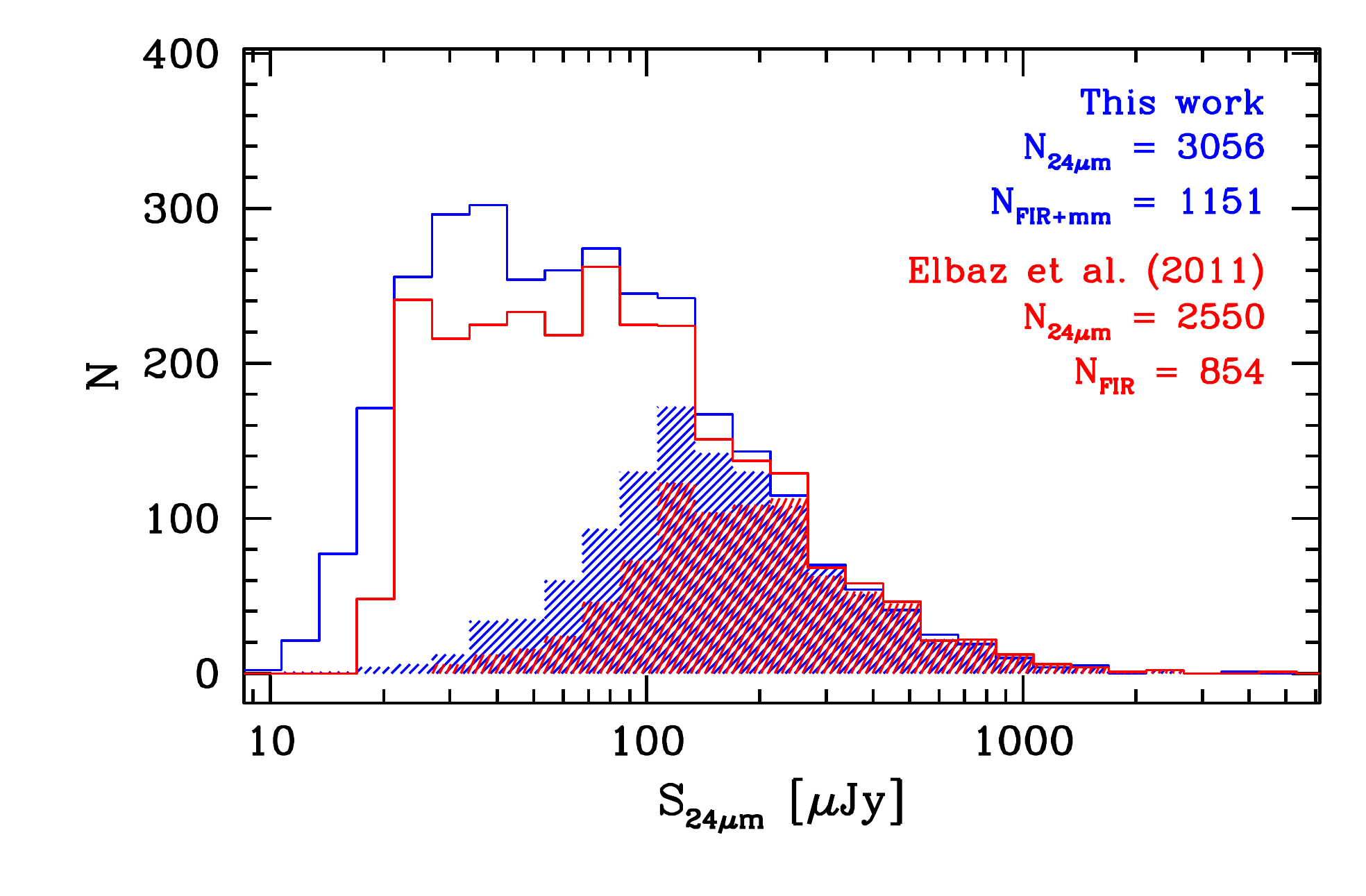}
\includegraphics[width=0.45\textwidth, trim=0 10mm 0 8mm]{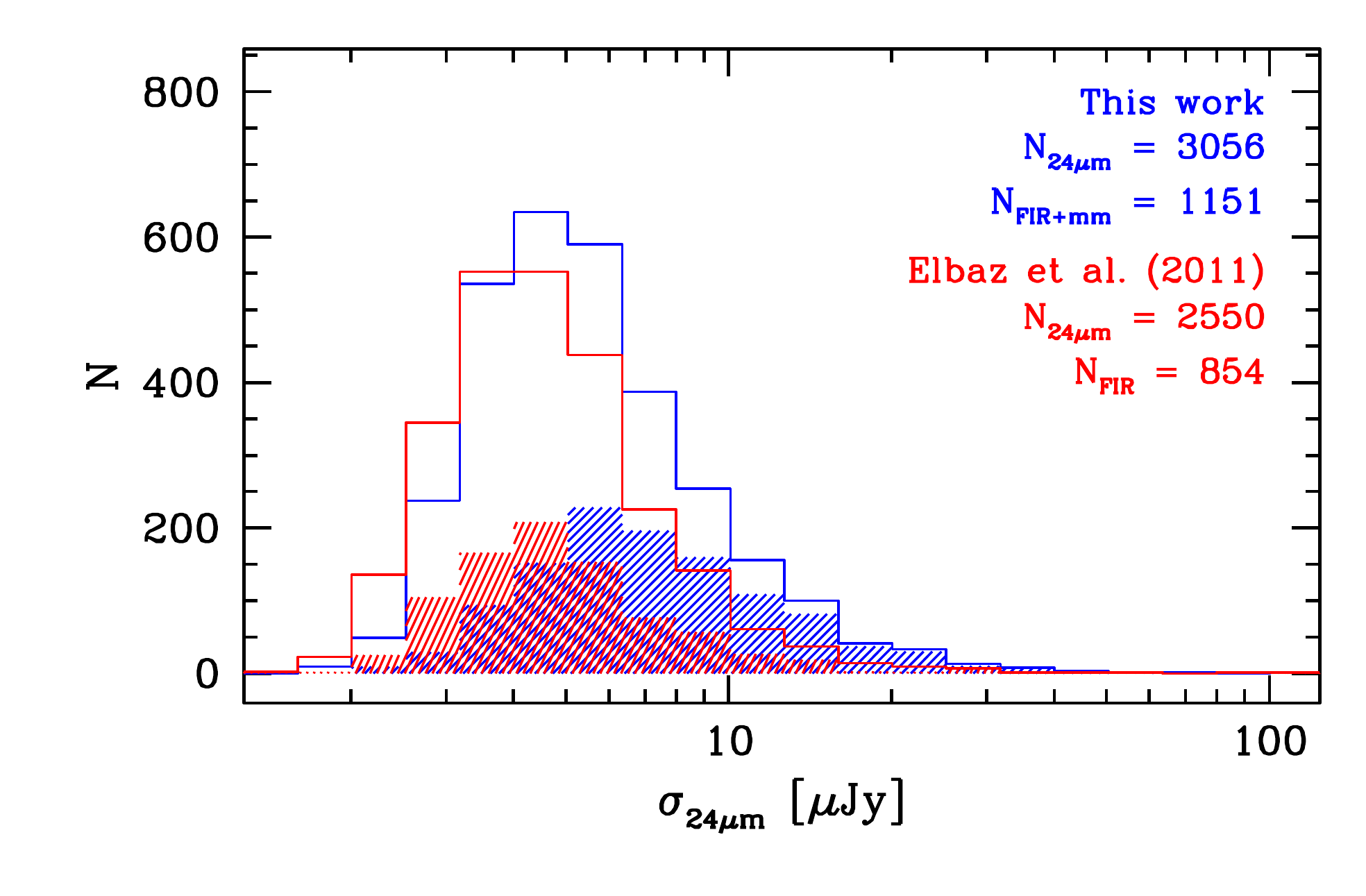}
\caption{%
    Comparison of histograms of the 24~$\mu$m flux densities (left) and flux density uncertainties (right) between this work and that of \citet{Elbaz2011}. 
    The open areas of the histogram show all prior sources, while the shaded areas represent the sources that are significantly detected 
    (with $\mathrm{S/N}_{\textnormal{FIR+mm}} \ge 5$ for our sources, or $\mathrm{S/N}_{\mathrm{FIR}\,{(100\;\mathrm{to}\;500{\mu}\mathrm{m})}} \ge 5$ for \citealt{Elbaz2011} sources, see definition of $\mathrm{S/N}_{\textnormal{FIR+mm}}$ in Section~\ref{Section_z_SFR} and Equation~\ref{Equation_SNR}). 
    Results from this work are shown in blue, and for \citet{Elbaz2011} in red.
    In the left panel, the histograms of bright sources are consistent, but there are more sources from this work at fainter flux densities. This means that we have about 450 more candidate prior sources for the longer wavelength photometry than in \citet{Elbaz2011}. 
    The FIR to mm detected sources, which are the red and blue shaded areas, also show consistency for brighter fluxes but are more numerous in this work. 
    In the right panel, due to the simulation-based flux uncertainty correction (see Section~\ref{Section_Simulation_Correction_df_corr}), the uncertainties in this work are generally higher, even for the FIR and 24~$\mu$m detected sources in the shaded area. 
    The $N_{24\,{\mu}\mathrm{m}}$ is the number of $\mathrm{S/N}_{24\,{\mu}\mathrm{m}} \ge 3$ sources in each catalog, and $N_{\textnormal{FIR+mm}}$ is the number of FIR-to-mm (or FIR 100-to-500~$\mu$m for \citealt{Elbaz2011} catalog) sources. 
    \label{compare_f24_histogram_dzliu_daddi_elbaz}%
}
\end{figure*}

\begin{figure*}
\centering
\includegraphics[width=0.45\textwidth, trim=0 10mm 0 8mm]{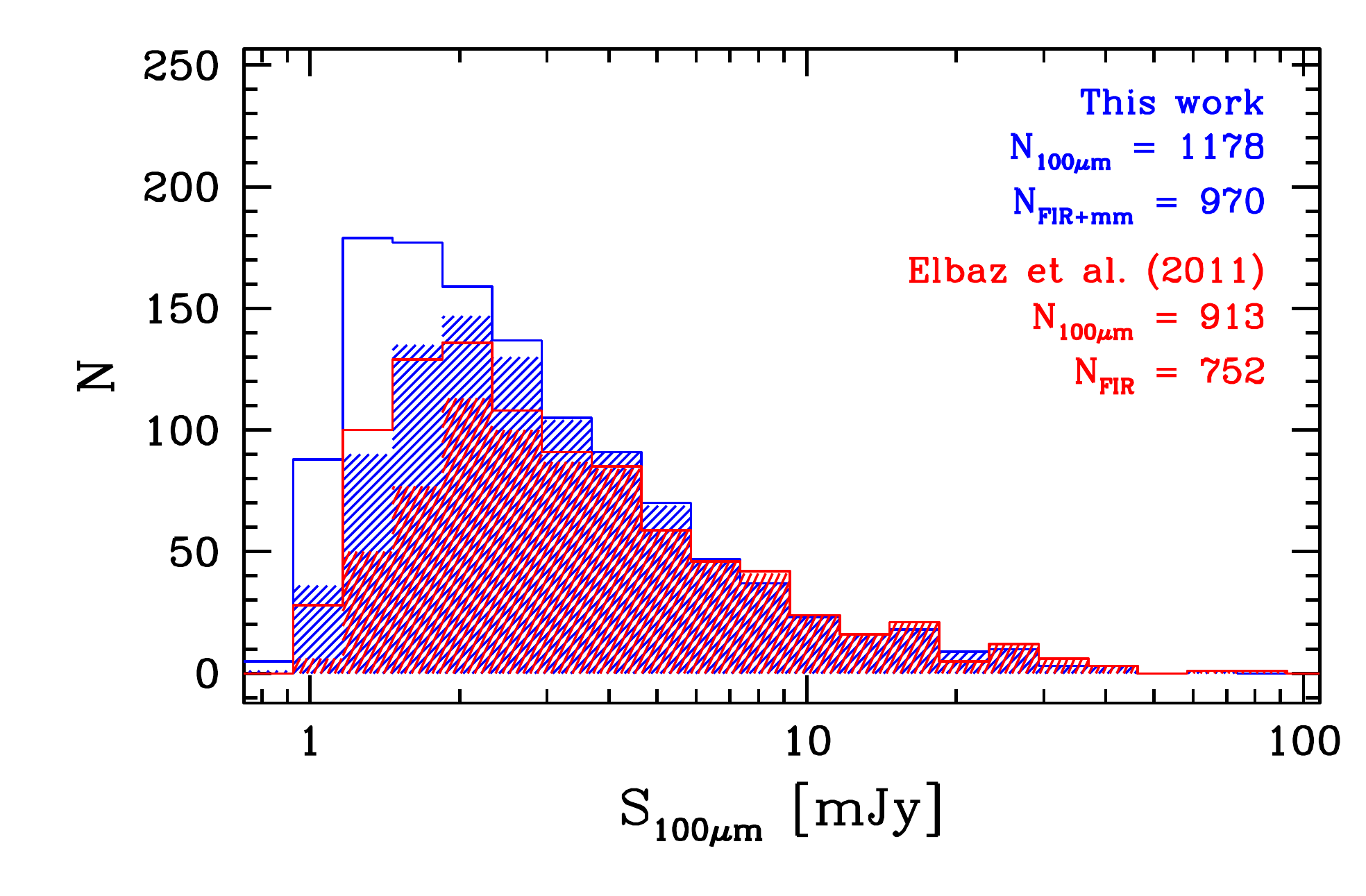}
\includegraphics[width=0.45\textwidth, trim=0 10mm 0 8mm]{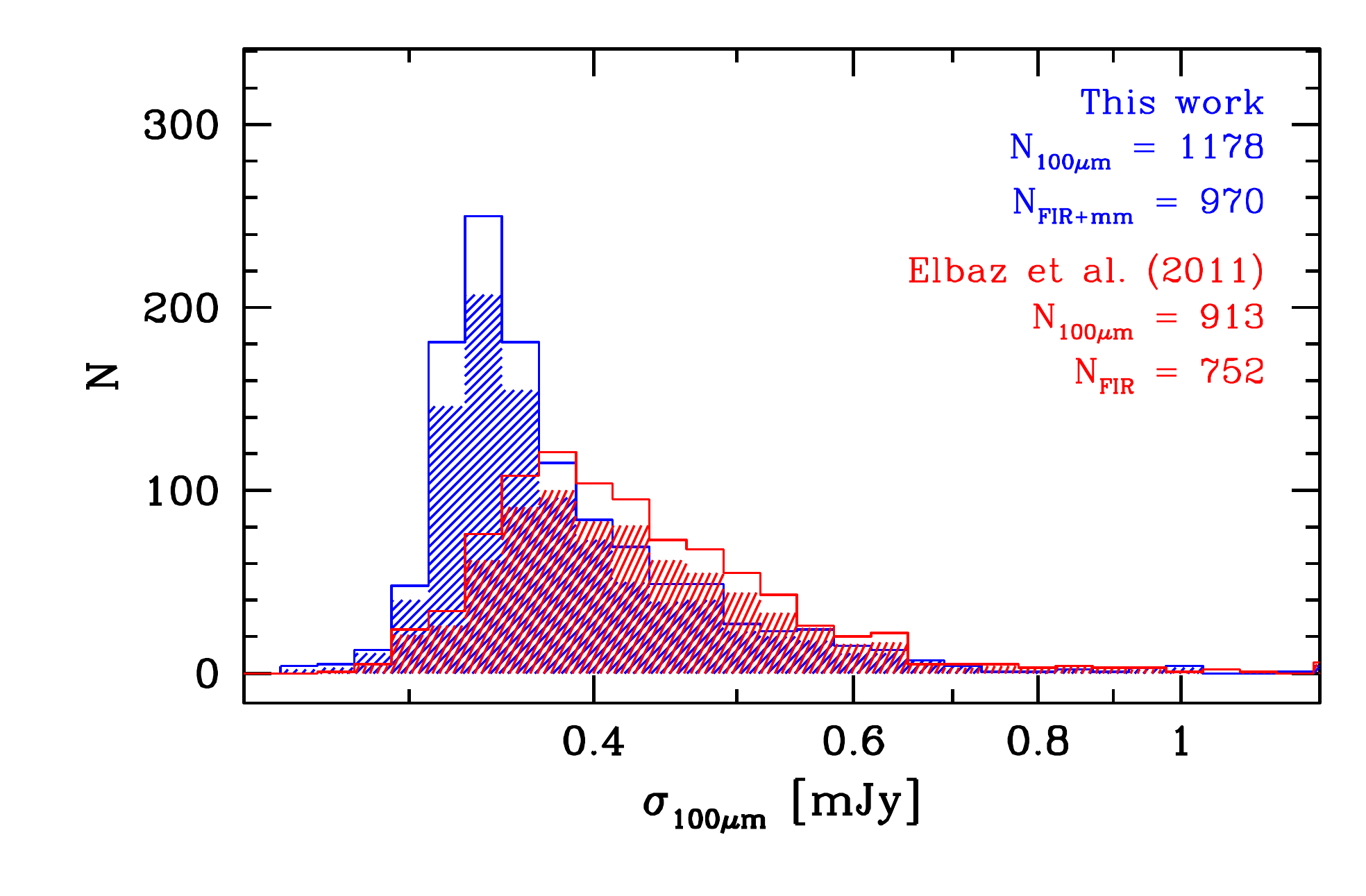}
\caption{%
    Comparison of histograms of the 100~$\mu$m flux densities (left) and flux density uncertainties (right) between this work and that of \citet{Elbaz2011}. Colors and histograms are defined in Fig.~\ref{compare_f24_histogram_dzliu_daddi_elbaz}. 
    In the left panel, this work has about 200 more sources and span a slightly deeper range of flux densities. 
    In the right panel, after applied the simulation-based flux uncertainty correction (see Section~\ref{Section_Simulation_Correction_df_corr}), the uncertainties in this work are lower than those in \citet{Elbaz2011} catalog, both because of the larger number of fainter sources in the left panel and the simulation-based corrections to the flux density uncertainties. 
    \label{compare_f100_histogram_dzliu_daddi_elbaz}%
}
\end{figure*}

\begin{figure*}
	\centering
	\includegraphics[width=0.45\textwidth, trim=0 10mm 0 8mm]{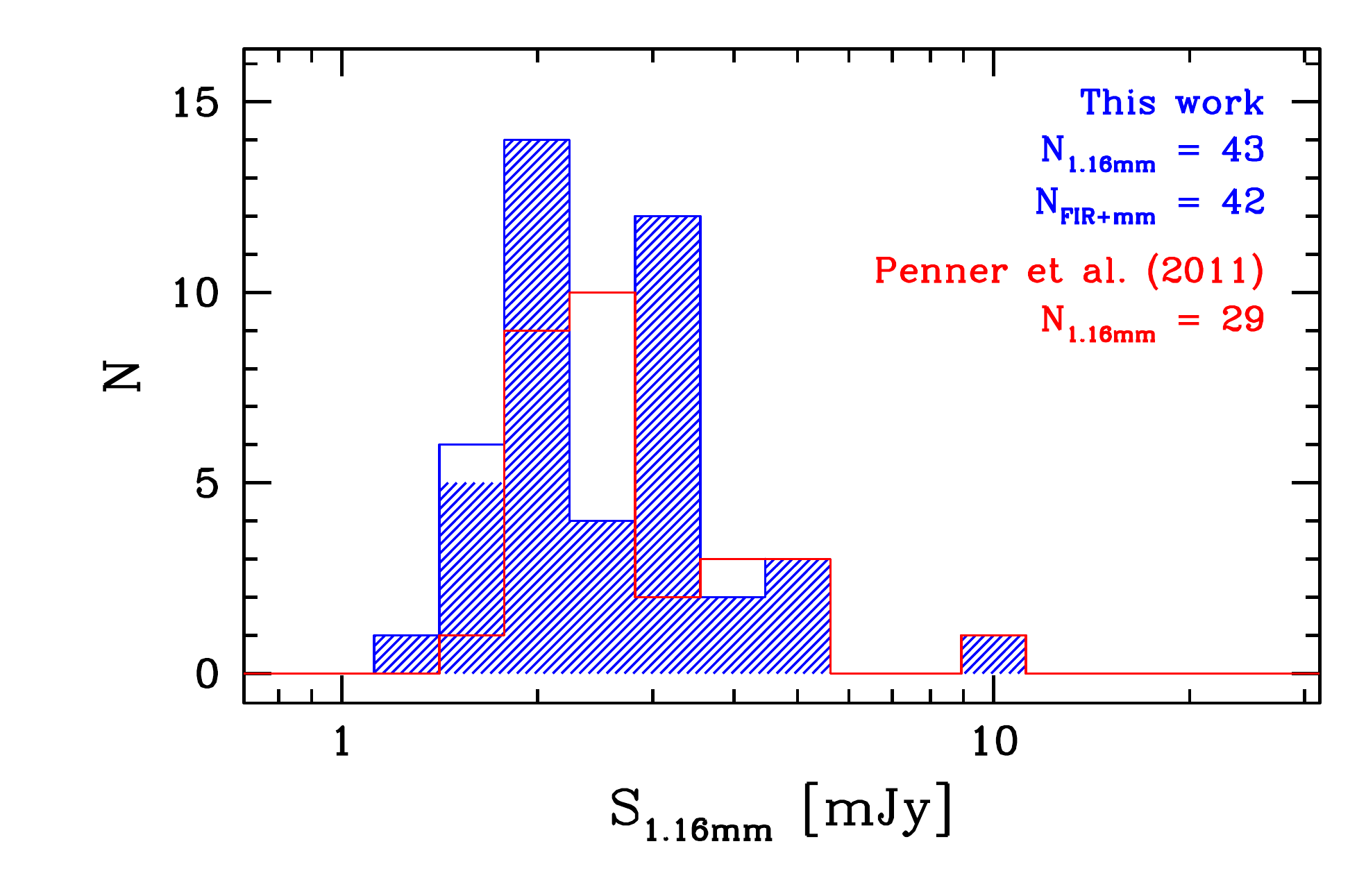}
	\includegraphics[width=0.45\textwidth, trim=0 10mm 0 8mm]{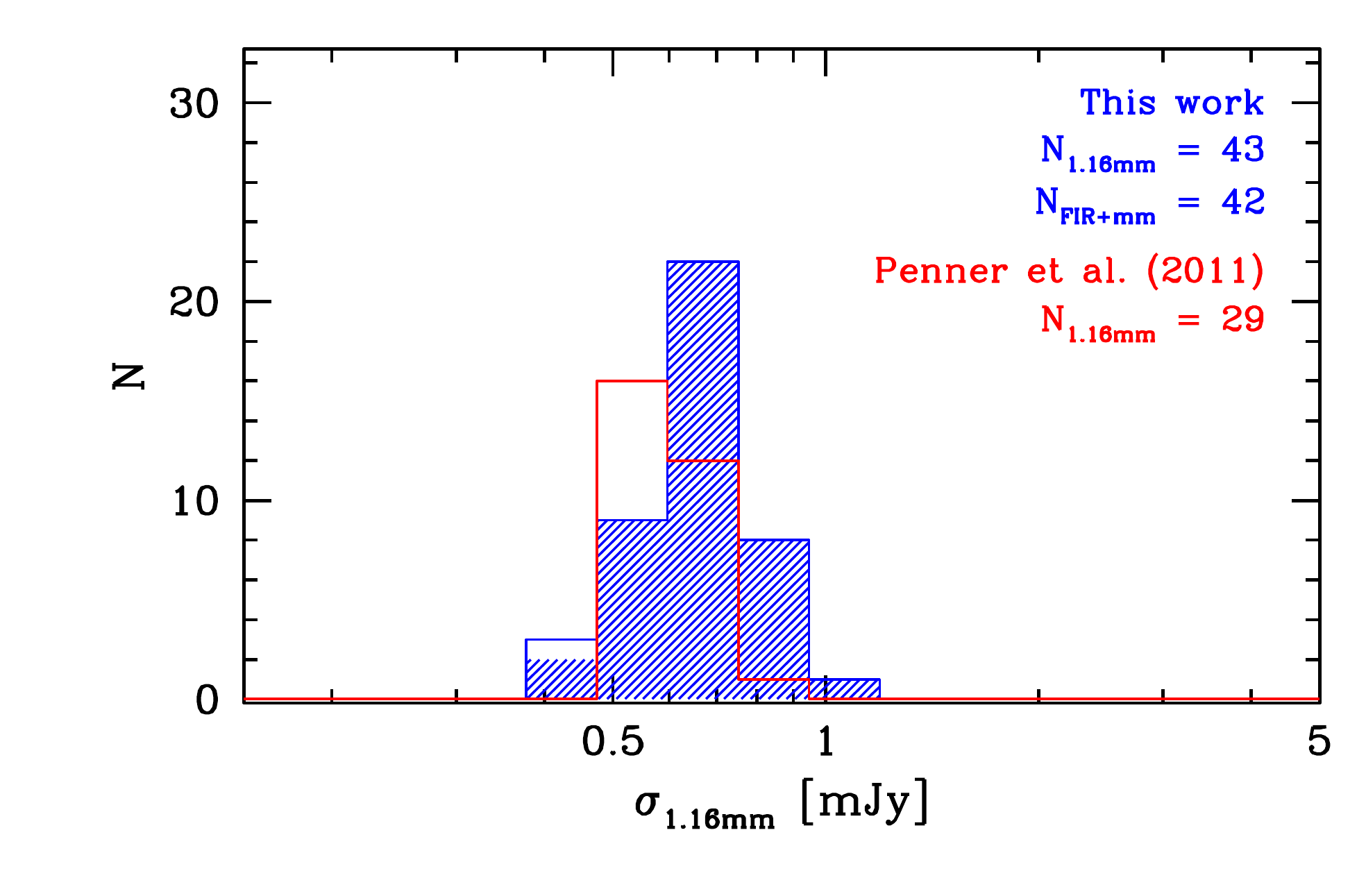}
	\caption{%
		Comparison of histograms of the 1.16~mm flux densities (left) and flux density uncertainties (right) between this work and that of \citet{Penner2011}. Colors and histograms are defined in Fig.~\ref{compare_f24_histogram_dzliu_daddi_elbaz}. The flux densities from \citet{Penner2011} are deboosted values when available. Most 1.16~mm-detected sources are also FIR+mm-detected sources. 
		In the left panel, the present work reports significant detections for about 10 more sources %
		over a similar range in flux density.
		In the right panel, after applying the simulation-based correction (see Section~\ref{Section_Simulation_Correction_df_corr}), the flux density uncertainties in this work are slightly larger than those in the \citet{Penner2011} catalog. 
		$N_{1.16\,\mathrm{mm}}$ is the number of $\mathrm{S/N}_{1.16\,\mathrm{mm}}\ge3$ sources in each catalog, and $N_{\mathrm{FIR+SMM}}$ is the number of sources with PACS 100~$\mu$m to 1.16~mm $\mathrm{S/N}\ge5$ and $\mathrm{S/N}_{1.16\,\mathrm{mm}}\ge3$. 
		\label{compare_f1160_histogram_dzliu_penner}%
	}
\end{figure*}

When all photometric measurements have been derived for all bands, we run a final pass of global SED fitting from which we derive various galaxy properties. While a detailed discussion of extraction of physical parameters and their uncertainties from FIR galaxy SEDs (which would include AGN components, dust temperature, IR8, radio excess, etc.) is beyond the scope of this work,  for the sake of the scientific analysis presented in this paper (see Section~\ref{Section_Final_Catalog_Highz}) it is worth mentioning the derivation of the best fitting $L_\mathrm{IR}$, and also SED-driven photometric redshifts. The latter are in most cases a simple refinements of optical/near-IR photometric redshifts, corresponding to the redshift within the allowed range (close to the previous optical/near-IR photometric redshifts) where the best fit to the FIR+mm SED 
is obtained. 
However for objects with no previous optical/near-IR photometric redshift, we fit over the whole range $0<z<8$, and the SED fitting provides an estimate of their photometric redshift and uncertainty. In a future publication we will investigate the possible derivation of FIR/mm/radio photometric redshifts and their performance, improving over the work of \cite{Daddi2009GN20} and several others. 

The overall flow chart of our \superdb{} method is presented in Fig.~\ref{Figure_FlowChartDiagram}.

\vspace{1truecm}

%
%

\section{Quality checks and known limitations}
\label{Section_Quality_checks}


\subsection{Comparing SED predictions with measurements}
\label{Section_Comparing_SED_Predictions}

An important step in our procedure is to predict fluxes for sources in a given band before actually performing measurements in that dataset. Sources with predicted fluxes (plus twice flux prediction uncertainty) brighter than $S_{\mathrm{cut}}$ are retained to be fitted with \galfit{}. 
For those fitted sources, we can now compare their measured fluxes to the SED predictions in order to investigate the presence of any potential bias in those predictions (and by extension, also the predictions for the faint/excluded/subtracted sources). 


These comparisons are shown in Fig.~\ref{Fig_Galsed_SED_flux_vs_Galfit_flux}. In the left panels we compare $S_{\mathrm{predict}}$ and $S_{\mathit{galfit}}$, and in the right panels we show the difference $(S_{\mathit{galfit}}-S_{\mathrm{predict}})$ divided by the total uncertainty, 
plotted versus $S_{\mathrm{predict}}$. The error bars represent the combined uncertainties of ${\sigma}_{{\mathit{galfit}}}$ and ${\sigma}_{{\mathrm{predict}}}$. 

$S_{\mathrm{predict}}$ and $S_{\mathit{galfit}}$ generally agree well with each other. There is a small fraction of sources with $S_{\mathrm{predict}}$ lower than $S_{\mathit{galfit}}$ at the faint end, as seen in the left panels (but less obvious in the right panels when actual flux uncertainties are taken into account). This small bias is driven by the requirement to have a significant \galfit{} detection in order to be shown in the plot, and vanishes when this selection bias is taken into account.

\subsection{Goodness of SED fitting and measurement errors}
\label{chi2}

\REVISED{
Another important overall check of our procedure is evaluating the goodness of fit of the global SEDs that we derived with our adopted template libraries, to verify at the same time that our templates are appropriate and that the error bars are sensible. We concentrate here on the FIR+mm range between 100--1100$\mu$m which is the range most affected by blending that we are interested to test, and also because mid-IR and radio deviations from our SEDs might be expected because galaxies can be more complicated than the models (AGN components with different shapes/slopes, strength of PAHs, etc). 
As we have already shown with examples of SEDs (see e.g., Fig.~\ref{Plot_SED_GN20}), the reduced $\chi^2$ values are generally reasonable and close to unity. 
We verified this aspect more systematically for the whole sample, considering the overall IR detection $\mathrm{S/N}$ ratio (see Eq.~\ref{Equation_SNR} in Section~\ref{Section_Final_Catalog_Highz}). For the 
518
galaxies with $\mathrm{S/N}_{\rm FIR+mm}>10$ we find a median $\chi_r^2=1.0$, while a median $\chi_r^2=0.8$ is found for the $5<\mathrm{S/N}_{\rm FIR+mm}<10$ sources. 
About 2\% of all galaxies in either sample 
cannot be fit by our models, with $\chi_r^2>4$. 
In most cases this appears to be due to an erroneous or problematic redshift (photometric or spectroscopic), either due to an error, or because of blending/superposition between foreground and background objects, either due to gravitational lensing or simply line-of-sight alignment. In some cases it is quite clear that only part of the FIR/mm emission is due to the galaxy under consideration as the IRAC prior, and that an additional component comes from a higher redshift object that contributes additional flux at longer wavelengths. Many of these could be scientifically interesting cases (e.g., lenses or the high redshift background galaxies). 
}

\begin{figure*}
	\centering
	\includegraphics[width=0.40\textwidth]{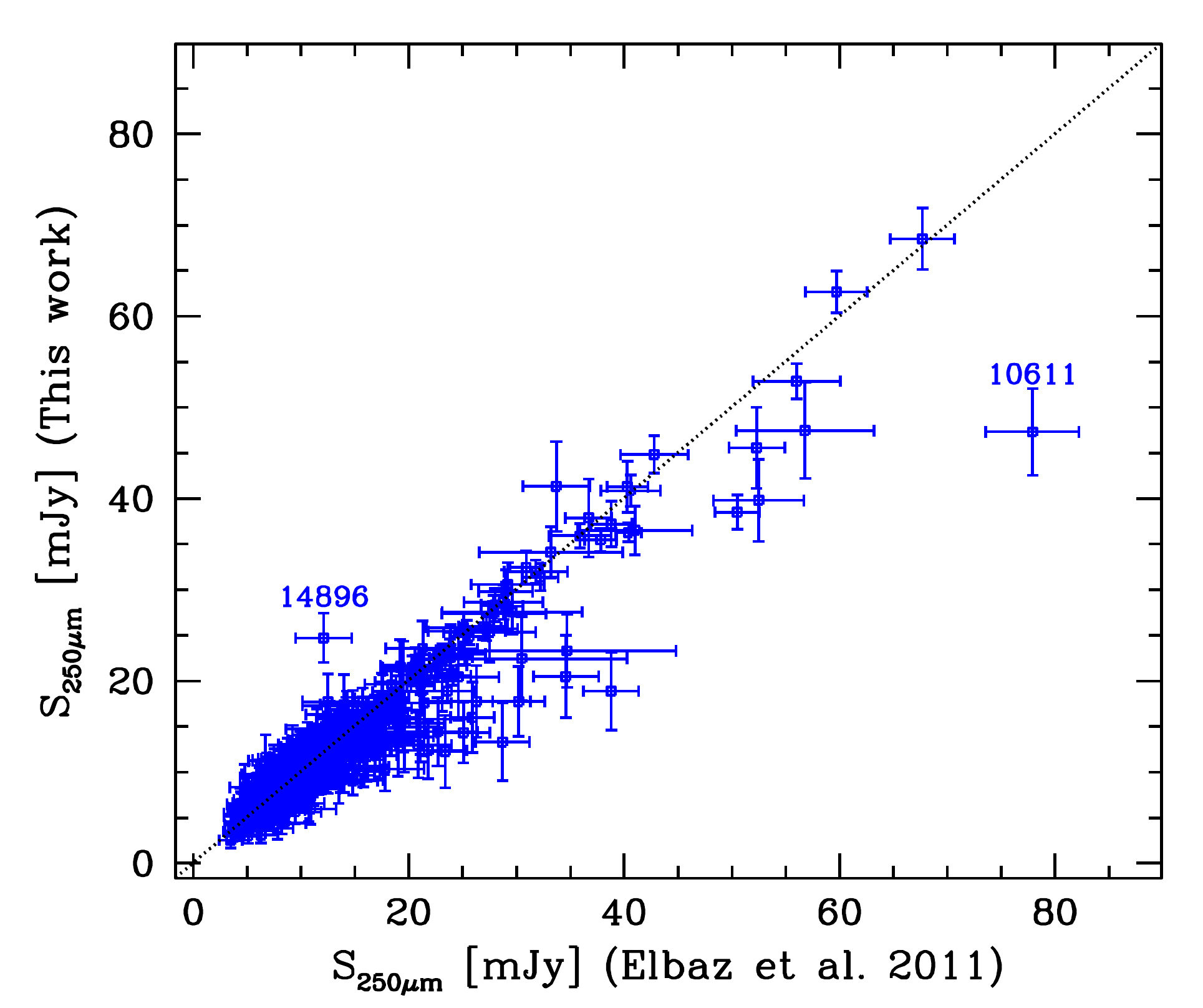}
	\includegraphics[width=0.40\textwidth]{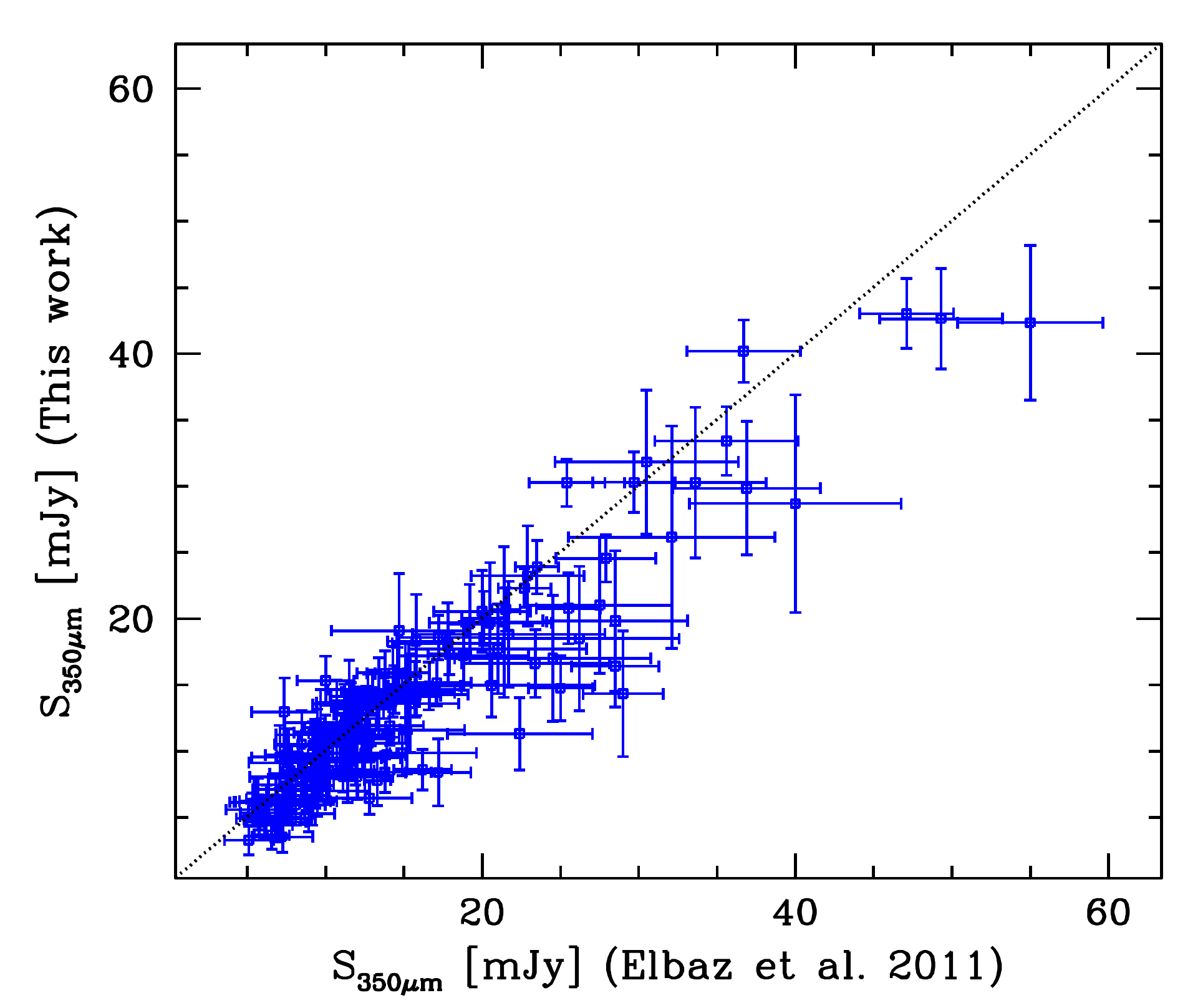}
	\includegraphics[width=0.40\textwidth]{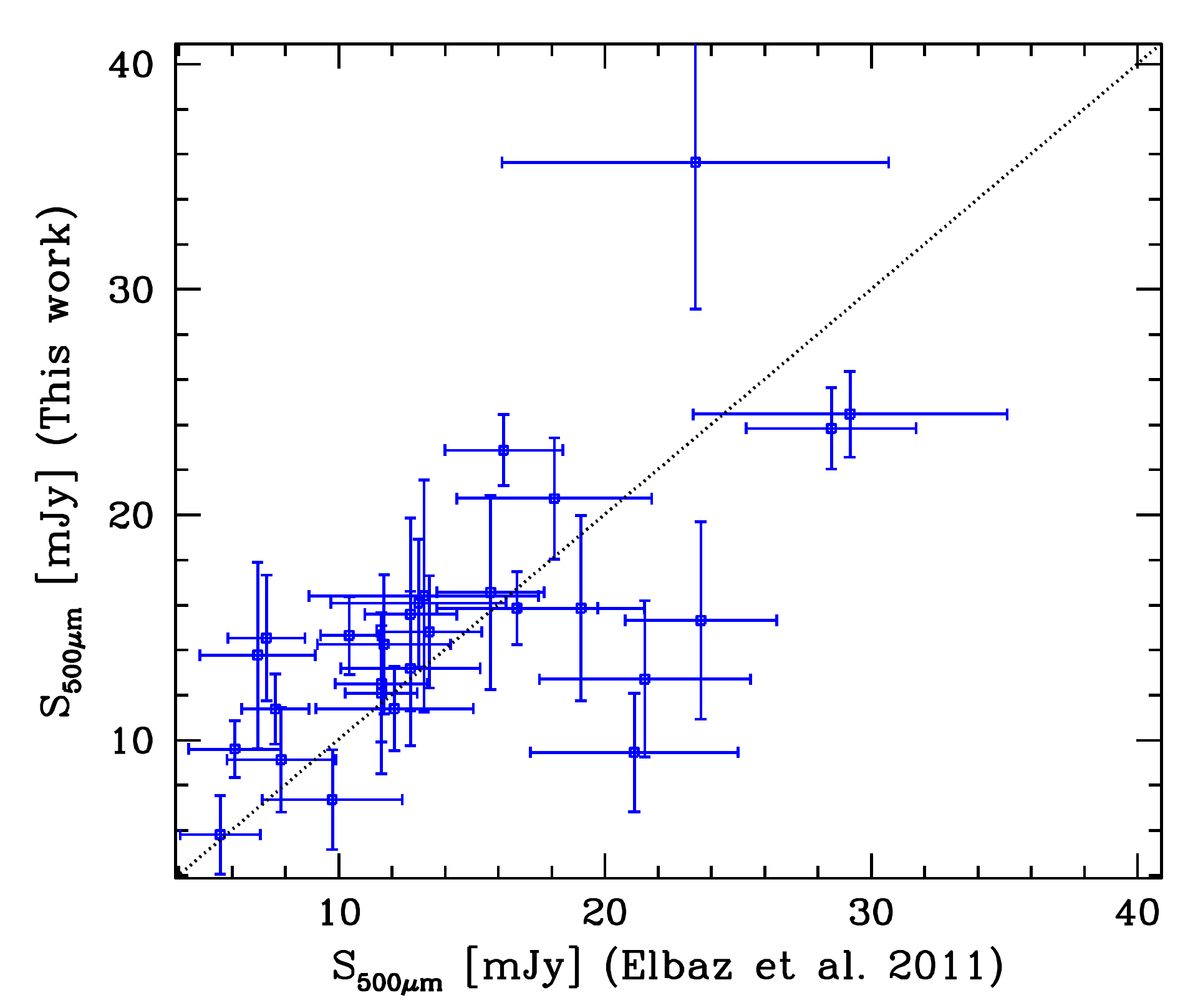}
	\includegraphics[width=0.40\textwidth]{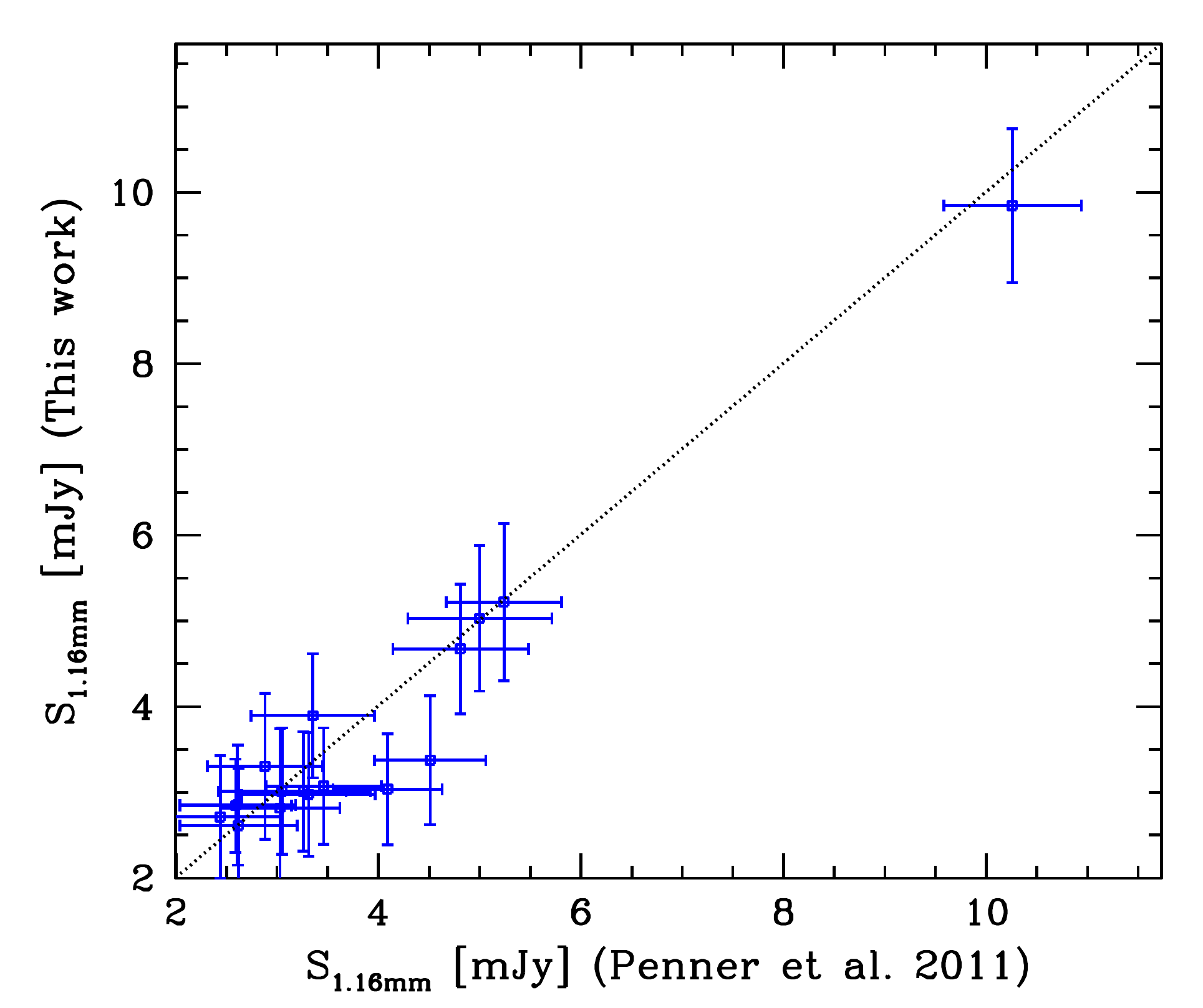}
	\caption{%
		Comparisons between flux density measurements in this work and literature catalogs at 250, 350, 500~$\mu$m and 1.16~mm. Literature fluxes for the three SPIRE bands are taken from \citet{Elbaz2011}, and the literature 1.16~mm flux densities are from \citet{Penner2011}. The figures show flux comparisons for sources detected in both our catalog and in previous work and cross-matched based on position.  \citet{Elbaz2011} used PACS-detected sources as priors for SPIRE photometry, while in this work we use SED fitting to the the near-infrared to PACS, SPIRE and radio photometry to determine the prior sources for SPIRE flux extraction, and we subtract the faint source flux contribution from the observed data.  Therefore our flux density measurement tend to be smaller than the literature values especially when a source is blended with neighbors.  Two most extreme outliers are marked with their IRAC ID in the first panel. These are ID~14896 (\citet{Elbaz2011} ID~2132) and ID~10611 (\citet{Elbaz2011} ID~1512). We discuss ID~14896 in detail in Section~\ref{Section_Compare_Measurements} and the next figures. Although the offset for ID~10611 has the opposite sense, we have checked this source has a very similar blending issue. 
		\label{Plot_f_compare_full_panels_v20160426}%
	}
\end{figure*}

\begin{figure*}
	\centering
	\includegraphics[width=0.32\textwidth, trim=0 2mm 0 0]{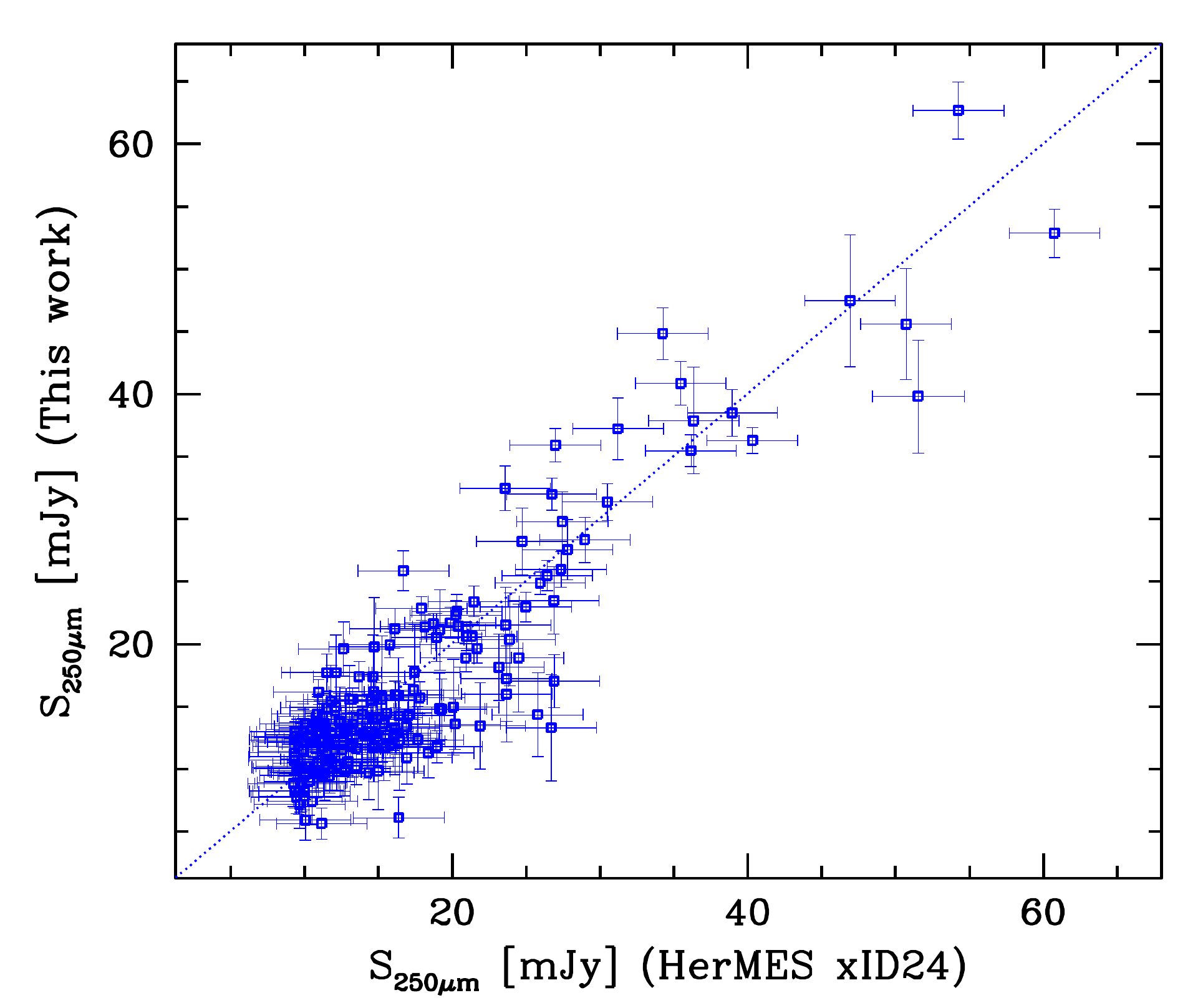}
	\includegraphics[width=0.32\textwidth, trim=0 2mm 0 0]{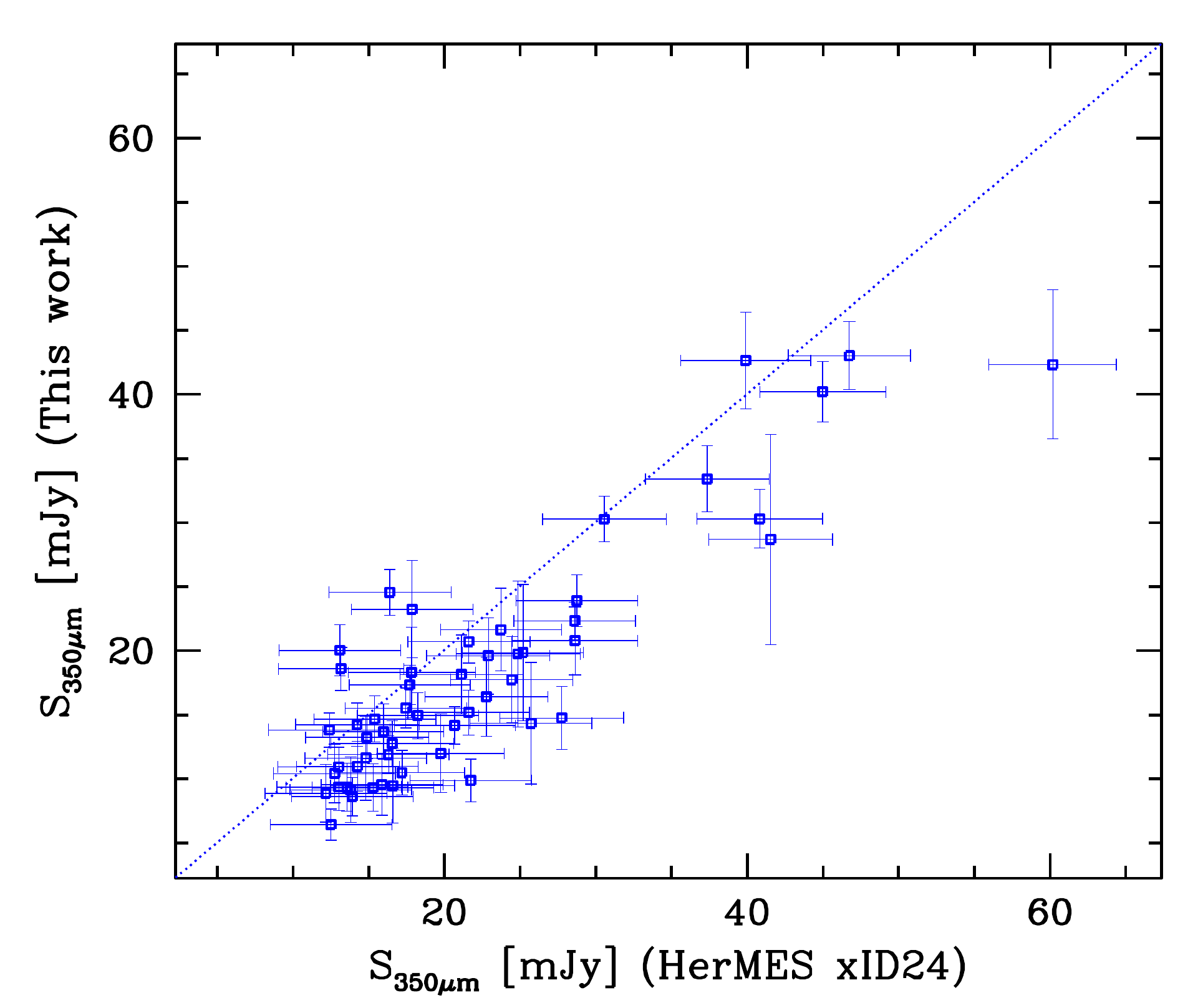}
	\includegraphics[width=0.32\textwidth, trim=0 2mm 0 0]{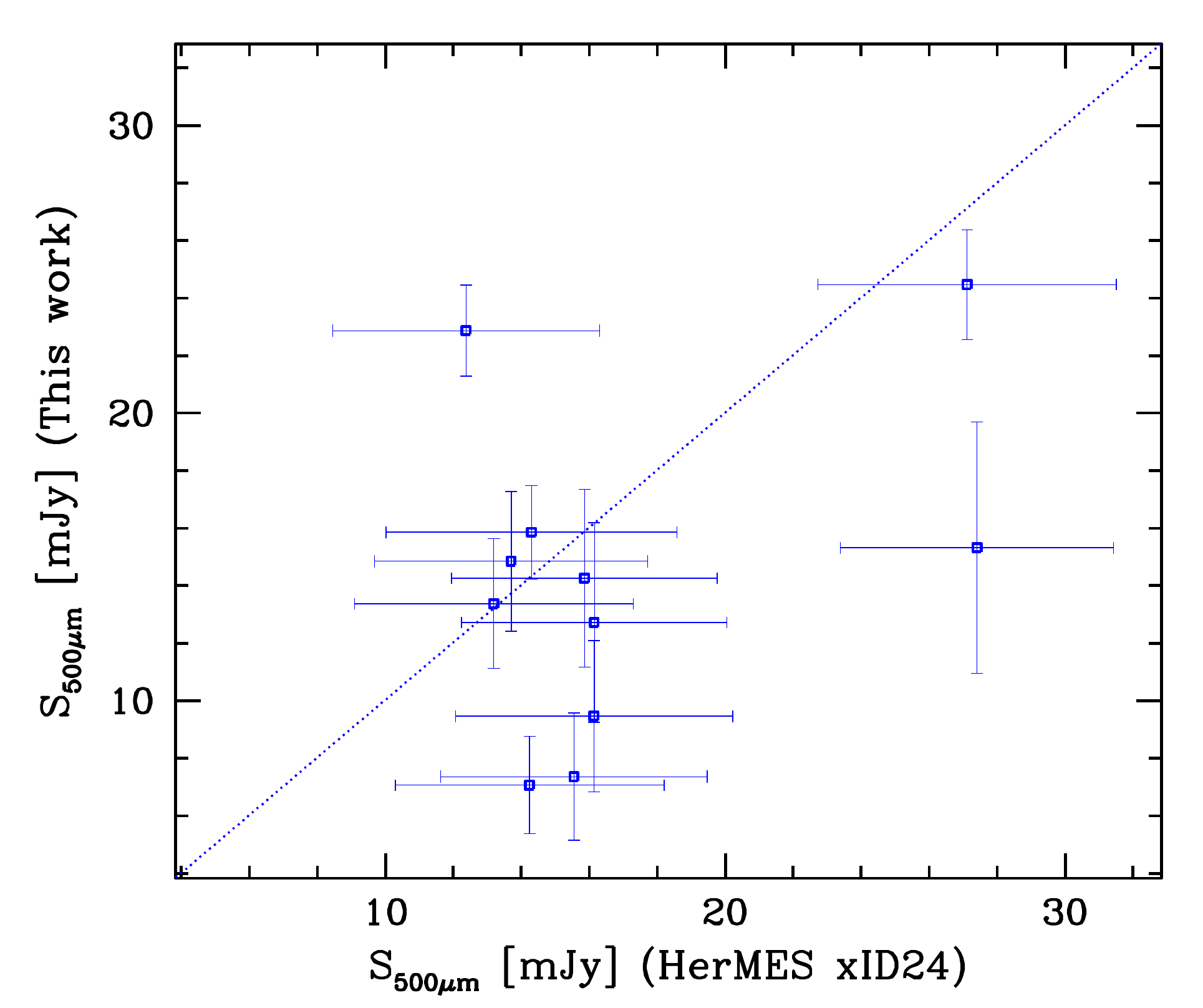}
	\caption{%
		\REREVISED{%
		Flux density measurement comparisons for sources with $\mathrm{S/N} \ge 3$ in both this work and the HerMES catalog (DR3; \citealt{Roseboom2010}). The three panels show SPIRE 250, 350 and 500~$\mu$m measurements, respectively. 
		} 
		\label{Plot_f_compare_HerMES}%
	}
\end{figure*}

\begin{figure}[ht]
\centering
\includegraphics[width=0.48\textwidth]{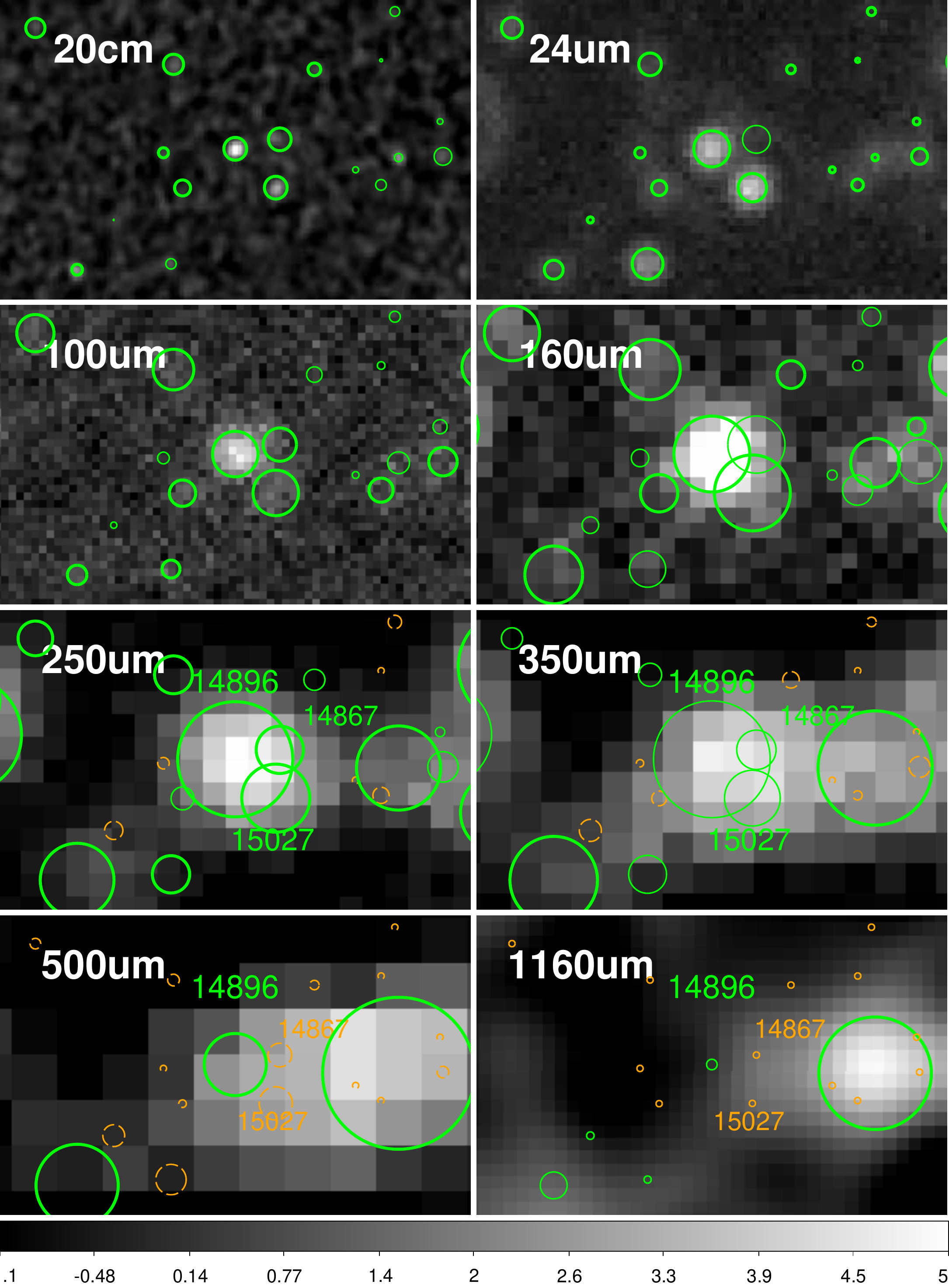}
\caption{%
	Multi-wavelength cutouts around the outlier ID~14896, which has a significant lower flux in our catalog than in that of \citet{Elbaz2011}. The discrepancy arises because \citet{Elbaz2011} assign more flux to ID 15027 than in our catalog. Each circle represents a 24+radio prior source (Section~\ref{Section_Initial_IRAC_Catalog}).  As in Fig.~\ref{Figure_Cutouts_GN20}, green circles are the fitted prior sources, while orange circles are the excluded ones. The circle sizes scale with flux (see caption of Fig.~\ref{Figure_Cutouts_GN20}). 
	The SEDs of the three heavily blended sources, ID~14896, ID~14867 and ID~15027, are shown in Fig.~\ref{Plot_SED_ID14896}. 
	%
	\label{Figure_Cutouts_ID14896}
}
\end{figure}

\begin{figure}[ht]
\centering
\includegraphics[width=0.33\textwidth, trim=0 12mm 0 0, clip]{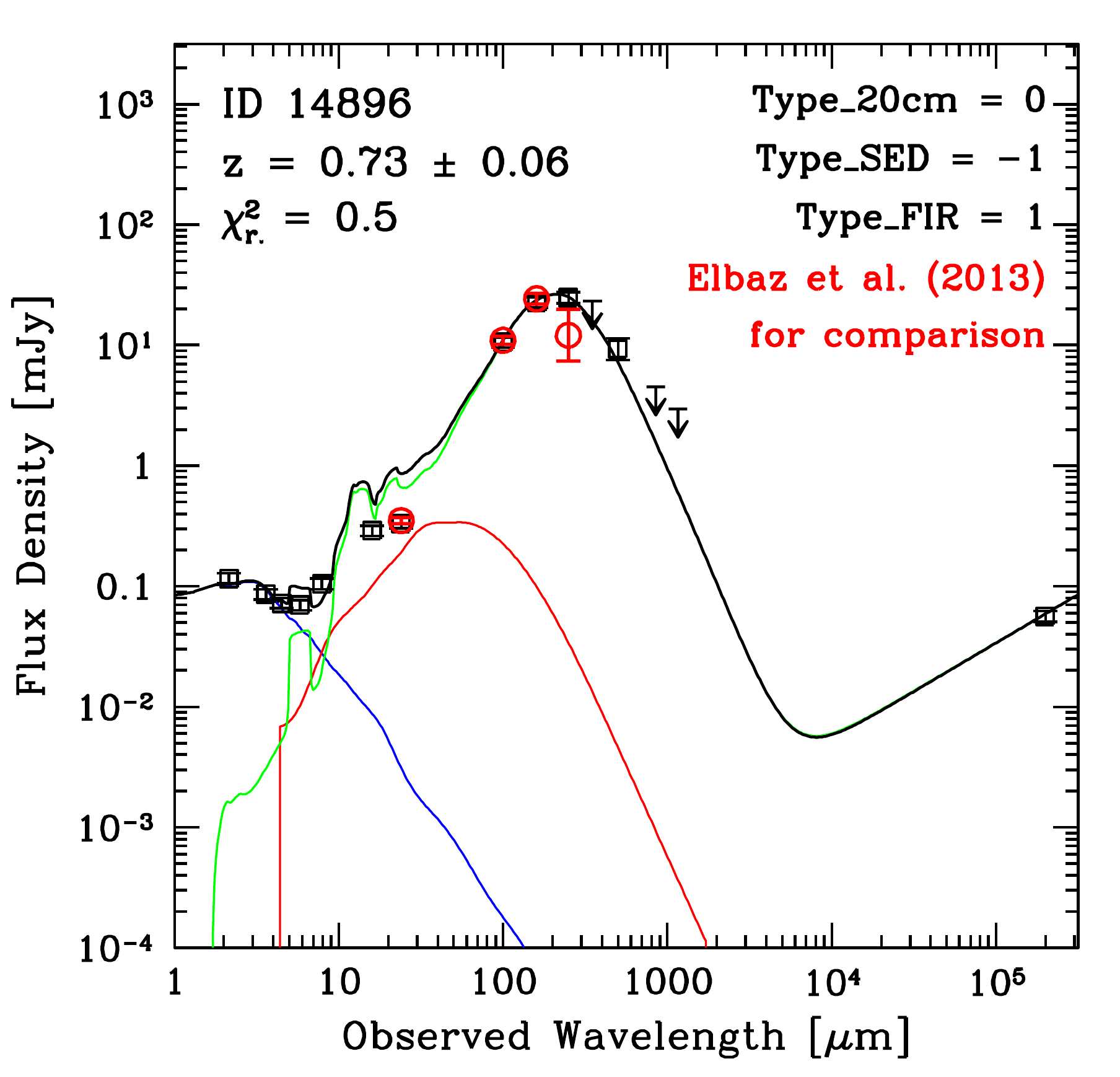}
\includegraphics[width=0.33\textwidth, trim=0 12mm 0 6mm, clip]{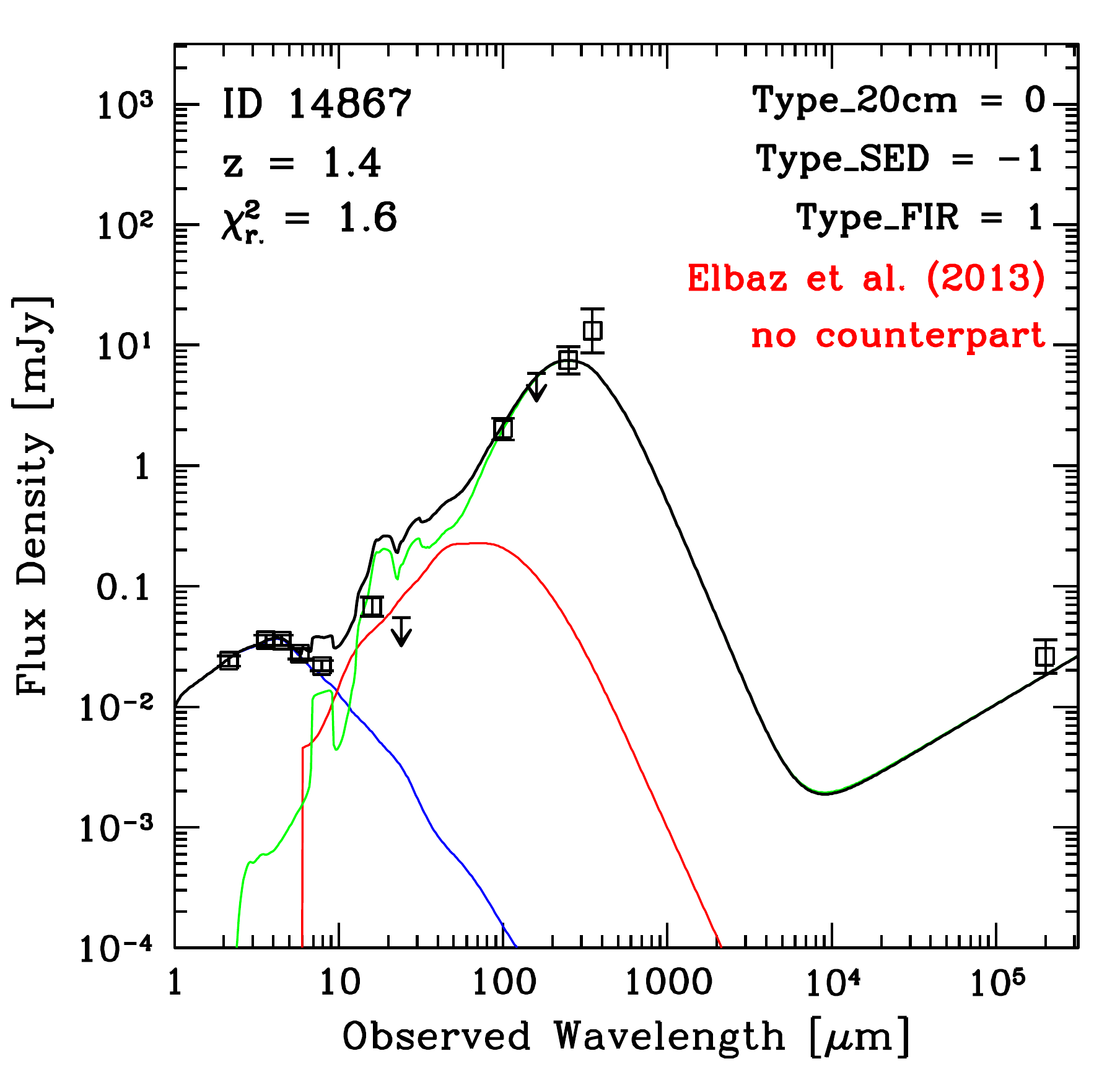}
\includegraphics[width=0.33\textwidth, trim=0 0mm 0 6mm, clip]{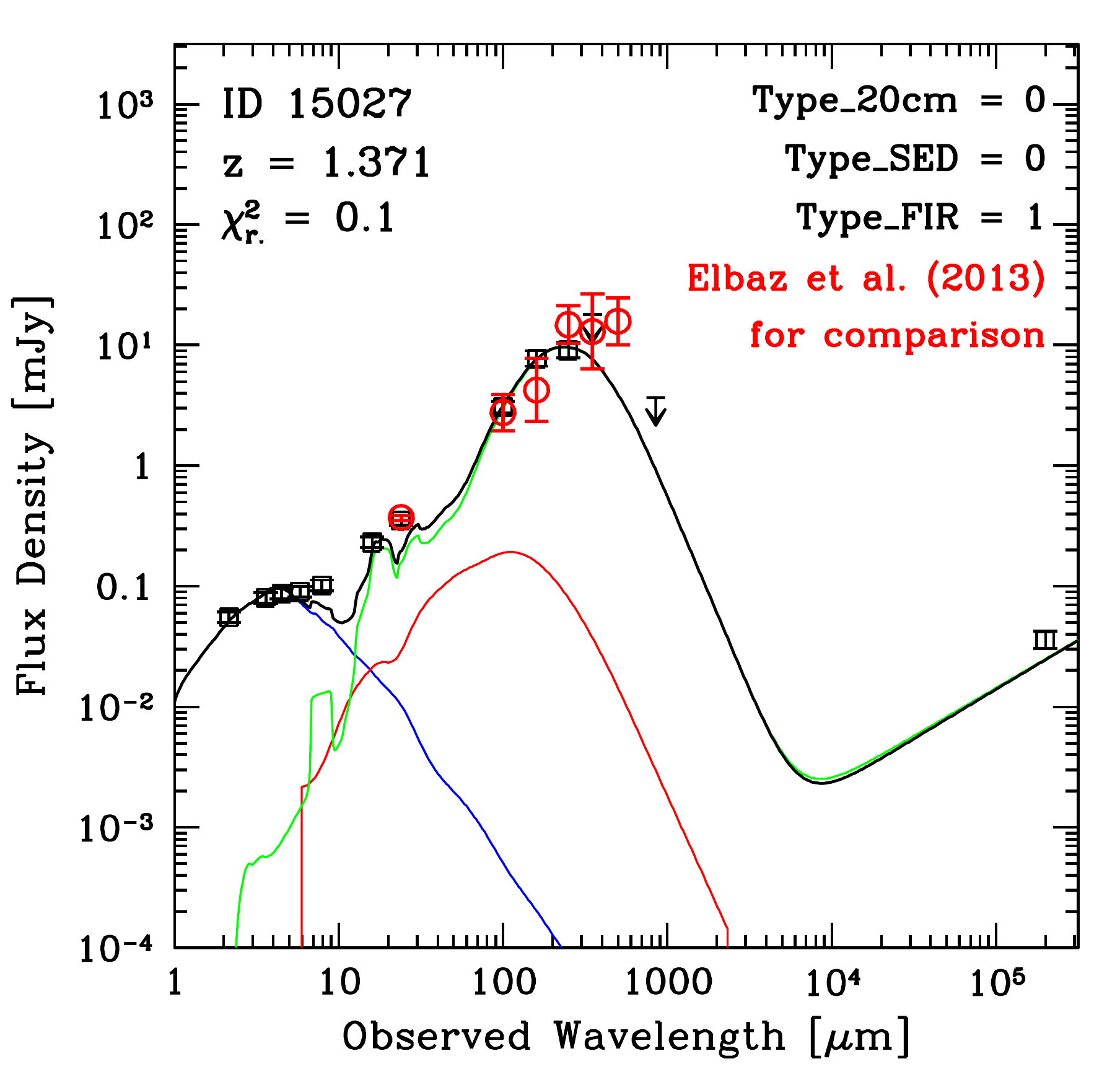}
\caption{%
	The SEDs of 3 sources highlighted in Fig.~\ref{Figure_Cutouts_ID14896}, to help explain the discrepancy between our photometry and that of \citet{Elbaz2011}. In the panels for ID 14896 and ID 15027 panels, we show red data points from \citet{Elbaz2011} for comparison. ID 14867 has no counterpart in \citet{Elbaz2011}. The best-fit redshift and \REVISED{reduced} $\chi^2$ are shown in the upper left corner of each panel. When the source has a spectroscopic redshift, we fix the redshift to that value, and no error is quoted. ID~14867 and 15027 both have spectroscopic redshifts, and our deblended photometry for the latter appears to agree better with the model SEDs than that of \citet{Elbaz2011}. 
	More details of the SED fitting are given in Section~\ref{Section_SED_Fitting}. See also the caption to Fig.~\ref{Plot_SED_GN20}. 
	\label{Plot_SED_ID14896}
} 
\end{figure}


\subsection{Comparison to literature results}
\label{Section_Compare_Measurements}

Here we compare our 24~$\mu$m, PACS and SPIRE flux measurements and uncertainties with those of \cite{Elbaz2011} \REVISED[]{and HerMES \citep[DR3;][]{Roseboom2010}\footnote{\url{http://hedam.lam.fr/HerMES/}}}, and our 1.16~mm \REVISED[]{catalog} with that of \citet{Penner2011}. 

In Fig.~\ref{compare_f24_histogram_dzliu_daddi_elbaz}, we compare our 24~$\mu$m flux and uncertainty histograms with the \citet{Elbaz2011} catalog. In the left panel, the flux histogram of $\mathrm{S/N}_{24\,{\mu}\mathrm{m}}\ge3$ sources in this work is shown in blue, and that of \citet{Elbaz2011} in red. The shaded area under each histogram indicates the sources with FIR+mm multi-band combined $\mathrm{S/N}_{FIR+mm}\ge5$ and $\mathrm{S/N}_{24\,{\mu}\mathrm{m}}\ge3$, where $\mathrm{S/N}_{FIR+mm}$ is defined as per Equation~\ref{Equation_SNR} in Section~\ref{Section_z_SFR}. \citet{Elbaz2011} do not include (sub-)mm bands, i.e., for their sources the combined $\mathrm{S/N}$ only involves 100~$\mu$m to 500~$\mu$m photometry. For the bright sources, the flux histograms from both works are consistent, while for fainter sources, the flux histogram from this work shows more prior sources. We have about 450 more 24~$\mu$m-detected candidate prior sources for the longer wavelength photometry than those used in \citet{Elbaz2011}. The shaded area, or the FIR to mm and 24~$\mu$m detected sources, also have a larger number of detections in our catalog. The newly detected sources are fainter. Notice how IR-detected sources fall off very rapidly below 100~$\mu$Jy, suggesting that our prior sample is quite complete and possibly redundant for \textit{Herschel} sources at typical redshifts (although we might still be somewhat incomplete for the highest redshifts, as previously noted in Section~\ref{Section_The_24_Radio_Catalog}). 
The right panel shows flux uncertainties, corrected based on simulations (see Section~\ref{Section_Simulation_Correction_df_corr}). They are generally higher than those in \citet{Elbaz2011}, but should have less bias and be more representative of the true uncertainties in the measurements (see Section~\ref{Section_Simulation_Correction_df_corr} and Section~\ref{Section_Simulation_Performance}). For the FIR and 24~$\mu$m-detected sources in the shaded area, our flux uncertainties are also generally higher. 

In Fig.~\ref{compare_f100_histogram_dzliu_daddi_elbaz}, we compare the 100~$\mu$m flux and uncertainty histograms from this work to those from the catalog of \citet{Elbaz2011}. In the left panel, our work has about 200 more sources and extends to a slightly \MINORREREVISED[deeper flux]{fainter flux} range. 
In the right panel, after applying the simulation-based flux uncertainty correction (see Section~\ref{Section_Simulation_Correction_df_corr}), our flux uncertainties are lower than those in \citet{Elbaz2011}.  This is due both to the greater number of fainter sources in the left panel, and to our simulation-based corrections to the flux uncertainties. 

Other PACS and SPIRE bands have similar histograms to that at 100$\,\mu$m and hence are not shown here. The comparisons of our 1.16~mm flux measurements and uncertainties with those of \citet{Penner2011} are shown in Fig.~\ref{compare_f1160_histogram_dzliu_penner}. We use their deboosted fluxes when possible. The number of detected sources is small at this wavelength, but our catalog has 
13
more detections. The sources span a similar range of flux in both works. After applying the simulation-based flux uncertainty corrections (see Section~\ref{Section_Simulation_Correction_df_corr}), the flux uncertainties in this work are slightly higher than those in \citet{Penner2011}. Similar to the case at 24~$\mu$m, we believe that the difference is mainly due to the fact that our corrections yield more reasonable flux uncertainties (based on the simulation results). 

The one-to-one flux comparisons for the three SPIRE bands and at 1.16~mm are presented in Fig.~\ref{Plot_f_compare_full_panels_v20160426} \REVISED{and Fig.~\ref{Plot_f_compare_HerMES}}. 
We cross-match common sources (within a limiting separation of \MINORREREVISED[5$''$ because of the large \textit{Herschel} PSFs]{1$''$}) and compare the fluxes \REVISED{from this work to those from \citet{Elbaz2011} or the HerMES catalog}. 
The majority of the sources show consistent fluxes, but some sources in the \citet{Elbaz2011} catalog have higher SPIRE fluxes, which is due to source blending. The two most extreme outliers are marked with their IRAC IDs in the first panel. They are ID~14896 \citep[ID~2132 in][]{Elbaz2011} and ID~10611 \citep[ID~1512 in][]{Elbaz2011}. ID~14896 is one of very few sources that have higher fluxes in our catalog than in that of \citet{Elbaz2011}. This is due to the complex blending situation; \citet{Elbaz2011} assign most of the flux to a nearby source but little to this object. 
Fig.~\ref{Figure_Cutouts_ID14896} shows the multi-band image cutouts around this source. The three sources: ID~14896, ID~14867 and ID~15027, become blended at 160~$\mu$m and longer wavelengths. Their SEDs are shown in Fig.~\ref{Plot_SED_ID14896}. For comparison, the red data points report values from \citet{Elbaz2011} for ID~14896 and ID~15027. ID~14867 has no counterpart in \citet{Elbaz2011}. 
In the third panel, ID~15027, with a relatively low spectroscopic redshift $z = 1.371$, is not likely to have SPIRE flux as bright as the red data points. In the first panel, the 20~cm and 500~$\mu$m fluxes of ID~14896 suggest that its 250~$\mu$m flux is not likely to be as low as that reported by \citet{Elbaz2011}. Thus in our opinion, for this case our deblending leads to more reasonable results. 
For the other extreme case mentioned above (ID~10611 and its surrounding sources), we have also checked their image cutouts and SEDs. The blending situation is very similar to that described above, and in the catalog of \citet{Elbaz2011} most of the SPIRE flux is attributed to ID~10611 (e.g., like ID~15027). Thus the detailed figures and SEDs are not shown here.

\REREVISED{
    Similar considerations apply to the HerMES catalogs (see Fig.~\ref{Plot_f_compare_HerMES}). In this case, we note that there are large differences with our measurements at 350 and 500$\mu$m, substantially larger than the uncertainties from both catalogs.
    Generally, we find that many sources in the HerMES catalog have their flux densities overestimated, likely because of unresolved blending. The number of single-band detected sources is almost doubled in our catalog compared to the HerMES catalog, for all SPIRE bands. The flux density uncertainties in HerMES catalog  (using confusion noise computed from the inverted covariance matrix and the PSF matrix, see \citealt{Roseboom2010,Roseboom2012}) tend to be larger than ours. 
}

\subsection{Effects of varying the prior density}

\begin{figure}[ht]
	\begin{center}
		\includegraphics[width=0.48\textwidth]{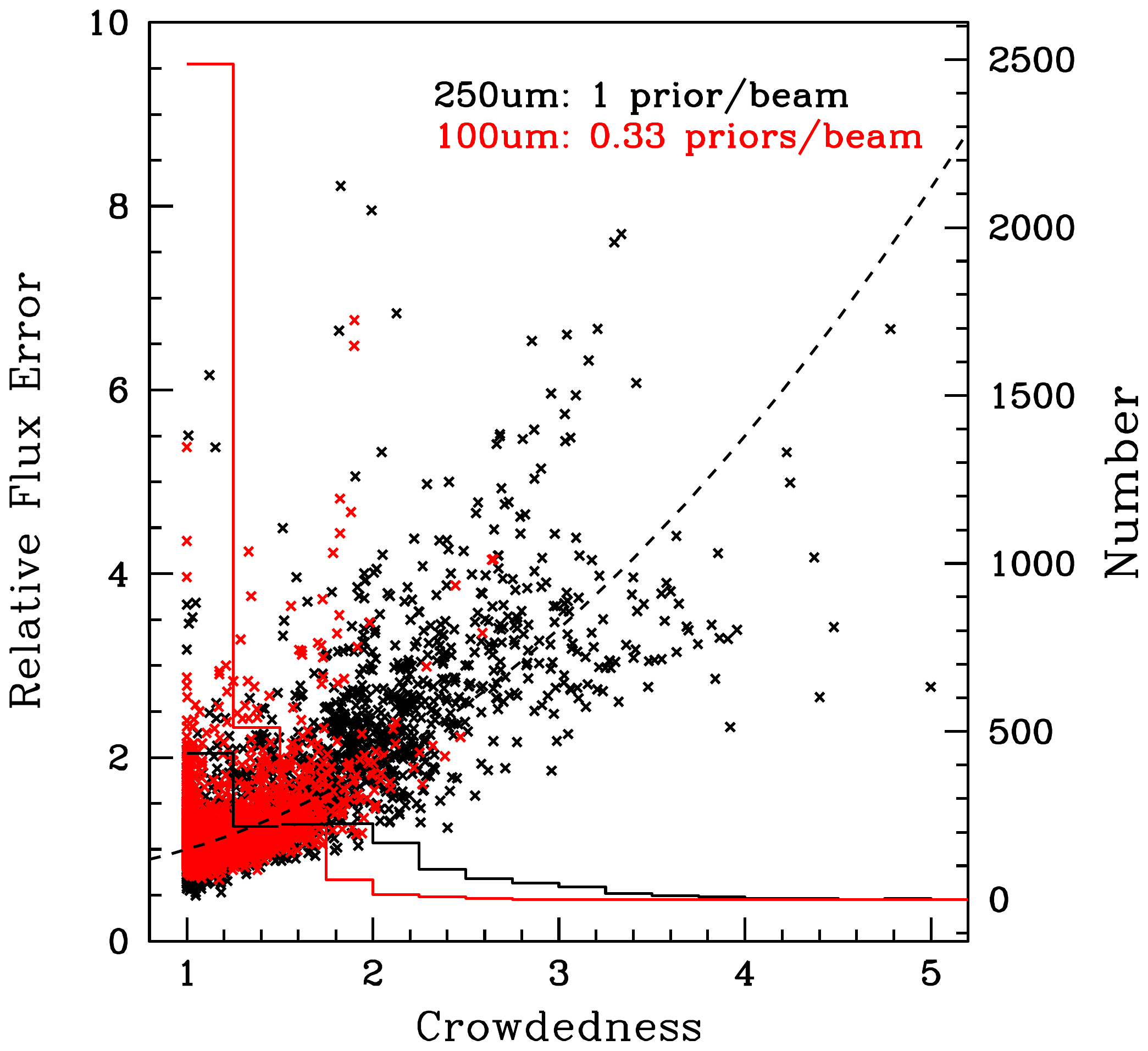}
		\caption{%
			Relative flux error as a function of \crowdedness{} (normalized for low values of \crowdedness), with values indicated on the left-Y axis. The results at 100~$\mu$m are shown in red, while those at 250~$\mu$m are in black. The dashed line is a reasonable fit to both data sets, scaling as
			(\crowdedness)$^{0.5}$. Histograms of \crowdedness{} values, with values shown on the right-Y axis. At 250~$\mu$m we are fitting a smaller number of priors that are in more crowded regions on average. 
			\label{Figure_priors}
		}
	\end{center}
\end{figure}

Our method attempts to use all available information on galaxies in the region of sky under consideration in order to define an ideal set of priors to fit FIR/(sub)mm data, using the smallest possible number of useful sources. Increasing the number of priors will increase photometric errors for all galaxies on average because of increased blending. This effect is exemplified in Fig.~\ref{Figure_priors}, where we show the normalized flux errors for the 100~$\mu$m and 250~$\mu$m bands as a function of the \crowdedness{} parameter. Despite these two bands having quite different numbers of priors per beam (0.3 and 1 respectively, see also Fig.~\ref{Fig_Galsed_Plot_Number_per_Beam}), the normalized behavior of the flux error versus \crowdedness{} is similar for the two bands. To first order, the relative flux error scales proportionally to \crowdedness{}, although a second order polynomial fit is a better description for values of \crowdedness$>2$. At 250~$\mu$m, as a result of the much larger beam, we are fitting a smaller absolute number of priors (see Fig.~\ref{Figure_priors} and Table~\ref{Table_1}), which on average have higher values of \crowdedness{}. The median value for the \crowdedness{} is 1.64 (1.05) at 250~$\mu$m (100~$\mu$m), and the median relative flux error is similarly larger 1.63 (vs.\ 1.02). The increase of the typical `noise' is the price to pay for fitting all reasonable priors. We note that this effect is not so dramatic, and errors are still quite acceptable, even if we are dealing with a density of priors approaching 1 per beam (in possible contradiction with results and discussions in \citealt{Scudder2016} and \citealt{Karim2013}).

\begin{figure}[ht]
	\begin{center}
		\includegraphics[width=0.42\textwidth, trim=0 0 -20mm 0]{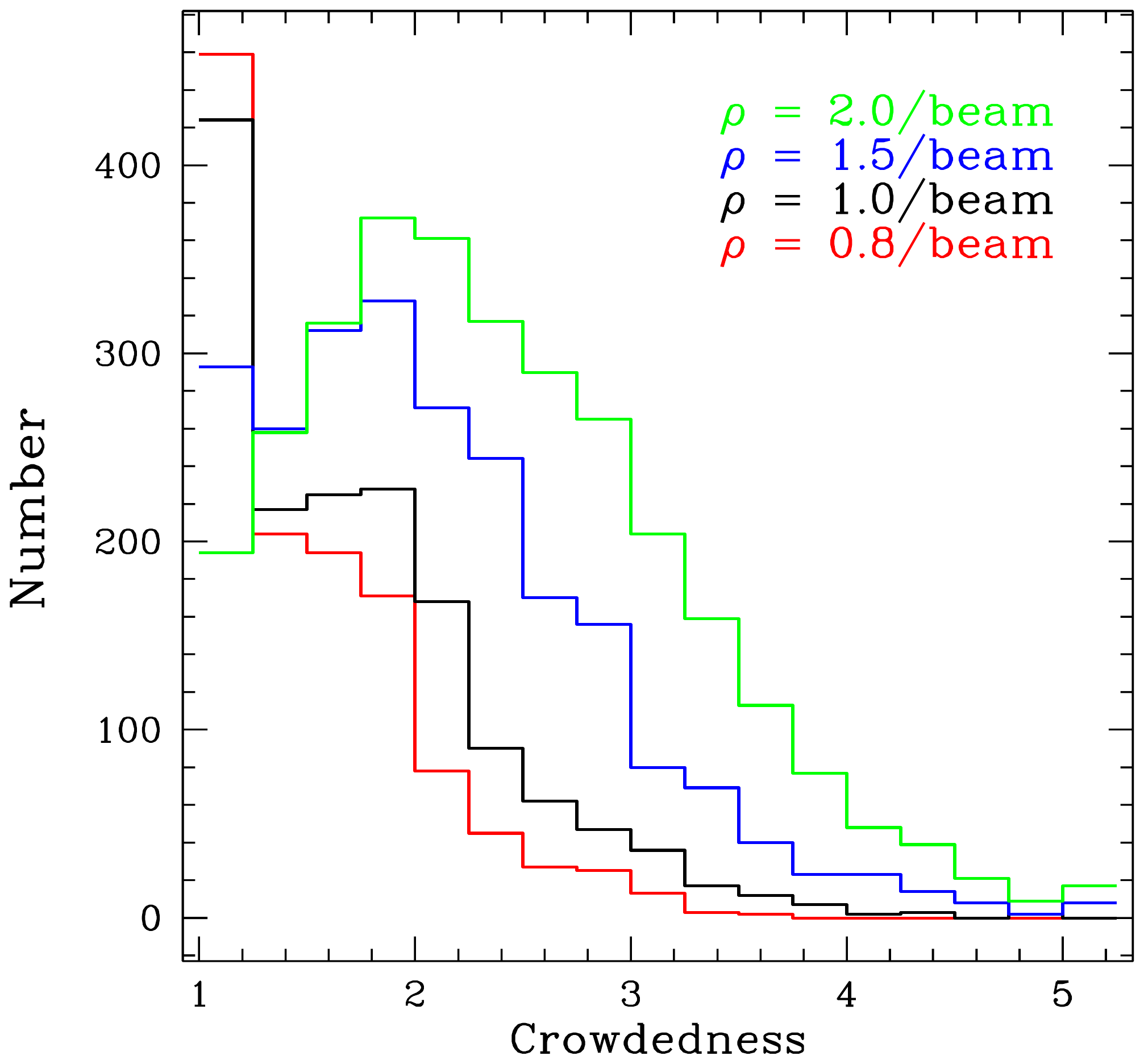}
		\caption{%
			Histograms of the number of priors versus \crowdedness{} for different choices of prior source density $\rho_{\mathrm{beam}}$ from 0.8 to 2.0, computed here for 250~$\mu$m and varying the choice of $S_{\mathrm{cut}}$. 
			\label{Figure_priorshist}
		}
	\end{center}
\end{figure}

Fig.~\ref{Fig_Galsed_cumulative_number_function} shows that, given the information in hand, it is not possible to reduce much further the number density of priors without failing to fit objects that could be potentially fairly bright in the FIR/mm imaging data.  For example, raising $S_{\mathrm{cut}}$ from 3 to 6$\sigma$ would reduce $\rho_{\mathrm{beam}}$ only from 1 to $\sim0.8$. On the other hand, if we were to reduce $S_{\mathrm{cut}}$ to 1.5$\sigma$ we would reach $\rho_{\mathrm{beam}}$ of about 2. It is interesting at this point to evaluate what the impact of these choices would be. Fig.~\ref{Figure_priorshist} shows the distribution of \crowdedness{} values for the different cases. Reducing the source density from 1.0 to 0.8 per beam has a fairly minor effect on the implied \crowdedness{}, and hence on the noise. Increasing the source density much beyond 1 has a much stronger effect: the fraction of sources in reasonable isolation, with \crowdedness~$<1.5$, drops from 42\% ($\rho_{\mathrm{beam}}=1.0$) to 15\% ($\rho_{\mathrm{beam}}=2.0$), while the sources with poor measurements due to blending with \crowdedness~$>2.5$ rise from 12\% ($\rho_{\mathrm{beam}}=1.0$) to 41\% ($\rho_{\mathrm{beam}}=2.0$). 
Therefore if we were to accept more sources for fitting, including some of those below $S_{\mathrm{cut}}=3\sigma$, this would result in a significantly increase in the noise and poorer performance for all objects. Staying within the limit $\rho_{\mathrm{beam}}\simlt 1$, as we have done for SPIRE in this paper, appears to be a reasonable choice.

\subsection{Robustness to source subtraction errors}

We are subtracting faint source models from the images before running the photometric analysis.
This procedure is prone to errors, as we might being over- (or under-)estimating the flux values to subtract. 
We expect that any such error would be automatically taken into account by our simulation-based approach. For example, if we were over-subtracting the fluxes of many faint sources, this would create, to first order, a negative (false) background, which we would detect and correct with our bias measurements, which in practice define the effective background in the data. To second order, the over-subtracted sources in this example, after zero-level determination, would result in increased background fluctuations and noise, because they are not likely to be spatially homogeneous, and they will have a variety of flux values. This extra noise would be reflected in our simulation-based calibration of uncertainties and thus taken into account in our error bars.

%
%
%
%

\subsection{Known limitations}

We are aware of a number of possible remaining problems that might still affect our flux measurements and their uncertainties.

One is the presence of additional sources (most often at high redshifts) which could significantly contribute flux in the data  but which are not in our prior list. Such objects could spuriously boost the inferred flux (and hence lead to underestimated errors) of other priors. Judging from our simulations, this effect 
does not seem to be significant, since the distribution of $(S_{\mathrm{in}}-S_{\mathrm{out}})/{\sigma}$ closely resembles a Gaussian with width equal to 1 (see Fig.~\ref{Fig_Galsim_df_corr_SPIRE} and Appendix~\ref{Section_Simulation_Performance}).

Also, a fraction of photometric redshifts or even spectroscopic redshifts could be incorrect. Experience from various surveys have shown that a few percent of galaxies may have substantially incorrect redshift measurements or estimates, even when using values classified as reliable, as we do here. 
Note that in order to accept a spectroscopic redshift from the literature, we require that it is in agreement within 10\% in $(1+z)$ with the photometric redshift, to avoid possible wrong redshifts.  
Using a wrong redshift can affect SED predictions.  To some extent, allowing flexibility in dust temperatures (or, more specifically, in $\left<U\right>$ values) moderates the impact because of the redshift-temperature degeneracy, especially if the redshifts are not catastrophically wrong. In this case, due to SED flux prediction \MINORREREVISED[errors]{uncertainties}, there could be a few objects that are retained for \galfit{} fitting instead of being dropped.  This would modestly increase \crowdedness{}, but this is probably a negligible effect.  Some objects might be erroneously dropped from fitting while still being bright enough for detection in the data. The latter cases could give rise to sources that would be detectable in the residual images 
%
if they are isolated. If they fall atop or near other priors, they may corrupt their photometry, biasing recovered fluxes high. 
Of course, in all cases, the impact of wrong redshifts would extend beyond simple photometric accuracy to the derivation of wrong bolometric $L_{\mathrm{IR}}$ and SFR, thus affecting the science. 

Finally, another limitation is that we are using \galfit{} with a constant background. Angular variations in the background would act as increased noise. Carefully accounting for background variations could improve photometric uncertainties, but is limited by our knowledge of PSF wings, and of course by the very large beams in the SPIRE and (sub)mm imaging. We attempted a correction of background variations at 24~$\mu$m and 16~$\mu$m, where the beam is still fairly small and manageable. Dealing with this for PACS and SPIRE imaging is beyond the scope of the present work.

\subsubsection{Some considerations about simulation methodology}
\REVISED{%
We carry out our simulations (Section~\ref{Section_Simulation}) independently for each band, adding objects randomly into the real image data with many Monte Carlo \MINORREREVISED[iterations]{repetitions}. An alternative approach, which we have not yet employed, might simultaneously and consistently simulate source fluxes over their full SEDs, then create fully simulated deep survey images in all of the bands, then carry out a full photometric analysis on these simulated data.  We recall, however, that we are also performing flux measurements one band at a time, and assigning \MINORREREVISED[errors]{uncertainties} to them in a reliable way at each band, as verified by our simulations. We use band-to-band information only to select which galaxies to fit and which to exclude, and we do this in a very conservative way because, in any case, we fit many more galaxies down to fainter fluxes than we can actually measure. Errors or limitations of this procedure would eventually factor into the \MINORREREVISED[error]{uncertainty} budget of the flux measurements, which is inferred from our simulations. Therefore it appears that simultaneous modeling of galaxies in all bands, while perhaps useful, might not be a critical limitation of this work in its current version.}

\REVISED{%
However, we recognize that there is interesting additional information that could be obtained by a full multi-band image simulation. In particular, such a simulation could permit an estimate for the amount of correlation\MINORREREVISED[]{s} among flux uncertainties between adjacent wavelengths, induced by the flux correlation of the galaxy SEDs themselves and their slowly rolling shapes (regardless whether the signal is from primary galaxies or from a superposition of those in the background, e.g., from confusion). Neglecting this might actually spuriously enhance the significance of source detection, as performed in the next sections by coadding the $\mathrm{S/N}$ over several bands. We use a fairly conservative threshold to limit the impact of this problem. We will return to this issue with an appropriate statistical approach in future work.}

\REVISED{%
Also, a full multi-band image simulation, including clustering \citep[e.g.,][]{Bethermin2017} and extended to very faint galaxies, could allow us to address the issue of the nature of ``additional sources'' (Section~\ref{Section_Additional_Sources_In_Residual}), i.e., to have information on which fraction of them might be spurious superpositions of several faint galaxies. This is also left to future work.  
}


\vspace{1truecm}



\section{Results and Discussion}
\label{Section_Final_Catalog_Highz}

In this final section we use the \MINORREREVISED[FIR/(sub)mm]{FIR+mm} catalog developed in this work to study the IR emission of high redshift galaxies and to provide an estimate of the SFR density in the Universe, up to high redshift.

In order to evaluate whether a source is detected in the FIR+mm \MINORREREVISED[]{catalog,} we adopt a simple approach considering the combined $\mathrm{S/N}$ over PACS, SPIRE, 850~$\mu$m and 1.16~mm bands:

\begin{equation}
\label{Equation_SNR}
\begin{split}
\mathrm{(S/N)}_{\textnormal{FIR+mm}}^{2}
= \mathrm{(S/N)}_{100\,{\mu}\mathrm{m}}^{2}
+ \mathrm{(S/N)}_{160\,{\mu}\mathrm{m}}^{2}
+ \mathrm{(S/N)}_{250\,{\mu}\mathrm{m}}^{2}
+ \\
\mathrm{(S/N)}_{350\,{\mu}\mathrm{m}}^{2}
+ \mathrm{(S/N)}_{500\,{\mu}\mathrm{m}}^{2}
+ \mathrm{(S/N)}_{850\,{\mu}\mathrm{m}}^{2}
+ \mathrm{(S/N)}_{1.16\,\mathrm{mm}}^{2}
\end{split}
\end{equation}


%
Our $(\mathrm{S/N})_\mathrm{FIR+mm}^{2}$ definition is equivalent to a Chi-Squared field whose expectation value (in the presence of pure noise) for 7 bands being added, is equal to \REVISED[$\sqrt{7} \approx 2.6$]{7,} and the probability to have $\mathrm{S/N}_\mathrm{FIR+mm} \ge 5$ is $7 \times 10^{-4}$ for pure Gaussian random fields \citep{Bloomfield2017}. In such an ideal case \REVISED{(but see Section~\ref{Section_Quality_checks})}, this would produce at most a couple of spurious IR detection from our starting pool of prior positions. 
Among the 3306 24+radio sources, 
1109
are detected with $\mathrm{S/N}_\mathrm{FIR+mm} \ge 5$, and 
518
are very significantly detected with $\mathrm{S/N}_\mathrm{FIR+mm} \ge 10$.


While we have applied our photometric technique to the whole GOODS-N area, \REVISED[]{in some of the remainder sections (e.g., for the derivation of the SFR density, Sections~\ref{Section_UV_unattenuated_SFR}--\ref{Section_Dropouts})} we focus our analysis on objects that lie in the central 134 arcmin$^2$ region (hereafter \goodArea{}), avoiding the outer perimeter where the instrumental noise at 24~$\mu$m starts to rise, and hence the 24~$\mu$m prior sources will be less complete.
This \goodArea{} contains 862 $\;\mathrm{S/N}_\mathrm{FIR+mm} \ge 5 \,$ sources and 
427 $\;\mathrm{S/N}_\mathrm{FIR+mm} \ge 10 \,$ sources. 


%
%
\begin{figure*}
	\begin{center}
		\includegraphics[width=0.95\textwidth]{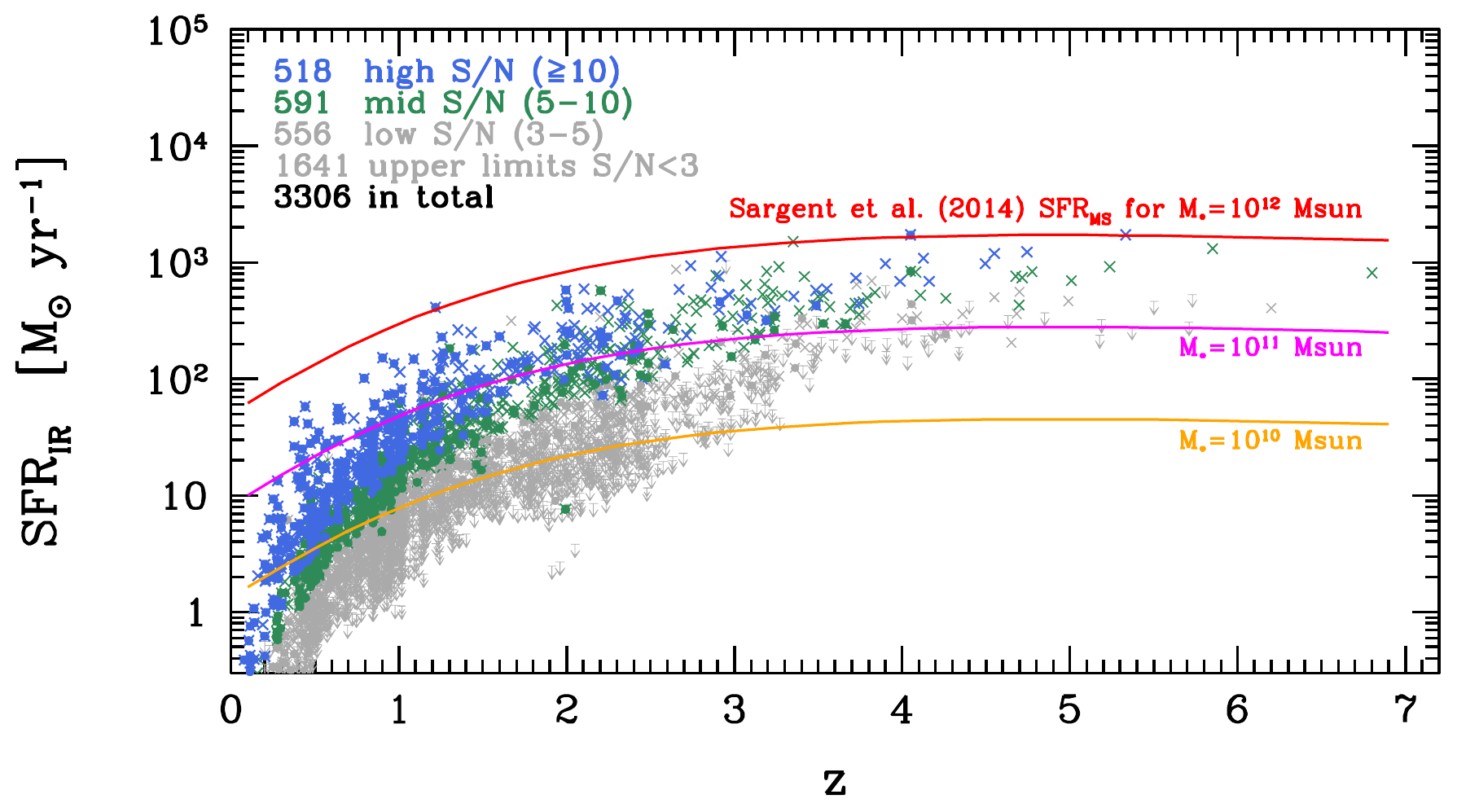}
		\caption{%
			Star formation rate (SFR) versus redshift for all 3306 sources in the 24+radio catalog (Section~\ref{Section_The_24_Radio_Catalog}). SFRs are computed from the integrated 8-1000$\,{\mu}\mathrm{m}$ infrared luminosities derived from their FIR+mm SEDs, $\mathrm{SFR}=L_{\mathrm{IR}}/1\times10^{10}\,\mathrm{M}_{\odot}\,\mathrm{yr}^{-1}$, assuming a Chabrier IMF \citep{Chabrier2003}. Colors indicate the combined $\mathrm{S/N}$ over the FIR+mm bands (see Equation~\ref{Equation_SNR}). 
			\REVISED{We highlight sources with spectroscopic redshifts with dots at the center of the cross symbols.} 
			We show the empirical tracks of the MS galaxy SFR as a function of redshift at three representative MS stellar masses: $M_{*}=10^{12}\,\mathrm{M}_{\odot}$, $10^{11}\,\mathrm{M}_{\odot}$ and $10^{10}\,\mathrm{M}_{\odot}$, according to \citet{Sargent2014}, who compiled literature data and derived empirical MS correlations (i.e., SFR as a function of $M_{*}$ and $z$). 
			\label{Figure_z_SFR}%
		}
	\end{center}
\end{figure*}

%
%
%
\begin{table*}

\begin{center}

\caption{ %
    GOODS-N ``Super-deblended'' Photometry Catalog (Example) %
    \label{Table_2} %
}

\begin{tabular*}{0.95\textwidth}{ @{\extracolsep{\fill}} l c c l c r r r c r r r r r r }

    \hline

        ID & 
        R.A. & 
        Decl. & 
        $z_{\mathrm{IR}}$ & 
        $z_{\mathrm{type}}$ & 
        $S_{500}$ & 
        $\sigma_{500}$ & 
        $\mathrm{S/N}_{\mathrm{IR}}$ & 
        $\log{M_{*}}$ & 
        $\mathrm{SFR}_{\mathrm{total}}$ & 
        $\mathrm{SFR}_{\mathrm{IR}}$ & 
        $\sigma_{\mathrm{SFR}_{\mathrm{IR}}}$ & 
        ${\mathrm{goodArea}}$ & 
        $T_{\mathrm{20cm}}$ & 
        $T_{\mathrm{SED}}$ \\ 
        
        (1) & 
        (2) & 
        (3) & 
        (4) & 
        (5) & 
        (6) & 
        (7) & 
        (8) & 
        (9) & 
        (10) & 
        (11) & 
        (12) & 
        (13) & 
        (14) & 
        (15) \\ 
    
    \hline

  564 & 189.2994995 & 62.3700371 & 4.055 & s & 44.0 & 5.8 & 19.9 & 11.25 & 1812.4 & 1726.0 & 86.6 & 1 & 0 & 1\\
  11499 & 189.2606049 & 62.2172966 & 4.50 & p & 10.4 & 1.8 & 18.2 & 11.44 & 983.5 & 977.4 & 53.7 & 1 & 0 & -1\\
  2592 & 189.3670044 & 62.3222733 & 4.74 & p & 14.5 & 2.3 & 16.3 & 11.37 & 1238.9 & 1231.0 & 75.4 & 1 & 0 & -1\\
  15289 & 188.9900818 & 62.1734276 & 3.075 & s & 8.7 & 2.3 & 15.4 & 8.06 & 386.1 & 354.9 & 25.5 & 0 & 0 & 1\\
  16332 & 189.2442017 & 62.1585388 & 3.55 & p & 11.3 & 3.8 & 14.9 & \nodata & 659.8 & 594.8 & 163.5 & 1 & 0 & 0\\
  3532 & 189.3077393 & 62.3073006 & 3.99 & p & 9.1 & 2.2 & 14.8 & 11.46 & 689.1 & 687.9 & 190.2 & 1 & 1 & -1\\
  9053 & 189.0358124 & 62.2431564 & 3.48 & p & 14.7 & 1.8 & 14.4 & 11.26 & 579.2 & 576.1 & 88.5 & 1 & 0 & -1\\
  4500 & 189.4094086 & 62.2934532 & 3.190 & s & 14.6 & 2.5 & 14.1 & 10.26 & 450.7 & 317.7 & 22.5 & 1 & 0 & 1\\
  18911 & 189.1133881 & 62.1015778 & 4.55 & p & 31.8 & 3.9 & 14.1 & \nodata & 1323.3 & 1193.0 & 100.4 & 0 & 0 & 0\\
  4990 & 189.1330261 & 62.2873993 & 4.16 & p & 12.6 & 2.3 & 13.8 & 10.32 & 698.5 & 687.5 & 49.8 & 1 & 0 & 1\\
  14914 & 188.9597931 & 62.1782951 & 5.33 & p & 20.1 & 2.6 & 13.6 & 12.55 & 1760.0 & 1724.0 & 132.1 & 0 & 0 & -1\\
  15213 & 189.0117798 & 62.1743240 & 3.24 & p & 5.9 & 2.9 & 12.3 & 10.45 & 399.7 & 384.2 & 148.9 & 1 & 0 & 0\\
  17624 & 189.0368347 & 62.1344681 & 3.36 & p & 12.1 & 4.0 & 11.7 & 11.46 & 512.1 & 511.2 & 88.0 & 0 & 0 & -1\\
  16810 & 189.0555725 & 62.1504669 & 3.05 & p & 10.3 & 1.6 & 11.4 & 10.84 & 338.3 & 327.8 & 71.1 & 0 & 0 & -1\\
  13107 & 189.1156006 & 62.1996002 & 3.49 & p & 9.2 & 1.9 & 11.3 & 11.07 & 433.2 & 427.9 & 128.4 & 1 & 0 & -1\\
  130 & 189.3960114 & 62.3908005 & 4.13 & p & 23.7 & 2.5 & 11.0 & 10.67 & 1109.8 & 1085.0 & 98.5 & 0 & 0 & 1\\
  3827 & 189.5537872 & 62.3028259 & 3.90 & p & 29.0 & 3.4 & 10.8 & \nodata & 1083.8 & 977.1 & 264.0 & 0 & 0 & 0\\
  16121 & 189.1436768 & 62.1616745 & 3.74 & p & 12.2 & 2.1 & 10.6 & 11.24 & 454.0 & 446.2 & 42.1 & 1 & 0 & -1\\
  17381 & 189.2108612 & 62.1394119 & 3.10 & p & 3.8 & 9.3 & 10.5 & 10.68 & 491.3 & 445.9 & 160.4 & 1 & 0 & 0\\
  18603 & 189.0650787 & 62.1119728 & 3.72 & p & 12.6 & 6.6 & 10.4 & 11.04 & 742.8 & 737.7 & 163.4 & 0 & 0 & 0\\

    \hline

    \vspace{-0.5ex}

\end{tabular*}

\begin{minipage}{0.95\textwidth}
    
    Table~\ref{Table_2} is published in its entirety in the machine-readable format. 
    A portion is shown here for guidance regarding its form and content. 
    Here we show a few example columns and a few example $\mathrm{S/N}_{\mathrm{FIR+mm}} \ge 10$ sources at $z \ge 3$ (sorted by the $\mathrm{S/N}_{\mathrm{FIR+mm}}$). 
    Column (1), (2) and (3) are the IRAC catalog from GOODS-\textit{Spitzer} Legacy Program 
    (see Section~\ref{Section_Initial_IRAC_Catalog}). 
    Column (4) \REREVISED[the spectroscopic redshift when available, otherwise the FIR+mm SED fitting photometric redshift]{is our IR-to-radio SED redshift, which is spectroscopic redshift if available, otherwise the SED fitted photometric redshift, see Section~\ref{Section_SED_Fitting}}.  
    \REREVISED[]{Column (5) indicates whether $z_{\mathrm{IR}}$ is spectroscopic redshift (``s'') or photometric redshift (``p''). The references for each spectroscopic redshifts are fully listed in the released machine-readable catalog.} 
    Column (6) and (7) are flux density and uncertainty in unit of mJy.  
    Column (8) is the FIR+mm combined S/N in Equation~\ref{Equation_SNR}.  
    Column (9) is the stellar mass from 3D-HST \citet{Skelton2014} or \citet{Pannella2015},  
    converted to Chabrier IMF when needed.  
    Column (10) is the total SFR, adding UV-unattenuated SFR (Section~\ref{Section_UV_unattenuated_SFR}) to the dust-obscured SFR. 
    Column (11) and (12) are the dust-obscured SFR and uncertainty from FIR+mm SED fitting  
    (see Fig.~\ref{Figure_z_SFR}  
    caption for deriving SFR from $L_{\mathrm{IR}}$).  
    Column (13) is the ``goodArea'' parameter,  
    equals 1 if the source is in the inner lower r.m.s. area in 24~$\mu$m  
    (has deeper prior catalog for surrounding sources),  
    otherwise 0 (has shallower or even incomplete prior catalog for surrounding sources).  
    Column (14) is the ``Type\_20cm'' parameter,  
    equals 1 if the source is classified as radio excess and  
    its radio data point is not fitted in SED fitting, otherwise 0.  
    Column (15) is the ``Type\_SED'' parameter,  
    equals 1 if the source is classified as starburst type  
    (its SED is fixed to using starburst templates),  
    or -1 if classified as main-sequence type  
    (its SED is fixed to using main-sequence templates), 
    otherwise 0 (allowing fitting all SED templates). 
    
\end{minipage}

\end{center}

\end{table*}


\subsection{Redshift, \REVISED{SFR and stellar mass} distributions}
\label{Section_z_SFR}

In Fig.~\ref{Figure_z_SFR}, we show the 
SFR versus redshift for all 3306 24+radio sources in the full GOODS-N field.
The SFR are derived from the pure dust component IR luminosity over 8--1000~$\mu$m, excluding any AGN torus components as derived from our SED fitting (see Section~\ref{Section_SED_Fitting}). 
Among all the detected sources (henceforth, the FIR+mm sources), 
106
sources have $z\ge 2.5$  (FIR+mm SED based photometric redshift  or spectroscopic redshift when available), 
70
have $z \ge 3$, and 
20 
have $z \ge 4$. Within the \goodArea{} we find 
71
sources at $z\ge 2.5$, 
49
at $z \ge 3$ and 
14
at $z\ge 4$.

In Fig.~\ref{Figure_z_SFR} we indicate the SFRs of typical MS galaxies at three fiducial stellar masses ($M_{*}=10^{12}\,\mathrm{M}_{\odot}$, $10^{11}\,\mathrm{M}_{\odot}$ and $10^{10}\,\mathrm{M}_{\odot}$ respectively), computed with the Eq.~(A1) of \cite{Sargent2014}. 
This FIR+mm sample has a non-uniform distribution in both SFR and stellar mass. At lower redshift (e.g., $z<1$), the limits of the SFR and stellar mass both rise quickly. 
At $z<1$, the GOODS-N FIR+mm data can detect emission from MS galaxies with $M_{*}<10^{10}\,\mathrm{M}_{\odot}$, but at $z>1$ all detected galaxies are more massive. 
At higher redshift (e.g., $z>3$), the sample is clearly biased towards high SFRs and stellar masses ($M_{*}>10^{11}\,\mathrm{M}_{\odot}$). Both the effective limits of SFR and stellar mass have risen by one to two orders of magnitude compared to those at $z\sim1$.


We present our FIR+mm catalog in Table~\ref{Table_2}. The full machine-readable version can be accessed via the on-line journal; here we only show a few important columns whose descriptions are in the caption.



\subsection{UV unattenuated SFRs}
\label{Section_UV_unattenuated_SFR}

\begin{figure*}
\centering
\includegraphics[width=0.49\textwidth, trim=0 3mm 0 3mm]{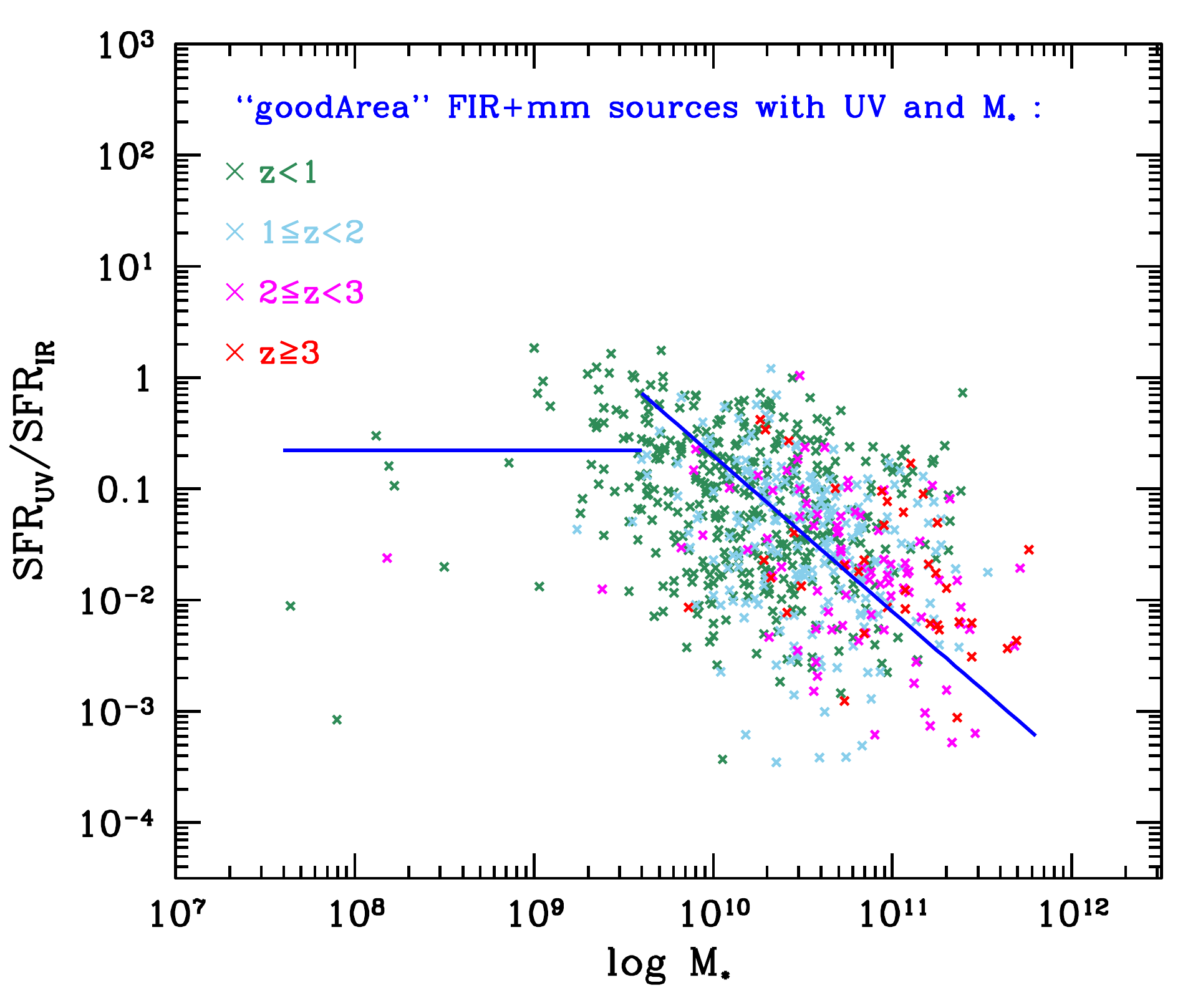}
\includegraphics[width=0.49\textwidth, trim=0 3mm 0 3mm]{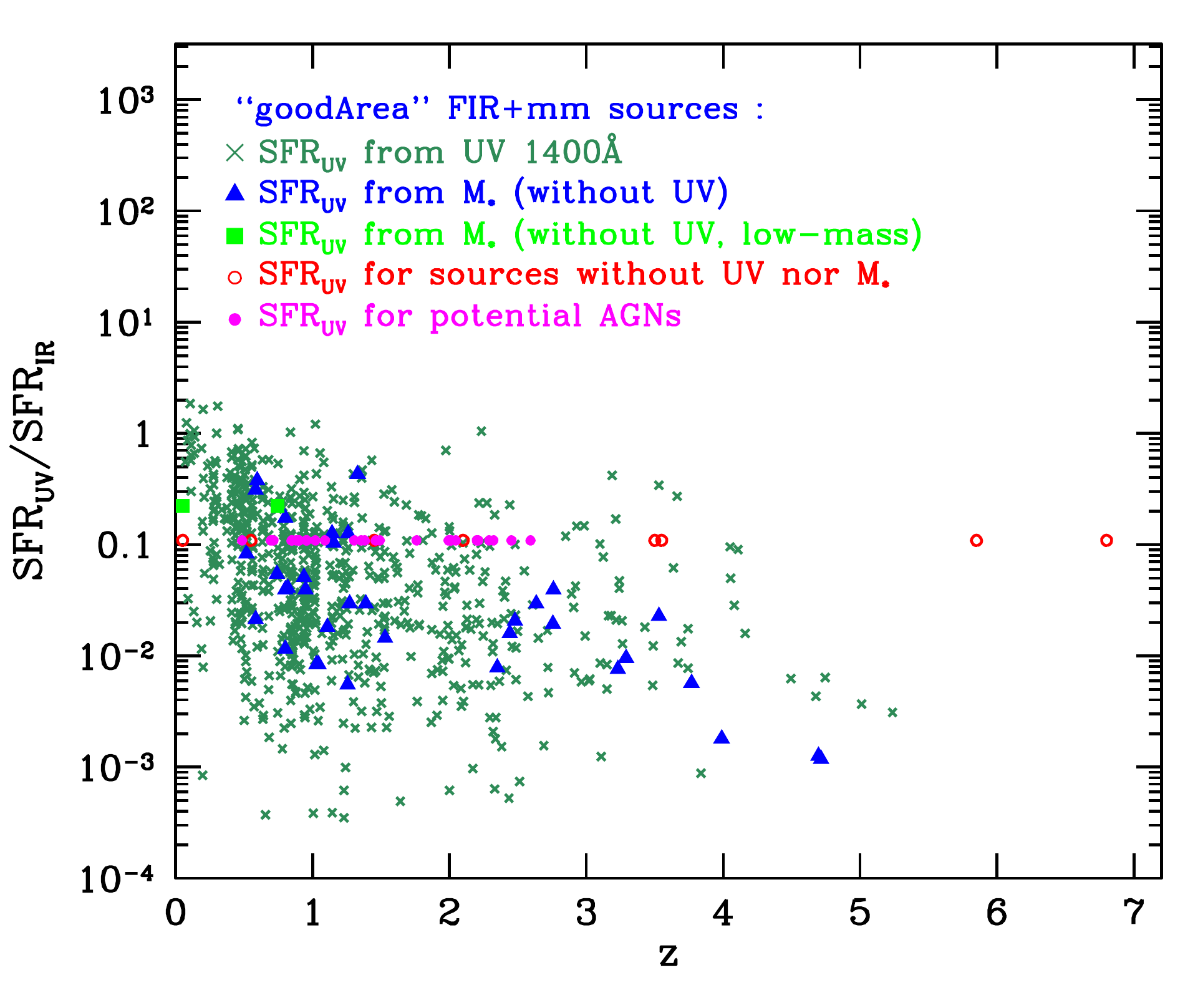}
\caption{%
The ratio of UV-uncorrected SFR ($\mathrm{SFR}_\mathrm{UV}$) and dust-obscured SFR ($\mathrm{SFR}_\mathrm{IR}$) versus stellar masses and redshift, for our FIR+mm $\mathrm{S/N}\ge5$ galaxies within the \goodArea{}. In the left panel, 
\MINORREREVISED[data points are sources with both rest-frame UV continuum and stellar mass available, and they are colored by redshifts]{data points indicate sources for which both UV continuum fluxes and stellar masses are available, and are color-coded by redshift}. 
In the right panel, data points are all FIR+mm sources in the \goodArea{}, and colors indicate different $\mathrm{SFR}_\mathrm{UV}$ estimation methods, depending on the availability of rest-frame UV continuum, $M_{*}$ and $\mathrm{SFR}_\mathrm{IR}$ (see text in Section~\ref{Section_UV_unattenuated_SFR}). $\mathrm{SFR}_\mathrm{IR}$ are obtained from our IR-to-radio SED fitting as described in Section~\ref{Section_z_SFR}. 
\label{Figure_SFR_UV_SFR_IR}}
\end{figure*}

\begin{figure*}
\centering
\includegraphics[width=0.30\textwidth, trim=4.0cm 2.5cm 0 0]{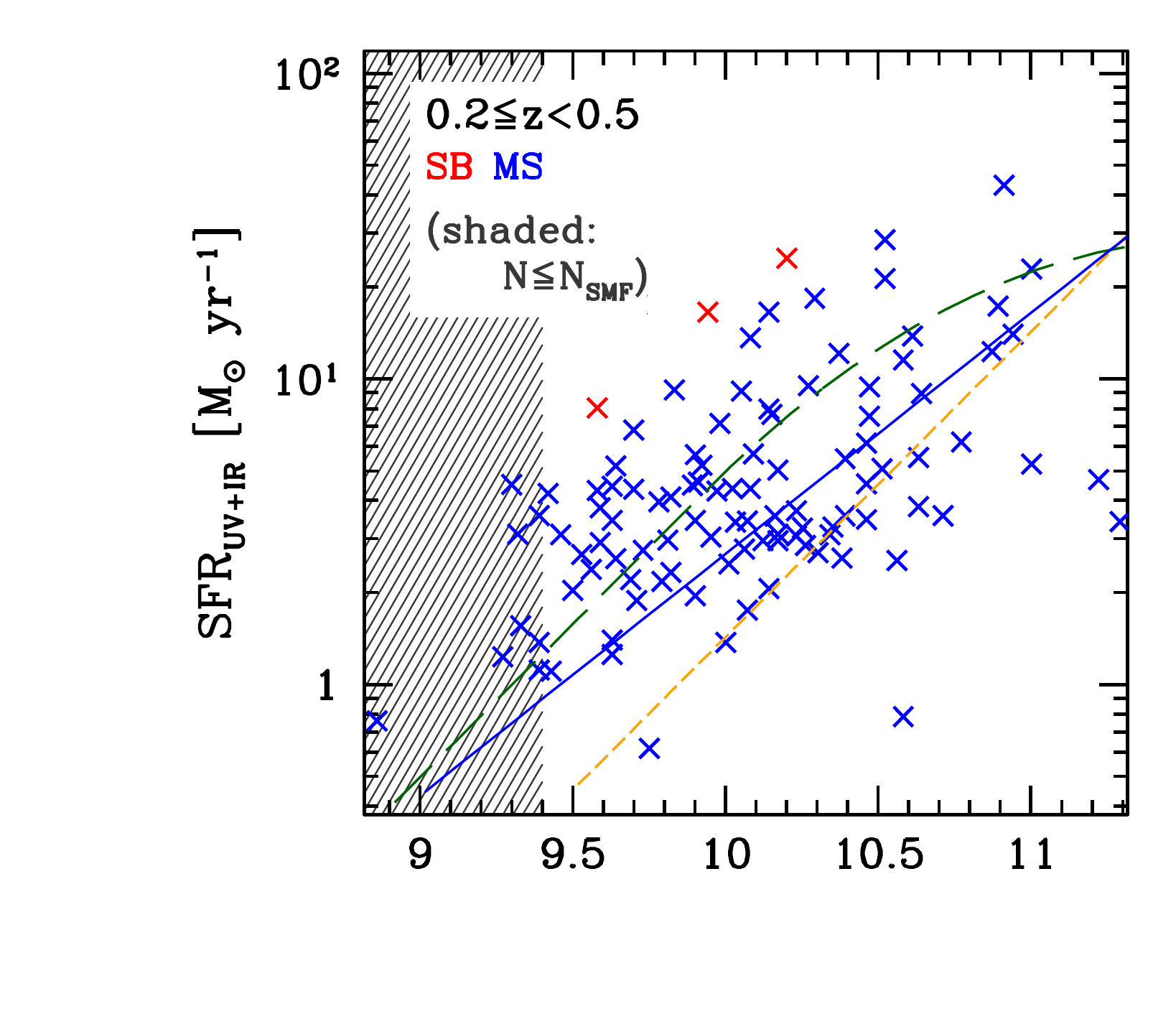}
\includegraphics[width=0.30\textwidth, trim=4.0cm 2.5cm 0 0]{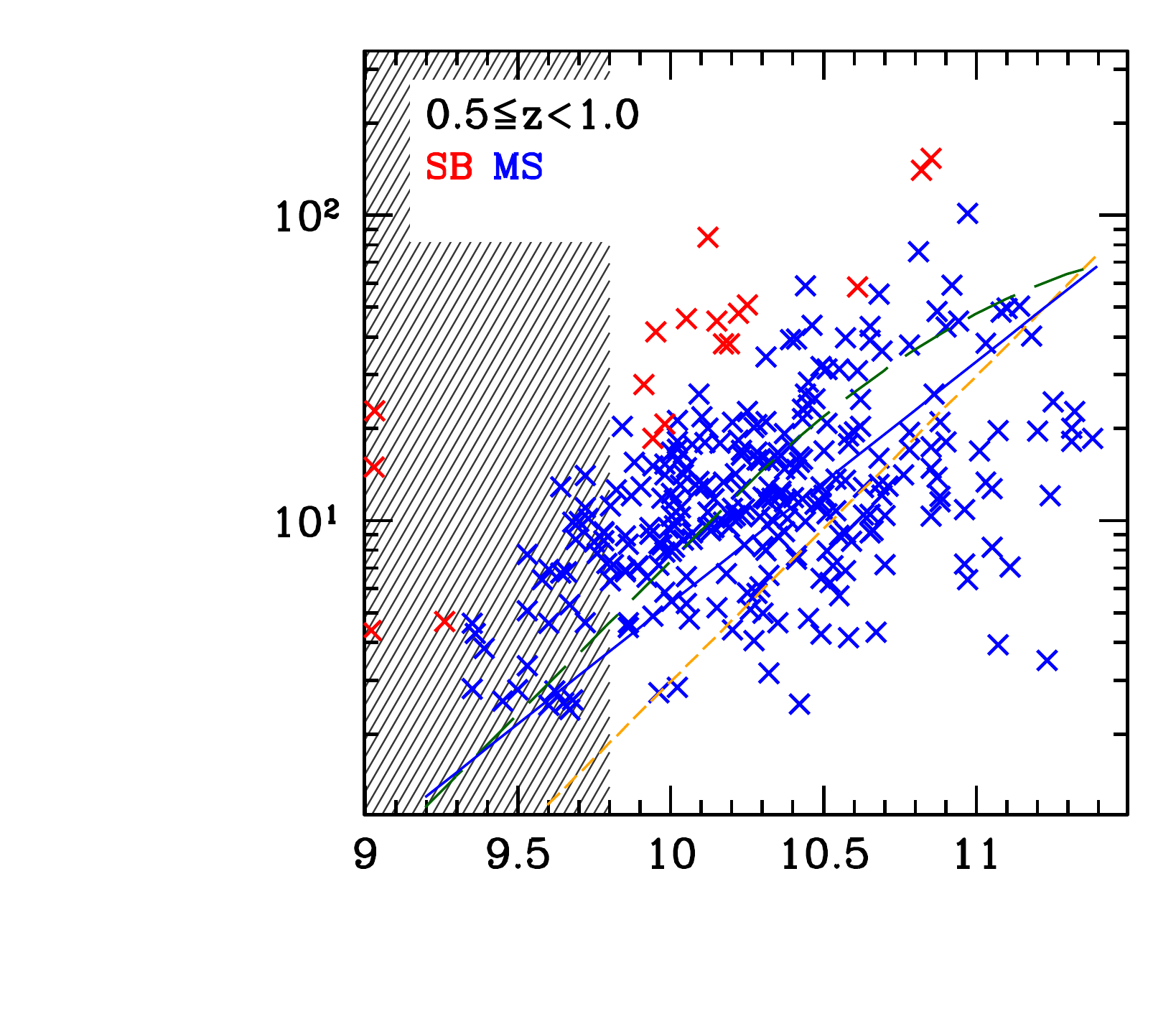}
\includegraphics[width=0.30\textwidth, trim=4.0cm 2.5cm 0 0]{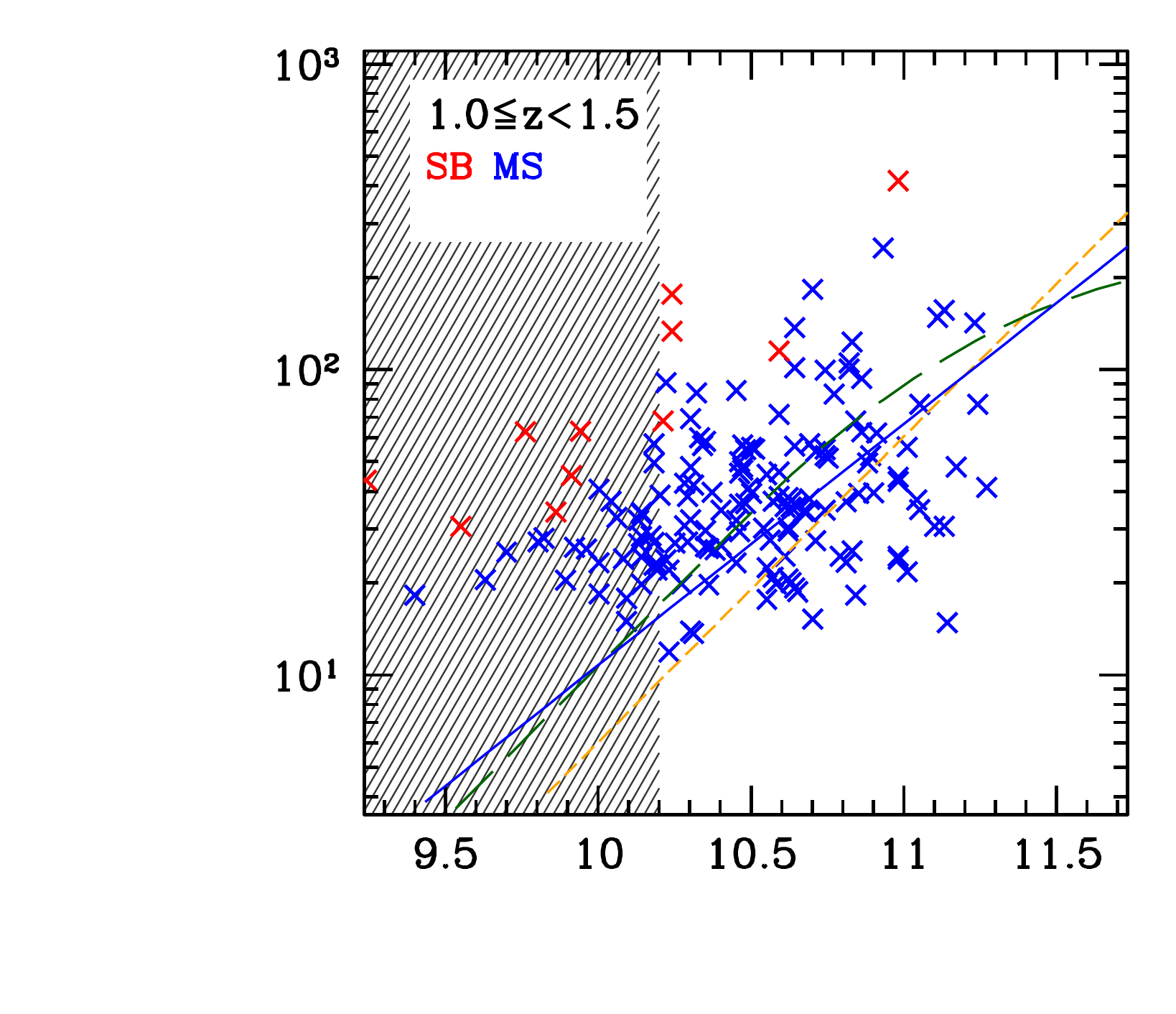}
\includegraphics[width=0.30\textwidth, trim=4.0cm 2.5cm 0 0]{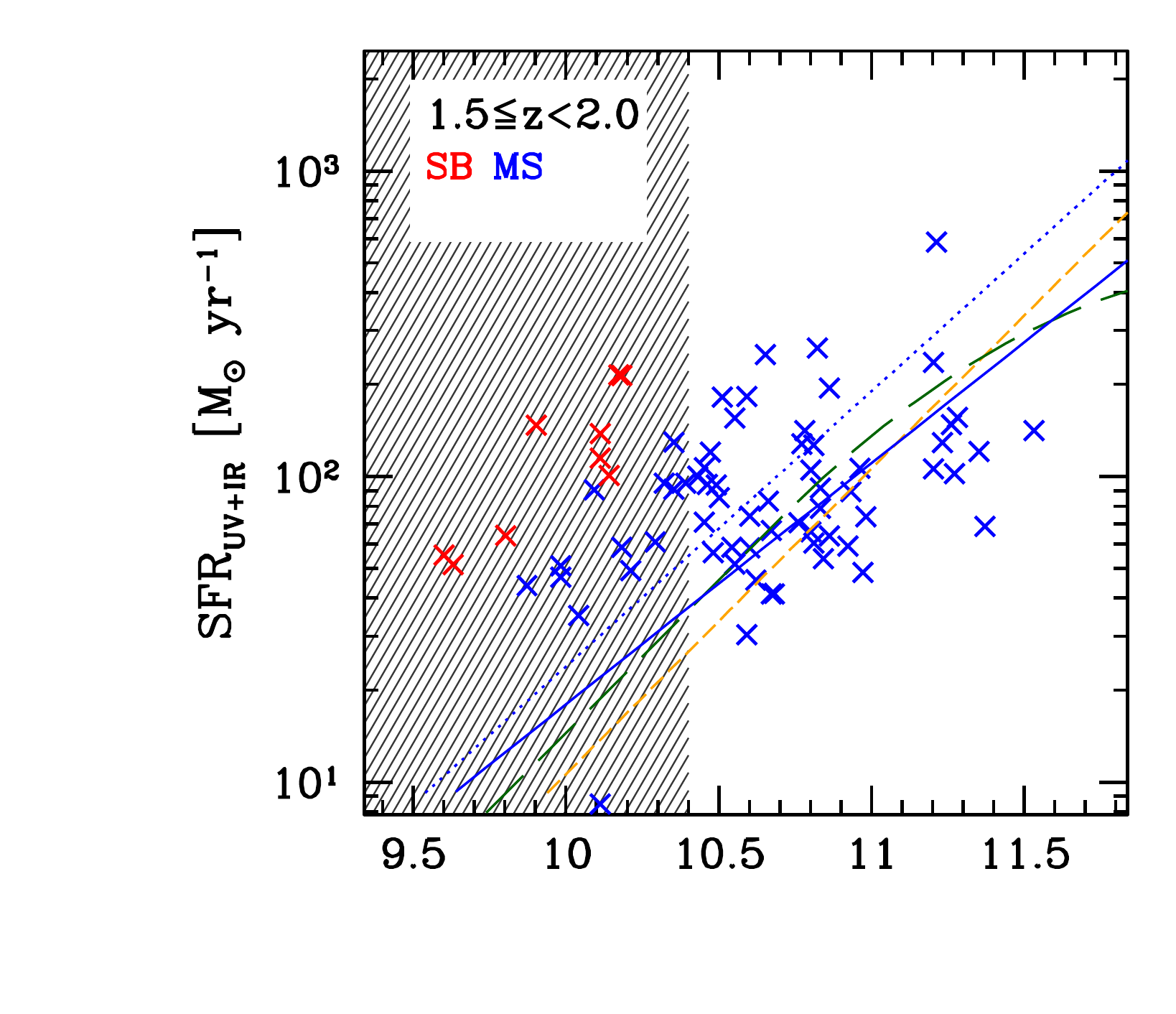}
\includegraphics[width=0.30\textwidth, trim=4.0cm 2.5cm 0 0]{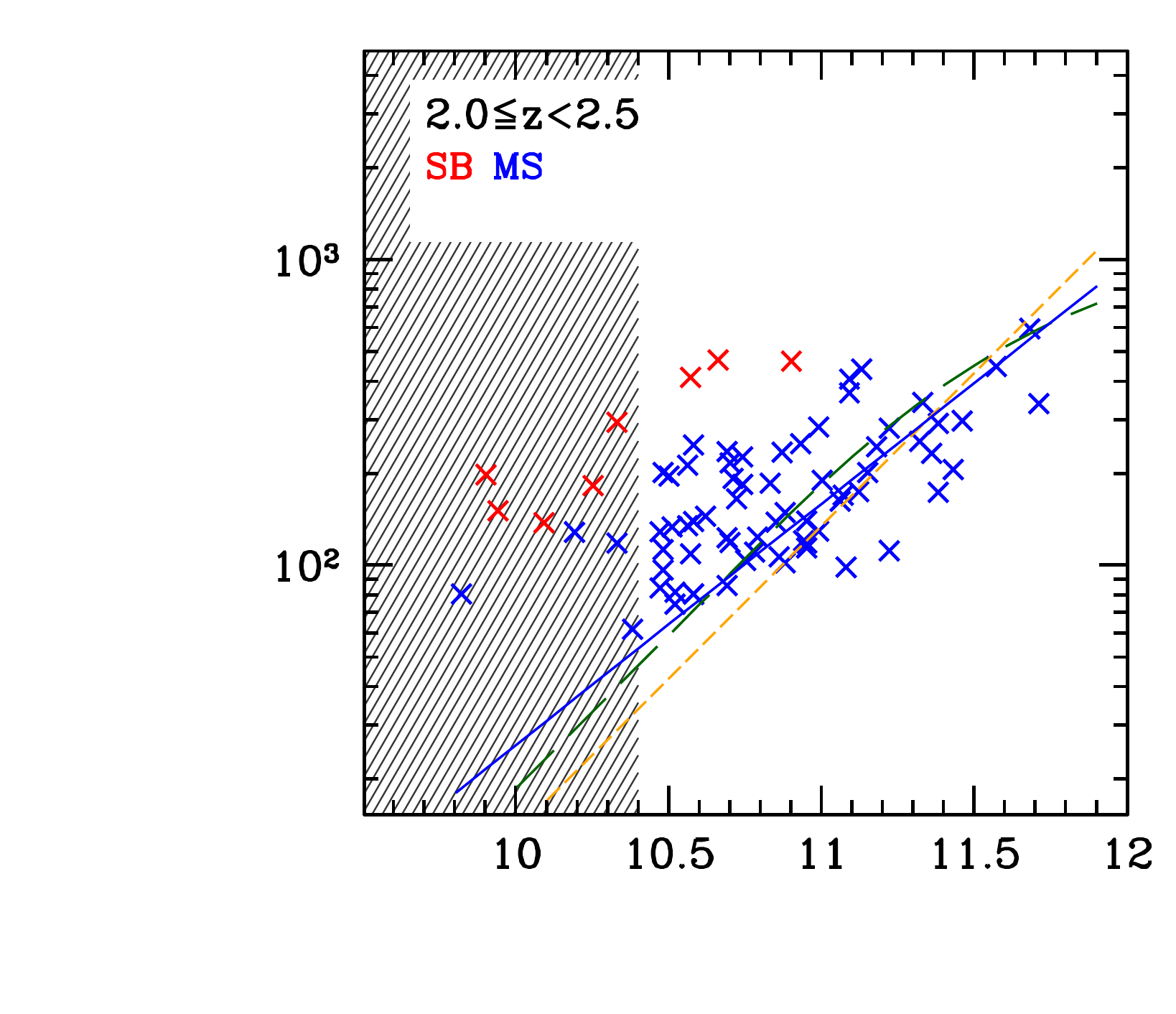}
\includegraphics[width=0.30\textwidth, trim=4.0cm 2.5cm 0 0]{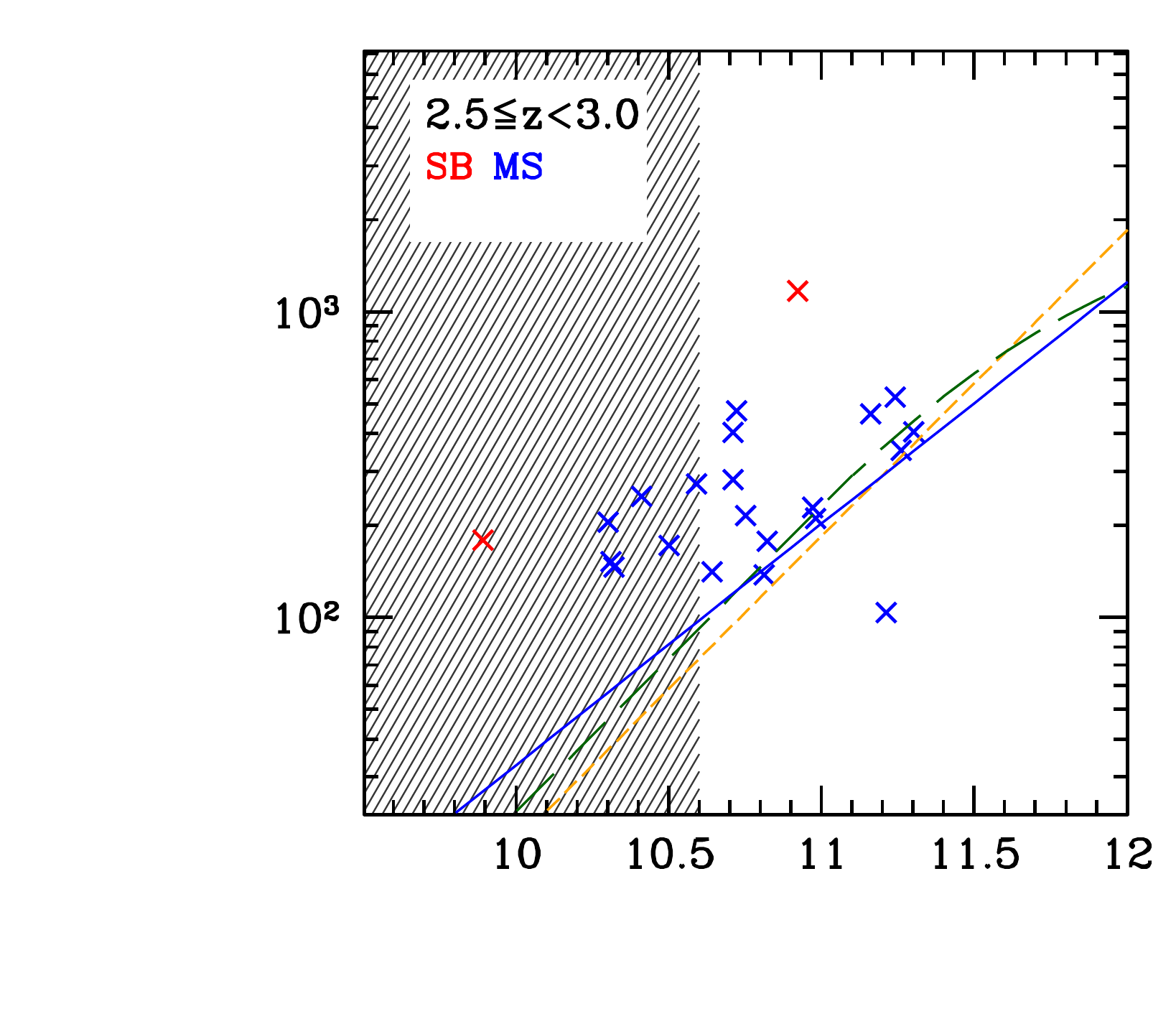}
\includegraphics[width=0.30\textwidth, trim=4.0cm 0.5cm 0 0]{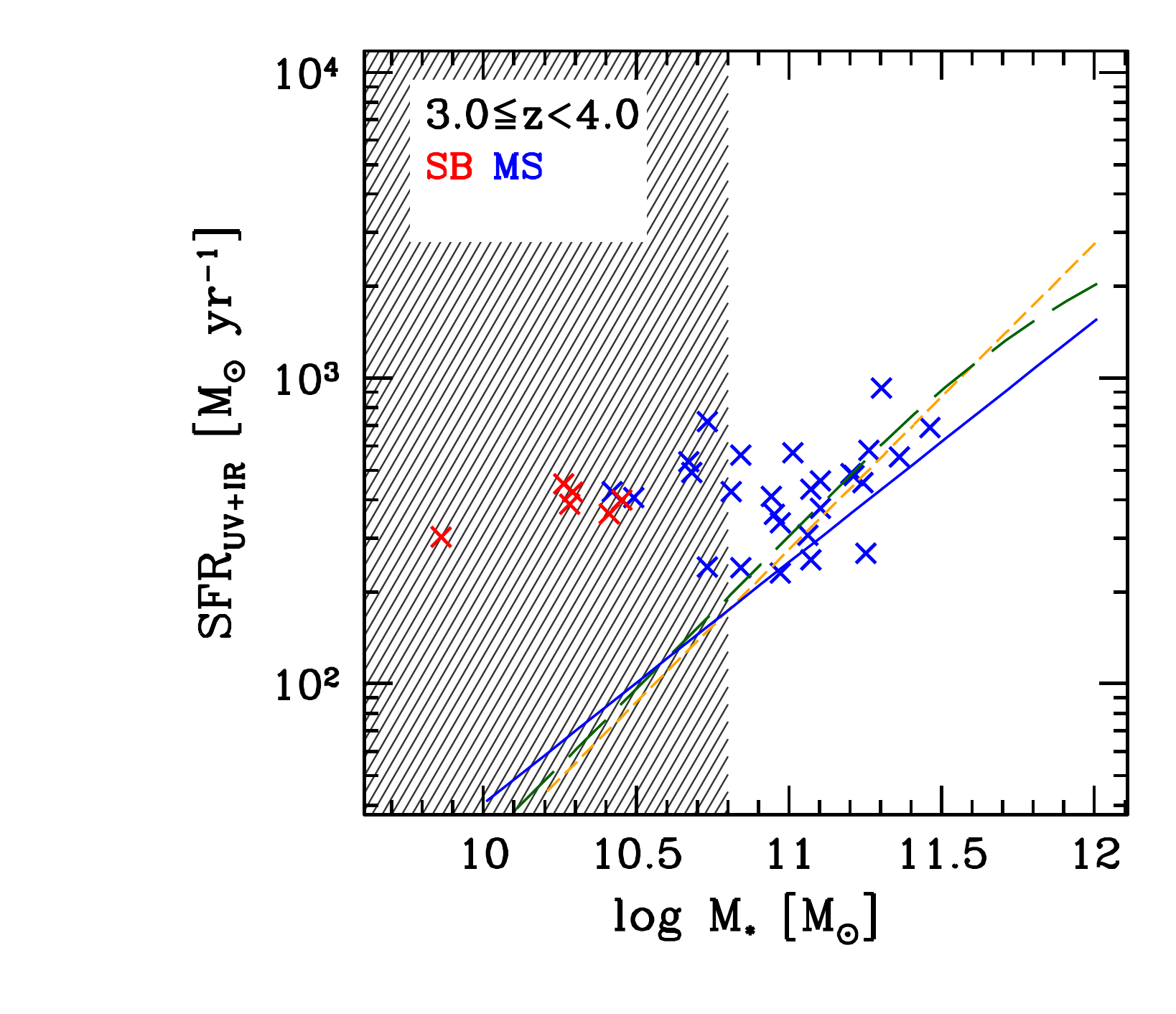}
\includegraphics[width=0.30\textwidth, trim=4.0cm 0.5cm 0 0]{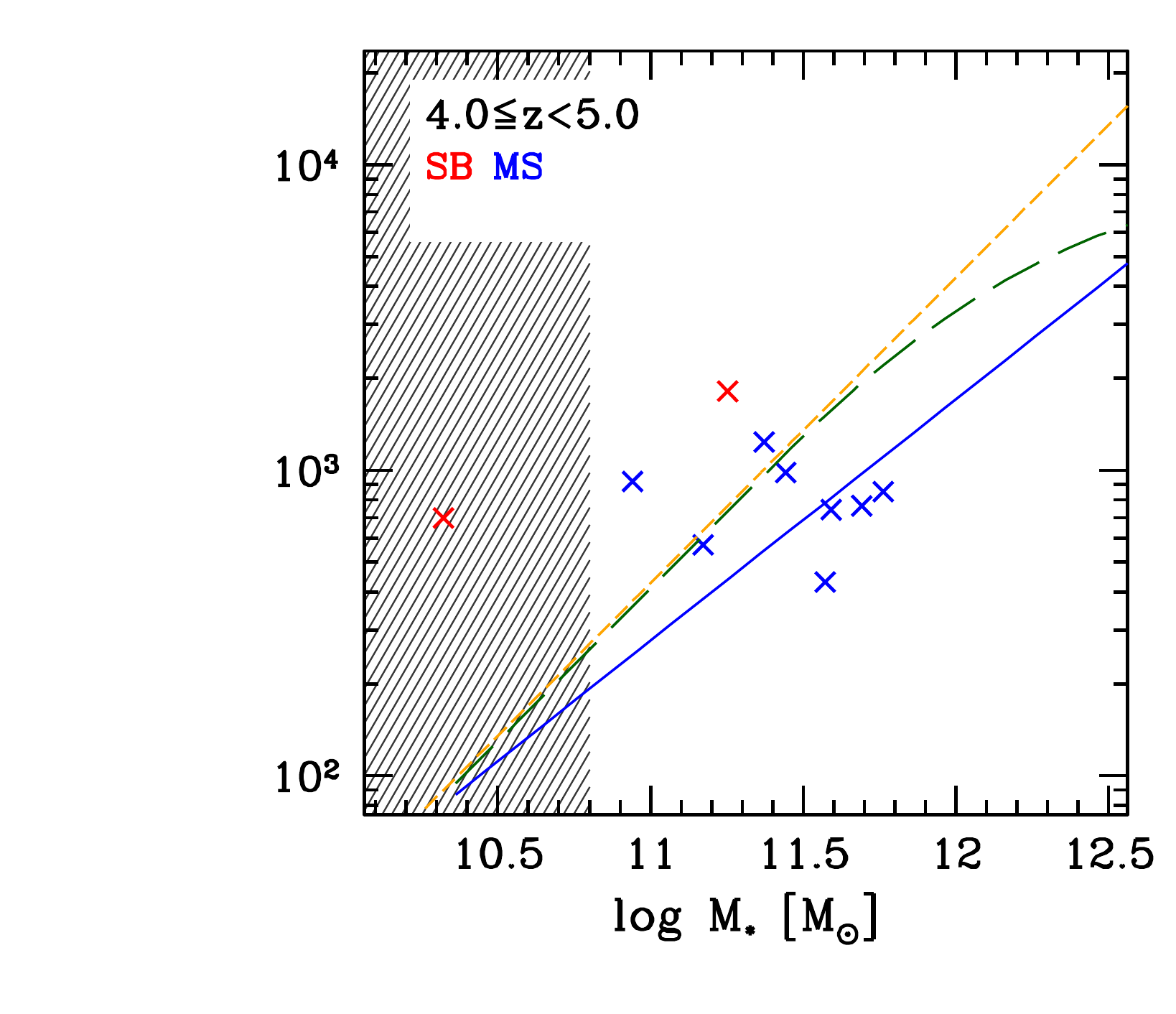}
\includegraphics[width=0.30\textwidth, trim=4.0cm 0.5cm 0 0]{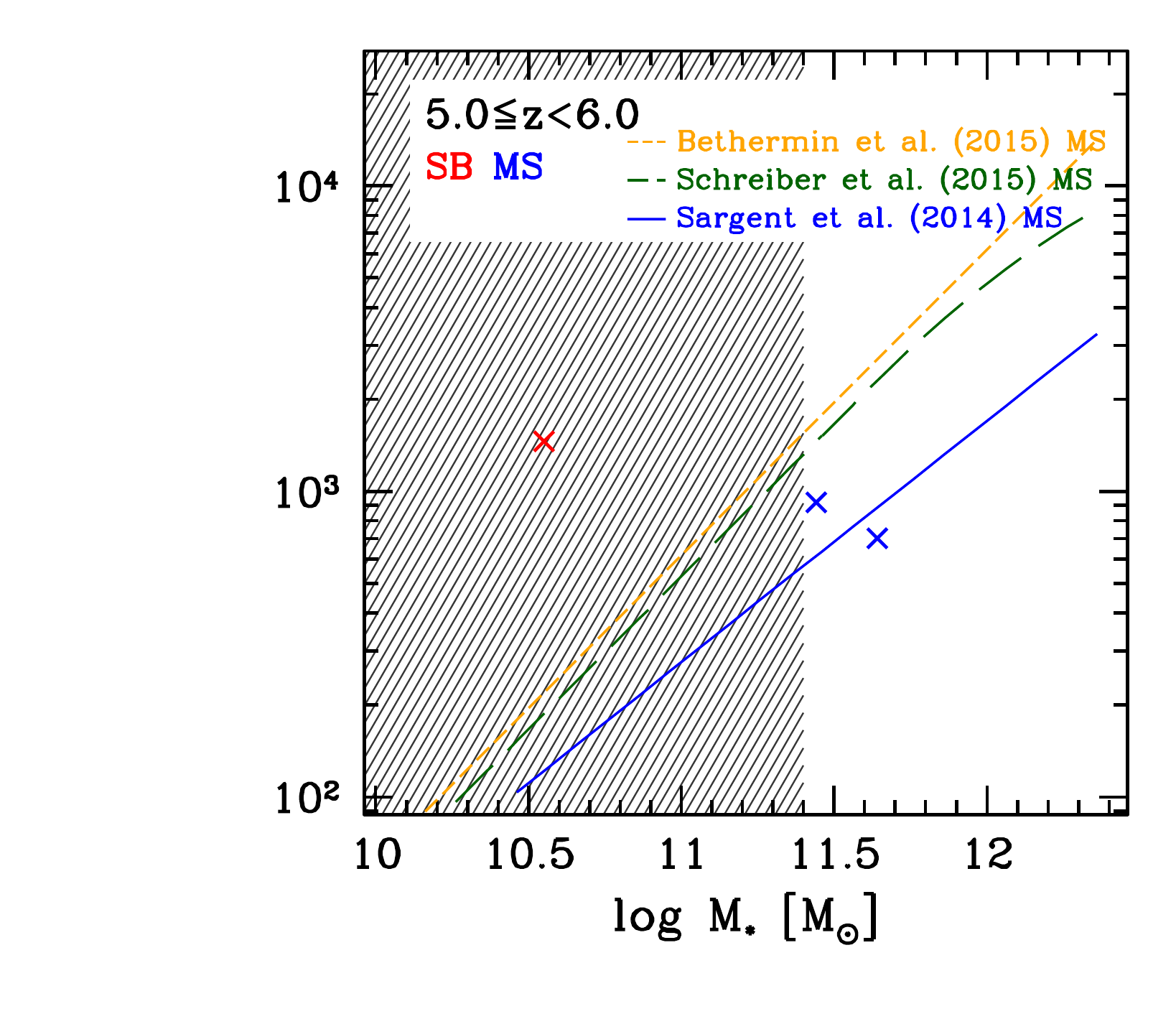}
\caption{%
    The 
    dust-obscured
    \REVISED{plus unobscured}
    SFR versus stellar mass $M_{*}$ for our FIR+mm sample in 9 redshift bins: $0.2 \le z_{\mathrm{phot}}<0.5$, $0.5 \le z_{\mathrm{phot}}<1.0$, $1.0 \le z_{\mathrm{phot}}<1.5$, $1.5 \le z_{\mathrm{phot}}<2.0$, $2.0 \le z_{\mathrm{phot}}<2.5$, $2.5 \le z_{\mathrm{phot}}<3.0$, $3.0 \le z_{\mathrm{phot}}<4.0$, $4.0 \le z_{\mathrm{phot}}<5.0$ and $5.0 \le z_{\mathrm{phot}}<6.0$.  SFR and $M_{*}$ are computed assuming a Chabrier IMF \citep{Chabrier2003}. 
    Red symbols are sources classified as SBs during our SED fitting (Section~\ref{Section_SED_Fitting}) according to their SFR, $\mathrm{S/N}$ of SFR and the distance to the MS relation of \citet{Sargent2014}, while blue symbols are all other sources. The shaded area indicates where our FIR+mm sample becomes incomplete (see the evaluation of incompleteness in Section~\ref{Section_Mstar_Histograms}) according to our stellar mass histogram analysis in Fig.~\ref{Plot_Mstar_Histogram_z_all_panels}. 
    Three literature MSs are indicated in each panel:  the blue solid line from \citet{Sargent2014}, the green long-dash line from \citet{Schreiber2015}, and the yellow dashed line from \citet{Bethermin2015}. 
    Note that the discrepancies between the three empirical MSs are small at lower redshifts, but not for the highest redshifts, where the difference can be as large as a factor of 3. 
    \label{Plot_Mstar_SFR_z}%
}
\end{figure*}

\REVISED{%
In order to estimate the total SFRs in our IR-detected galaxies we need to add the unattenuated SFR that can be directly recovered in the rest frame UV ($\mathrm{SFR}_\mathrm{UV}$). 
We thus cross-matched the 3D-HST catalog for the rest-frame UV flux densities of the FIR+mm galaxies ($\mathrm{S/N}_{\mathrm{FIR+mm}}\ge5$\MINORREREVISED[ and within \goodArea{}]{}). 94\% of them have rest-frame 1400~$\angstrom$ flux densities from the 3D-HST catalog. 
The $\mathrm{SFR}_\mathrm{UV}$ is in general a small contribution (10\% typically) compared to $\mathrm{SFR}_\mathrm{IR}$ for our IR selected sample. 
Fig.~\ref{Figure_SFR_UV_SFR_IR} shows the ratio of $\mathrm{SFR}_\mathrm{UV}$ and $\mathrm{SFR}_\mathrm{IR}$ versus stellar masses and redshifts (and in \goodArea{}). The left panel shows sources with rest-frame 1400~$\angstrom$ flux densities and stellar masses, colored by their redshifts. The right panel shows all FIR+mm sources (also in \goodArea{}), with different colors indicating the method adopted to estimate $\mathrm{SFR}_\mathrm{UV}$, as discussed below. 
Method (1): if a galaxy is in the 3D-HST catalog (matched within  1.0$''$ radius) and has rest-frame 1400~$\angstrom$ flux densities, then we use the \cite{Kennicutt1998ARAA} calibration: $\mathrm{SFR}_\mathrm{UV} = 1.4 \times 10^{-28} \times 4 \pi (\frac{\mathrm{d_{L}}}{\mathrm{[cm]}})^2 (\frac{S_{1400\angstrom}}{\mathrm{[erg\,s^{-1}\,cm^{-2}\,Hz^{-1}]}})$. We also check that using instead the rest-frame 1700~$\angstrom$ flux densities  (also provided by 3D-HST)  leads to no more than 0.1~dex difference in average. 
Method (2): if a galaxy has no rest-frame 1400~$\angstrom$ but has a stellar mass, then we use a simple correlation between $\mathrm{SFR}_\mathrm{UV}/\mathrm{SFR}_\mathrm{IR}$ and stellar mass (as shown in the left panel of Fig.~\ref{Figure_SFR_UV_SFR_IR}, analogous to the stellar mass--attenuation relation that is fairly constant with redshift -- \citealt{Pannella2015}) that we obtained with the galaxies in method (1): $\log_{10} (\mathrm{SFR}_\mathrm{UV}/\mathrm{SFR}_\mathrm{IR}) = {13.3-1.4 \log_{10}(M_{*})}$. This relation appears only to be valid for $\log_{10}(M_{*})\ge{9.6}$, and does not hold at very low stellar masses, where IR detected objects are mainly SB sources (e.g., see the right panel of Fig.~\ref{Plot_Mstar_SFR_z}). For the only 2 sources with $\log_{10}(M_{*})<{9.6}$ and no UV continuum, we just adopted the average $\mathrm{SFR}_\mathrm{UV}/\mathrm{SFR}_\mathrm{IR}$ value of the $\log_{10}(M_{*})<{9.6}$, UV-continuum-available sample ($\mathrm{SFR}_\mathrm{UV}/\mathrm{SFR}_\mathrm{IR} = 0.22$). 
Finally, for 8 galaxies that do not have UV continuum nor stellar mass, as well as for a small sample of $\sim$28 objects where UV continuum is most likely dominated by the emission of an AGN \MINORREREVISED[]{(e.g., indicated by the mid-IR AGN SED fits and optical/near-IR images)}, we adopted the median 
$\mathrm{SFR}_\mathrm{UV}/\mathrm{SFR}_\mathrm{IR}$ of the whole sample ($=0.11$).
%
In the following sections, we add $\mathrm{SFR}_\mathrm{UV}$ and $\mathrm{SFR}_\mathrm{IR}$ to obtain the total SFR ($\mathrm{SFR}_\mathrm{UV+IR}$) for each galaxy, and use the total SFR for all further analysis. 
}

%
%


\subsection{Stellar mass and SFR properties}
\label{Section_Mstar_SFR}

%

In Fig.~\ref{Plot_Mstar_SFR_z} we compare our sample with empirical MS correlations in redshift bins up to $z = 6$. Only \goodArea{} FIR+mm sources with stellar masses from 3D-HST or \citet{Pannella2015} are shown here;  7 FIR+mm sources lack stellar mass estimates and are not shown. 
%
The differences between the three MSs considered here (from \citealt{Sargent2014}, \citealt{Schreiber2015} and \citealt{Bethermin2015}) are relatively large at both $z<1.5$ and $z>4$, but the different estimates are more consistent at $1.5<z<4$.  
In the highest redshift bins, the  \citet{Sargent2014} MS has the lowest SFR normalizations. The difference between these three MS relations \REVISED{(in some cases their extrapolations)} can reach a factor of 3 at $M_{*}\sim10^{11.5}\,\mathrm{M}_{\odot}$ and $z\sim$4--6, where global uncertainties and lack of direct IR data in the previous derivations still prevent a solid assessment of the MS level. 

The red symbols in each panel indicate sources classified as strong starbursts during the SED fitting (Section~\ref{Section_SED_Fitting}), relative to the MS of \citet{Sargent2014}. 
They are relatively rare, and mostly have masses $M_{*} \lesssim 10^{10.5}\,\mathrm{M}_{\odot}$. 
The fraction of SBs in this sample is 9.8\% at $1<z<4$, intermediate between the fraction of 30-50\% for submillimeter galaxy-like selection at the highest luminosities and the fraction of 2\% to 3\% derived for a sample at $1.5<z<2.5$ selected by stellar mass (complete to $M_{*} \ge 10^{10}\,\mathrm{M}_{\odot}$; \citealt{Rodighiero2011}~\footnote{Their definition of SB is the same as we adopted in this work, i.e., $\log{\mathrm{SFR}}-\log{\mathrm{SFR}_\mathrm{MS}}>0.6$ dex}). 
%
%

The shaded area in each panel indicates the stellar mass range over which the sample is incomplete, evaluated by analyzing stellar mass functions (SMFs) for star-forming galaxies as described in the next section. 
Our FIR+mm sample should have a high degree of completeness at the characteristic stellar mass $M_{\mathrm{star}}^{*}$ in the 4 lowest redshift bins, where
\REVISED{
	$M_{\mathrm{star}}^{*}$ for star-forming galaxies is estimated to be in the range of $10^{10.3}$--$10^{10.9}\,\mathrm{M}_{\odot}$ at $0.2<z<2.5$ according to \citet{Davidzon2017}, \citet{Ilbert2013} and \citet{Muzzin2013}.
}

At $z>3$, estimates of $M_{\mathrm{star}}^{*}$ fall in the range $10^{10.74-11.56}\,\mathrm{M}_{\odot}$ (\citealp{Ilbert2013}; \citealp{Muzzin2013}; \citealp{Grazian2015}; \citealp{Davidzon2017}). At these redshifts, the FIR+mm sample can only probe a small number of galaxies at the high-mass end of the SMF.

%
%
\begin{figure*}
\centering
\includegraphics[width=0.32\textwidth, trim=0 1cm 0 1.8cm]{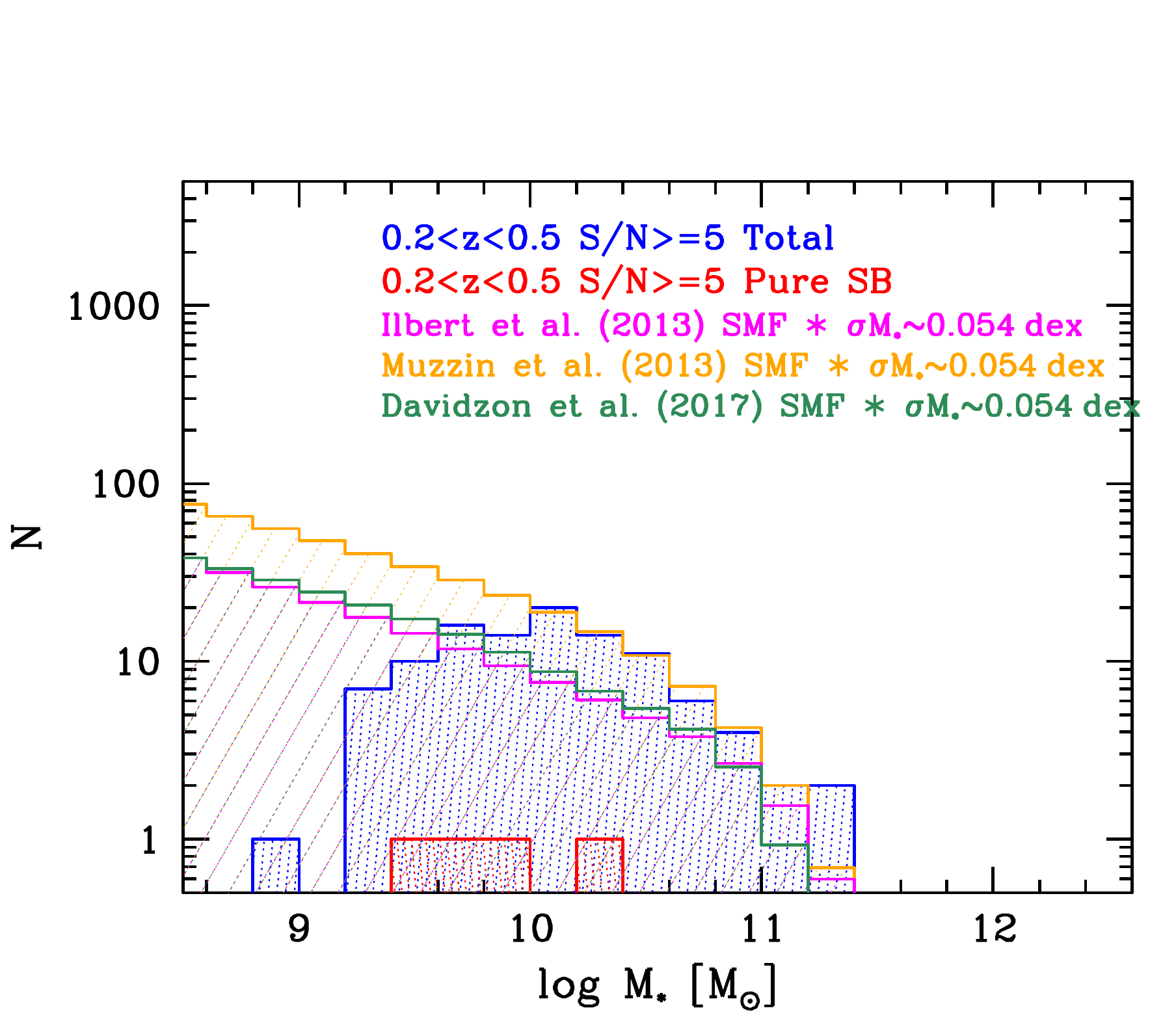}
\includegraphics[width=0.32\textwidth, trim=0 1cm 0 1.8cm]{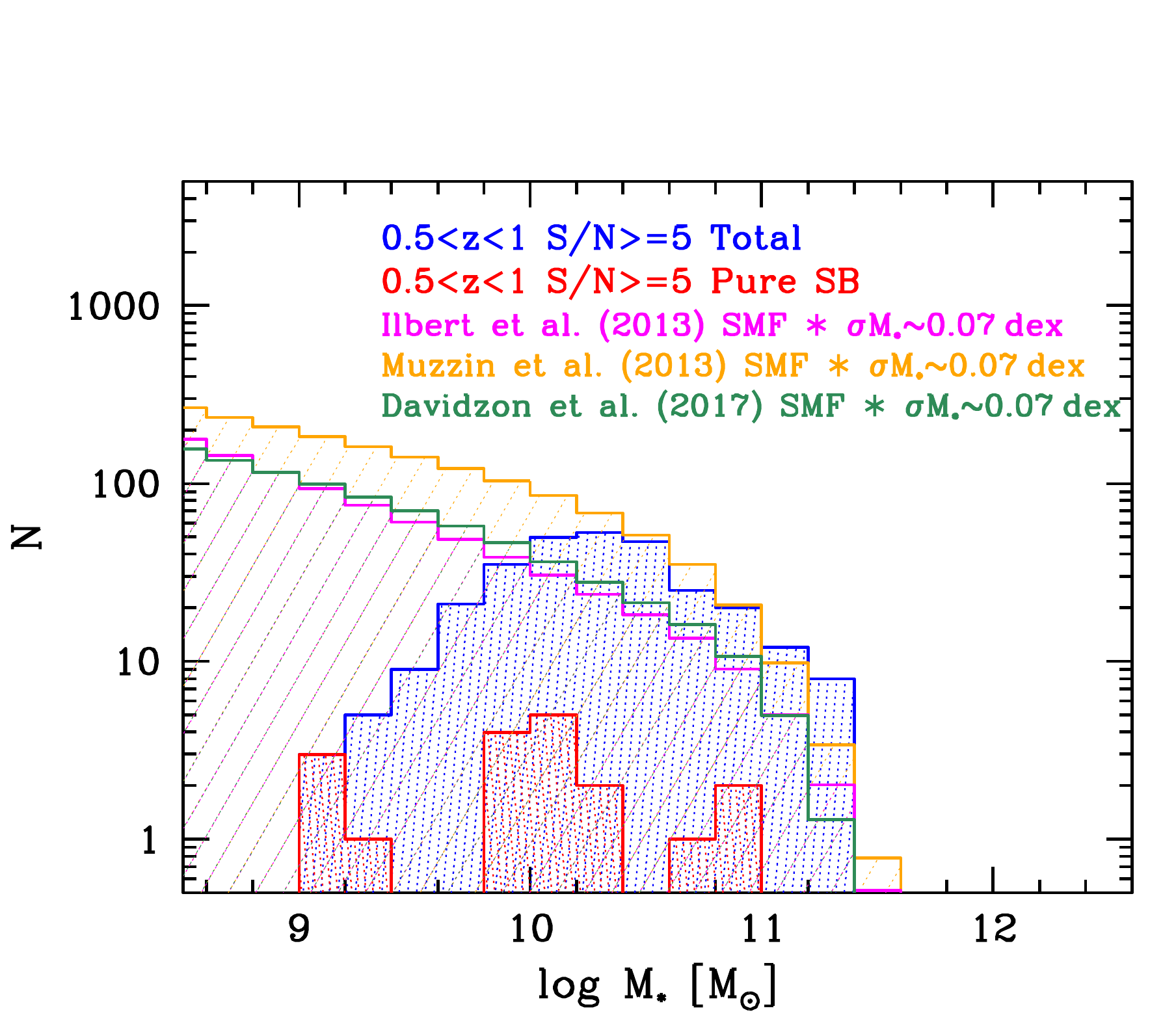}
\includegraphics[width=0.32\textwidth, trim=0 1cm 0 1.8cm]{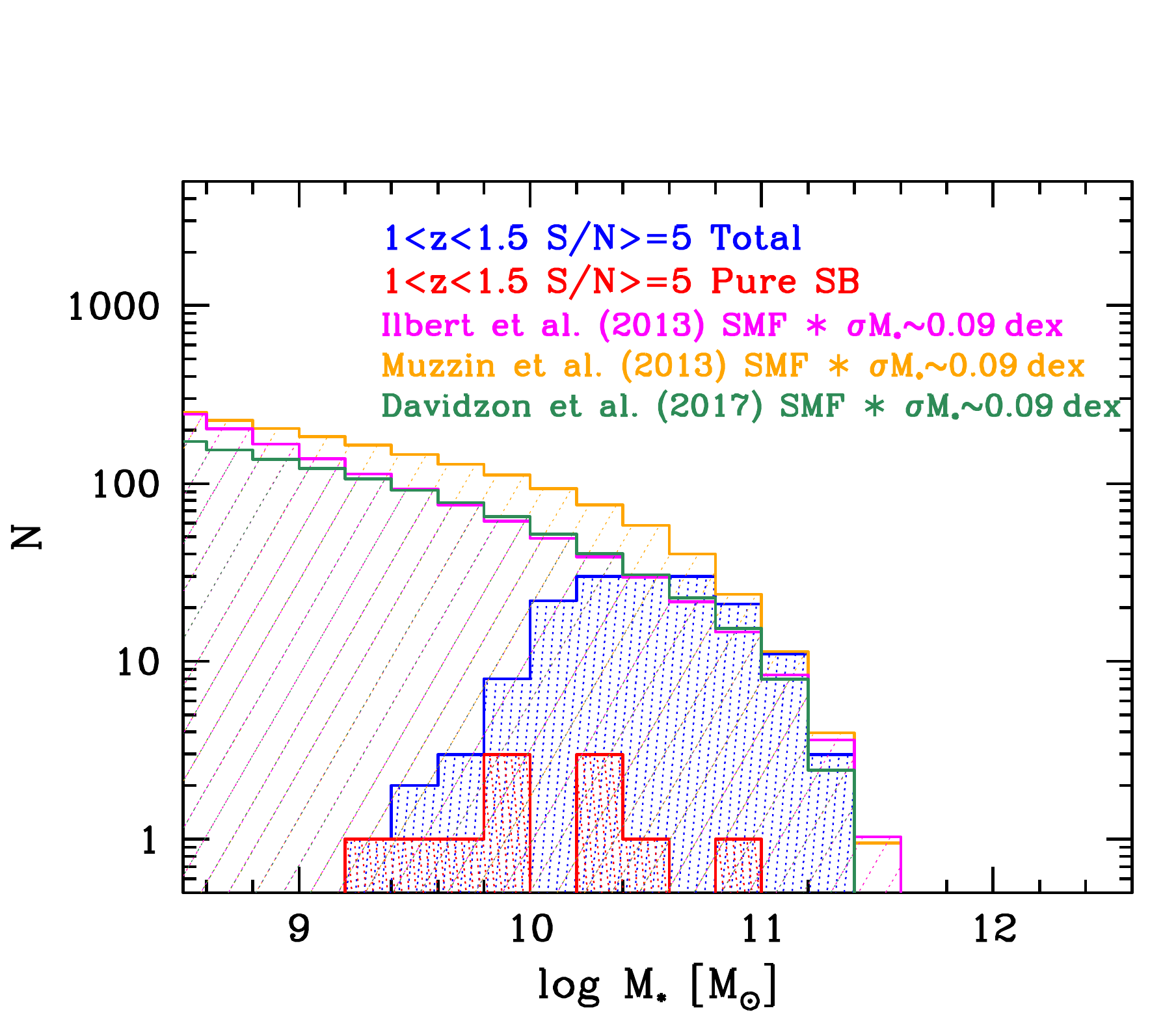}
\includegraphics[width=0.32\textwidth, trim=0 1cm 0 1.8cm]{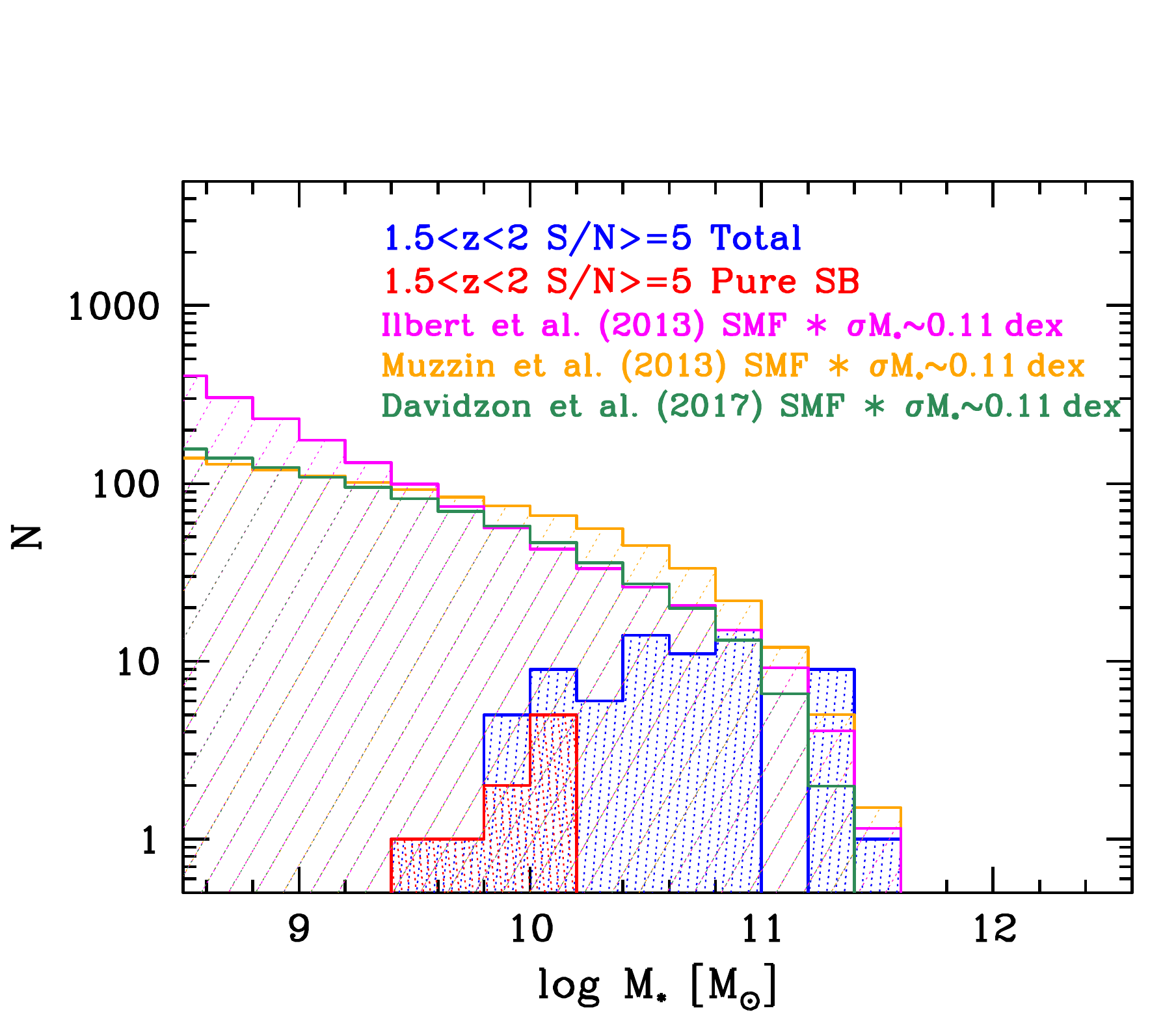}
\includegraphics[width=0.32\textwidth, trim=0 1cm 0 1.8cm]{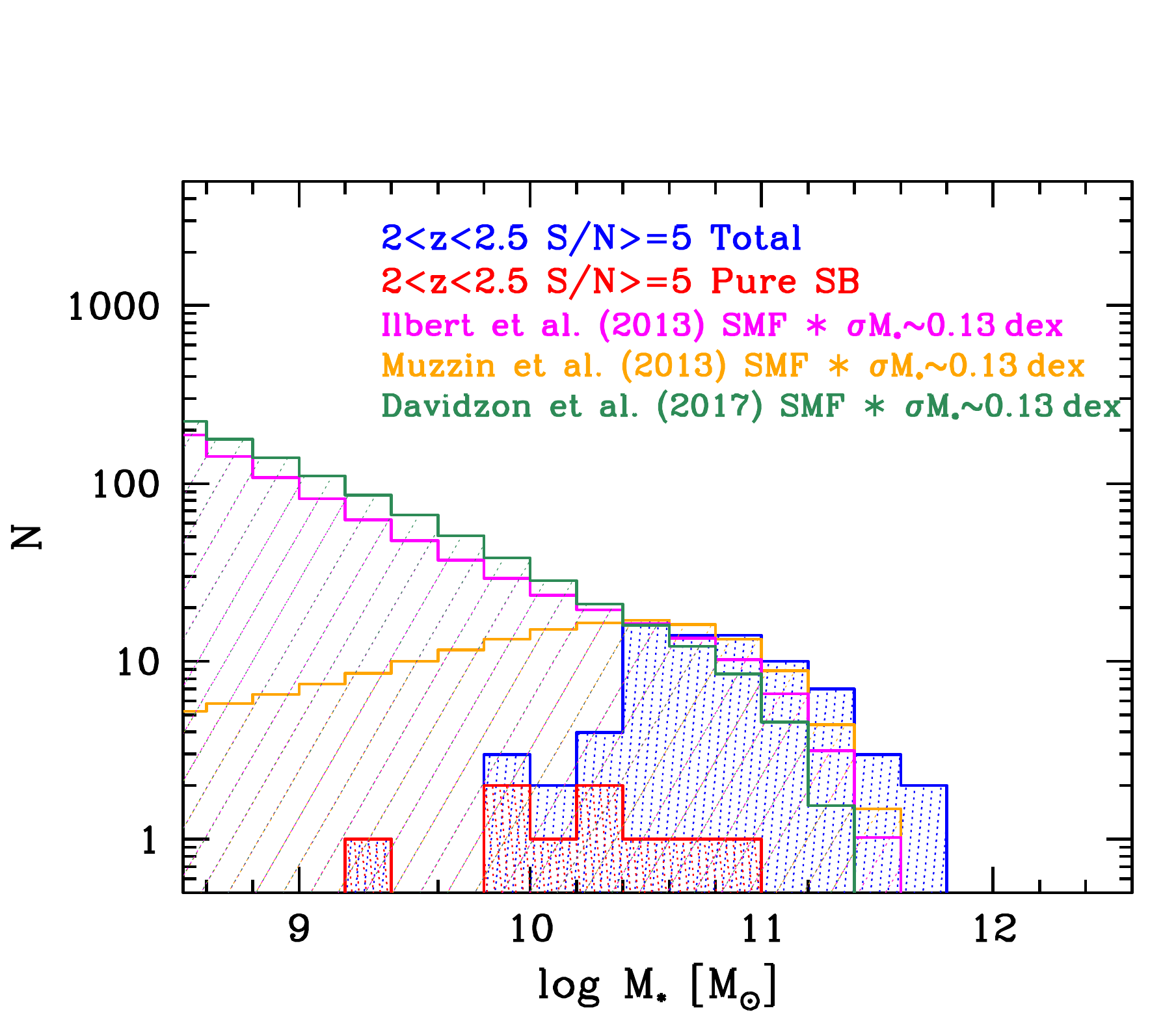}
\includegraphics[width=0.32\textwidth, trim=0 1cm 0 1.8cm]{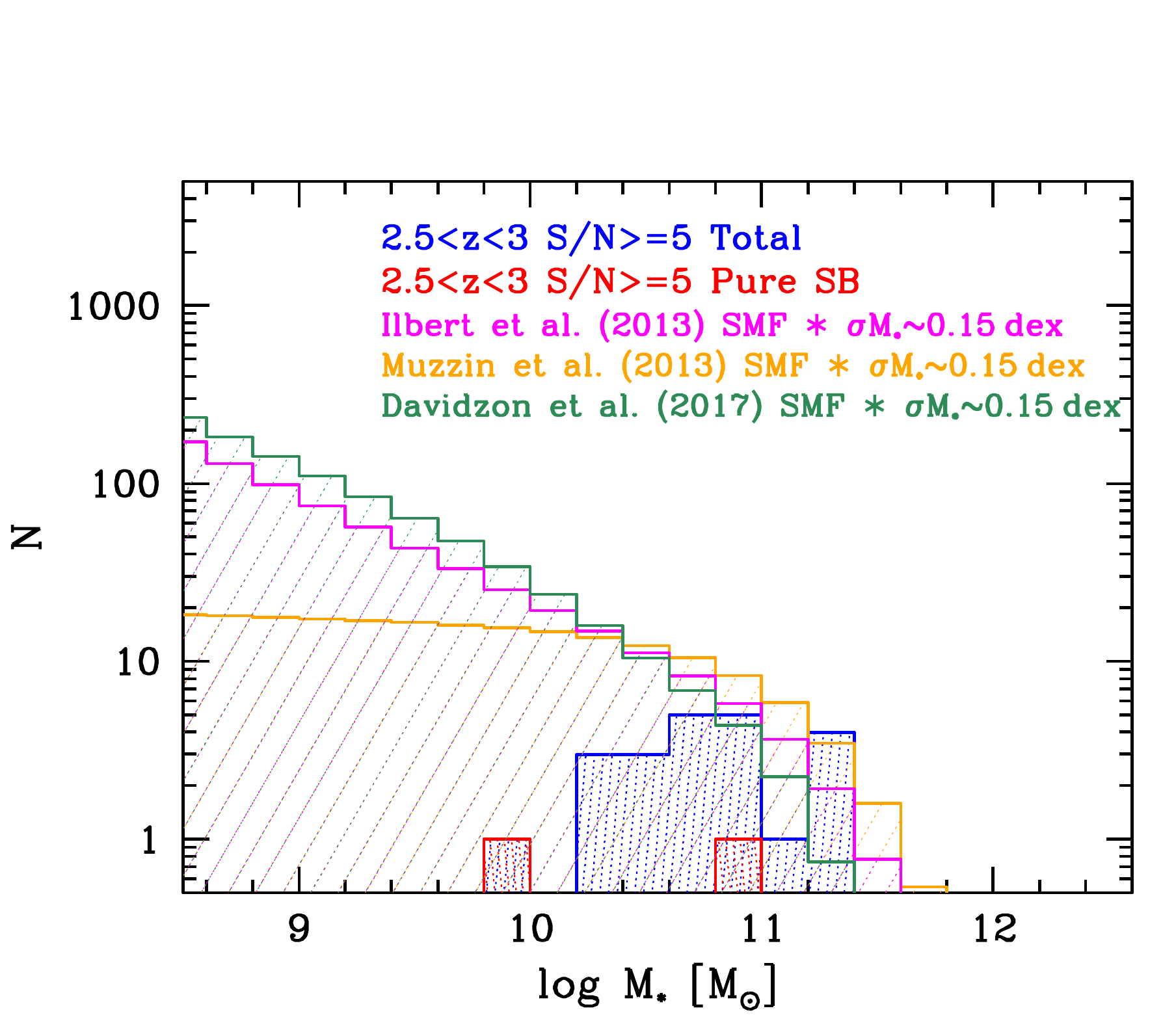}
\includegraphics[width=0.32\textwidth, trim=0 0.2cm 0 1.8cm]{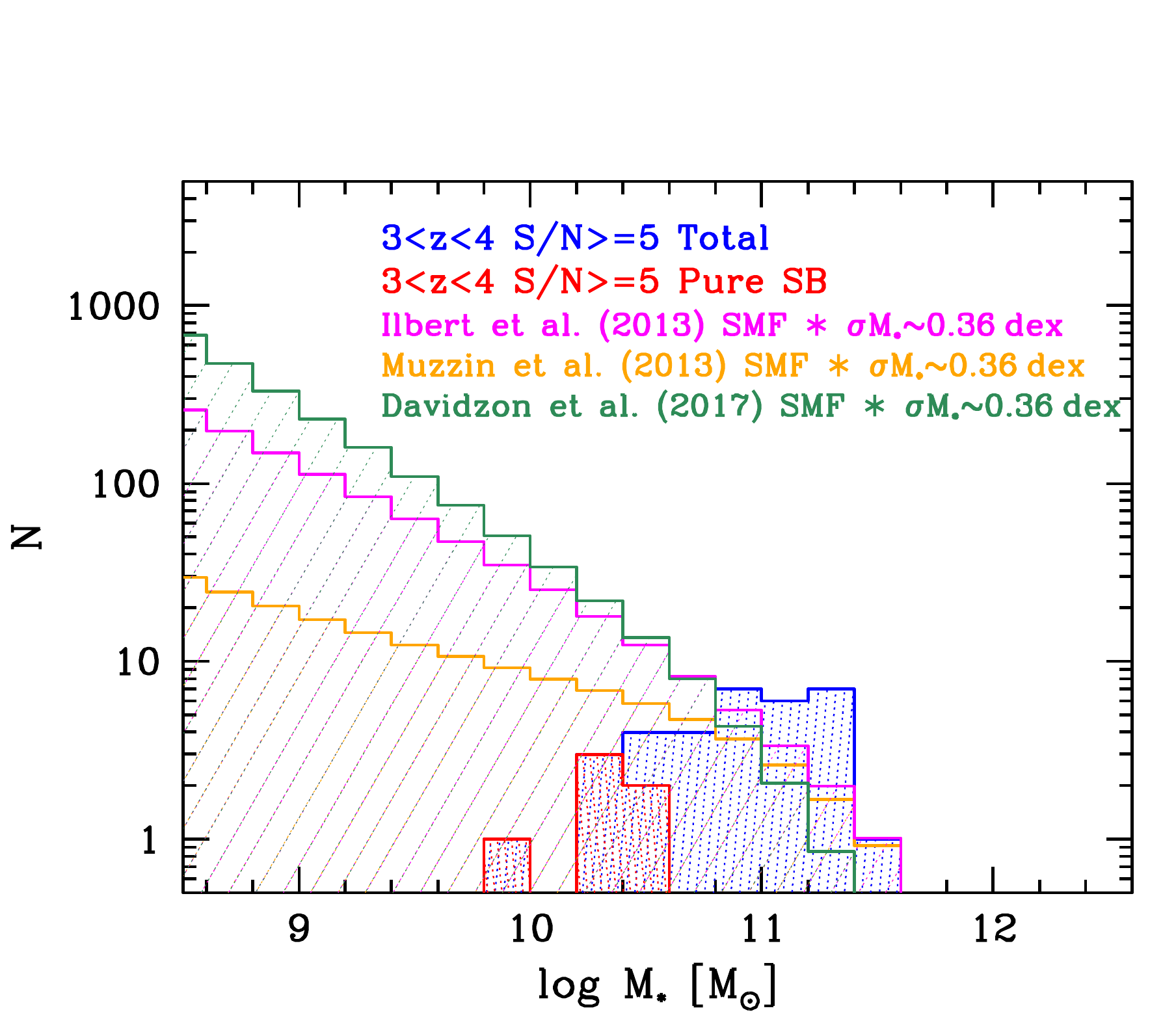}
\includegraphics[width=0.32\textwidth, trim=0 0.2cm 0 1.8cm]{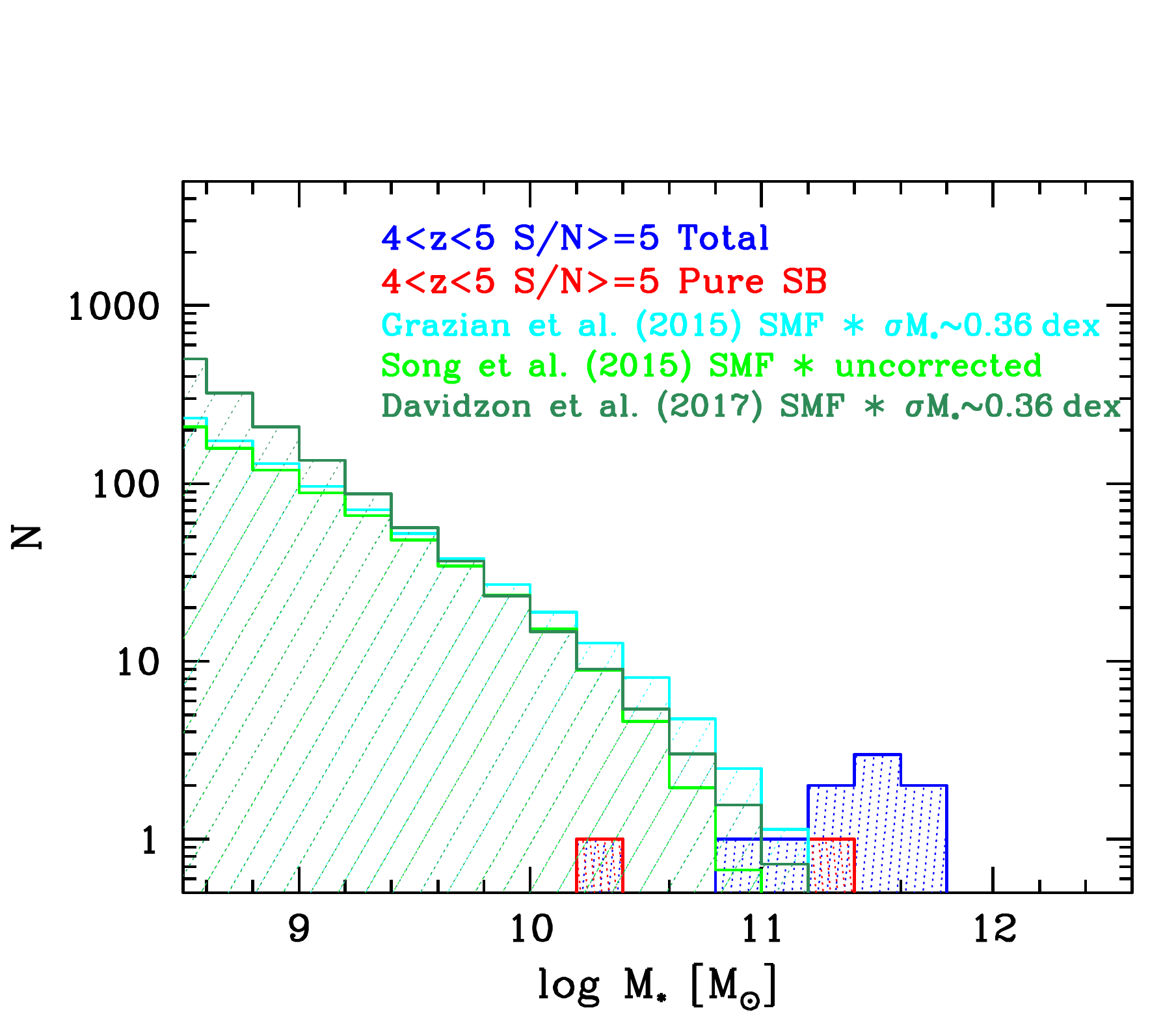}
\includegraphics[width=0.32\textwidth, trim=0 0.2cm 0 1.8cm]{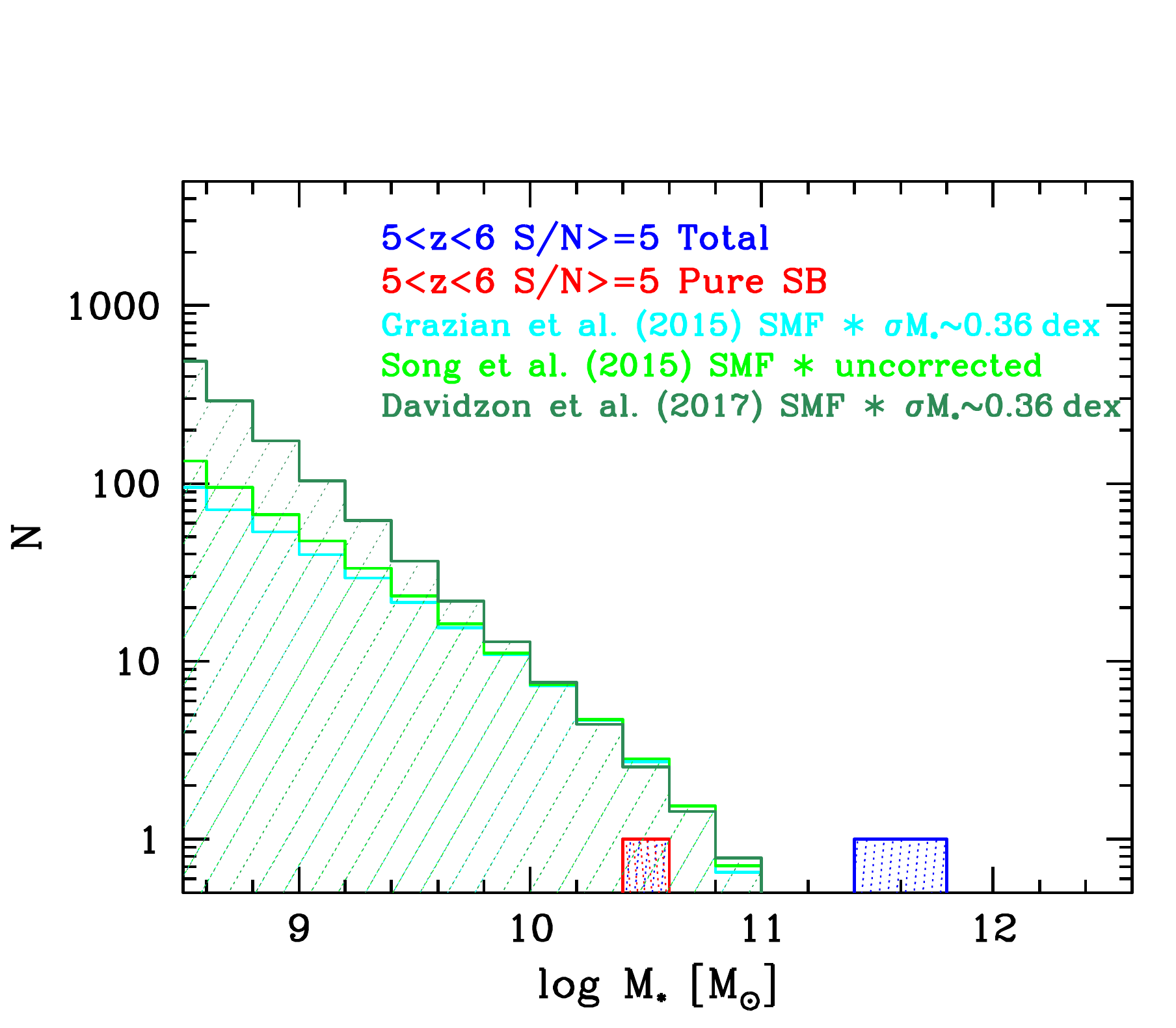}
\caption{%
    Stellar mass histograms of FIR+mm detected sources within the 134 arcmin$^2$ \goodArea{} in 9 redshift bins (blue histograms). \REVISED{This figure illustrates how we estimate the incompleteness of our sample 
at each redshift.}
    The red histograms indicate sources classified as pure starbursts (SB) in our SED fitting. 
    The empirical stellar mass functions (SMFs) of \citet{Ilbert2013}, \citet{Muzzin2013}, \citet{Grazian2015}, \citet{Song2016} and \citet{Davidzon2017} are shown as the magenta, yellow, cyan, light green and dark green histograms, respectively. 
    These SMFs are derived for star-forming galaxies; they have been \MINORREREVISED[converted to]{expressed for} a Chabrier IMF when necessary, and convolved with typical stellar mass uncertainties ($\sigma_{M_{*}}$) as indicated by the labels in each panel, following Appendix A of \citet{Ilbert2013} (except for \citet{Song2016}, which is the directly observed SMF). 
    The SMFs of \citet{Ilbert2013} and \citet{Muzzin2013} extend up to redshift 4, while that of \citet{Grazian2015}, \citet{Song2016} and \citet{Davidzon2017} probe higher redshifts. \citealt{Davidzon2017} does not fully cover the highest redshift bin; there we show a simple linear redshift extrapolation. 
    \label{Plot_Mstar_Histogram_z_all_panels}%
}
\end{figure*}

%
%
\begin{figure*}
\centering
\includegraphics[width=0.32\textwidth, trim=0 1cm 0 2cm]{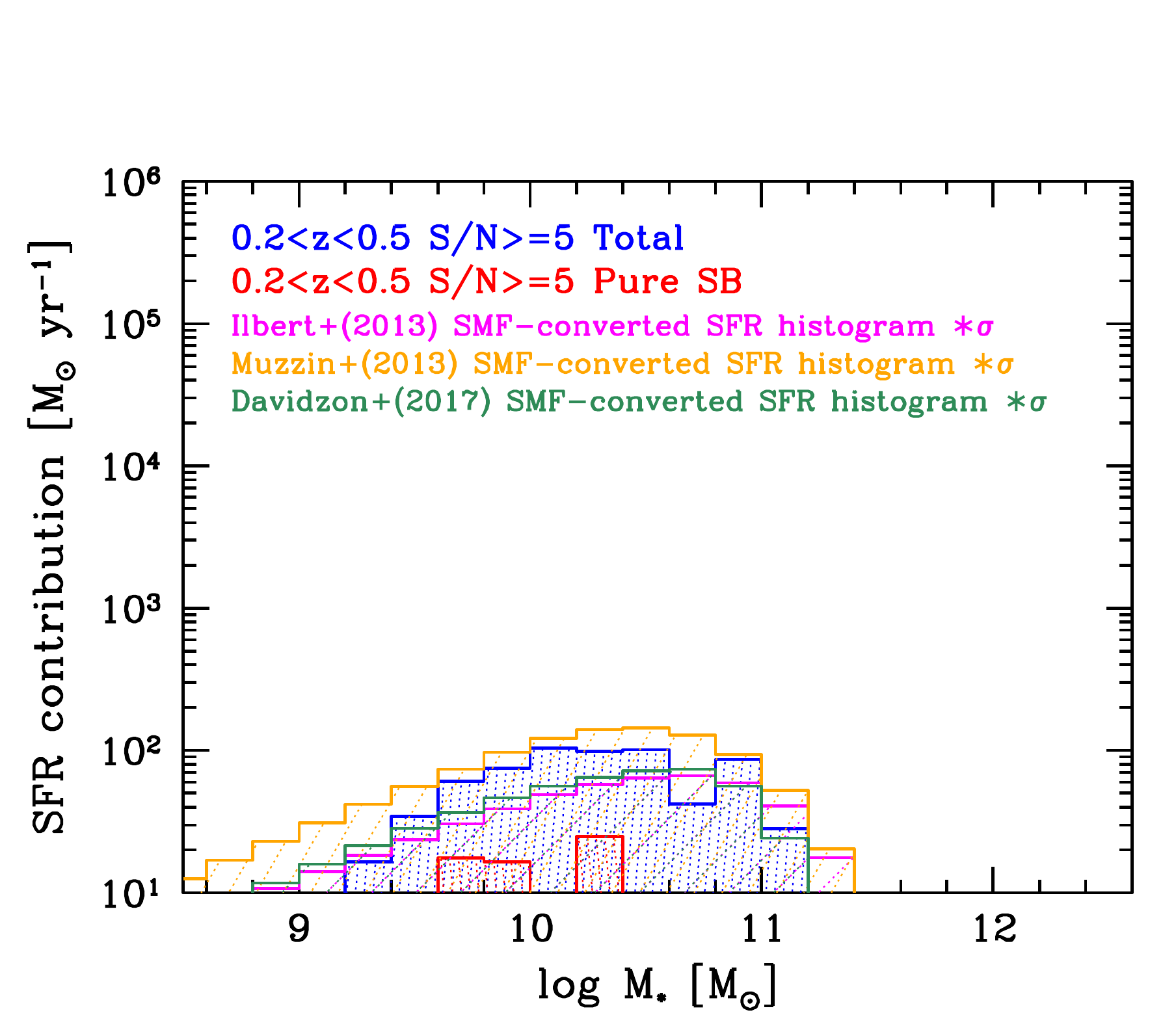}
\includegraphics[width=0.32\textwidth, trim=0 1cm 0 2cm]{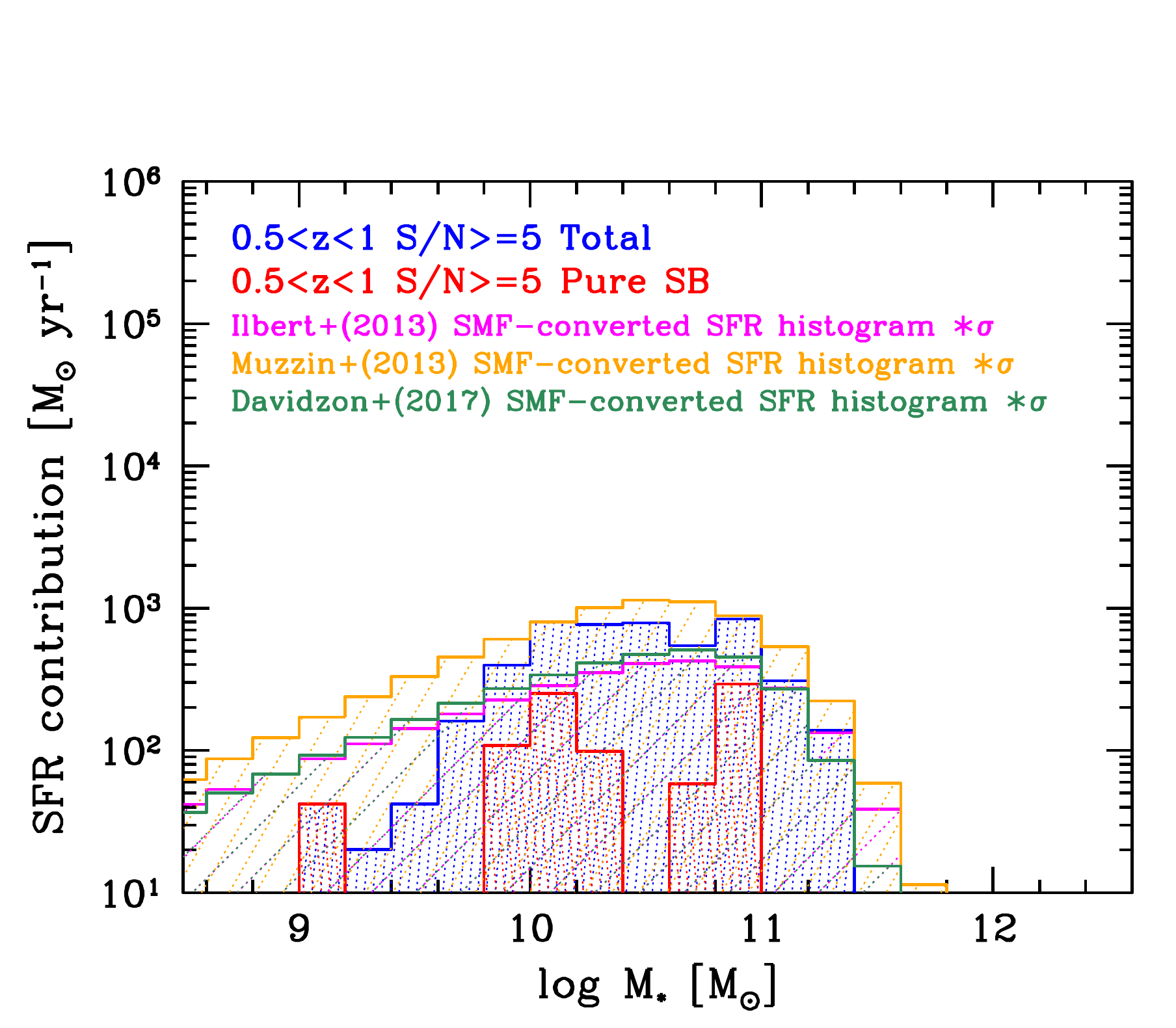}
\includegraphics[width=0.32\textwidth, trim=0 1cm 0 2cm]{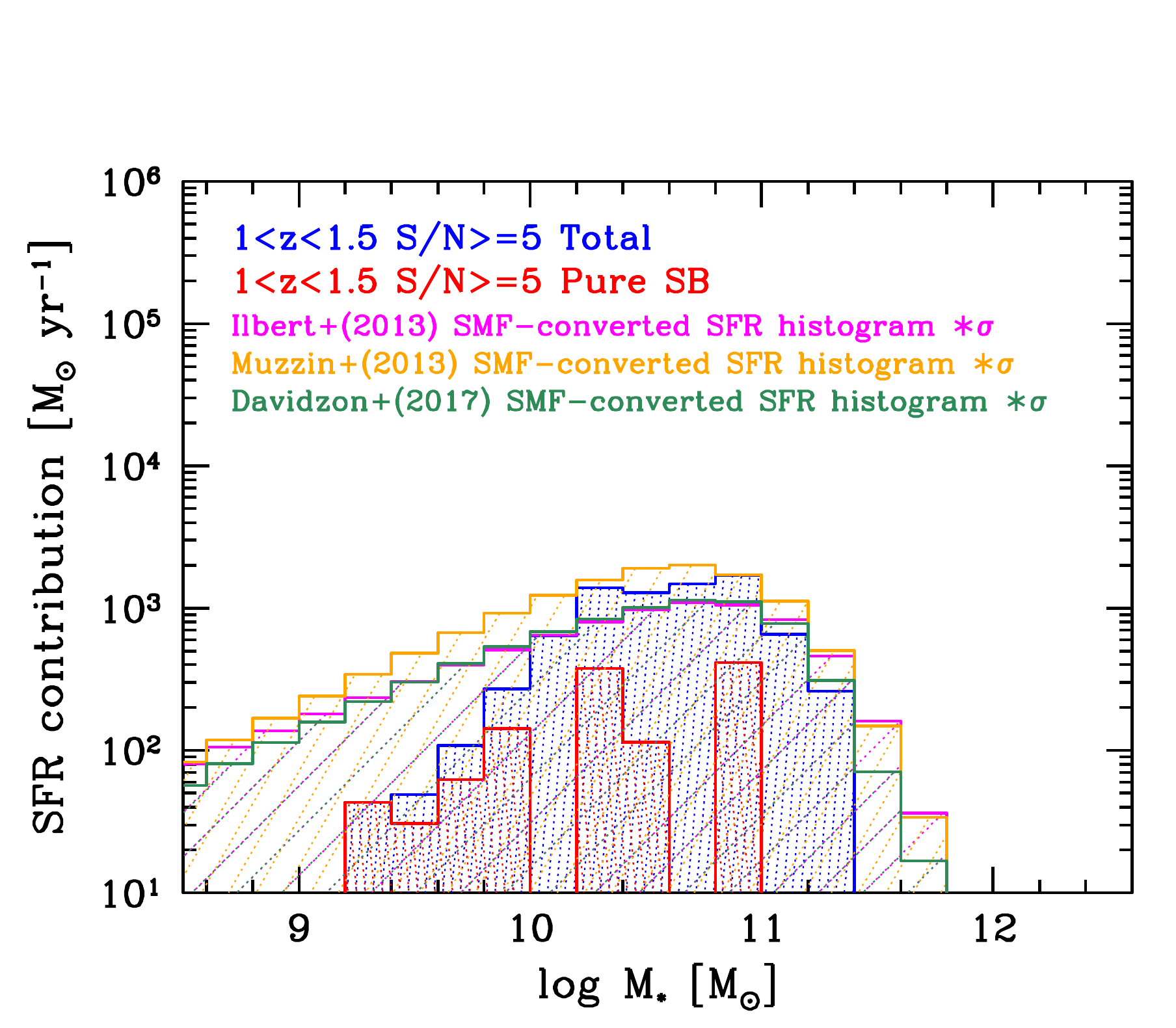}
\includegraphics[width=0.32\textwidth, trim=0 1cm 0 2cm]{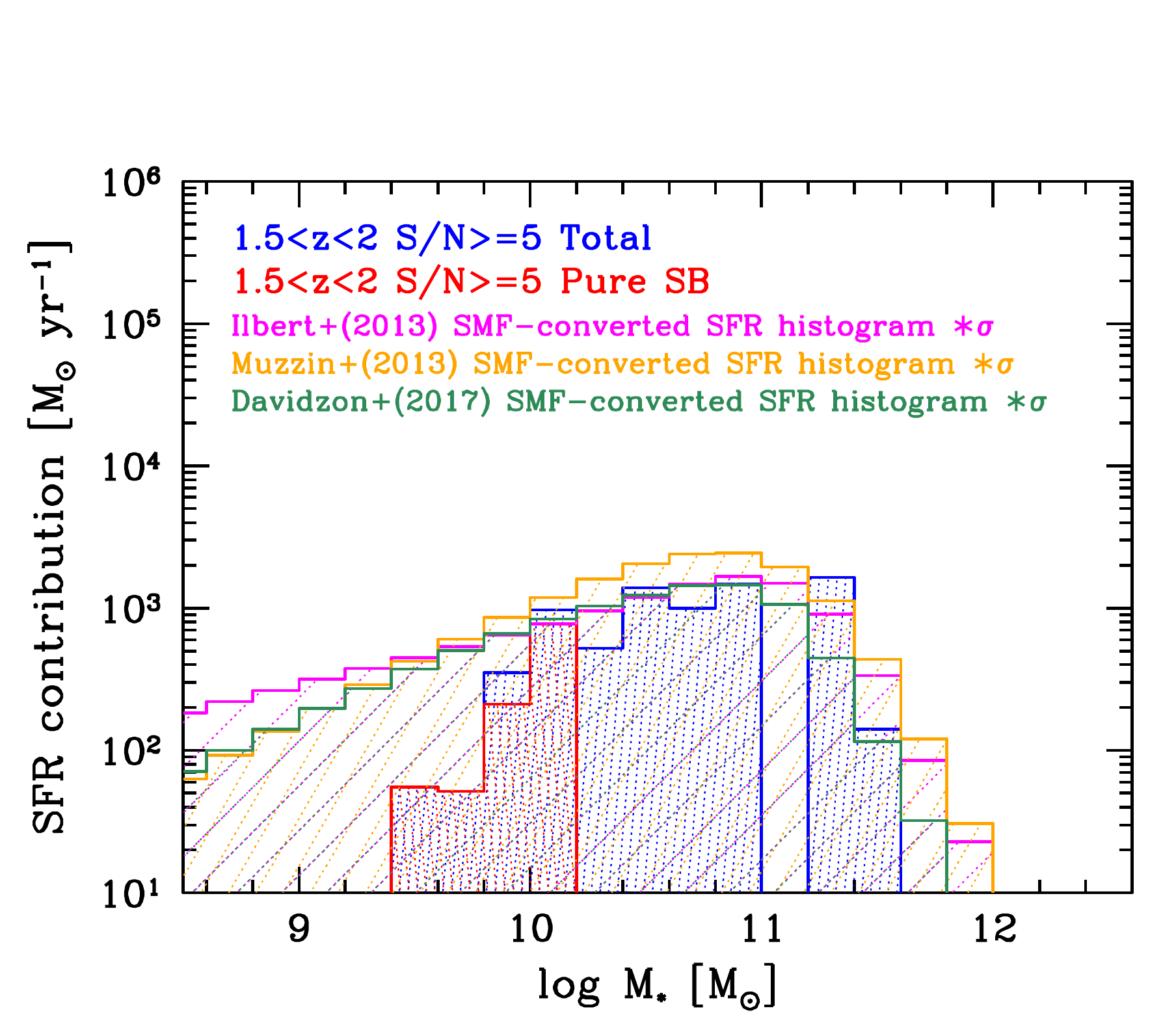}
\includegraphics[width=0.32\textwidth, trim=0 1cm 0 2cm]{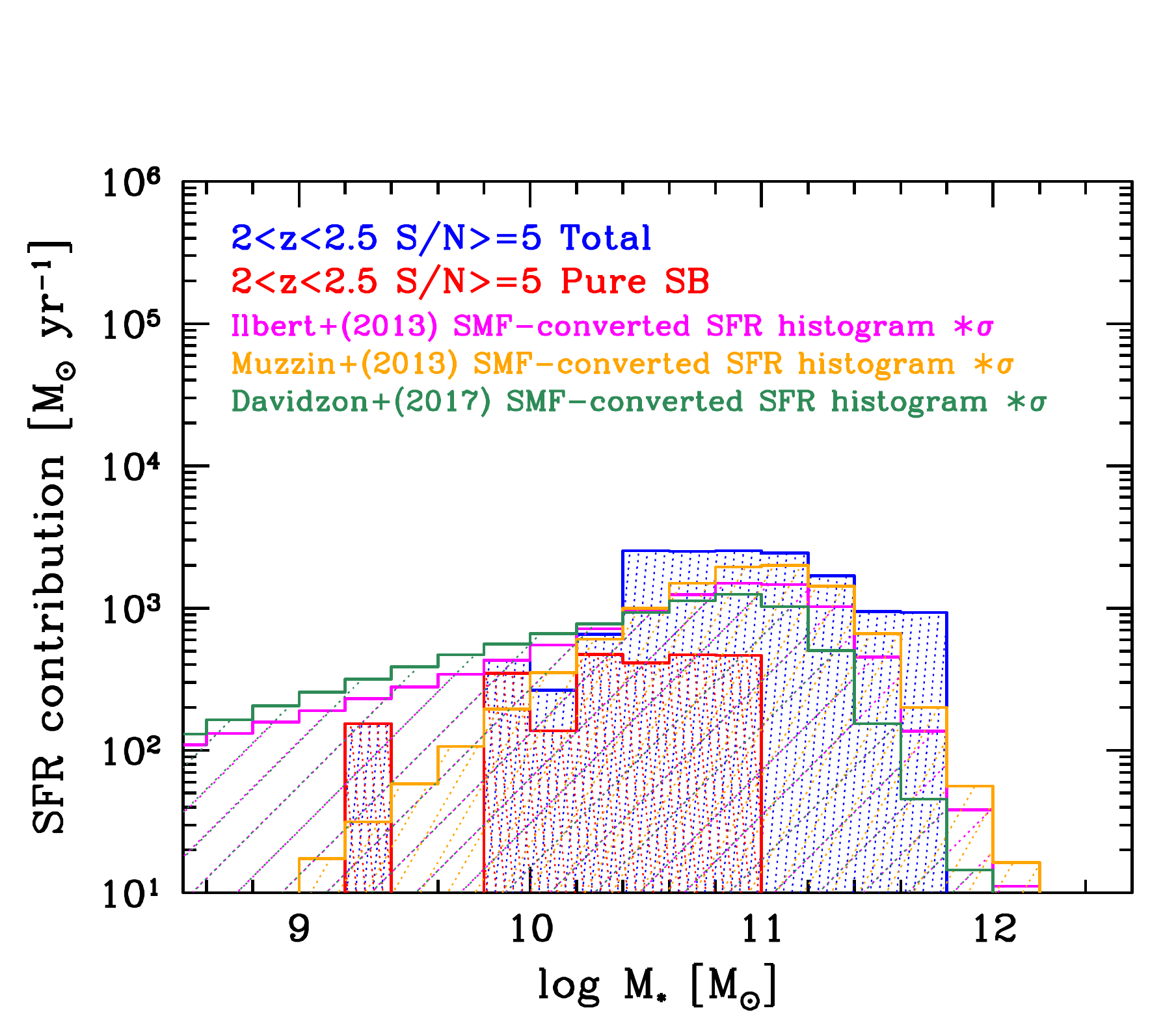}
\includegraphics[width=0.32\textwidth, trim=0 1cm 0 2cm]{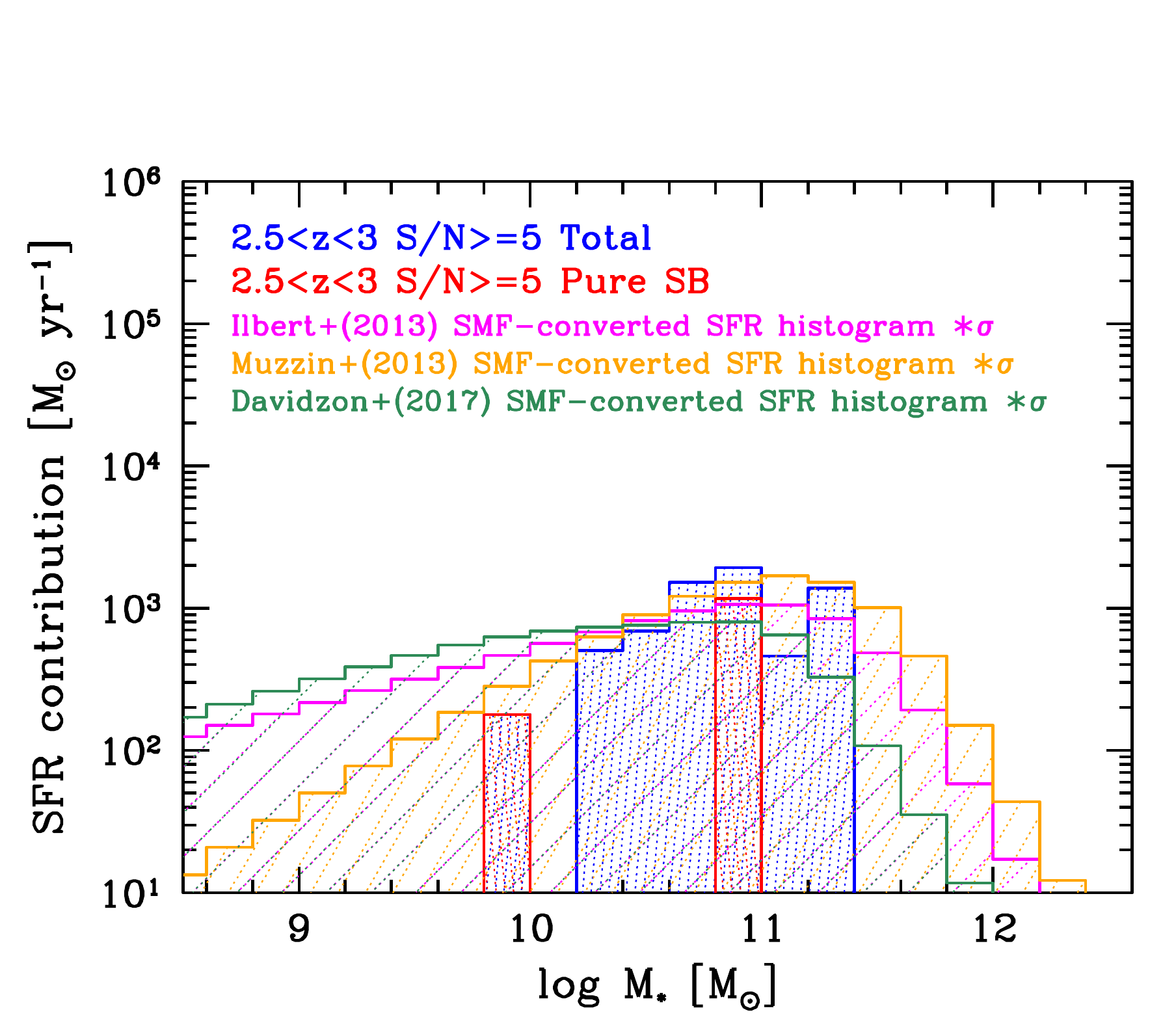}
\includegraphics[width=0.32\textwidth, trim=0 0 0 2cm]{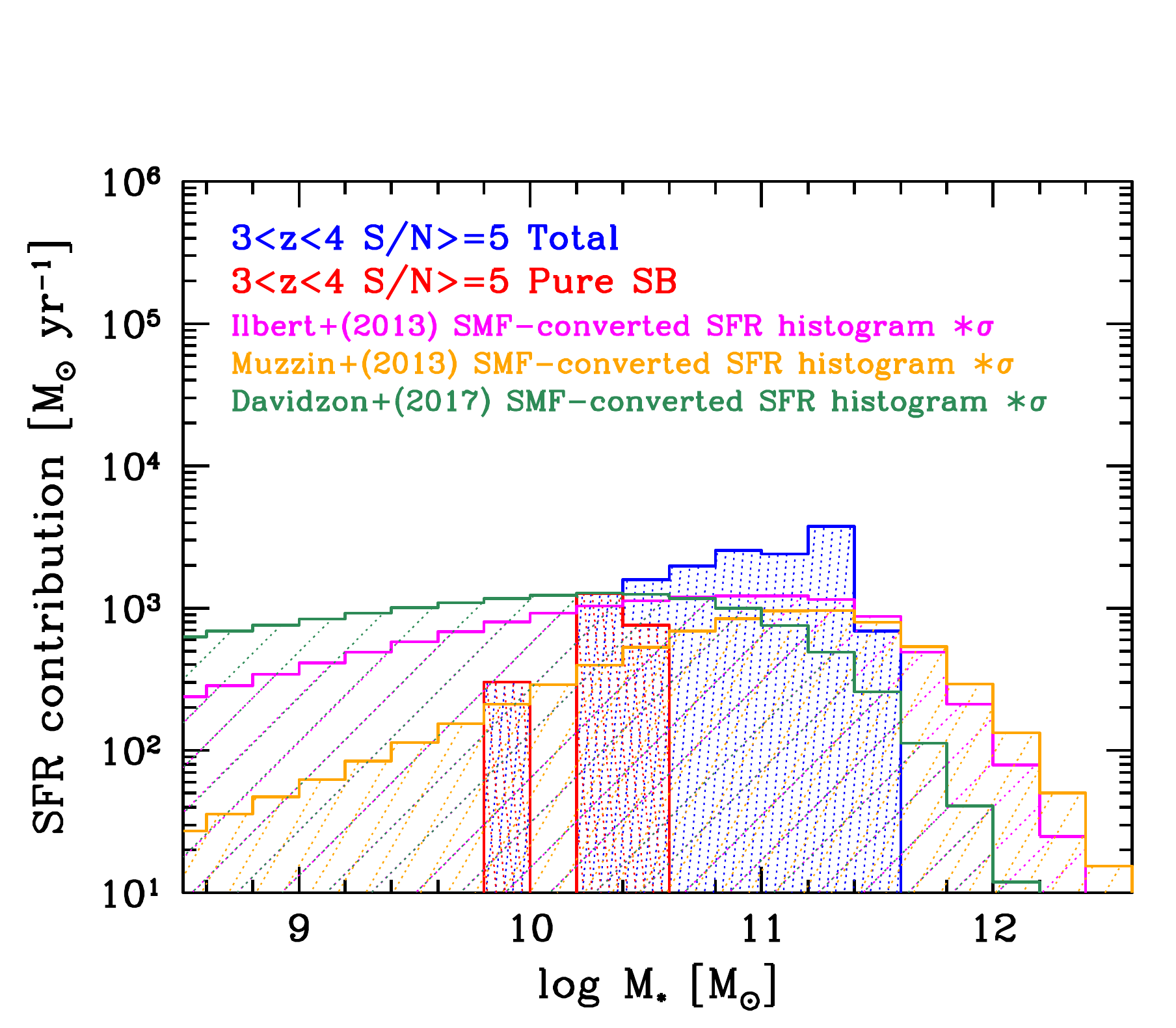}
\includegraphics[width=0.32\textwidth, trim=0 0 0 2cm]{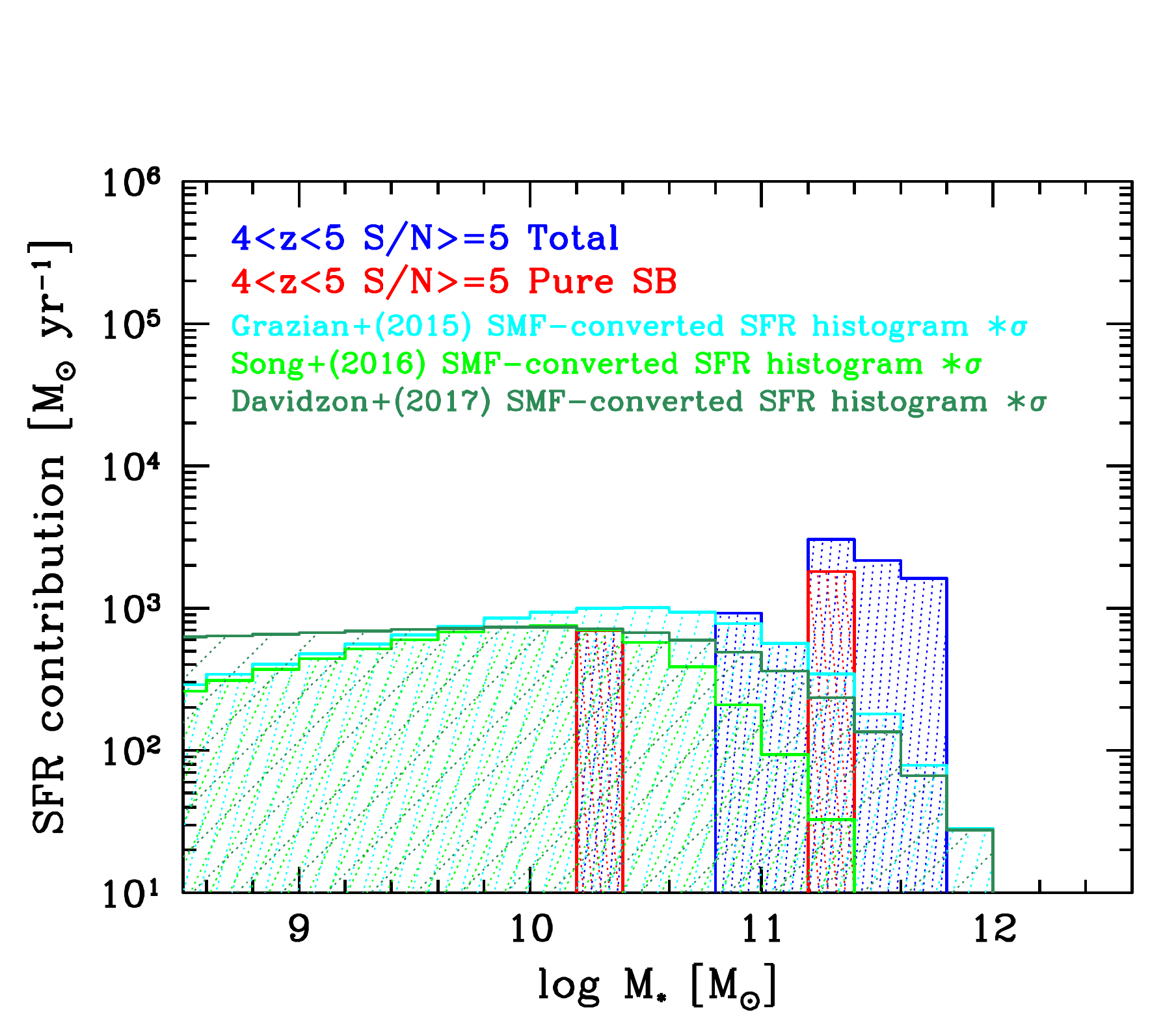}
\includegraphics[width=0.32\textwidth, trim=0 0 0 2cm]{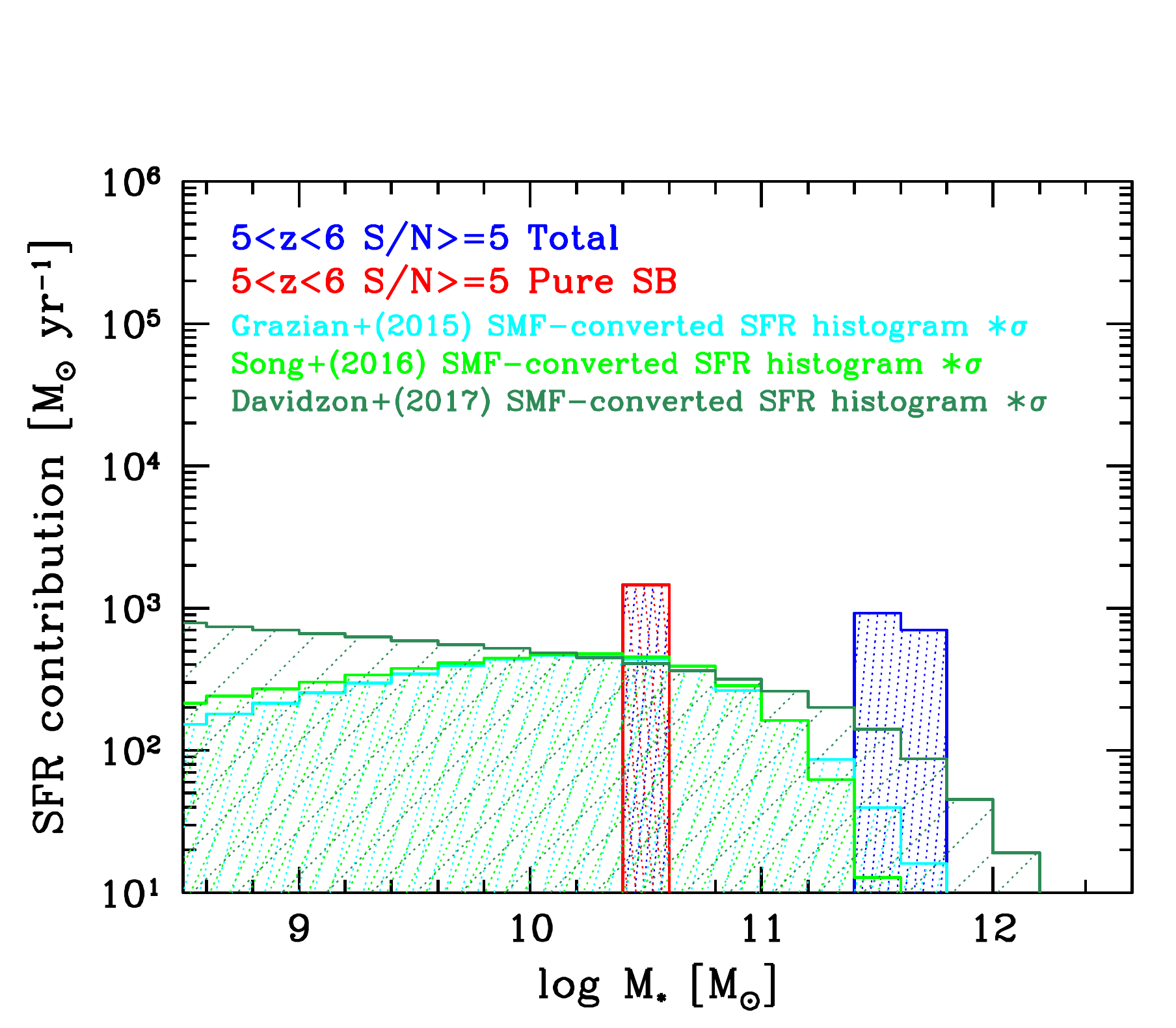}
\caption{%
    \REVISED{%
    As in Fig.~\ref{Plot_Mstar_Histogram_z_all_panels}, but showing the measured SFR contribution from each mass bin.}
    We convert the SMF histograms into SFR histograms by multiplying by the specific-SFR defined by MS of \citet{Sargent2014} 
    (see Section~\ref{Section_CSFRD}). This figure shows the direct comparison between the data and the empirical expectations from the SMFs without any renormalization. The re-normalized comparison, used to assess completeness correction, can be seen in Fig.~\ref{Plot_Mstar_SFR_Contribution_Renormalized}.
    \label{Plot_Mstar_SFR_Contribution_z_bin_all_panels}%
}
\end{figure*}


\subsection{Comparing with literature stellar mass functions (SMFs)}
\label{Section_Mstar_Histograms}

To evaluate the incompleteness of our FIR+mm sample
we consider sources in each redshift bin, further dividing them into bins of stellar masses in Fig.~\ref{Plot_Mstar_Histogram_z_all_panels}, and compare their distributions with various star-forming galaxy SMFs from the literature.
%

The literature SMFs are convolved to a common stellar mass uncertainty ($\sigma_{M_{*}}$, see labels in each panel) in order to account for measurement errors when comparing with our observed stellar mass distributions (blue and red histograms in the figure, representing the full FIR+mm sample and the SB sub-sample respectively). This is done by first using the Eddington bias-corrected (i.e., intrinsic) SMFs in these literature, then convolving them with a common $\sigma_{M_{*}}$ (which is redshift-dependent), following the approach described in Appendix~A of \citet{Ilbert2013} (except for \citealt{Song2016} where the authors have not made that correction). We multiply by the \goodArea{} co-moving volume to convert SMF (in units of number per volume) to the absolute number of galaxies within each stellar mass bin and each redshift bin. 


The literature SMFs do not always agree with each other, \REVISED{especially at the low-mass end and at high redshifts}, but in general they agree with our observed histograms up to the $3<z<4$ bin without doing any fitting. \cite{Muzzin2013} presented two best-fits to the SMF at each redshift: one with fixed slope at the low-mass end, the other with non-fixed slope. Their fixed-slope SMFs agree better with the other two $z<4$ SMFs in the literature, while the non-fixed-slope SMFs show large discrepancies at the low-mass end compared to other SMFs at $z > 2$. In order to probe the variety of SMFs, we show their non-fixed-slope SMFs in the figures. 
Note that the SMF of \citet{Davidzon2017} does not fully cover our highest redshift bin. We have linearly extrapolated \MINORREREVISED[results from]{} their SMFs at $3.5<z<4.5$ and $4.5<z<5.5$ to our bin at $5<z<6$ for purpose of comparison. Because their SMF is even higher at $4.5<z<5.5$ than at $3.5<z<4.5$, the extrapolated SMF is also higher. 

To evaluate the stellar mass range (shown as the shaded-area in Fig.~\ref{Plot_Mstar_SFR_z}) over which our sample becomes incomplete in each redshift bin, we estimate at which value of stellar mass the number of observed galaxies in our sample starts to be substantially smaller than the number predicted by each SMF.

In the 3 bins with the highest redshifts in Fig.~\ref{Plot_Mstar_Histogram_z_all_panels}, our FIR+mm sample tends to have more galaxies at the highest masses than would be predicted by the literature SMFs. In the $3.0<z<4.0$ panel, the excess is not large, but in the $4.0<z<5.0$ panel, our FIR+mm sample has 8 more sources than is predicted by the SMFs of \citet{Grazian2015} and \citet{Song2016}, or 6 more sources than predicted by the \citet{Davidzon2017} SMF. Note that in the GOODS-N field there is a well known $z=4.055$ proto-cluster which includes three FIR+mm sources detected in our catalogs: GN20, GN20.2a and GN20.2b \citep{Daddi2009GN20,Tan2014}. 
Thus small number statistics \MINORREREVISED[are]{is} already an important factor at these high redshifts. 

Moreover, \citet{Grazian2015} discussed the importance of cosmic variance \REVISED{(i.e., clustering of galaxies)} between the GOODS-S and UDS fields in their Fig.~3 \REVISED{and Section 4.2.4}. The SMF derived from UDS sources is much higher than the SMF derived from GOODS-S sources at the high-mass end;  the UDS SMF agrees better with the SMFs of \citet{Ilbert2013} and \citet{Muzzin2013} at similar redshifts. Hence cosmic variance may also be responsible for the excess in the highest redshift bins found in this work. 

Our forthcoming work in the 2 square degree COSMOS field using the same approach (S. Jin et al., 2017, in preparation) will provide results that are less affected by cosmic variance in the high-mass and high-redshift bins.



\subsection{Comparing with SFR histograms} 
\label{Section_SFR_Histograms}

In Fig.~\ref{Plot_Mstar_SFR_Contribution_z_bin_all_panels}, we analyze the SFR contribution of our sample of galaxies in bins of stellar mass. 
We divide our sources into several stellar mass bins 
and then compute the sum of SFRs in each bin. 

For comparison, we use SMFs from the literature, first multiplying by the co-moving volume to obtain the absolute number of galaxies, then multiplying by the SFR per unit mass predicted by the MS correlation to get the SFR contribution for each stellar mass bin (\MINORREREVISED[$\int \mathrm{SMF} \times \mathrm{SFR}_{\mathrm{MS}}$]{SMF-converted SFR}). In this way, we are able to see how the SFR of this FIR+mm sample is distributed as a function of stellar mass, and to estimate what fraction of the co-moving SFR density this sample contributes to the whole population of star-forming galaxies at each redshift up to 6. Here we show the results computed assuming the MS correlation of \citet{Sargent2014}, but the main results do not change substantially if the MSs of \citet{Bethermin2015} or \citet{Schreiber2015} were used instead. 

The shape of these SMF-converted SFR histograms usually displays a peak around $M_{\mathrm{star}}^{*}$.
In the three lowest redshift bins, the predicted SFR distribution based on the \citet{Muzzin2013} SMF has the best agreement with our observations, but implies a higher SFR contribution from the most massive galaxies ($M_{*}>10^{11.0}\,\mathrm{M}_{\odot}$), where we might be incomplete due to the limited volume of our survey. The SFR distribution predicted from the \citet{Davidzon2017} SMF matches our data well at the high-mass end, but is too low overall. The prediction based on the \citet{Ilbert2013} SMF falls in between the other two.

The FIR+mm sample is already accounting for most of the global SFR density at these redshifts.
In the intermediate redshift bins ($1.5<z<4.0$), the \citet{Ilbert2013} prediction gives the best fit to the normalization. The \citet{Muzzin2013} distribution gives a better fit to the shape at the low-mass end, which is probably due to the different stellar mass limit in \citet{Muzzin2013} (e.g., $M_{*,\mathrm{95\%\,lim.}} = 10^{10.54}\,\mathrm{M}_{\odot}$ at $2<z<2.5$) compared to that in \citet{Ilbert2013} (e.g., $M_{*,\mathrm{90\%\,lim.}} = 10^{10.01}\,\mathrm{M}_{\odot}$ at $2<z<2.5$), keeping in mind the current difficulties in determining the low-mass end slope of SMFs (see Appendix~C of \citealt{Muzzin2013}). 
In the two highest redshift bins, the SFR distribution predicted from the \citet{Davidzon2017} SMF seems to be the closest one to the observed SFR histogram, 
but the SFR excess of our FIR+mm sample is still obvious.

​

%
%
\begin{figure*}
\begin{center}
\includegraphics[width=0.8\textwidth, trim=0 0 0 0]{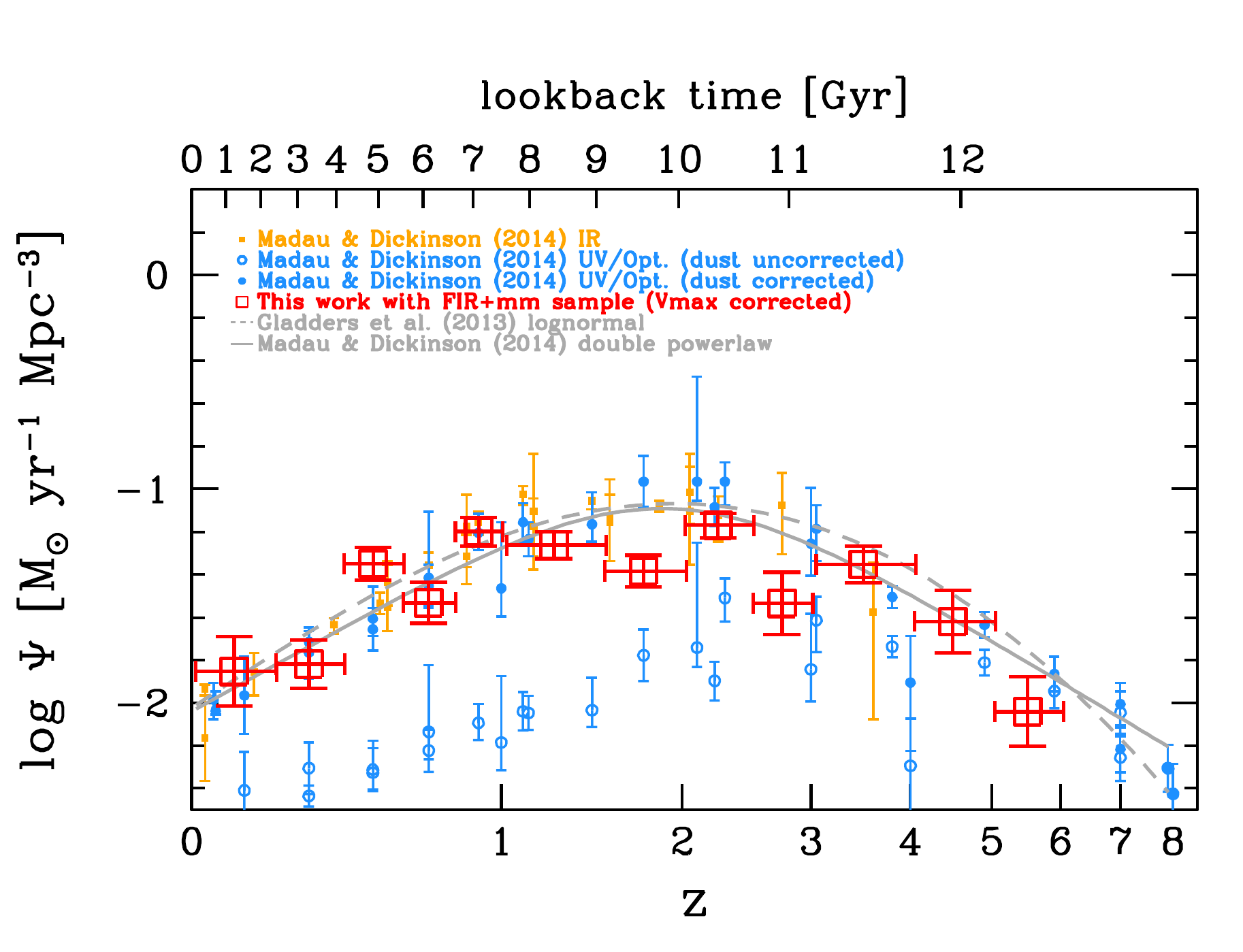}
\caption{%
    The co-moving SFR densities ($\Psi$) of our FIR+mm sample (red symbols) compared with values from the literature. 
    The horizontal error bars represent the redshift bin ranges, and vertical error bars indicate the uncertainties. 
    The measurement details (i.e., 1/$V_{max}$ and bootstrapping methods) are discussed in Section~\ref{Section_CSFRD}. 
    The yellow solid squares are the IR-based measurements from \citet{Madau2014a}. Blue filled circles are the \MINORREREVISED[UV/optical-based]{rest-frame UV-based} dust attenuation corrected measurements, while the blue open circles are \MINORREREVISED[UV/optical-]{rest-frame UV} measurements with no dust attenuation correction. 
    The dashed gray curve shows the best-fit log-normal function of \citet{Gladders2013}, while the solid gray curve is the best-fit double power-law function given in \citet{Madau2014a}. 
    \label{Plot_Cosmic_SFR}%
}
\end{center}
\end{figure*}

%
%
\begin{figure}
\begin{center}
\includegraphics[width=0.49\textwidth, trim=0 6mm 0 10mm]{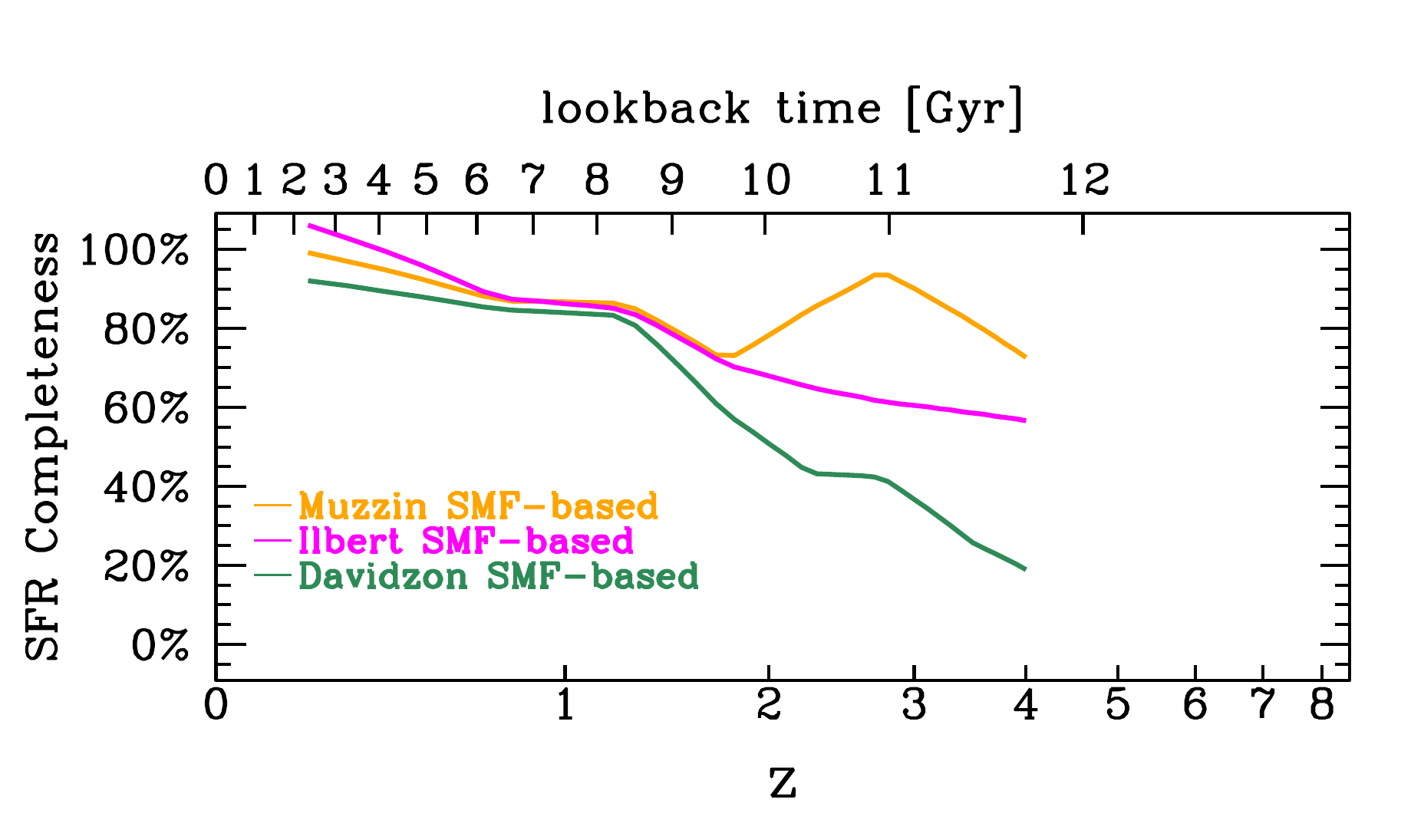}
\caption{%
    The completeness in SFR for our FIR+mm sample, estimated based on fitting/re-normalizing the SMFs. See details in Section~\ref{Section_CSFRD_Correction} and Appendix Fig.~\ref{Plot_Mstar_SFR_Contribution_Renormalized}. 
    \label{Plot_Cosmic_SFR_Correction}%
}
\end{center}
\end{figure}

%
%
\begin{figure*}
\begin{center}
\includegraphics[width=0.8\textwidth, trim=0 0 0 0]{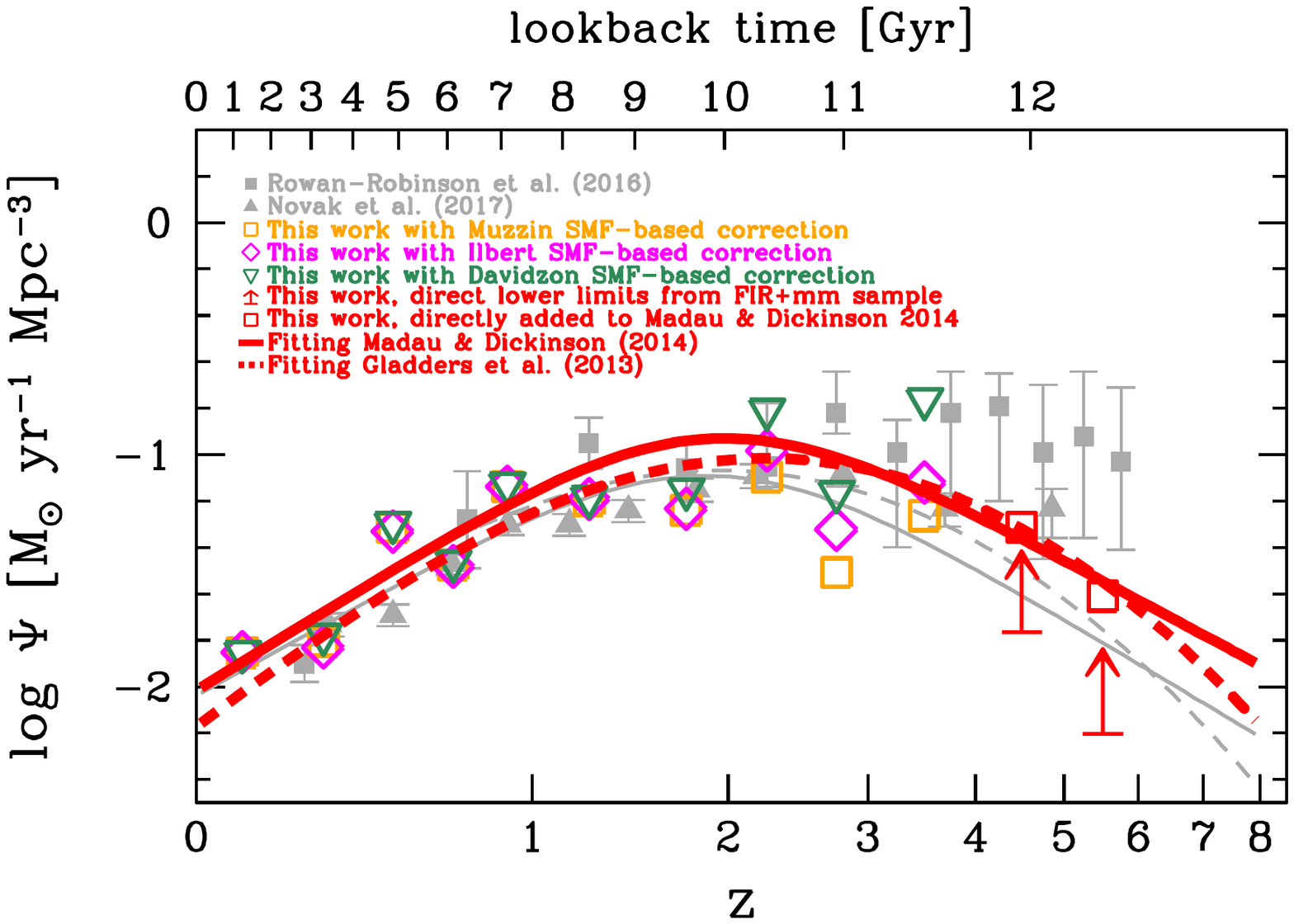}
\caption{%
    The incompleteness-corrected cosmic SFR densities based on our FIR+mm sample. We estimate the incompleteness using several different SMFs from the literature (\citealp{Muzzin2013}; \citealp{Ilbert2013}; \citealp{Davidzon2017}; \citealp{Grazian2015}; \citealp{Song2016}) and assume the MS correlation fitted by \cite{Sargent2014}. 
    At $z<4$, we show the incompleteness-corrected cosmic SFR density derived from each SMF-based correction using different colors and symbols (see labels in the figure). 
    At $z>4$, we show lower limits as well as the summed total cosmic SFR density that combines our FIR+mm sample and the UV-based samples compiled by \cite{Madau2014a}. 
    See details in Section~\ref{Section_CSFRD_Correction}. 
    The two gray curves are the same as in Fig.~\ref{Plot_Cosmic_SFR}, but here we also show two red thick curves \REVISED[]{(solid curve --- double power-law function as in \cite{Madau2014a}; dashed curve --- log-normal function as in \cite{Gladders2013})} which are our new best-fit to all the colored \REVISED{(i.e., green, magenta, yellow and red)} data points (details in the text). 
    The grey data points show other results from the recent literature derived from far-IR or radio data extending out to $z = 5$ to 6 (\citealp{RowanRobinson2016}; \citealp{Novak2017}).
    All values assume a Chabrier IMF (or are converted to Chabrier IMF when needed). 
    \label{Plot_Cosmic_SFR_Corrected}%
}
\end{center}
\end{figure*}

\begin{table*}

\begin{center}

\caption{ %
    GOODS-N FIR+mm Star Formation Rate Densities %
    \label{Table_3} %
}

\begin{tabular*}{0.75\textwidth}{ @{\extracolsep{\fill}} c c c c c c c c }

        \hline
        
        $z_{\mathrm{low}}$                              &
        $z_{\mathrm{high}}$                             &
        $\Psi_{\mathrm{SFR,obs},\textnormal{UV+IR}}$    &
        $\Psi_{\mathrm{SFR,obs,\textnormal{IR-only}}}$  &
        ${\sigma}_{\Psi_{\mathrm{SFR,obs}}}$            &
        $\Psi_{\mathrm{SFR},\mathrm{corr.}}$            &
        SMF$_{\mathrm{corr.}}$                          &
        $f_{\mathrm{comp.}}$                            \\[2pt]
        
        
        \hline
        
            0.0 &     0.2 &  -1.852 & -2.00 &   0.164 &  -1.852 &       Muzzin &        100.0\% \\ 
            \multicolumn{5}{c}{}                      &  -1.852 &       Ilbert &        100.0\% \\ 
            \multicolumn{5}{c}{}                      &  -1.852 &     Davidzon &        100.0\% \\ 
            0.2 &     0.4 &  -1.818 & -1.93 &   0.114 &  -1.805 &       Muzzin &         97.0\% \\ 
            \multicolumn{5}{c}{}                      &  -1.830 &       Ilbert &        102.8\% \\ 
            \multicolumn{5}{c}{}                      &  -1.776 &     Davidzon &         90.7\% \\ 
            0.4 &     0.6 &  -1.349 & -1.42 &   0.077 &  -1.315 &       Muzzin &         92.5\% \\ 
            \multicolumn{5}{c}{}                      &  -1.331 &       Ilbert &         96.0\% \\ 
            \multicolumn{5}{c}{}                      &  -1.293 &     Davidzon &         88.0\% \\ 
            0.6 &     0.8 &  -1.531 & -1.56 &   0.096 &  -1.476 &       Muzzin &         88.1\% \\ 
            \multicolumn{5}{c}{}                      &  -1.482 &       Ilbert &         89.3\% \\ 
            \multicolumn{5}{c}{}                      &  -1.462 &     Davidzon &         85.4\% \\ 
            0.8 &     1.0 &  -1.199 & -1.22 &   0.067 &  -1.137 &       Muzzin &         86.8\% \\ 
            \multicolumn{5}{c}{}                      &  -1.137 &       Ilbert &         86.8\% \\ 
            \multicolumn{5}{c}{}                      &  -1.125 &     Davidzon &         84.2\% \\ 
            1.0 &     1.5 &  -1.263 & -1.30 &   0.062 &  -1.199 &       Muzzin &         86.4\% \\ 
            \multicolumn{5}{c}{}                      &  -1.191 &       Ilbert &         84.8\% \\ 
            \multicolumn{5}{c}{}                      &  -1.183 &     Davidzon &         83.2\% \\ 
            1.5 &     2.0 &  -1.383 & -1.41 &   0.075 &  -1.239 &       Muzzin &         71.8\% \\ 
            \multicolumn{5}{c}{}                      &  -1.233 &       Ilbert &         70.8\% \\ 
            \multicolumn{5}{c}{}                      &  -1.150 &     Davidzon &         58.5\% \\ 
            2.0 &     2.5 &  -1.170 & -1.19 &   0.057 &  -1.098 &       Muzzin &         84.8\% \\ 
            \multicolumn{5}{c}{}                      &  -0.983 &       Ilbert &         65.0\% \\ 
            \multicolumn{5}{c}{}                      &  -0.806 &     Davidzon &         43.3\% \\ 
            2.5 &     3.0 &  -1.534 & -1.54 &   0.146 &  -1.509 &       Muzzin &         94.4\% \\ 
            \multicolumn{5}{c}{}                      &  -1.322 &       Ilbert &         61.5\% \\ 
            \multicolumn{5}{c}{}                      &  -1.161 &     Davidzon &         42.3\% \\ 
            3.0 &     4.0 &  -1.353 & -1.38 &   0.086 &  -1.263 &       Muzzin &         81.3\% \\ 
            \multicolumn{5}{c}{}                      &  -1.120 &       Ilbert &         58.6\% \\ 
            \multicolumn{5}{c}{}                      &  -0.762 &     Davidzon &         25.7\% \\ 
            4.0 &     5.0 &  -1.619 & -1.64 &   0.145 &  -1.313 &         MD14 &    $\sim$49.4\% \\ 
            5.0 &     6.0 &  -2.040 & -2.07 &   0.164 &  -1.610 &         MD14 &    $\sim$37.1\% \\ 
        
        \hline
        
        \vspace{-0.5ex}

\end{tabular*}

\begin{minipage}{0.75\textwidth}

    The measured SFR density ($\Psi_{\mathrm{SFR,\,obs.},\textnormal{UV+IR}}$ and $\Psi_{\mathrm{SFR,obs.},\textnormal{IR-only}}$, 
    in unit of $\mathrm{M}_{\odot}\,\mathrm{yr}^{-1}\,\mathrm{Mpc}^{-3}$) 
    and uncertainty (${\sigma}_{\Psi_{\mathrm{SFR,obs.}}}$) for the observed FIR+mm galaxies in this work 
    as described in Section~\ref{Section_CSFRD}, 
    as well as the incompleteness-corrected cosmic SFR density 
    ($\Psi_{\mathrm{SFR,\,corr.}}$) and the completeness percentage 
    in SFR ($f_{\mathrm{comp.}}$, i.e., 
    $=10^{(\Psi_{\mathrm{SFR,\,obs.},\textnormal{UV+IR}}-\Psi_{\mathrm{SFR,\,corr.}})} \times 100\%$) 
    as described in Section~\ref{Section_CSFRD_Correction}. 
    We use literature SMFs and the \citet{Sargent2014} MS correlation to 
    estimate the completeness percentage, therefore each SMF leads to a 
    corrected value of $\Psi_{\mathrm{SFR,\,corr.}}$ at each redshift bin. 
    %
    %
    The SMF$_{\mathrm{corr.}}$ column indicates which SMF is used (\citealt{Muzzin2013}; \citealt{Ilbert2013}; \citealt{Grazian2015}; \citealt{Song2016}; \citealt{Davidzon2017}), \REVISED{%
     except for the last two rows at $z>4$, where it becomes prohibitive to estimate any incompleteness correction. 
     Given the evidence that our FIR+mm sample does not substantially overlap with the UV-based LBG samples at $z>4$ 
     (see Section~\ref{Section_Dropouts}), we correct the $z>4$ cosmic SFR densities by adding the observed SFR densities 
     from our FIR+mm sample to the UV-based, attenuation-corrected SFR densities from LBG samples (i.e., the values reported in \citealt[][MD14]{Madau2014a}). 
     The percentages $\sim$49.4\% and $\sim$37.1\% thus represent the SFR ratio of our sample to the sum of our sample and the LBGs.
    }

\end{minipage}

\end{center}

\end{table*}


\subsection{Comparing with literature cosmic SFR densities}
\label{Section_CSFRD}

%

At redshifts $z > 3$, our understanding of the cosmic SFR density is still largely limited to UV- and optical-based studies. Direct knowledge of dust-obscured star formation at such high redshift, i.e., from the FIR to mm data, mainly comes from a small number of rarest and most ultraluminous galaxies, leaving considerable uncertainty about true SFR densities at those high redshifts \citep{Madau2014a}. 
Recent attempts to directly measure dust-obscured SFR densities from FIR and submm data have been pushed 
\REVISED{to redshift $3<z<4$, e.g., by \citet{Gruppioni2013} and \citet{Burgarella_2013}, and to redshift 6, e.g., by \citet{RowanRobinson2016} and \citet{Bourne2017}.}

In this and the next section, we first directly compute the inferred SFR densities from our FIR+mm \goodArea{} sample up to redshift 6 (Fig.~\ref{Plot_Cosmic_SFR}) and compare with results from the literature, then estimate the incompleteness corrections using literature SMFs (Fig.~\ref{Plot_Cosmic_SFR_Correction}). Finally, we repeat the comparison after correcting for incompleteness in our sample (Fig.~\ref{Plot_Cosmic_SFR_Corrected}). 
%


We use the non-parametric 1/$V_{max}$ method \citep{Schmidt1968} to estimate the SFR density of our FIR+mm sample in each redshift bin in order to eliminate the \REVISED[]{Malmquist bias}. 
For each source, we compute the farthest redshift $z_{max}$ at which it can still be detected in our catalog (i.e., $\mathrm{S/N}_{\textnormal{FIR+mm}} \ge 5$). Then for each redshift bin (lower and higher boundaries are $z_1$ and $z_2$), we compute the co-moving volume for each source ($V_{max}$) from $z_1$ to ${min}(z_{max},z_2)$ \citep[i.e., the smaller value, e.g.,][]{Pozzetti2003}. Then the SFR density of this sample is the sum of all individual $\mathrm{SFR}/V_{max}$ in each redshift bin. 
To account for more realistic uncertainties, we compute $\mathrm{SFR}/V_{max}$ in a bootstrapping method: we randomly select the same number of sources in each redshift bin, allowing duplication, then repeat 1000 times to compute the mean and rms (i.e., the uncertainty of the measured SFR density). 

Fig.~\ref{Plot_Cosmic_SFR} shows the SFR densities directly measured from our observed sample, and compares them to previous results  
from the literature \citep{Madau2014a, Gladders2013}.
The fluctuations from bin to bin at $0<z<3$ (where we expect a smooth redshift dependence)
are significant at the $\sim2\sigma$ level and
suggest total uncertainties in our determinations of order of $\sim$0.2~dex, \REVISED[probbaly]{probably} dominated by cosmic variance
(although note that the bins do not have equal co-moving volumes). 

\subsection{Incompleteness corrections for cosmic SFR densities}
\label{Section_CSFRD_Correction}

\REVISED{
	We estimate the completeness of our measured SFR density from the FIR+mm sample to the total cosmic SFR density in each redshift bin in the following way, with the help of the SFR histograms in bins of stellar masses. 
	We assume that SMF can predict the total cosmic SFR density based on the assumption of MS correlation (e.g., \citealt{Sargent2014}), thus the SMF-\MINORREREVISED[predicted]{converted} SFR histogram should be able to fit the observed SFR histogram when the FIR+mm sample is complete at certain stellar mass bin, for example as indicated by Fig.~\ref{Plot_Mstar_SFR_Contribution_z_bin_all_panels}. 
	The fitting is done by adjusting the normalization of SMF-\MINORREREVISED[predicted]{converted} SFR histogram (see Fig.~\ref{Plot_Mstar_SFR_Contribution_Renormalized}), so as to achieve a minimum $\chi^2$ in the stellar mass range where the FIR+mm sample is at least 50\%-complete in stellar mass. The 50\%-complete mass ($M_{*,\,50\%\,lim.}$) is determined by the renormalized version of Fig.~\ref{Plot_Mstar_Histogram_z_all_panels} (Fig.~\ref{Plot_Mstar_SFR_Contribution_Renormalized} in the Appendix)\MINORREREVISED[, where]{:} we calculate above which stellar mass the number of FIR+mm galaxies starts to be larger than half of what is predicted by SMF, i.e., $N_{\mathrm{FIR+mm}} \ge 0.5 \times N_{\mathrm{SMF}}$. 
} 

Note that, as mentioned at the end of Section~\ref{Section_UV_unattenuated_SFR}, the directly observed SFRs for our FIR+mm sample are SFR$_{\textnormal{UV+IR}}$. We emphasize that the aim of the incompleteness correction is to account for the SFR$_{\textnormal{UV+IR}}$ from fainter/lower-mass galaxies that are not included in our sample. Although one might consider correcting the observed IR-only SFRs in a same fashion, we caution that this may not be appropriate if fainter/lower-mass galaxies have less dust attenuation, as is observed at $z < 3$ (e.g., \citealt{Pannella2015}).

We then integrate the renormalized SMF-\MINORREREVISED[predicted]{converted} SFR histogram 
down to $10^{8}\,\mathrm{M}_{\odot}$  (as was done in \citealt{Ilbert2013} and \citealt{Muzzin2013})
to obtain an estimate of the total cosmic SFR density.  The ratio of the total observed SFR to the total cosmic SFR gives the completeness fraction ($f_{\mathrm{comp.}}$). 

In Fig.~\ref{Plot_Cosmic_SFR_Correction}, we present the cosmic SFR completeness fractions versus redshift, estimated using three different SMFs from the literature.  We present the incompleteness-corrected total SFR densities in Fig.~\ref{Plot_Cosmic_SFR_Corrected}.

At $z<2$, the completeness of our sample drops linearly with $\log(1+z)$, and the three SMF-based results are fully consistent. However, at $z > 2$, the three SMF-based estimates differ by large factors, largely due to the different low-mass slopes of the literature SMFs.
The SMF from \citet{Muzzin2013} has a shallower slope at low masses, and fits our observed histograms much better, leading to a completeness factor of about 80\% at $z\sim3$. The SMFs from \citet{Ilbert2013} and \citet{Davidzon2017} have steeper low-mass slopes, and hence lead to much larger incompleteness corrections (see Table~\ref{Table_3}).


At $z>4$, the completeness fractions that we estimate from the renormalized SMFs are very low (see Appendix Fig.~\ref{Plot_Mstar_SFR_Contribution_Renormalized}), ranging from $\sim 0.2\%$ to $\sim 5\%$. 
If we do not re-normalize the SMFs but just integrate the SMF-based SFR histogram (i.e., as shown in Fig.~\ref{Plot_Mstar_SFR_Contribution_z_bin_all_panels}), we would get a total cosmic SFR density that is fully consistent with the fit by \cite{Madau2014a} at $z\sim5$, and would not reflect the excess SFR that we see in the observed SFR histograms.
The main reason is that at $z>4$ our FIR+mm sample is almost entirely disjoint from the UV-based samples that have traditionally been used to estimate the cosmic SFRD, as we will demonstrate in the next sections.  Thus to estimate the total cosmic SFR density at this high redshift range, we need to add our \MINORREREVISED[]{directly observed} FIR+mm SFR density to the UV-based SFR density. In Fig.~\ref{Plot_Cosmic_SFR_Corrected}, we show the summed total cosmic SFR densities at $z>4$ as the red open squares. 
Our directly observed FIR+mm SFR densities are also shown as the red arrows \MINORREREVISED[]{for comparison}. 
We list all corrected cosmic SFR density values in Table~\ref{Table_3}. 


Our corrected measurements imply total SFR densities that are consistent with those reported in \cite{Madau2014a} at $0.5<z<2$,  but possibly up to 0.4~dex higher at $2<z<4$ where the results are less reliable and depend strongly on the assumed SMF. 
However, we also note that several recent FIR, (sub)mm and radio studies also seem to find similar trends. 
In Fig.~\ref{Plot_Cosmic_SFR_Corrected}, we show the recent cosmic SFR density measurements from \citet{RowanRobinson2016} and \cite{Novak2017} for comparisons. 
Based on much wider (20~$\mathrm{degree}^2$) but shallower FIR surveys, \citet{RowanRobinson2016} studied the FIR luminosity function with 3035 SPIRE 500$\,{\mu}$m sources. Their cosmic SFR densities (the gray filled squares) are much higher than the UV estimates at $3.5<z<6$ (i.e., the \citealt{Madau2014a} curve) by factors of $\sim 2-6 \times$. They assumed a fixed 500~$\mu$m luminosity function at $z>3.5$ (\MINORREREVISED[]{\citealt{Saunders_1990_LF} functional form with} fixed parameters $\alpha=1.2$ and $\sigma=0.60$), but whether such fixed form is applicable is still controversial. 
%
\cite{Novak2017} used deep VLA imaging data in the 2~deg$^2$ COSMOS field to derive and integrate the 3~GHz luminosity function (assuming pure luminosity evolution) up to $z = 5$ to 6, and estimated the cosmic SFR density assuming a redshift dependent FIR-radio correlation.  Their results are shown with gray triangles in Fig.~\ref{Plot_Cosmic_SFR_Corrected}\MINORREREVISED[)]{}, and are in good agreement with the \cite{Madau2014a} curve and the average of our incompleteness-corrected data points at $z<4$. However, at $z \sim 5$, their cosmic SFR density is $\sim 3 \times$ the value from \cite{Madau2014a}. 
%

\REVISED[%
Meanwhile, several works using HST optical/near-IR (e.g., $H_{\mathrm{F160W}}$) band catalog sources as priors and applying stacking techniques in (sub)mm image data {Bourne2017,Dunlop2017} found very consistent cosmic SFR densities to the {Madau2014a} values at $z\sim5$. 
]{%
    Recently, \cite{Bourne2017} and \cite{Dunlop2017} have investigated dust-obscured SFR densities by stacking submm and mm data for faint galaxies selected from deep HST imaging data.  \cite{Bourne2017} used SCUBA2 data at 850$\,\mu$m and 450$\,\mu$m covering 230~arcmin$^2$ in three of the CANDELS fields, stacking emission for galaxies selected from the 3D-HST catalog of \citet{Momcheva2016}. \cite{Dunlop2017} used ALMA 1.3~mm data covering 4.5~arcmin$^2$ in the HUDF, stacking for galaxies selected from the deep HUDF HST data.
    They convert the stacked single (sub)mm band flux into dust obscured SFRs  
    using submillimeter galaxy template SEDs, 
    then sum this with the raw UV SFRs, uncorrected for dust attenuation.
    At $z<2$, the total cosmic SFR densities that they derive agree well with the compilation of \cite{Madau2014a}, 
    similar to our own findings here.
    At higher redshift, their results are $\sim0.1-0.2$~dex larger than the \cite{Madau2014a} curve, but still $\sim0.2-0.3$~dex lower than the present work. It is possible that the most luminous IR--bright galaxies at $z>2$--3 were not part of their stacked samples, which were selected based on near-IR (1.6$\,\mu$m) HST data. 
}
\REVISED[This could likely be the fact that their sample lacks very dusty galaxies as in this work (which is IRAC, MIPS plus radio prior based).]{}

As a further step, we make a simple 
parametric fit to our incompleteness-corrected data points (i.e., all colored data points in Fig.~\ref{Plot_Cosmic_SFR_Corrected}) using the \cite{Madau2014a} double power-law and \citet[][see also \citealt{Abramson2016}]{Gladders2013} log-normal equations. 
\REVISED{From the $\chi^2$ distribution statistics, we compute the median value of each parameter within the $\chi^2_{\mathrm{min}}$ to $\chi^2_{\mathrm{min}}+d\chi^2$ range~\footnote{$d\chi^2=3.53$ for 3 parameter fitting, and $d\chi^2=4.72$ for 4 parameter fitting (e.g., \citealt{Press1992}, chapter 15.6).}, then obtain the} following newly fitted equations and parameters:
\begin{equation}
\begin{split}
\Psi = \Psi_0 \frac{(1+z)^{3.0}}{1+[(1+z)/2.9]^{5.6}} \ \mathrm{M}_{\odot}\,\mathrm{yr}^{-1}\,\mathrm{Mpc}^{-3} \\ 
\Psi_0 = 0.00587 \ \textnormal{(Chabrier IMF)} \\
\end{split}
\end{equation}
(the red thick solid curve in Fig.~\ref{Plot_Cosmic_SFR_Corrected}), and 
\begin{equation}
\begin{split}
\Psi = \frac{A_0}{t\sqrt{2 \pi \tau^2}} \exp{\left[\frac{-(\ln{(t)} - T_0)^2}{2 \tau^2}\right]} \ \mathrm{M}_{\odot}\,\mathrm{yr}^{-1}\,\mathrm{Mpc}^{-3} \\
t \ \textnormal{is the cosmic age in Gyr,} \ T_0 = 1.50, \ \tau = 0.66 \\
A_0 = 0.575 \ \textnormal{(Chabrier IMF)} \\
\end{split}
\end{equation}
(the red thick dashed curve in Fig.~\ref{Plot_Cosmic_SFR_Corrected}).

The new curves are similar to the previous fits at $z<1$. The double power-law (solid curve) is higher at $z\sim2$, but the log-normal (dashed curve) peaks at values similar to the original fits at $z\sim2$, where our SMF-\MINORREREVISED[predicted]{converted} SFR densities agree well with the literature data.  At $z>3$, both new curves can generally fit the new FIR/radio-based data points. 
But they diverge 
in the highest redshift bin, where \cite{RowanRobinson2016} measured a very high value but \MINORREREVISED[where]{} we derive a lower value of the FIR SFR density. 
\REVISED[]{%
There might still be some less extreme dusty galaxies, with lower infrared luminosities and SFRs, that are missed in both our FIR+mm sample and in UV-based samples. Such objects might be recovered by future ALMA/NOEMA surveys. 
}

\begin{figure*}
	\begin{center}
		\includegraphics[width=0.95\textwidth]{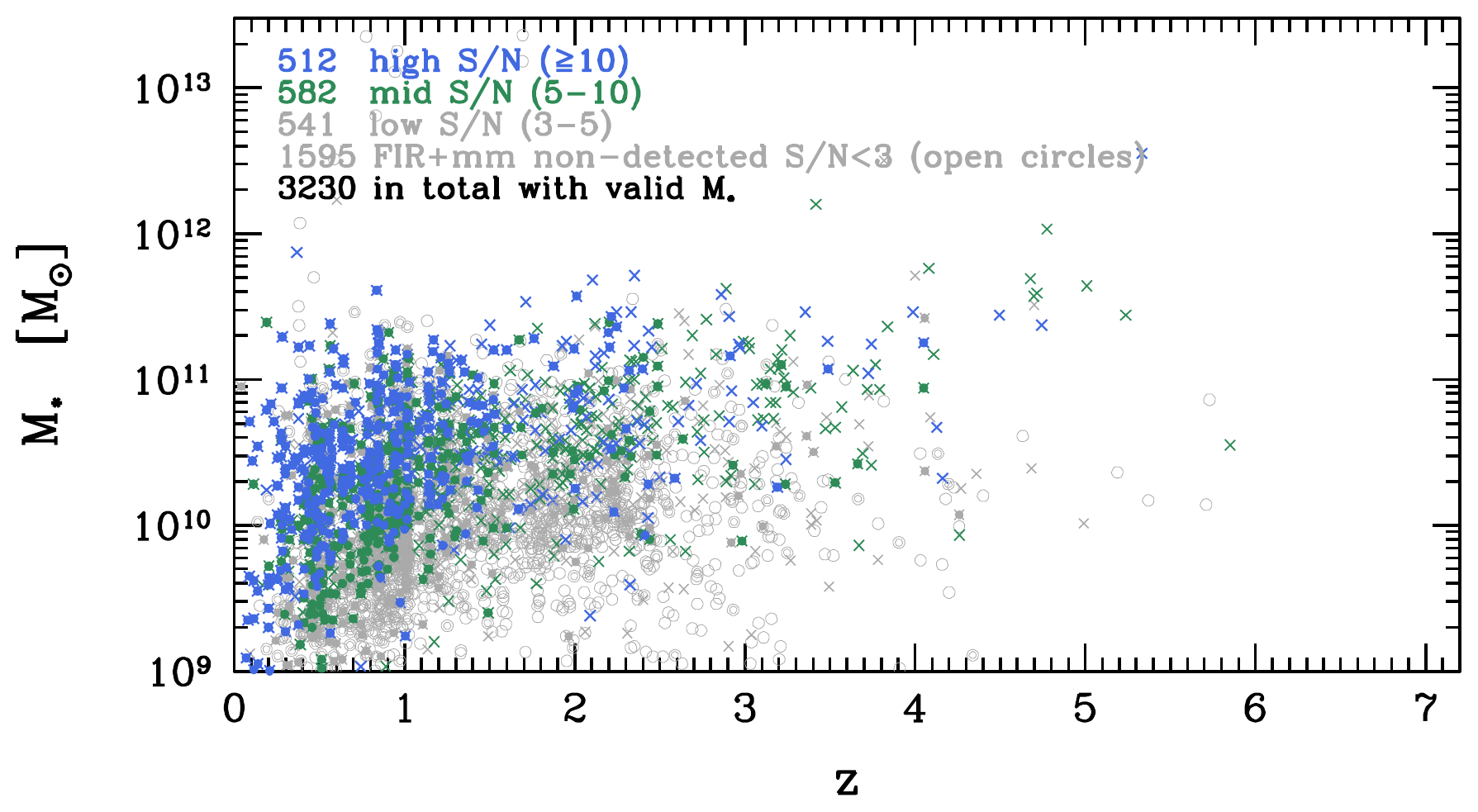}
		\caption{%
			Similar to Fig.~\ref{Figure_z_SFR} but showing  stellar masses versus redshift  for objects in our prior list. Only  sources having stellar mass estimates from optical/near-IR catalogs are shown:  the number of plotted sources is thus smaller here than in Fig.~\ref{Figure_z_SFR}. The colors  and symbol shapes coding is the same as in Fig.~\ref{Figure_z_SFR} and reflects the FIR+mm properties, not stellar masses (i.e., upper limits mean that the object was not detected at FIR+mm, not that $M^*$ is an upper limit, etc).
			\label{Figure_z_Mstar}%
		}
	\end{center}
\end{figure*}

\subsection{Uncertainties in the cosmic SFR density studies}
\label{Section_CSFRD_uncertainty}


Using our new \MINORREREVISED[far-IR]{FIR+mm} photometric catalog, the completeness-corrected cosmic SFR densities are quite well constrained up to redshifts $z\sim$~2--3.  Variations up and down around the average trend reflect small sample statistics together with cosmic variance in the relatively small GOODS-North field. 

At $z>3$, despite our concerted effort on photometry, the uncertainties are still much larger, not only due to cosmic variance but also due to other factors that we briefly discuss here. 
There are remaining limitations in the photometry even for bright sources due to residual blending. For example, we have  evaluated the incompleteness in recovering IR sources as a function of their SFR/masses and redshifts. Figs.~\ref{Plot_Mstar_Histogram_z_all_panels} and \ref{Plot_Mstar_SFR_Contribution_z_bin_all_panels} show that this is not likely to be a big issue at $z<3$ for bright galaxies (while we correct for faint galaxies), but this clearly becomes more important at $z>3$ (Fig.~\ref{Plot_Cosmic_SFR_Correction}). 
This is further exemplified by  the ``additional sources'' that we found and neglected in the SFRD study (Section~\ref{Section_Additional_Sources_In_Residual}).  Assuming that some of these are not spurious, some of them are likely to have redshifts $z>3$, as shown by the example of HDF850.1, which is not included in the SFRD calculation.
The limited sensitivity of the IR data also is a factor, permitting direct detections of sources only with SFR~$>500\,\mathrm{M_\odot\,yr^{-1}}$ at $z>4$. This may be far brighter than the threshold at which UV selection starts to include the large majority of galaxies, although the flux range for which this is true is currently unknown at $z>3$--4.  In any case, this may be another source of significant incompleteness, even for UV+IR samples at these redshifts.
Meanwhile the accuracy of photometric redshifts can also be a problem.  The photometric data used to estimate redshifts are very good for well-studied deep survey fields like GOODS-North, but catastrophic photometric redshift failures might still occur, e.g., in the case of red galaxies with apparent photometric redshifts $z > 3$--6. It is noteworthy that some of the most distant and luminous star forming galaxy candidates that we find, with $z>5$ (see next sections), are also embarrassingly massive \REVISED{(see Fig.~\ref{Figure_z_Mstar}).}
If their redshifts are correct, which must be tested with spectroscopy, these objects could pose interesting challenges for galaxy formation models, or may demonstrate limitations in our ability to derive accurate stellar masses for such objects (see Section~\ref{Section_z5} for more discussions about these sources).

\subsection{Comparing with Lyman-break galaxies}
\label{Section_Dropouts}

We have found a substantial population of IR-detected galaxies all the way out to redshifts $z=3$--6, and have estimated that their contribution to the SFR density of the Universe is significant and comparable to or larger than that derived from samples selected at UV rest-frame wavelengths. It is therefore interesting to discuss the possible overlap between galaxy samples selected at IR and UV wavelengths in order to evaluate whether they trace {\em independent} components of the cosmic SFR density (which should then be summed to derive the total), or if we are measuring the same population.


At $z > 3$, the 912\AA\ Lyman limit and 1216\AA\ Lyman~$\alpha$ forest spectral breaks are redshifted through optical passbands, causing the galaxies to become very faint and to ``drop out'' at bluer wavelengths while still being clearly detected in redder filters.   Optical color--color diagrams can then be used to select samples of Lyman break galaxies (LBGs) in specific redshift ranges based on these ``dropout'' color signatures.  For the HST ACS filters that were used to observe GOODS-N, most $B$-dropout LBGs fall into the redshift range $2.8 \le z \le 4.4$, while $V$-dropout LBGs are found to have $z \approx 4.4-6.5$ (\citealp{Giavalisco2004}; \citealp{Vanzella2009}; \citealp{Dahlen2010}). However, the Lyman break technique relies on detecting and measuring rest-frame UV light, and may therefore miss a fraction of dusty galaxies whose UV emission is strongly attenuated and reddened.  To investigate how dusty galaxies are missed in LBG samples, we compare our GOODS-N FIR+mm detected sources at $z = 2.8$--4.4 and $z = 4.4-5.5$ with the $B$- and $V$-dropout criteria (e.g., \citealp{Giavalisco2004}; \citealp{Dickinson2004}; \citealp{Stark2009}; \citealp{Vanzella2009}; \citealp{Bouwens2012}). 

\begin{figure}
\includegraphics[width=0.48\textwidth]{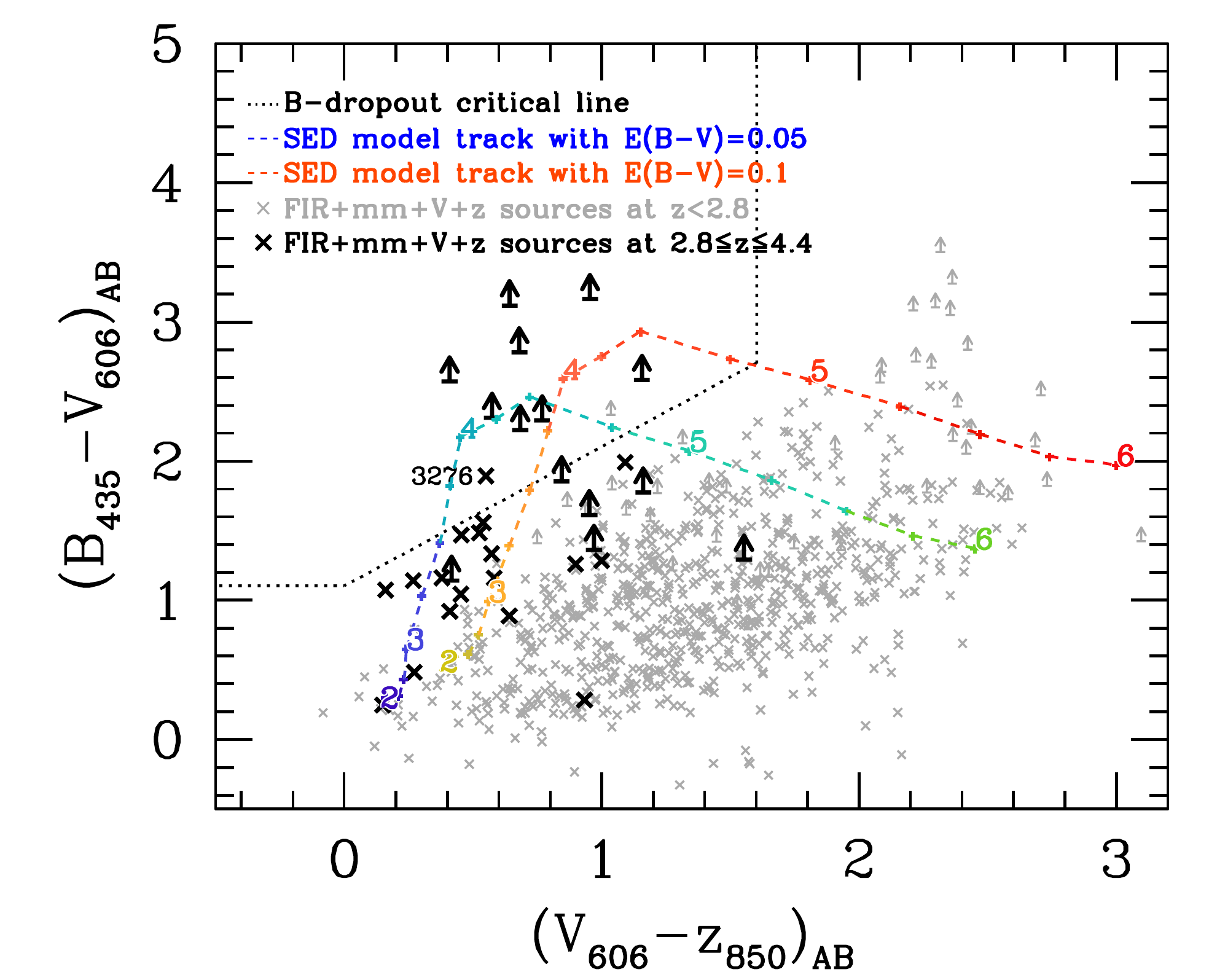}
\caption{%
Color-color diagram for identifying $B$-dropouts at $2.8 \le z \le 4.4$ (\citealp{Vanzella2009}; \citealp{Dahlen2010}). 
Data points are FIR+mm detected galaxies which are also detected at $V_{606}$ and $z_{850}$ bands. Gray points indicate galaxies with 
$z<2.8$, and black points indicate galaxies at 
$2.8 \le z \le 4.4$
\citep[e.g., Fig.~1 of][]{Vanzella2009}. 
Galaxies not detected in $B_{435}$ are shown as upward arrows (1$\,\sigma$ lower limits). 
The dotted line indicates the $B$-dropout criteria (Eq.~\ref{Equation_Bdropout}). 
Two colored curves indicate two model tracks calculated from our SED templates (Fig.~\ref{Fig_SED_Templates}), with E(B-V) $=0.05$ and $0.1$ as two examples. 
The small colored number at each node of the track indicates the redshift. 
\label{Figure_Dropouts_Bdropouts}
}
\end{figure}

\begin{figure}
\includegraphics[width=0.48\textwidth]{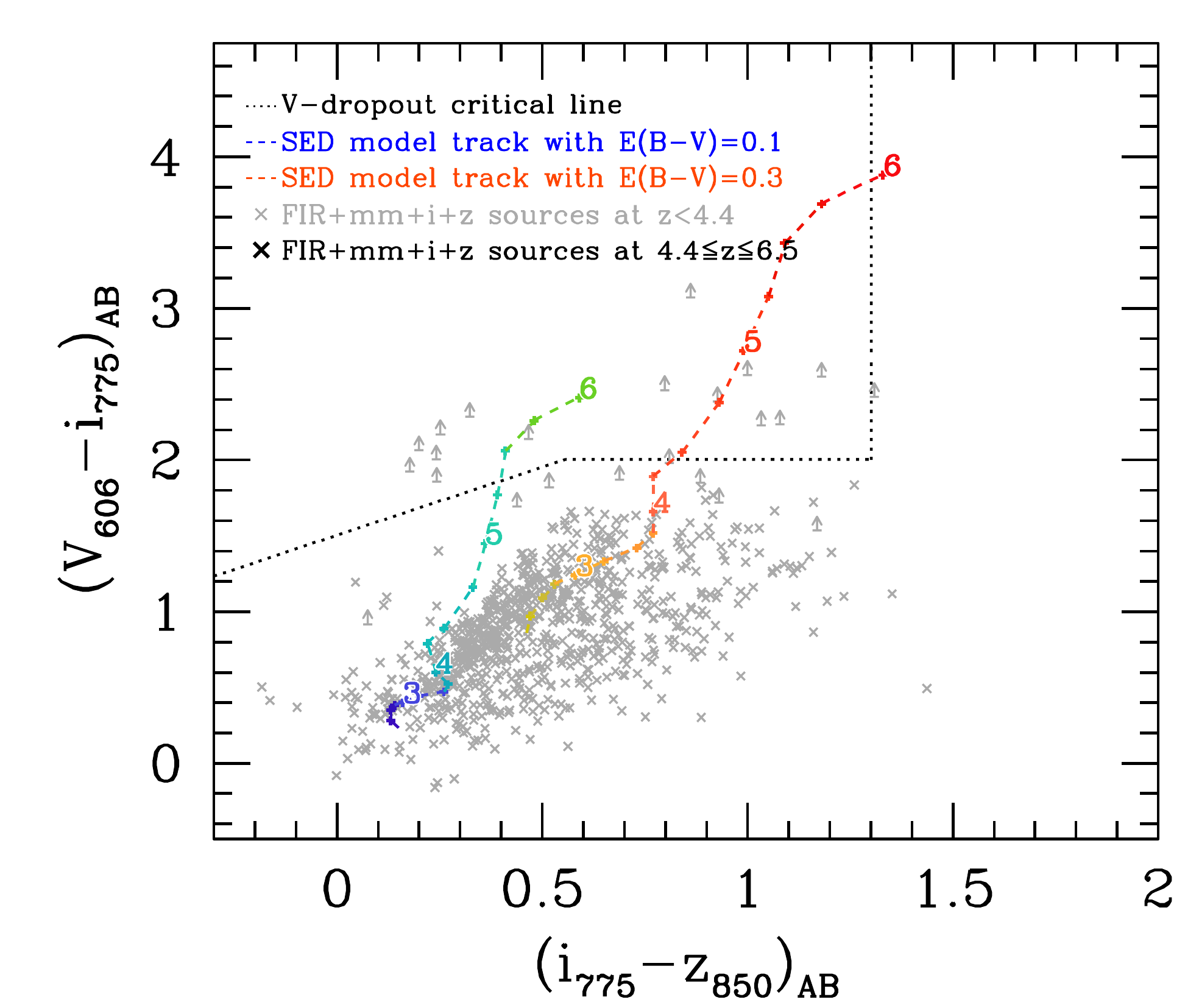}
\caption{%
\REVISED{Similar to Fig.~\ref{Figure_Dropouts_Bdropouts} but for $V$-dropouts at $4.4 \le z \le 6.5$ \citep{Vanzella2009}. No boldface point is seen here because none of the FIR+mm galaxies in our sample at $4.4<z<6.5$ are detected by HST in these optical bands.}
\label{Figure_Dropouts_Vdropouts}
}
\end{figure}

We use the HST optical to near-infrared photometry from \citet{Skelton2014} (cross-matched in Section~\ref{Section_Initial_IRAC_Catalog}) and use the color-color criteria from \cite{Vanzella2009} to select $B$- and $V$-dropout LBGs in GOODS-N: 
\begin{itemize}
%
%
\item $B$-dropout
\begin{equation}
\label{Equation_Bdropout}
\hspace{0.5cm}
\begin{cases}
\; (B_{435}-V_{606}) > (V_{606}-z_{850}) + 1.1 \phantom{AAAAAAAAA}\\
\; \mathrm{and} \; (B_{435}-V_{606}) \ge 1.1 \\
\; \mathrm{and} \; (V_{606}-z_{850}) \le 1.6 \\
\end{cases}
\end{equation}
%
%
\item $V$-dropout
\begin{equation}
\label{Equation_Vdropout}
\hspace{0.5cm}
\begin{cases}
\; 
\begin{cases}
\; (V_{606}-i_{775}) > 0.9\times(i_{775}-z_{850}) + 1.5 \\
\; \mathrm{or} \; (V_{606}-i_{775}) > 2.0 \\
\end{cases} \\
\; \mathrm{and} \; (V_{606}-i_{775}) > 1.2 \\
\; \mathrm{and} \; (i_{775}-z_{850}) < 1.3 \\
\; \mathrm{and} \; \mathrm{S/N}\,(B_{435}) < 2 \\
\mathrm{and} \; \mathrm{S/N}\,(i_{775}) \ge 3 \\
\end{cases}
\end{equation}
\end{itemize}

%
%

\begin{figure*}[htb]
\includegraphics[width=0.33\textwidth]{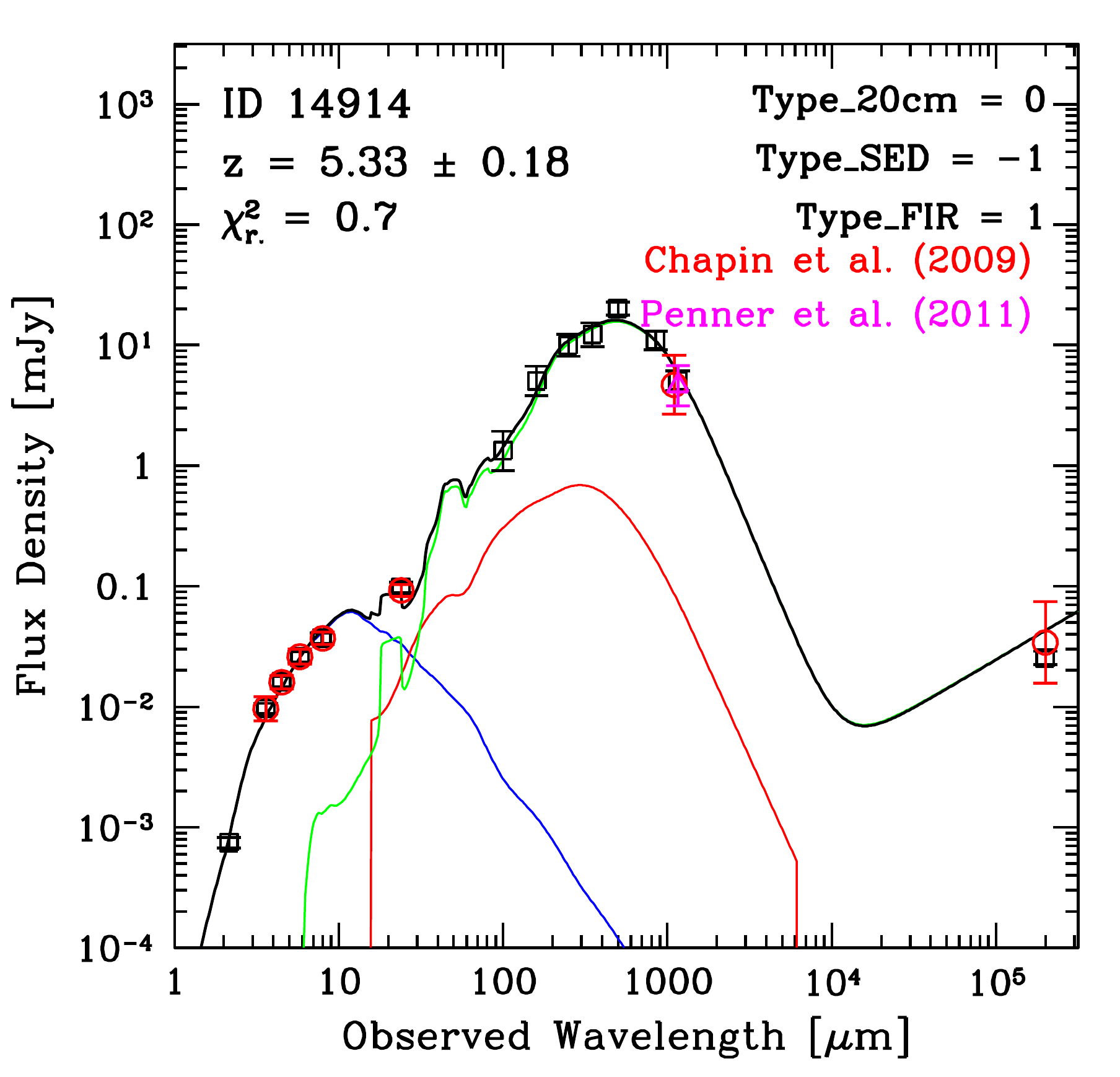}
\includegraphics[width=0.33\textwidth]{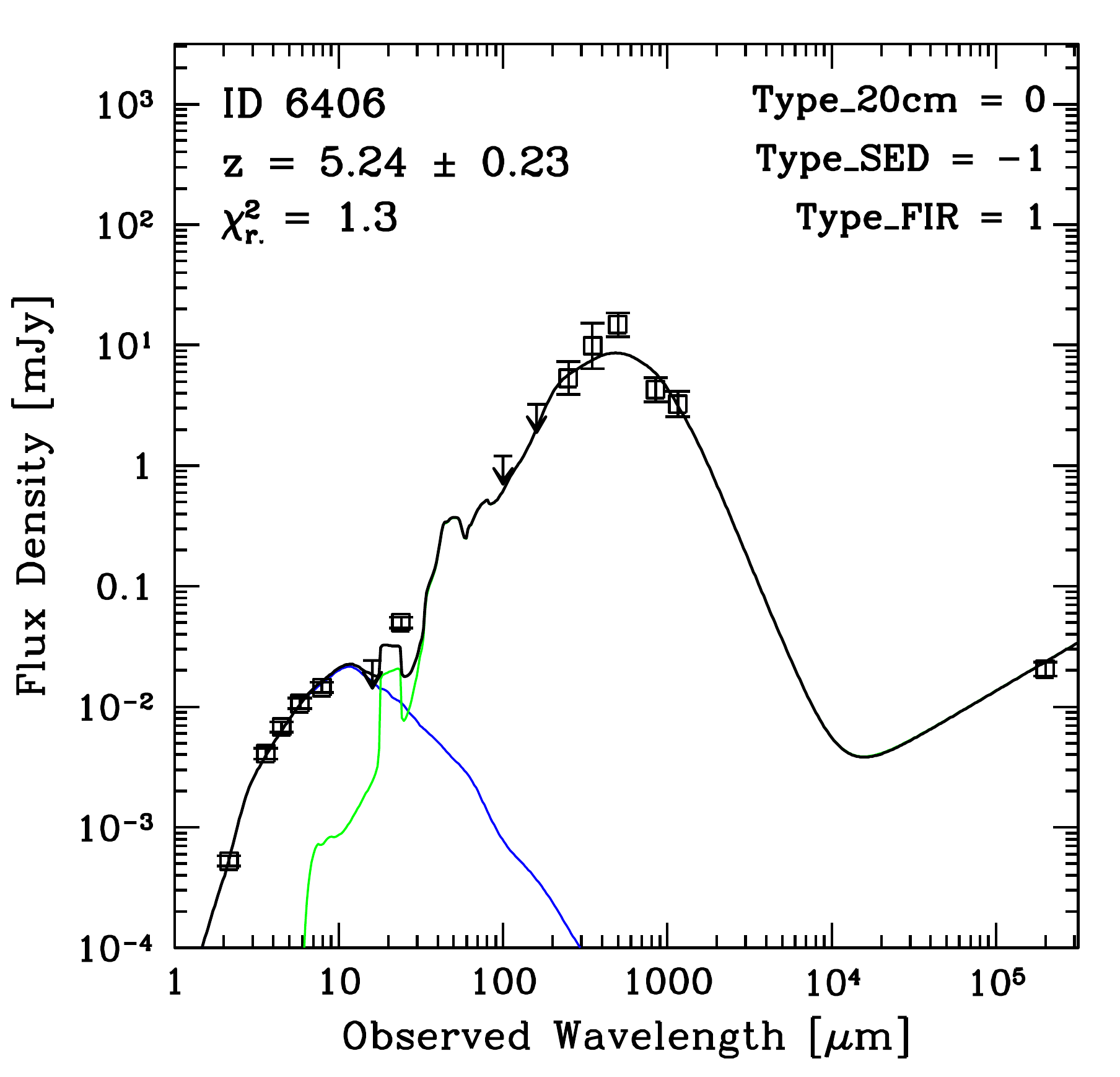}
\includegraphics[width=0.33\textwidth]{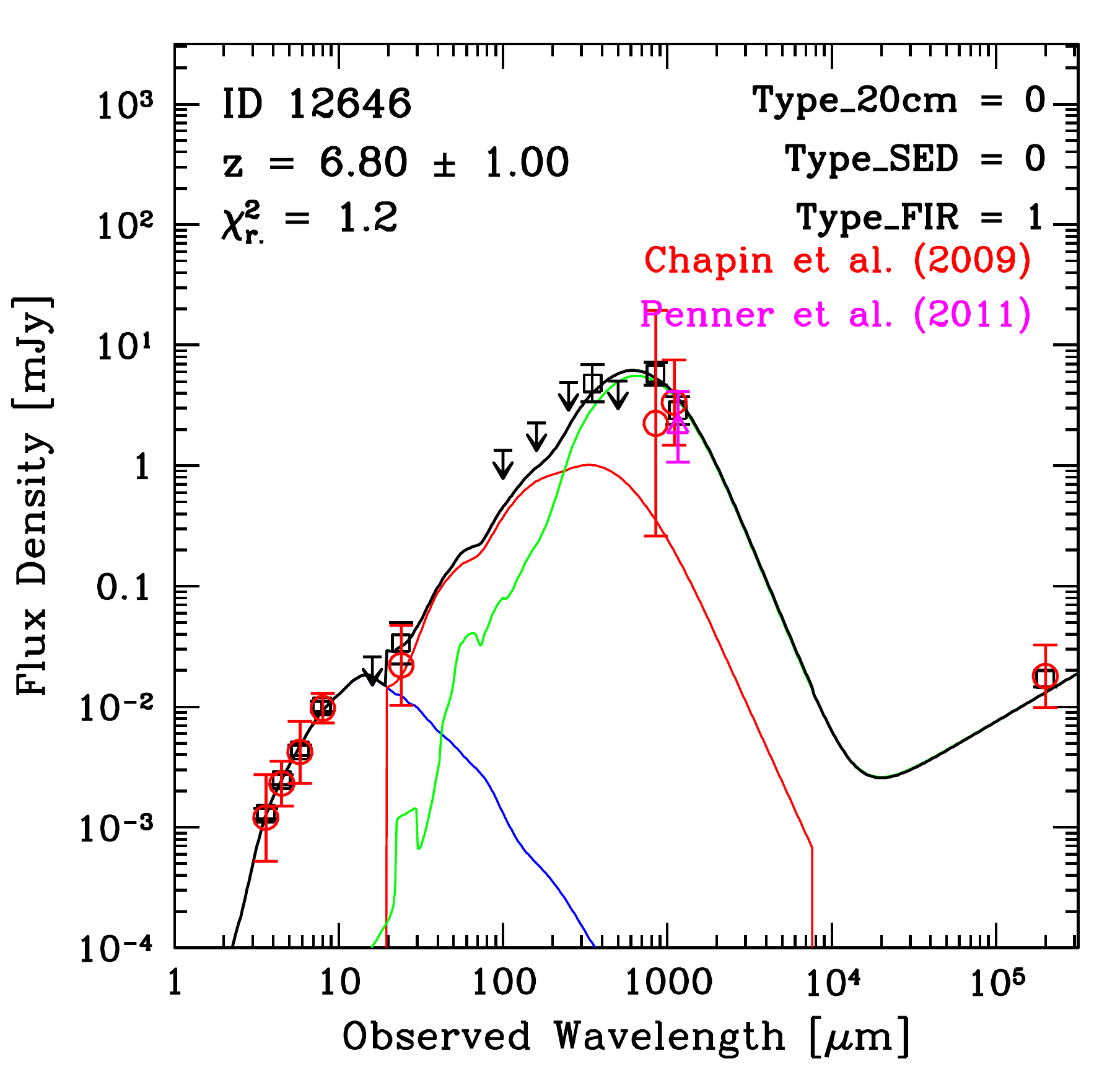}
\caption{%
The best-fit SEDs from the $\left<U\right>$-dependent SED fitting for the three $z \ge 5$ MS type FIR+mm galaxies: ID~14914, ID~6406 and ID~12646. The blue (near-IR wavelengths), red (if present at mid-IR wavelengths) and green (FIR wavelengths) curves represent the stellar, mid-infrared AGN (if present) and dust components. Parameters in the upper right corner are described in Section~\ref{Section_SED_Fitting}. 
\label{Plot_SEDs_z5}}
\end{figure*}

In Fig.~\ref{Figure_Dropouts_Bdropouts},   FIR+mm detected sources at any redshift that also have $V_{606}$ and $z_{850}$ \REVISED{3-sigma} detections 
are plotted in the $B_{435}-V_{606}$ versus $V_{606}-z_{850}$ diagram. The subscripts indicate the HST observing bands, i.e., F435W for $B$ band, F606W for $V$ band and F850LP for $z$ band. 
We also show two modeled tracks as the colored curves, which are calculated by redshifting SED templates (BC03+DL07, e.g., Fig.~\ref{Fig_SED_Templates}). Here we show two example model tracks with $E(B-V)=0.05$ and $0.1$. Heavier dust attenuation will make the model track shift toward redder colors.  Data points within the upper left region identified by the dotted lines indicate galaxies that meet the $B_{435}$-dropout color criteria. 

Among all the FIR+mm and $V_{606}$ and $z_{850}$ detected galaxies, 
%
%
%
%
%
%
%
%
we highlight in Fig.~\ref{Figure_Dropouts_Bdropouts} the 32 ones having $2.8 \le z \le 4.4$. 
Only 9 
of them can be identified as $B$-dropouts (we use 1~$\sigma$ limits, as in \citealt{Vanzella2009}).
In comparison, there are a total of 78 FIR+mm galaxies with photometric redshift within $2.8 \le z \le 4.4$ (59 have valid optical photometry in 3D-HST catalog, often thus with near-IR WFC3 detections but lacking ACS detections in the optical). 
Therefore, only half of our FIR+mm sample at $2.8<z<4.4$ has a significant HST optical counterpart in $V$ and $i$ to start with, and only about 10\% of the parent sample (20\% of those with optical counterparts) could be identified as $B$-dropouts at the depths of the current HST imaging. The \MINORREREVISED[optical detected]{optically-detected} \MINORREREVISED[]{FIR+mm} galaxies \MINORREREVISED[inside $2.8<z<4.4$ that fail detection as $B$-drops appear to do this]{with $2.8<z<4.4$ fail to be selected as $B$-dropouts} for a number of reasons: photometric noise affecting these faint detections, effect of photometric redshifts (some objects might be slight out of the boundary, in reality), lack of depth in the optical data (upper limits in the $B$-band are not always stringent enough).
In addition, even for the very small number of galaxies that can be selected as $B$-dropouts, the rest-frame UV-based SFR significantly underestimates the dust-obscured SFR. 
For example for ID~3276, a source that is detected in $B$ band and is within the $B$-dropout region in Fig.~\ref{Figure_Dropouts_Bdropouts} (with its ID number labeled in the figure), 
the optical/near-IR (rest-frame UV) based SFR, corrected for dust attenuation, is $\sim32\,\mathrm{M}_{\odot}\,\mathrm{yr}^{-1}$ (stellar mass $10^{11.06}\,\mathrm{M}_{\odot}$) in the 3D-HST catalog. For comparison, its FIR+mm based SFR is $256 \pm 37\,\mathrm{M}_{\odot}\,\mathrm{yr}^{-1}$ in our FIR+mm SED fitting.

In Fig.~\ref{Figure_Dropouts_Vdropouts}, we show the distribution of FIR+mm galaxies with $i_{775}$ and $z_{850}$ detections in the $V_{606}-i_{775}$ versus $i_{775}-z_{850}$ diagram. 
None of the 7 galaxies in our sample  with a photometric redshift within the $V$-dropout range of $4.4 \le z \le 6.5$ is detected in $i_{775}$ and $z_{850}$.
There are some galaxies within the $V$-dropout color selection region for which the photometric redshift estimate is $z < 4.4$.

All in all, there is very little overlap between our FIR+mm-selected sample of high redshift galaxies and dropout samples. For those few FIR+mm-bright objects that can be selected as LBGs, their UV-estimated SFRs  are largely under-estimated compared to those derived from their far-infrared luminosities.  We thus conclude that the SFR density contributions we have derived from IR data are largely independent from the ones estimated from UV-selected samples, at least at $z \gtrsim 3.5$.

\begin{figure*}[htb]
\centering
\includegraphics[width=0.96\textwidth, trim=0 0 0 0, clip]{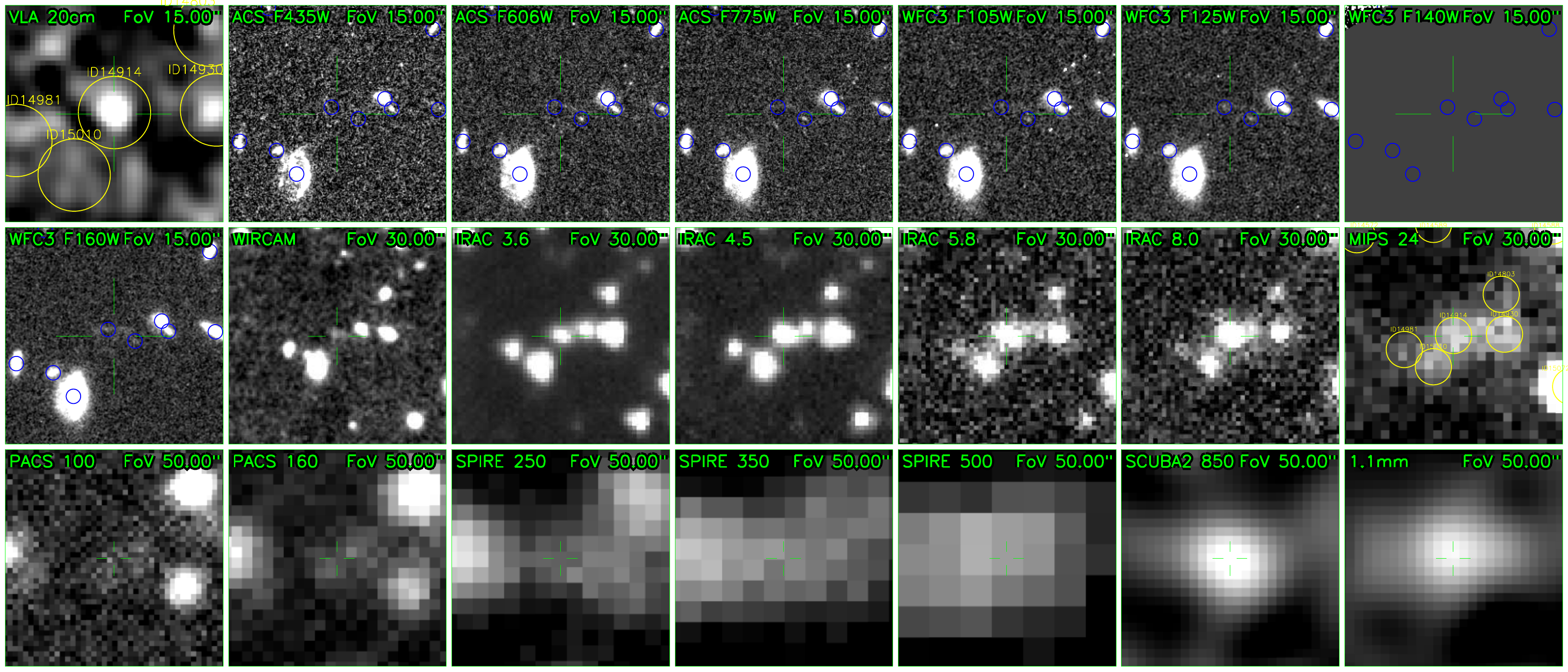}
\caption{%
The multi-wavelength cutouts of ID~14914. The 
instrument and bandpass 
of each cutout image is labelled in each panel. Yellow circles in the first panel and the MIPS 24 panel are our 24+radio sources and have a diameter of 5$''$. Blue circles in the HST panels are the 3D-HST sources and have a diameter of 0.5$''$. 
All panels are centered on the IRAC position (as indicated by the green cross) of the source under examination (ID~14914). The field of view is also indicated in each panel. 
\label{Plot_cutouts_ID14914}}
\end{figure*}

\begin{figure*}[htb]
\centering
\includegraphics[width=0.95\textwidth, trim=0 0 0 0, clip]{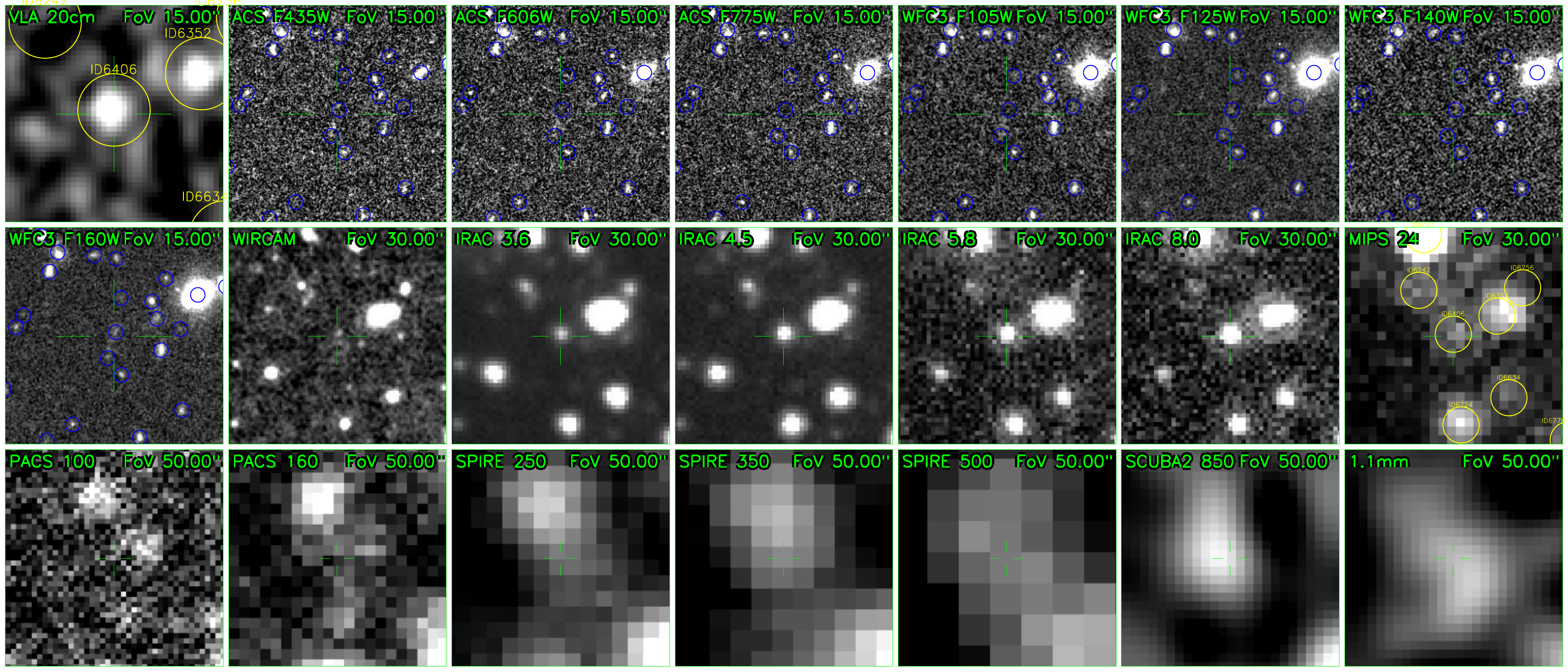}
\caption{%
The multi-wavelength cutouts of ID~6406. See the caption in Fig.~\ref{Plot_cutouts_ID14914}. 
\label{Plot_cutouts_ID6406}}
\end{figure*}


\begin{figure*}[htb]
\centering
\includegraphics[width=0.95\textwidth, trim=0 0 0 0, clip]{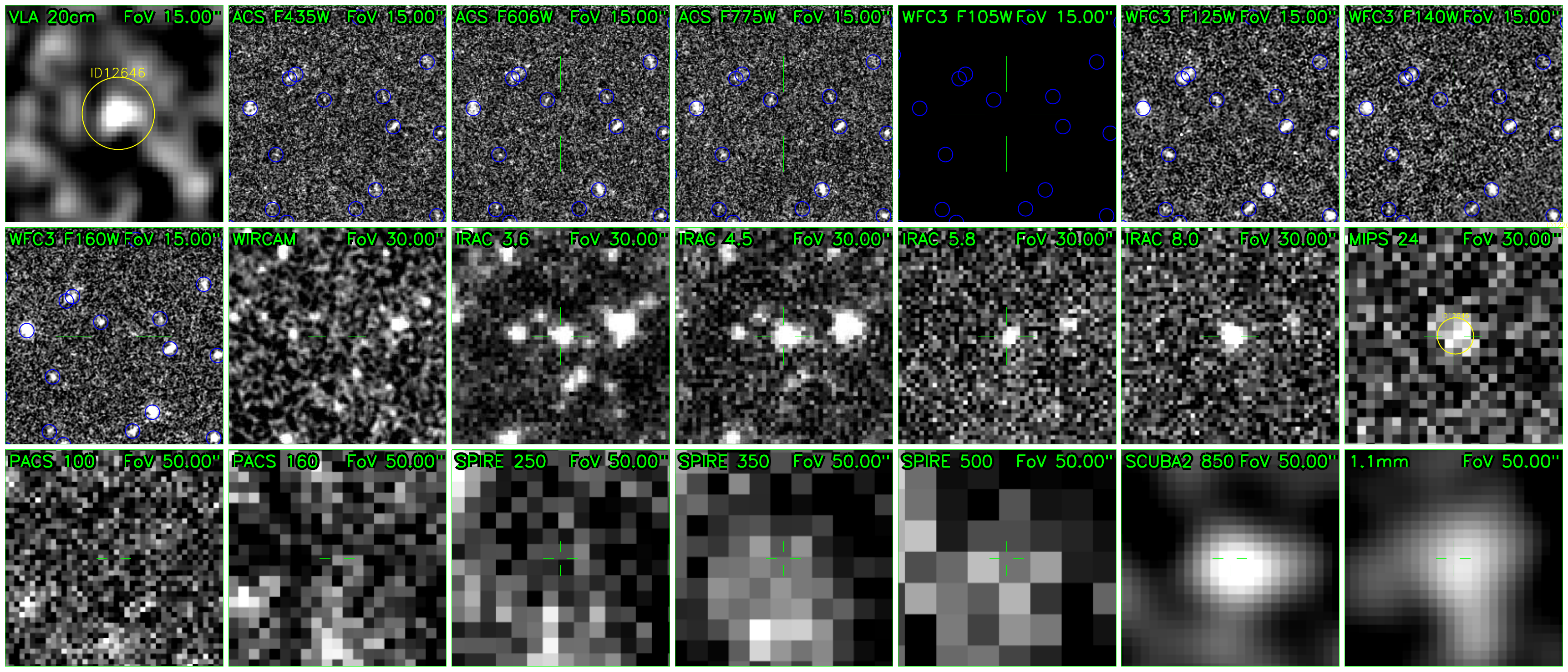}
\caption{%
The multi-wavelength cutouts of ID~12646. See the caption in Fig.~\ref{Plot_cutouts_ID14914}. 
\label{Plot_cutouts_ID12646}}
\end{figure*}

\subsection{A sample of candidate $z \ge 5$ galaxies}
\label{Section_z5}


In our FIR+mm catalog, we find three sources that have FIR+mm SED photometric redshifts $z \ge 5$, and whose SFRs and stellar masses appear to be consistent with the MS at these redshifts (at least within the current, large uncertainties on its locus). 
These are ID~14914 (which is not in the \goodArea{}), ID~6406 
and ID~12646 (the right-most blue data point in Fig.~\ref{Figure_z_SFR}, our highest redshift candidate, with 
$z\sim6.8$). We discuss the properties of these three particularly interesting sources in some details below. 

ID~14914 is detected in all FIR+mm bands. Multi-wavelength SEDs and cutout images are presented in 
Figures~\ref{Plot_SEDs_z5}~and~\ref{Plot_cutouts_ID14914}. 
Weak extended emission can be seen in the HST ACS and WFC3 images. Two sources from the 3D-HST catalogs are very close to ID~14914. We cross-matched the closer (0.585$''$) 3D-HST source, to the north-east, as the counterpart. It is clearly detected by IRAC, as well as in the (sub)mm and at 20~cm. Blending is not significant. The 3D-HST ID number and other properties derived from our analysis are listed below: 

\begin{equation}
\begin{split}
\mathrm{ID}{\,}_{\textnormal{IRAC}} &= 14914 \\
\mathrm{ID}{\,}_{\textnormal{3D-HST}} &= 7338 \\
z{\,}_{\textnormal{FIR+mm}} &= 5.33 \pm 0.18 \\
z{\,}_{\textnormal{3D-HST}} &= 5.57 \ (5.335 - 5.90) \ \ \text{\footnotemark}\\
M_{*,\textnormal{3D-HST}} &= 10^{12.55} \ \mathrm{M}_{\odot}\\
M_{*} &\approx 10^{11.68} \ \mathrm{M}_{\odot} \ \ \text{\footnotemark}\\
\mathrm{SFR}_{\textnormal{3D-HST}} &= 436.5 \ \mathrm{M}_{\odot}\,\mathrm{yr}^{-1}\\
\mathrm{SFR}_{\textnormal{FIR+mm}} &= 1724 \pm 132.1 \ \mathrm{M}_{\odot}\,\mathrm{yr}^{-1} \\
\left<U\right> &\approx 45 \\
\end{split}
\end{equation}
\addtocounter{footnote}{-1}
\footnotetext{The photometric redshift range in brackets indicate the 1~$\sigma$ (68\%) lower and upper confidence limits.}

\noindent This source is also identified as AzGN04 and GN1200.12 in the 1.1--1.2~mm catalogs of \cite{Perera2008}, \cite{Greve2008} and \cite{Penner2011}. We show the photometry of \cite{Chapin2009} and \cite{Penner2011} in Fig.~\ref{Plot_SEDs_z5} for comparison (as red circles and magenta triangle respectively). Our measurements are in good agreement with these previous values. 

\REVISED[(Moved from footnote 32 to here:) ]{%
	Note that the 3D-HST catalog \citep{Skelton2014} reports a stellar mass of $M_{*,\textnormal{3D-HST}} = 10^{12.55}\,\mathrm{M}_{\odot}$ for this object, at a photometric redshift $z=5.57$. However, from our SED fitting, or simply by comparing IRAC fluxes with other, similar sources, such as GN20, we find a much lower stellar mass at a similar redshift, $z=5.33$. \footnote{GN20 has a peak stellar (over IRAC wavelengths) flux density of $\sim$35~$\mu$Jy, redshift $z=4.055$, and stellar mass $M_{*}=10^{11.28}\,\mathrm{M}_{\odot}$ (see Fig.~\ref{Plot_SED_GN20}).  Scaling these values to $z=5.33$ with $d_{L}^2\;(1+z)^{-1}$ and to an IRAC peak flux of $\sim$60~$\mu$Jy (see Fig.~\ref{Plot_SEDs_z5}), we would infer $M_{*} \approx 10^{11.68}\,\mathrm{M}_{\odot}$.} 
	This value is not far from our own SED fitting results, and it seems more reasonable than the 3D-HST value. 
}
\REVISED{Also note that this source is not in the \goodArea{}, thus it was not included in our cosmic SFR density analyses.}

ID~6406 is detected in the long-wavelength FIR and (sub)mm bands. Image cutouts are presented in Fig.~\ref{Plot_cutouts_ID6406}. 
It is almost invisible in ACS images,  but is detected in the near-IR WFC3, and becomes clear in the IRAC images, as well as the (sub)mm and radio data. 
Its 3D-HST and FIR+mm properties are listed below: 

\begin{equation}
\begin{split}
\mathrm{ID}{\,}_{\textnormal{IRAC}} &= 6406 \\
\mathrm{ID}{\,}_{\textnormal{3D-HST}} &= 26337 \\
z{\,}_{\textnormal{FIR+mm}} &= 5.24 \pm 0.23 \\
z{\,}_{\textnormal{3D-HST}} &= 5.47 \ (3.725 - 5.82)\\
M_{*,\textnormal{3D-HST}} &= 10^{11.44} \ \mathrm{M}_{\odot}\\
\mathrm{SFR}_{\textnormal{3D-HST}} &= 0.1 \ \mathrm{M}_{\odot}\,\mathrm{yr}^{-1} \\
\mathrm{SFR}_{\textnormal{FIR+mm}} &= 918 \pm 109 \ \mathrm{M}_{\odot}\,\mathrm{yr}^{-1} \\
\left<U\right> &\approx 50 \\
\end{split}
\end{equation}

This source has no millimeter counterpart in \cite{Perera2008}, \cite{Greve2008} and \cite{Penner2011}. The closest mm source is AzGN31 in \cite{Perera2008} catalog, at a distance of 4.9$''$, and ID~33 in \cite{Penner2011} catalog, at a distance of 8.8$''$.  Recent NOEMA mm follow-up observations detect a source that accurately corresponds with the IRAC and VLA 20~cm position (D.\ Liu et al. 2018, in preparation).

Multi-wavelength cutouts of ID~12646 are presented in Fig.~\ref{Plot_cutouts_ID12646}. 
There is no obvious detection in ACS or WFC3 images at the exact IRAC position. 
The closest 3D-HST source is 3D-HST ID~12669 at a photometric redshift $z = 4.65$ (4.248 -- 5.06), but this is located 1.52$''$ away from the IRAC position. Given the radio localization, this is unlikely to be the counterpart of ID~12646. ID~12646 is also not matched to any source in the catalog of \cite{Pannella2015}, hence we do not have a stellar mass for it. The emission starts to be seen in the WIRCAM $K_s$ image, but has an irregular morphology and low $\mathrm{S/N}$ ratio. It is fairly bright in IRAC images and bright at (sub)mm wavelengths, but with some offset from the emission peak. It matches AzGN10 in \cite{Perera2008}, \cite{Chapin2009} and \cite{Penner2011}. 
Its physical properties are listed below: 
\begin{equation}
\begin{split}
\mathrm{ID}{\,}_{\textnormal{IRAC}} &= 12646 \\
z{\,}_{\textnormal{FIR+mm}} &= 6.8 \pm 1.0 \\
M_{*} &\approx 10^{11.44} \ \mathrm{M}_{\odot} \ \ \text{\footnotemark}\\
\mathrm{SFR}_{\textnormal{FIR+mm}} &= 814 \pm 249 \ \mathrm{M}_{\odot}\,\mathrm{yr}^{-1} \\
\left<U\right> &\approx 31 \\
\end{split}
\end{equation}
\footnotetext{Similarly, by scaling GN20 to a peak stellar flux of $\sim$23~$\mu$Jy 
	over 
	IRAC wavelengths and 
    a photometric redshift of 6.8, we would infer $M_{*} \approx 10^{11.44}\,\mathrm{M}_{\odot}$.
}


The photometry of \cite{Chapin2009} and \cite{Penner2011} is also shown in the ID~12646 panel of Fig.~\ref{Plot_SEDs_z5}. The measurements generally agree with each other. To estimate the photometric redshift, \cite{Chapin2009} adopted a simple method which infers redshift as a function of IRAC and 24~$\mu$m flux according to \cite{Pope2006}, and derived a photometric redshift $z \sim 3$. Note that using their equation, GN20 (ID~564 in this work) with a spectroscopic redshift of 4.055 would have a photometric redshift $z \sim 2.6$. 


While these sources are interesting  candidates for the most distant IR-detected galaxies in GOODS-N, we emphasize \MINORREREVISED[how]{that} we still only have photometric redshift estimates. In particular, the optical/near-IR SEDs are very red, faint at optical wavelengths and rising at $>$~3~$\mu$m. It is hard to obtain precise redshifts in these cases. \MINORREREVISED[Note, however,]{However, note} that the redshifts estimated from the FIR\MINORREREVISED[]{+mm} SEDs agree well with the optical/near-IR photometric redshifts in all cases, while using templates with $\left<U\right>\sim$~30--40, lower than what is expected for MS galaxies at $z>4$ \citep{Bethermin2015}.
Because of the dust temperature versus redshift degeneracies, using warmer templates would further increase the estimated FIR\MINORREREVISED[]{+mm} redshifts. We cannot exclude that these sources are actually SB galaxies, and their inferred $\left<U\right>$ values would be appropriate in that case. However, even in that case, substantially lower redshifts would require very cold dust SEDs, which seems unlikely. 

Spectroscopic observations will be essential to measure redshifts for these objects, and to \MINORREREVISED[help]{} understand the reasons behind the different SED fitting results. Due to the dusty nature of these FIR+mm galaxies, optical/near-IR spectroscopic observations will likely not be as efficient as for LBGs. Mid-IR spectroscopic surveys with JWST, and (sub)mm spectral line scans with ground-based interferometric arrays, will be the most efficient ways to measure redshifts. 



\vspace{1truecm}

\section{Summary}

In this work, we have presented detailed \superdb{} photometry for the exceptionally deep far-infrared to (sub-)millimeter imaging data that are available in the GOODS-North field. 
To overcome the heavy blending problems introduced by the large PSFs of these data, especially at \textit{Herschel} SPIRE and (sub)mm wavelengths, 
we have developed a new method in which we choose the prior sources to use in fitting for the photometric measurements, and the ones to be ``frozen'' and subtracted at each FIR/mm band. 
In this method, we run SED fitting to predict the flux density of each source for each band, then determine the critical flux value for choosing an actual prior source list at each band by considering both the number density and the expected flux detection limit. 

The number of fitted sources for each band can therefore be kept to reasonable values of $\lesssim 1$ per PSF beam area. For the sources that are not fit, which are hopelessly faint at that wavelength, we eliminate their flux \MINORREREVISED[]{contribution} by modeling their images and subtracting them from the observed data. In this way, the problem of flux boosting of fitted sources can also be largely reduced. 

We generated Monte-Carlo simulations to verify our photometric measurements, and to generate statistical correction recipes for obtaining reasonable measurement uncertainties with a nearly Gaussian behavior. The corrections are linked to three measurable parameters in three-step recipes. Thus for real sources, we can correct flux biases and uncertainties for each source based on its measured parameters. The final uncertainties follow well-behaved statistics, which are important for obtaining more reasonable physical properties from SED fitting. 

Finally, we identify a list of 
1109
IRAC catalog sources with combined $\mathrm{S/N} \ge 5$ over the FIR and (sub)mm bands, including 
70
detections at $z \ge 3$. Comparing with the empirical SFR--$M_{*}$ main-sequence relation at these redshifts, the majority of these dusty galaxies  appear to be classified as ``normal'' MS galaxies, although we caution that the MS level is not yet very well known at these highest redshifts. 



We compute the co-moving SFR densities that are directly observed for our detected FIR+mm sources using the $1/V_{max}$ method plus a boot-strapping analysis.
We estimate the completeness in SFR for our FIR+mm sample from the SMF-converted SFR histograms. 
At $z \le 2$, values estimated using different stellar mass functions from the literature are similar and consistent, but they vary considerably at $z \sim 2-4$ due to the existence of significant disagreements between faint-end SMF slopes in the literature.  These are even worse at $z \sim 4-6$, where we effectively detect only the few most massive galaxies, including, notably, galaxies in the GN20 protocluster at $z = 4.05$. 

We find that the completeness-corrected cosmic SFR densities agree well with the standard literature results from \citet{Madau2014a} and \citet{Gladders2013} at $z \le 2$. 
At $z \sim 2-4$, the corrections vary strongly with the adopted SMFs, but we suggest that the total SFR density 
could be higher than the literature SFR densities by as much as a factor of 3. 
At $z>4$, 
we cannot reliably estimate the incompleteness, so we report only the directly-observed SFR densities in Fig.~\ref{Plot_Cosmic_SFR}, and that value minus the 1$\,\sigma$ uncertainty as 
\REVISED[very conservative]{} lower limits \REVISED[]{to the dust-obscured cosmic SFR} 
in Fig.~\ref{Plot_Cosmic_SFR_Corrected}. 


We find that our estimates for co-moving SFR densities are largely independent from those derived from UV-selected galaxies, and thus should be in principle summed with the UV-derived values, at least for $z \gtrsim 3$ and only for the directly-measured quantities at the bright end, as our sample does not obviously overlap with $B$- and $V$-dropout LBGs.

We finally present the discovery of three galaxies expected to be at $z>5$, including one at $z\sim6.8$, that are the most distant candidates in our GOODS-N sample. Their stellar masses are seemingly very large, a peculiarity shared with much of the sample at $z>4$ that merits further investigations.



\acknowledgments

We are grateful to the referee for a detailed and helpful report that resulted in an improved presentation of this paper. 
We thank Iary Davidzon for helpful discussion on stellar mass functions and Daniel Stern 
and the late Hyron Spinrad for unpublished spectroscopic redshifts that are used in this paper. 
DL and ED would like to thank the Chinese Academy of Sciences (CAS) --  Centre national de la recherche scientifique (CNRS) Joint PhD Program. 
Y.G. acknowledges support from the National Natural Science Foundation
of China (grants 11390373 and 11420101002) and the Chinese Academy
of Sciences' Key Research Program of Frontier Sciences. 
MTS acknowledges support from a Royal Society Leverhulme Trust Senior Research Fellowship (LT150041).
This work is based in part on observations made with Herschel, a European Space Agency Cornerstone Mission with significant participation by NASA.  Support for work by MED and HI was provided by NASA through an award issued by JPL/Caltech.
Their work was also supported by a NASA Keck PI Data Award, administered by the NASA Exoplanet Science Institute. Data presented herein were obtained at the W.\ M.\ Keck Observatory from telescope time allocated to the National Aeronautics and Space Administration through the agency's scientific partnership with the California Institute of Technology and the University of California. The Observatory was made possible by the generous financial support of the W.\ M.\ Keck Foundation.
The authors wish to recognize and acknowledge the very significant cultural role and reverence that the summit of Mauna Kea has always had within the indigenous Hawaiian community. We are most fortunate to have the opportunity to conduct observations from this mountain.



\vspace{2truecm}

\appendix

\section{Photometry Image Products}

Here we present the photometry image products for all the bands analyzed in this paper. 
In Fig.~\ref{galfit_100_FIT_goodsn_100_Map} (and all following figures in an analogous manner), the three images in the first row (top) demonstrate the process of subtraction of \textit{hopelessly faint} sources (see description in Section~\ref{Section_Prior_Extraction_Photometry}). 
From left to right, we show the original image for this band, the faint-source-model image that is constructed based on SED-predicted flux values, and the faint-source-subtracted image which is made by subtracting the middle panel from the left panel. 
The three images in the second row (bottom) demonstrate the prior-extraction photometry for this band. From left to right, they show the faint-source-subtracted image (identical to the top-right panel, but here marking sources retained to be fit with yellow circles), the best-fit-model image, and the residual image which is the left panel minus the middle panel. 
We run SExtractor on the residual image to detect additional sources for each band, shown with green circles (although, at this 100$\mu$m band we detect no  additional source). 
The first two rows of Figs.~\ref{galfit_160_FIT_goodsn_160_Map}~to~\ref{galfit_1160_FIT_goodsn_1160_Map} are identical to the panels shown in Fig.~\ref{galfit_100_FIT_goodsn_100_Map}. 
The third row in these later figures, from left to right, shows the faint-source-subtracted image, the best-fit-model image (this time including both the prior sources and additional sources detected in the residual maps), and the final residual image. 

All images have the same logarithmic scaling. The radii of the circle symbols are proportional to the square root of their measured source flux \MINORREREVISED[]{densities}, but capped to a maximum when $\mathrm{S/N}$ reaches 10.
Sources with $\mathrm{S/N} < 3$ are shown as dashed circles. 

\begin{figure}[hb]
\centering
\includegraphics[width=0.9\textwidth]{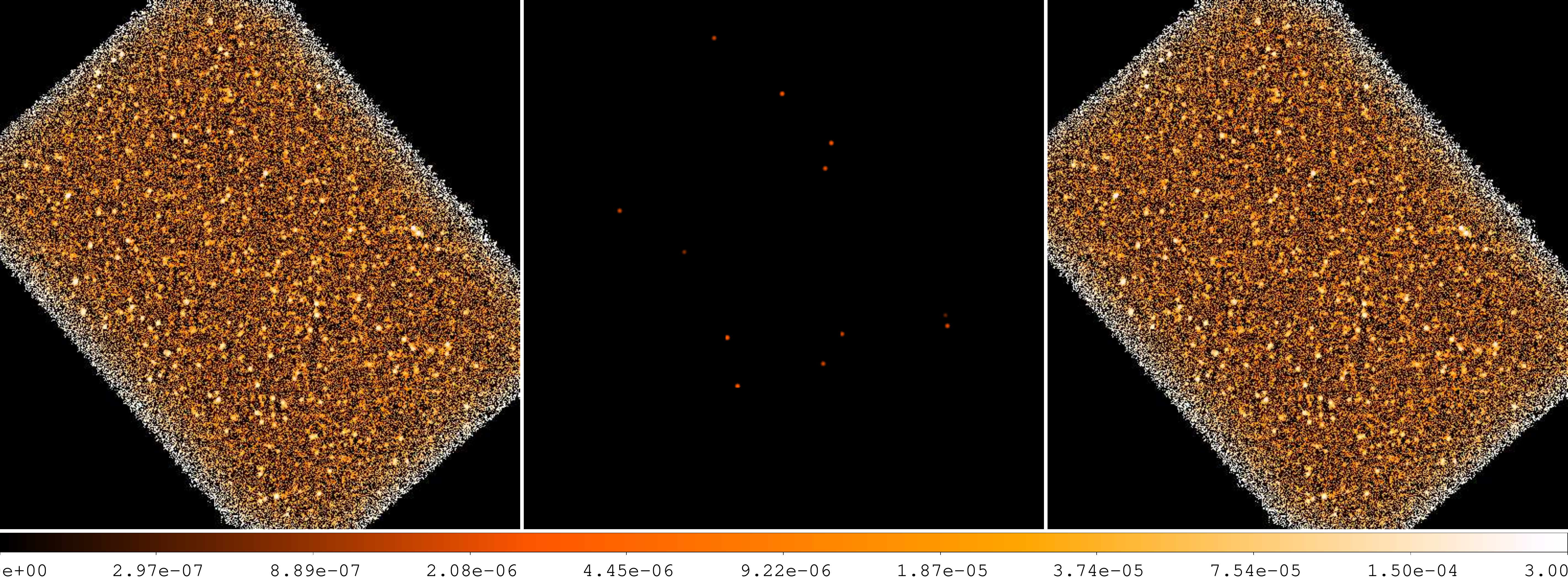}
\includegraphics[width=0.9\textwidth]{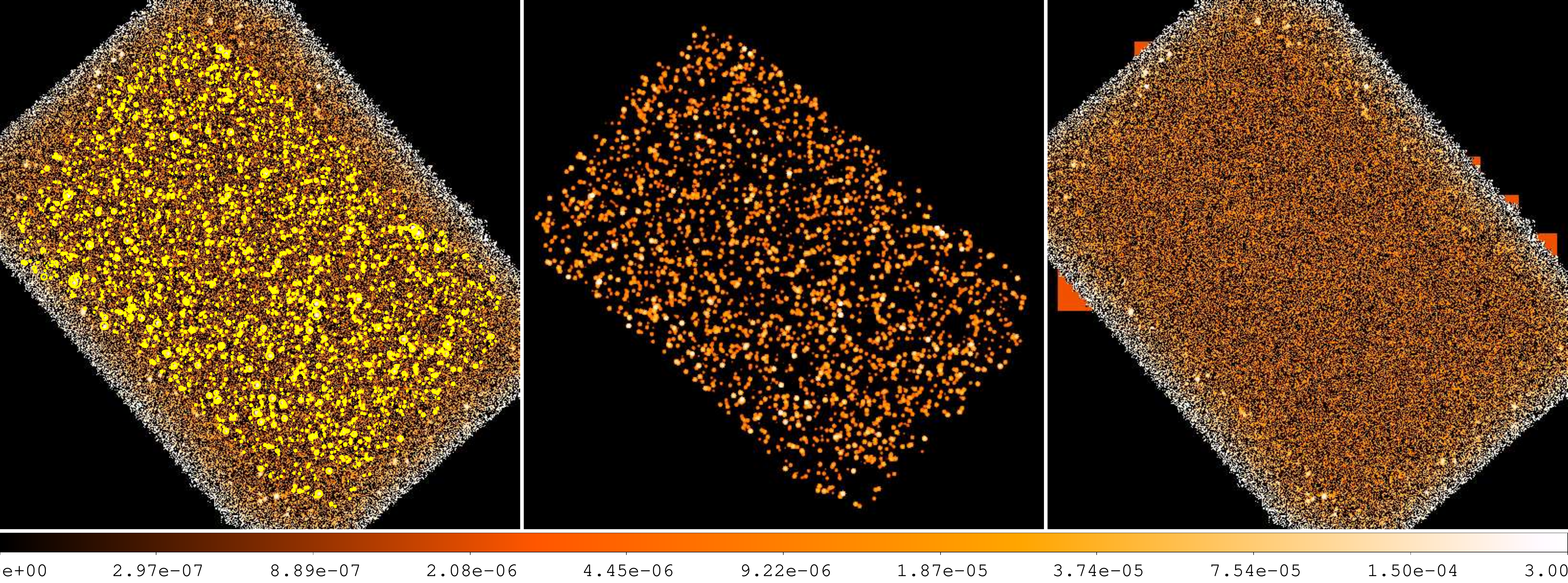}
\caption{%
    Photometry image products at 100~$\mu$m. See descriptions in the text. 
    \label{galfit_100_FIT_goodsn_100_Map}
}
\end{figure}

\begin{figure}
\centering
\includegraphics[width=0.9\textwidth]{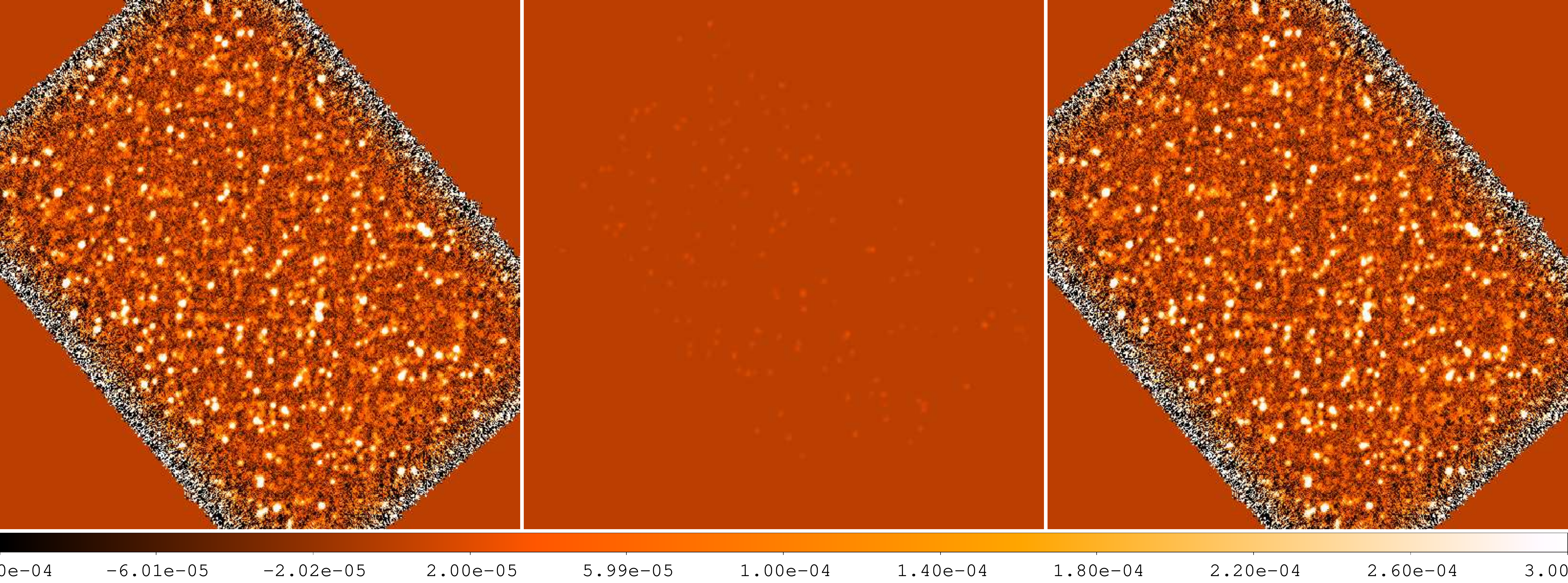}
\includegraphics[width=0.9\textwidth]{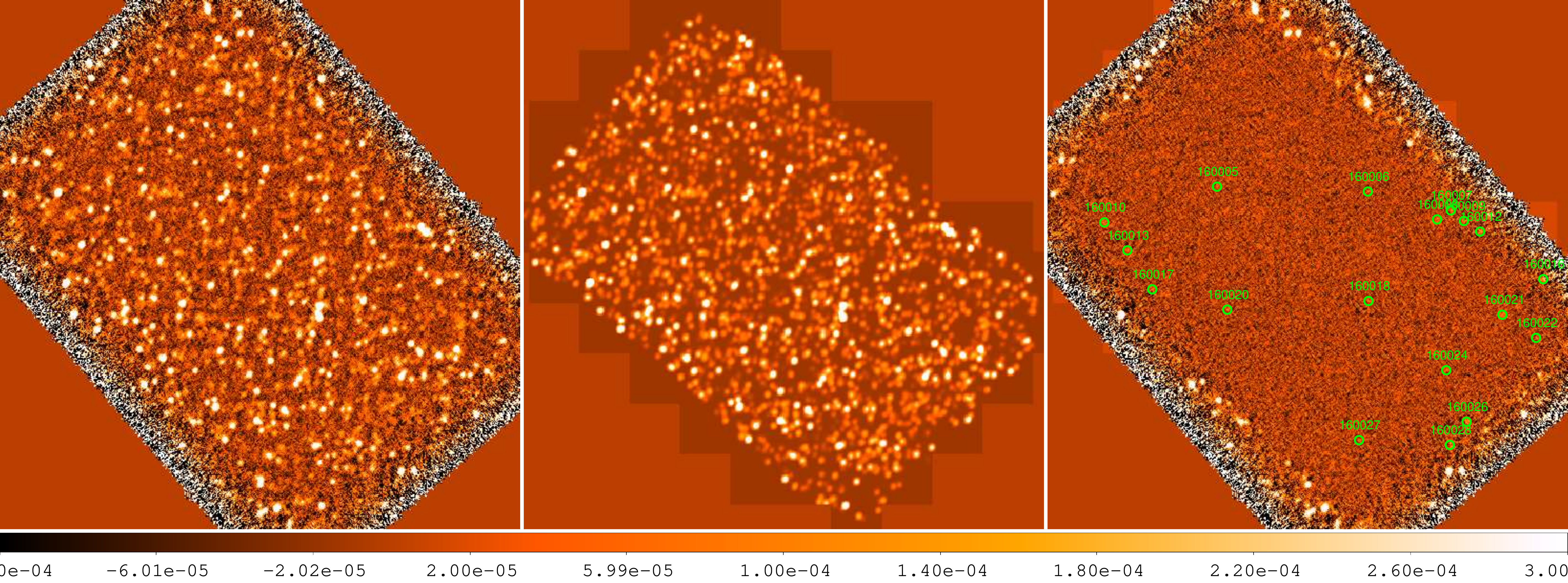}
\includegraphics[width=0.9\textwidth]{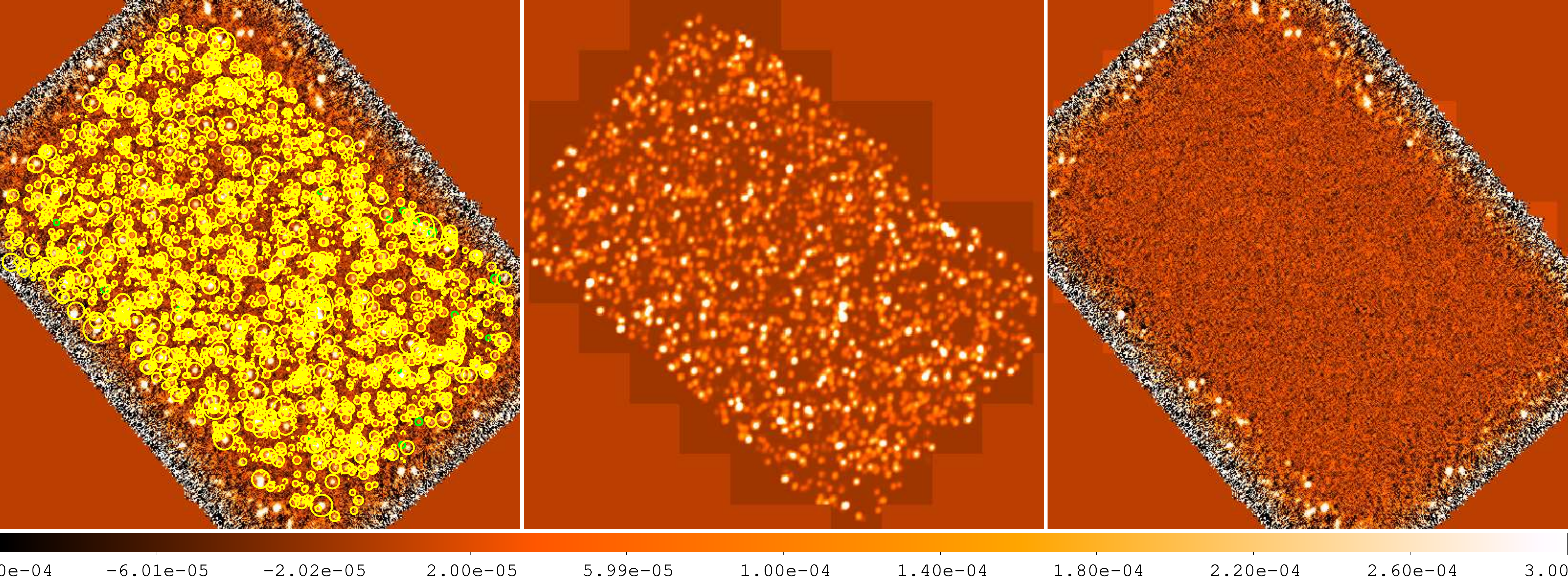}
\caption{%
    Photometry image products at 160~$\mu$m. See descriptions in the text. 
    \label{galfit_160_FIT_goodsn_160_Map}
}
\end{figure}

\begin{figure}
\centering
\includegraphics[width=0.9\textwidth]{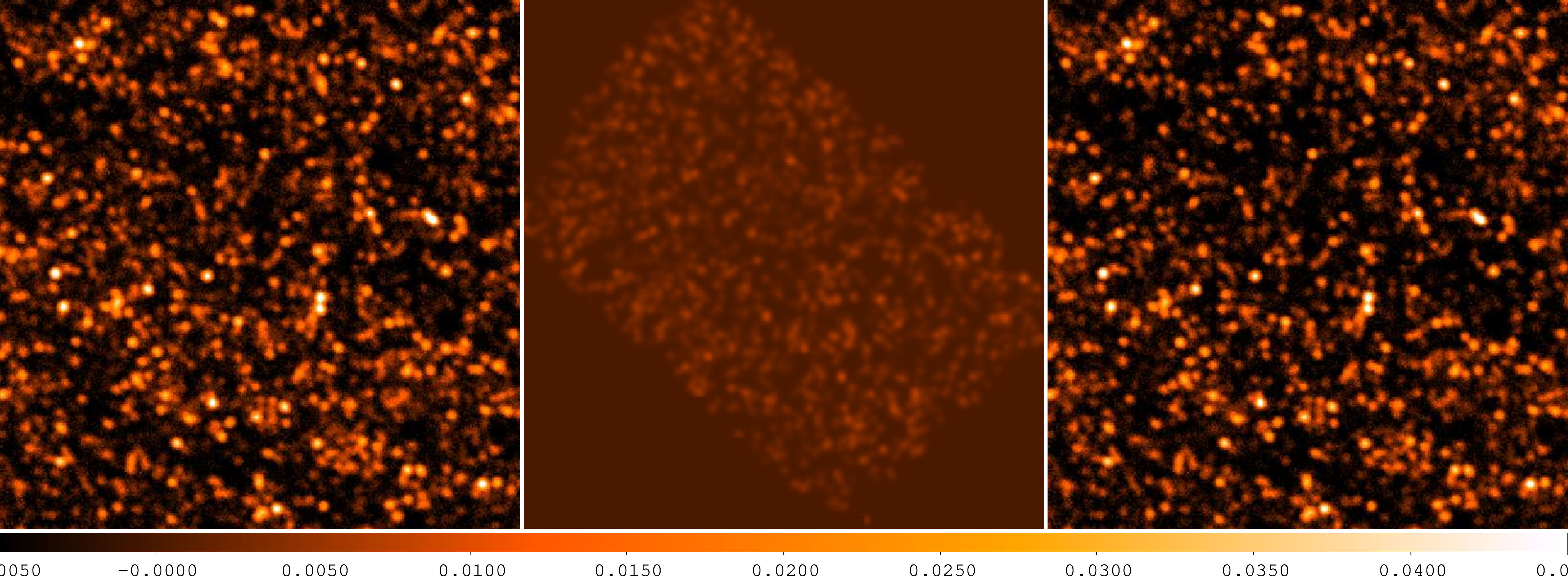}
\includegraphics[width=0.9\textwidth]{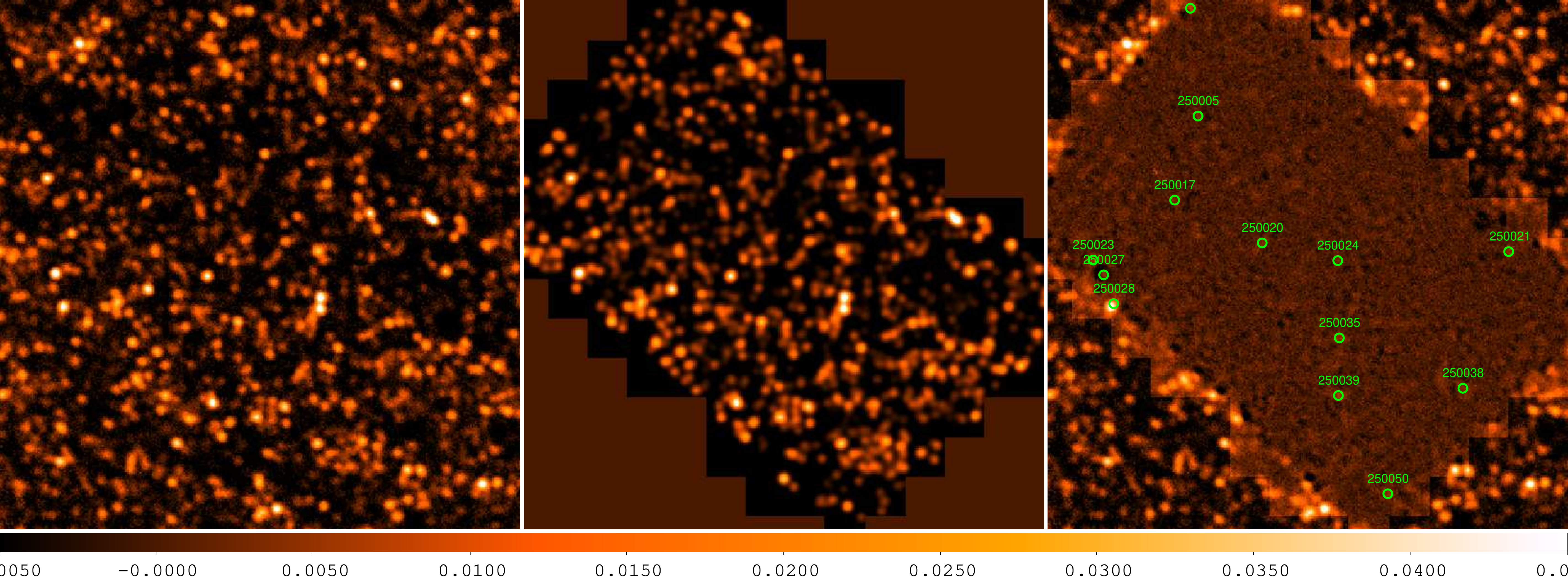}
\includegraphics[width=0.9\textwidth]{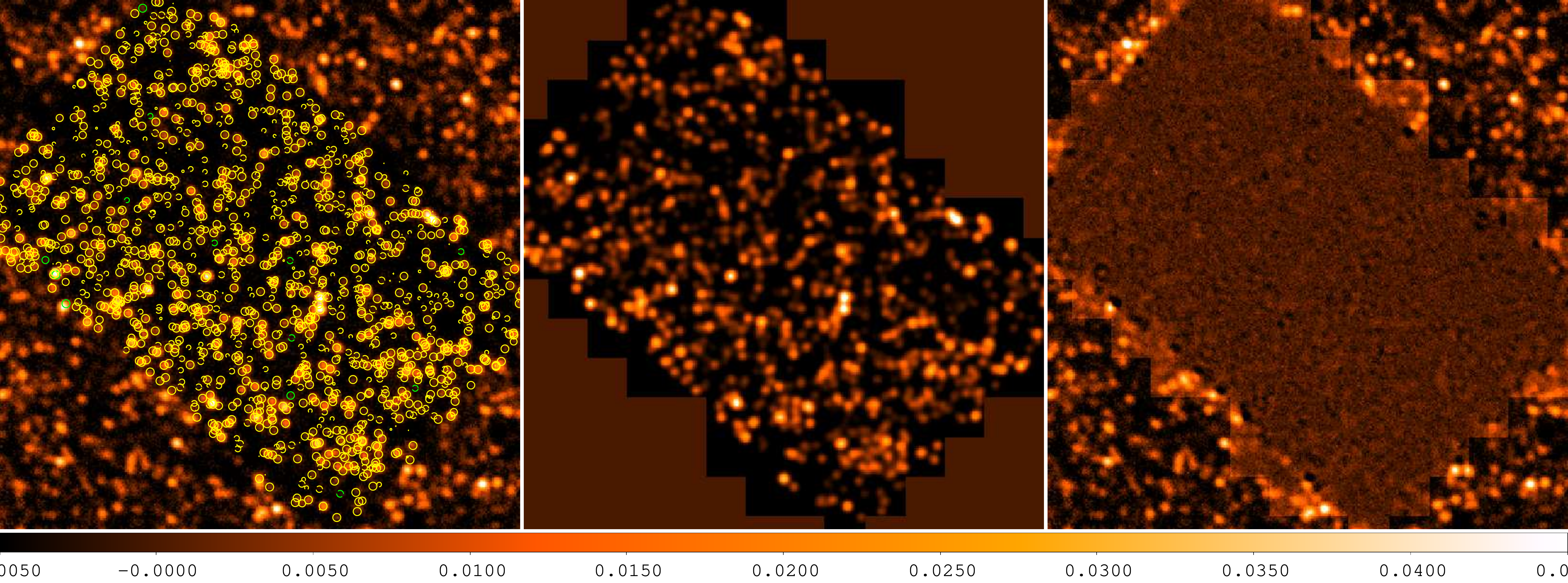}
\caption{%
    Photometry image products at 250~$\mu$m. See descriptions in the text. 
    \label{galfit_250_FIT_goodsn_250_Map}
}
\end{figure}

\begin{figure}
\centering
\includegraphics[width=0.9\textwidth]{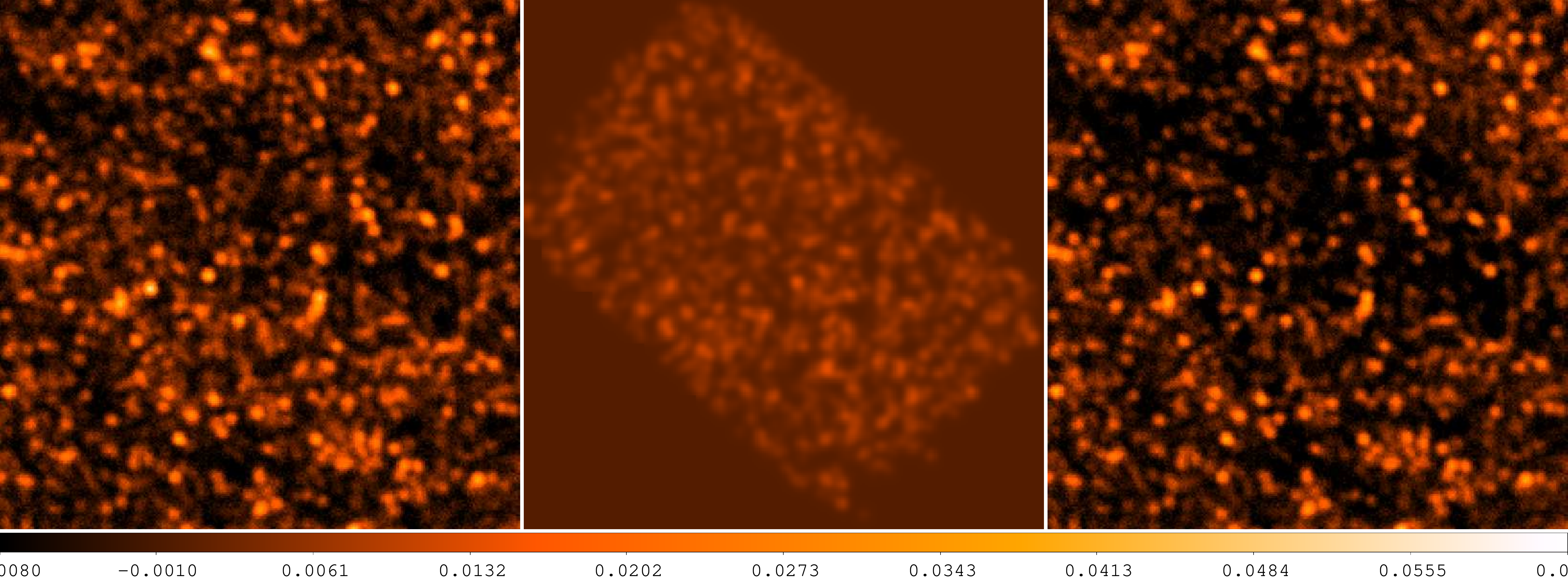}
\includegraphics[width=0.9\textwidth]{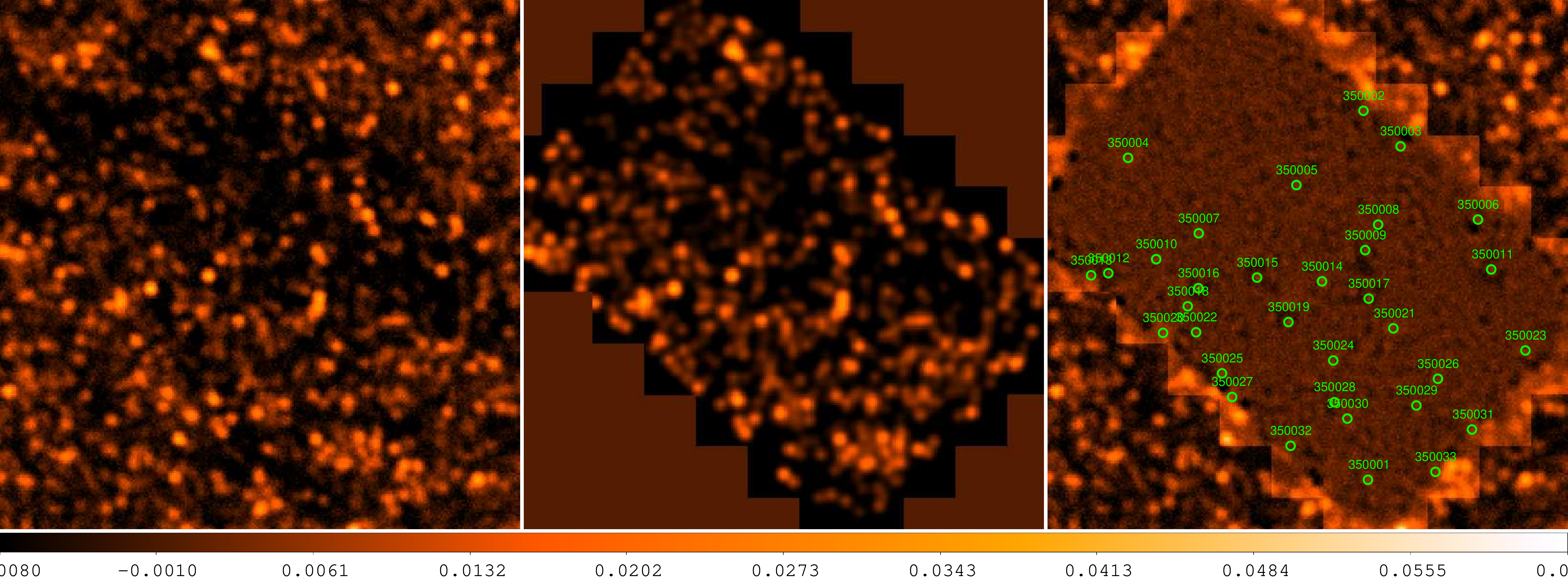}
\includegraphics[width=0.9\textwidth]{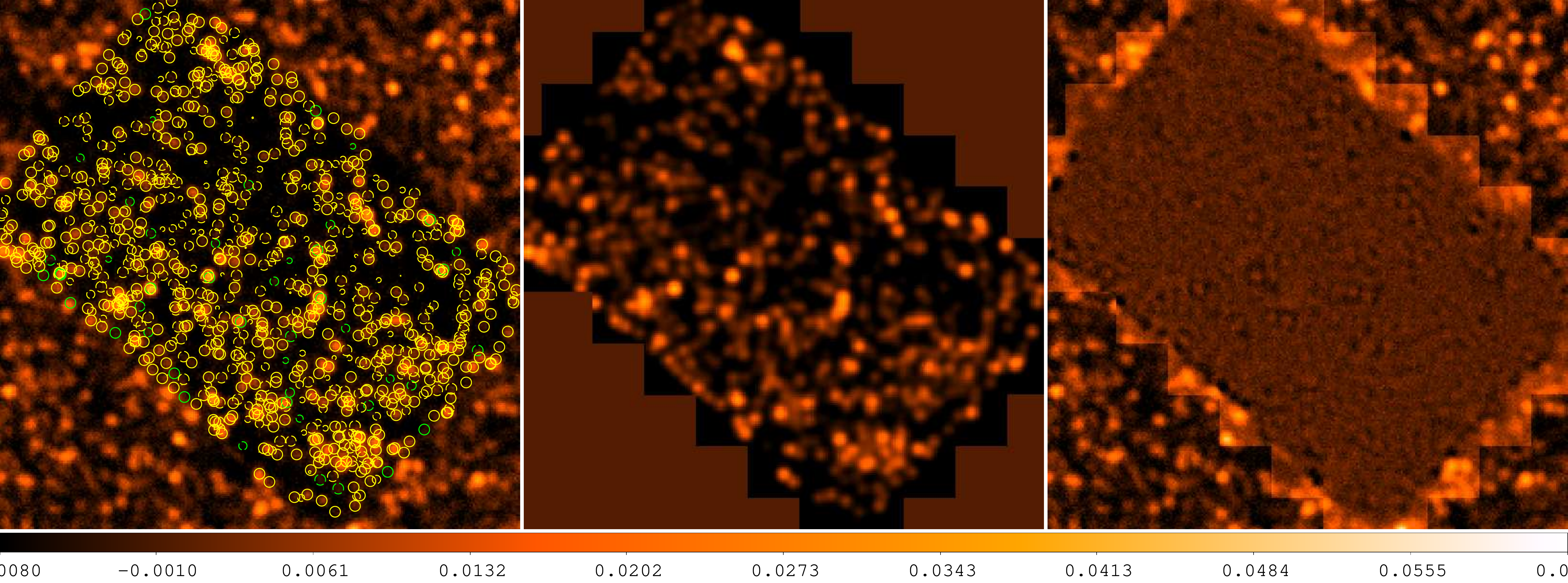}
\caption{%
    Photometry image products at 350~$\mu$m. See descriptions in the text. 
    \label{galfit_350_FIT_goodsn_350_Map}
}
\end{figure}

\begin{figure}
\centering
\includegraphics[width=0.9\textwidth]{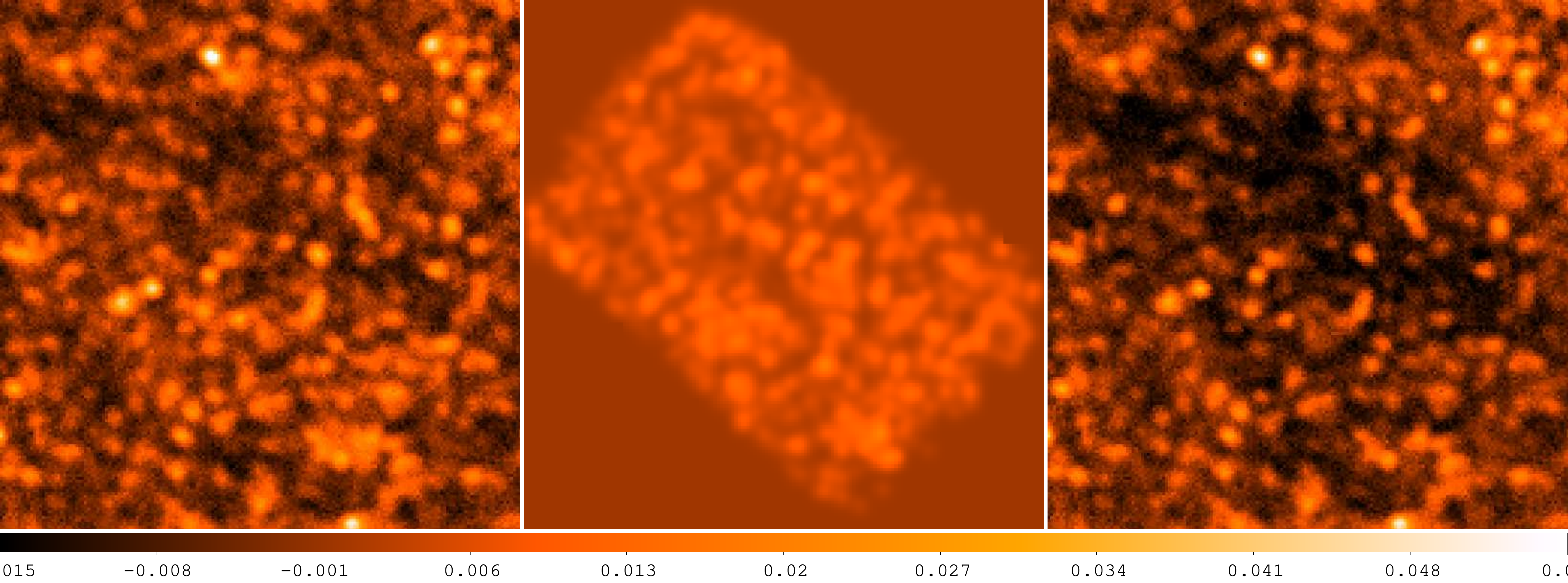}
\includegraphics[width=0.9\textwidth]{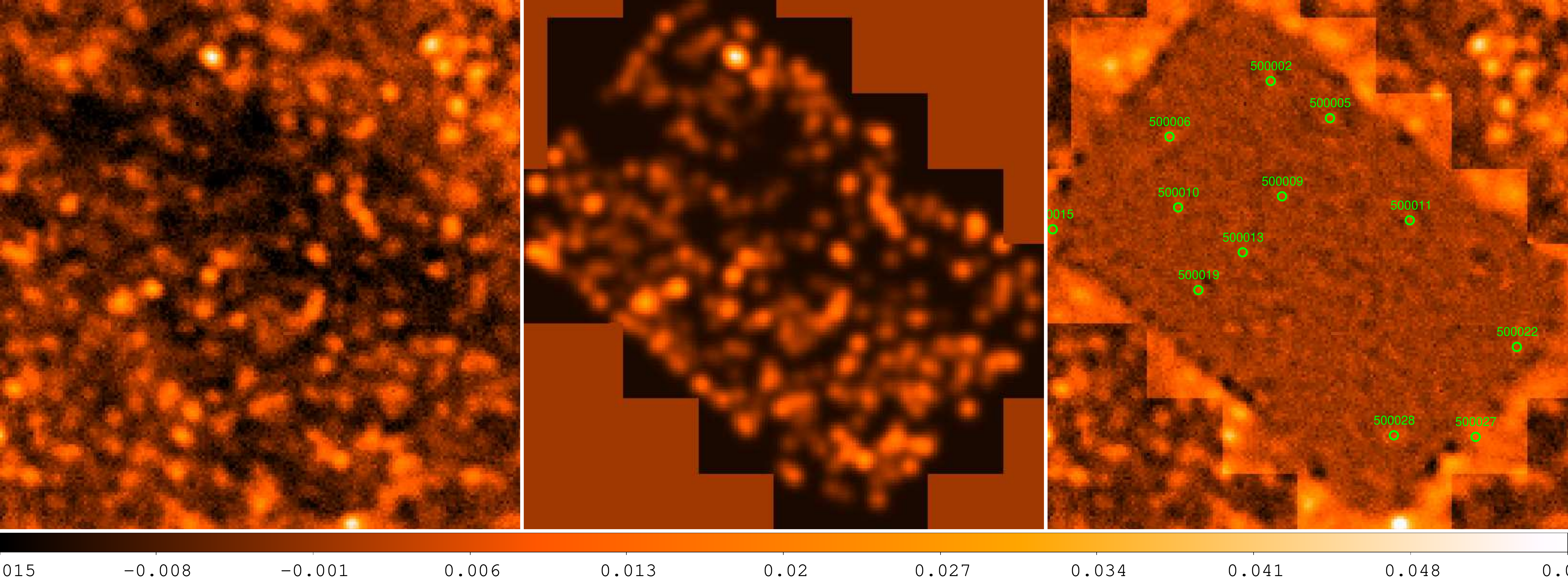}
\includegraphics[width=0.9\textwidth]{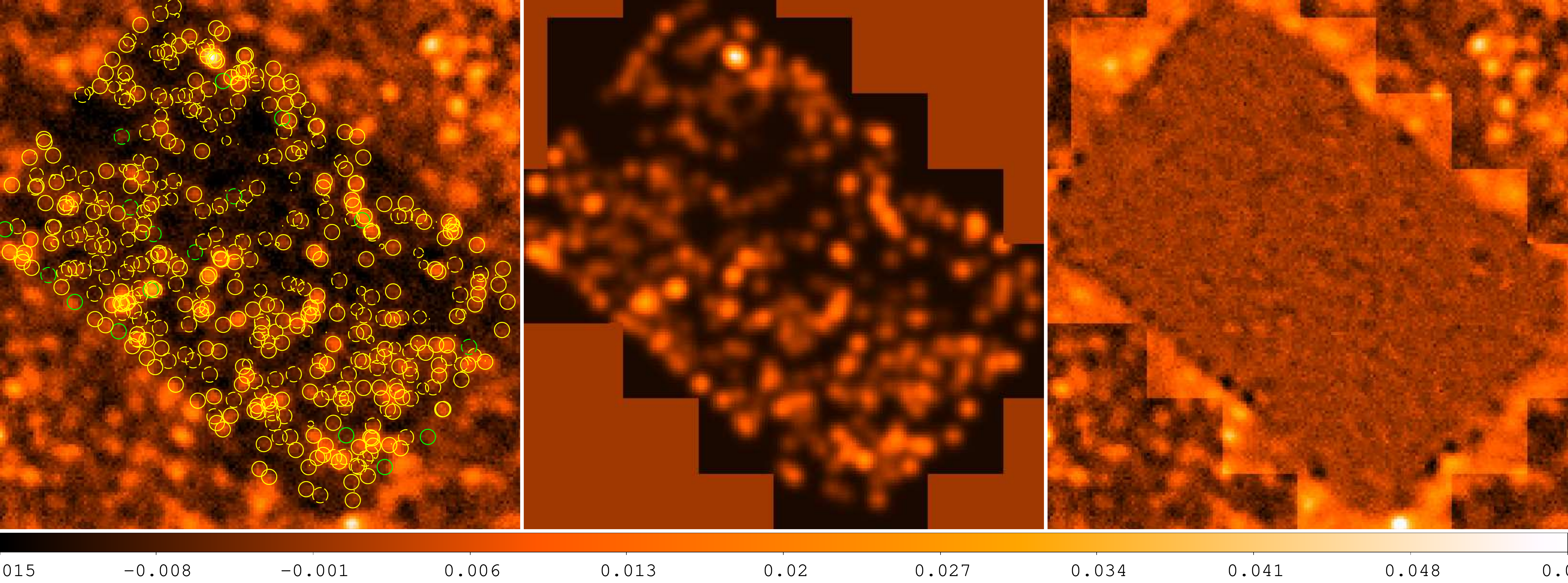}
\caption{%
    Photometry image products at 500~$\mu$m. See descriptions in the text. 
    \label{galfit_500_FIT_goodsn_500_Map}
}
\end{figure}

\begin{figure}
\centering
\includegraphics[width=0.9\textwidth]{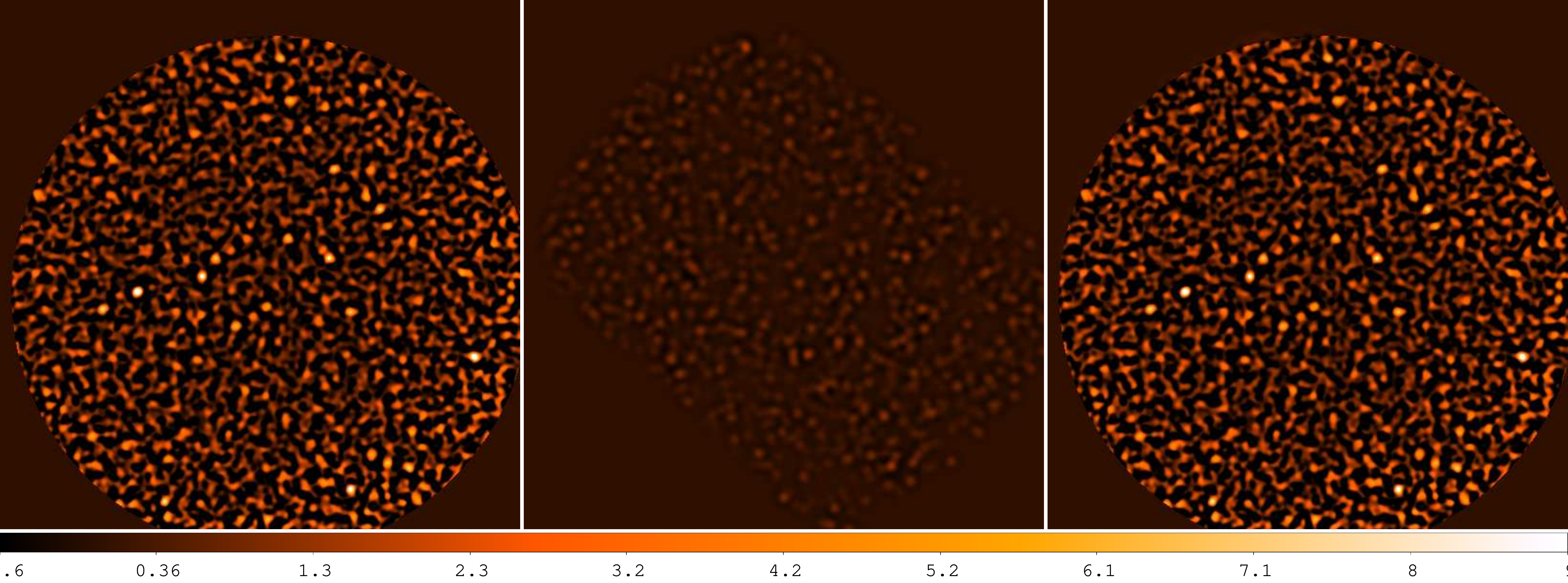}
\includegraphics[width=0.9\textwidth]{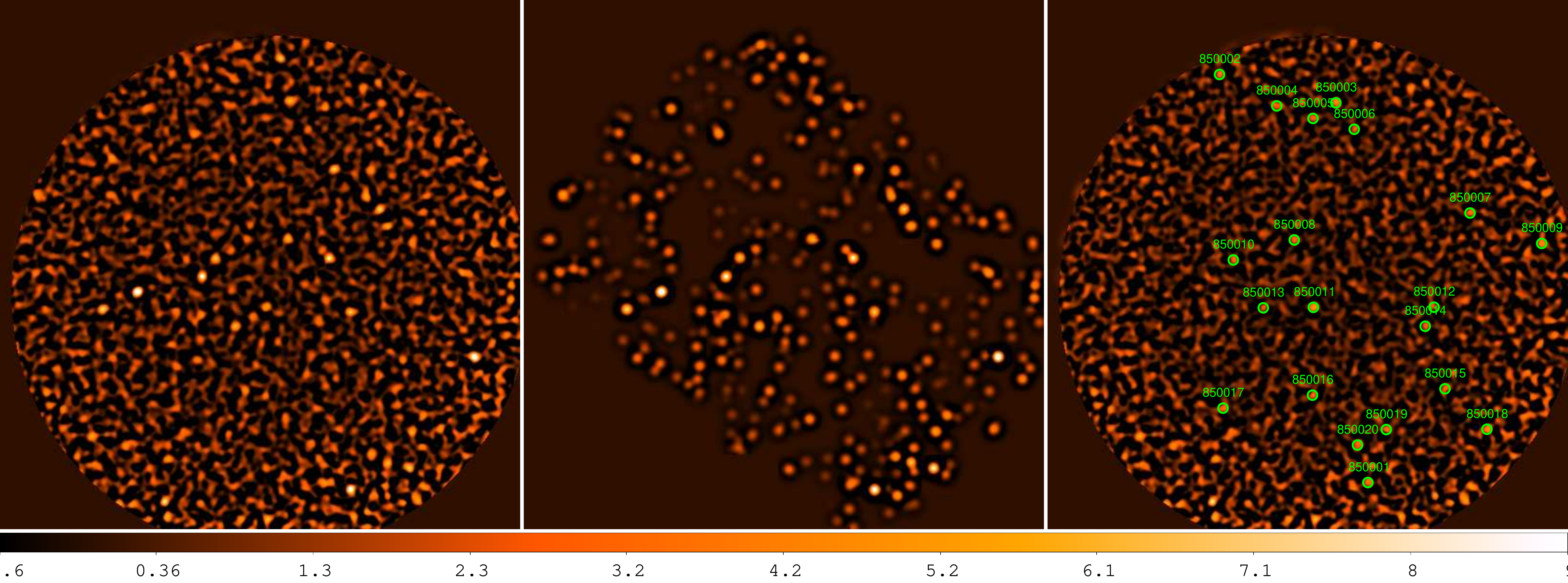}
\includegraphics[width=0.9\textwidth]{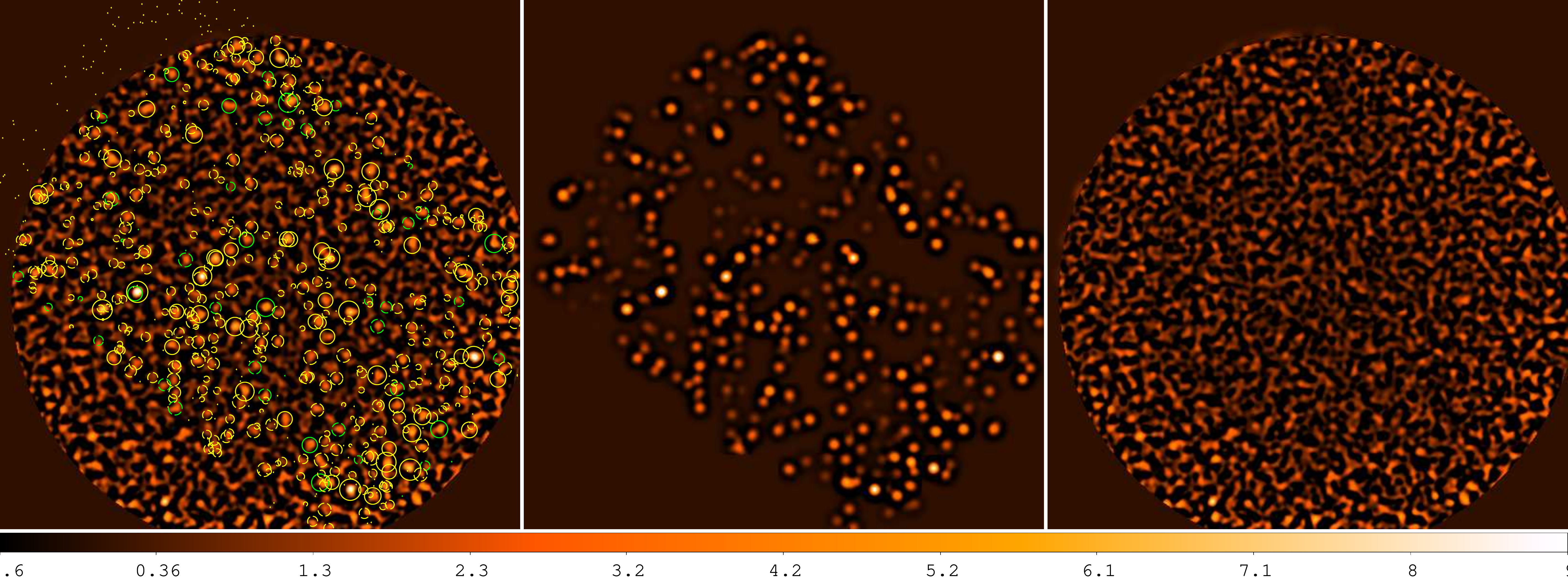}
\caption{%
    Photometry image products at 850~$\mu$m. See descriptions in the text. 
    \label{galfit_850_FIT_goodsn_850_Map}
}
\end{figure}

\begin{figure}
\centering
\includegraphics[width=0.9\textwidth]{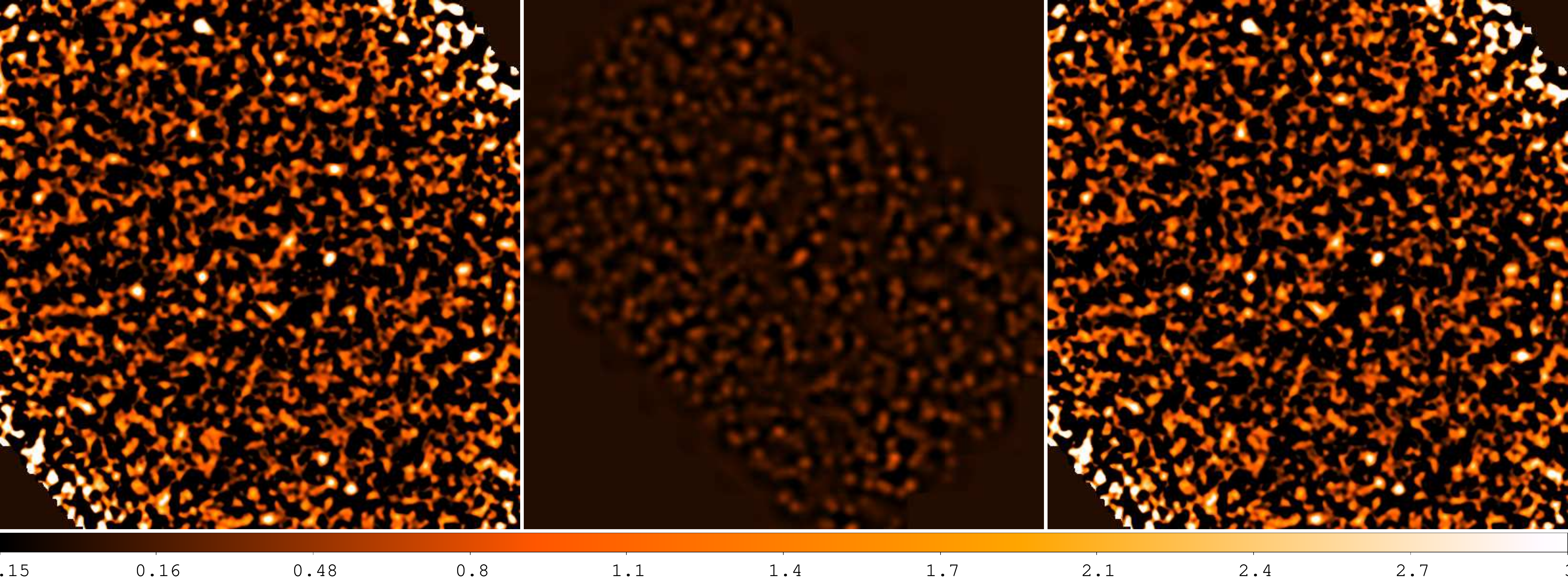}
\includegraphics[width=0.9\textwidth]{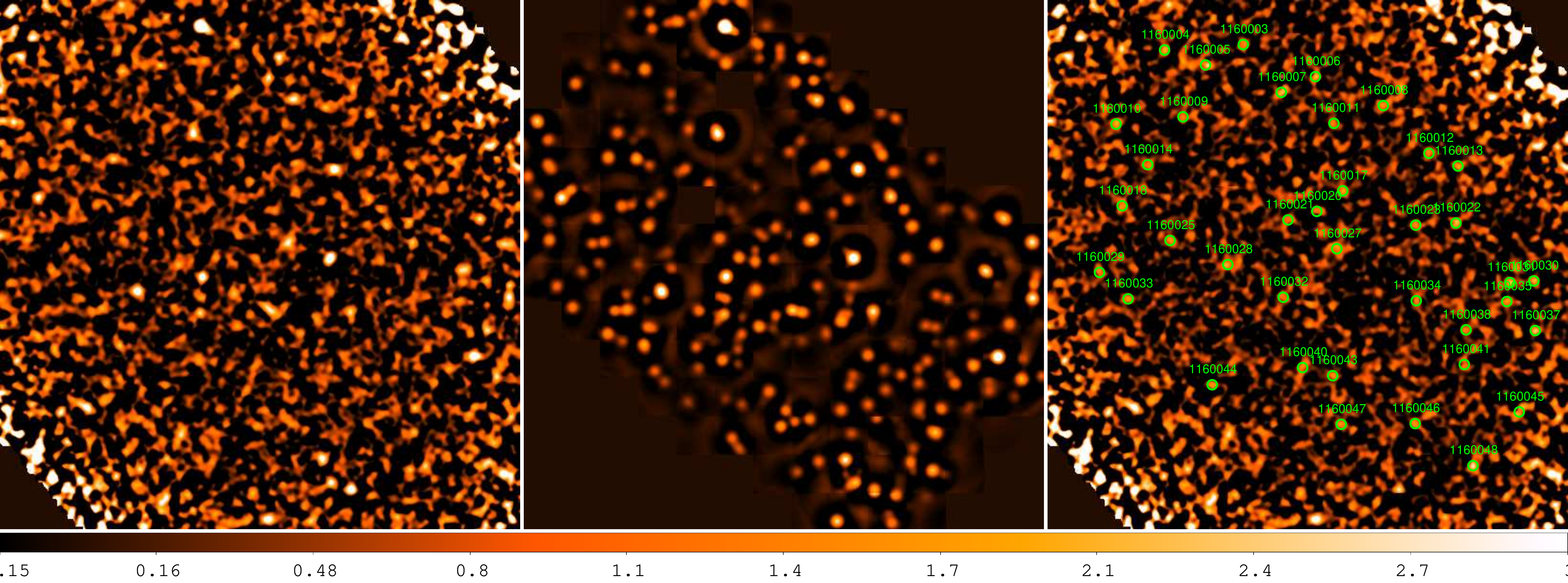}
\includegraphics[width=0.9\textwidth]{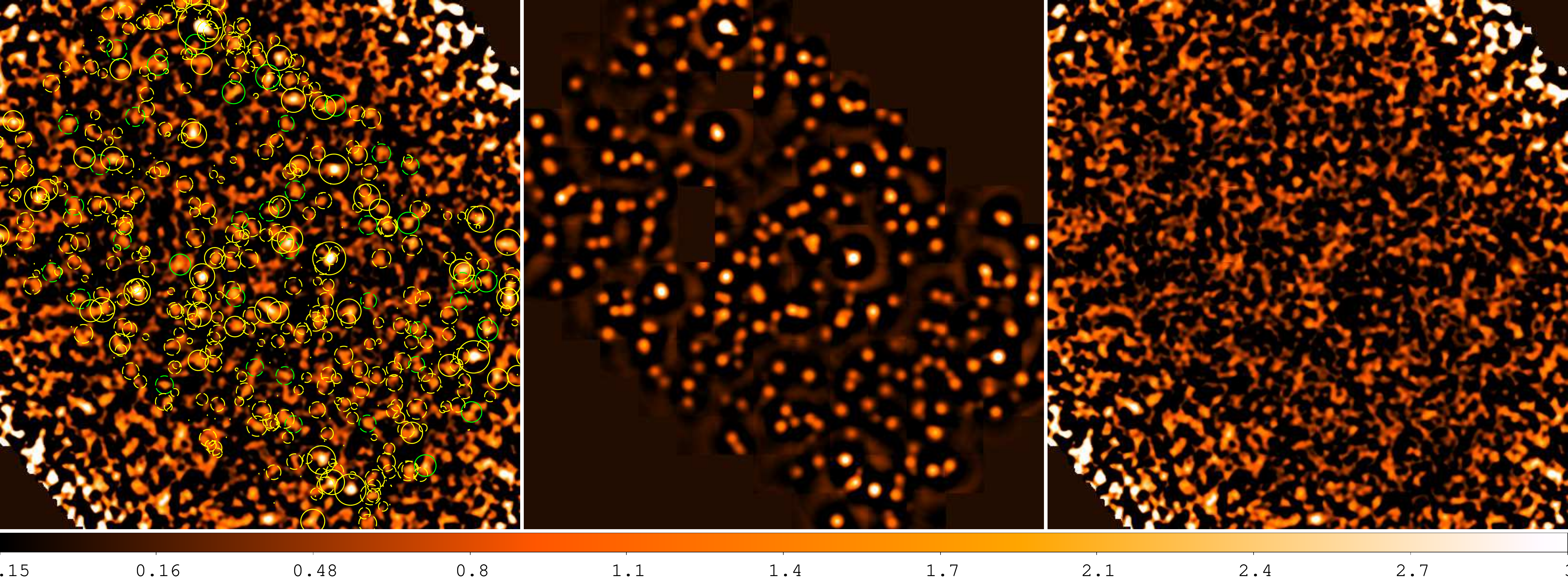}
\caption{%
    Photometry image products at 1.16~mm. See descriptions in the text. 
    \label{galfit_1160_FIT_goodsn_1160_Map}
}
\end{figure}

\clearpage

\section{Simulation Correction Analyses}
\label{Section_Simulation_Performance}

\begin{figure}[hb]
    \centering
    
    \begin{subfigure}[b]{\textwidth}\centering
    \includegraphics[height=2.6cm, trim=0 1cm 0 0]{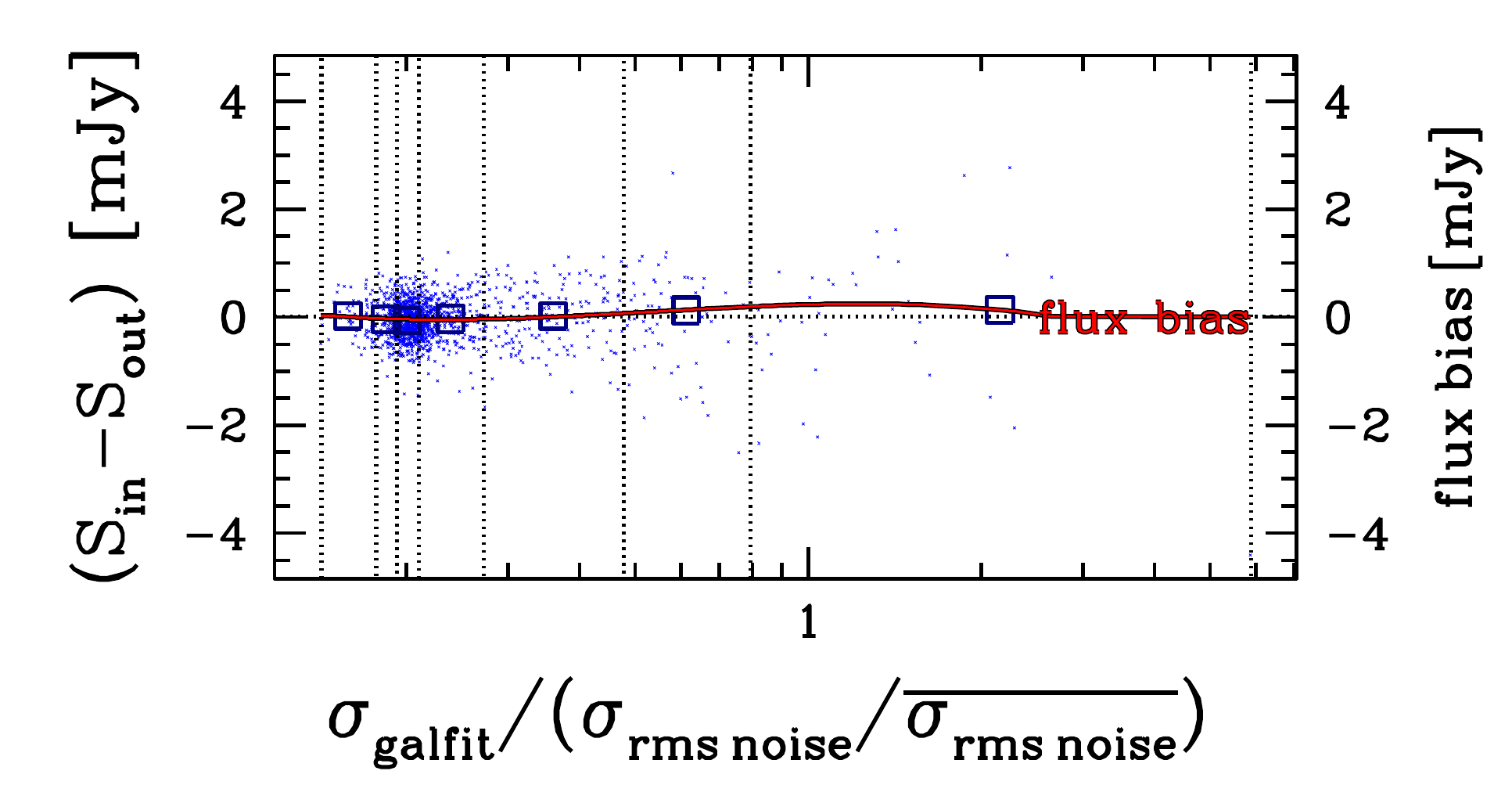}
    \includegraphics[height=2.6cm, trim=0 1cm 0 0]{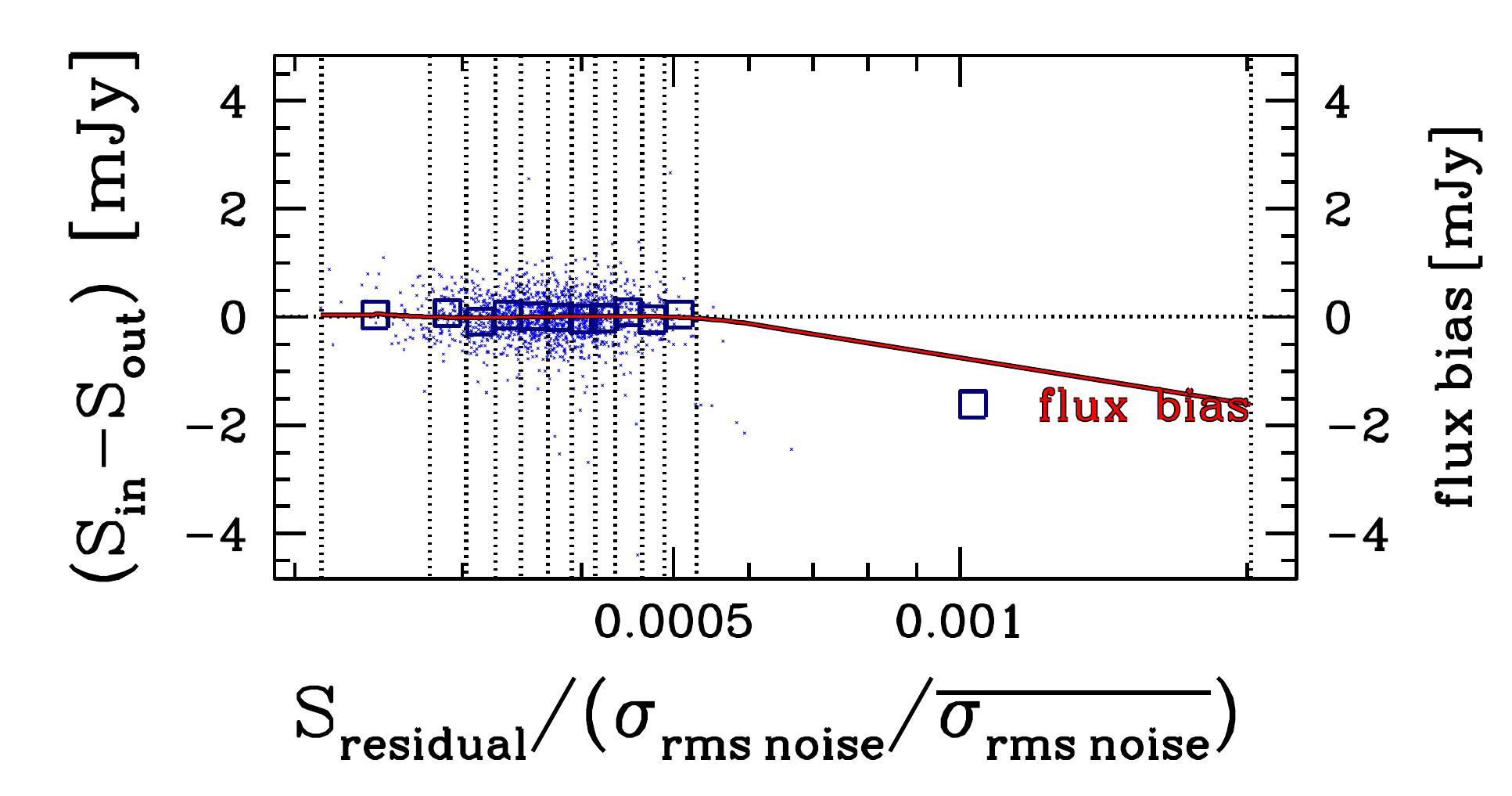}
    \includegraphics[height=2.6cm, trim=0 1cm 0 0]{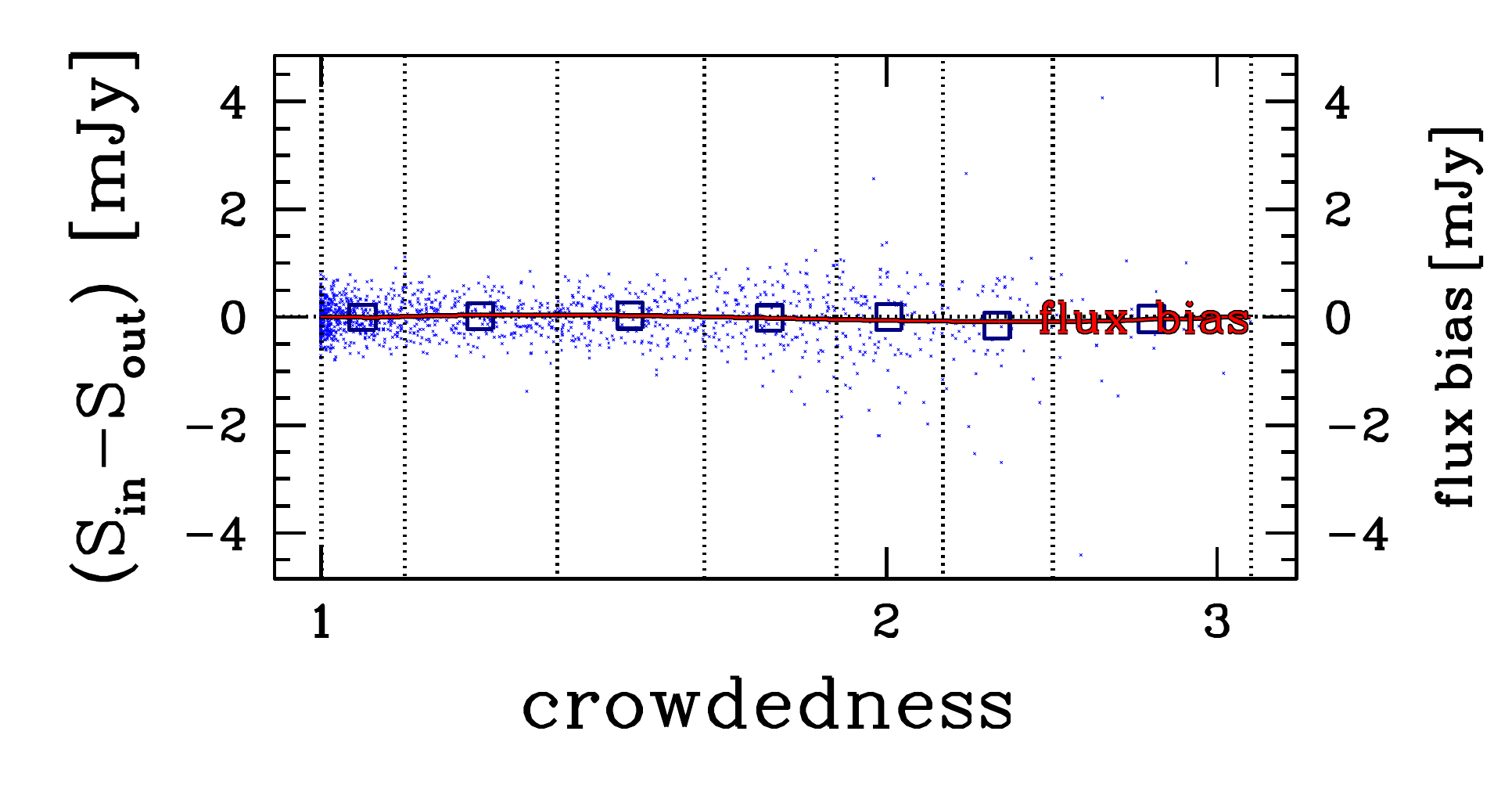}
    \end{subfigure}
    
    \begin{subfigure}[b]{\textwidth}\centering
    \includegraphics[height=2.6cm, trim=0 1cm 0 0]{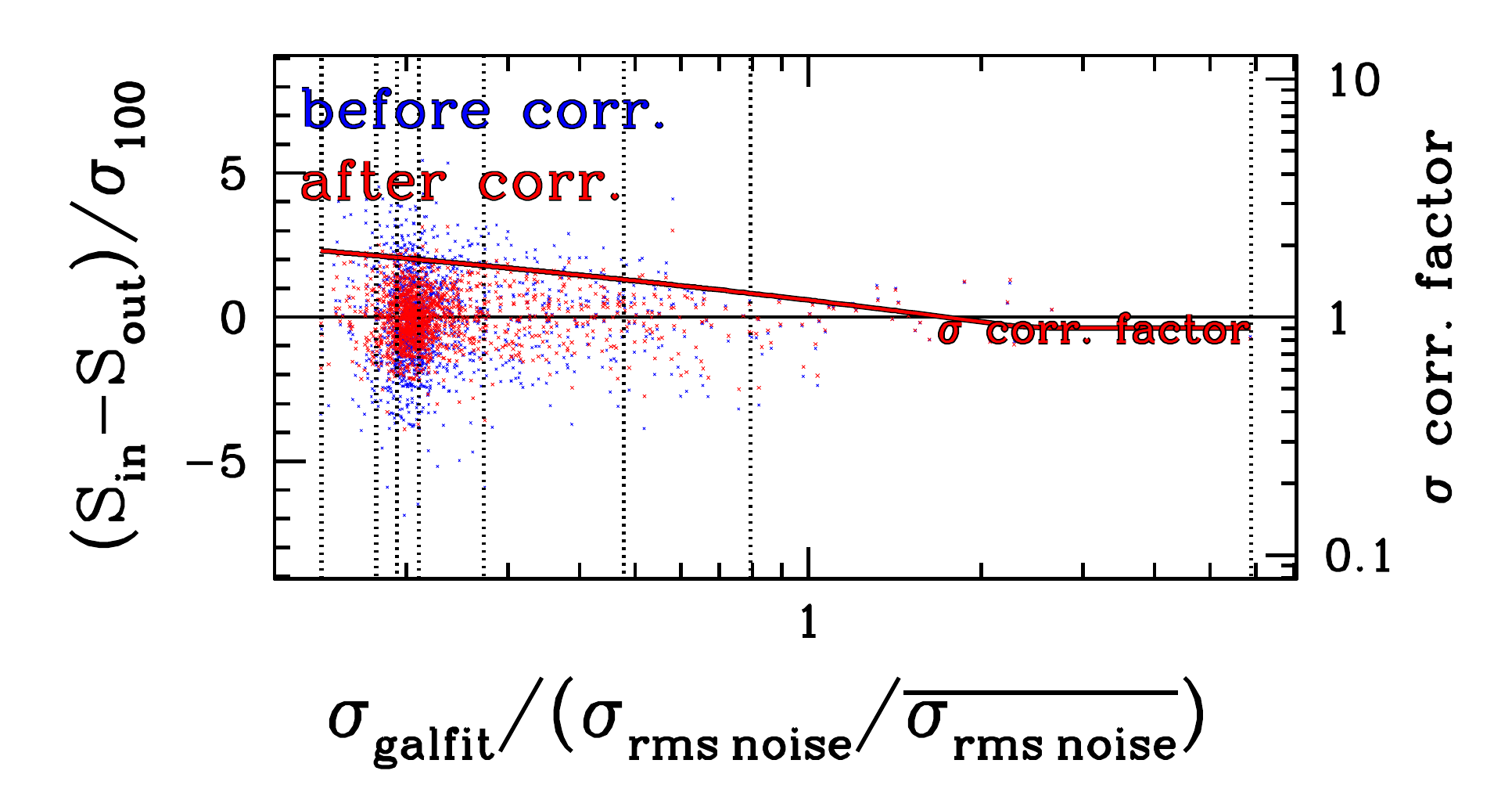}
    \includegraphics[height=2.6cm, trim=0 1cm 0 0]{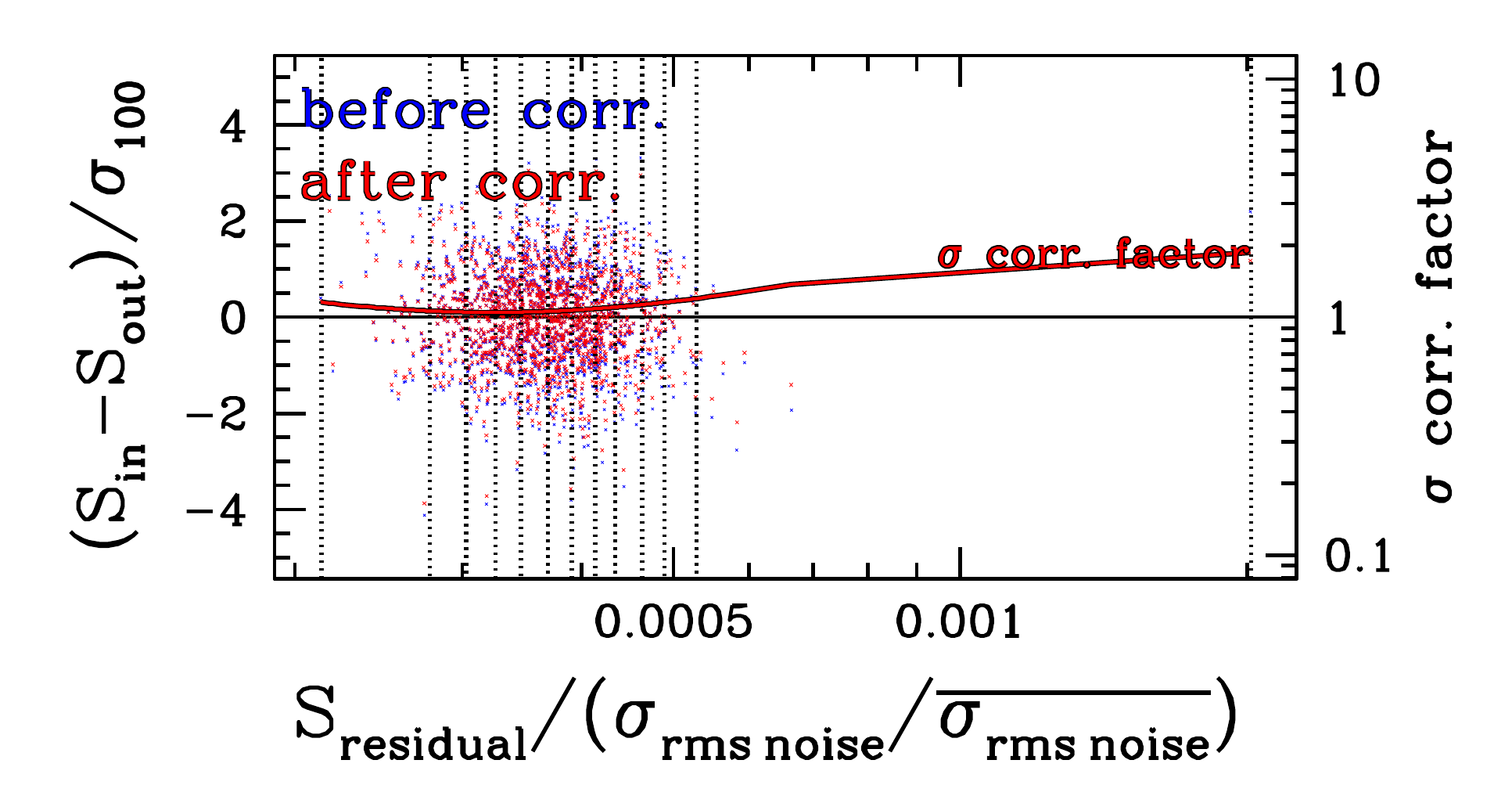}
    \includegraphics[height=2.6cm, trim=0 1cm 0 0]{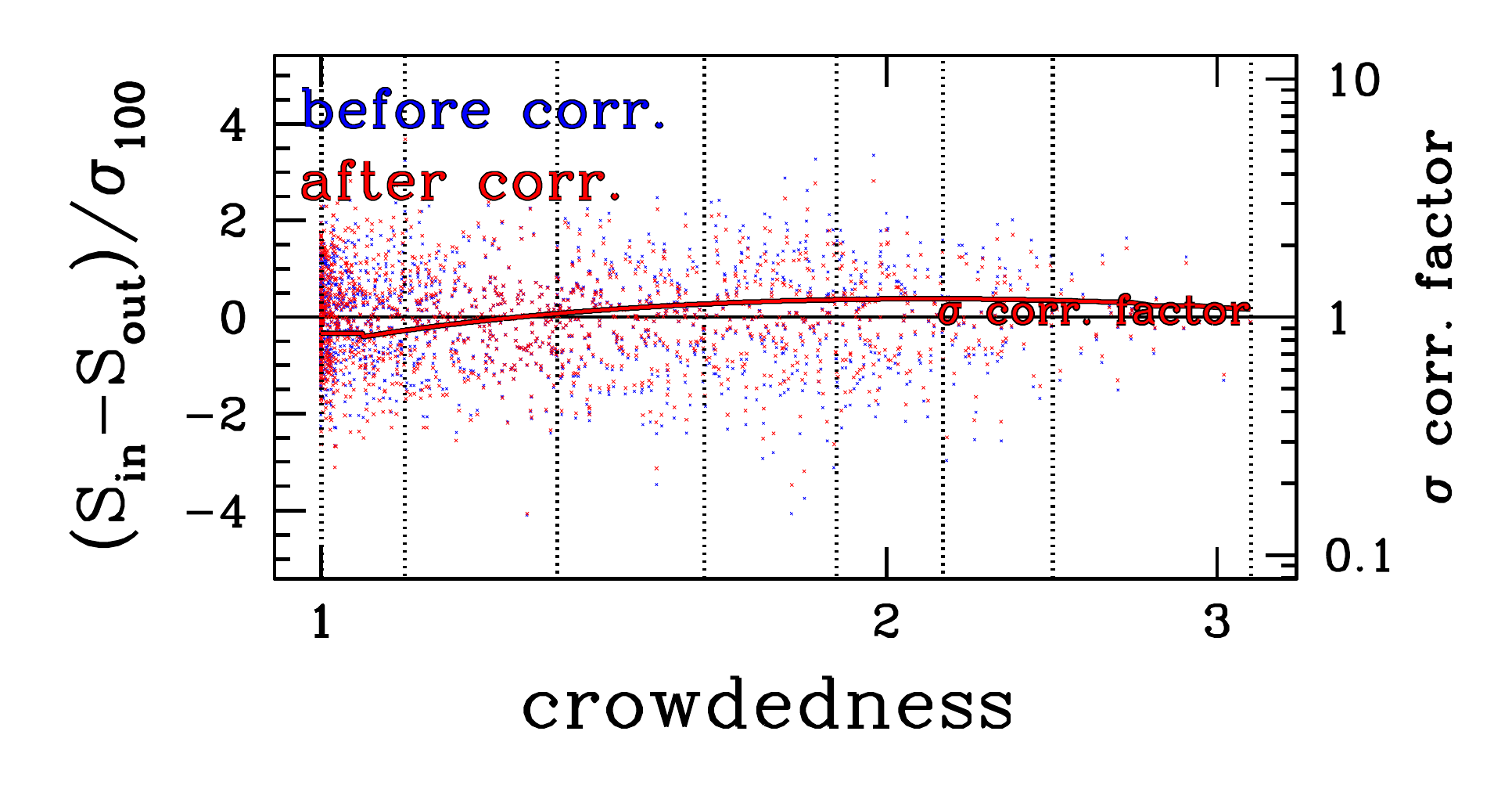}
    \end{subfigure}
    
    \begin{subfigure}[b]{\textwidth}\centering
    \includegraphics[height=2.6cm, trim=0 1cm -1.8cm 0]{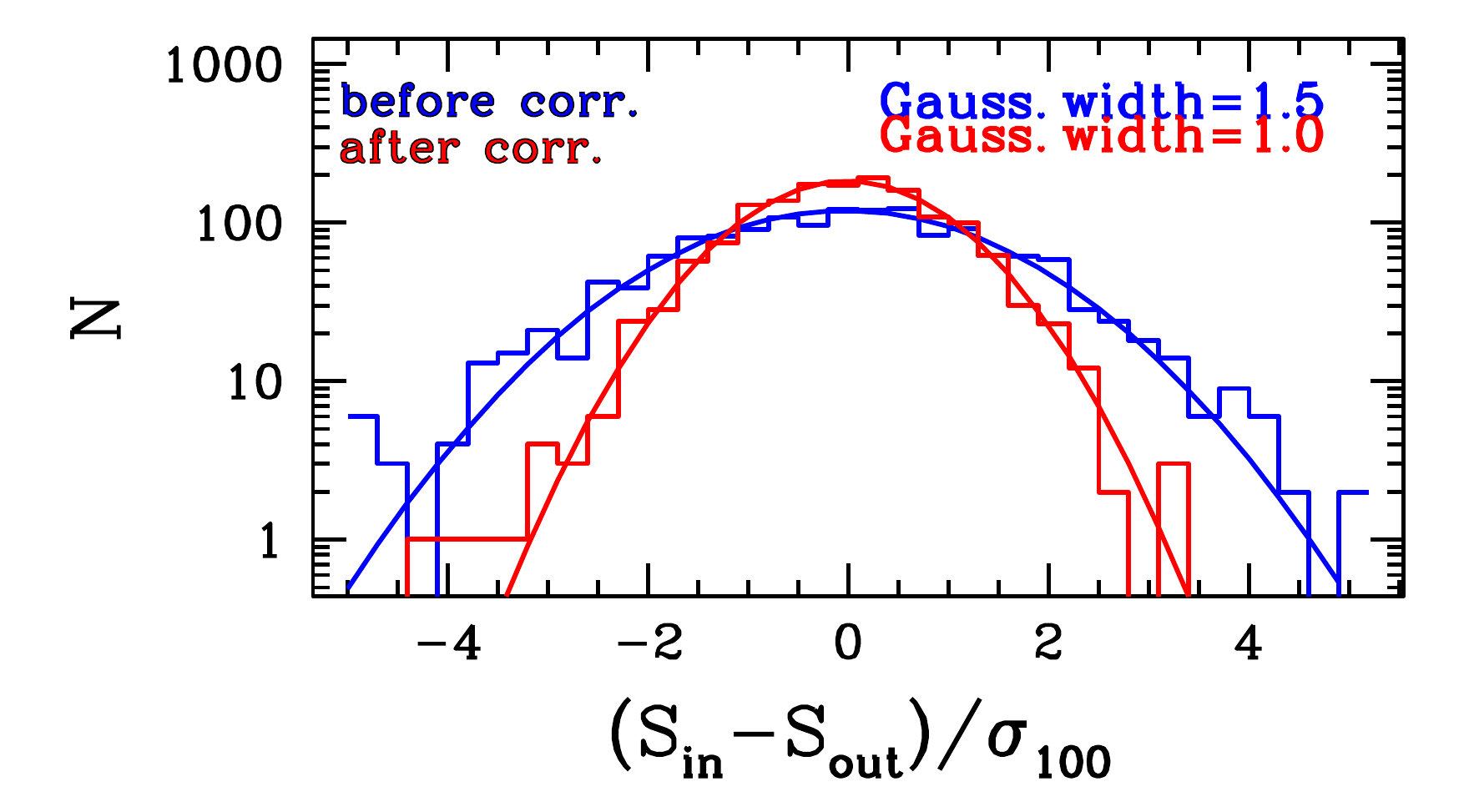}
    \includegraphics[height=2.6cm, trim=0 1cm -1.8cm 0]{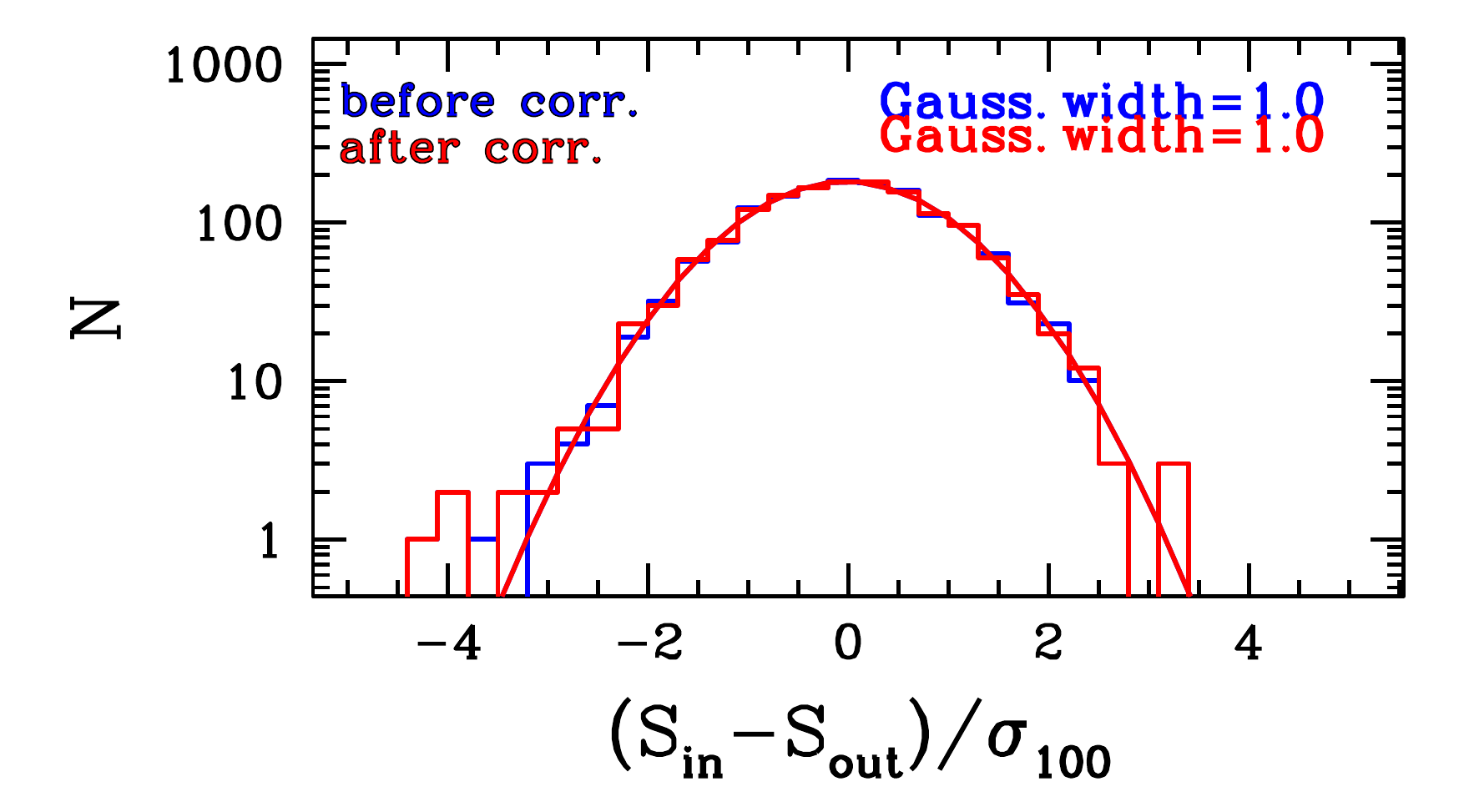}
    \includegraphics[height=2.6cm, trim=0 1cm -1.8cm 0]{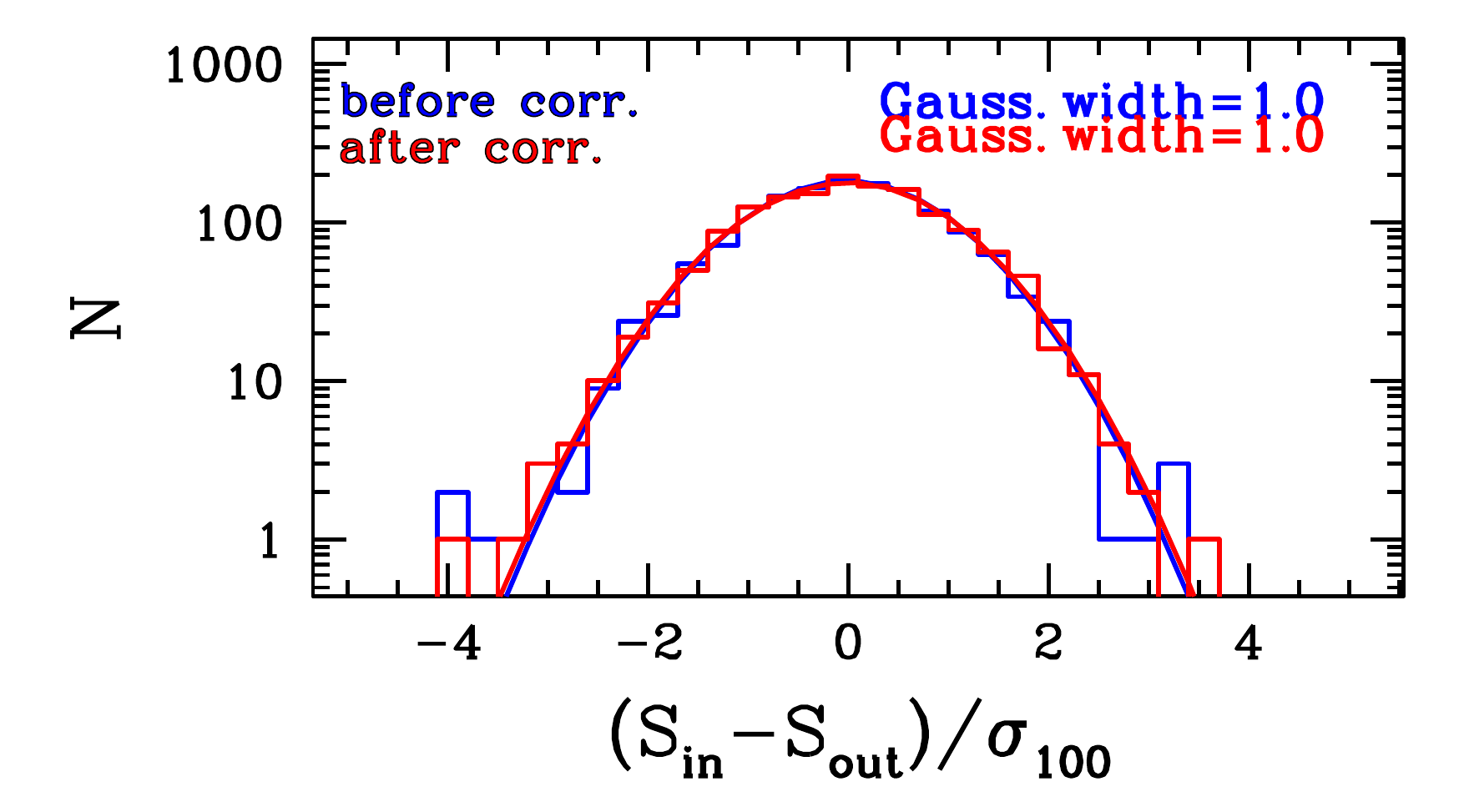}
    \end{subfigure}
    
    \begin{subfigure}[b]{\textwidth}\centering
    \includegraphics[height=2.6cm, trim=0 1cm -1.8cm 0]{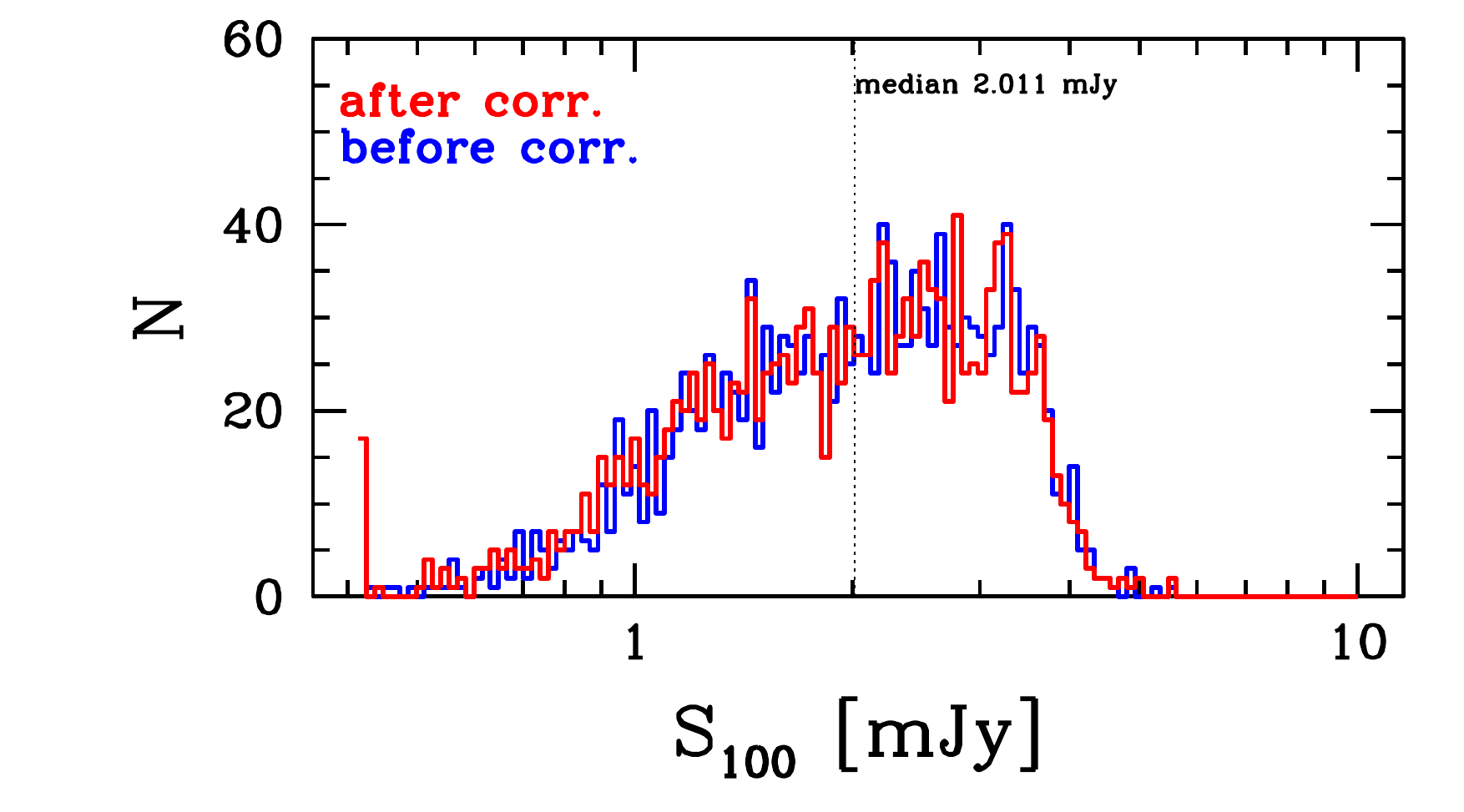}
    \includegraphics[height=2.6cm, trim=0 1cm -1.8cm 0]{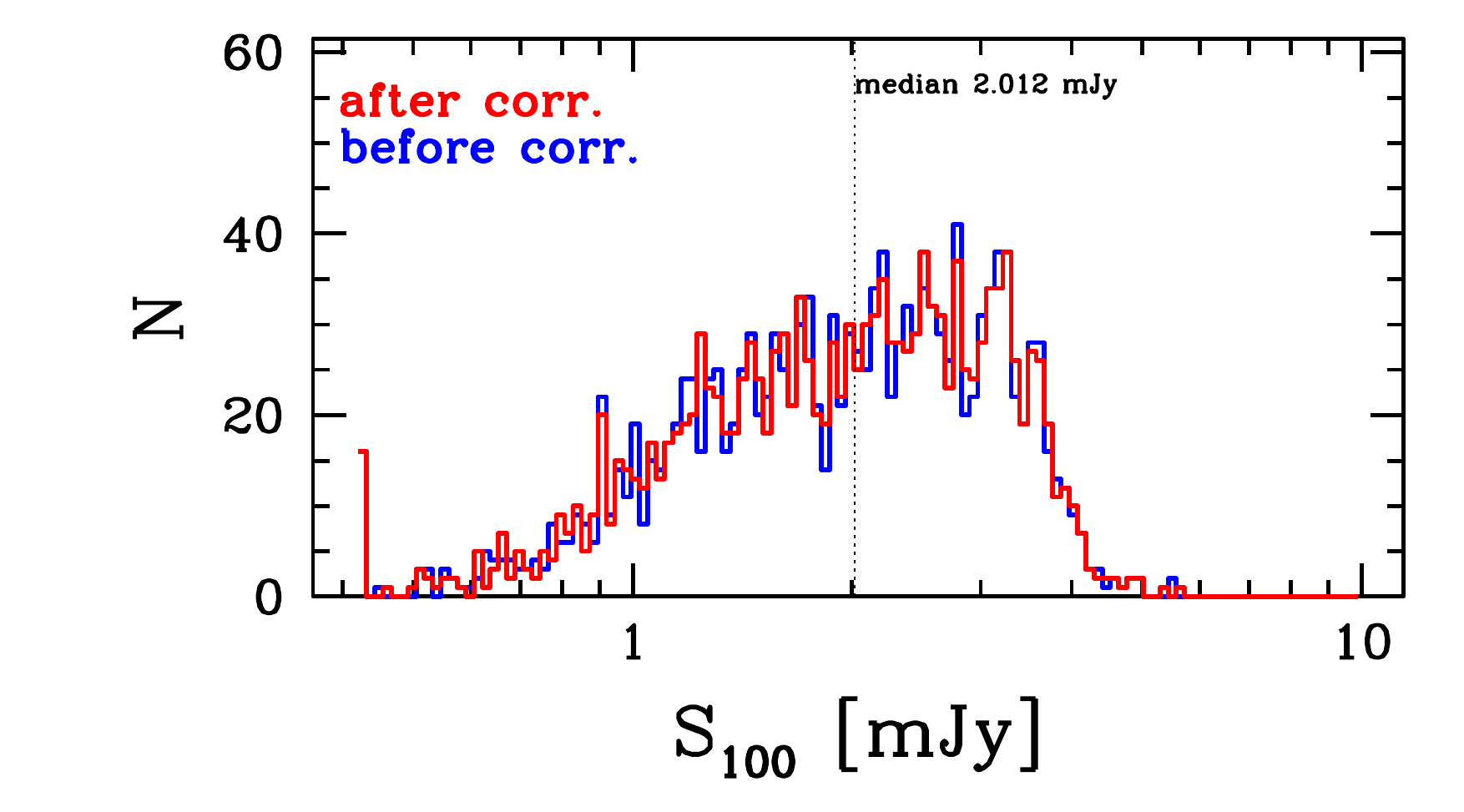}
    \includegraphics[height=2.6cm, trim=0 1cm -1.8cm 0]{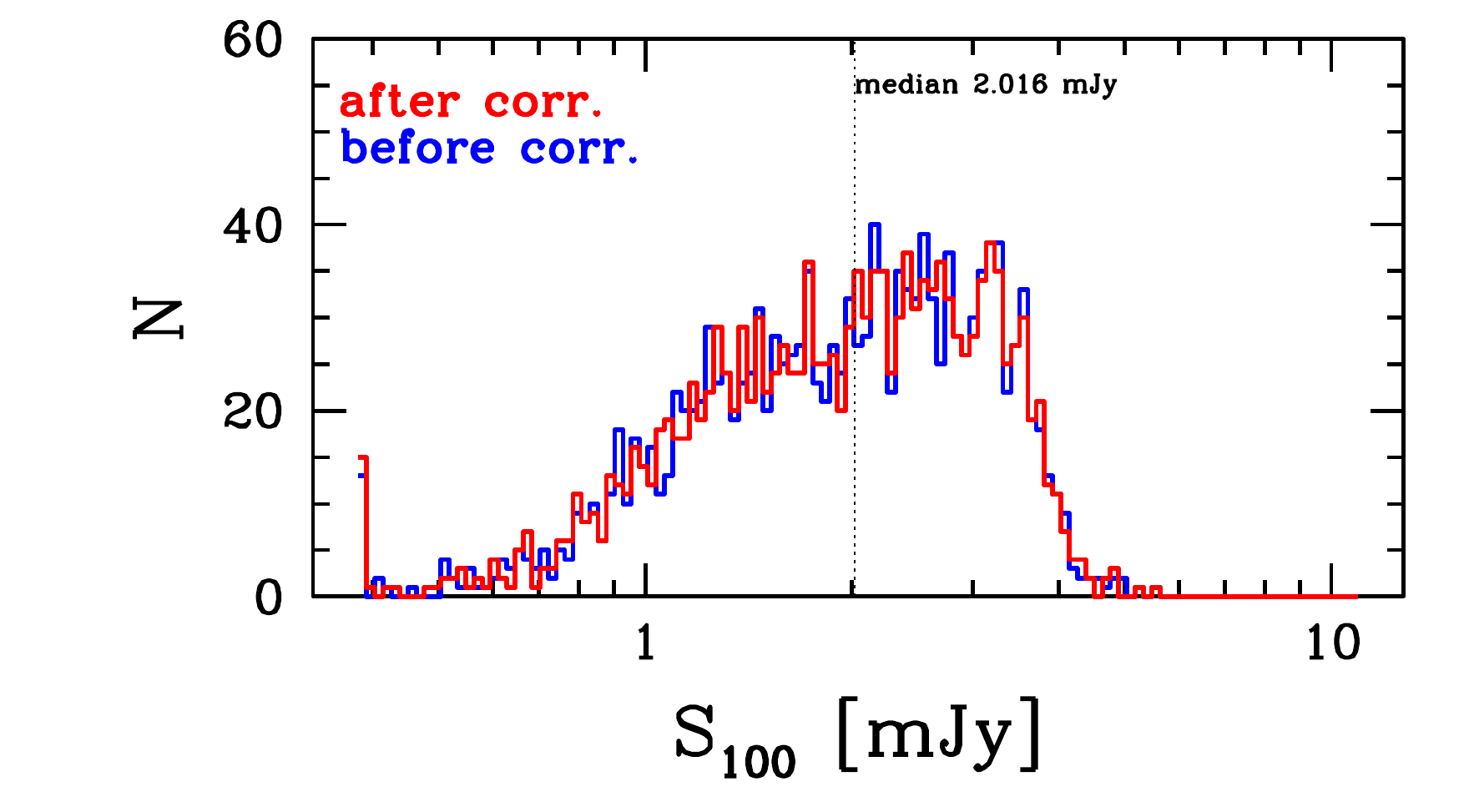}
    \end{subfigure}
    
    \begin{subfigure}[b]{\textwidth}\centering
    \includegraphics[height=2.6cm, trim=0 1cm -1.8cm 0]{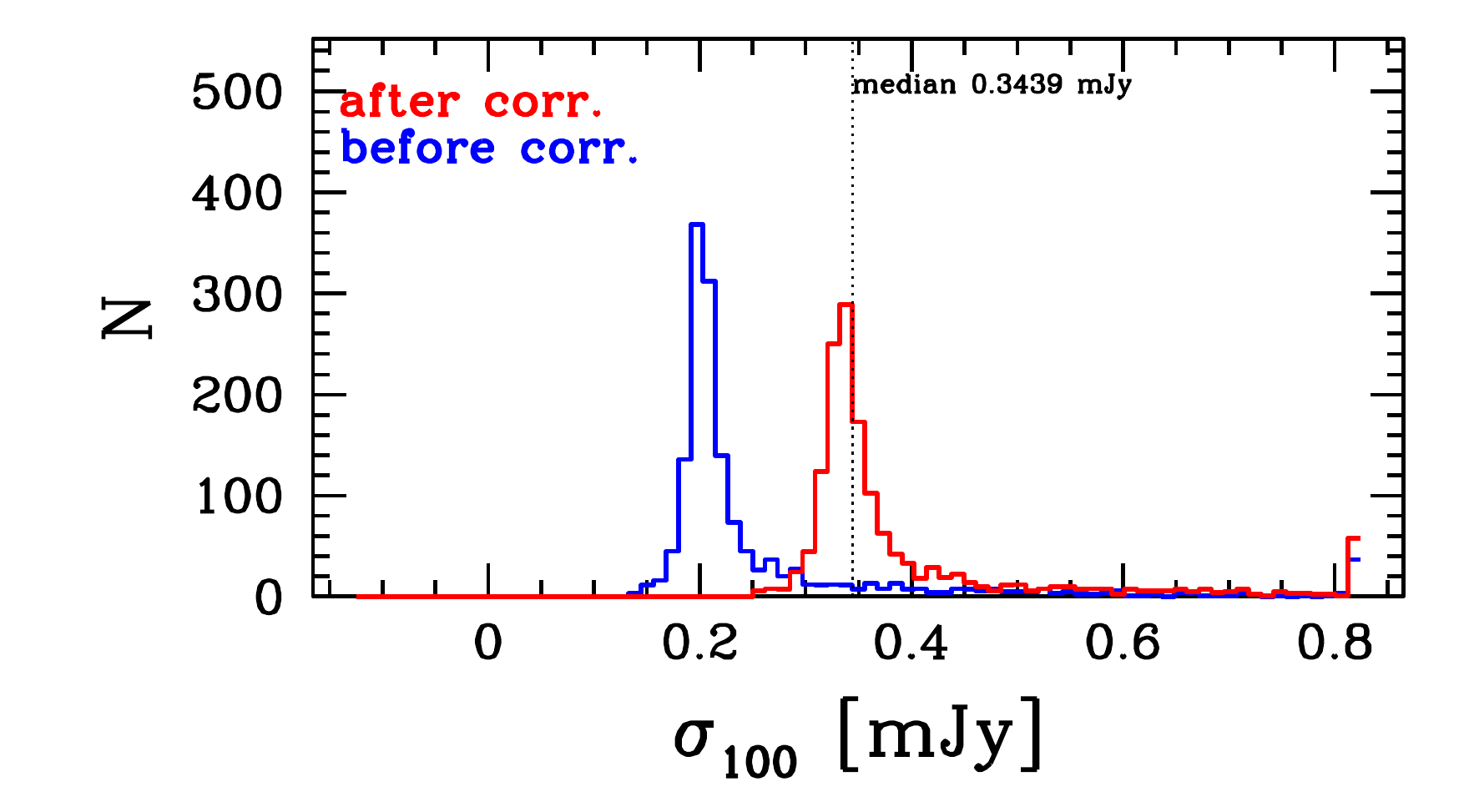}
    \includegraphics[height=2.6cm, trim=0 1cm -1.8cm 0]{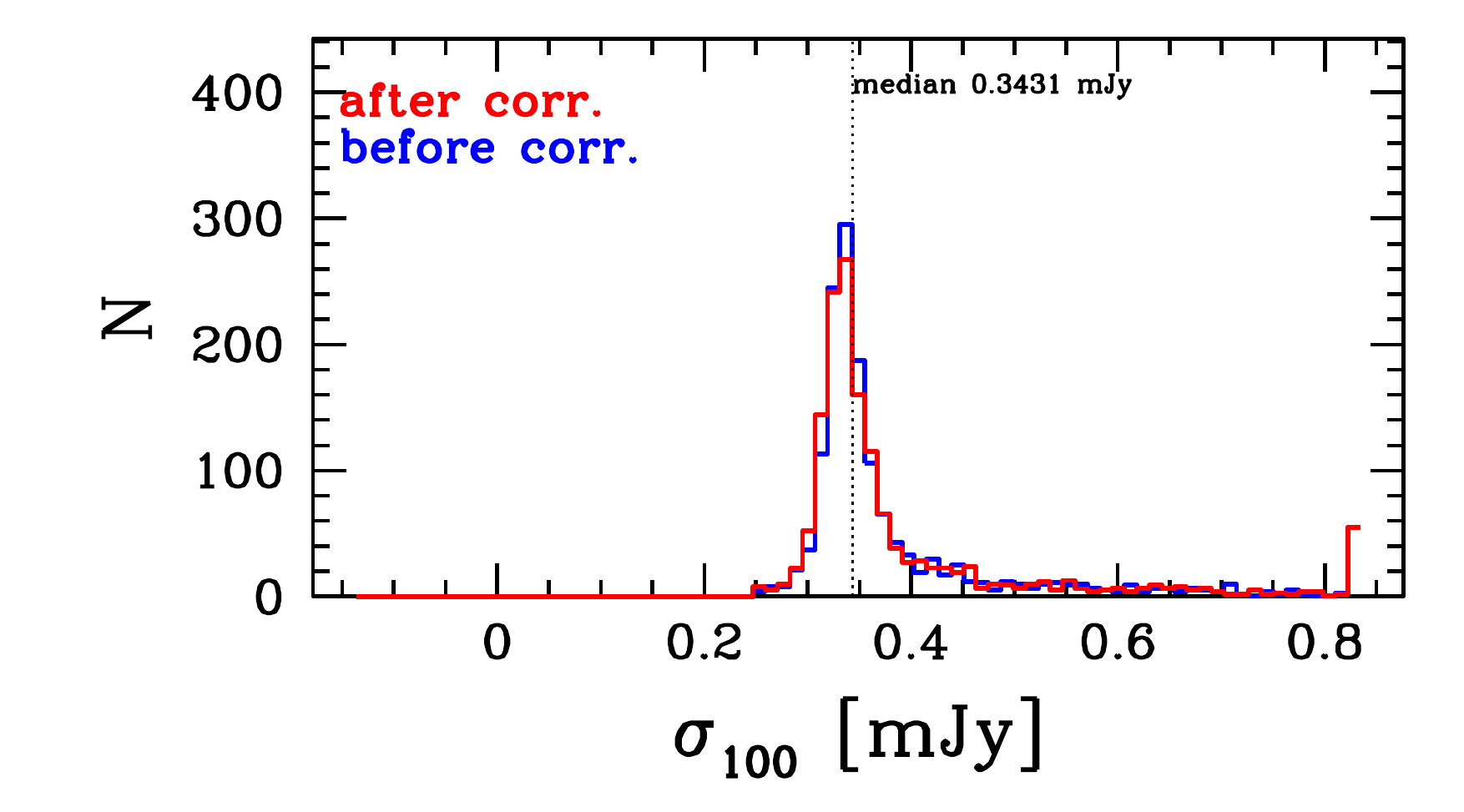}
    \includegraphics[height=2.6cm, trim=0 1cm -1.8cm 0]{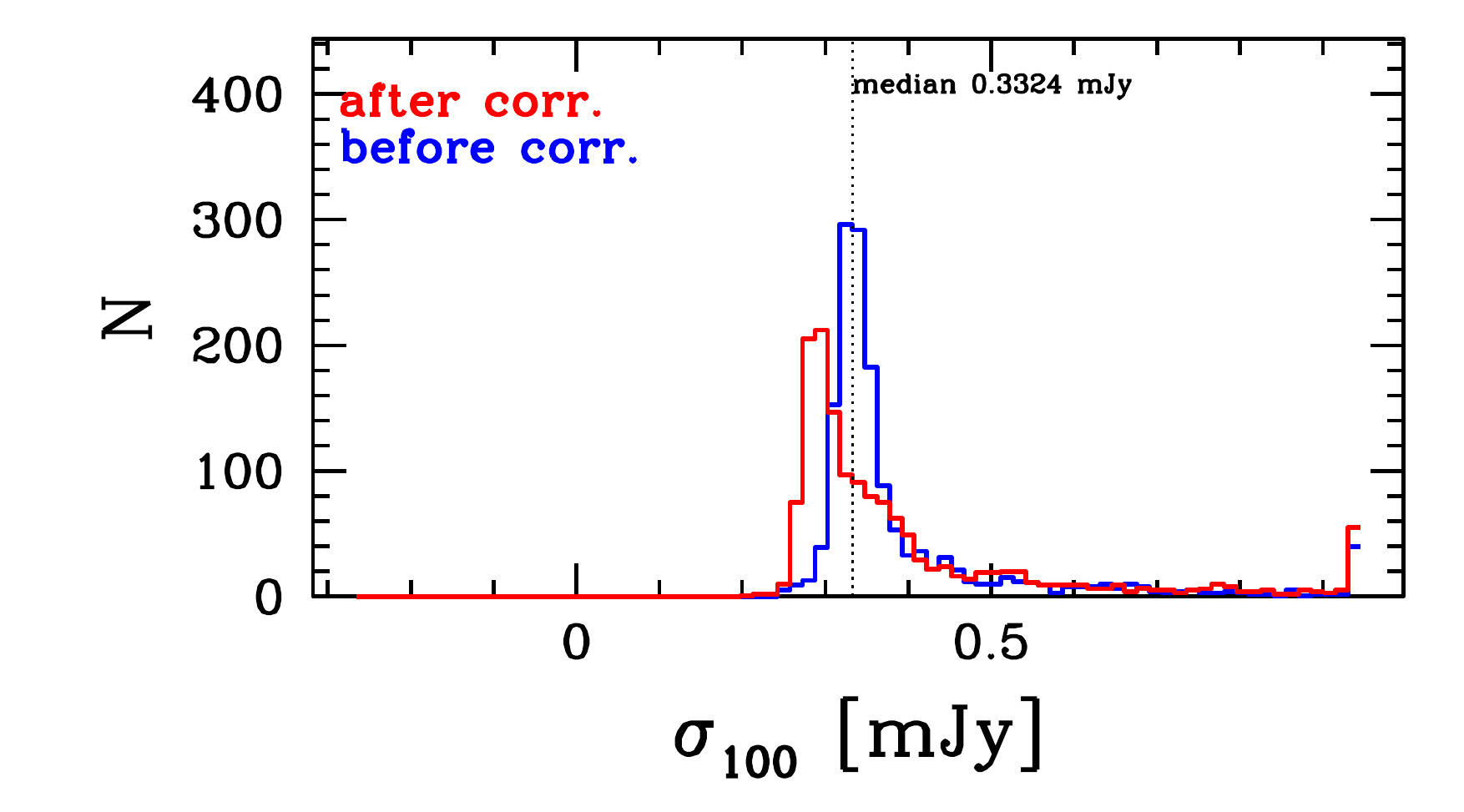}
    \end{subfigure}
    
    \caption{
        Simulation correction analyses at 100~$\mu$m. See descriptions in the text. 
        \label{Figure_galsim_100_bin}
    }
\end{figure}

Here we present figures reporting the results of our simulation-based correction recipes for all the bands that we analyzed.  In the main text, for simplicity, we have shown only the case of 350~$\mu$m, see Fig.~\ref{Fig_Galsim_df_corr_SPIRE}, and we do not duplicate that figure here.  For example, in Fig.~\ref{Figure_galsim_100_bin}, for the observations at 100~$\mu$m, we bin the simulation data points by three measurable parameters: the \galfit{} flux uncertainty normalized by the local rms noise at the source position (${\sigma}_\mathit{galfit}/{\sigma}_{\mathrm{rms\,noise}}$) in the first column, the residual flux within one PSF beam area on the residual image, normalized also by the local rms noise ($S_{\mathrm{residual}}/{\sigma}_{\mathrm{rms\,noise}}$) in the second column, and the \crowdedness{} parameter (see Section~\ref{Section_Simulation_Method}) in the third column. 
The bins used for the analysis are indicated by the dashed vertical lines in the first and second row images. 

In the first row, we analyze the difference between the input and output fluxes of each simulated source ($S_{\mathrm{in}}-S_{\mathrm{out}}$), which is used to correct the flux bias for our photometry. We fit a 3-order polynomial function to the bin-averaged ($S_{\mathrm{in}}-S_{\mathrm{out}}$) for each parameter, shown as the red curve.  

In the second row, we analyze the flux difference ($(S_{\mathrm{in}}-S_{\mathrm{out}})$) divided by the flux uncertainty (${\sigma}$) of each simulated source ($(S_{\mathrm{in}}-S_{\mathrm{out}})/{\sigma}$). The scatter in each bin is the correction factor to be applied to ${\sigma}$. Blue points show the uncorrected data points, while red points show the corrected values. After correction, the scatter of $(S_{\mathrm{in}}-S_{\mathrm{out}})/{\sigma}$ in each bin is very close to 1.0, indicating that the corrected ${\sigma}$ is statistically consistent with the scatter of $S_{\mathrm{in}}-S_{\mathrm{out}}$. We fit a 3-order polynomial function to the bin-averaged flux uncertainty correction factor for each parameter, as shown by the red curve, whose value can be read on the right axis. 

In the third row, we show the histograms of $(S_{\mathrm{in}}-S_{\mathrm{out}})/{\sigma}$ before and after the three-parameter corrections. Its shape, after correction (i.e., the red histogram), get closer to a well-behaved Gaussian distribution (i.e., symmetric and with a Gaussian width of 1.0), and is thus an improvement over the uncorrected one (i.e., the blue histogram). See the text in Sections~\ref{Section_Simulation_Correction_df_corr} for more detailed information. 

In the fourth and fifth rows, we show the histograms of flux densities, and of flux density uncertainties, respectively. In each panel, the blue histogram shows values before correction, and the red histogram shows the results after correction. From left to right, the panels are in the same three-parameter order. Generally the flux density changes are very small, while the corrections to the flux density uncertainties are larger with respect to the \galfit{} initial values.

\begin{figure}
    \centering
    
    \begin{subfigure}[b]{\textwidth}\centering
    \includegraphics[height=2.6cm, trim=0 1cm 0 0]{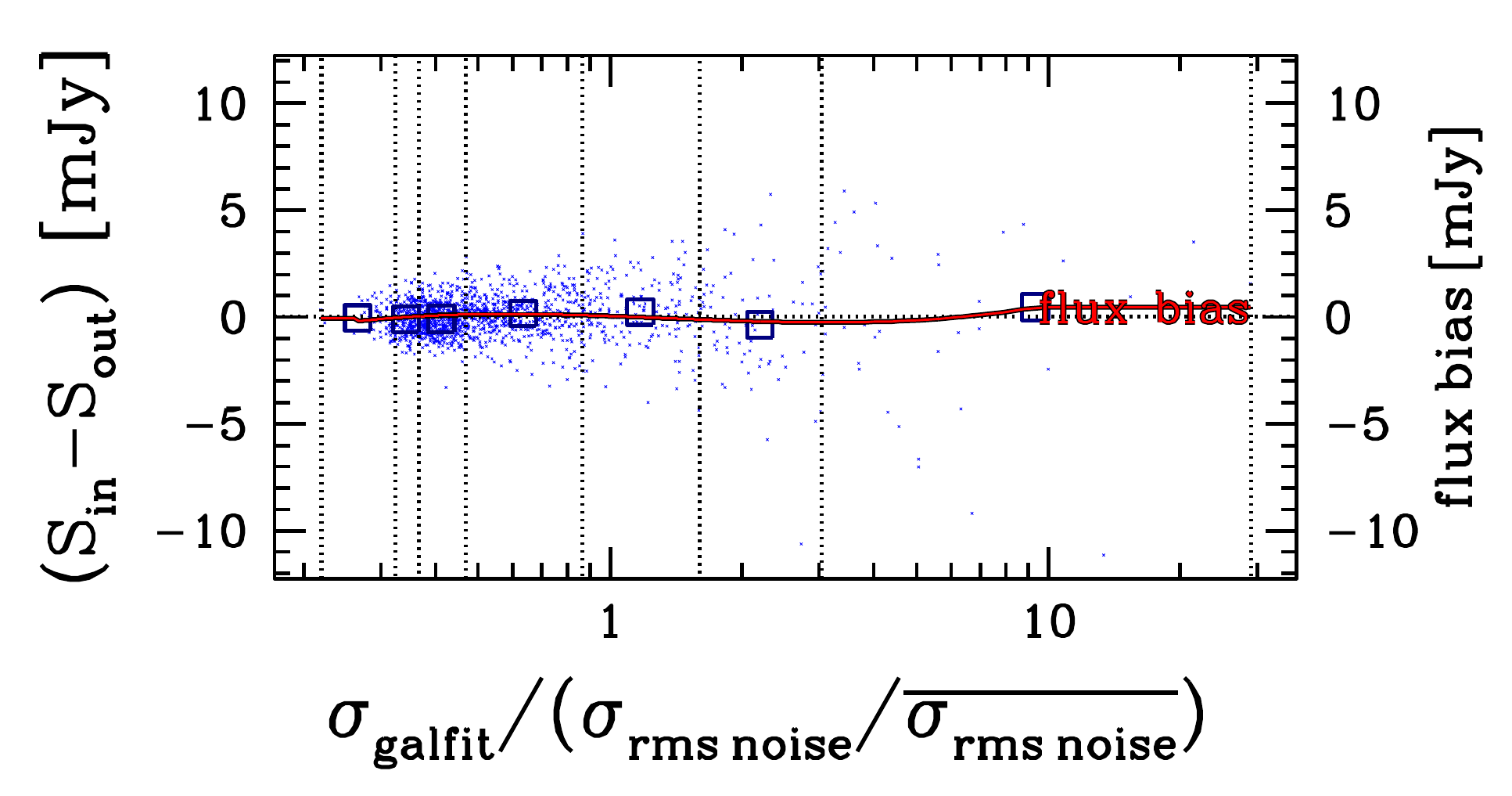}
    \includegraphics[height=2.6cm, trim=0 1cm 0 0]{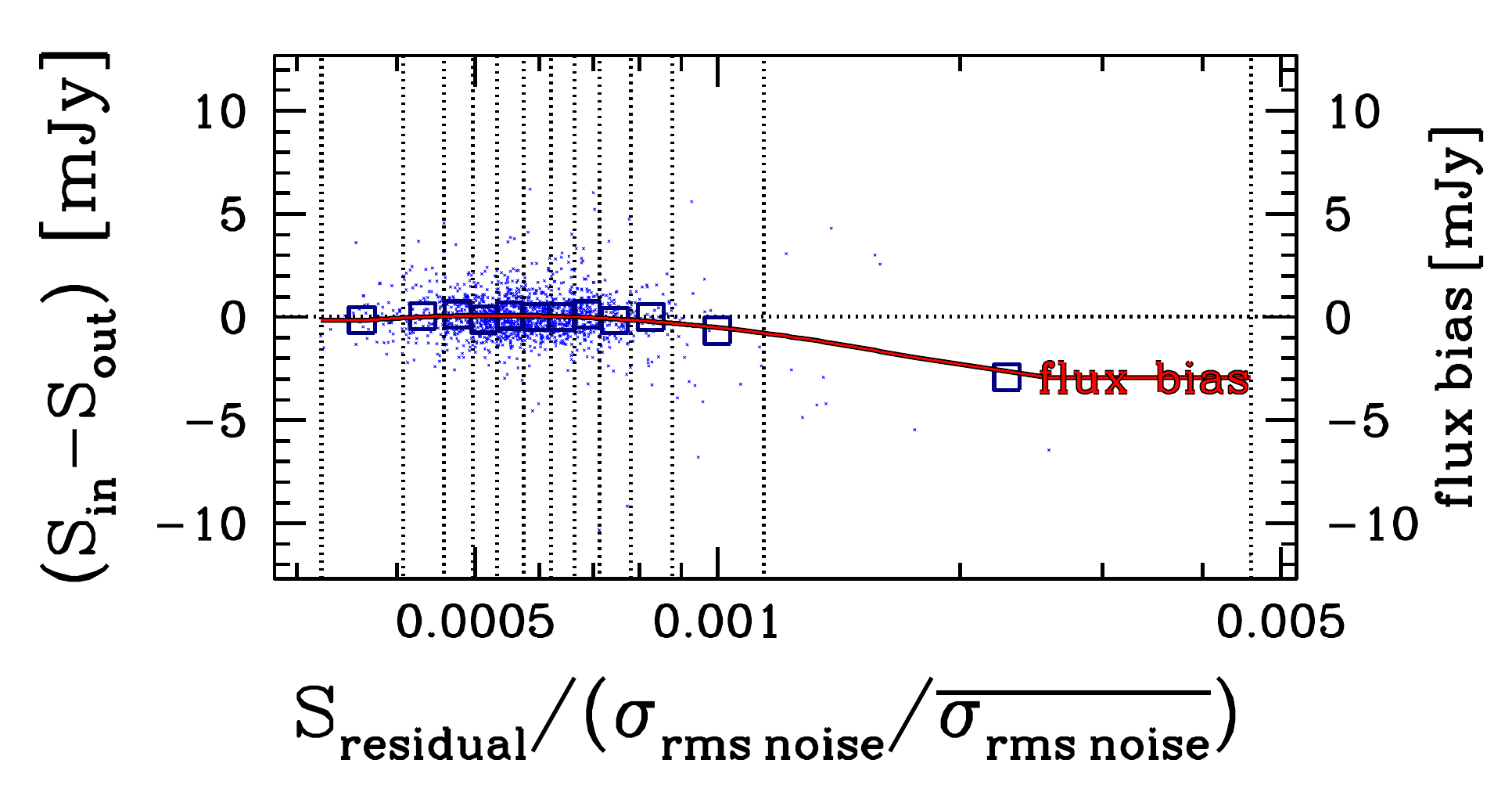}
    \includegraphics[height=2.6cm, trim=0 1cm 0 0]{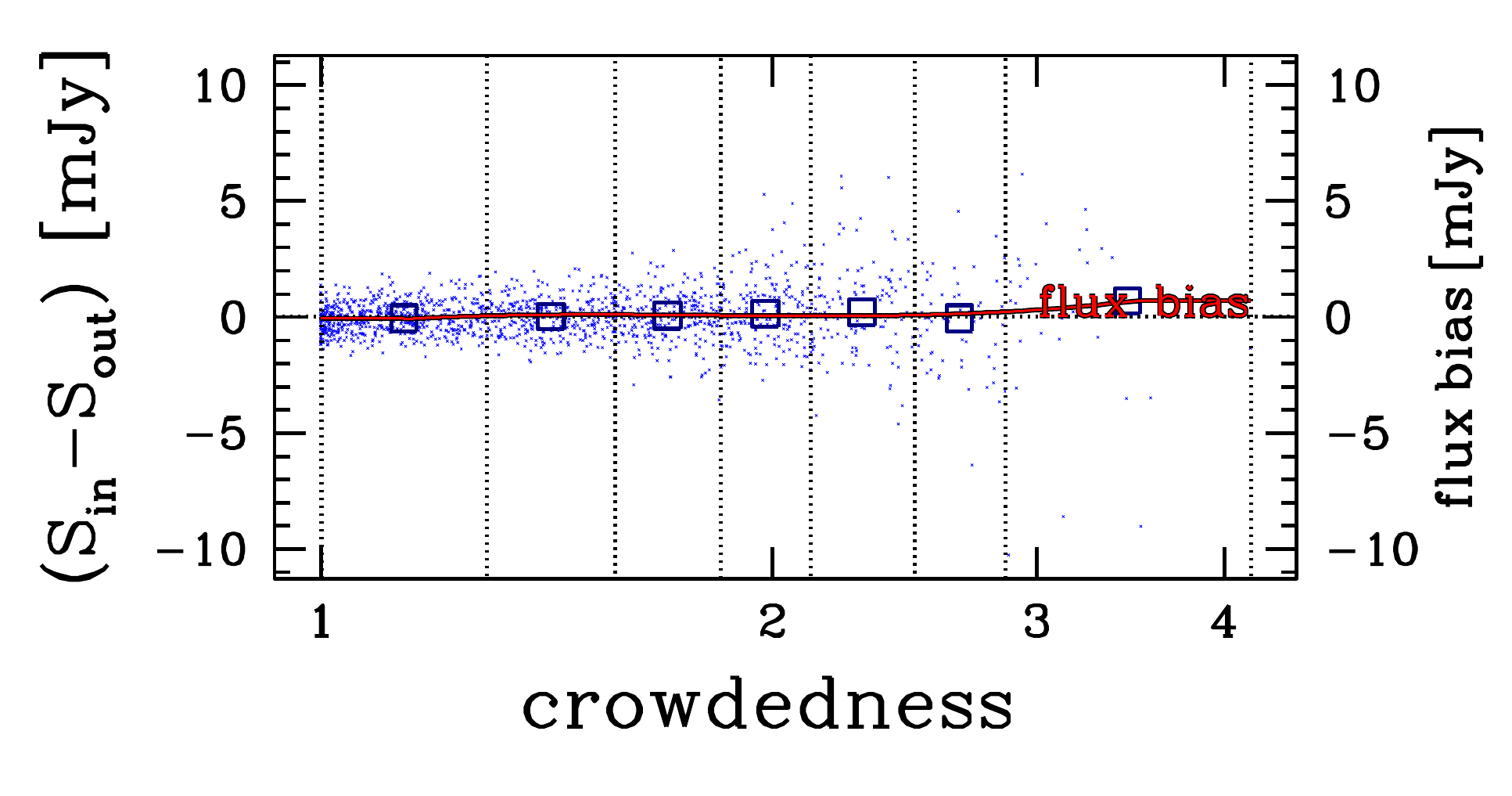}
    \end{subfigure}
    
    \begin{subfigure}[b]{\textwidth}\centering
    \includegraphics[height=2.6cm, trim=0 1cm 0 0]{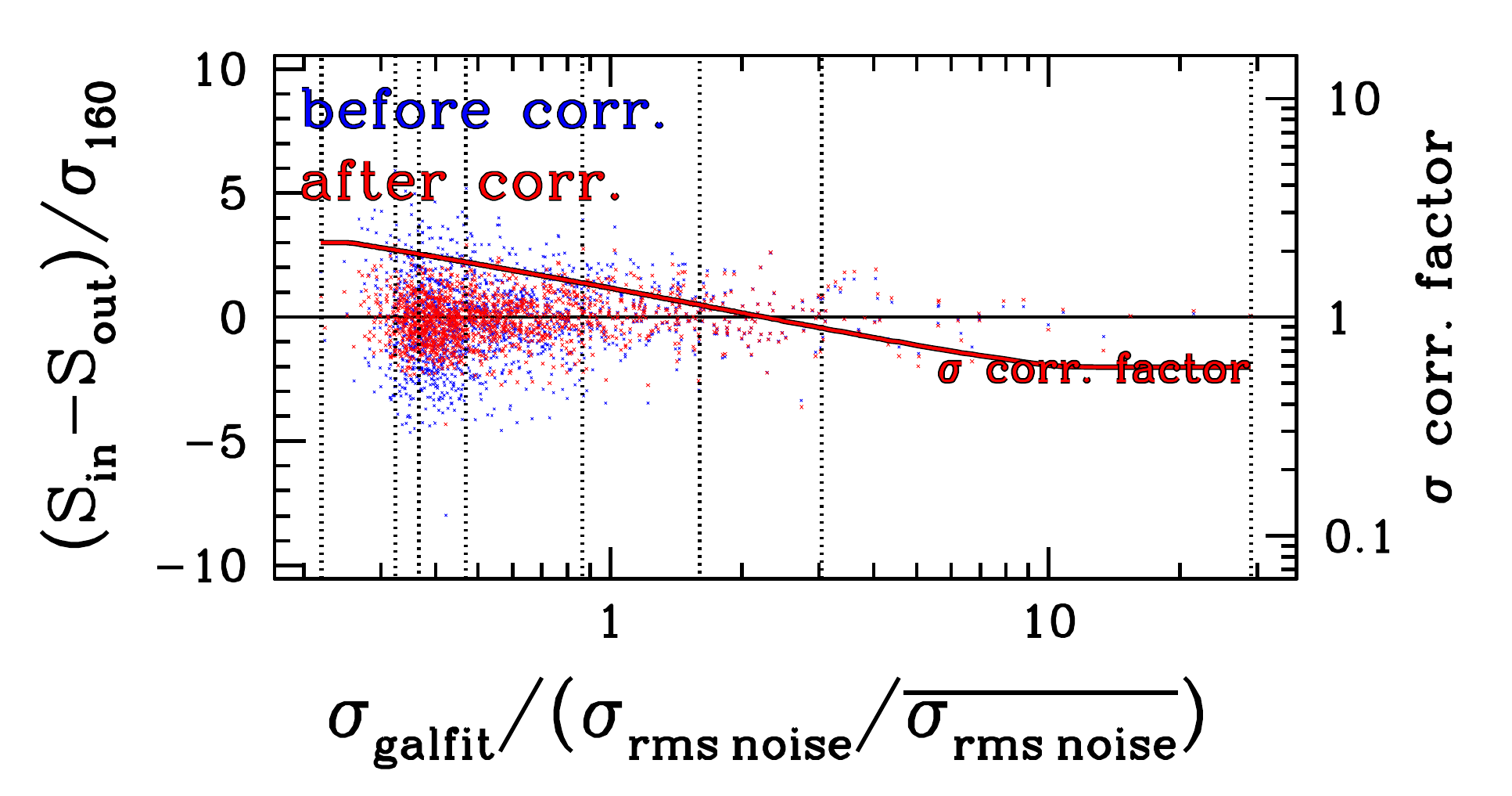}
    \includegraphics[height=2.6cm, trim=0 1cm 0 0]{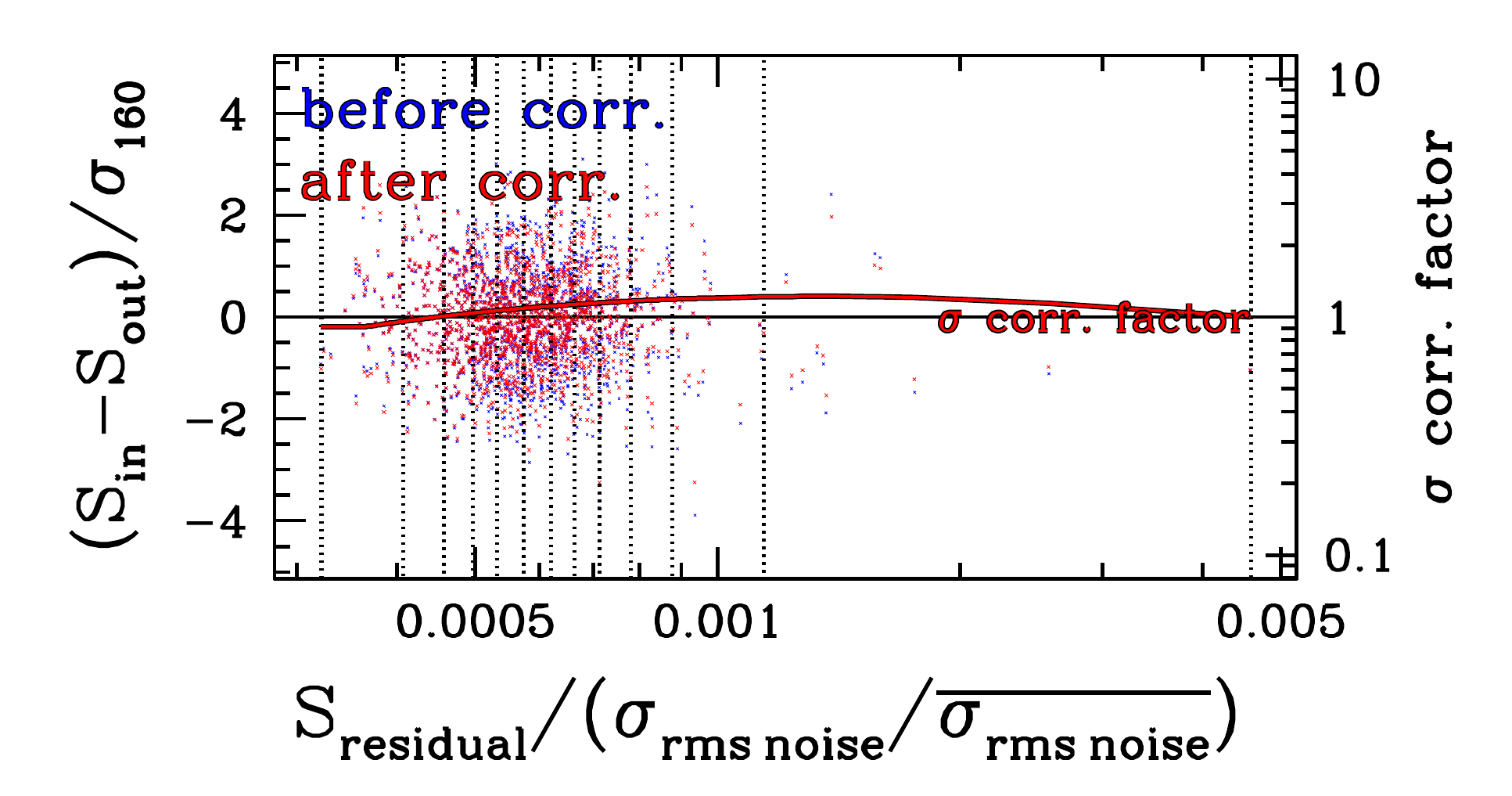}
    \includegraphics[height=2.6cm, trim=0 1cm 0 0]{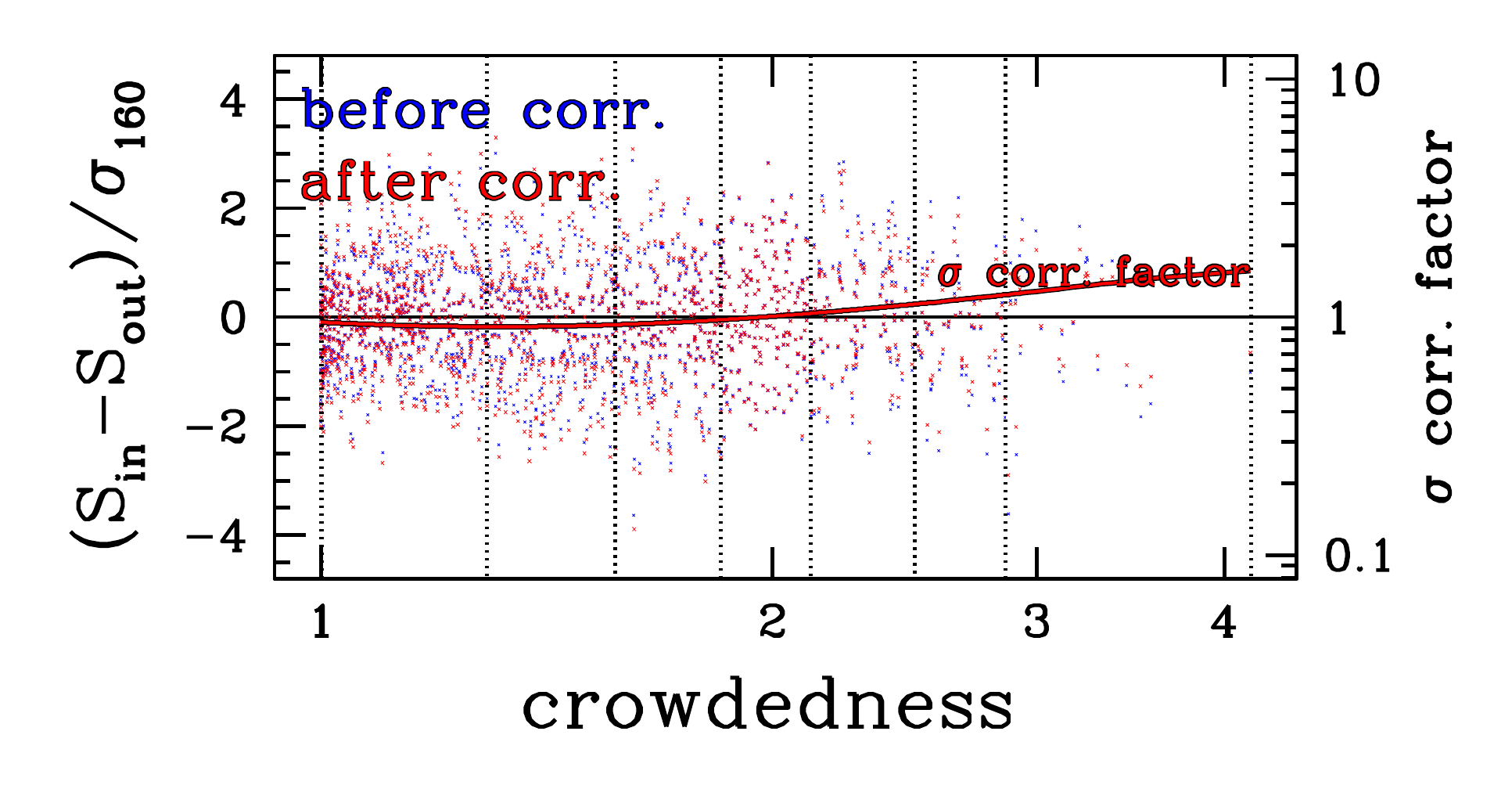}
    \end{subfigure}
    
    \begin{subfigure}[b]{\textwidth}\centering
    \includegraphics[height=2.6cm, trim=0 1cm -1.8cm 0]{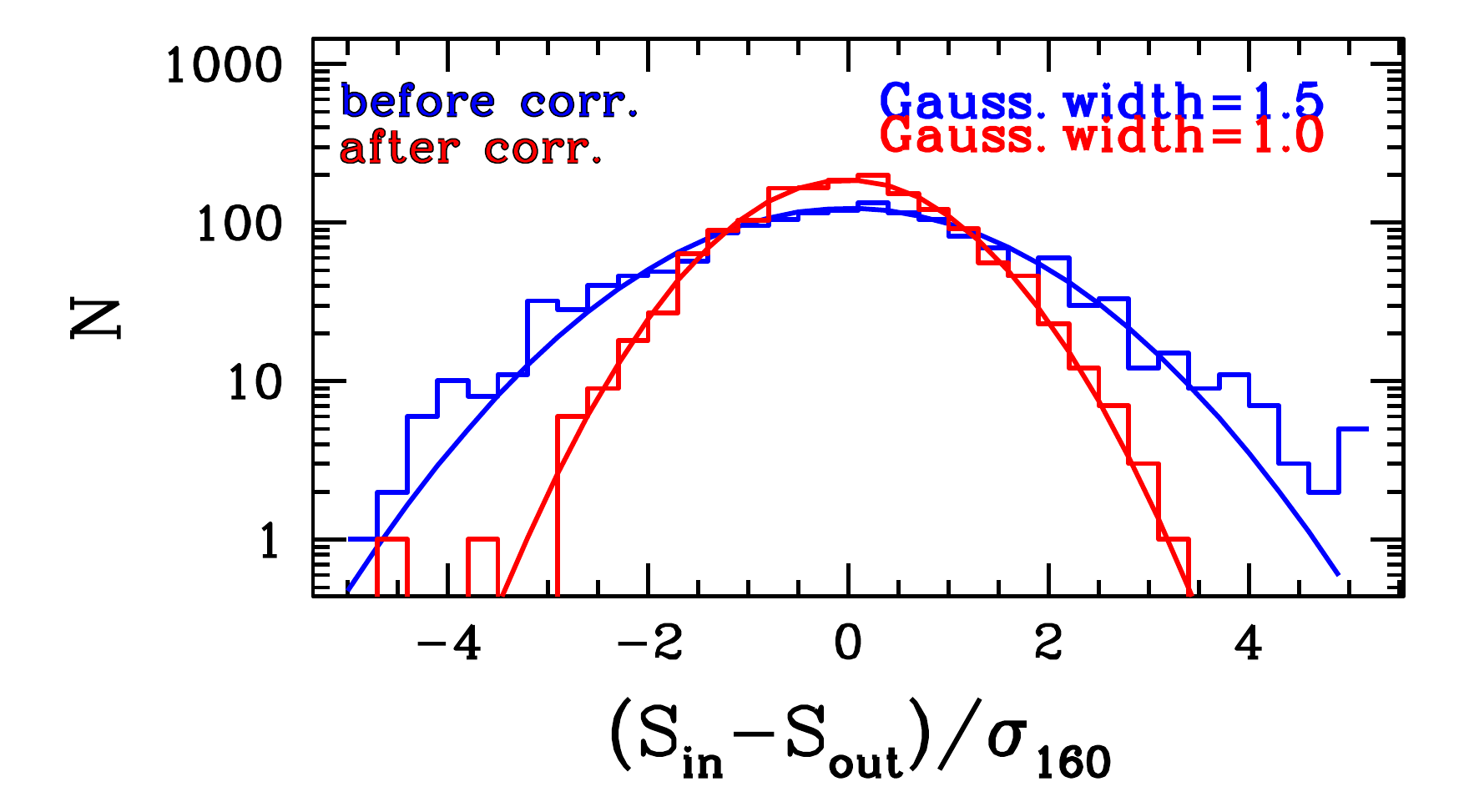}
    \includegraphics[height=2.6cm, trim=0 1cm -1.8cm 0]{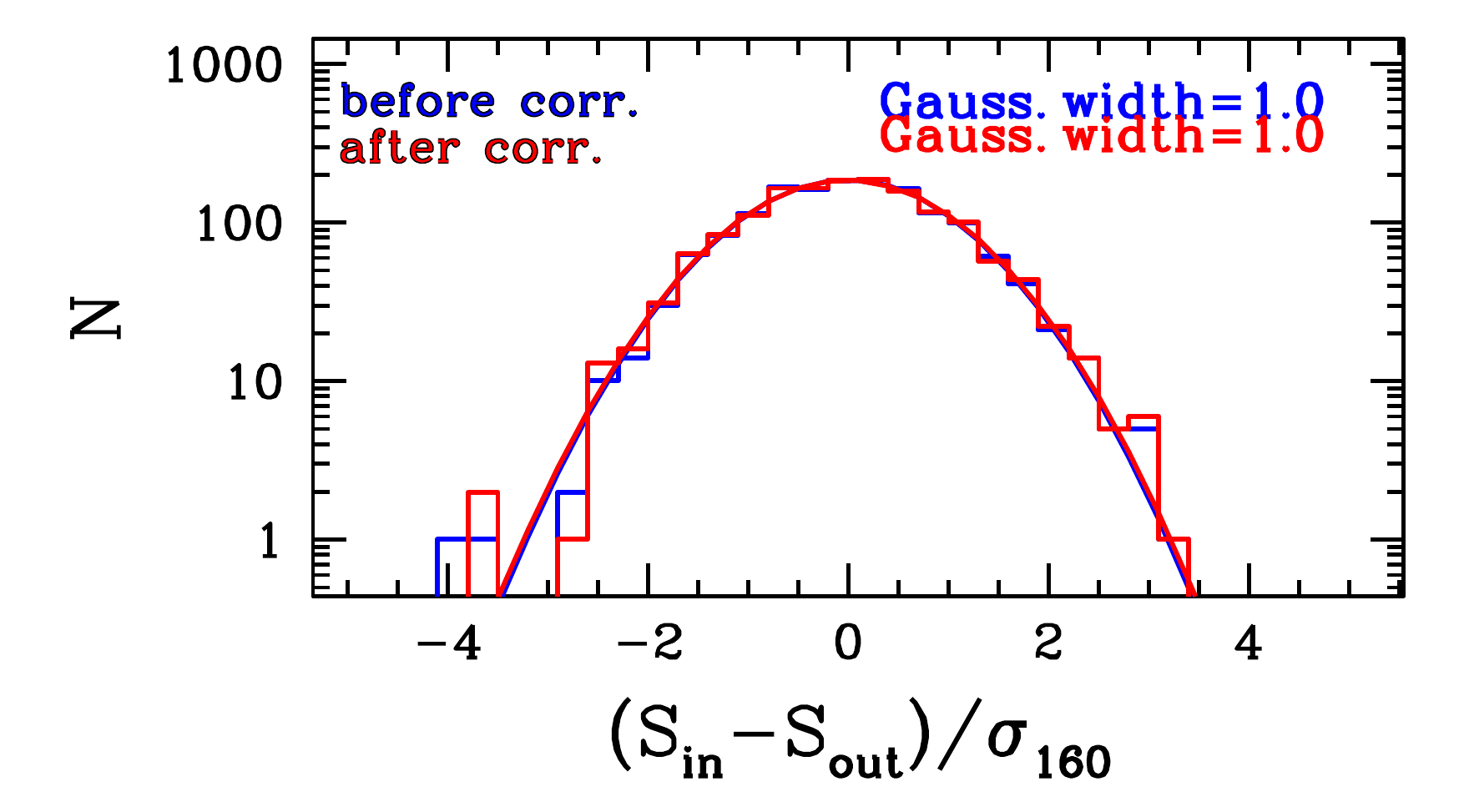}
    \includegraphics[height=2.6cm, trim=0 1cm -1.8cm 0]{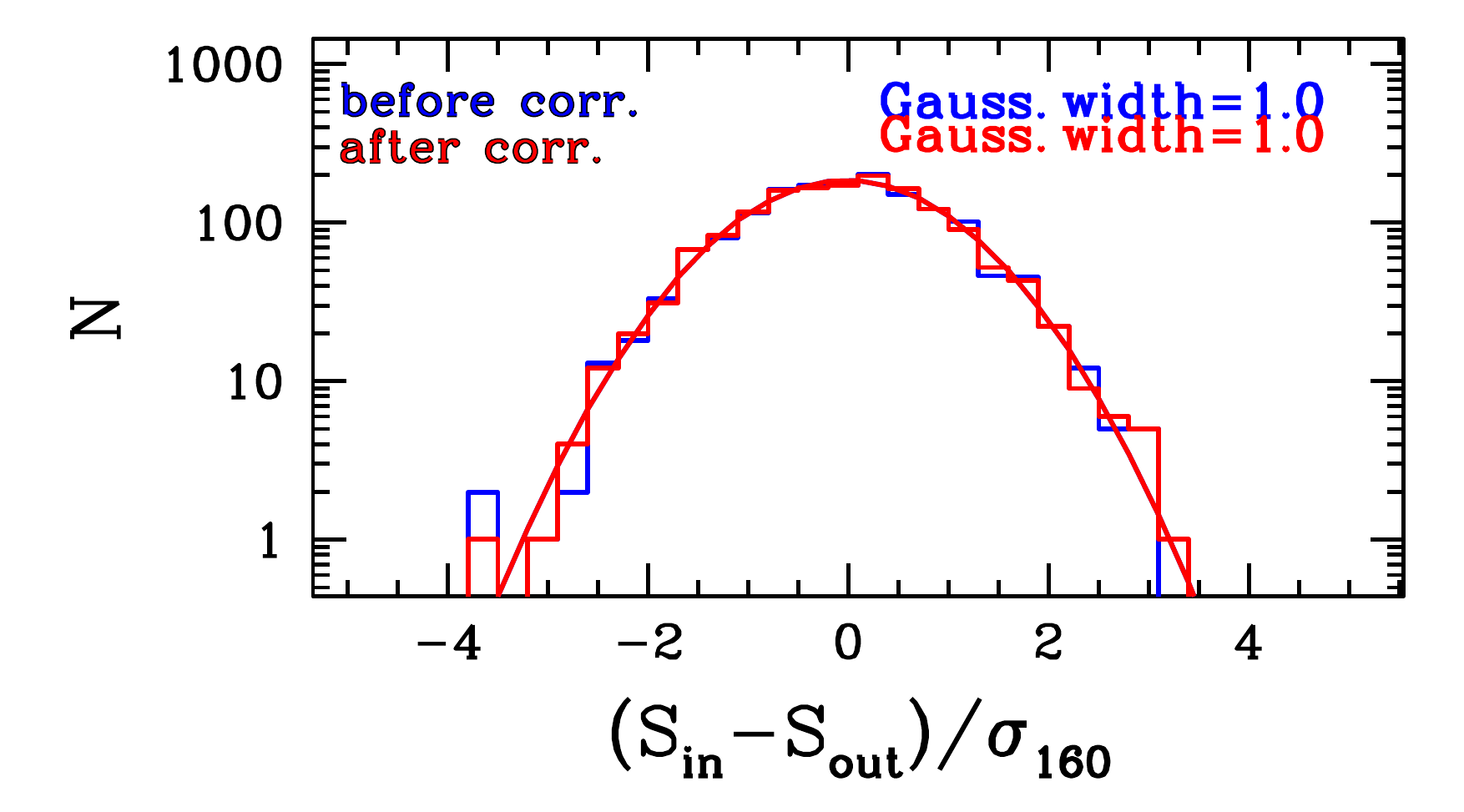}
    \end{subfigure}
    
    \begin{subfigure}[b]{\textwidth}\centering
    \includegraphics[height=2.6cm, trim=0 1cm -1.8cm 0]{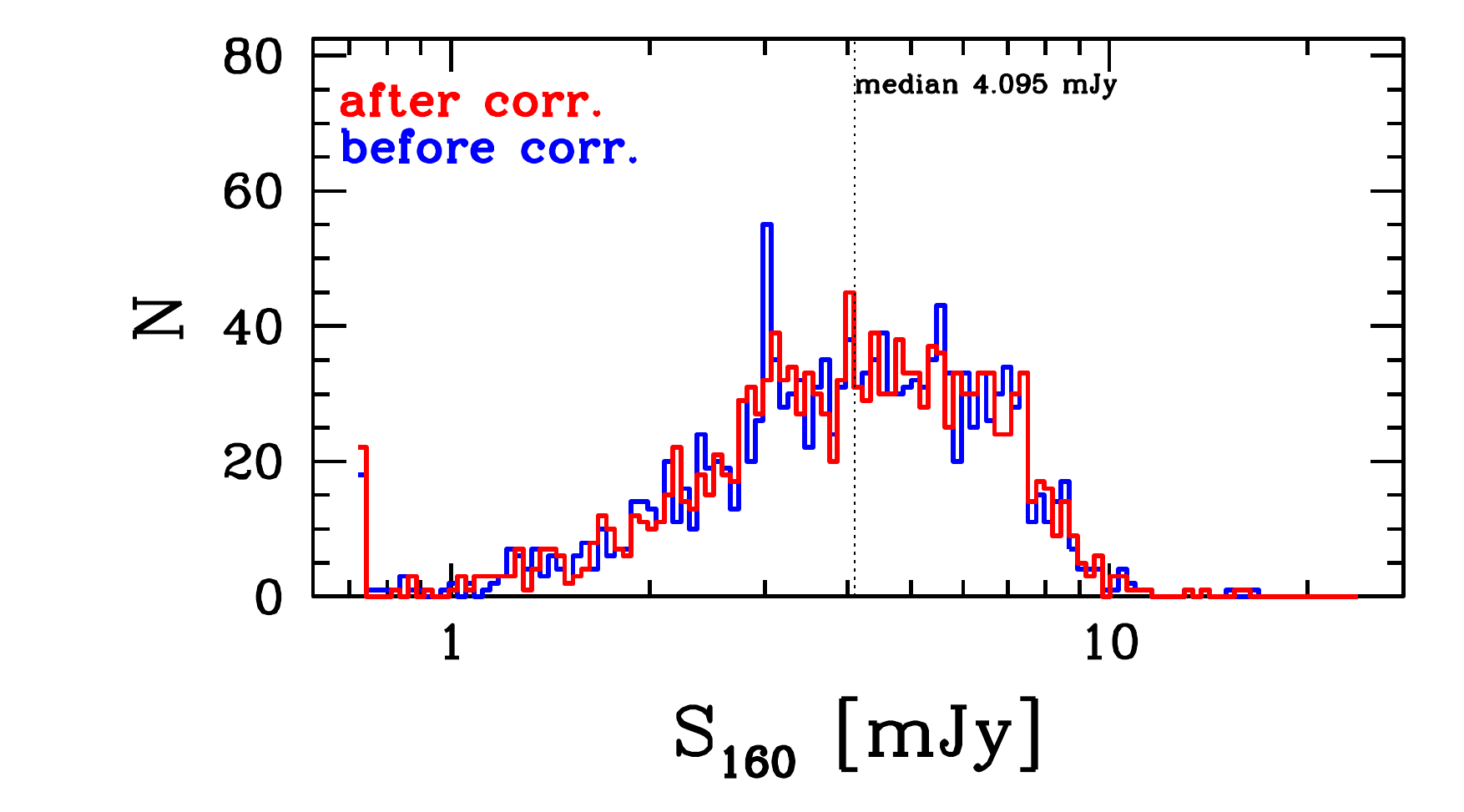}
    \includegraphics[height=2.6cm, trim=0 1cm -1.8cm 0]{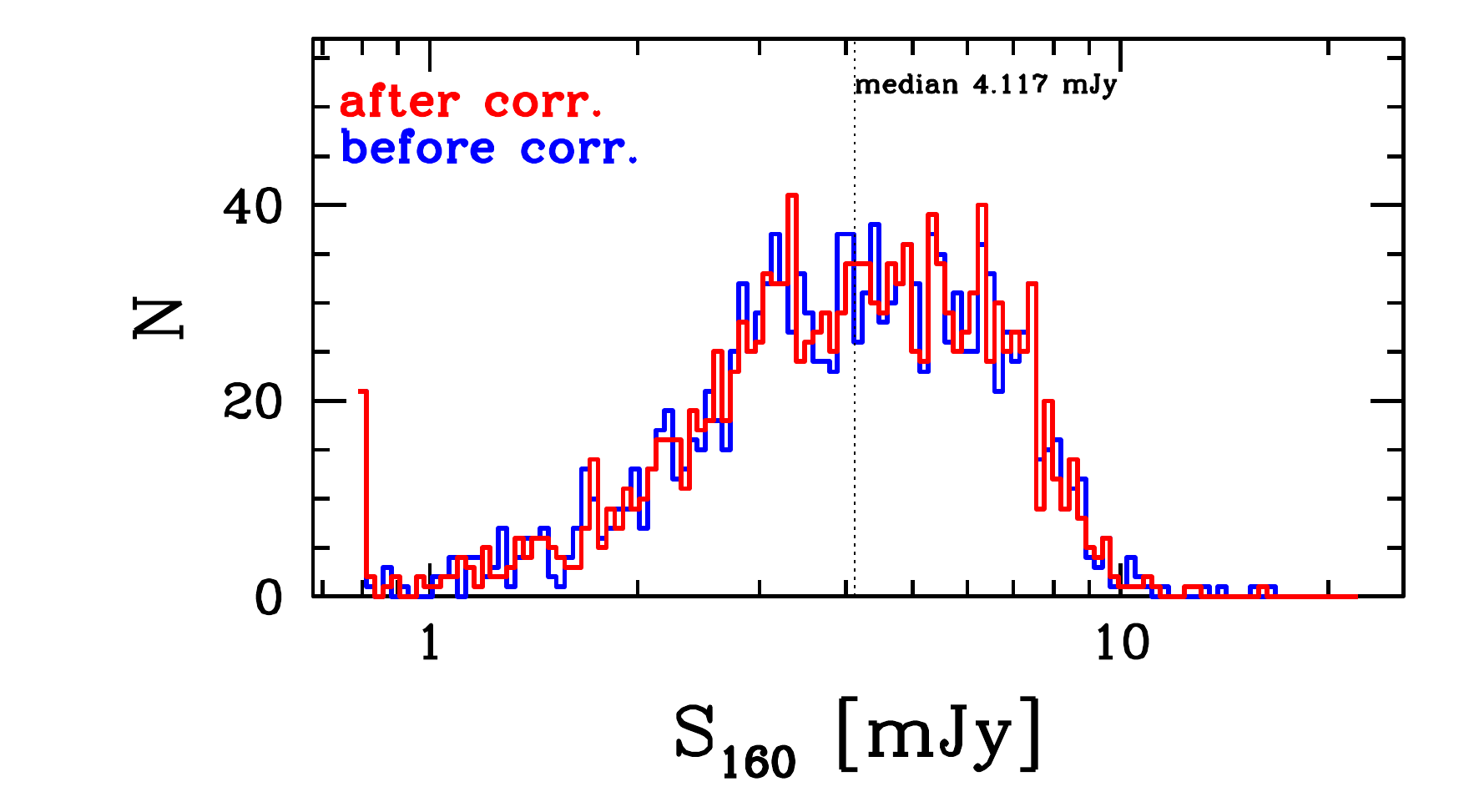}
    \includegraphics[height=2.6cm, trim=0 1cm -1.8cm 0]{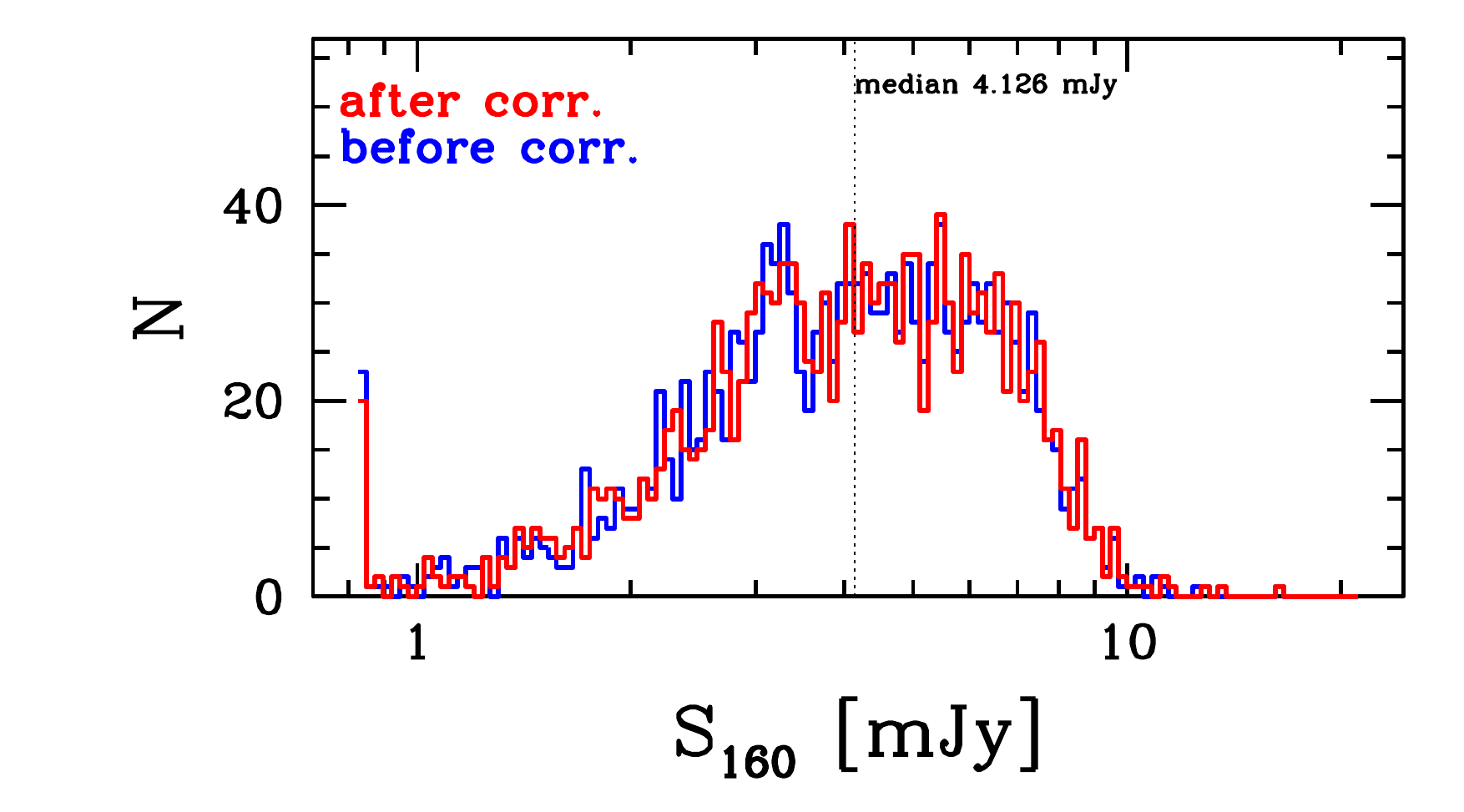}
    \end{subfigure}
    
    \begin{subfigure}[b]{\textwidth}\centering
    \includegraphics[height=2.6cm, trim=0 1cm -1.8cm 0]{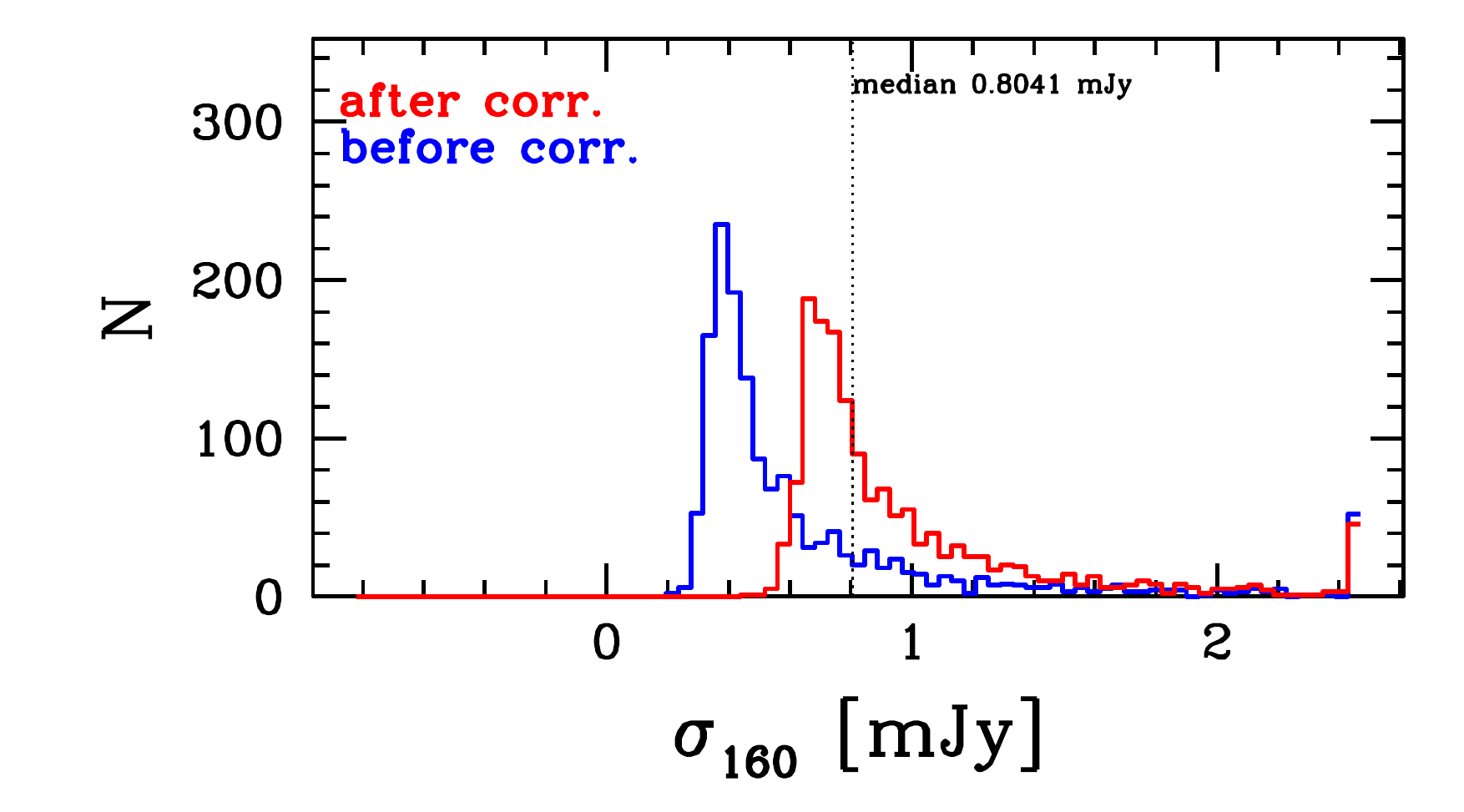}
    \includegraphics[height=2.6cm, trim=0 1cm -1.8cm 0]{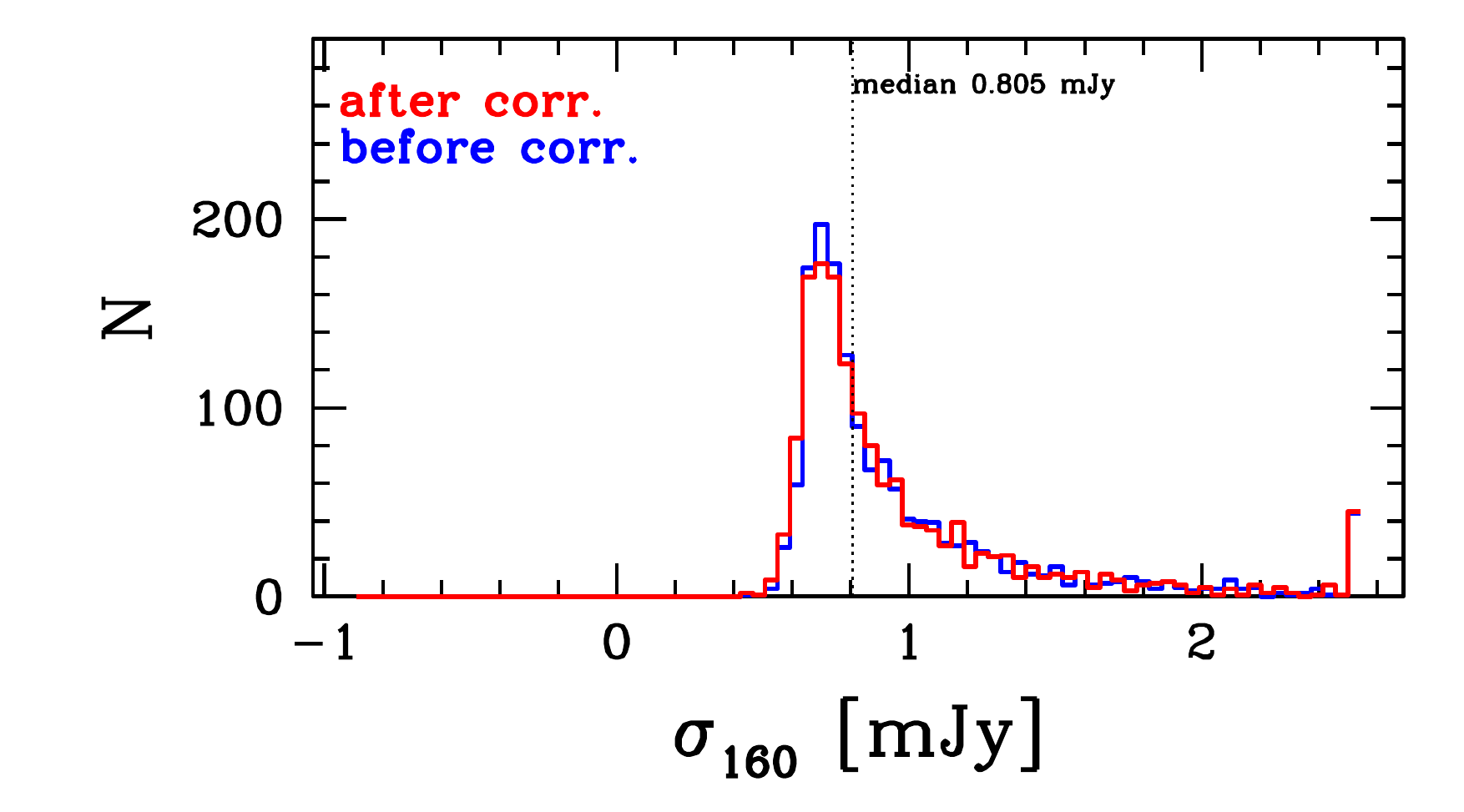}
    \includegraphics[height=2.6cm, trim=0 1cm -1.8cm 0]{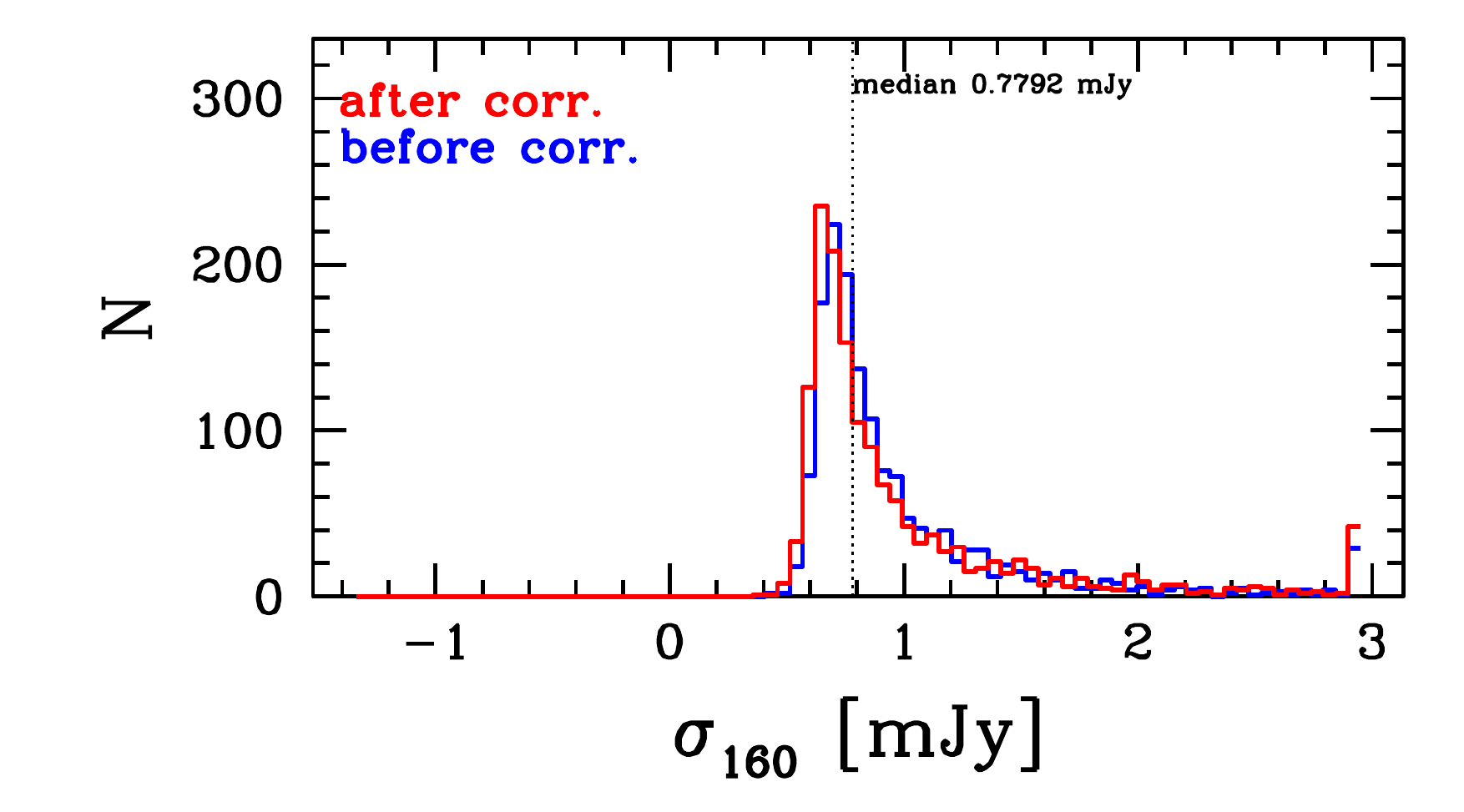}
    \end{subfigure}
    
    \caption{%
        Simulation correction analyses at 160~$\mu$m. See descriptions in the text. 
        \label{Figure_galsim_160_bin}
    }
\end{figure}

\begin{figure}
    \centering
    
    \begin{subfigure}[b]{\textwidth}\centering
    \includegraphics[height=2.6cm, trim=0 1cm 0 0]{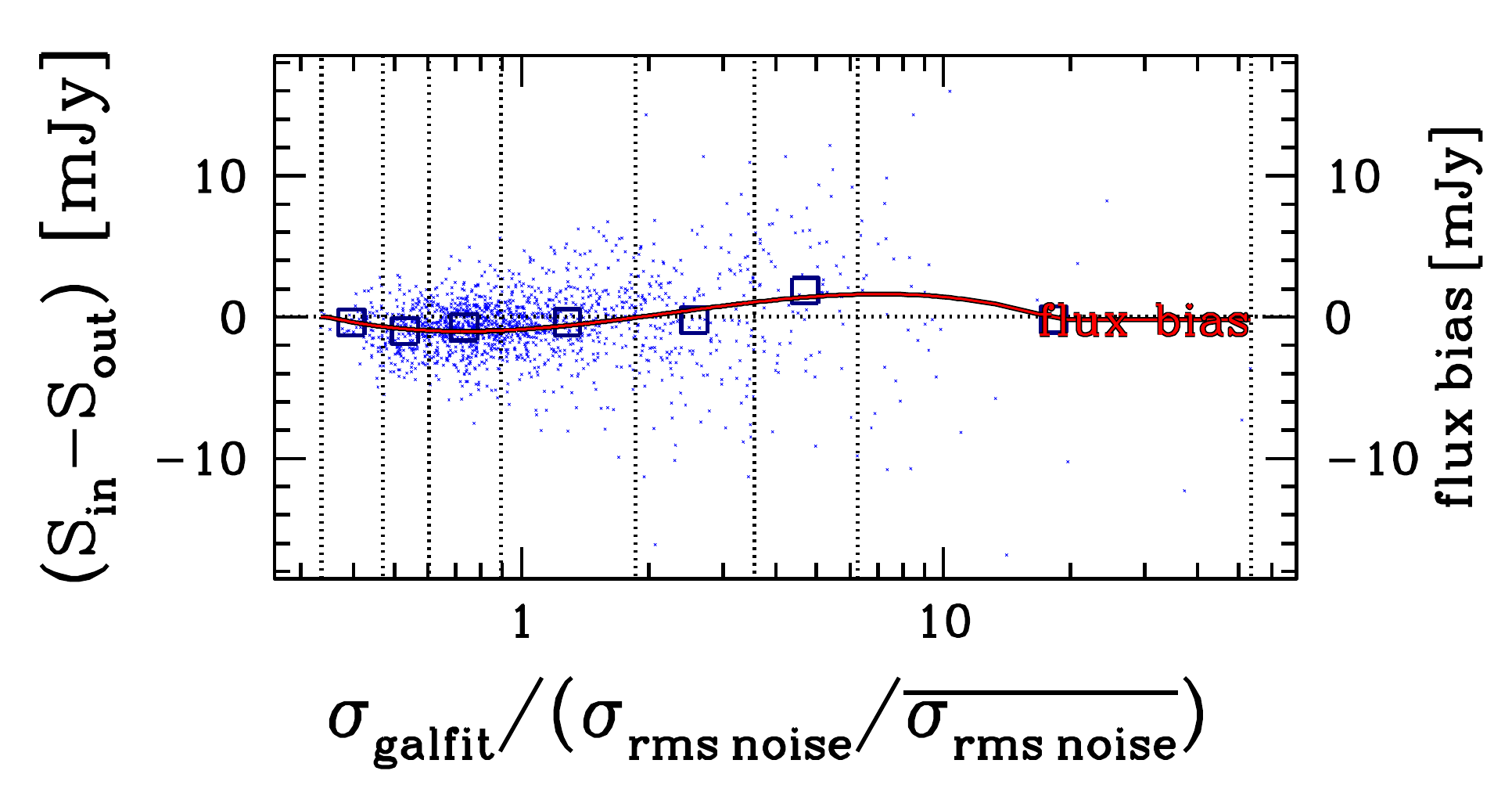}
    \includegraphics[height=2.6cm, trim=0 1cm 0 0]{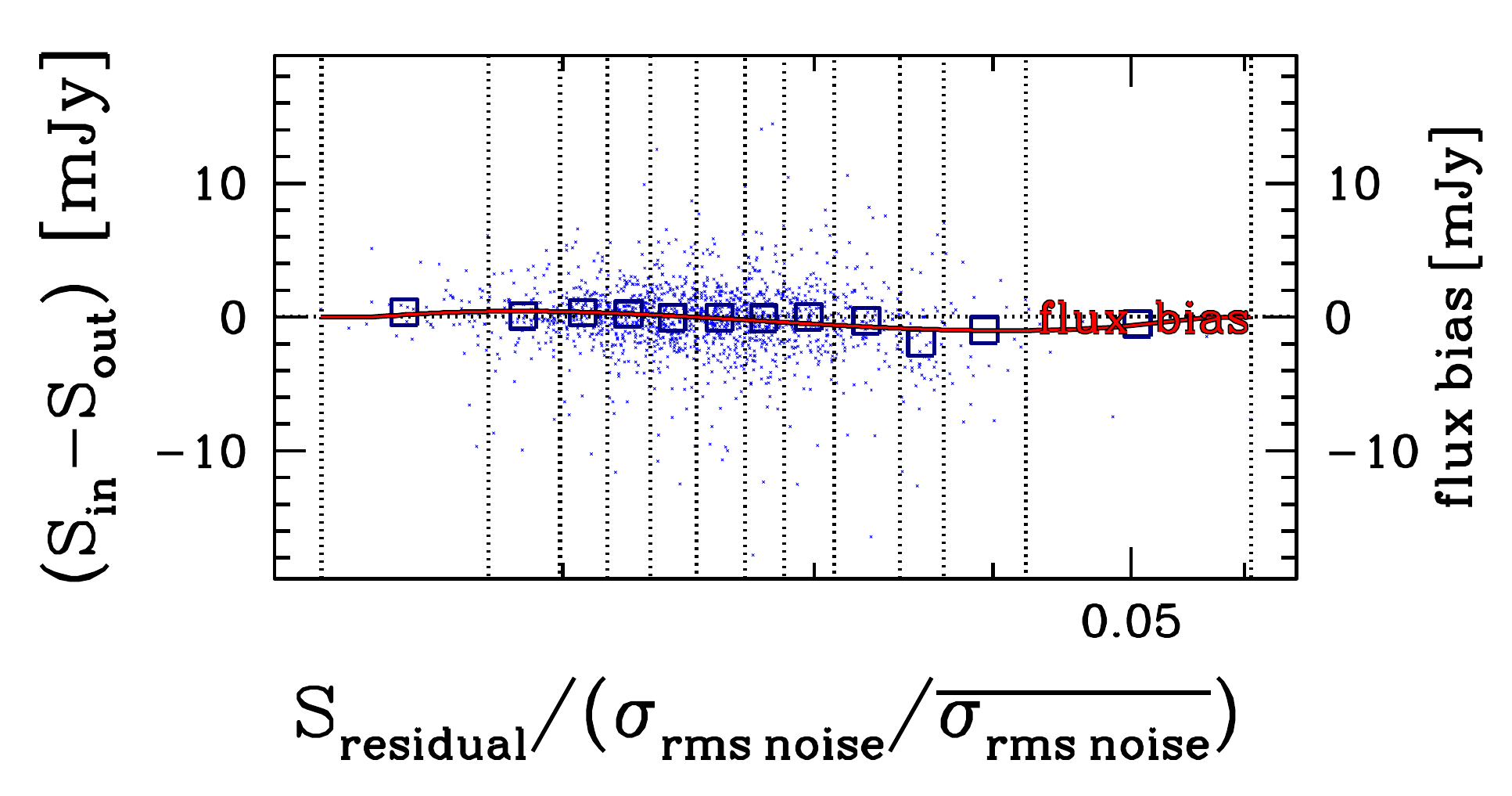}
    \includegraphics[height=2.6cm, trim=0 1cm 0 0]{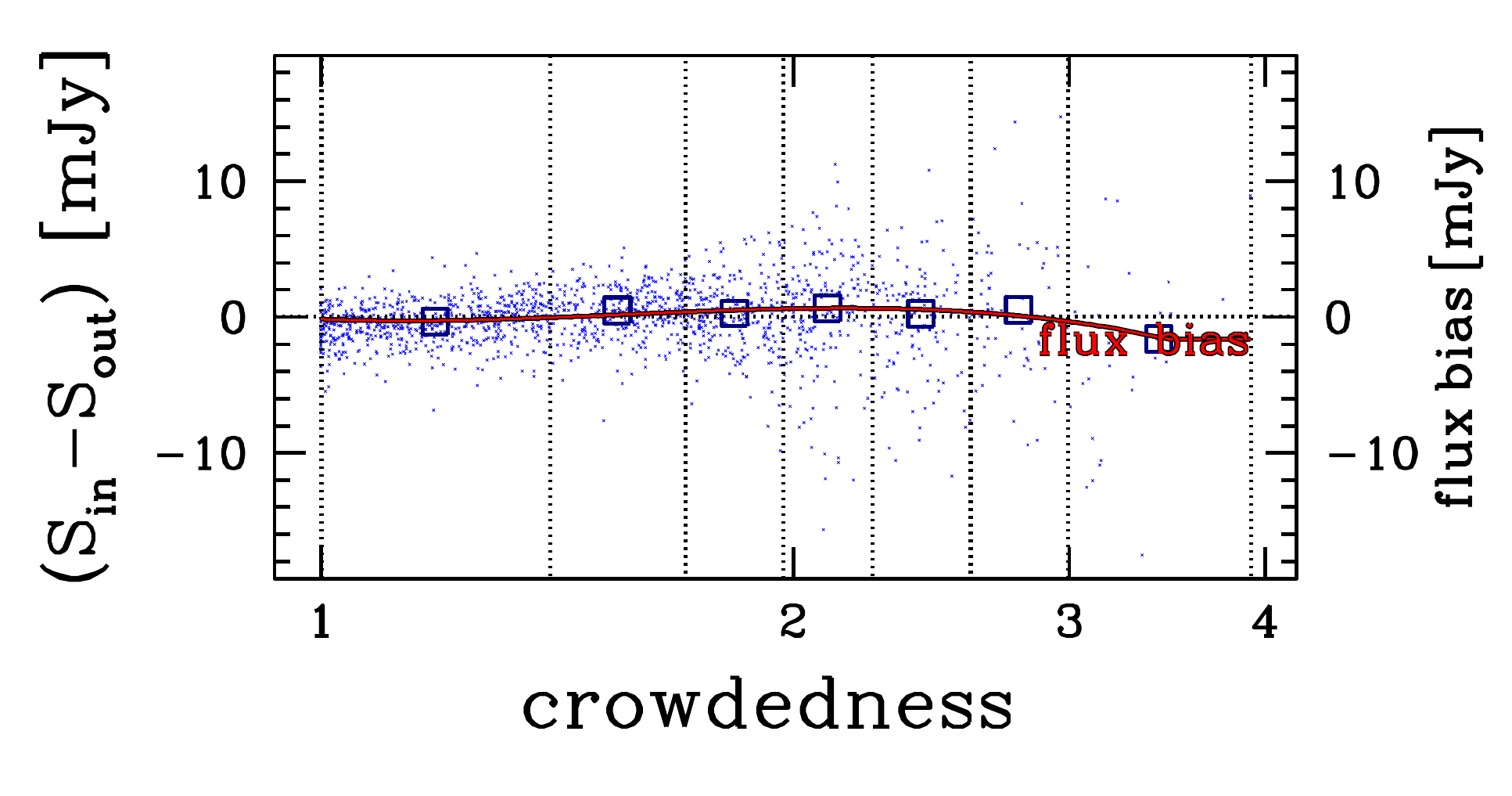}
    \end{subfigure}
    
    \begin{subfigure}[b]{\textwidth}\centering
    \includegraphics[height=2.6cm, trim=0 1cm 0 0]{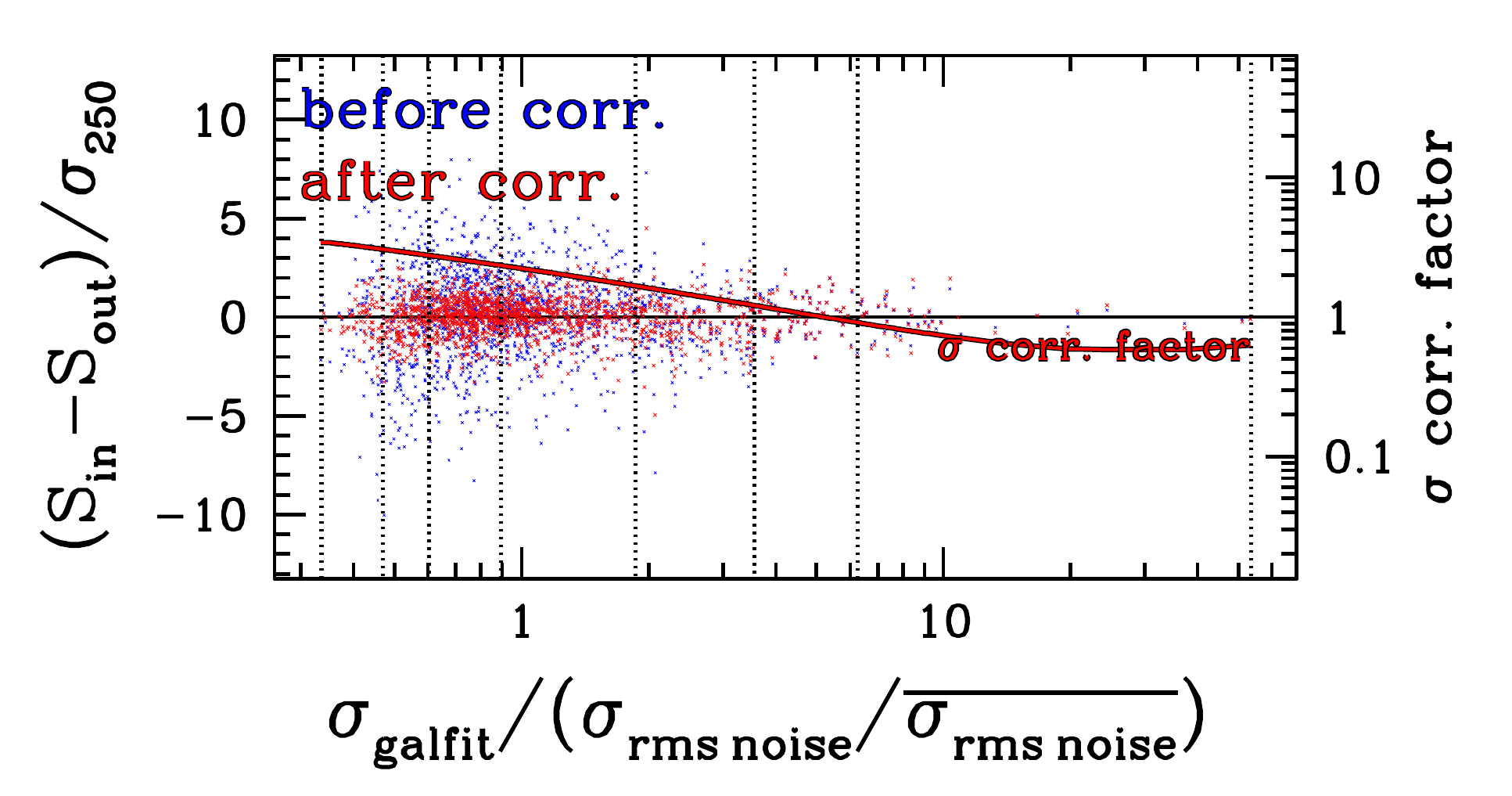}
    \includegraphics[height=2.6cm, trim=0 1cm 0 0]{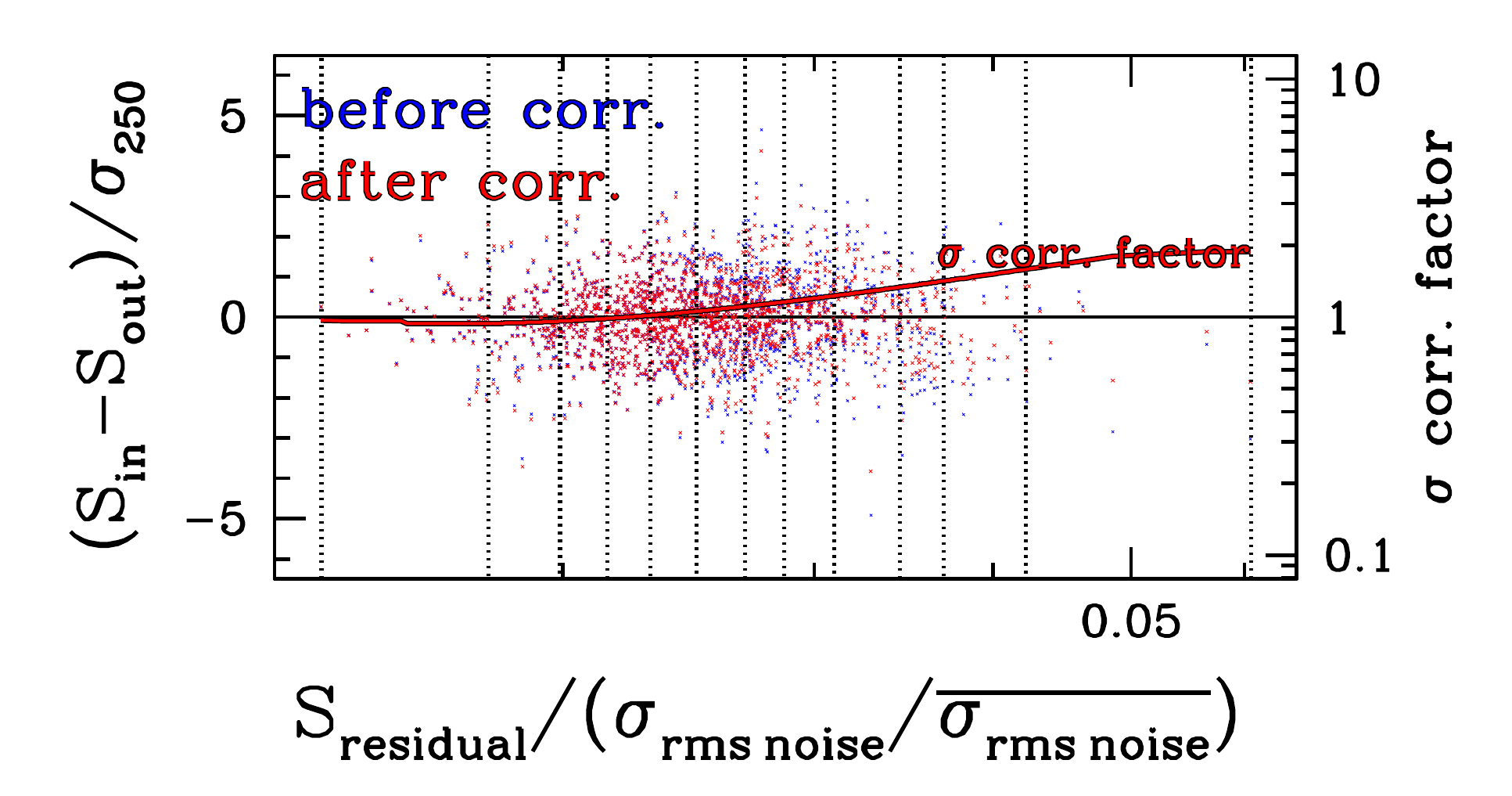}
    \includegraphics[height=2.6cm, trim=0 1cm 0 0]{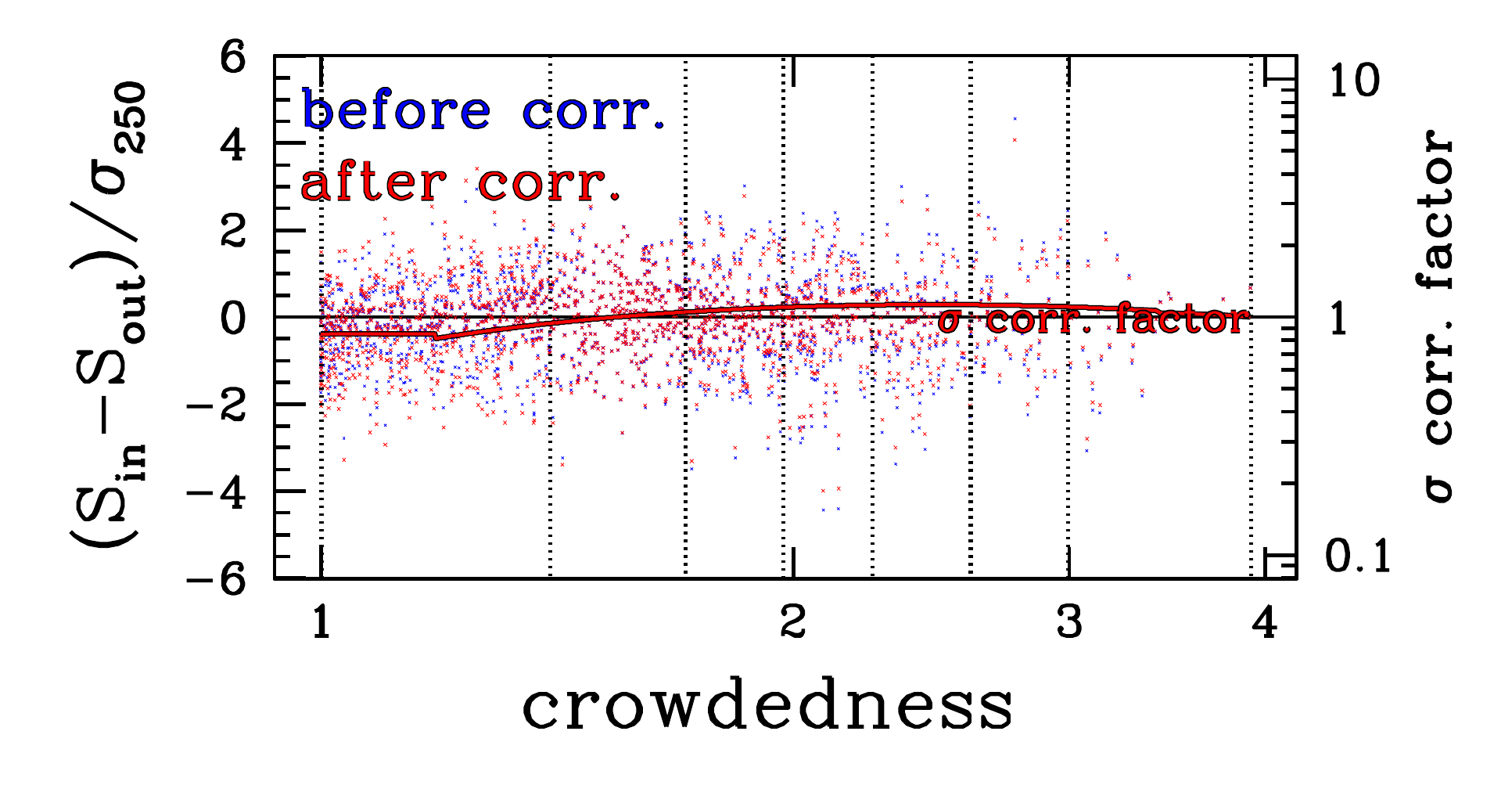}
    \end{subfigure}
    
    \end{figure}
    \begin{figure}\ContinuedFloat
    
    \begin{subfigure}[b]{\textwidth}\centering
    \includegraphics[height=2.6cm, trim=0 1cm -1.8cm 0]{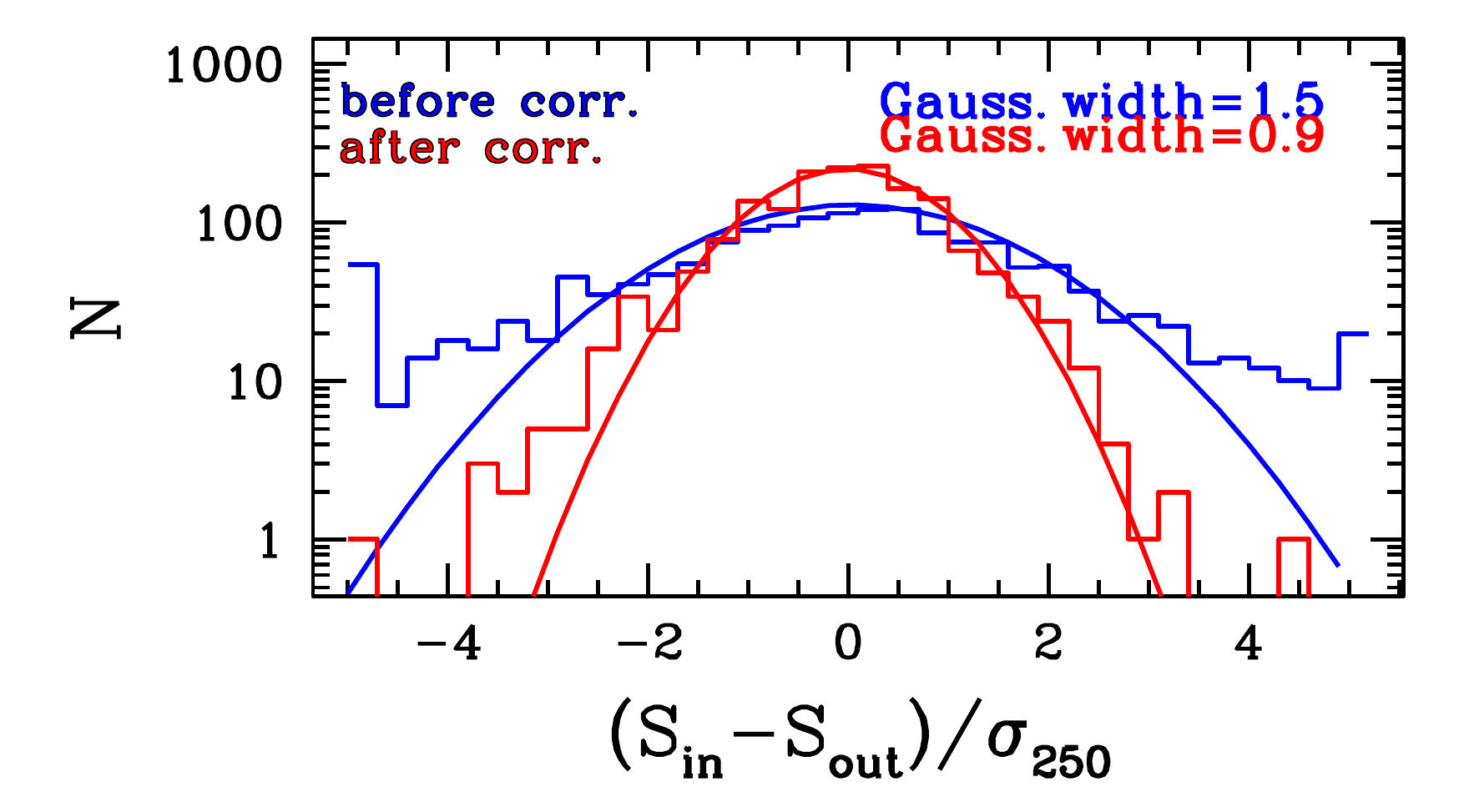}
    \includegraphics[height=2.6cm, trim=0 1cm -1.8cm 0]{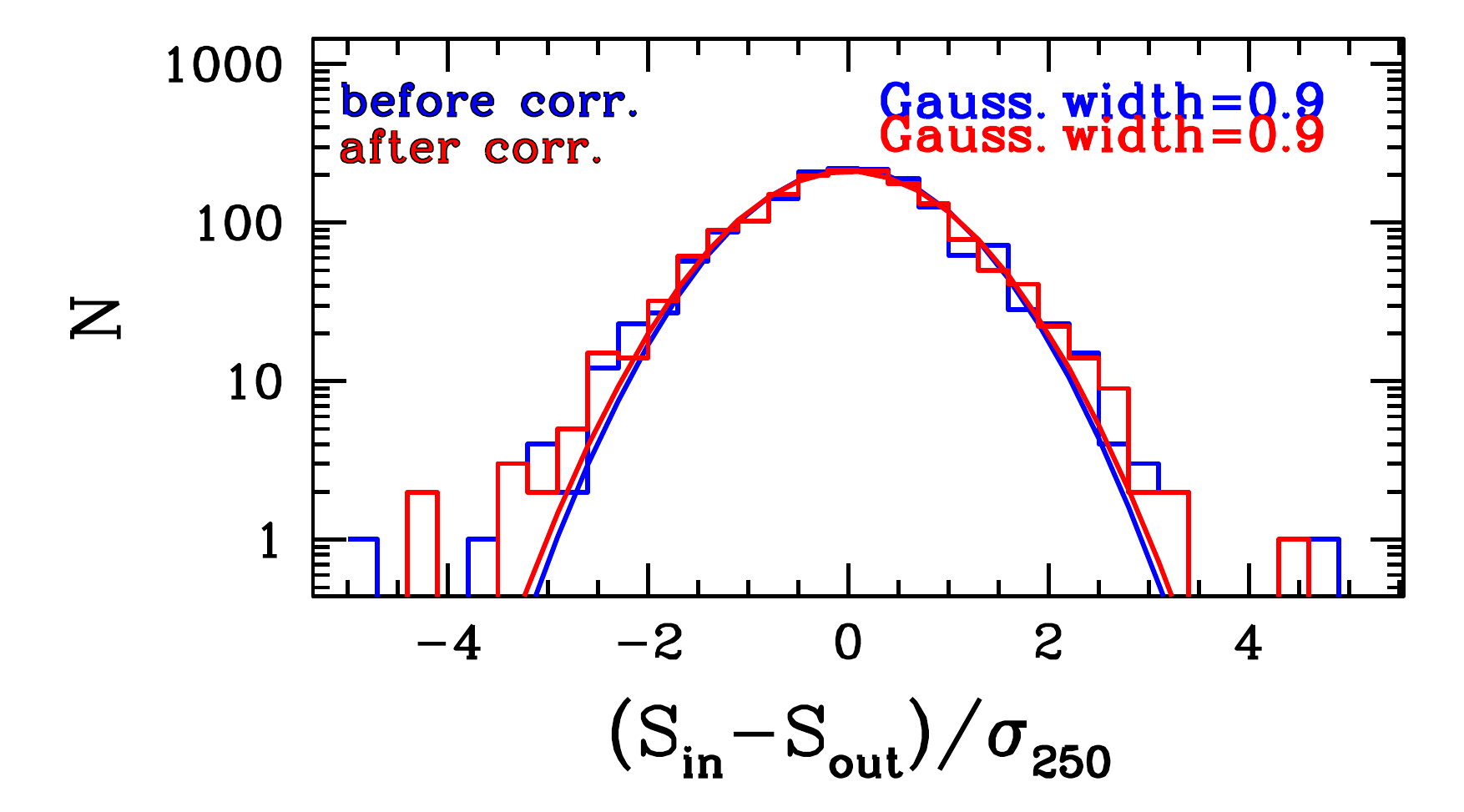}
    \includegraphics[height=2.6cm, trim=0 1cm -1.8cm 0]{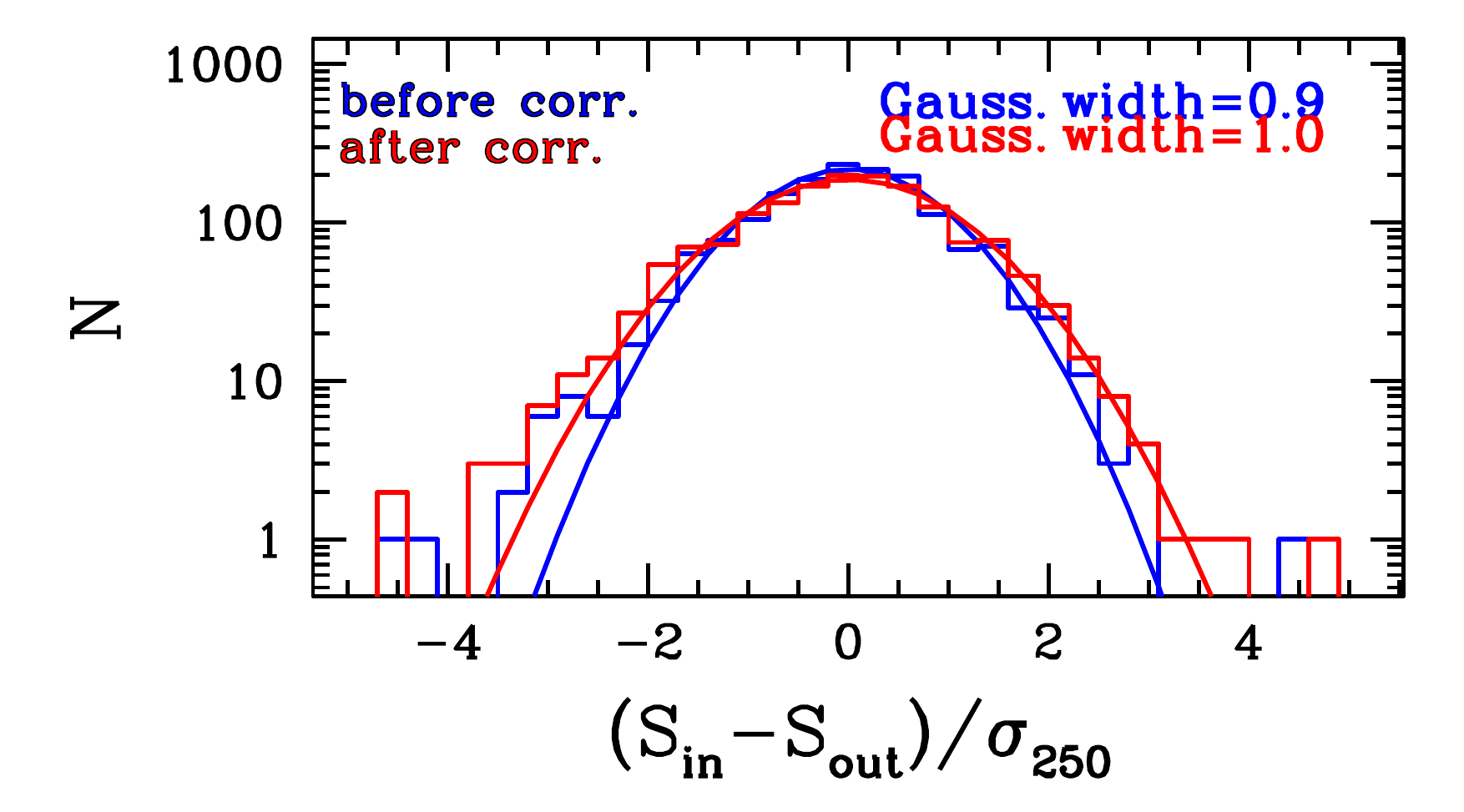}
    \end{subfigure}
    
    \begin{subfigure}[b]{\textwidth}\centering
    \includegraphics[height=2.6cm, trim=0 1cm -1.8cm 0]{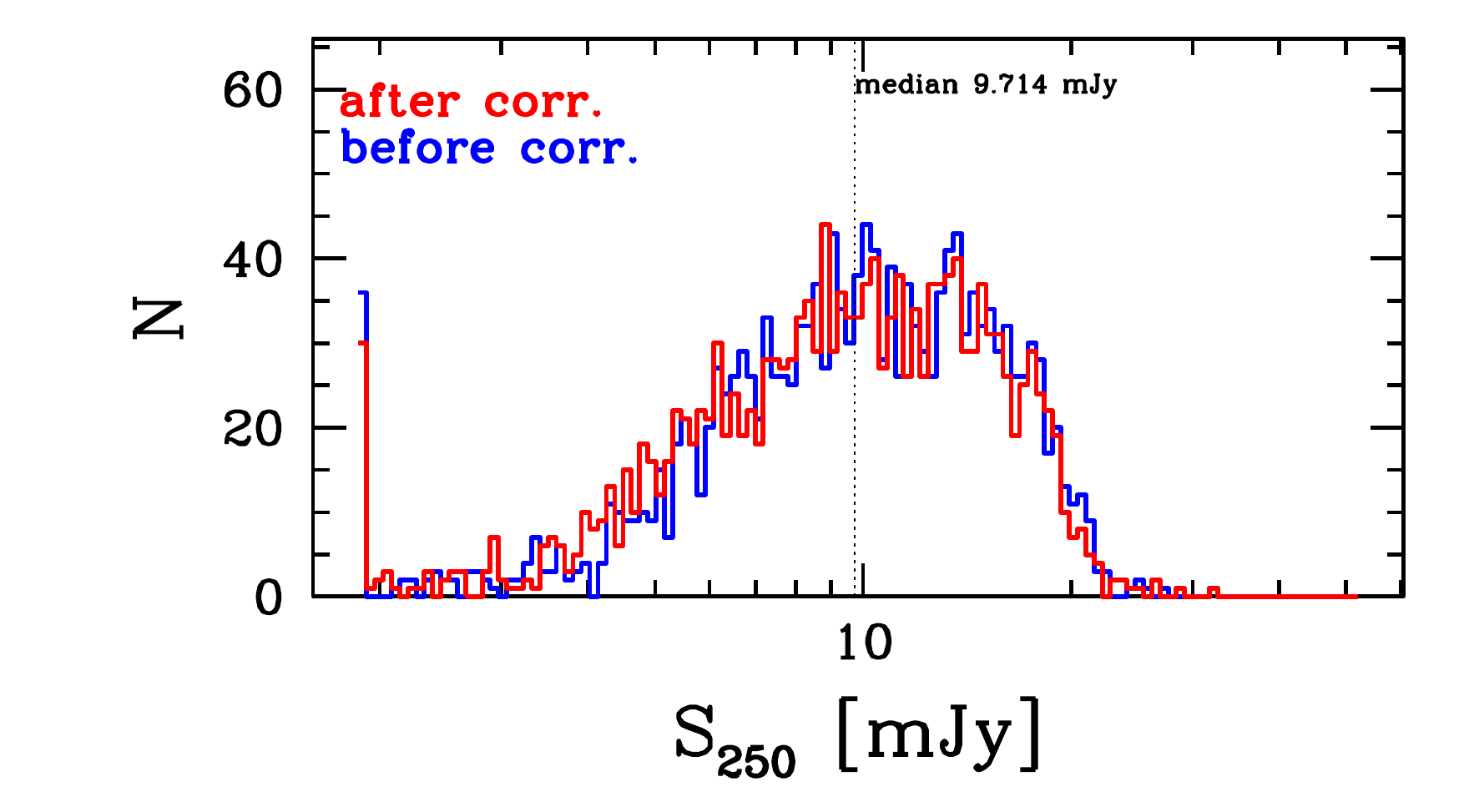}
    \includegraphics[height=2.6cm, trim=0 1cm -1.8cm 0]{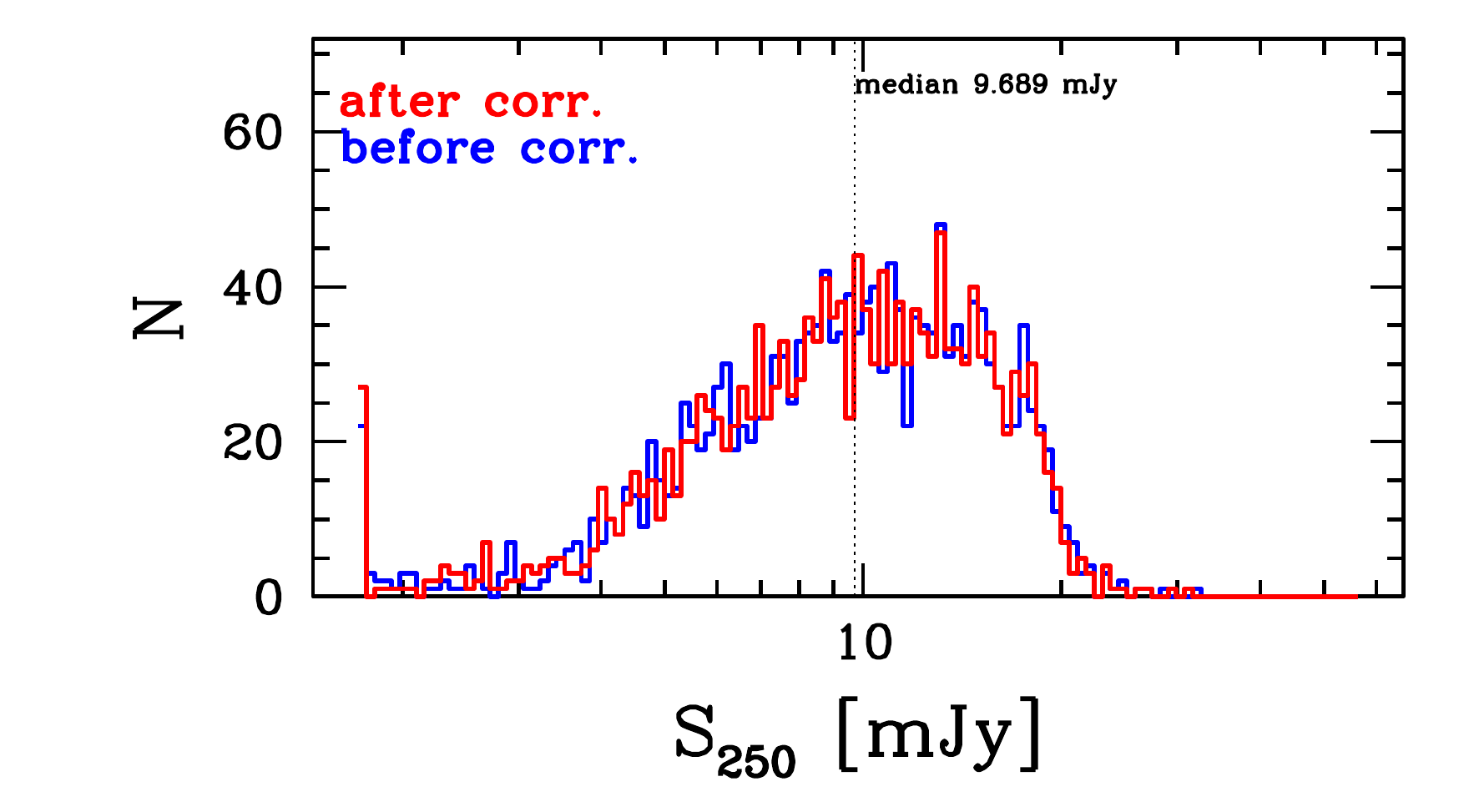}
    \includegraphics[height=2.6cm, trim=0 1cm -1.8cm 0]{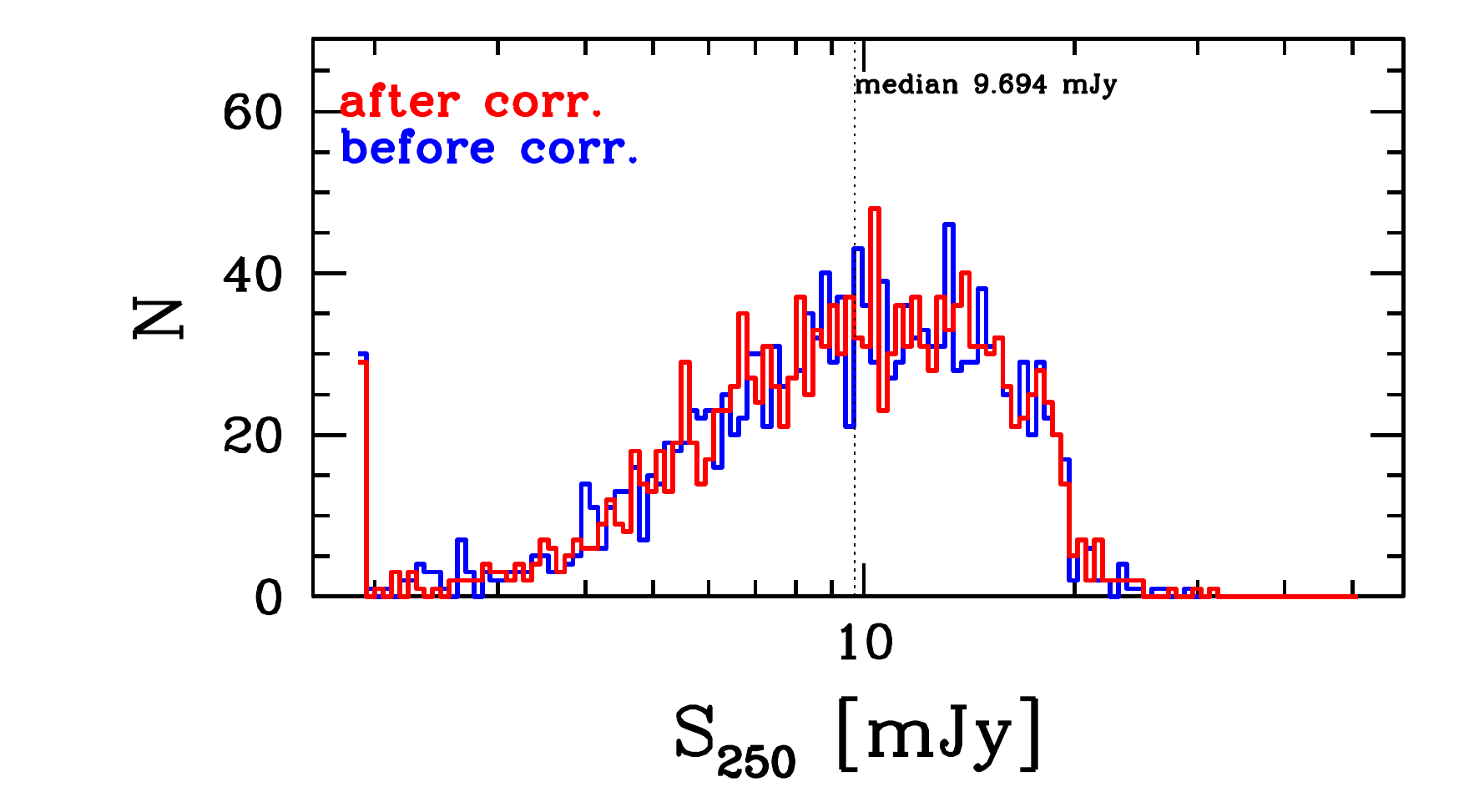}
    \end{subfigure}
    
    \begin{subfigure}[b]{\textwidth}\centering
    \includegraphics[height=2.6cm, trim=0 1cm -1.8cm 0]{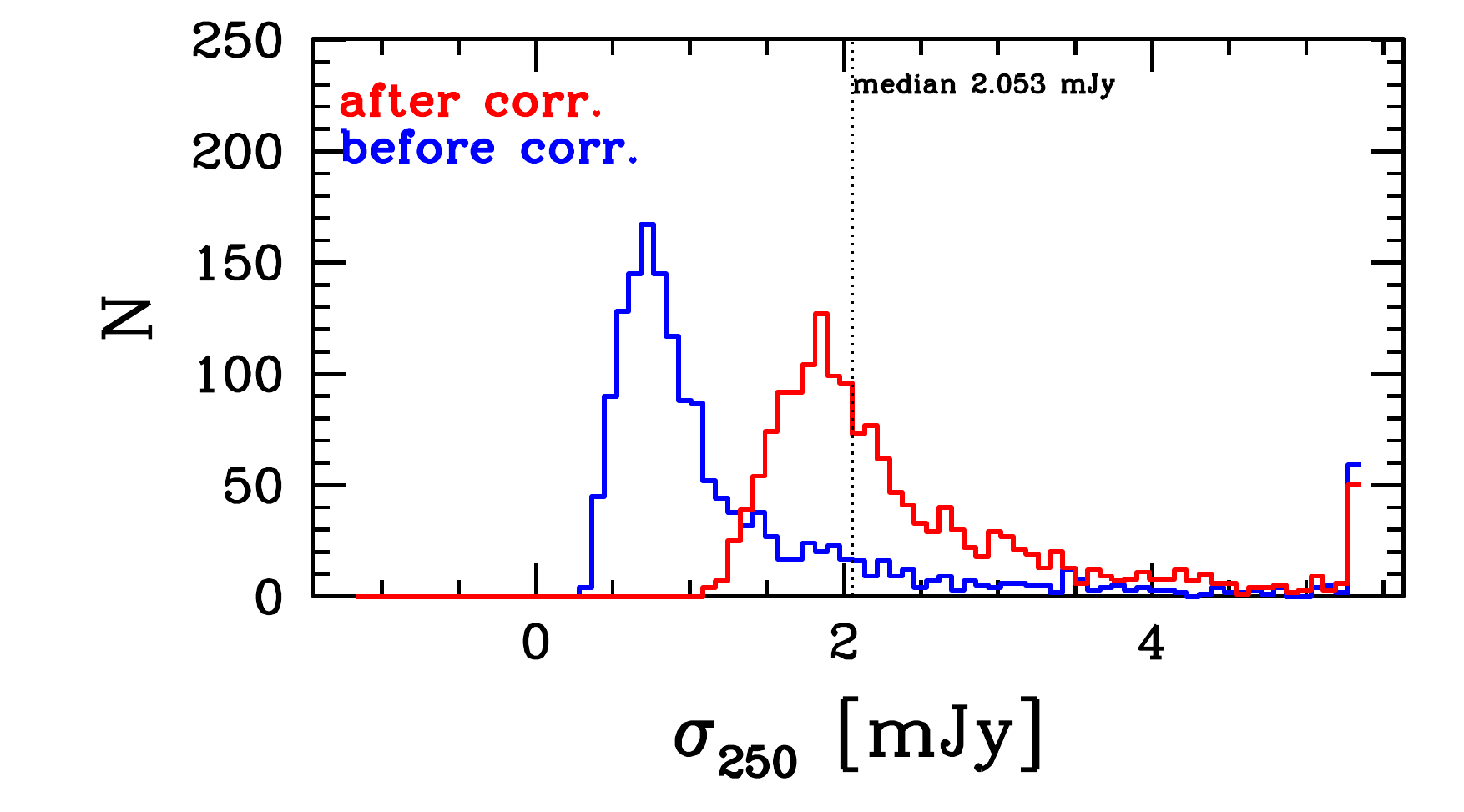}
    \includegraphics[height=2.6cm, trim=0 1cm -1.8cm 0]{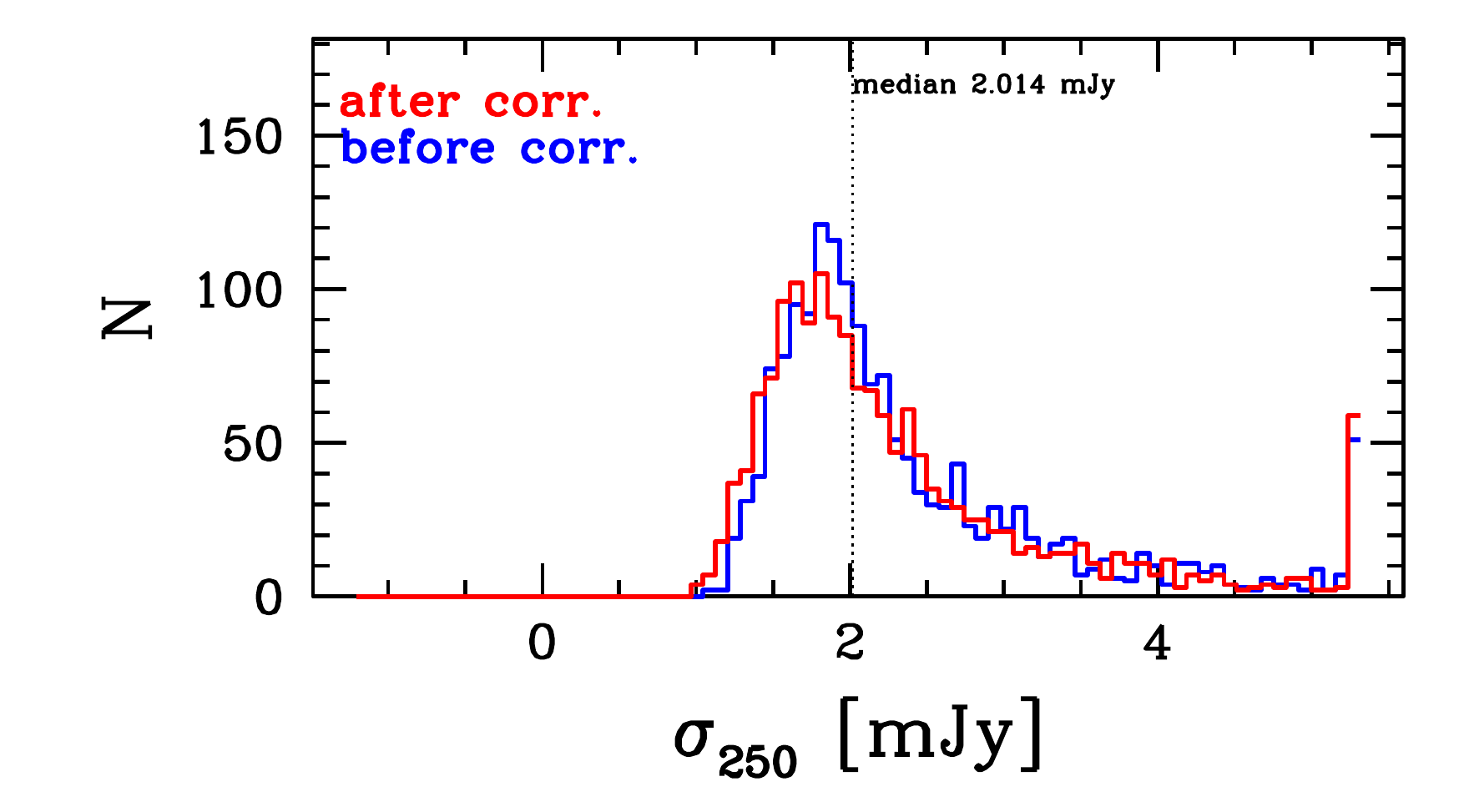}
    \includegraphics[height=2.6cm, trim=0 1cm -1.8cm 0]{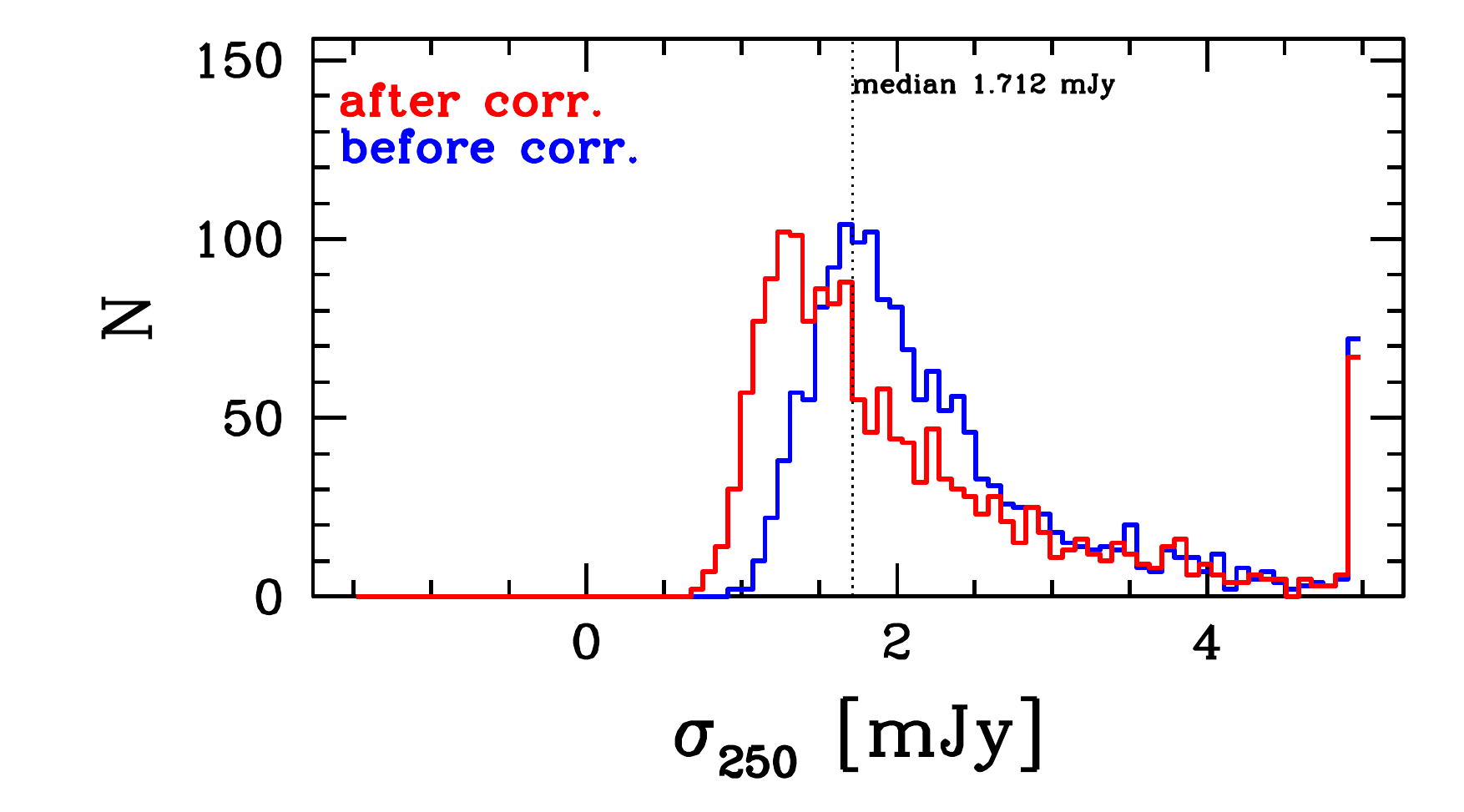}
    \end{subfigure}
    
    \caption{%
        Simulation correction analyses at 250~$\mu$m. See descriptions in the text. 
        \label{Figure_galsim_250_bin}
    }
\end{figure}


\begin{figure}
\centering

    \begin{subfigure}[b]{\textwidth}\centering
    \includegraphics[height=2.6cm, trim=0 1cm 0 0]{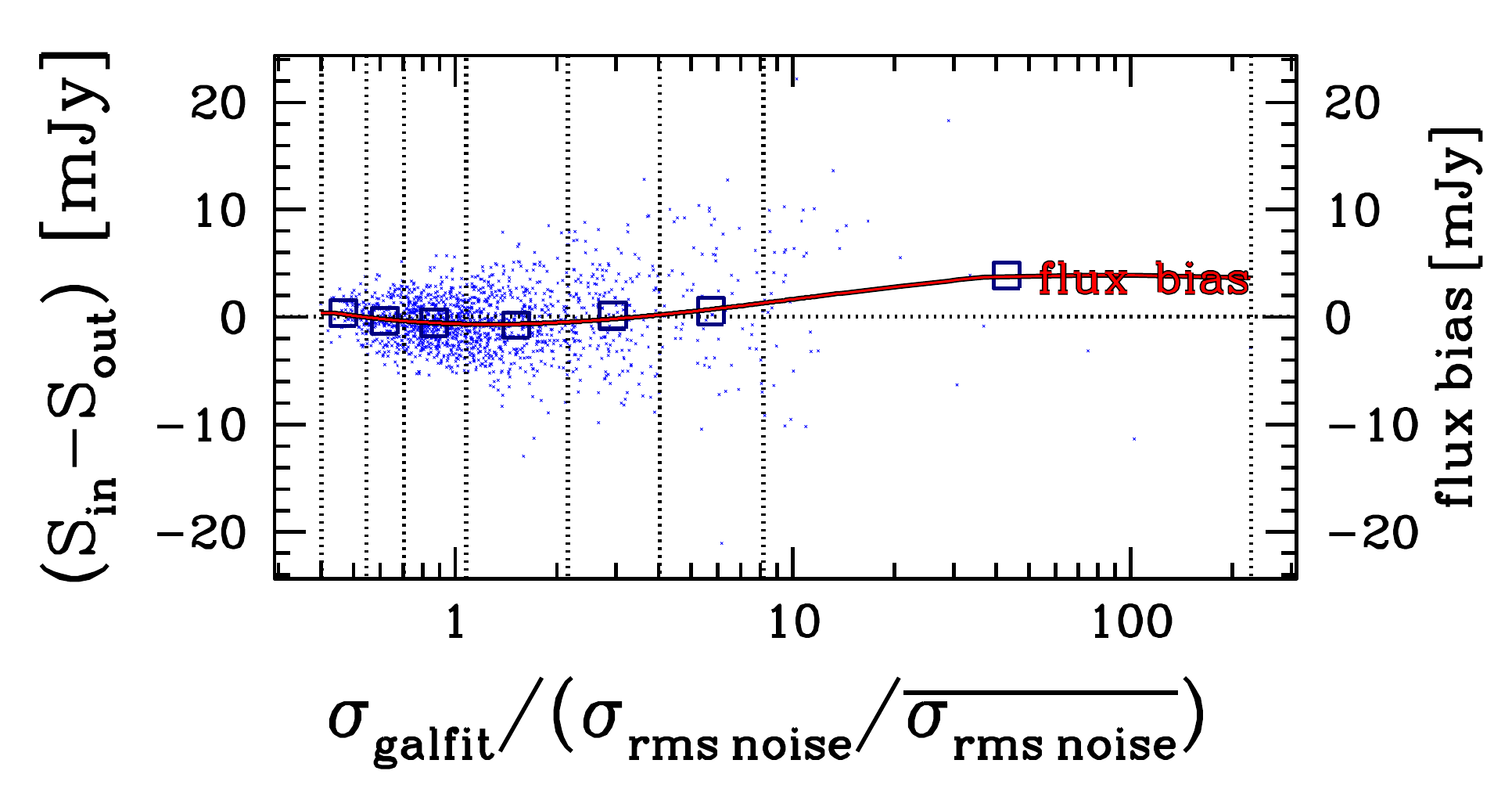}
    \includegraphics[height=2.6cm, trim=0 1cm 0 0]{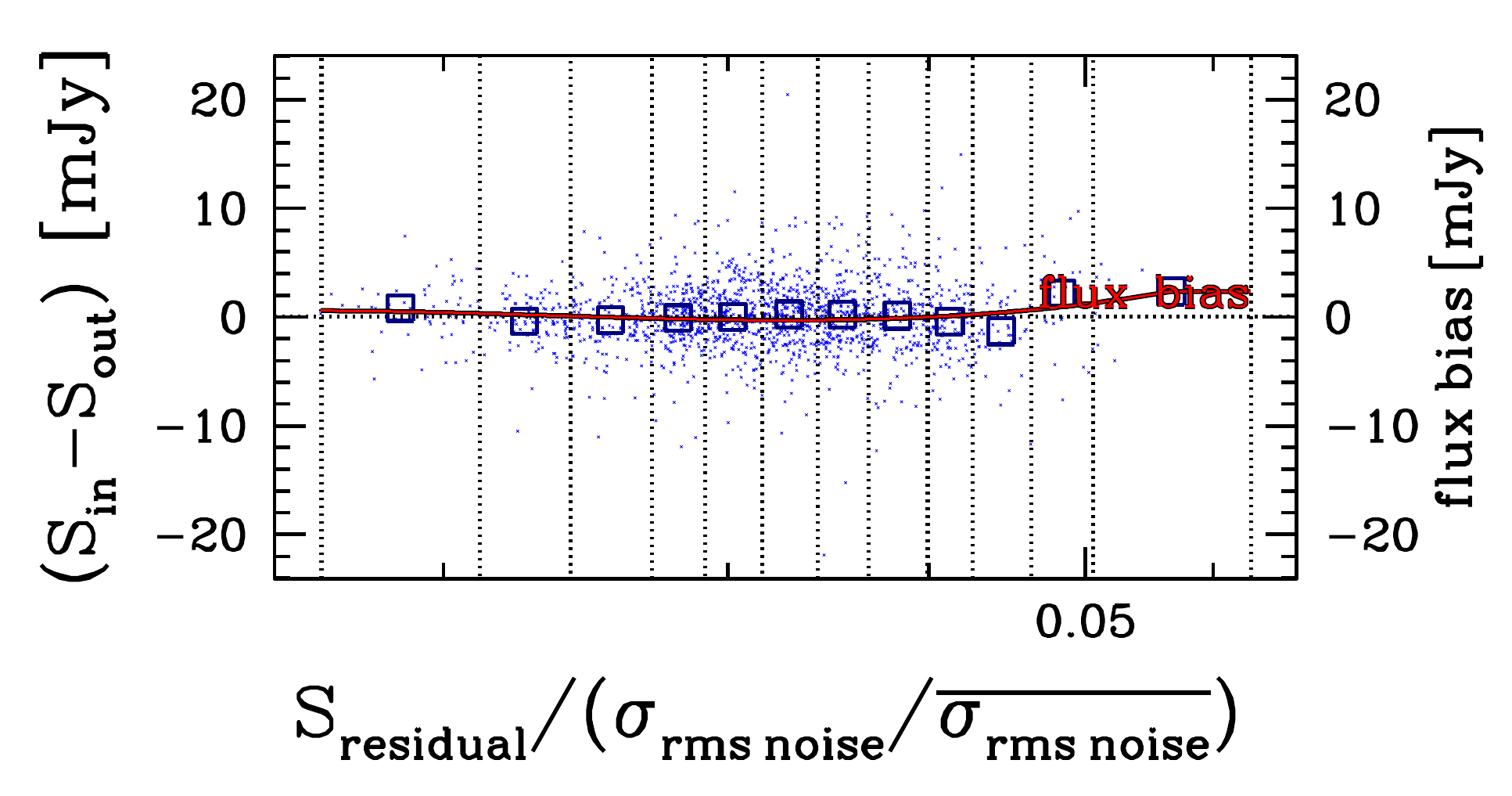}
    \includegraphics[height=2.6cm, trim=0 1cm 0 0]{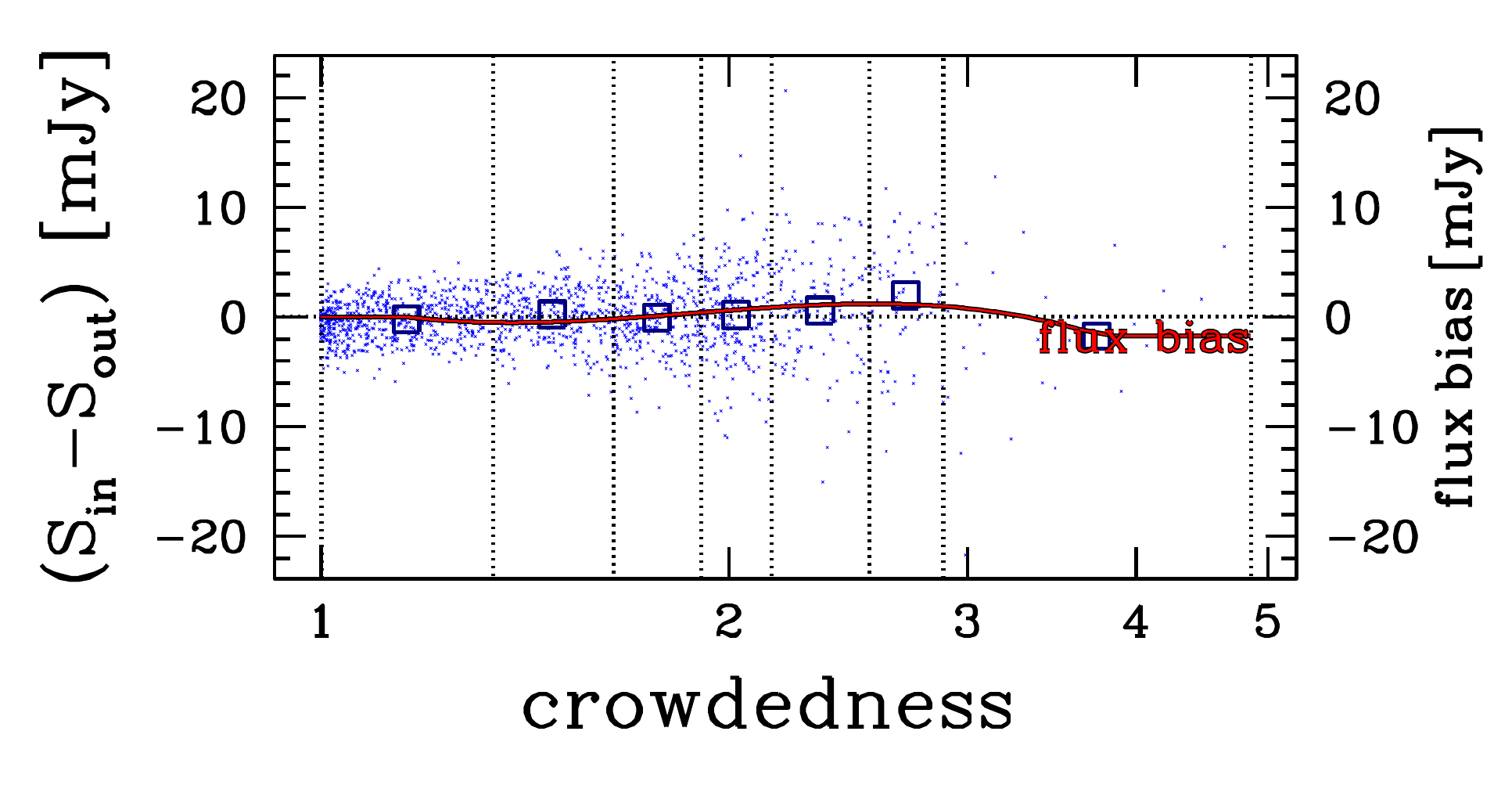}
    \end{subfigure}

    \begin{subfigure}[b]{\textwidth}\centering
    \includegraphics[height=2.6cm, trim=0 1cm 0 0]{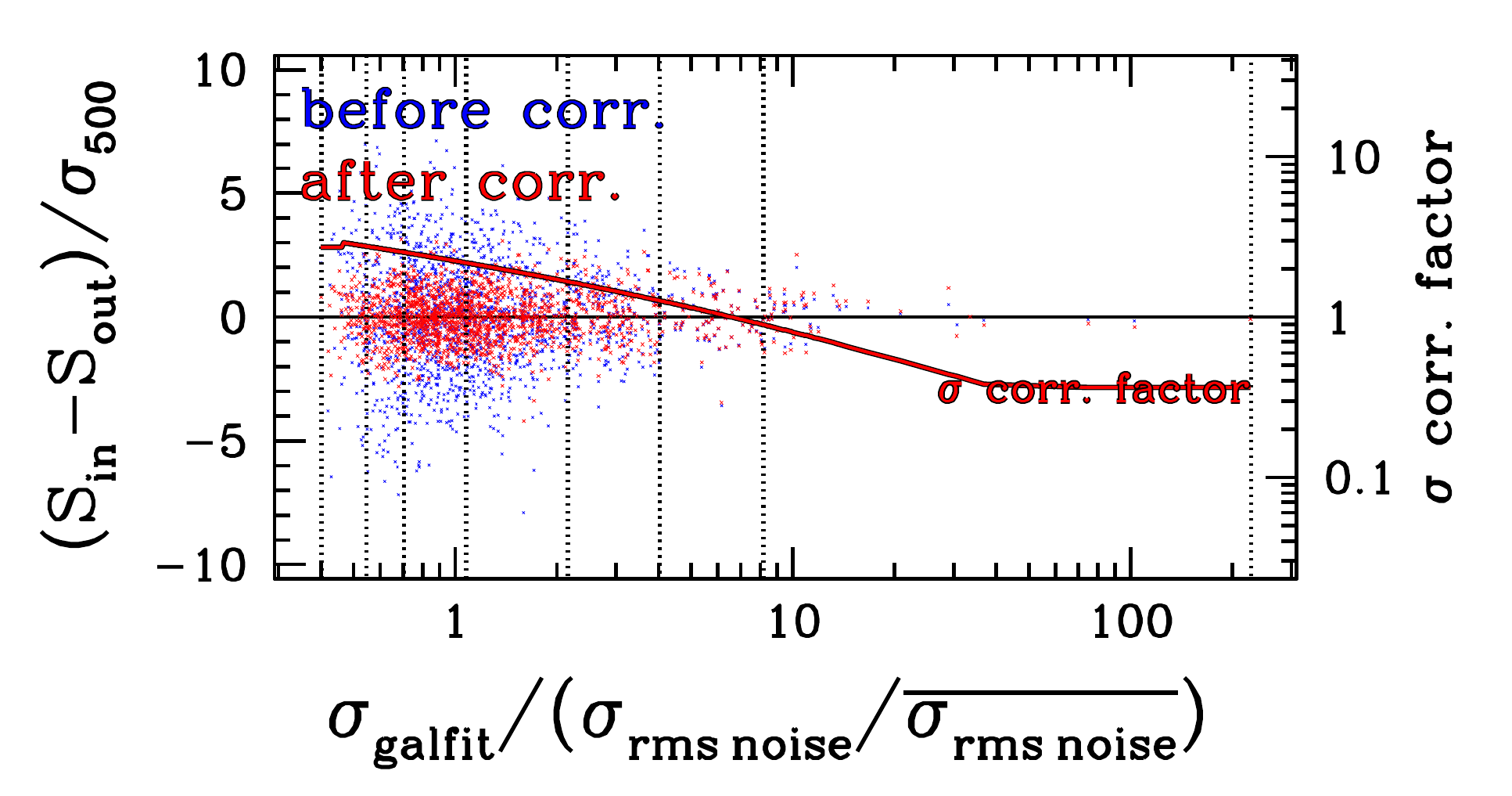}
    \includegraphics[height=2.6cm, trim=0 1cm 0 0]{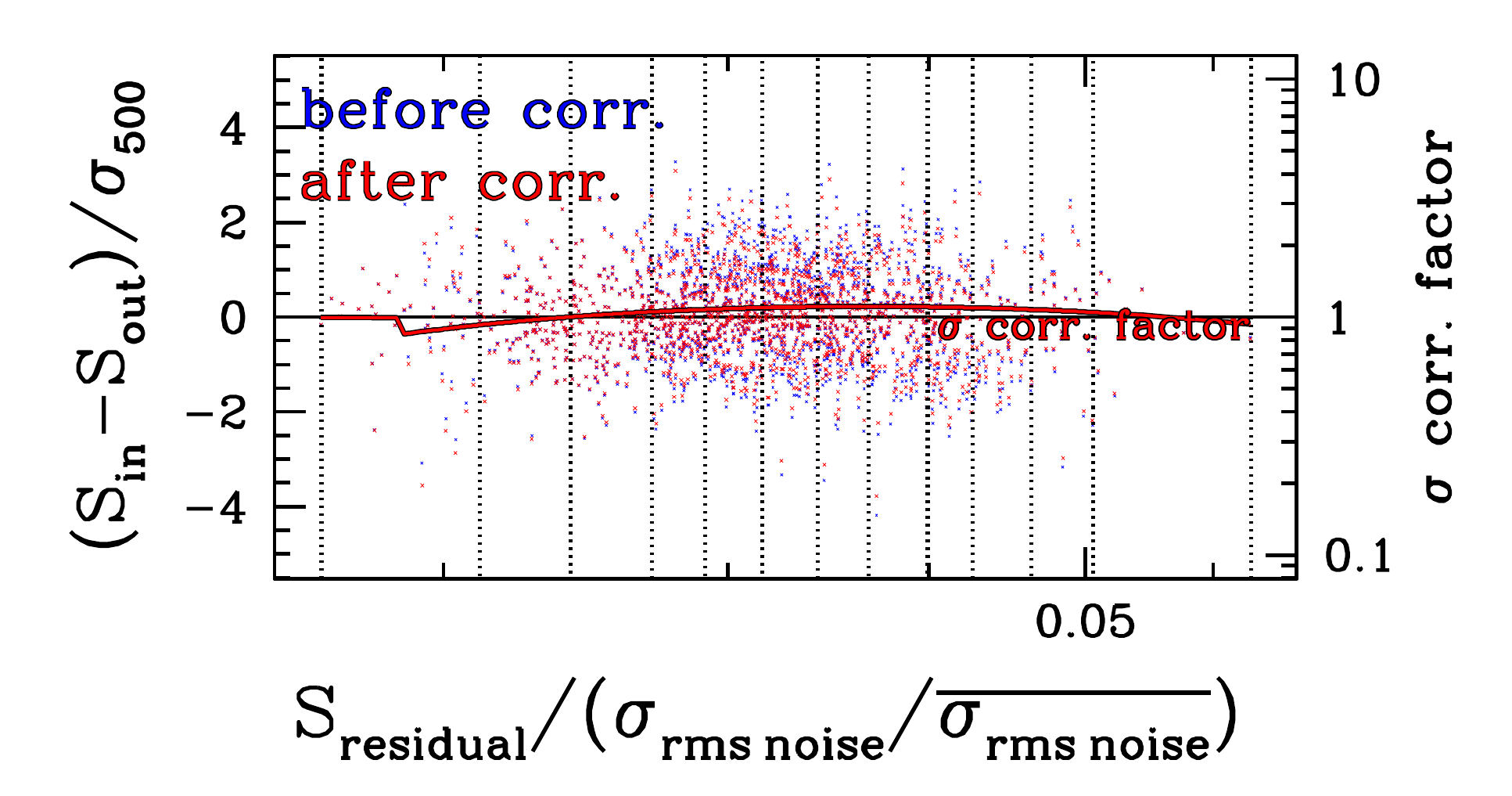}
    \includegraphics[height=2.6cm, trim=0 1cm 0 0]{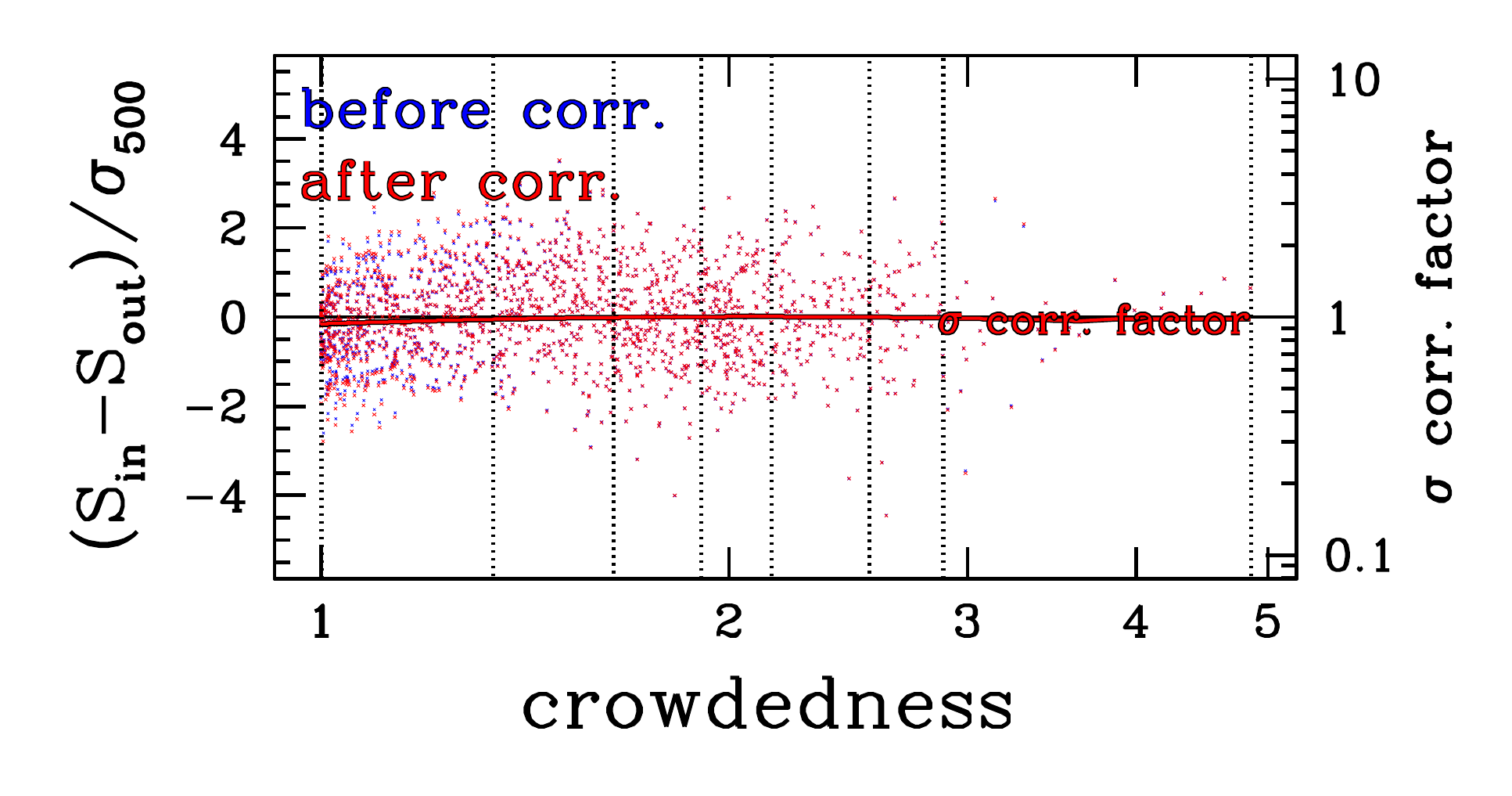}
    \end{subfigure}

    \begin{subfigure}[b]{\textwidth}\centering
    \includegraphics[height=2.6cm, trim=0 1cm -1.8cm 0]{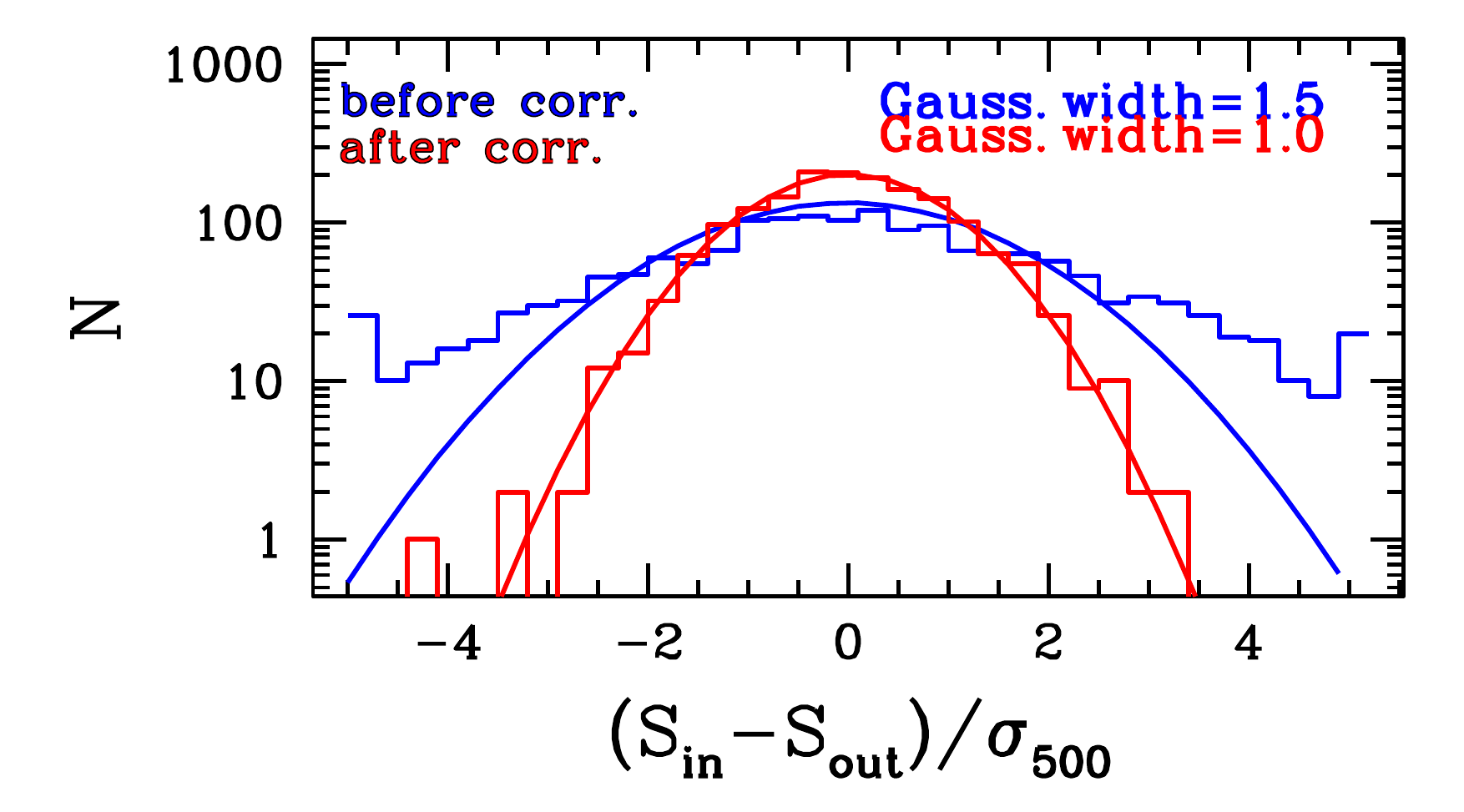}
    \includegraphics[height=2.6cm, trim=0 1cm -1.8cm 0]{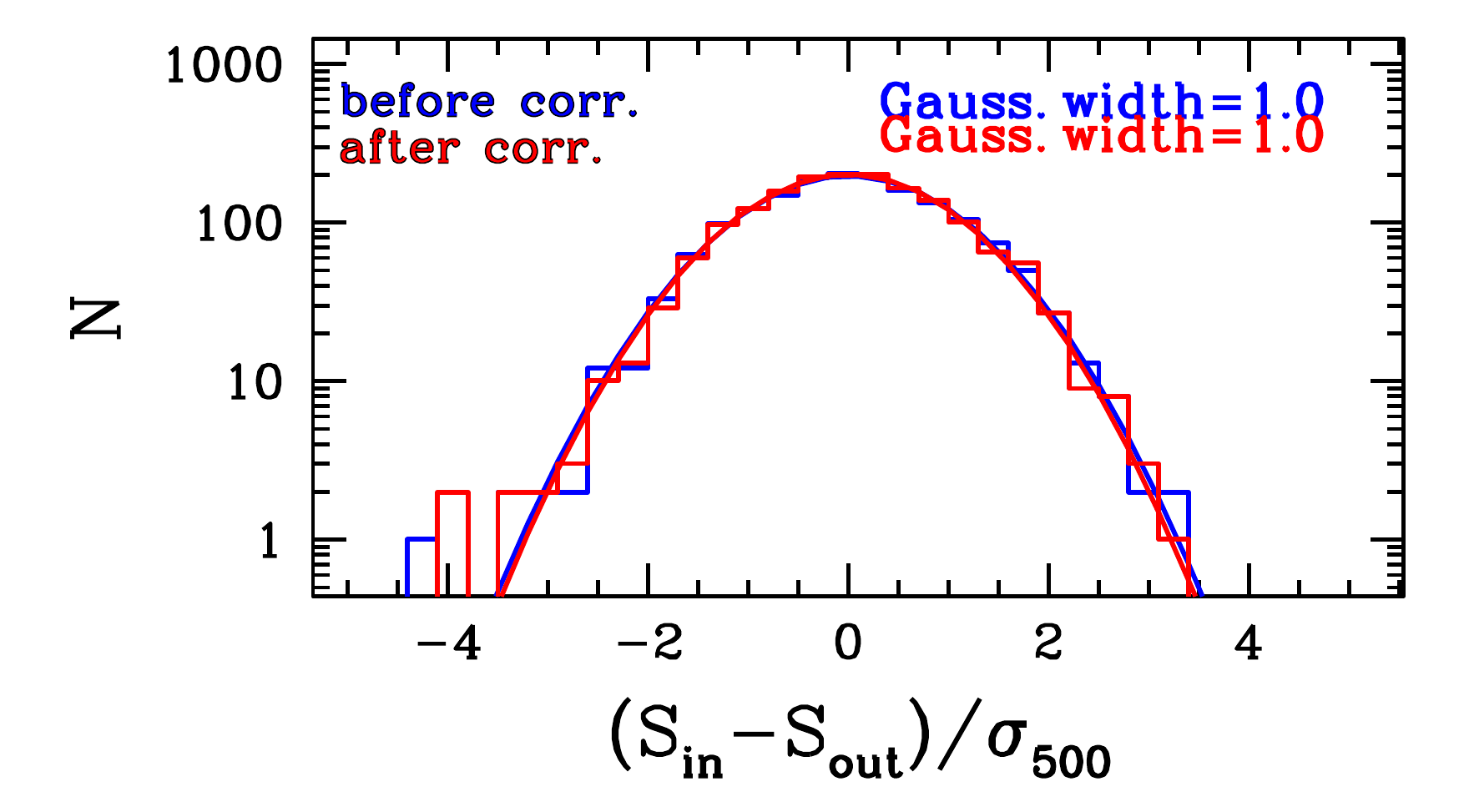}
    \includegraphics[height=2.6cm, trim=0 1cm -1.8cm 0]{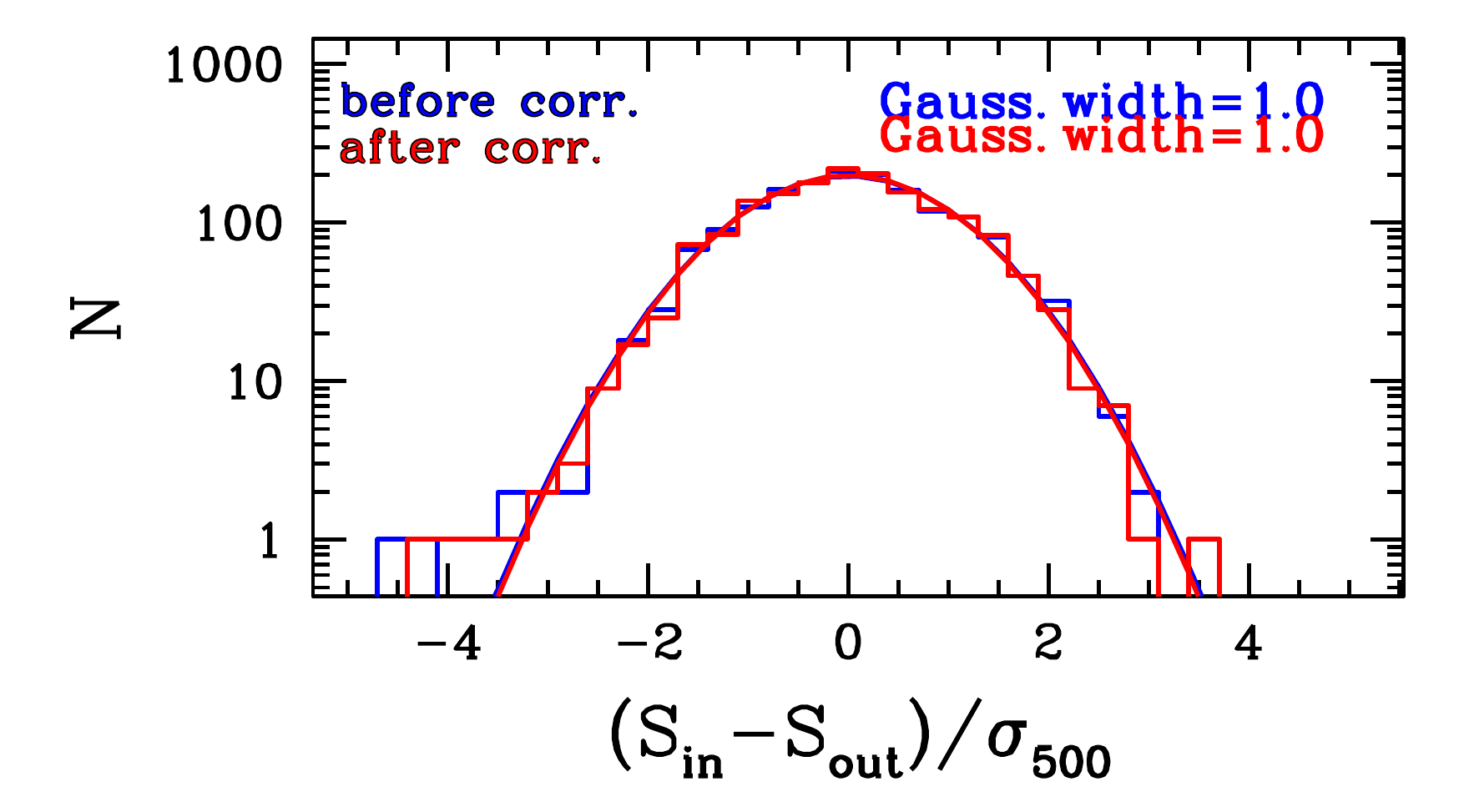}
    \end{subfigure}

    \begin{subfigure}[b]{\textwidth}\centering
    \includegraphics[height=2.6cm, trim=0 1cm -1.8cm 0]{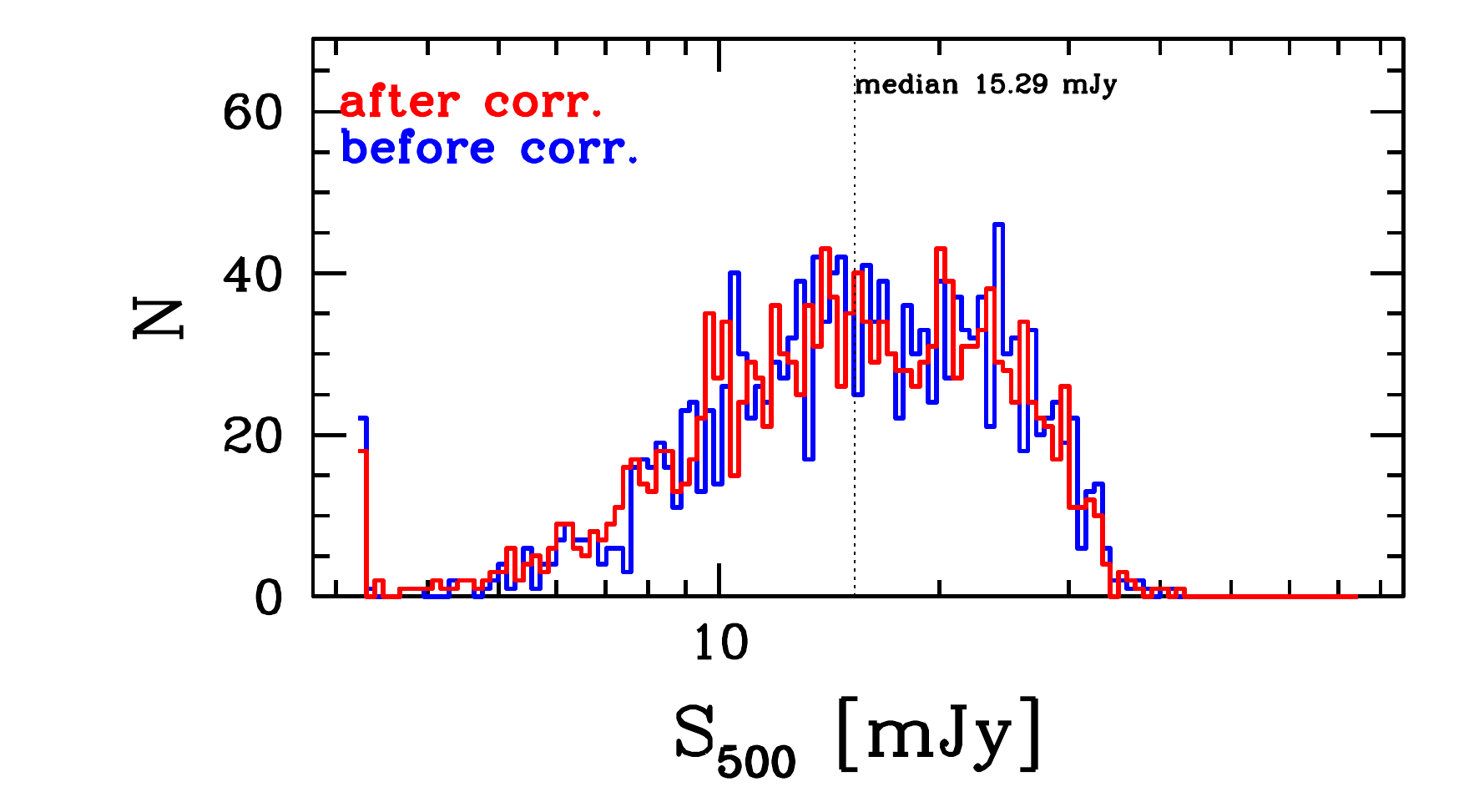}
    \includegraphics[height=2.6cm, trim=0 1cm -1.8cm 0]{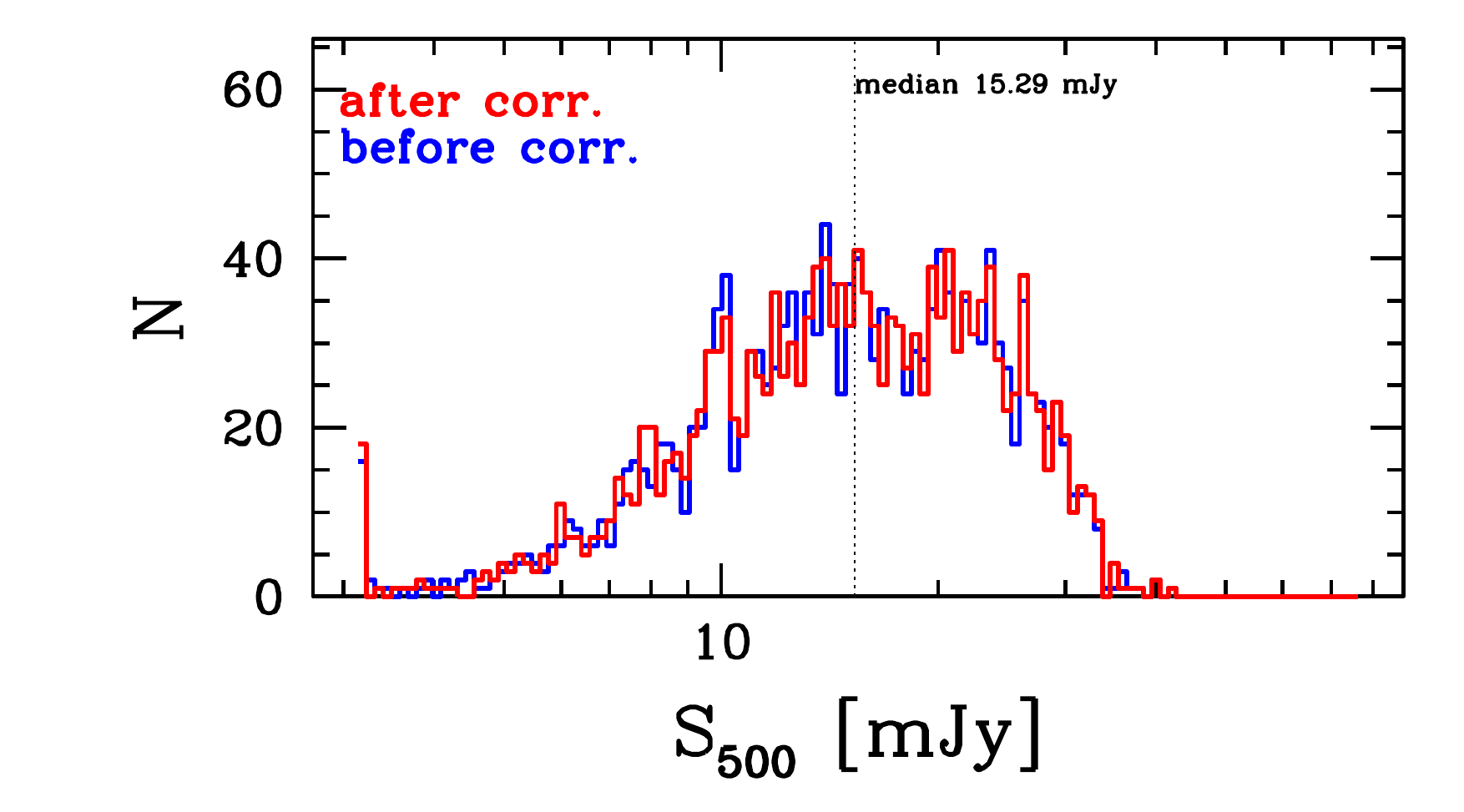}
    \includegraphics[height=2.6cm, trim=0 1cm -1.8cm 0]{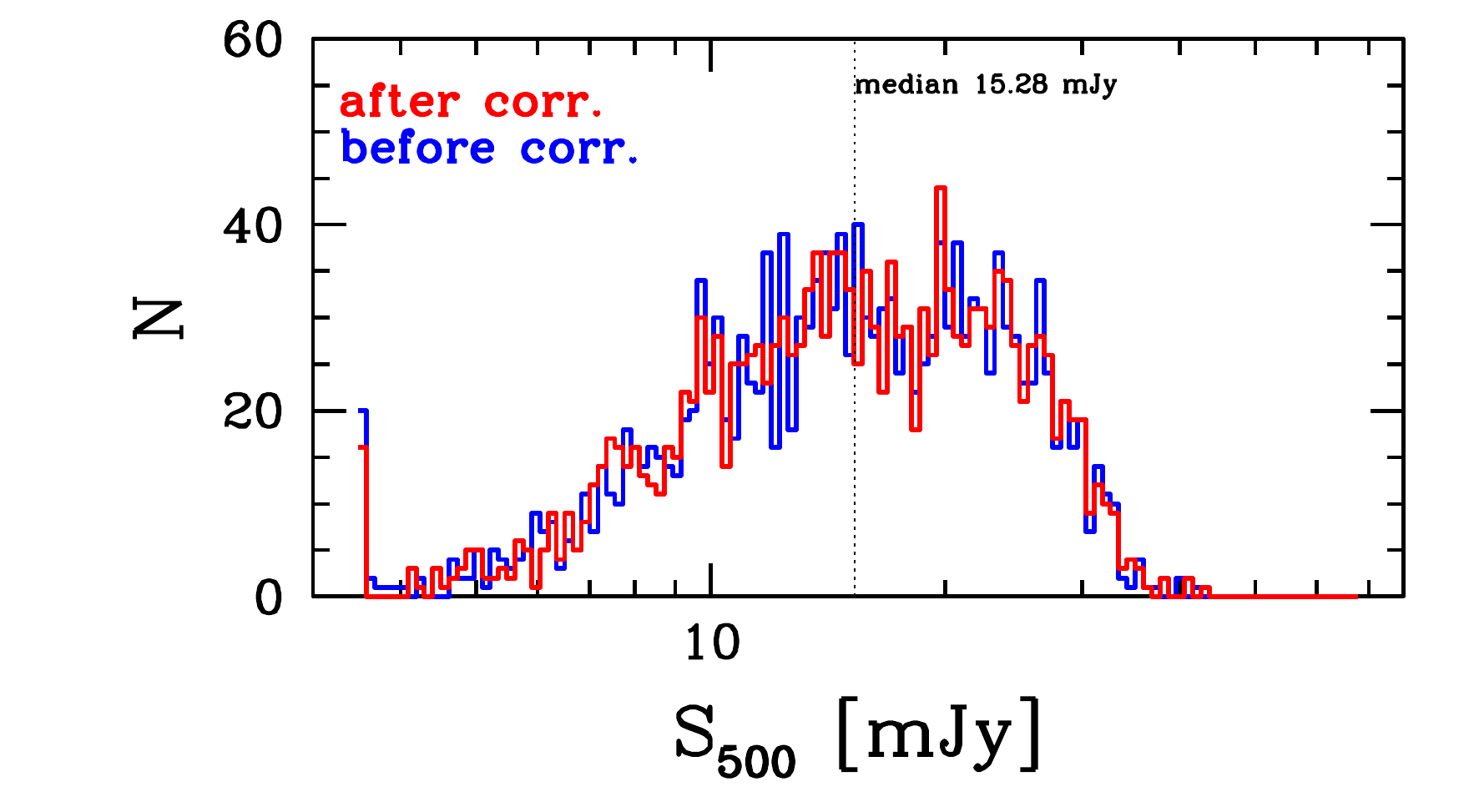}
    \end{subfigure}
    
    \end{figure}
    \begin{figure}\ContinuedFloat

    \begin{subfigure}[b]{\textwidth}\centering
    \includegraphics[height=2.6cm, trim=0 1cm -1.8cm 0]{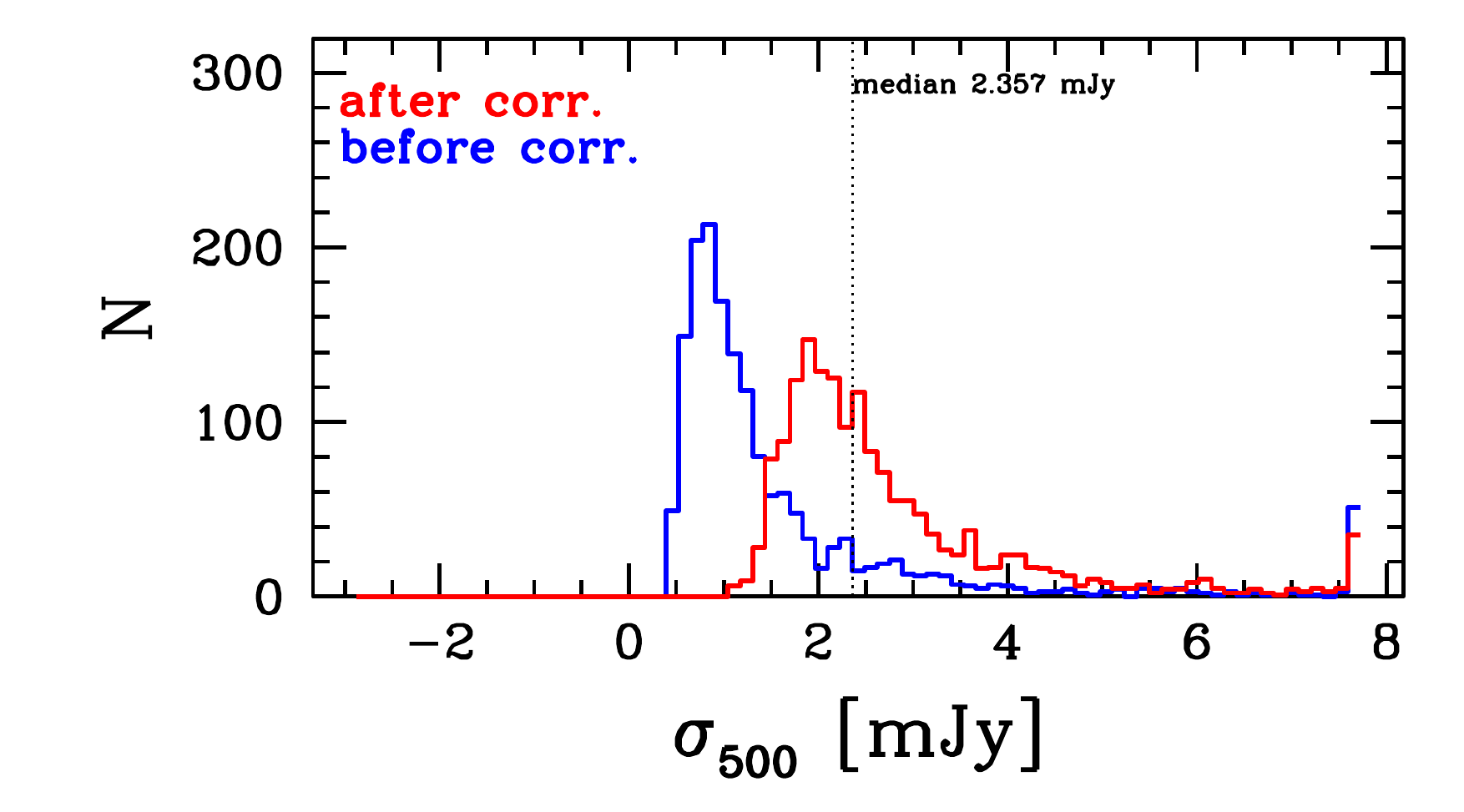}
    \includegraphics[height=2.6cm, trim=0 1cm -1.8cm 0]{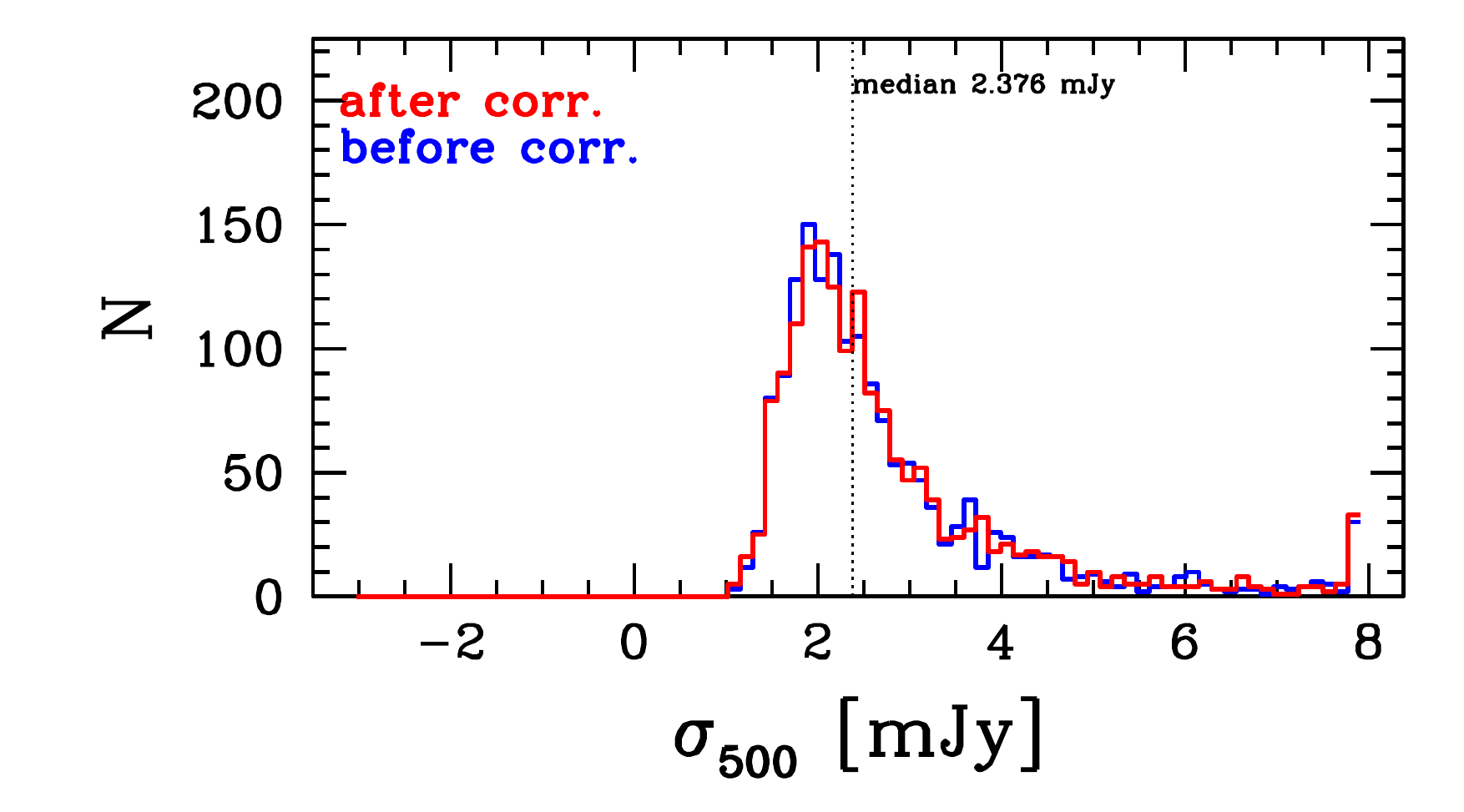}
    \includegraphics[height=2.6cm, trim=0 1cm -1.8cm 0]{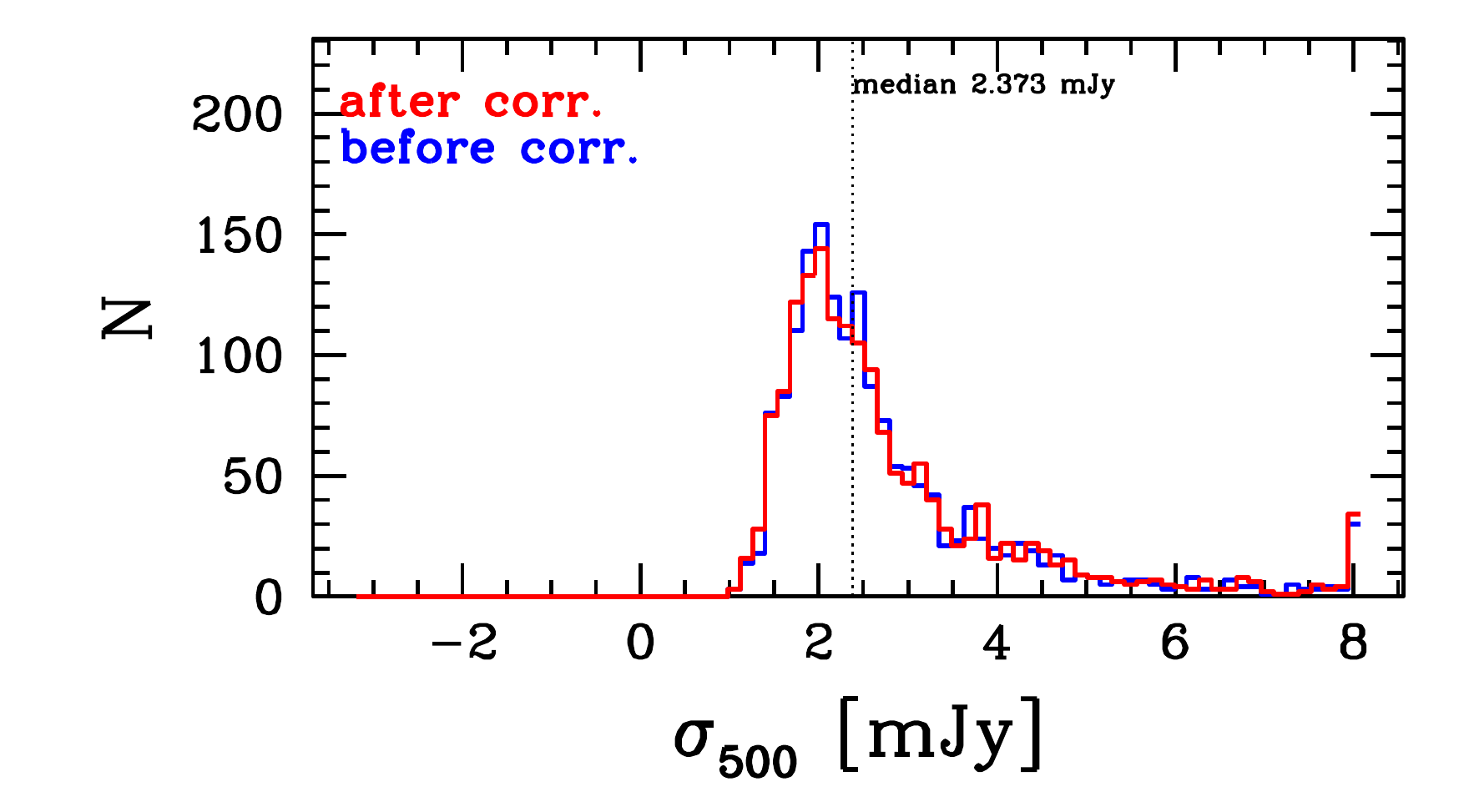}
    \end{subfigure}

\caption{%
    Simulation correction analyses at 500~$\mu$m. See descriptions in the text. 
        \label{Figure_galsim_500_bin}
}
\end{figure}

\begin{figure}
\centering

    \begin{subfigure}[b]{\textwidth}\centering
    \includegraphics[height=2.6cm, trim=0 1cm 0 0]{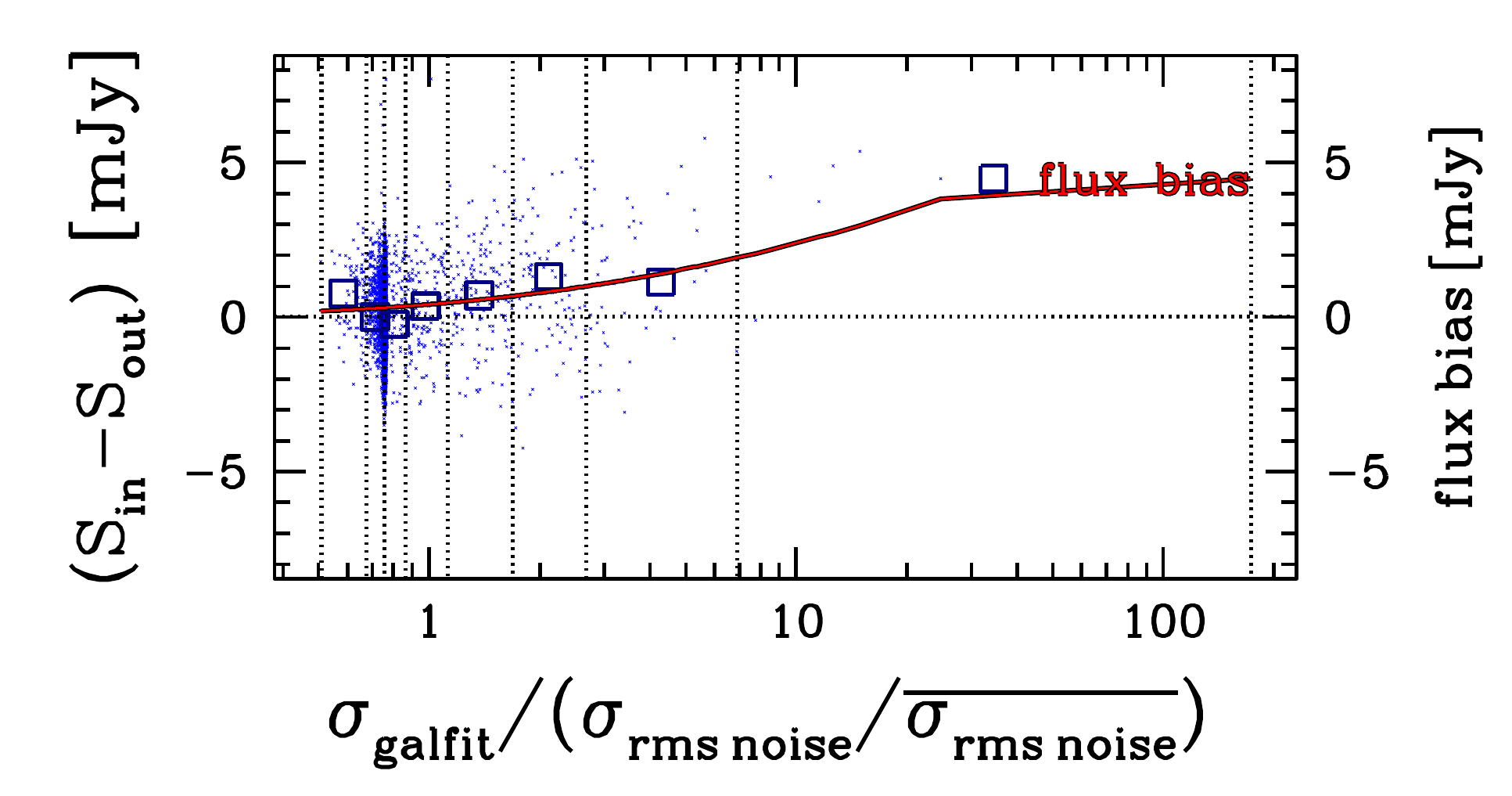}
    \includegraphics[height=2.6cm, trim=0 1cm 0 0]{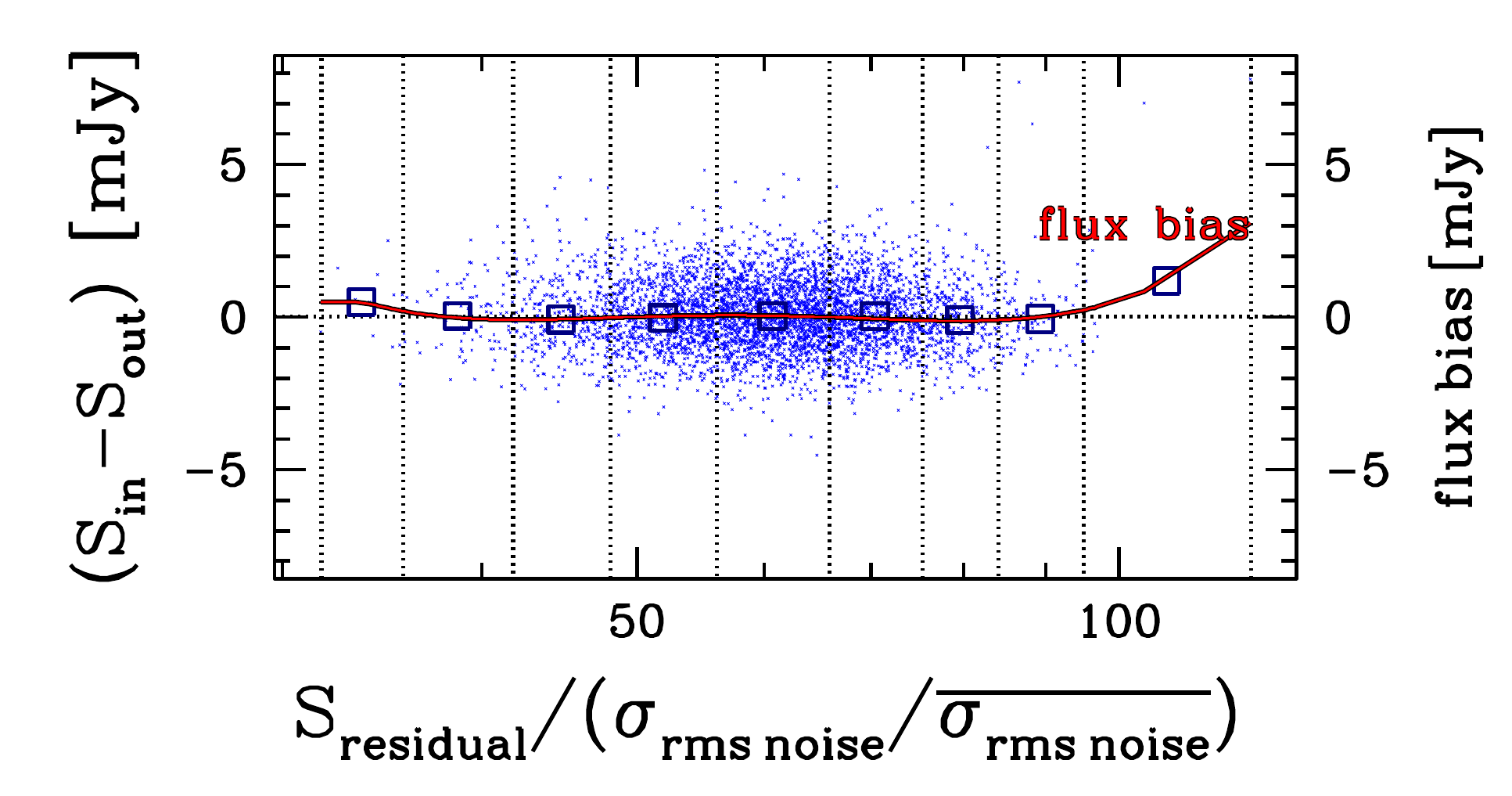}
    \includegraphics[height=2.6cm, trim=0 1cm 0 0]{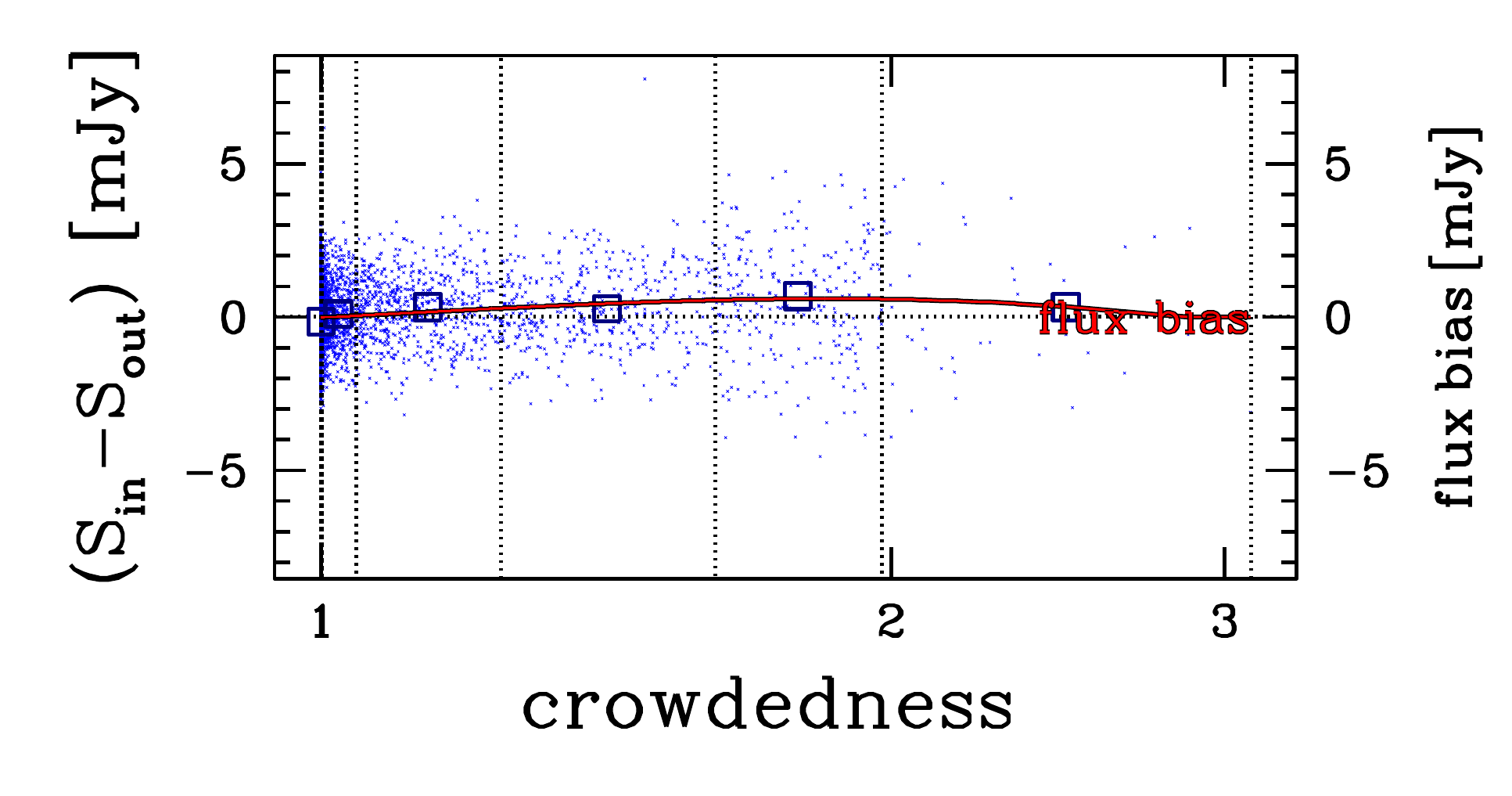}
    \end{subfigure}

    \begin{subfigure}[b]{\textwidth}\centering
    \includegraphics[height=2.6cm, trim=0 1cm 0 0]{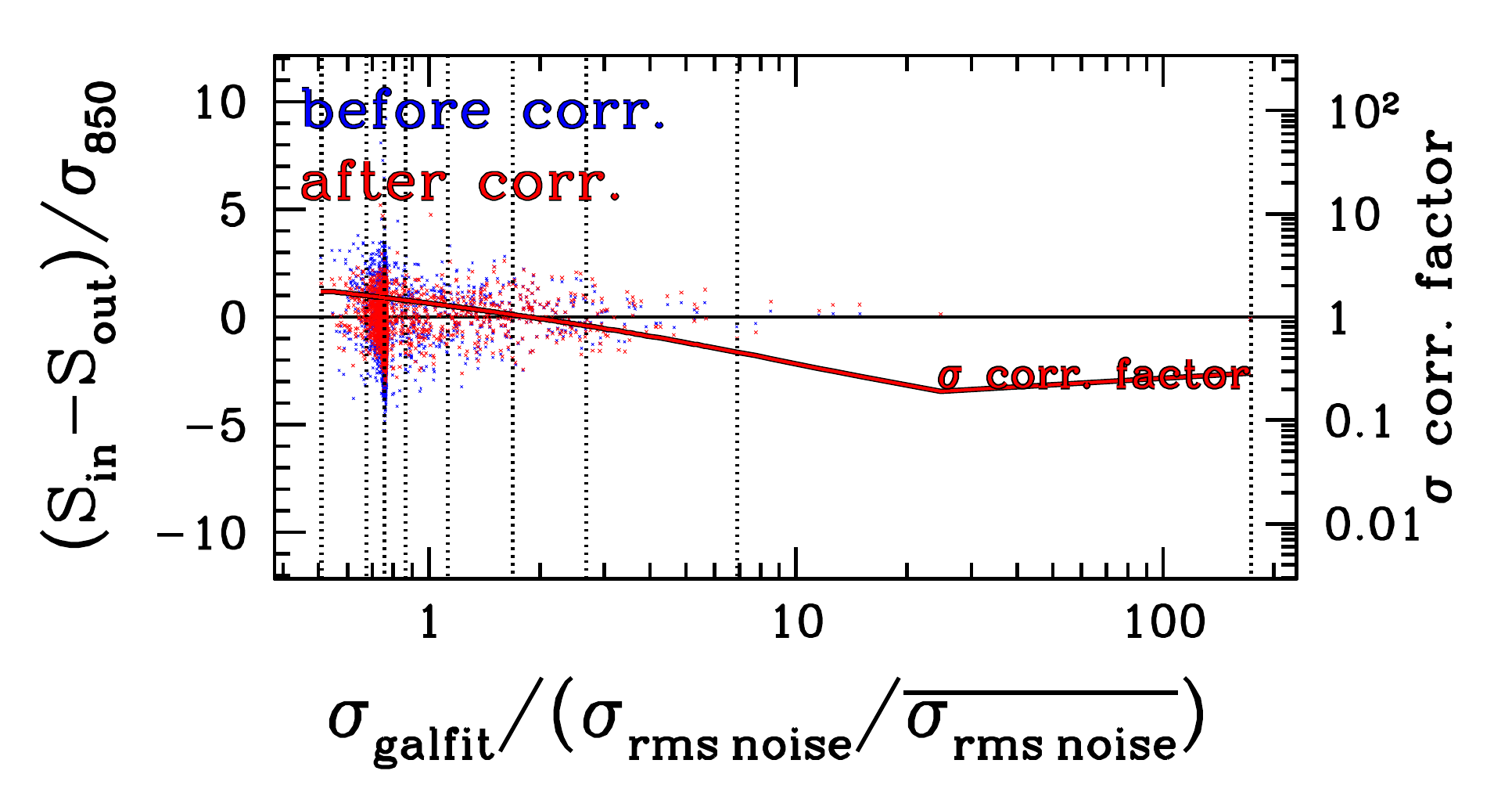}
    \includegraphics[height=2.6cm, trim=0 1cm 0 0]{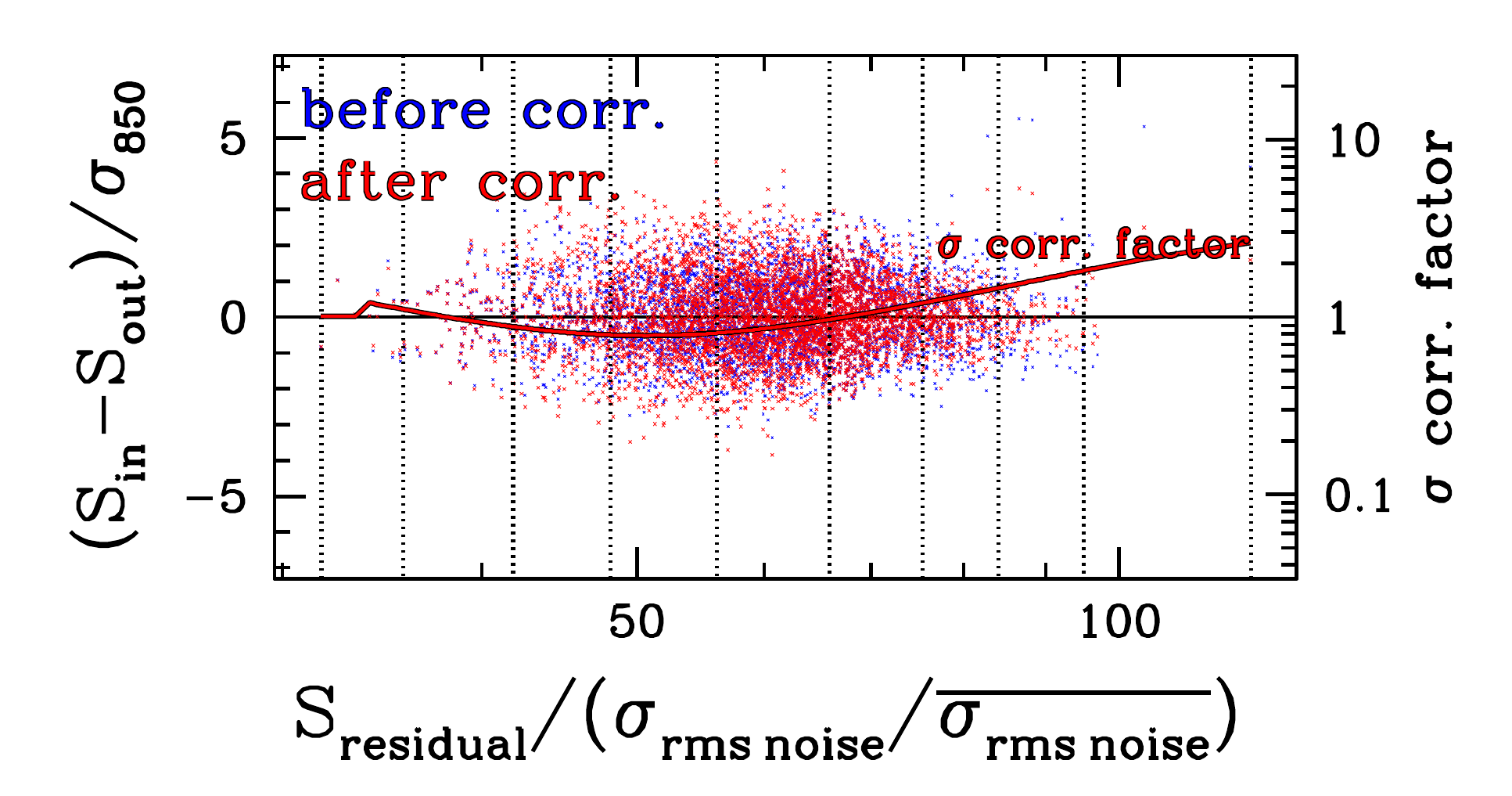}
    \includegraphics[height=2.6cm, trim=0 1cm 0 0]{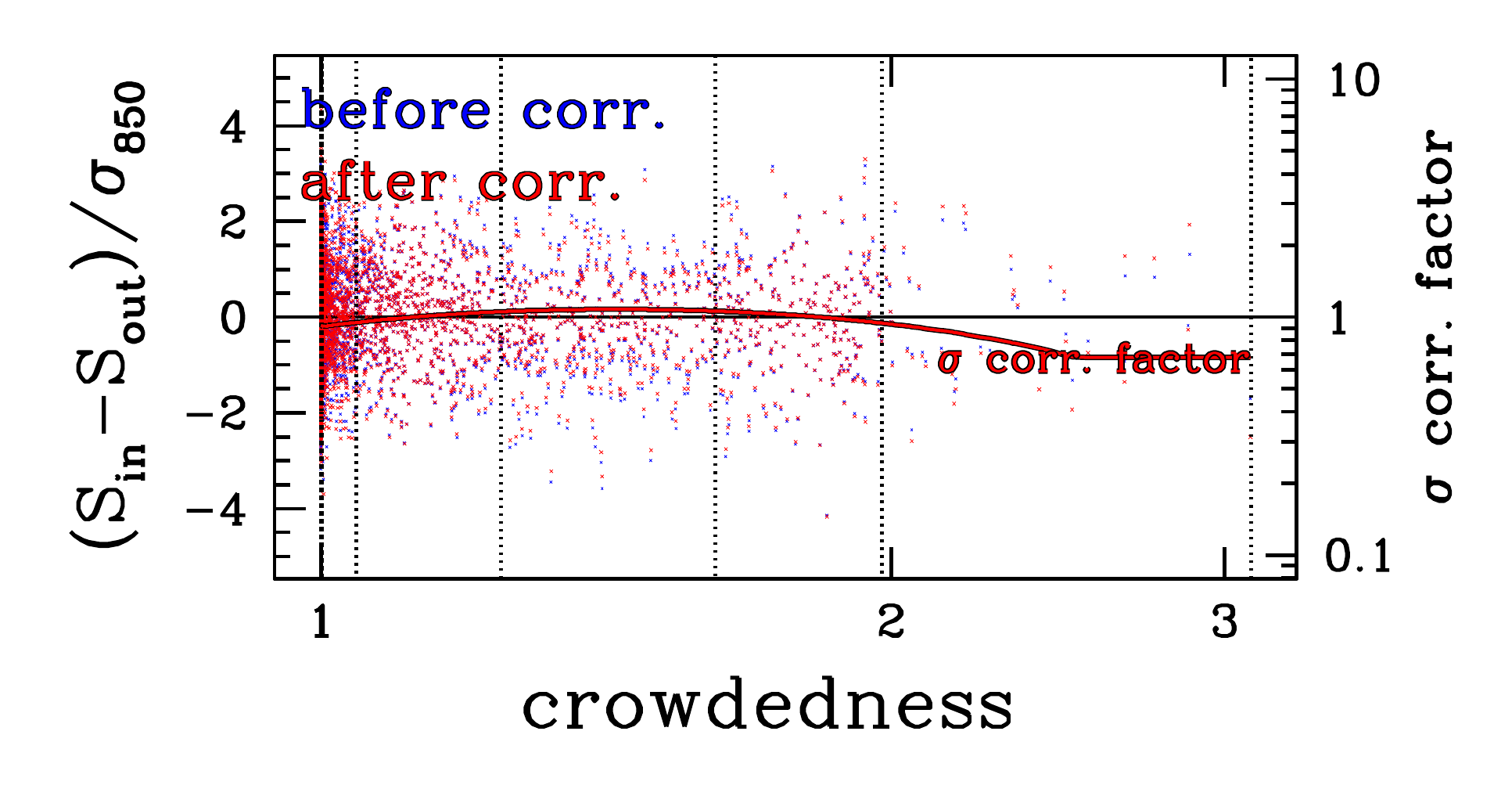}
    \end{subfigure}

    \begin{subfigure}[b]{\textwidth}\centering
    \includegraphics[height=2.6cm, trim=0 1cm -1.8cm 0]{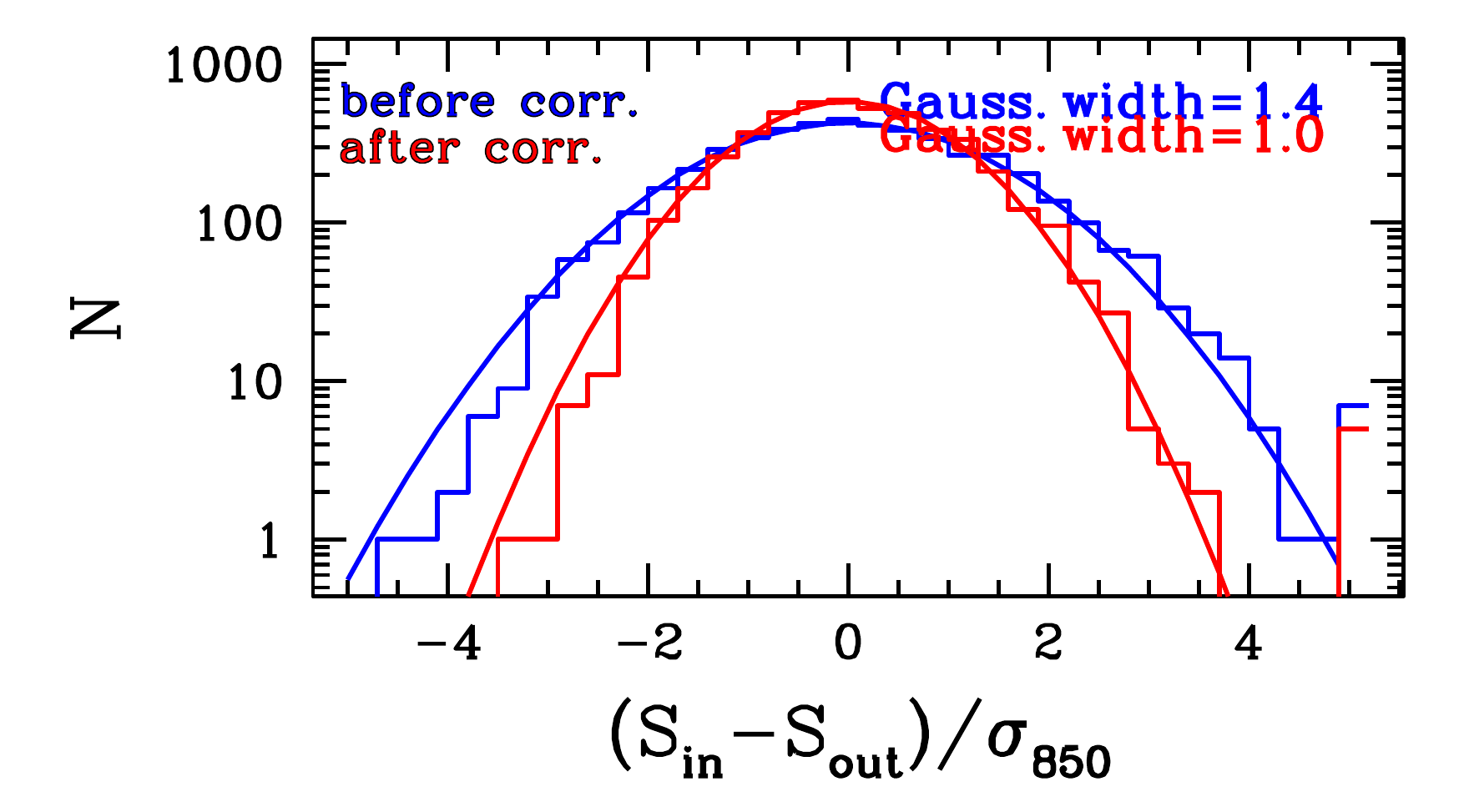}
    \includegraphics[height=2.6cm, trim=0 1cm -1.8cm 0]{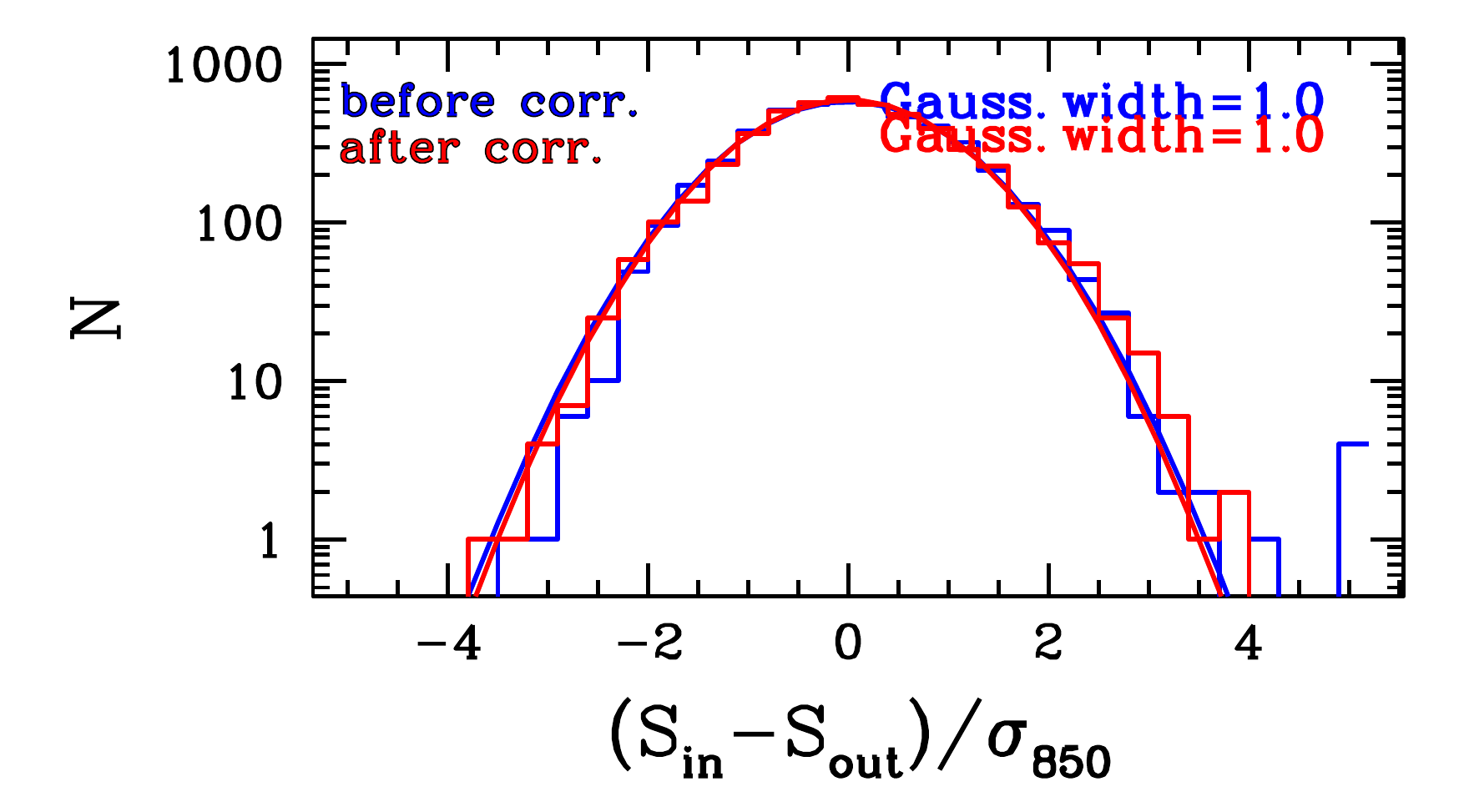}
    \includegraphics[height=2.6cm, trim=0 1cm -1.8cm 0]{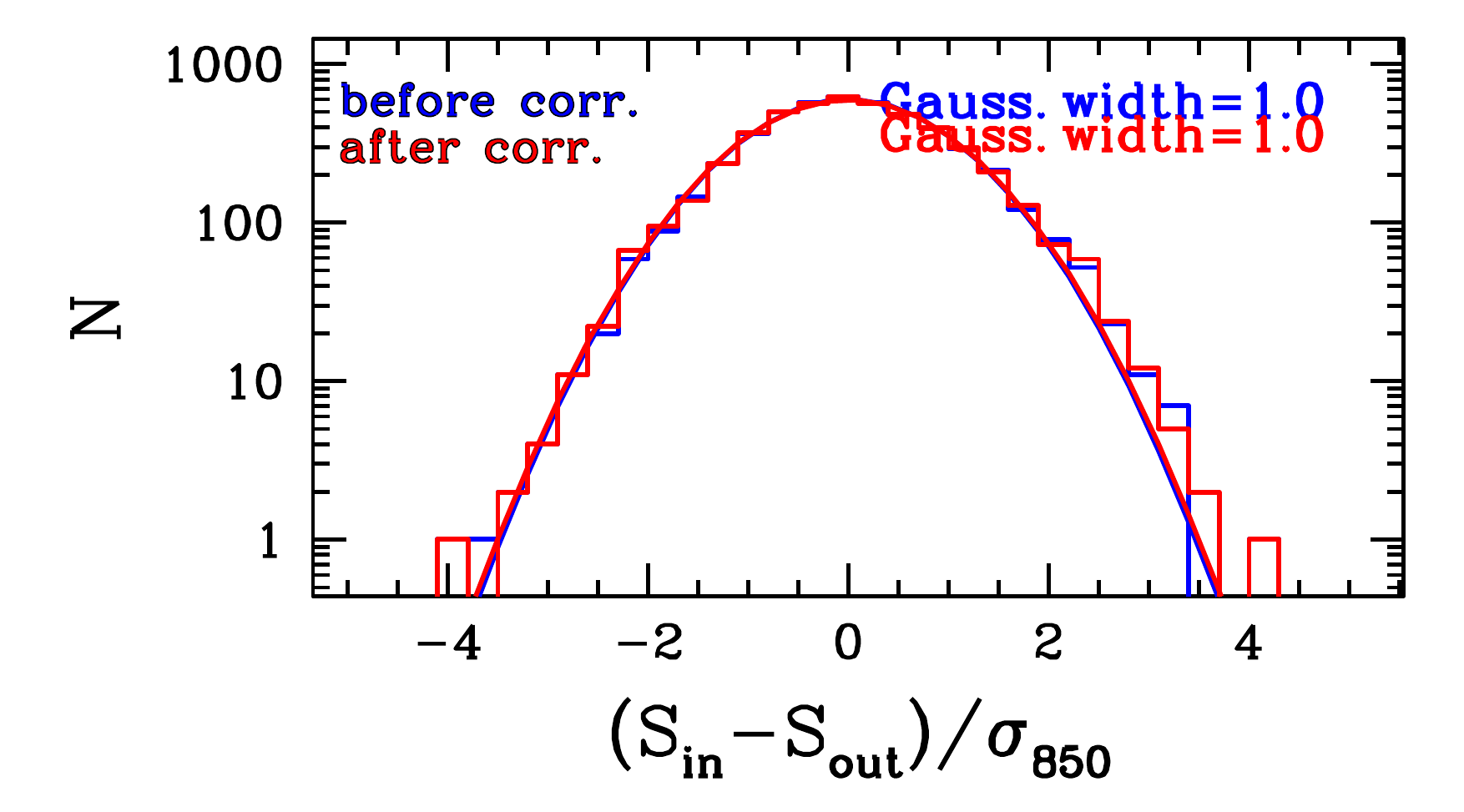}
    \end{subfigure}

    \begin{subfigure}[b]{\textwidth}\centering
    \includegraphics[height=2.6cm, trim=0 1cm -1.8cm 0]{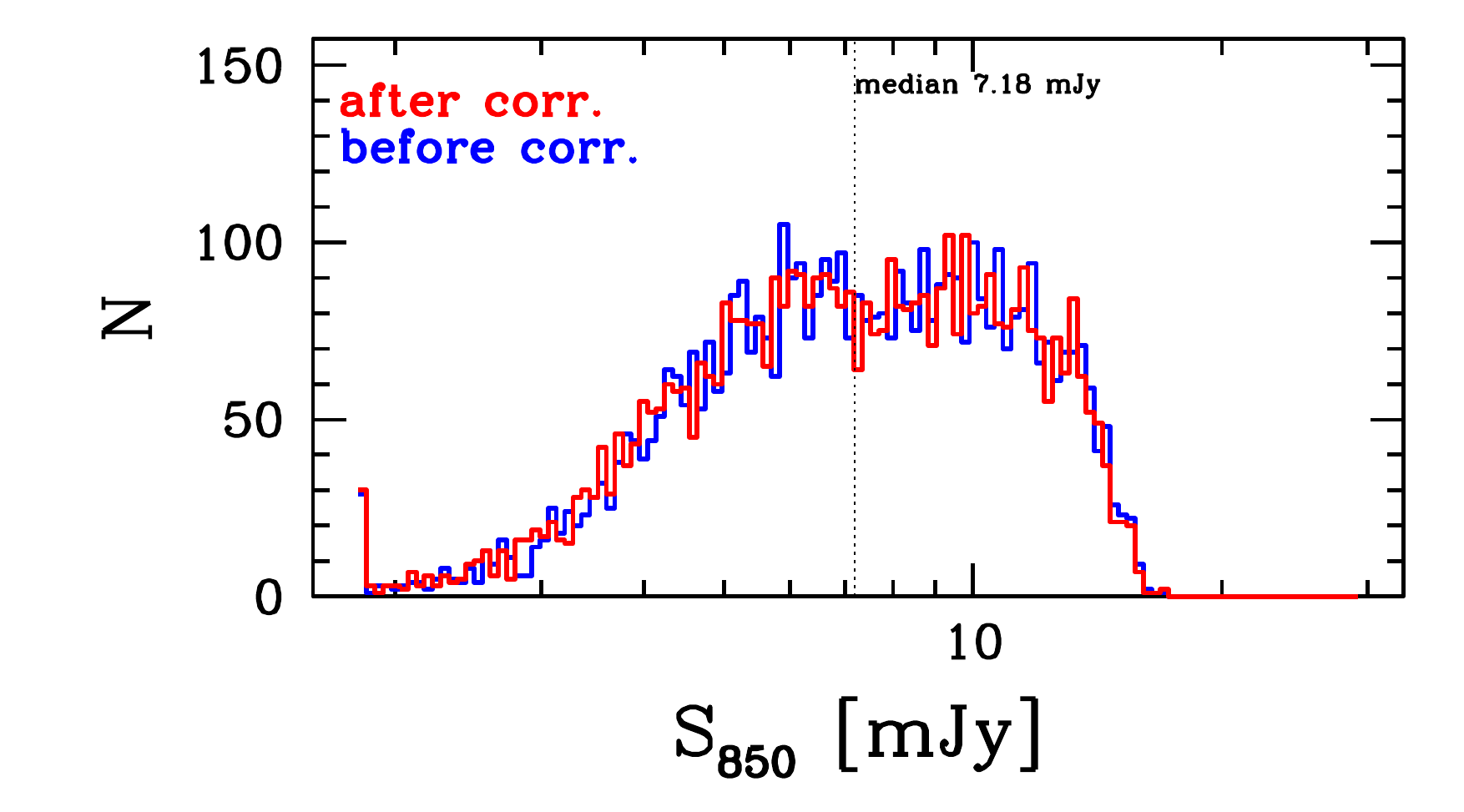}
    \includegraphics[height=2.6cm, trim=0 1cm -1.8cm 0]{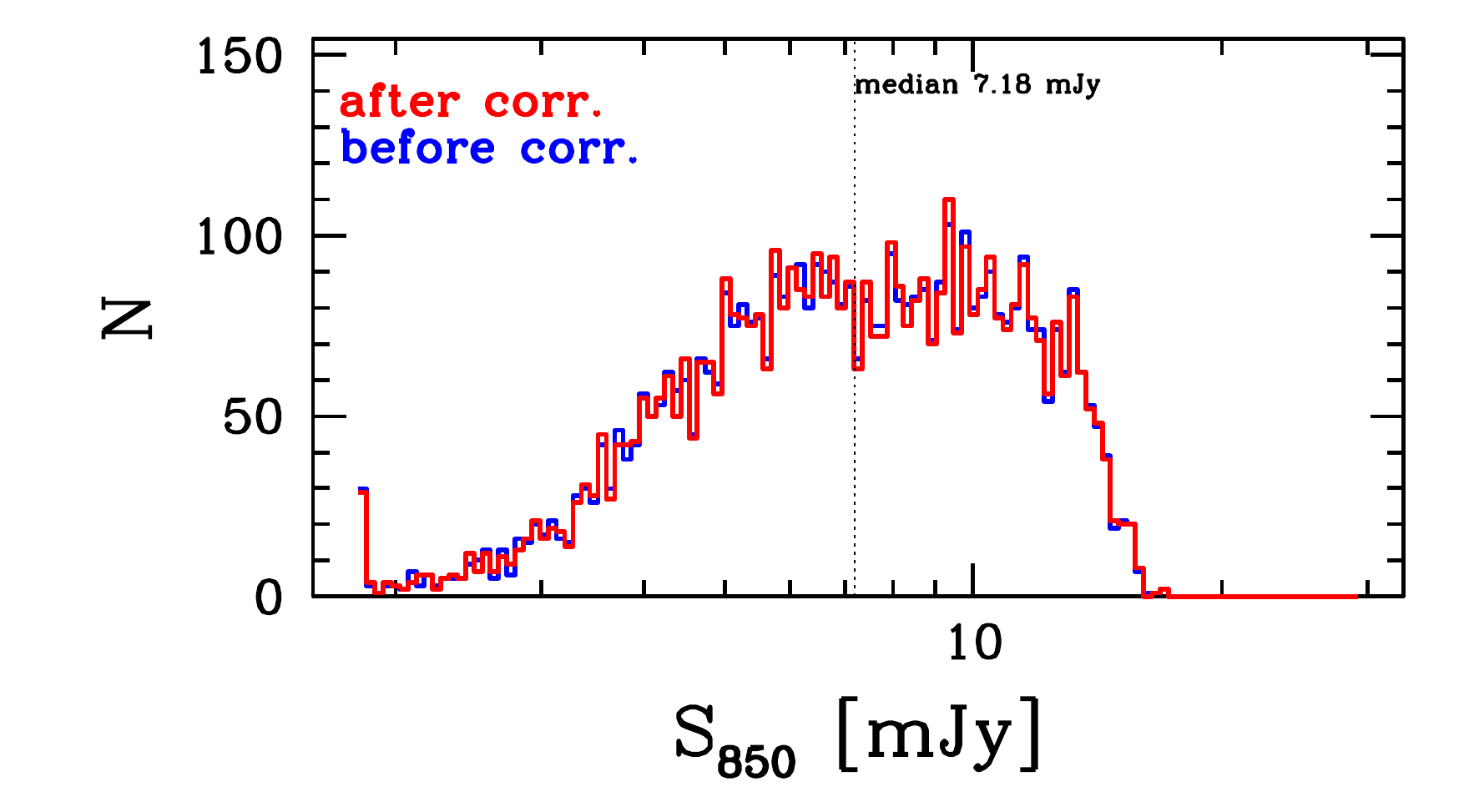}
    \includegraphics[height=2.6cm, trim=0 1cm -1.8cm 0]{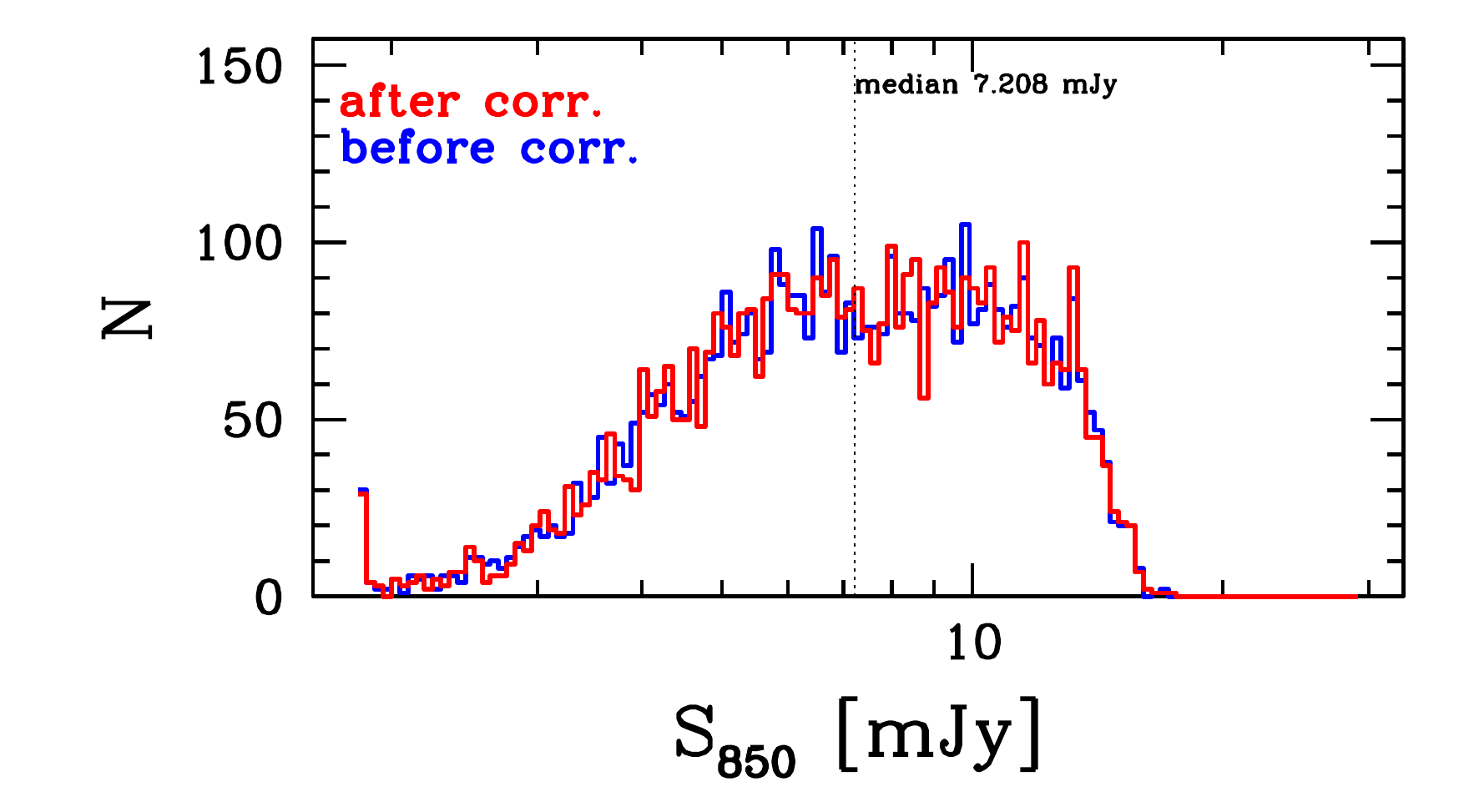}
    \end{subfigure}

    \begin{subfigure}[b]{\textwidth}\centering
    \includegraphics[height=2.6cm, trim=0 1cm -1.8cm 0]{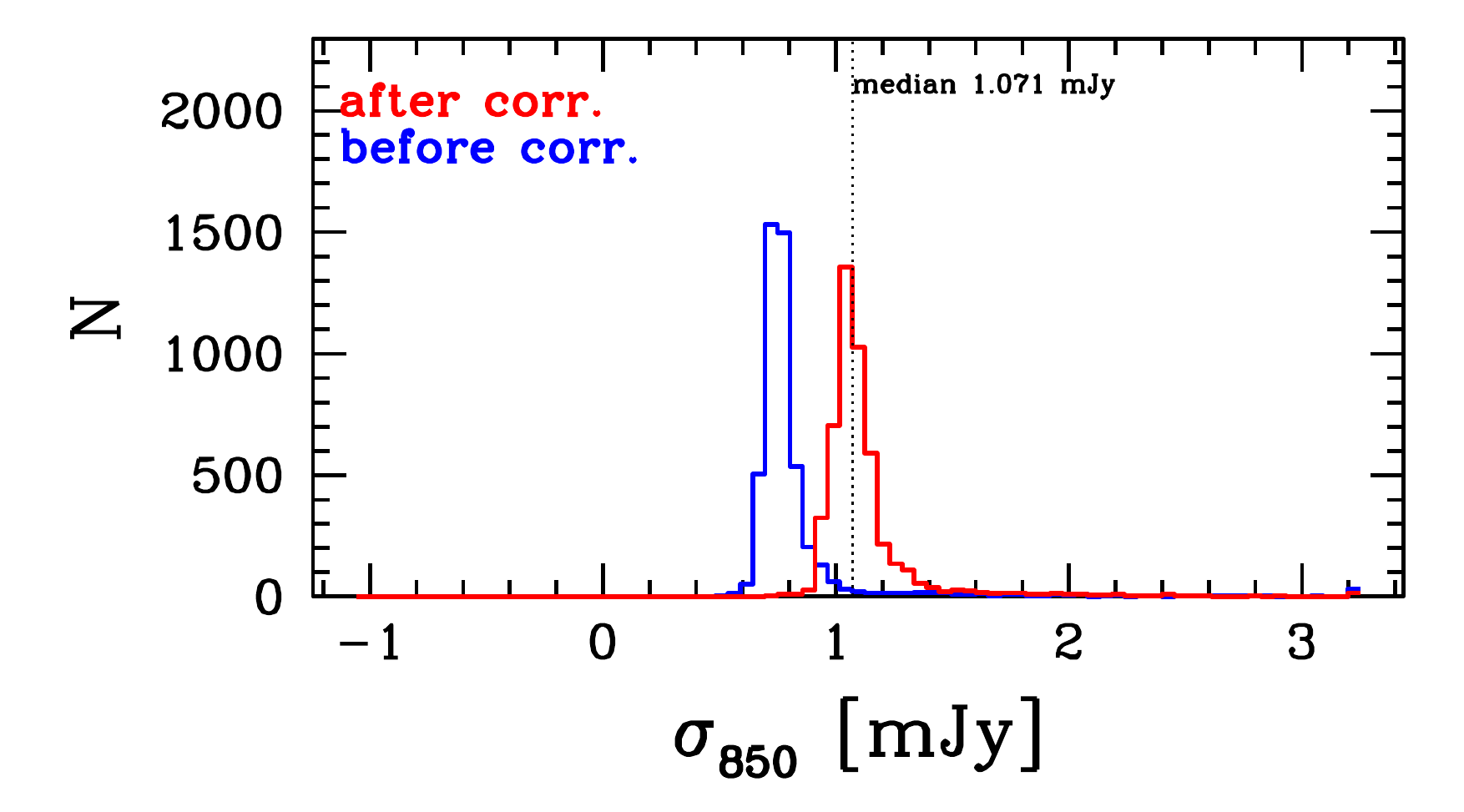}
    \includegraphics[height=2.6cm, trim=0 1cm -1.8cm 0]{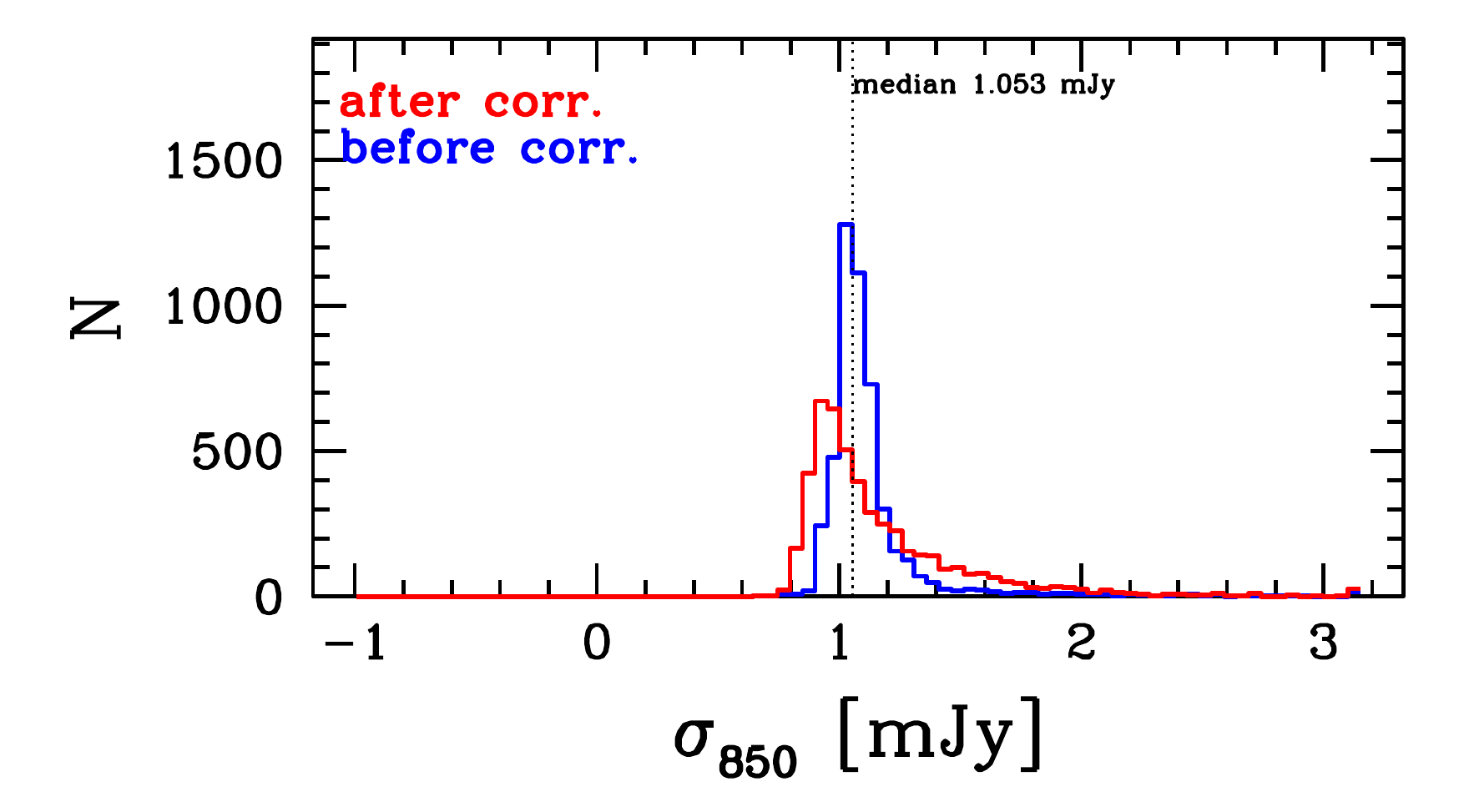}
    \includegraphics[height=2.6cm, trim=0 1cm -1.8cm 0]{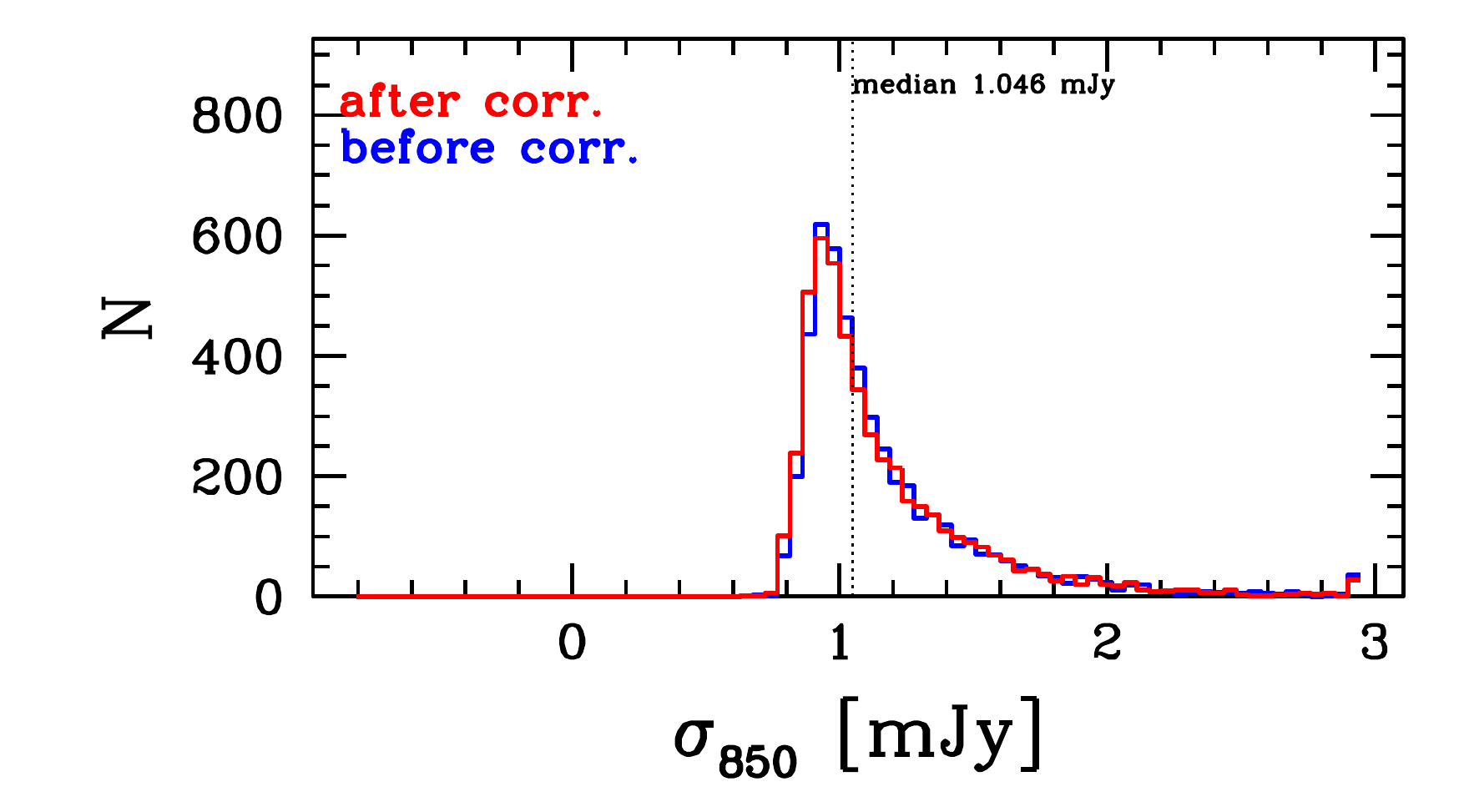}
    \end{subfigure}

\caption{%
    Simulation correction analyses at 850~$\mu$m. See descriptions in the text. 
    \label{Figure_galsim_850_bin}
}
\end{figure}

\begin{figure}
    \centering
    
    \begin{subfigure}[b]{\textwidth}\centering
    \includegraphics[height=2.6cm, trim=0 1cm 0 0]{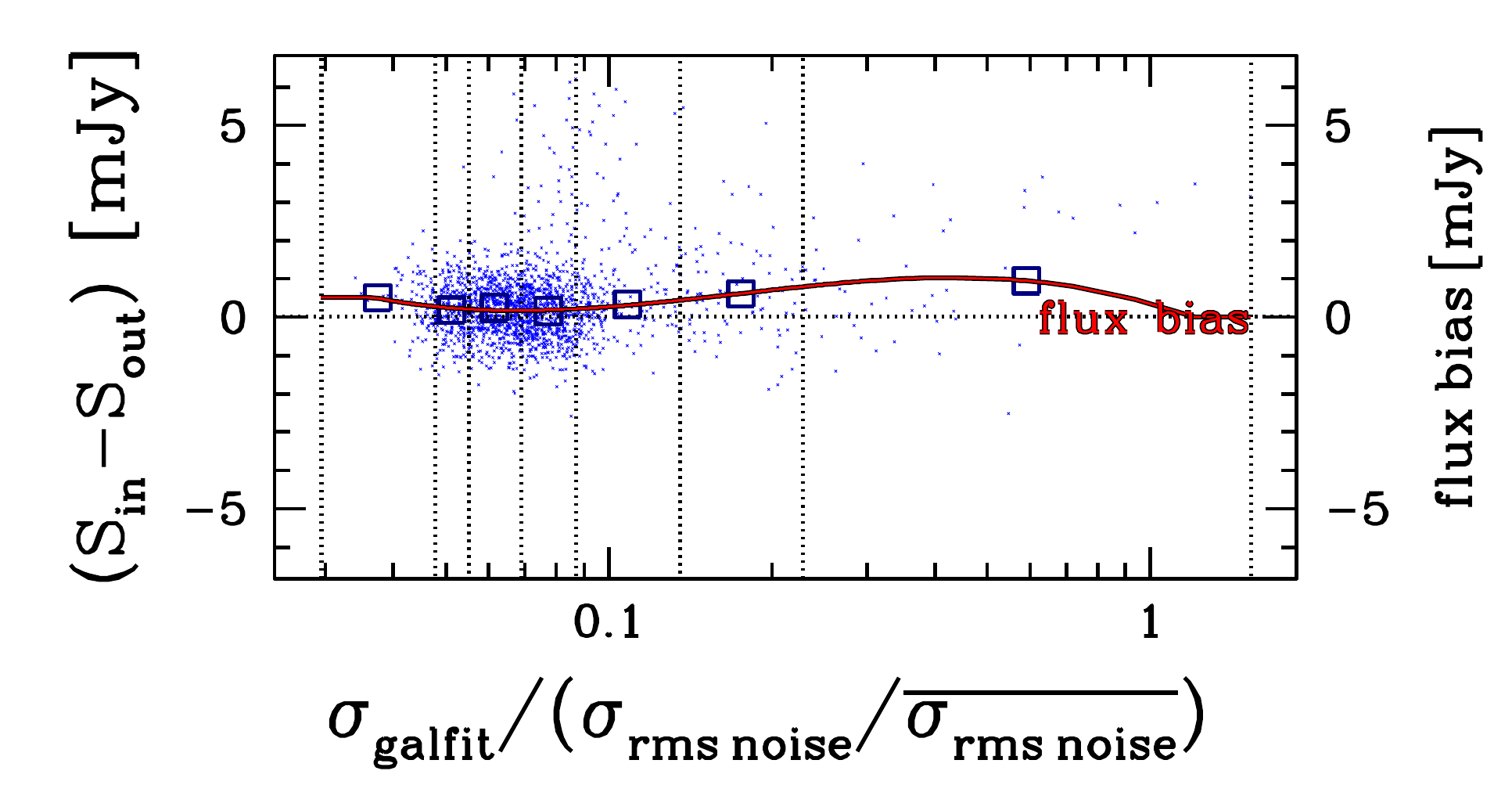}
    \includegraphics[height=2.6cm, trim=0 1cm 0 0]{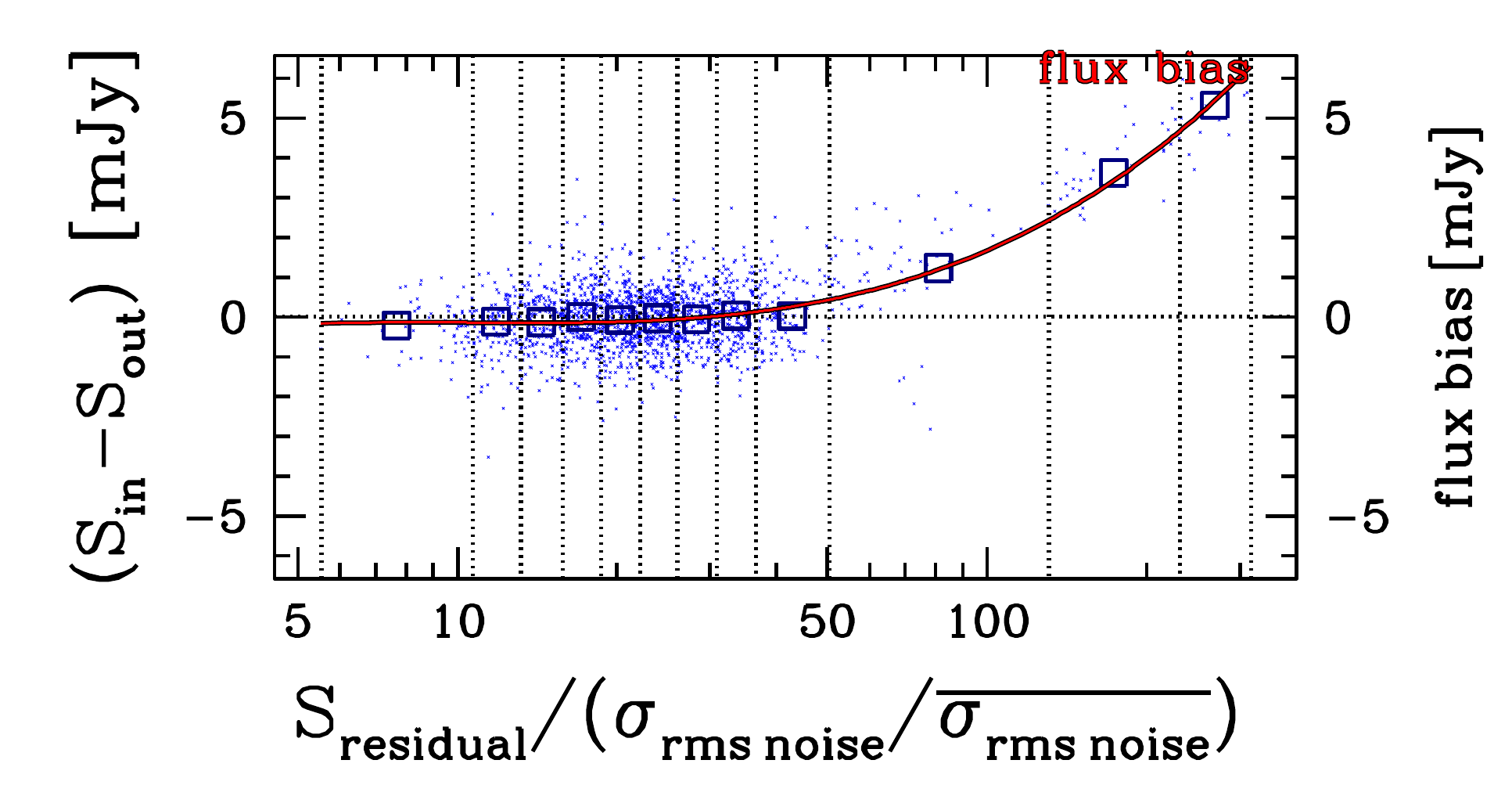}
    \includegraphics[height=2.6cm, trim=0 1cm 0 0]{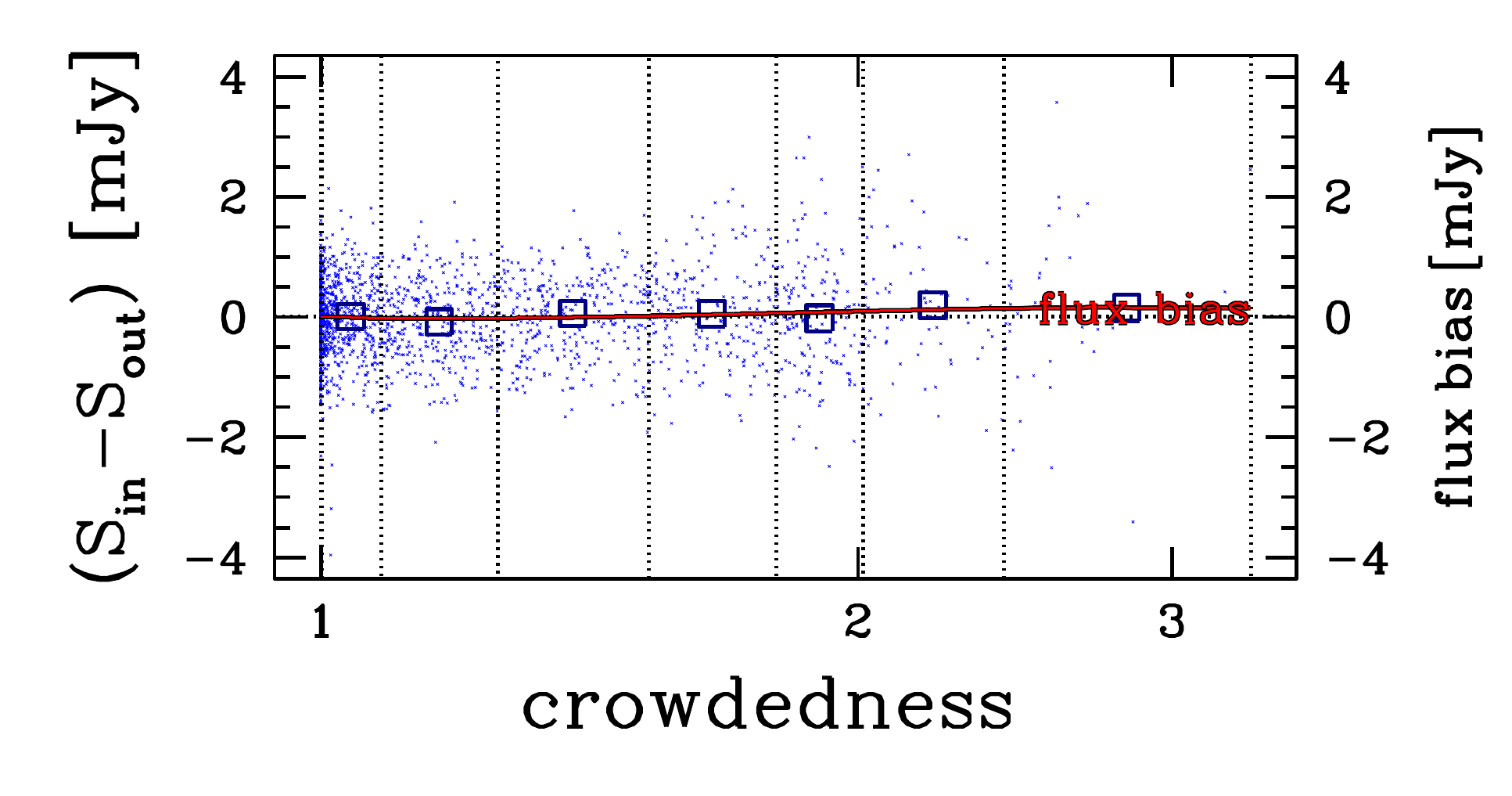}
    \end{subfigure}
    
    \end{figure}
    \begin{figure}\ContinuedFloat
    
    \begin{subfigure}[b]{\textwidth}\centering
    \includegraphics[height=2.6cm, trim=0 1cm 0 0]{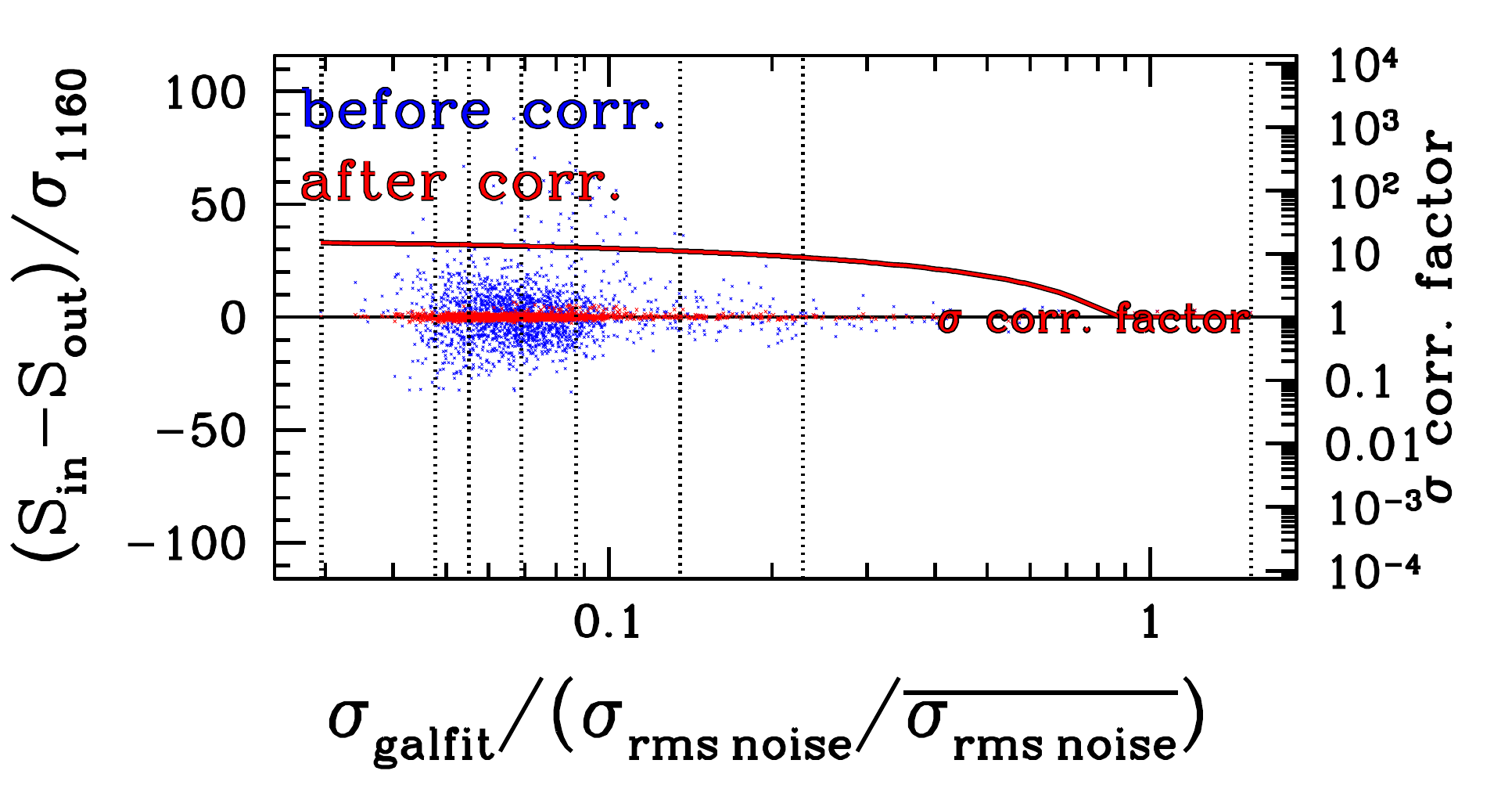}
    \includegraphics[height=2.6cm, trim=0 1cm 0 0]{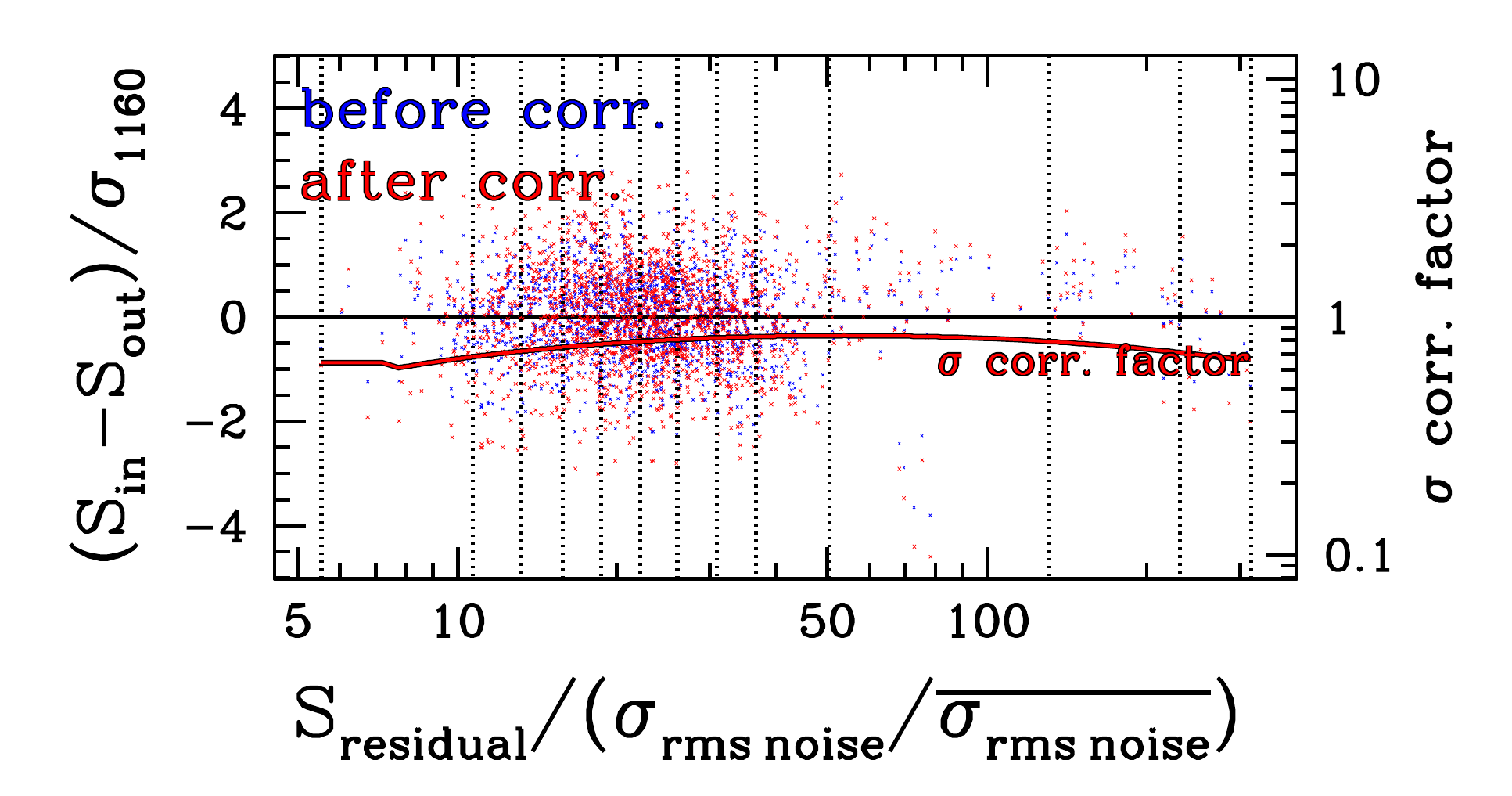}
    \includegraphics[height=2.6cm, trim=0 1cm 0 0]{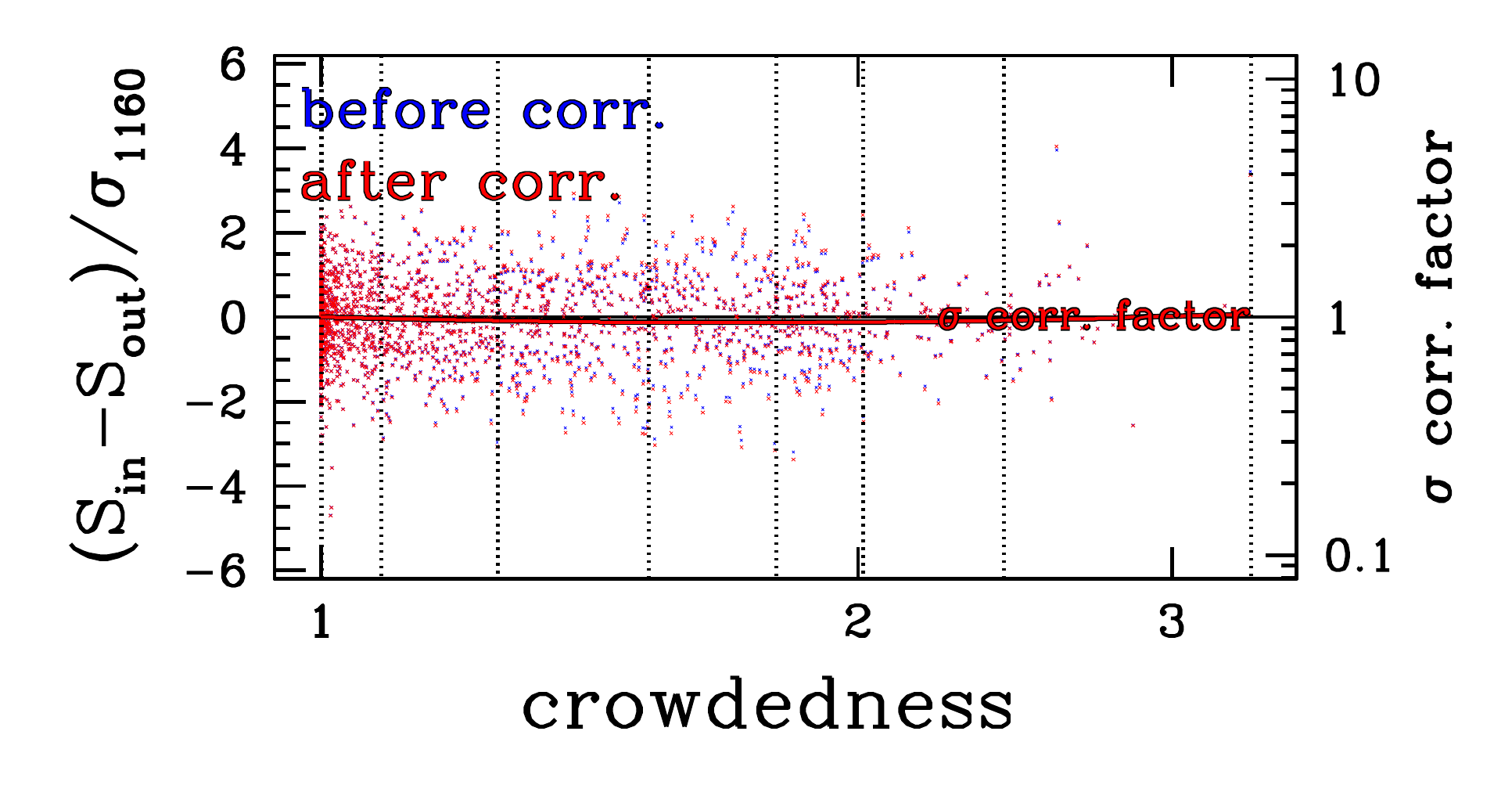}
    \end{subfigure}
    
    \begin{subfigure}[b]{\textwidth}\centering
    \includegraphics[height=2.6cm, trim=0 1cm -1.8cm 0]{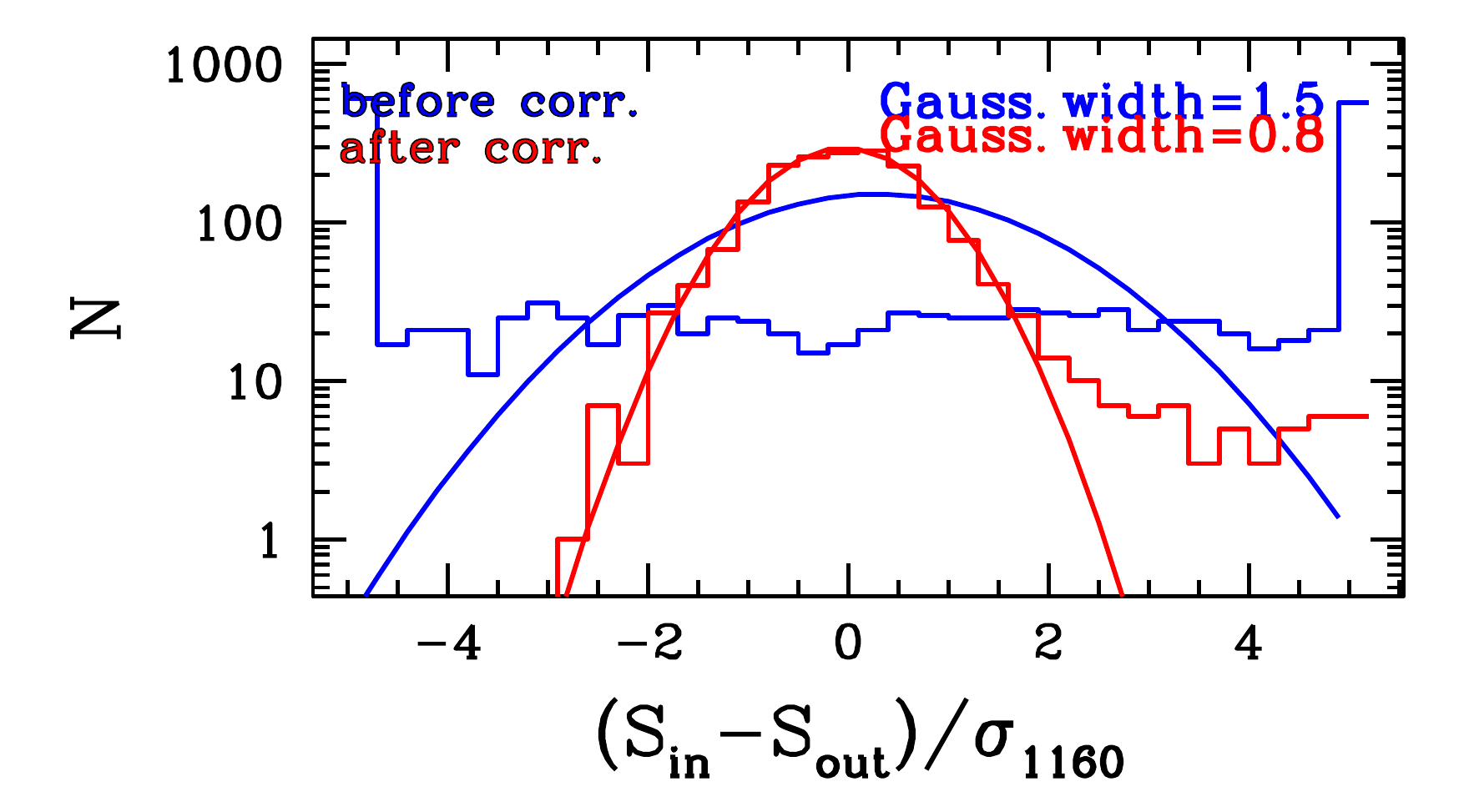}
    \includegraphics[height=2.6cm, trim=0 1cm -1.8cm 0]{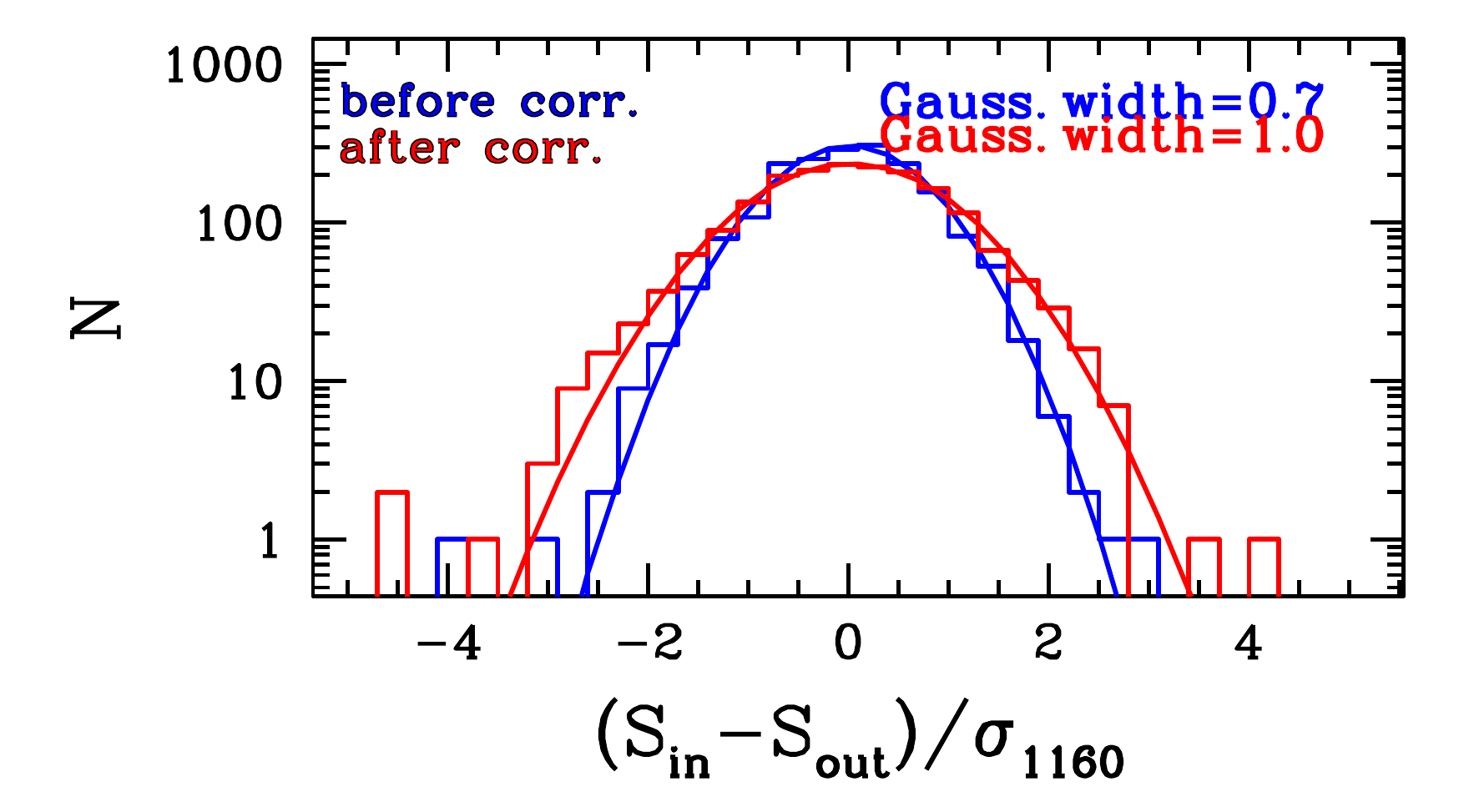}
    \includegraphics[height=2.6cm, trim=0 1cm -1.8cm 0]{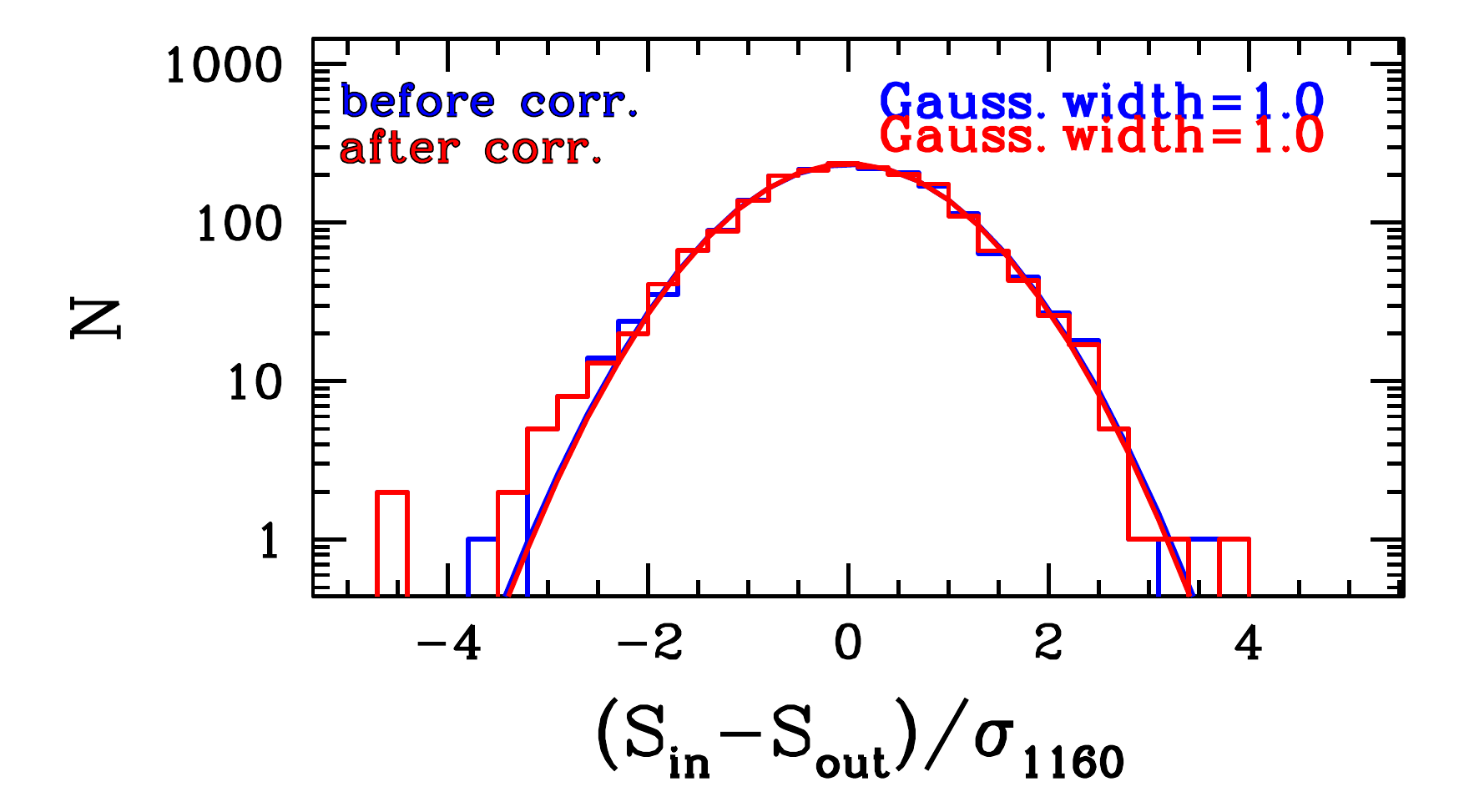}
    \end{subfigure}
    
    \begin{subfigure}[b]{\textwidth}\centering
    \includegraphics[height=2.6cm, trim=0 1cm -1.8cm 0]{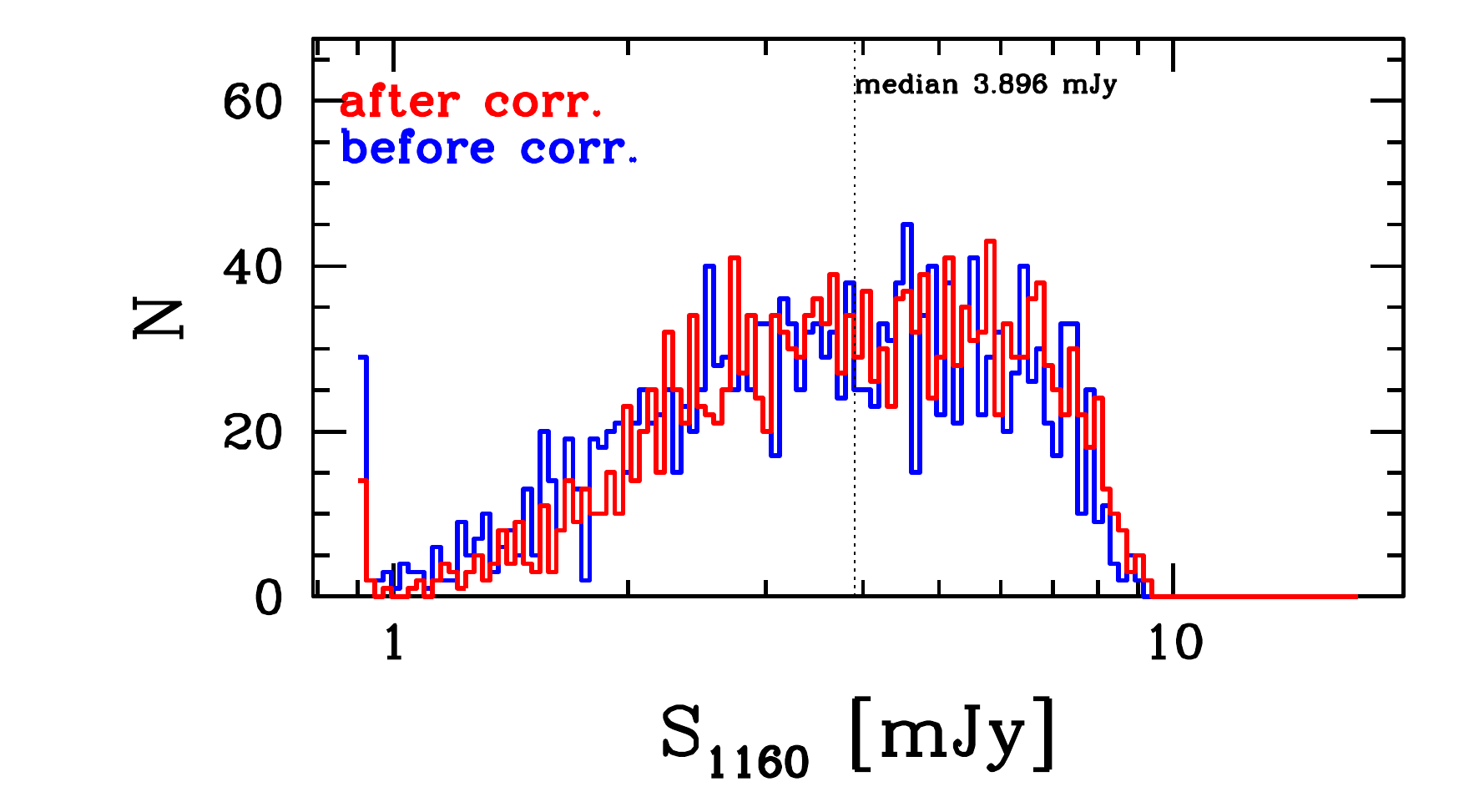}
    \includegraphics[height=2.6cm, trim=0 1cm -1.8cm 0]{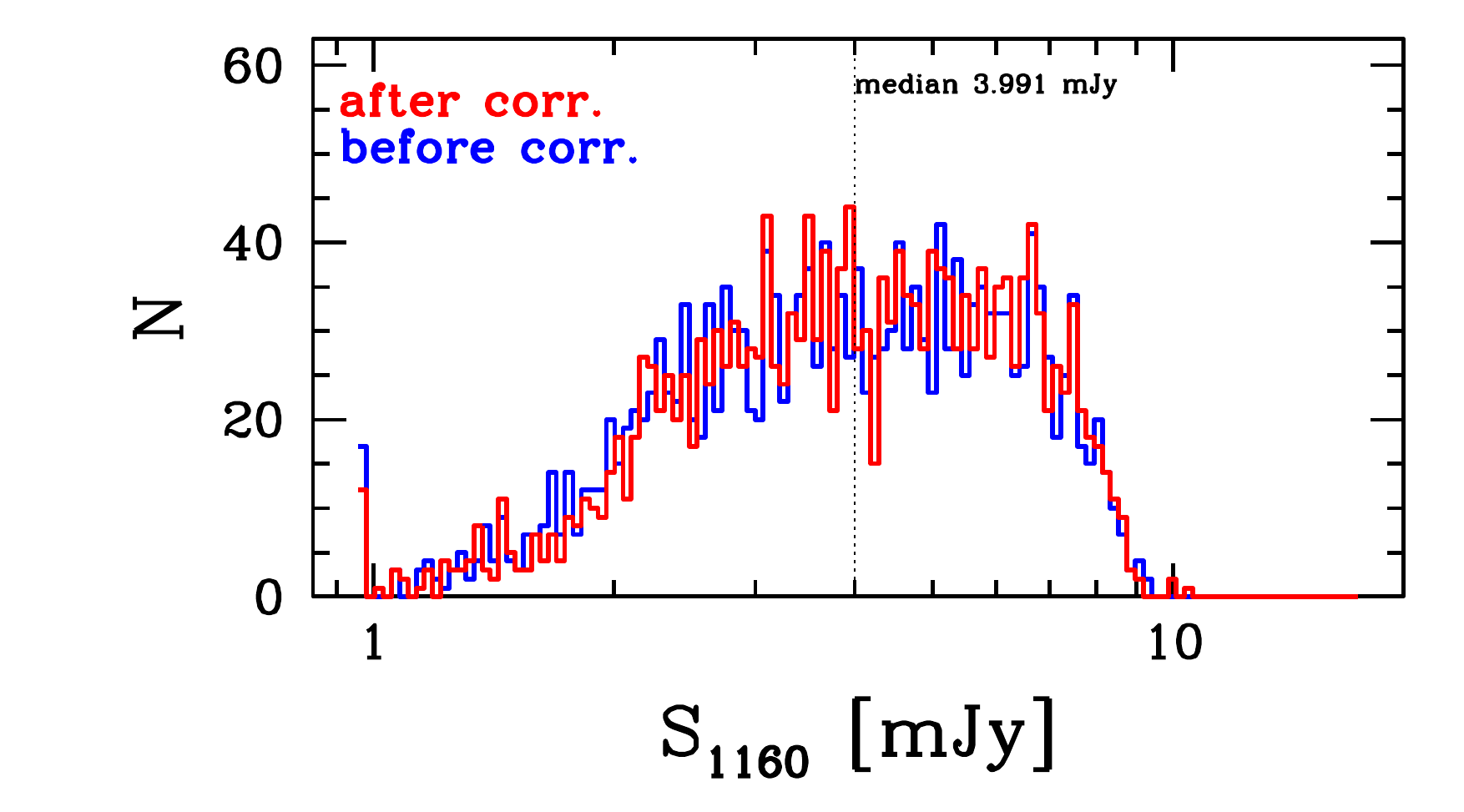}
    \includegraphics[height=2.6cm, trim=0 1cm -1.8cm 0]{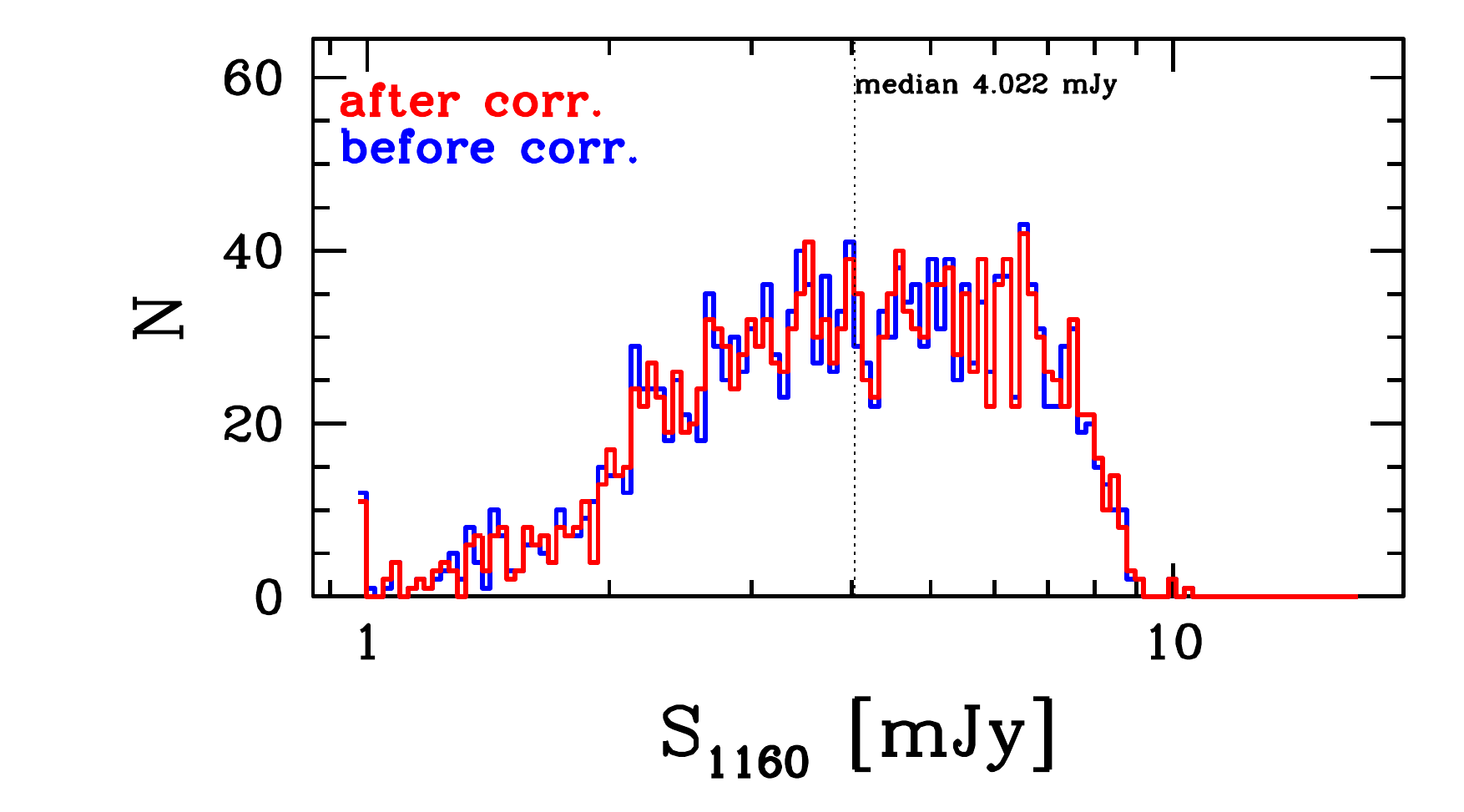}
    \end{subfigure}
    
    \begin{subfigure}[b]{\textwidth}\centering
    \includegraphics[height=2.6cm, trim=0 1cm -1.8cm 0]{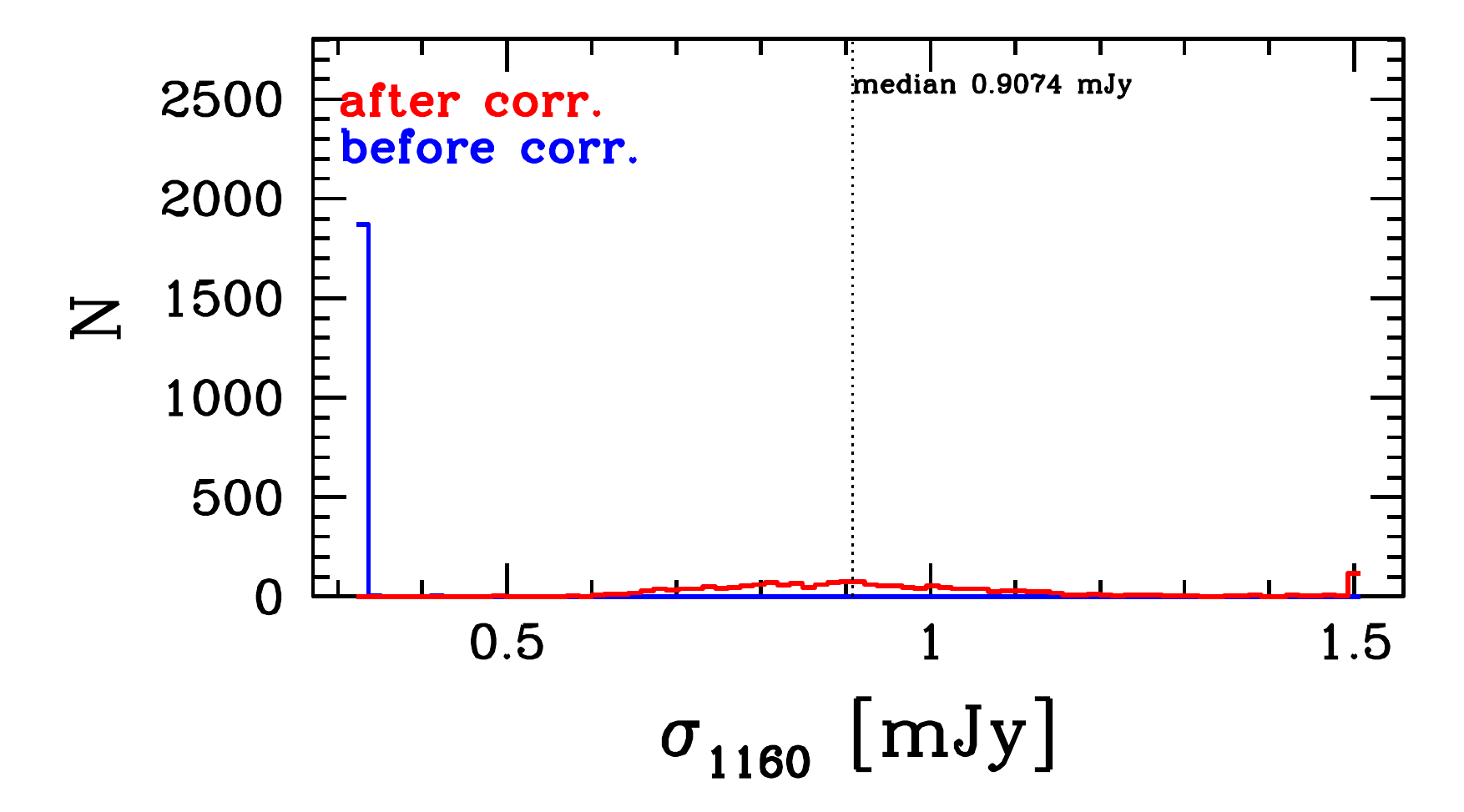}
    \includegraphics[height=2.6cm, trim=0 1cm -1.8cm 0]{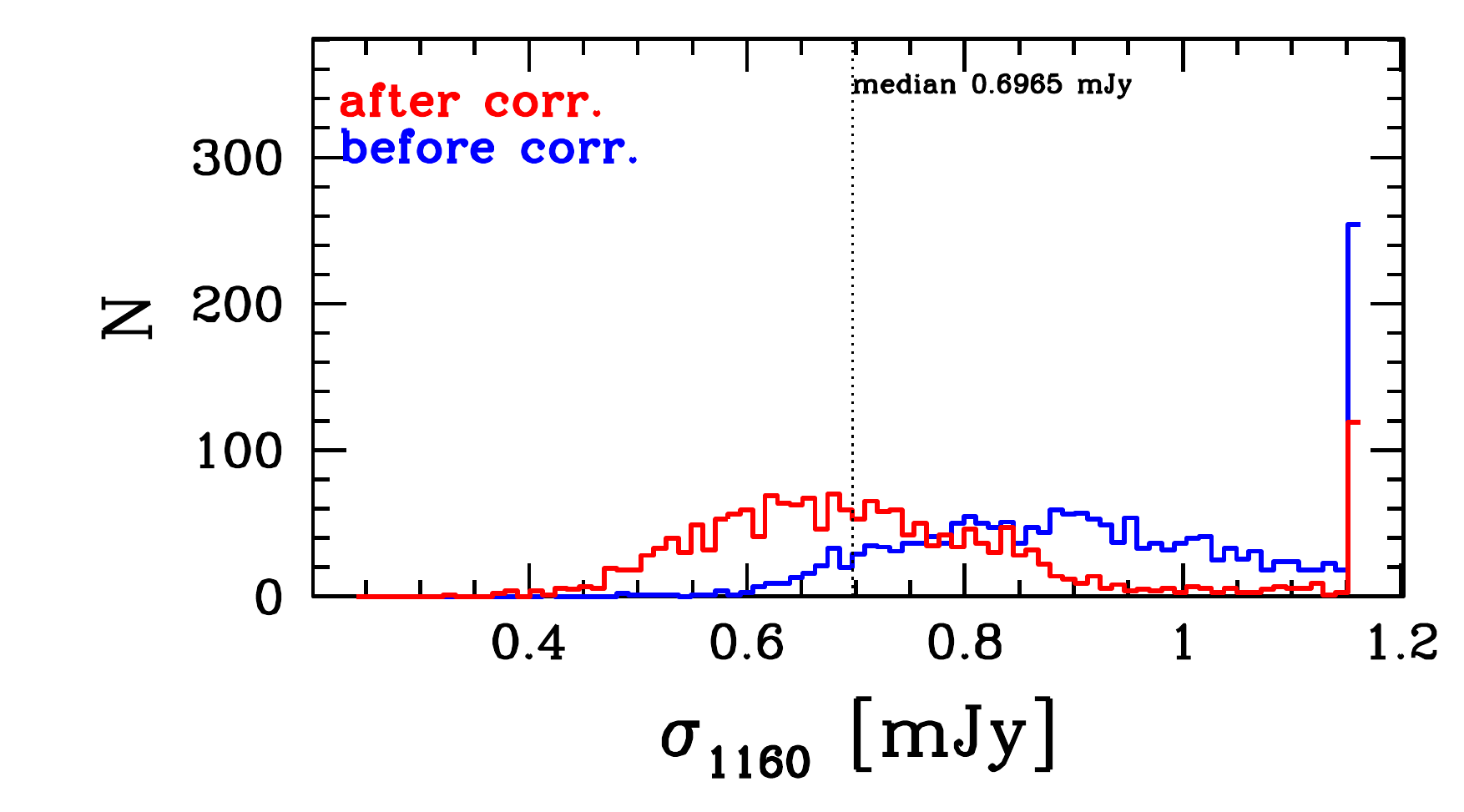}
    \includegraphics[height=2.6cm, trim=0 1cm -1.8cm 0]{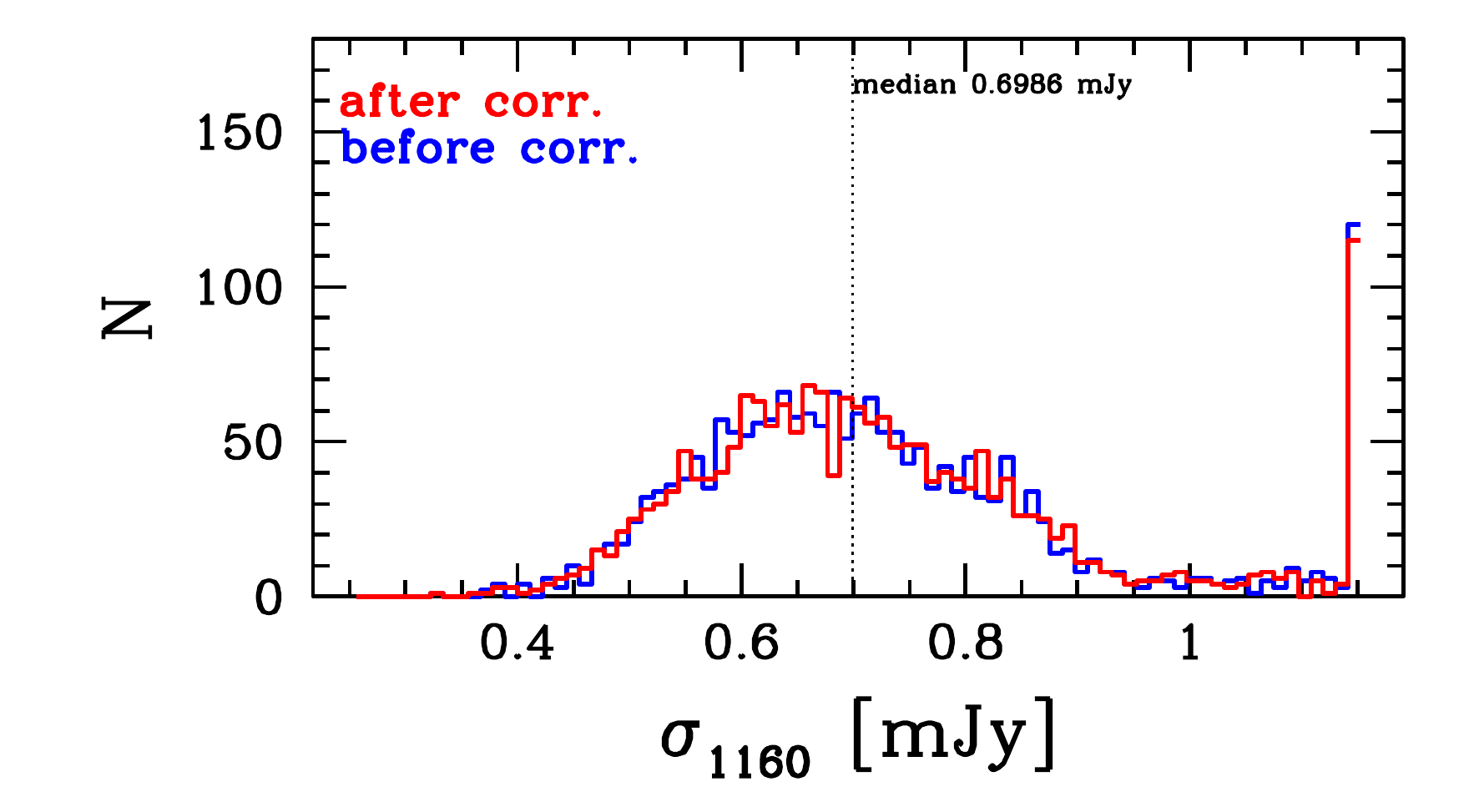}
    \end{subfigure}
    
\caption{%
    Simulation correction analyses at 1.16~mm. See descriptions in the text. 
    \label{Figure_galsim_1160_bin}
}
\end{figure}



\begin{figure}
    \centering
    
    \begin{subfigure}[b]{\textwidth}\centering
    \includegraphics[height=2.6cm, trim=0 1cm 0 0]{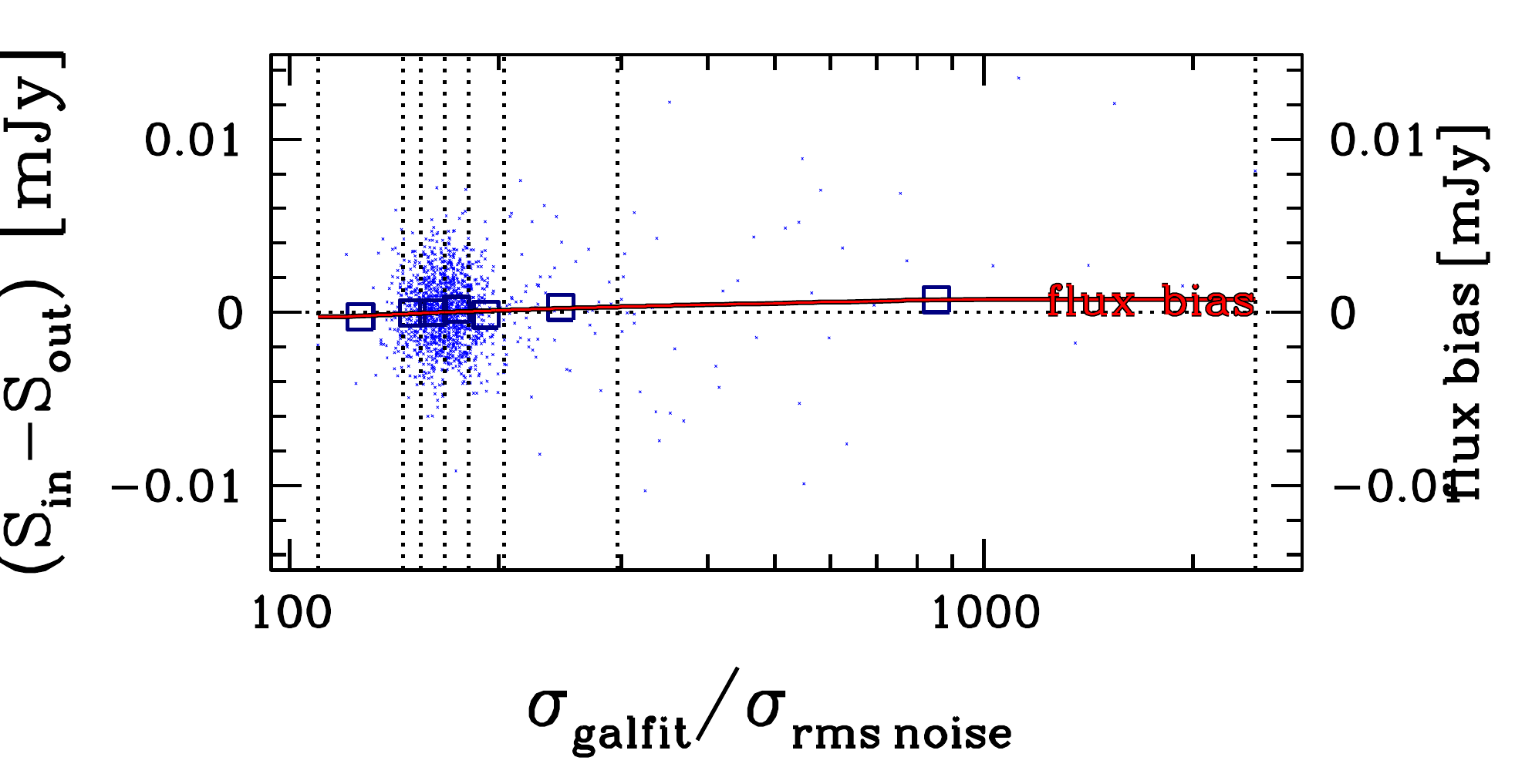}
    \includegraphics[height=2.6cm, trim=0 1cm 0 0]{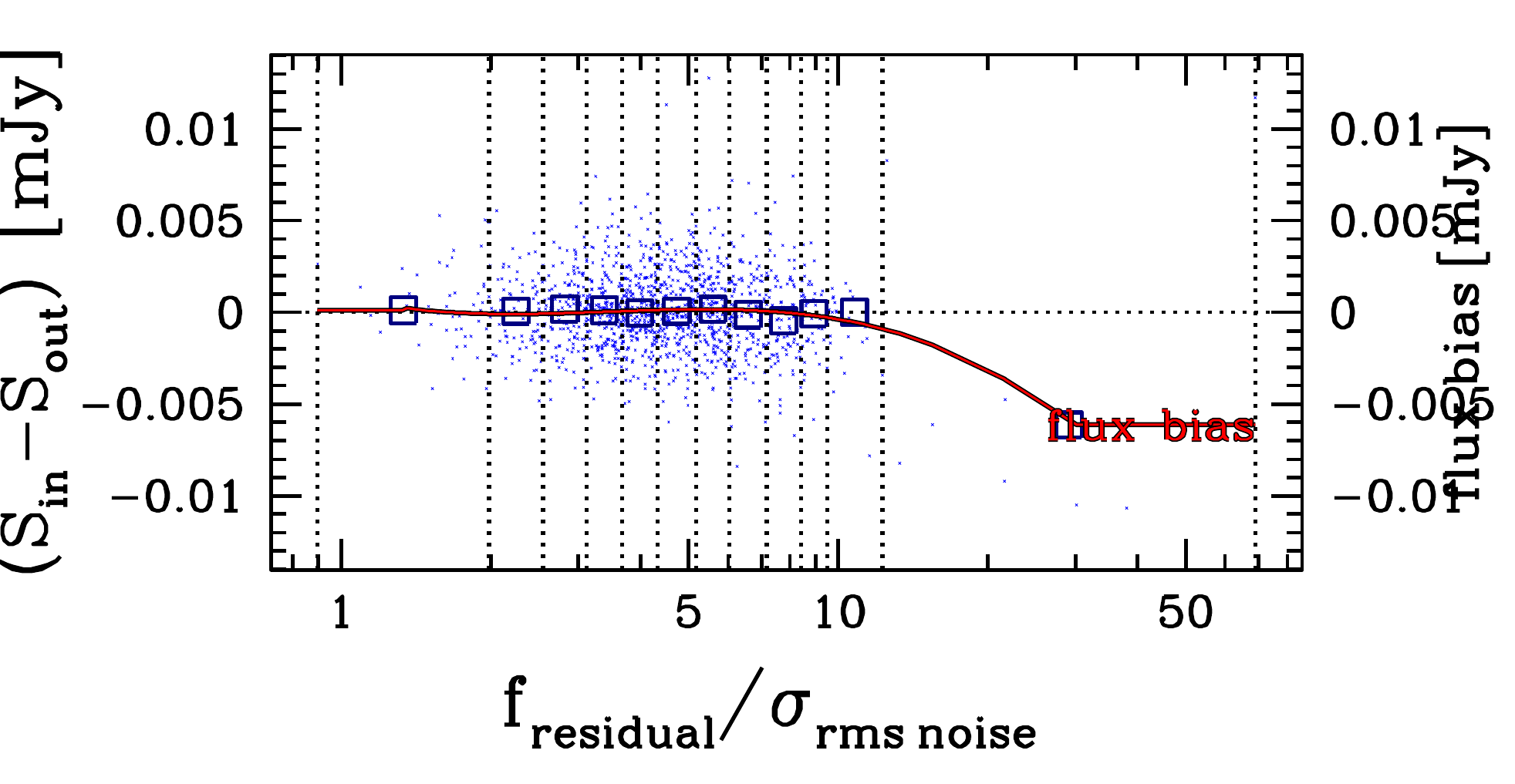}
    \includegraphics[height=2.6cm, trim=0 1cm 0 0]{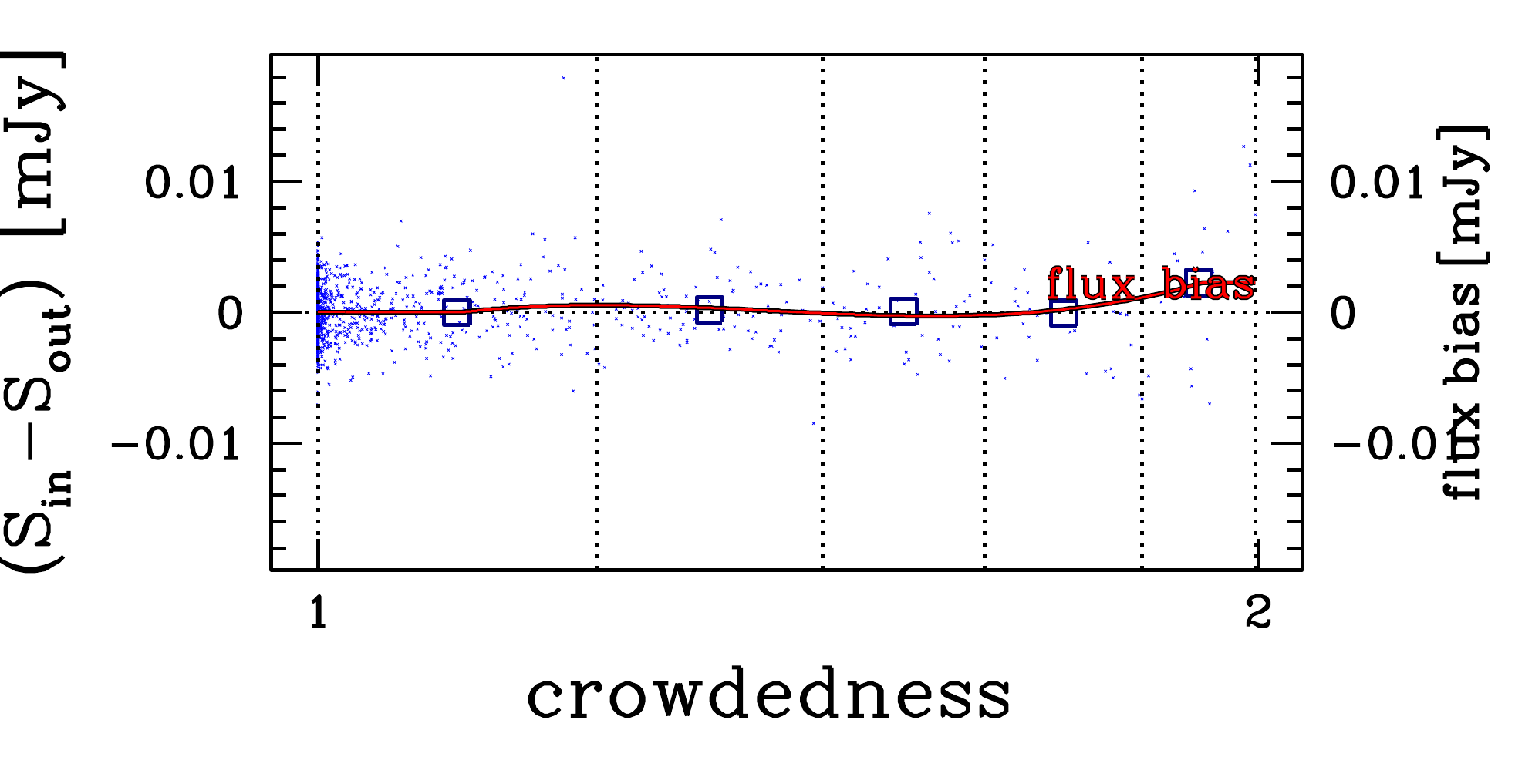}
    \end{subfigure}
    
    \begin{subfigure}[b]{\textwidth}\centering
    \includegraphics[height=2.6cm, trim=0 1cm 0 0]{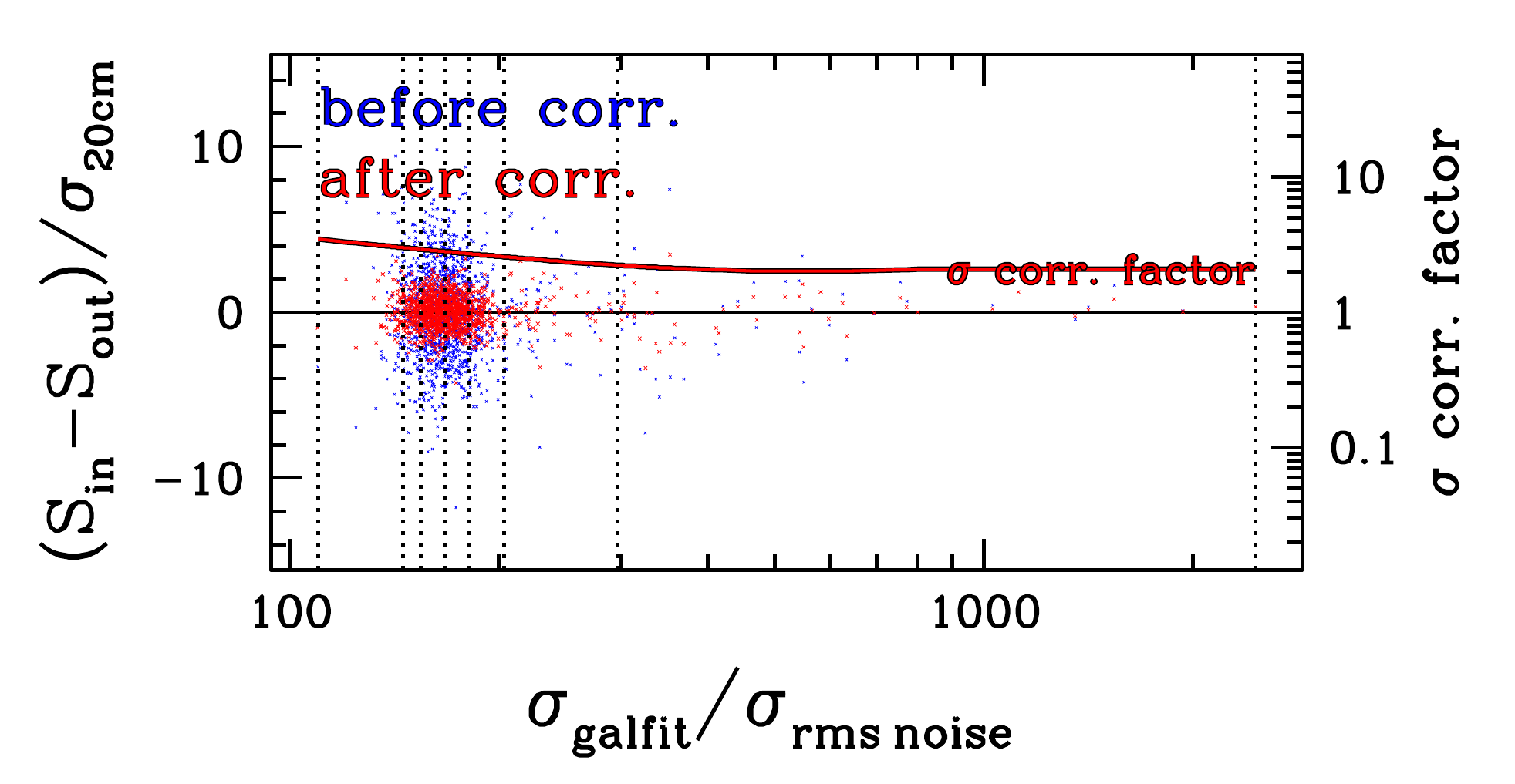}
    \includegraphics[height=2.6cm, trim=0 1cm 0 0]{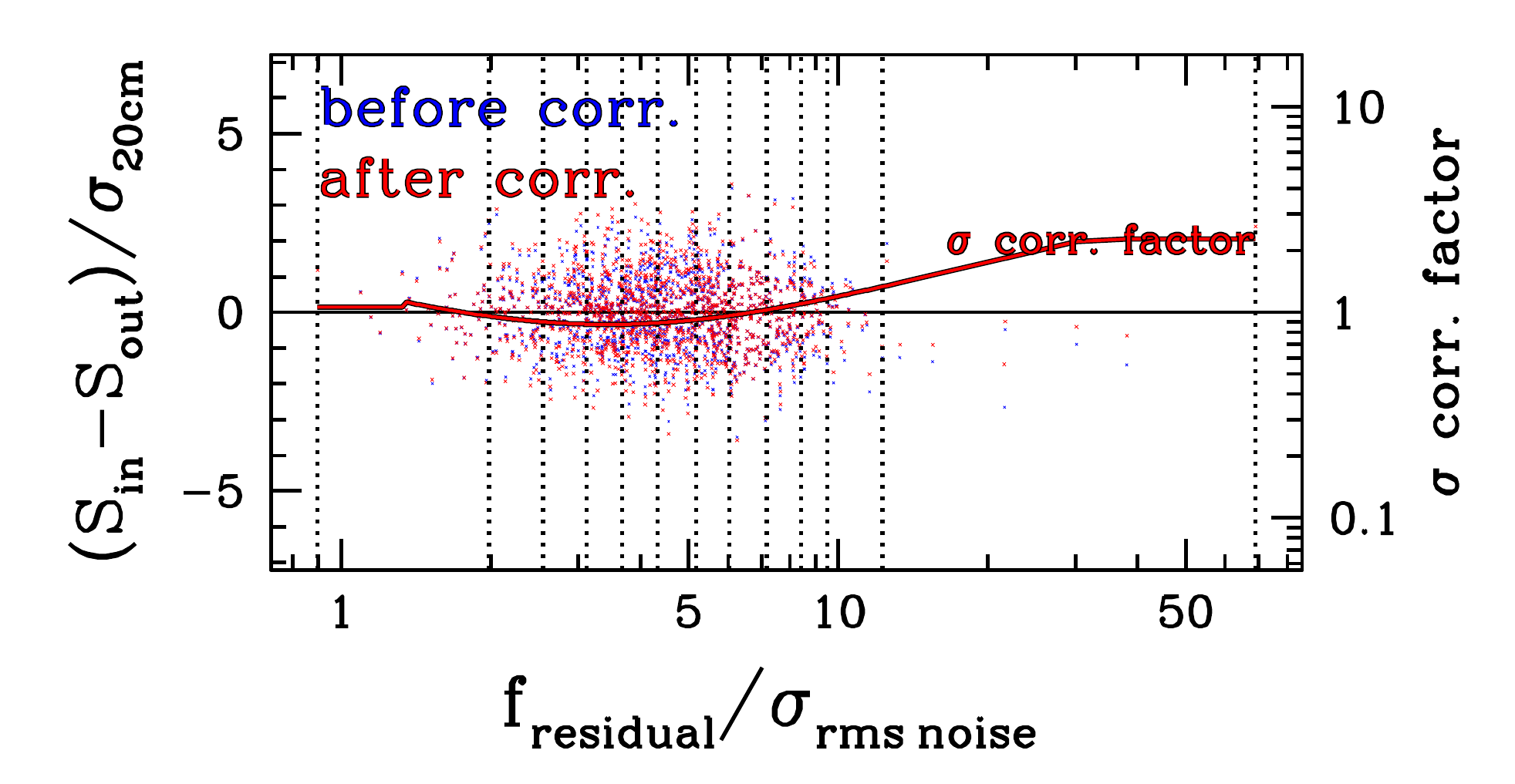}
    \includegraphics[height=2.6cm, trim=0 1cm 0 0]{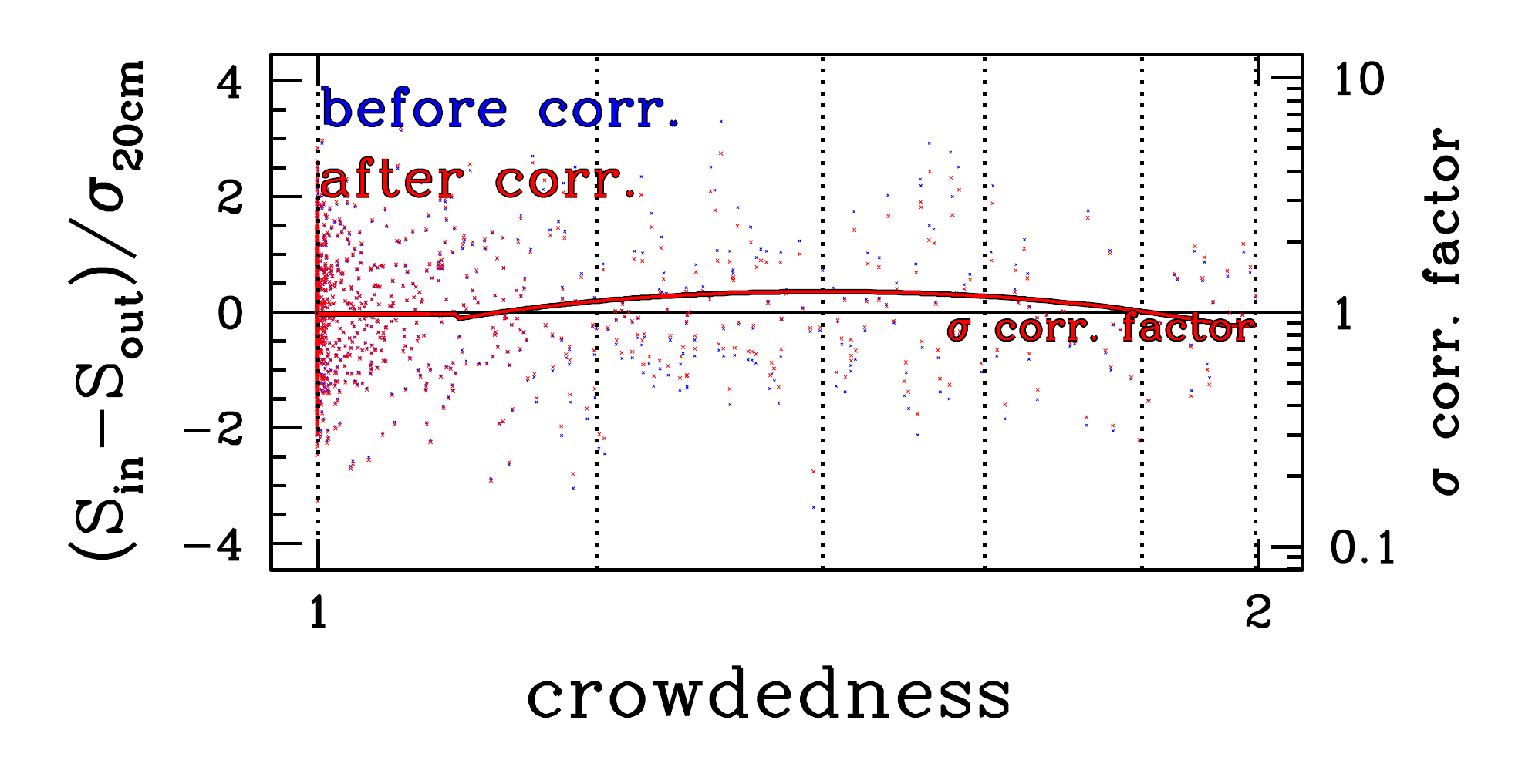}
    \end{subfigure}
    
    \begin{subfigure}[b]{\textwidth}\centering
    \includegraphics[height=2.6cm, trim=0 1cm -1.8cm 0]{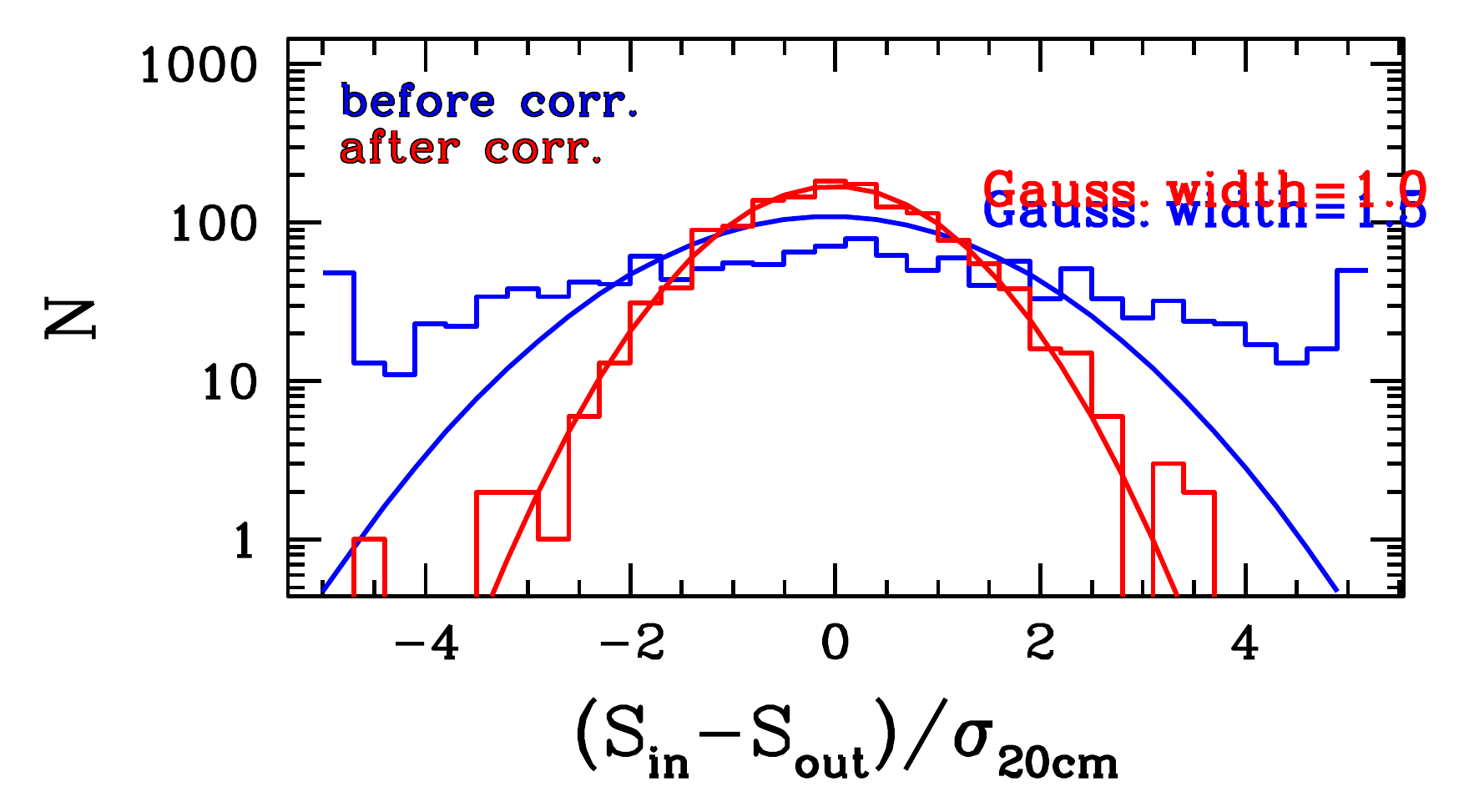}
    \includegraphics[height=2.6cm, trim=0 1cm -1.8cm 0]{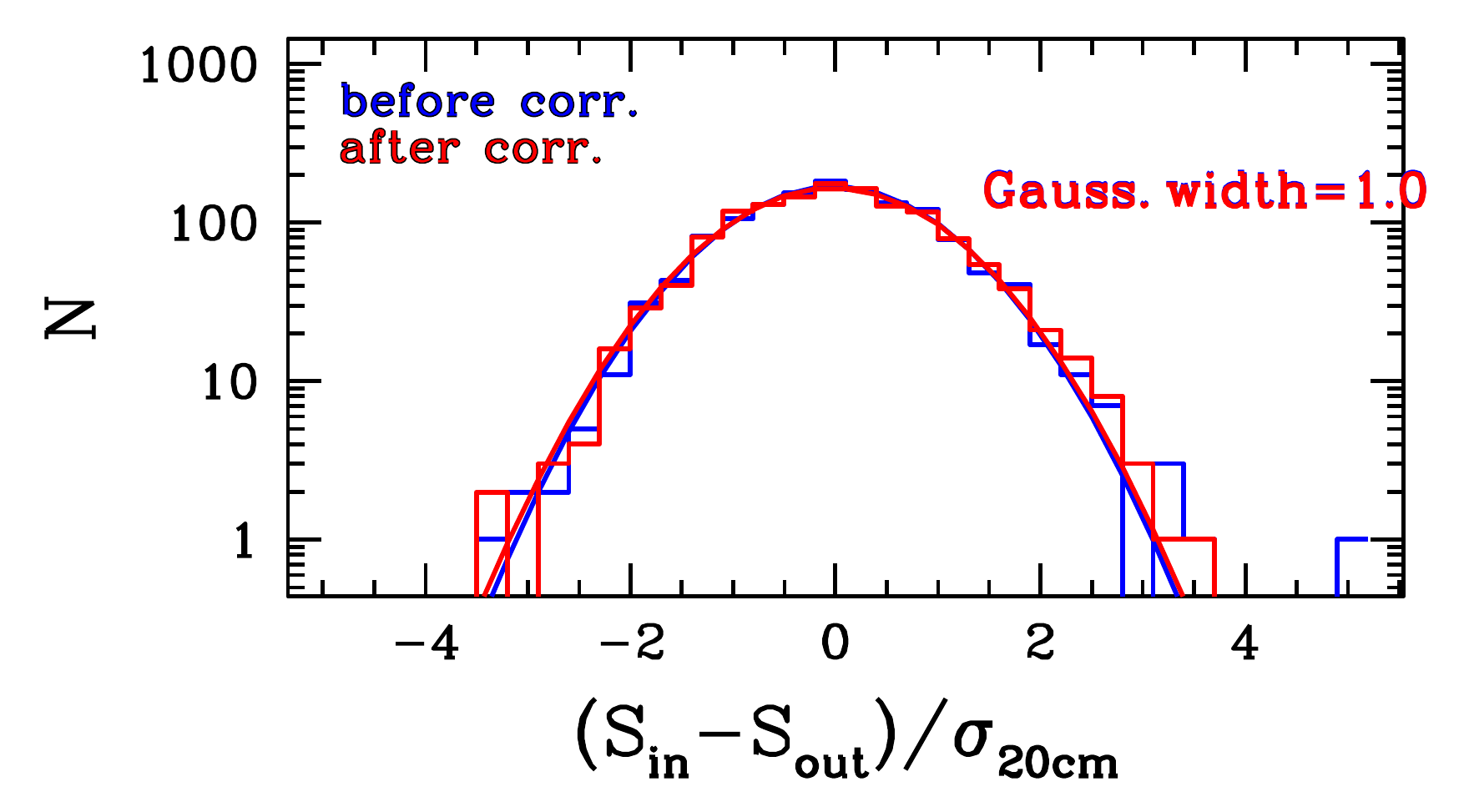}
    \includegraphics[height=2.6cm, trim=0 1cm -1.8cm 0]{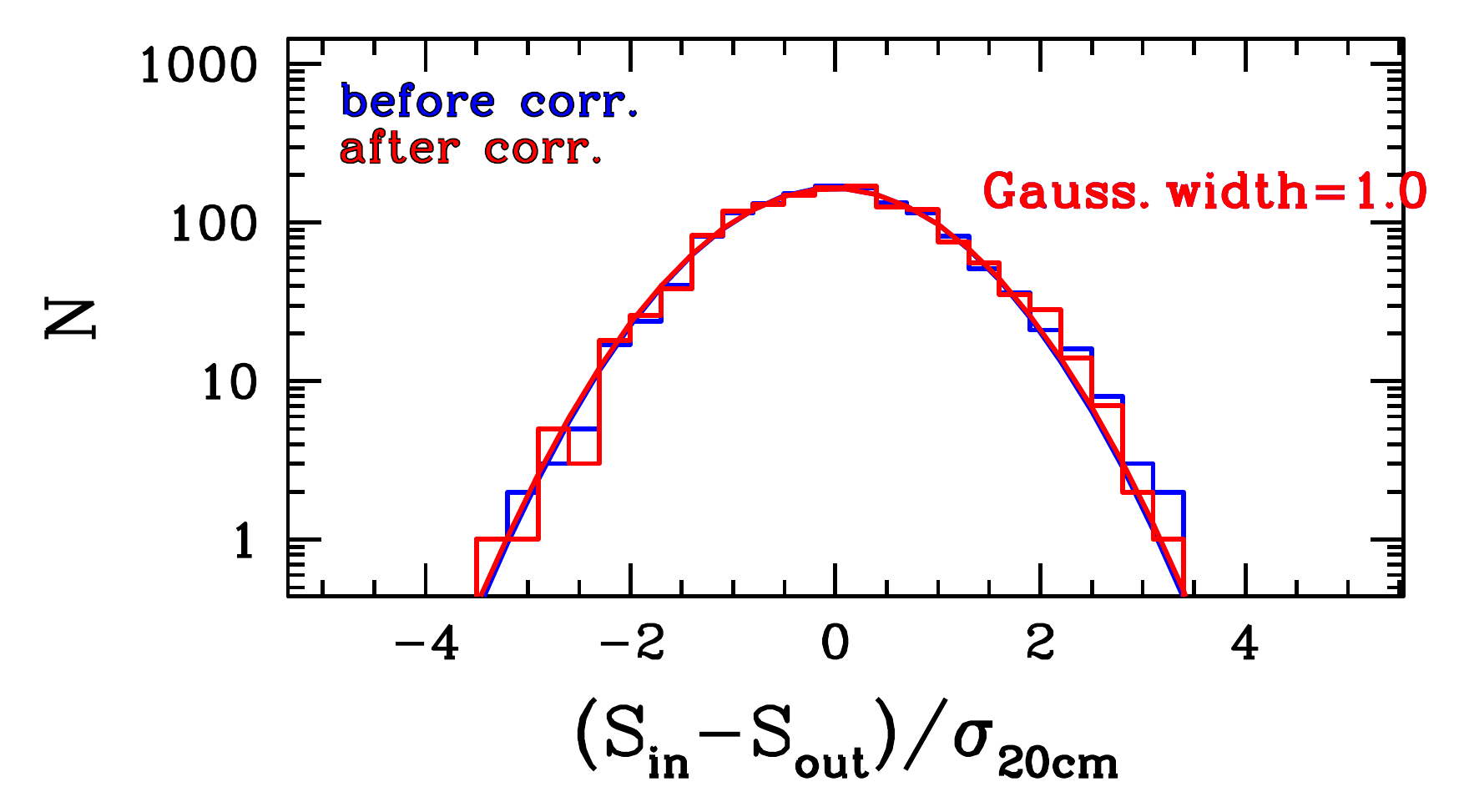}
    \end{subfigure}
    
    \end{figure}
    \begin{figure}\ContinuedFloat
    
    \begin{subfigure}[b]{\textwidth}\centering
    \includegraphics[height=2.6cm, trim=0 1cm -1.8cm 0]{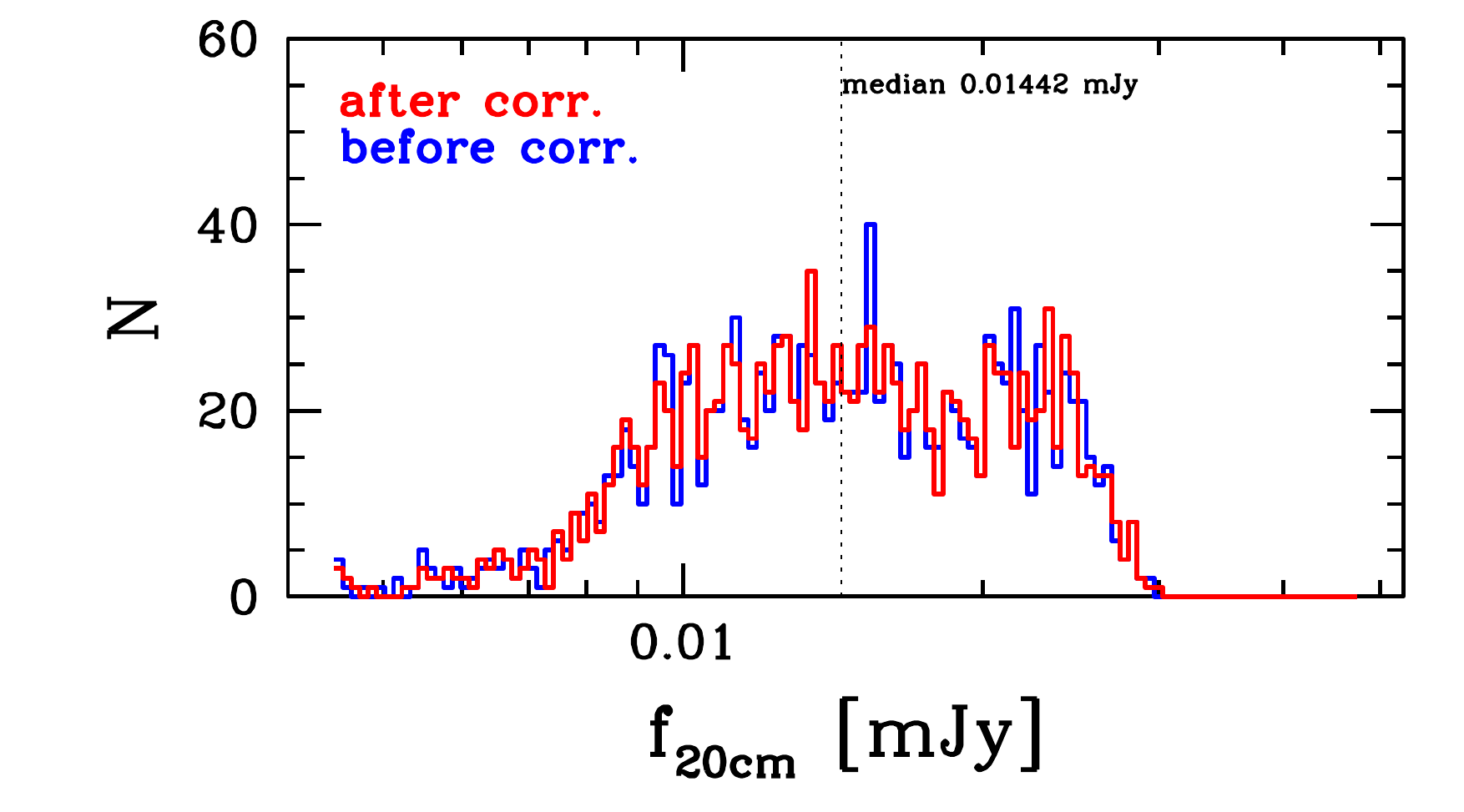}
    \includegraphics[height=2.6cm, trim=0 1cm -1.8cm 0]{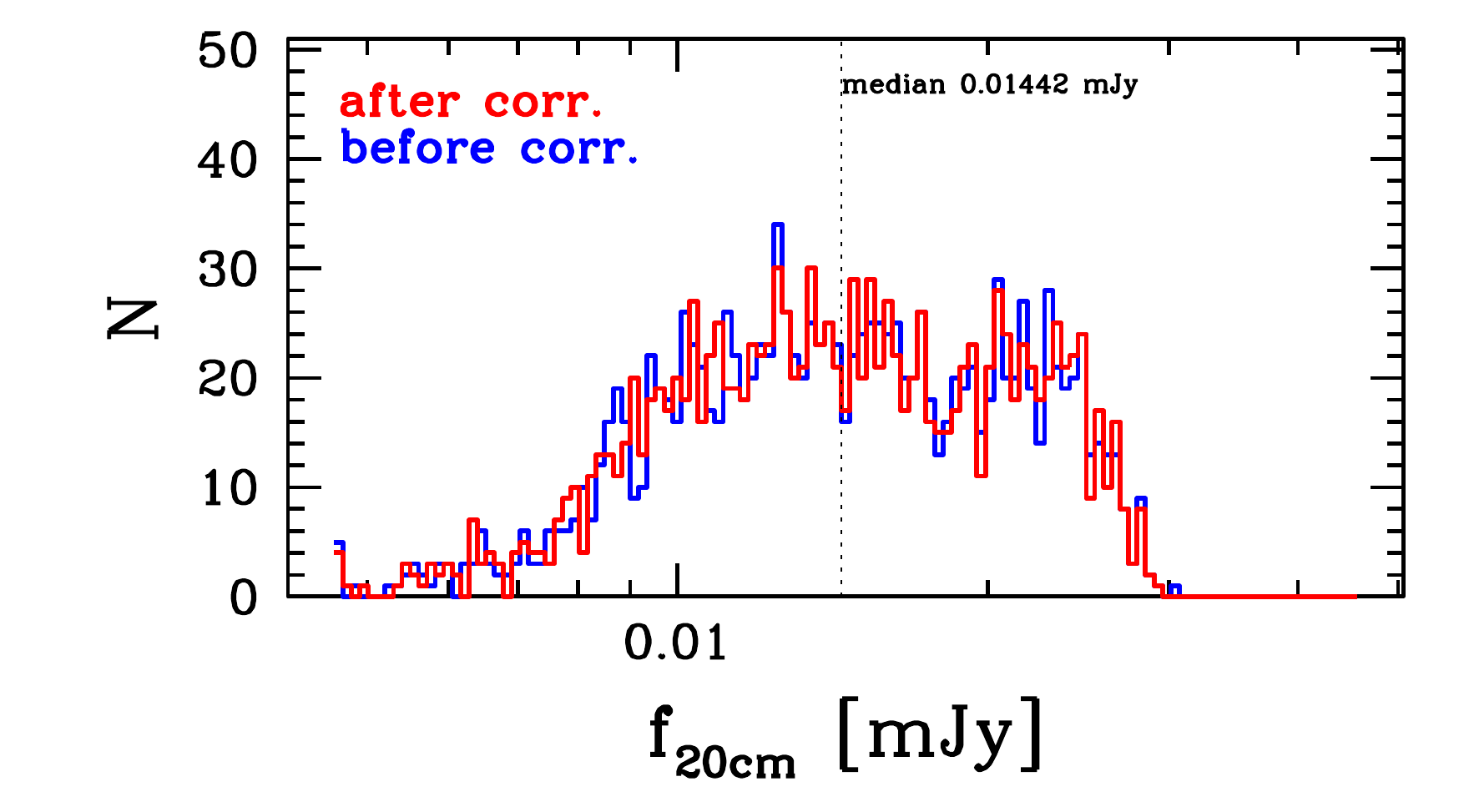}
    \includegraphics[height=2.6cm, trim=0 1cm -1.8cm 0]{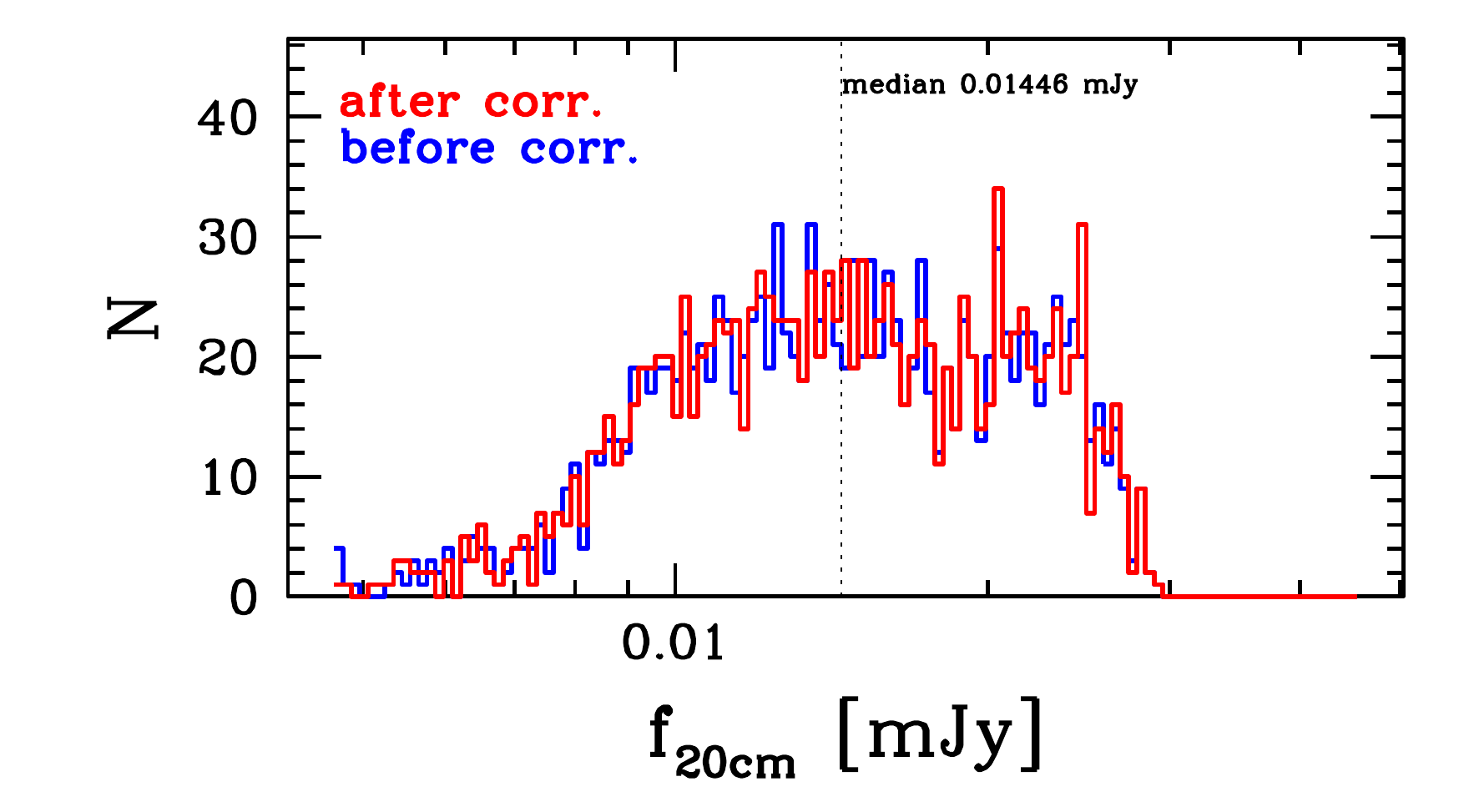}
    \end{subfigure}
    
    \begin{subfigure}[b]{\textwidth}\centering
    \includegraphics[height=2.6cm, trim=0 1cm -1.8cm 0]{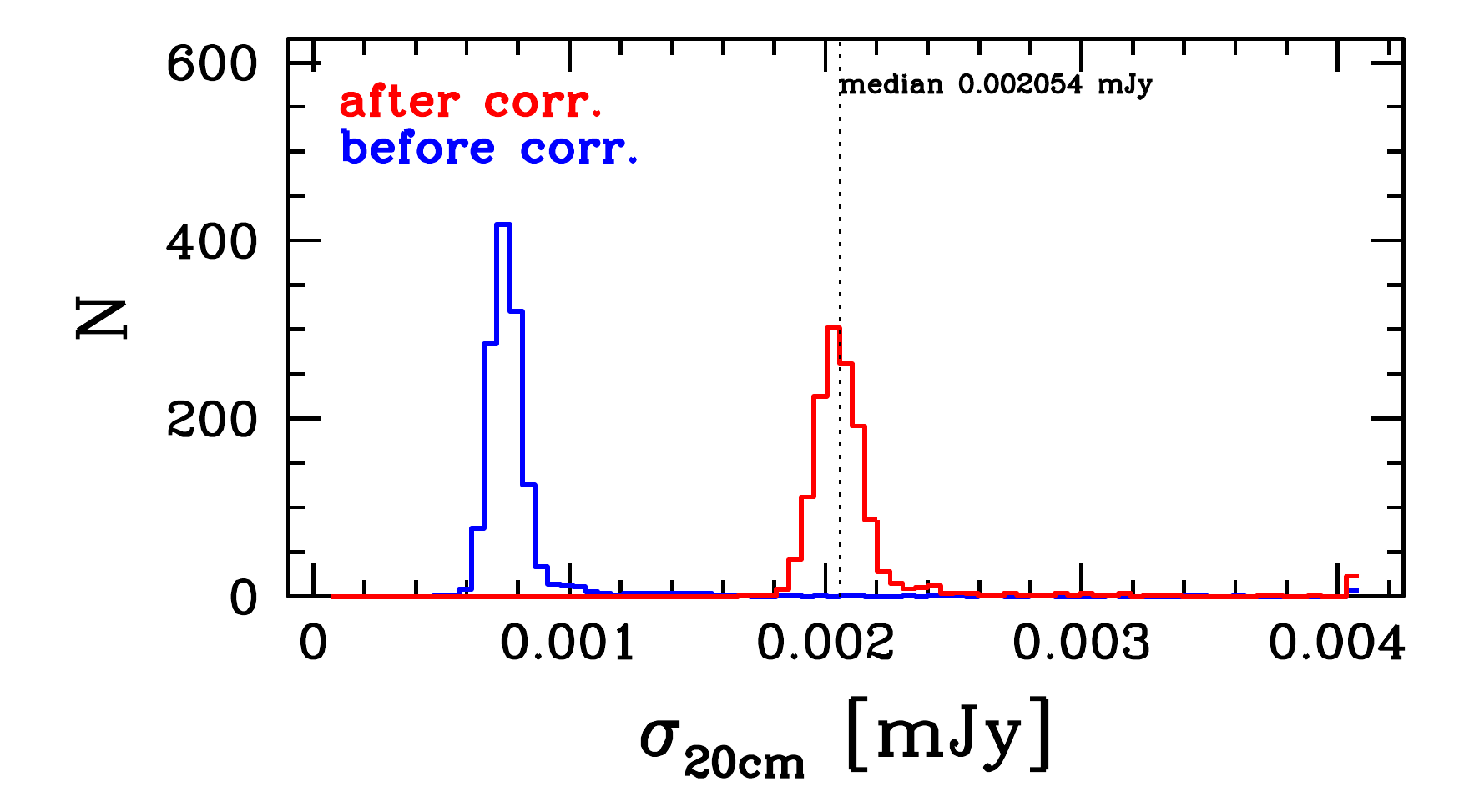}
    \includegraphics[height=2.6cm, trim=0 1cm -1.8cm 0]{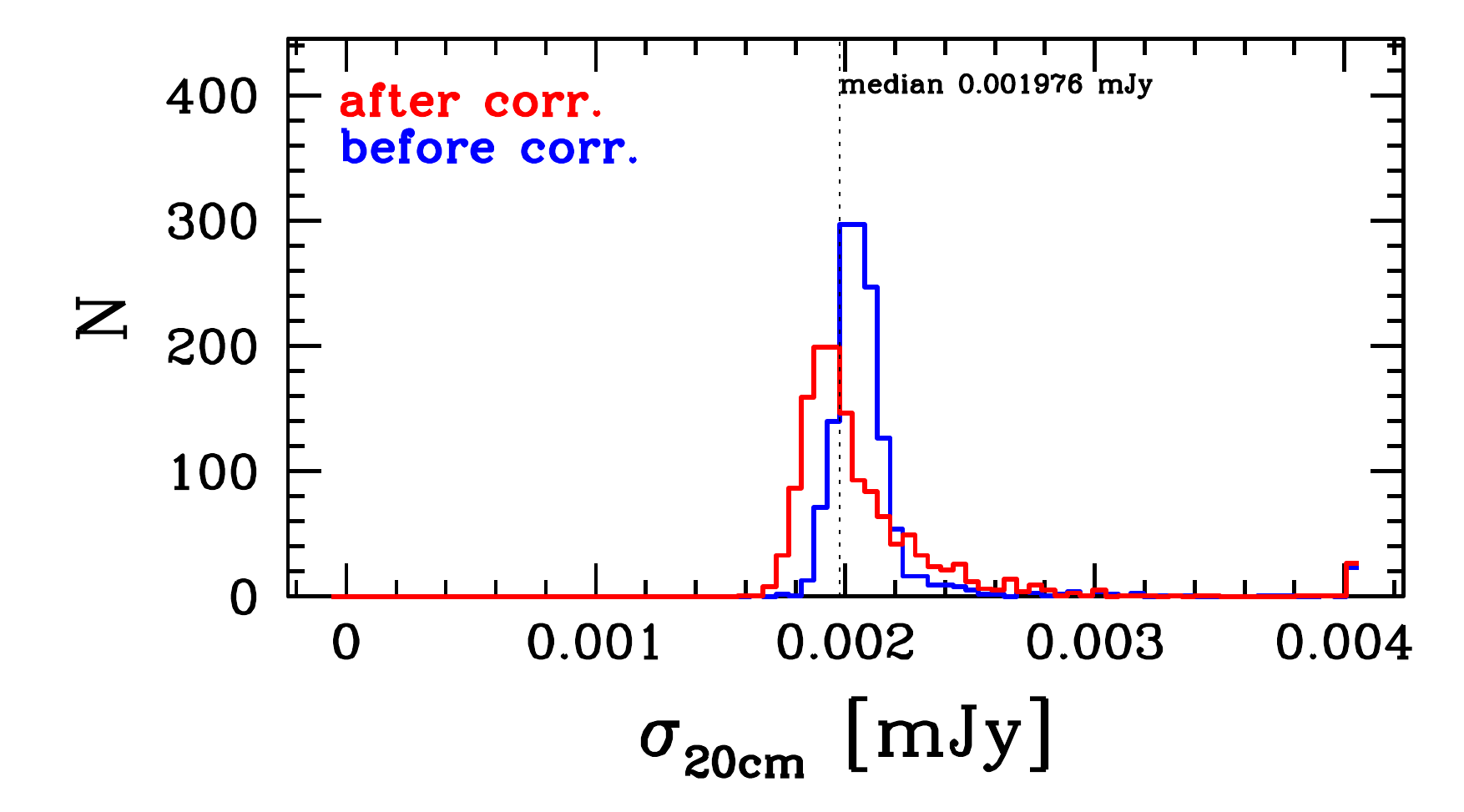}
    \includegraphics[height=2.6cm, trim=0 1cm -1.8cm 0]{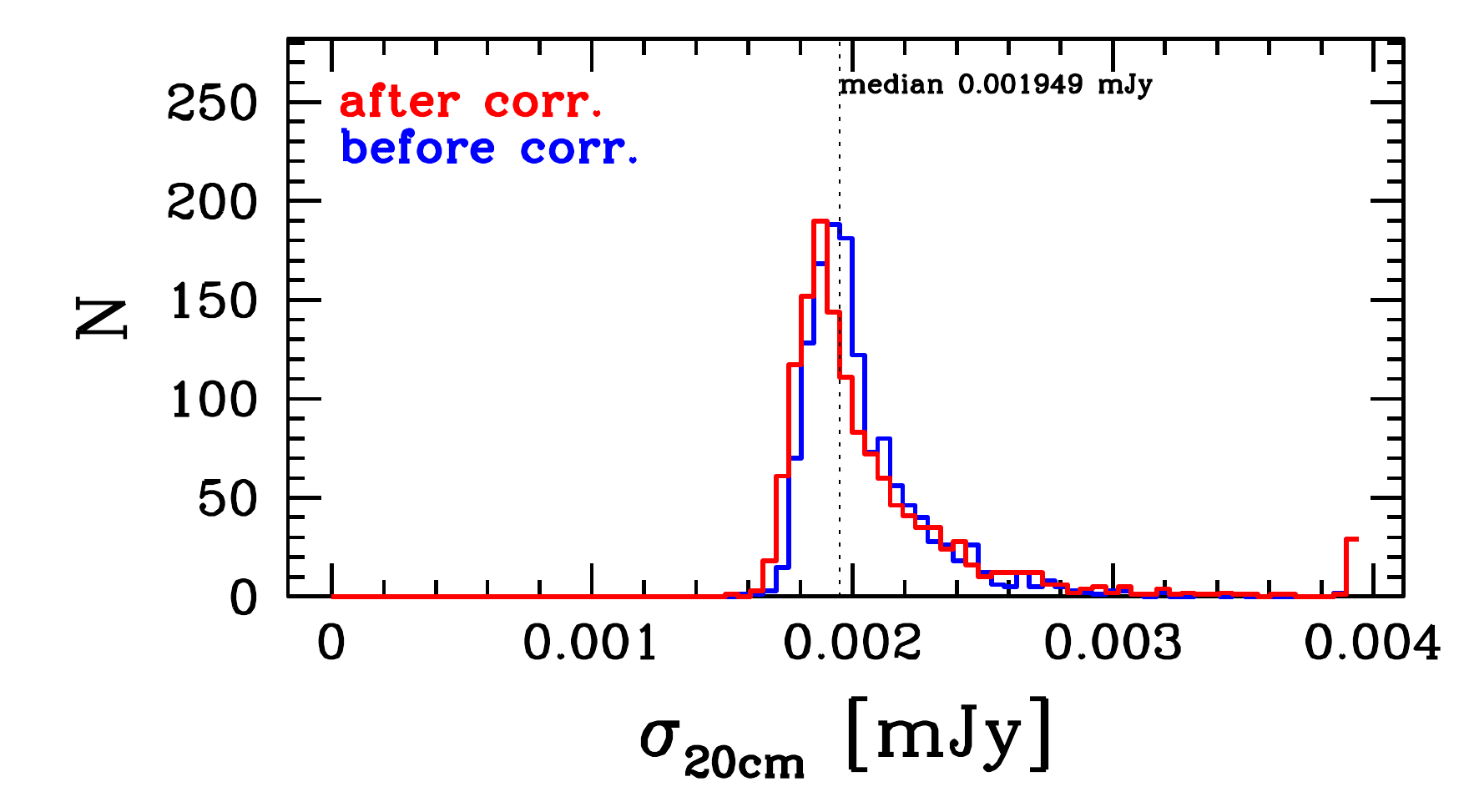}
    \end{subfigure}
    
    \caption{%
        Simulation correction analyses at 20~cm. See descriptions in the text. 
        \label{Figure_galsim_20cm_bin}
    }
\end{figure}



\begin{figure}
    \centering
    
    \begin{subfigure}[b]{\textwidth}\centering
    \includegraphics[height=2.6cm, trim=0 1cm 0 0]{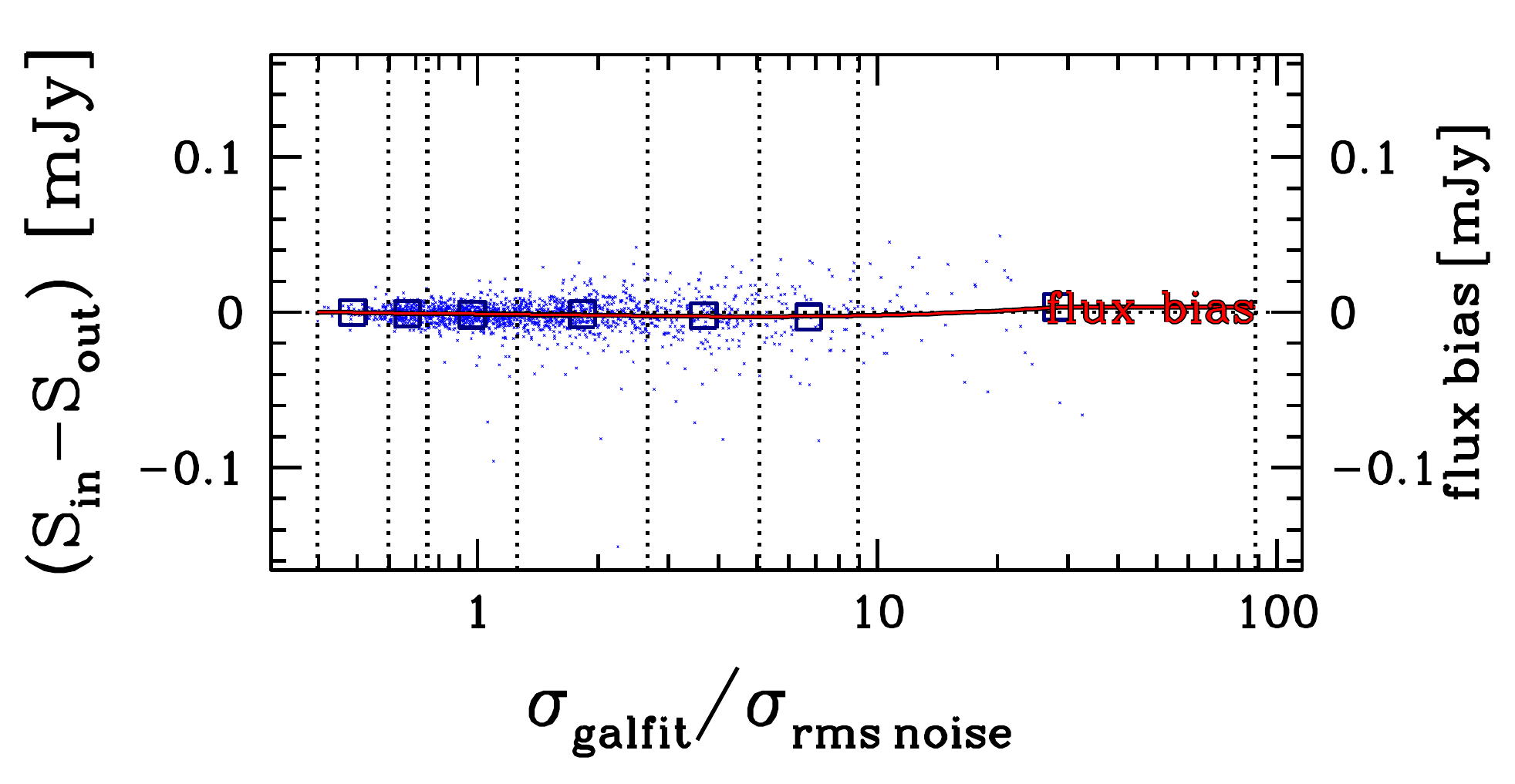}
    \includegraphics[height=2.6cm, trim=0 1cm 0 0]{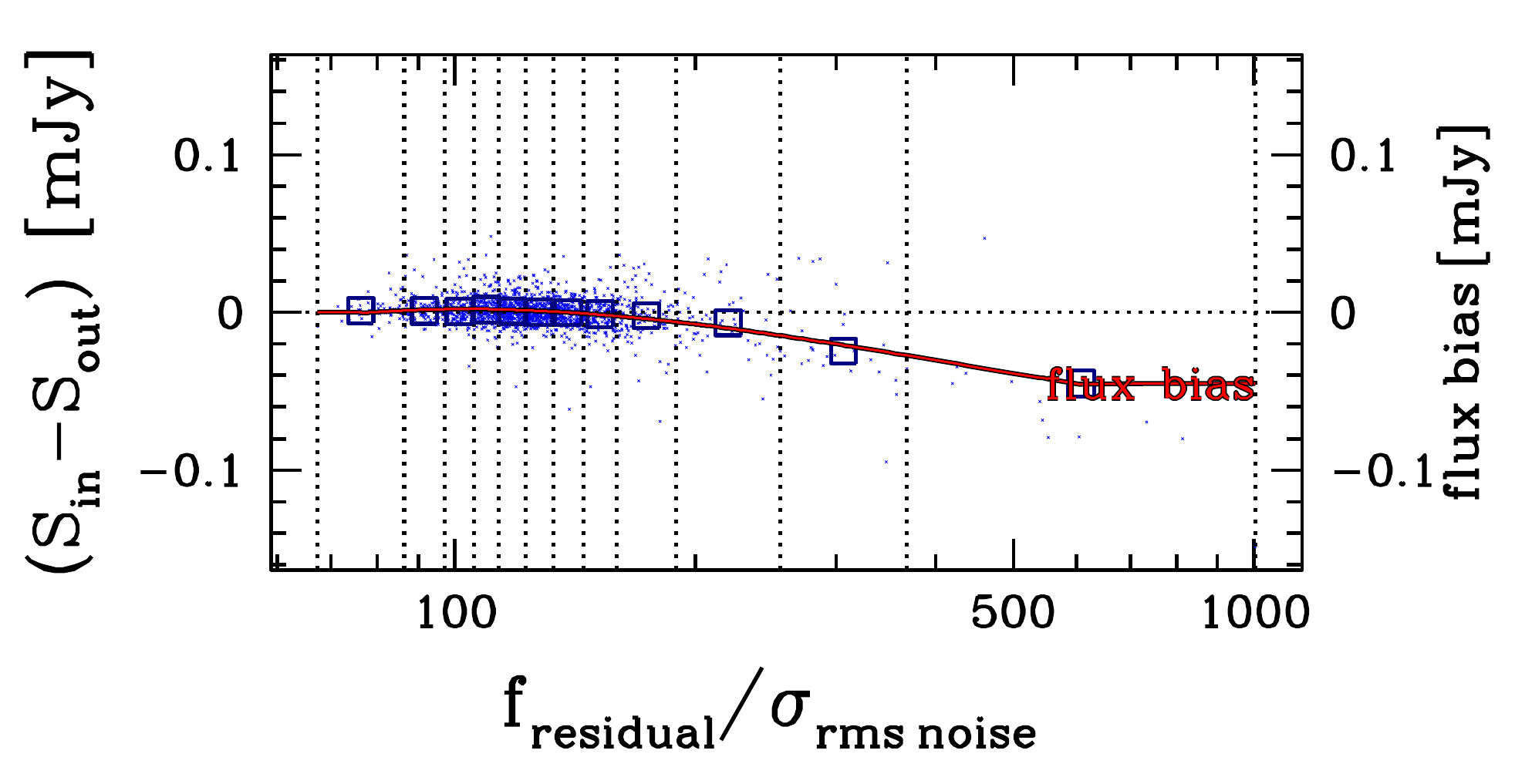}
    \includegraphics[height=2.6cm, trim=0 1cm 0 0]{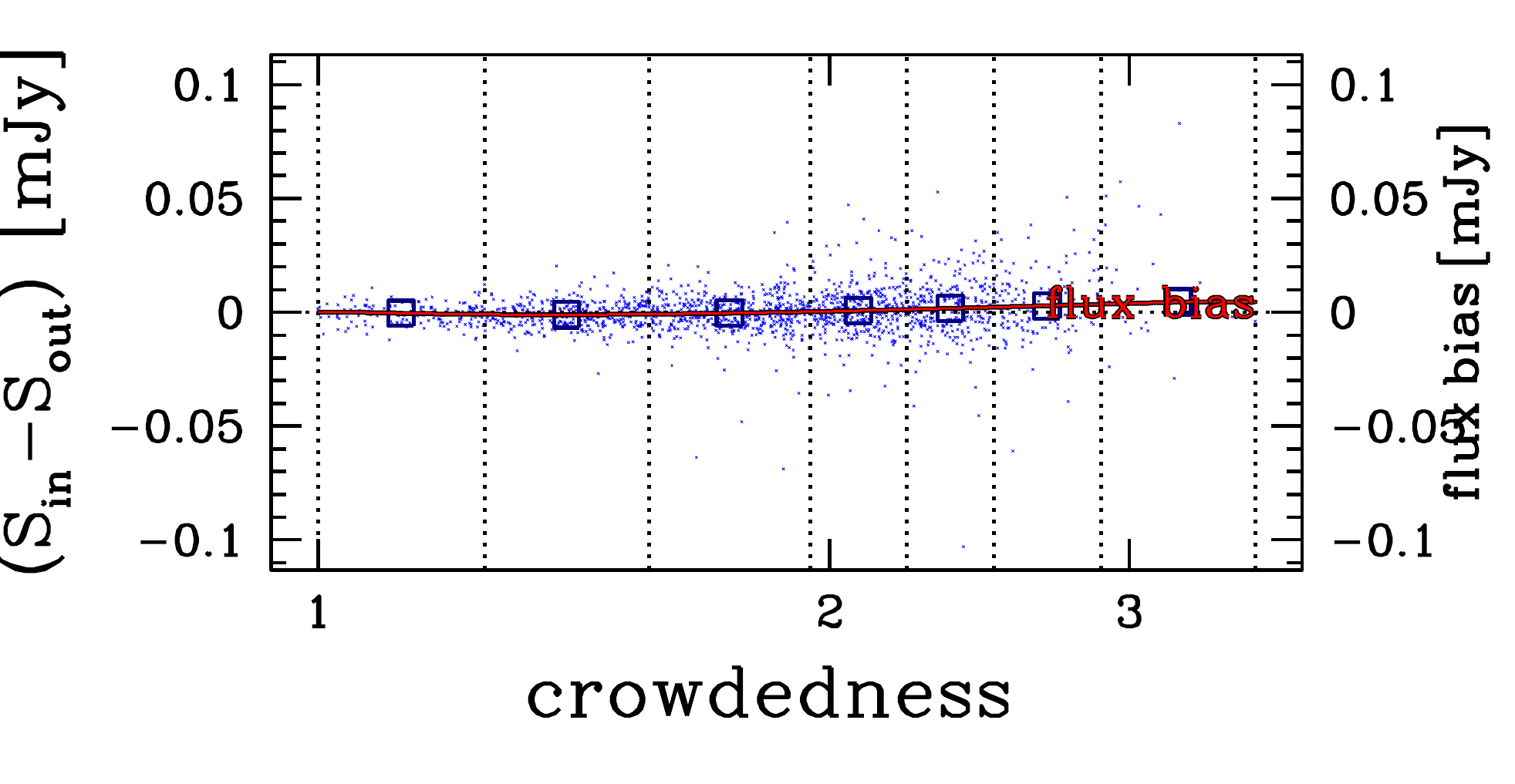}
    \end{subfigure}
    
    \begin{subfigure}[b]{\textwidth}\centering
    \includegraphics[height=2.6cm, trim=0 1cm 0 0]{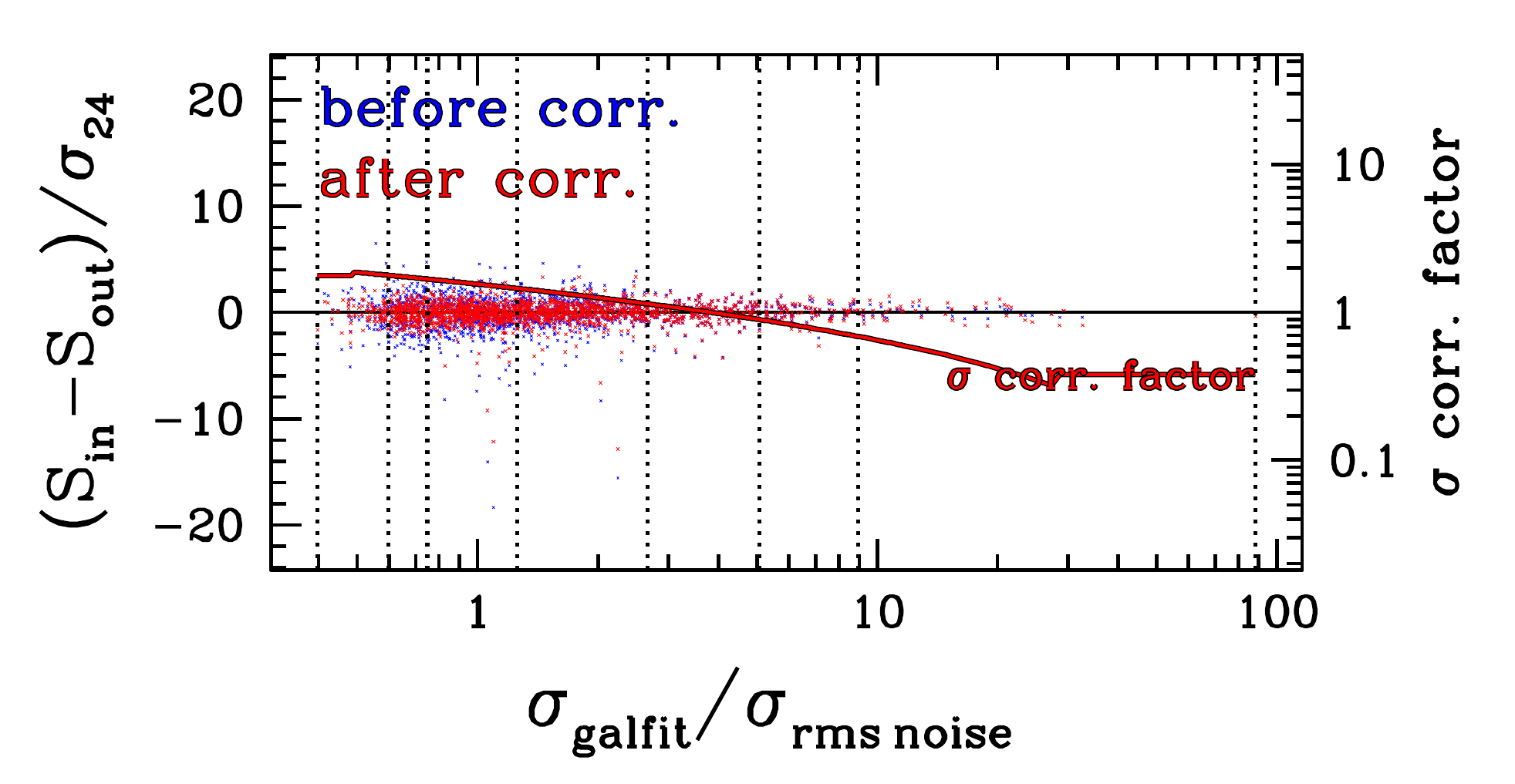}
    \includegraphics[height=2.6cm, trim=0 1cm 0 0]{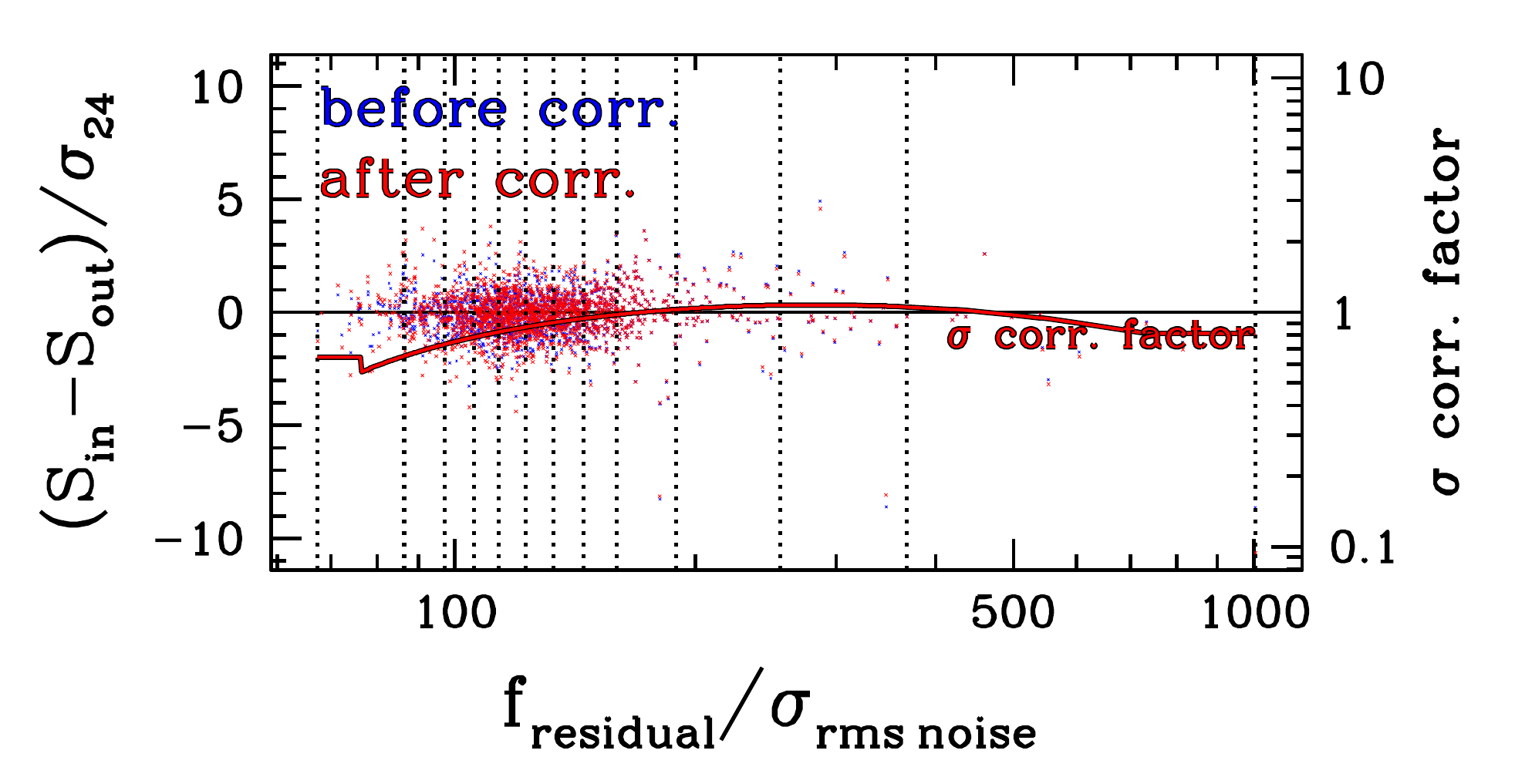}
    \includegraphics[height=2.6cm, trim=0 1cm 0 0]{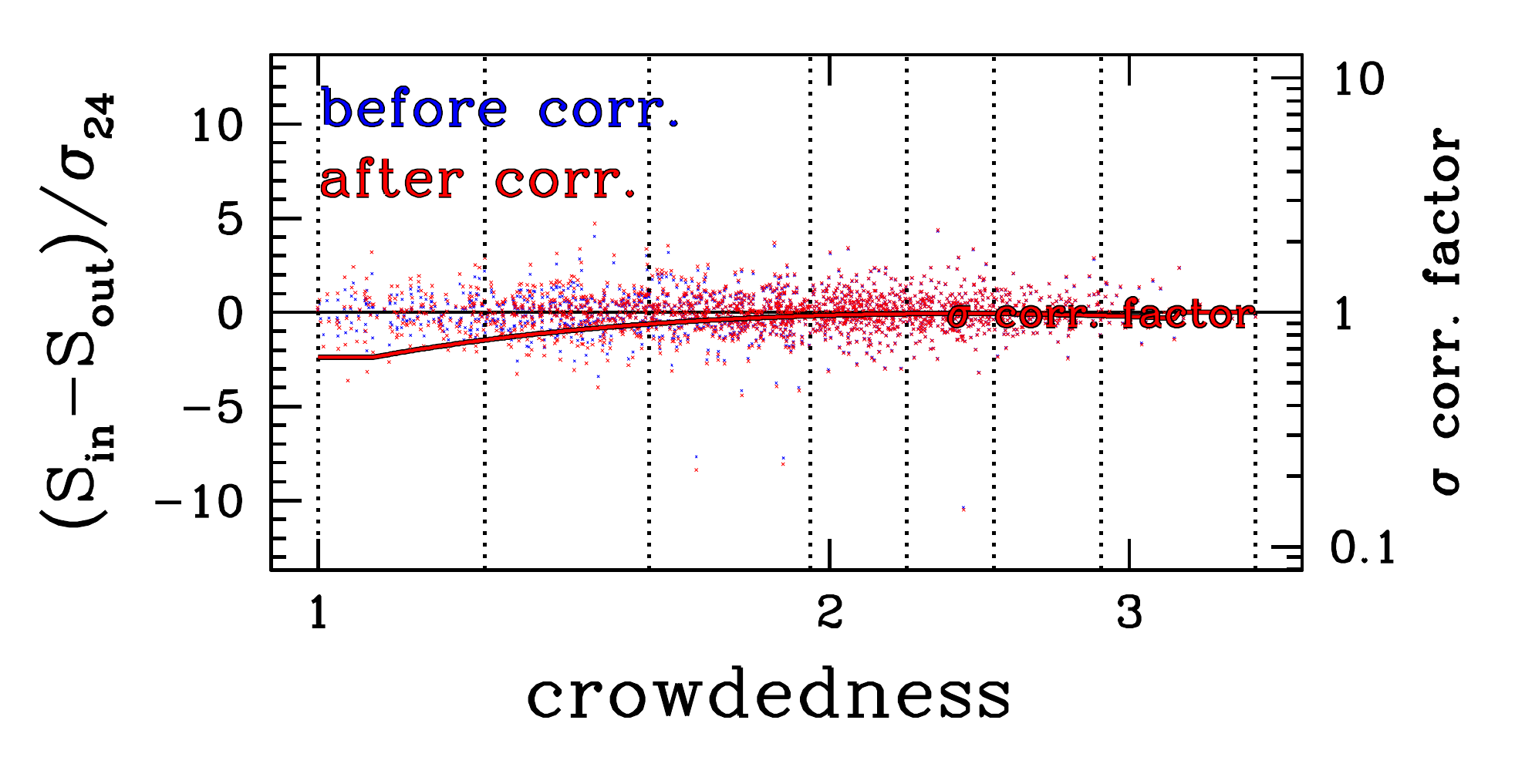}
    \end{subfigure}
    
    \begin{subfigure}[b]{\textwidth}\centering
    \includegraphics[height=2.6cm, trim=0 1cm -1.8cm 0]{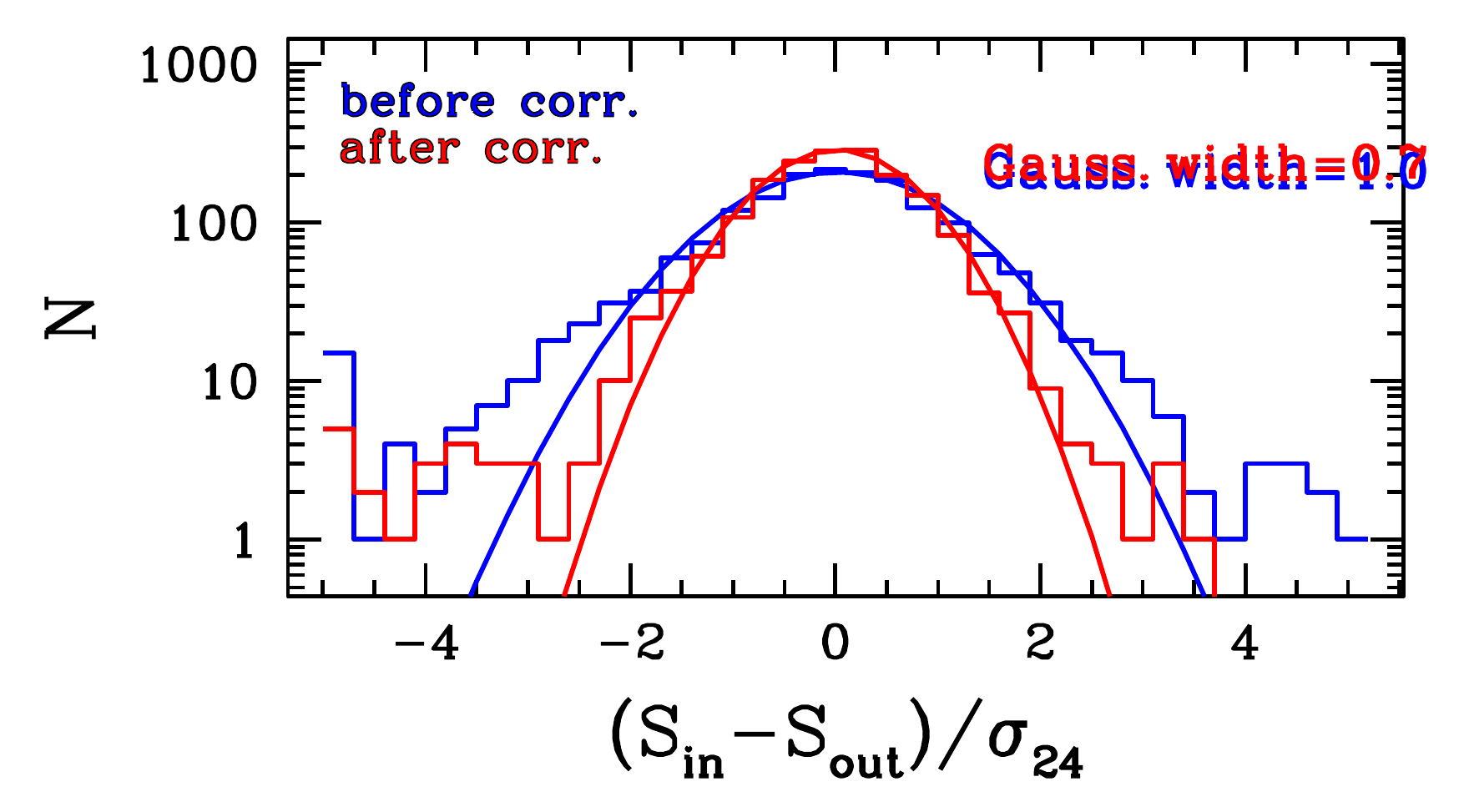}
    \includegraphics[height=2.6cm, trim=0 1cm -1.8cm 0]{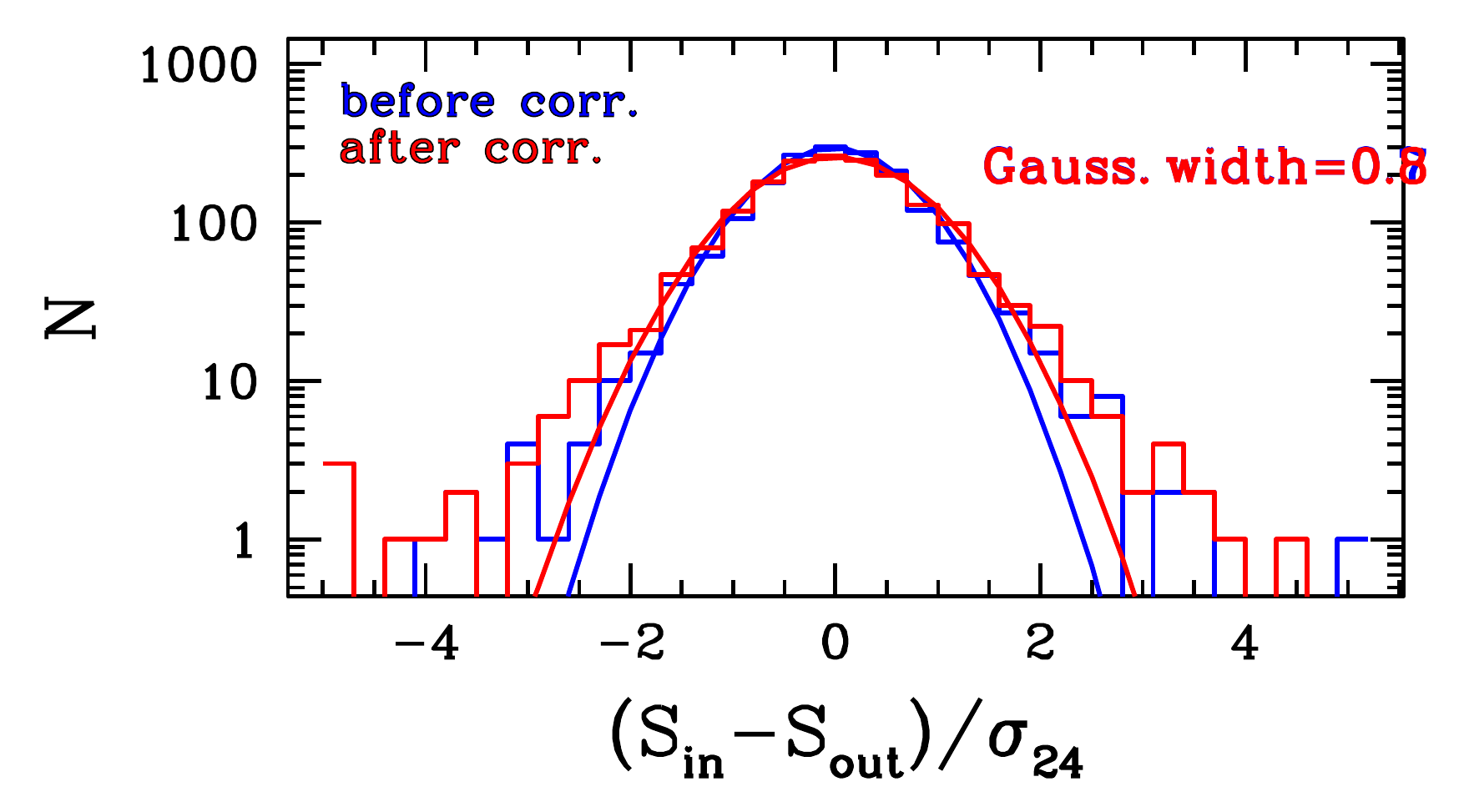}
    \includegraphics[height=2.6cm, trim=0 1cm -1.8cm 0]{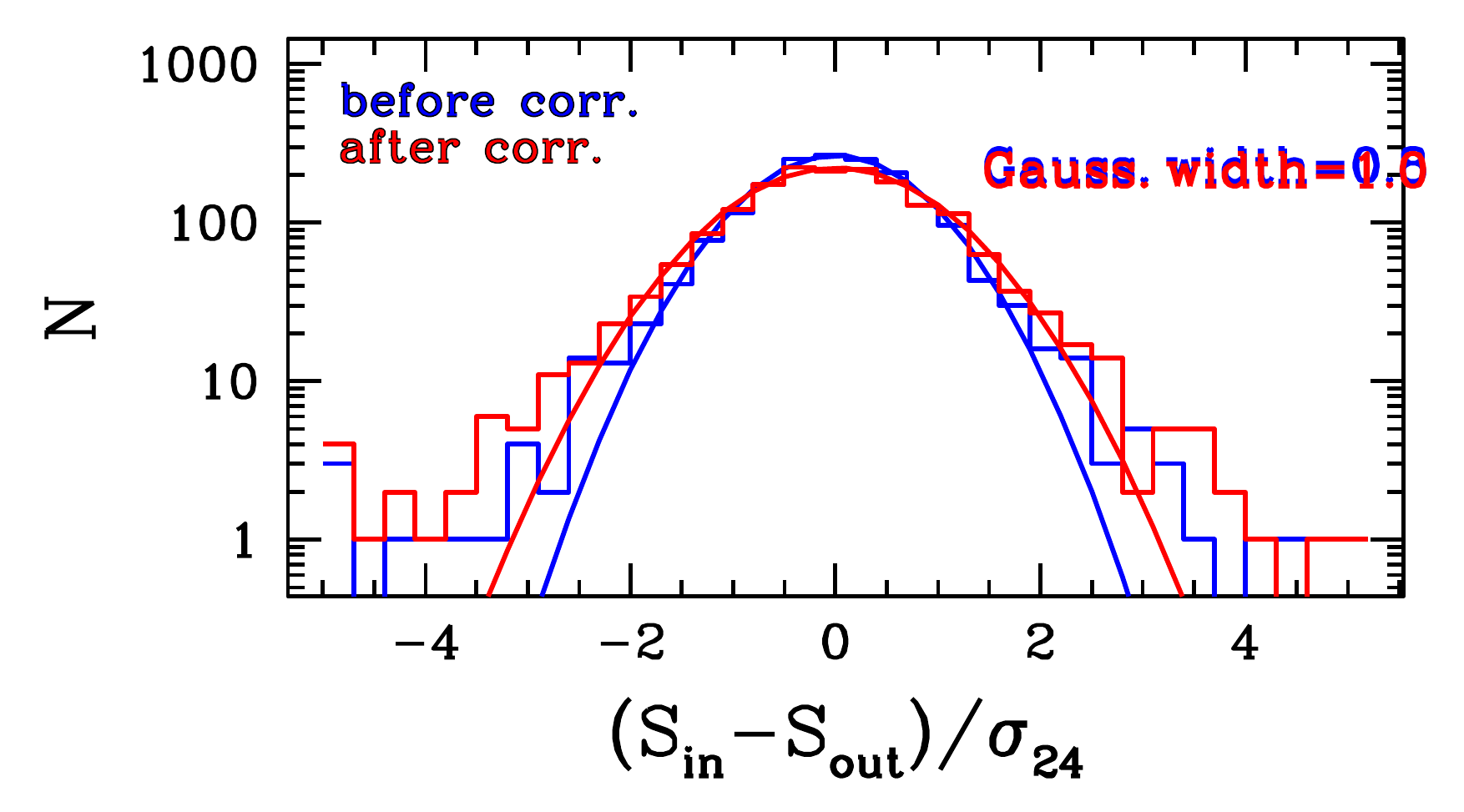}
    \end{subfigure}
    
    \begin{subfigure}[b]{\textwidth}\centering
    \includegraphics[height=2.6cm, trim=0 1cm -1.8cm 0]{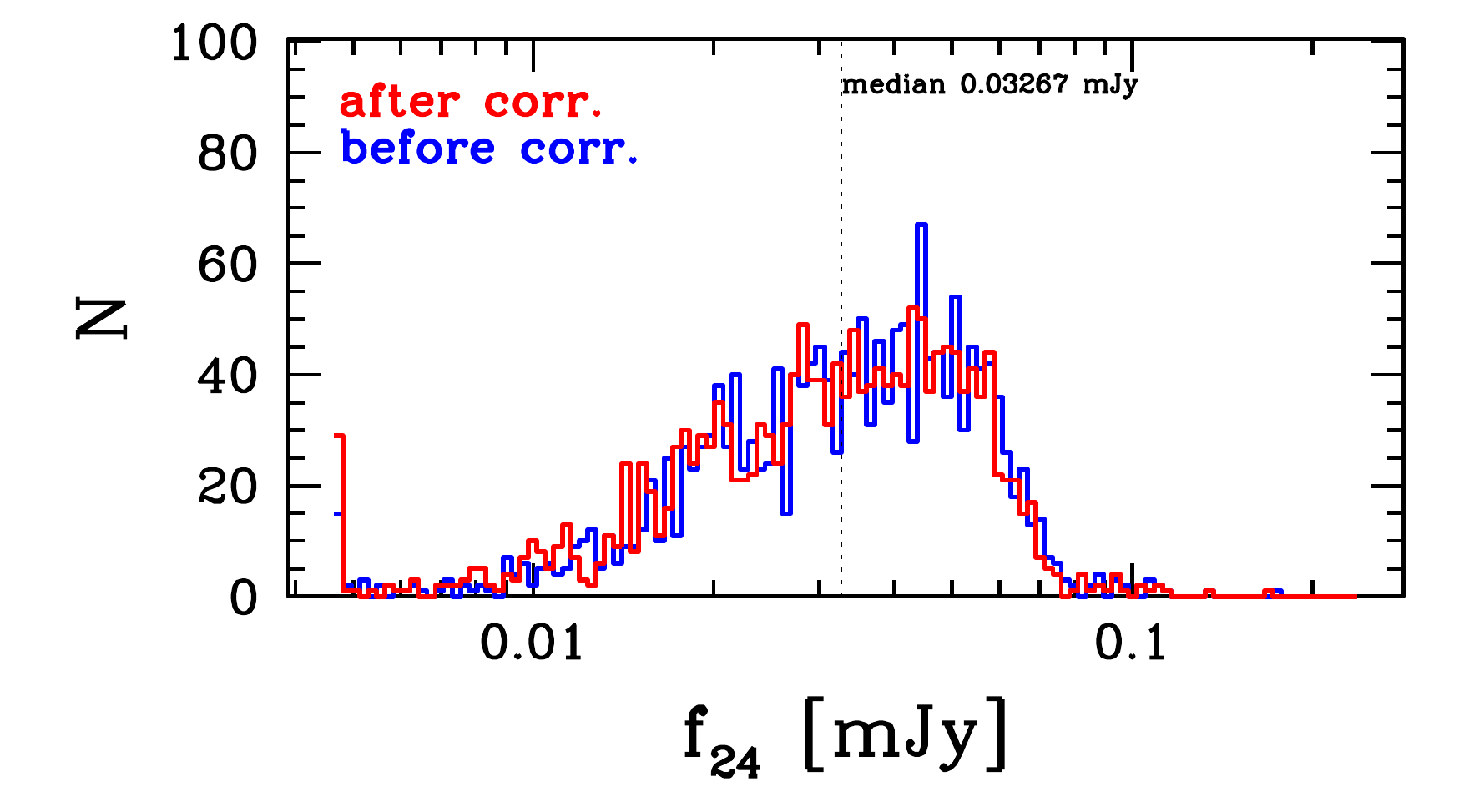}
    \includegraphics[height=2.6cm, trim=0 1cm -1.8cm 0]{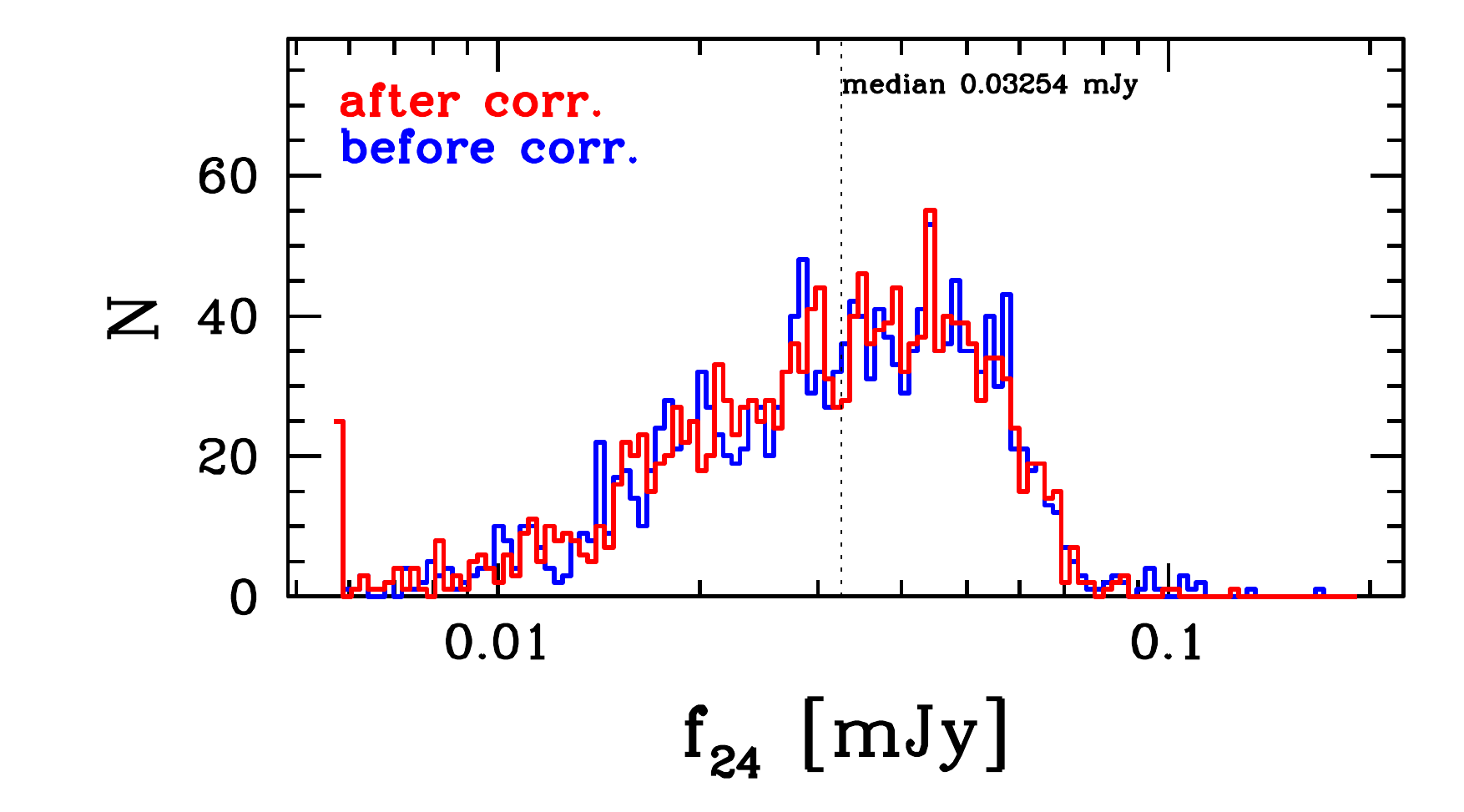}
    \includegraphics[height=2.6cm, trim=0 1cm -1.8cm 0]{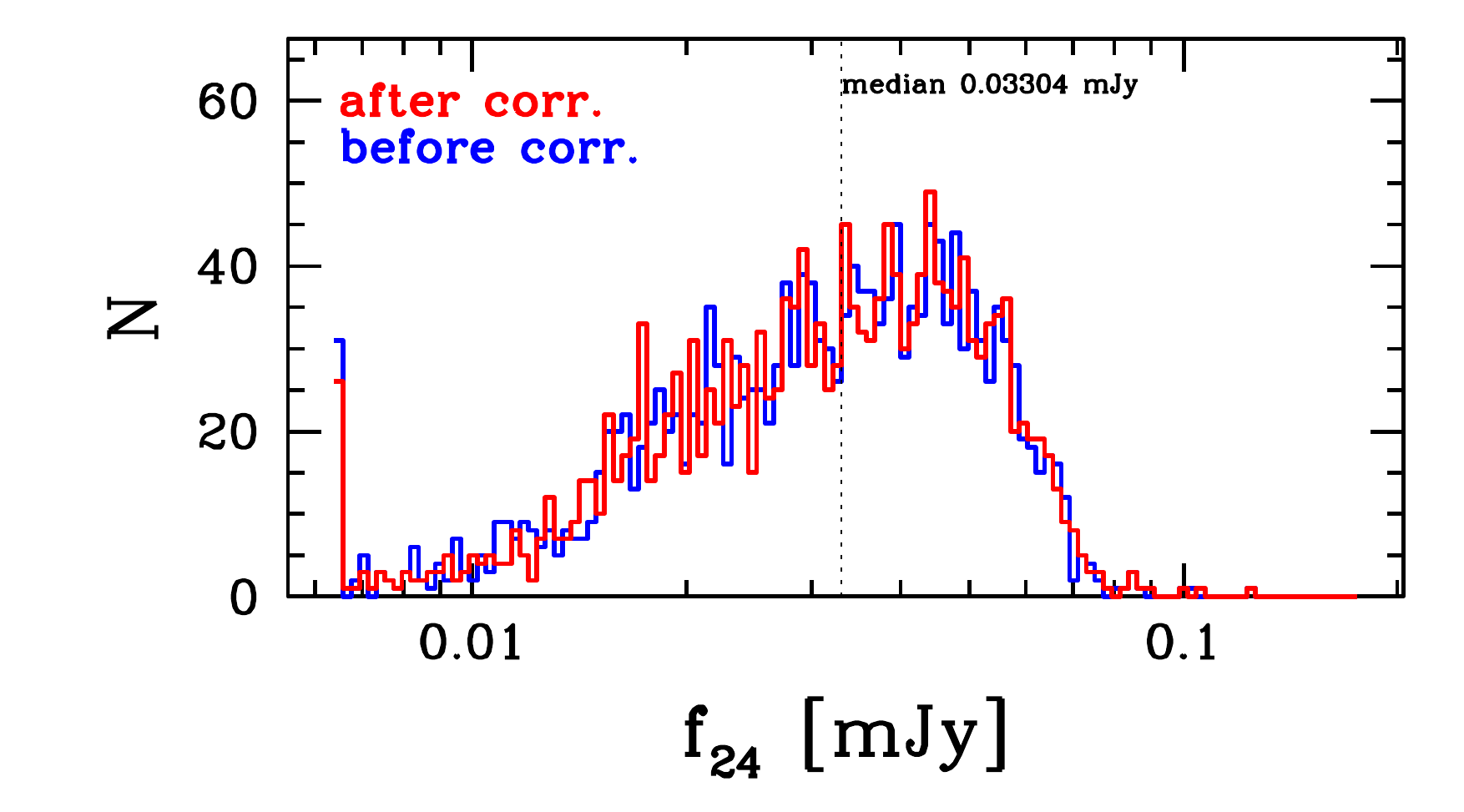}
    \end{subfigure}
    
    \begin{subfigure}[b]{\textwidth}\centering
    \includegraphics[height=2.6cm, trim=0 1cm -1.8cm 0]{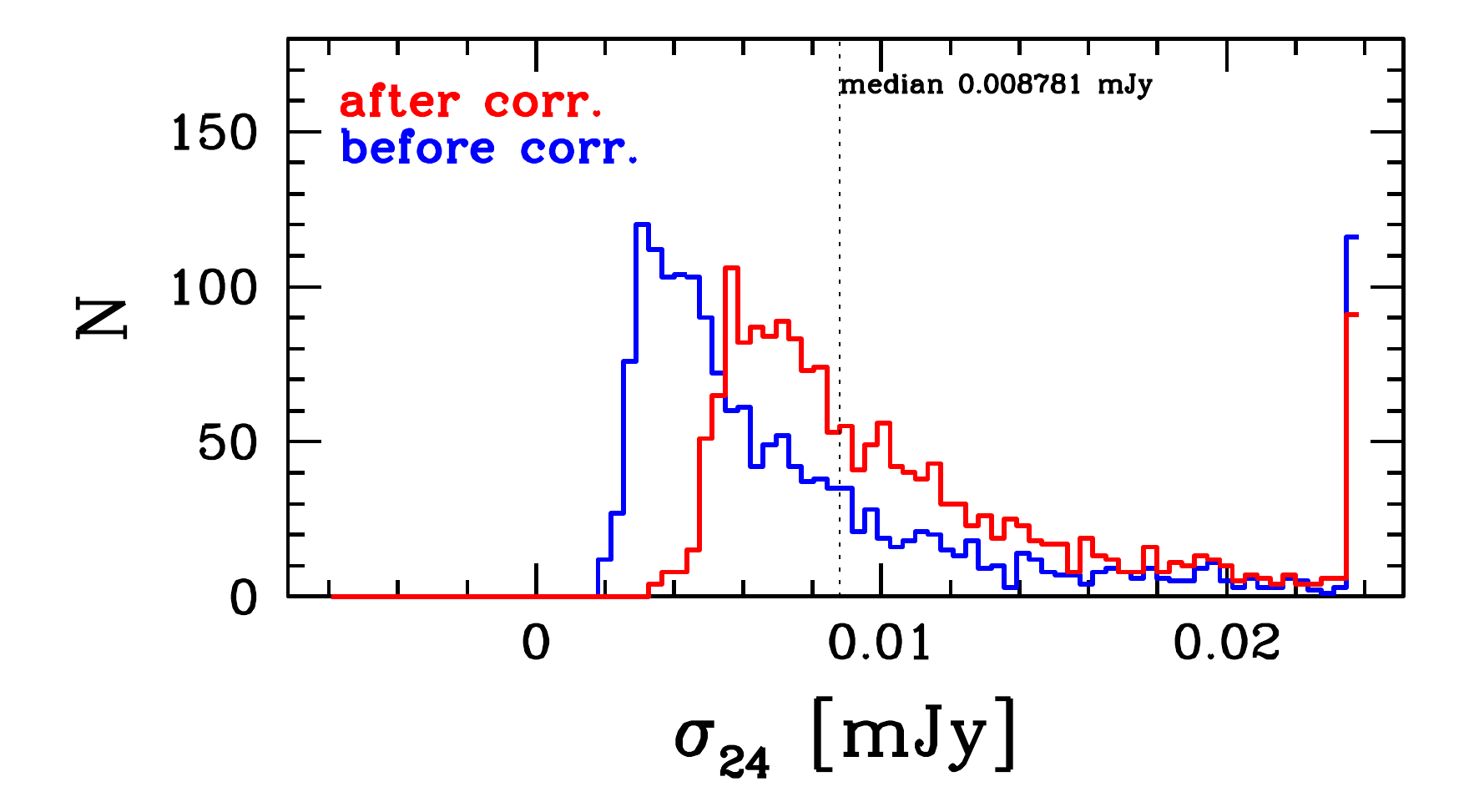}
    \includegraphics[height=2.6cm, trim=0 1cm -1.8cm 0]{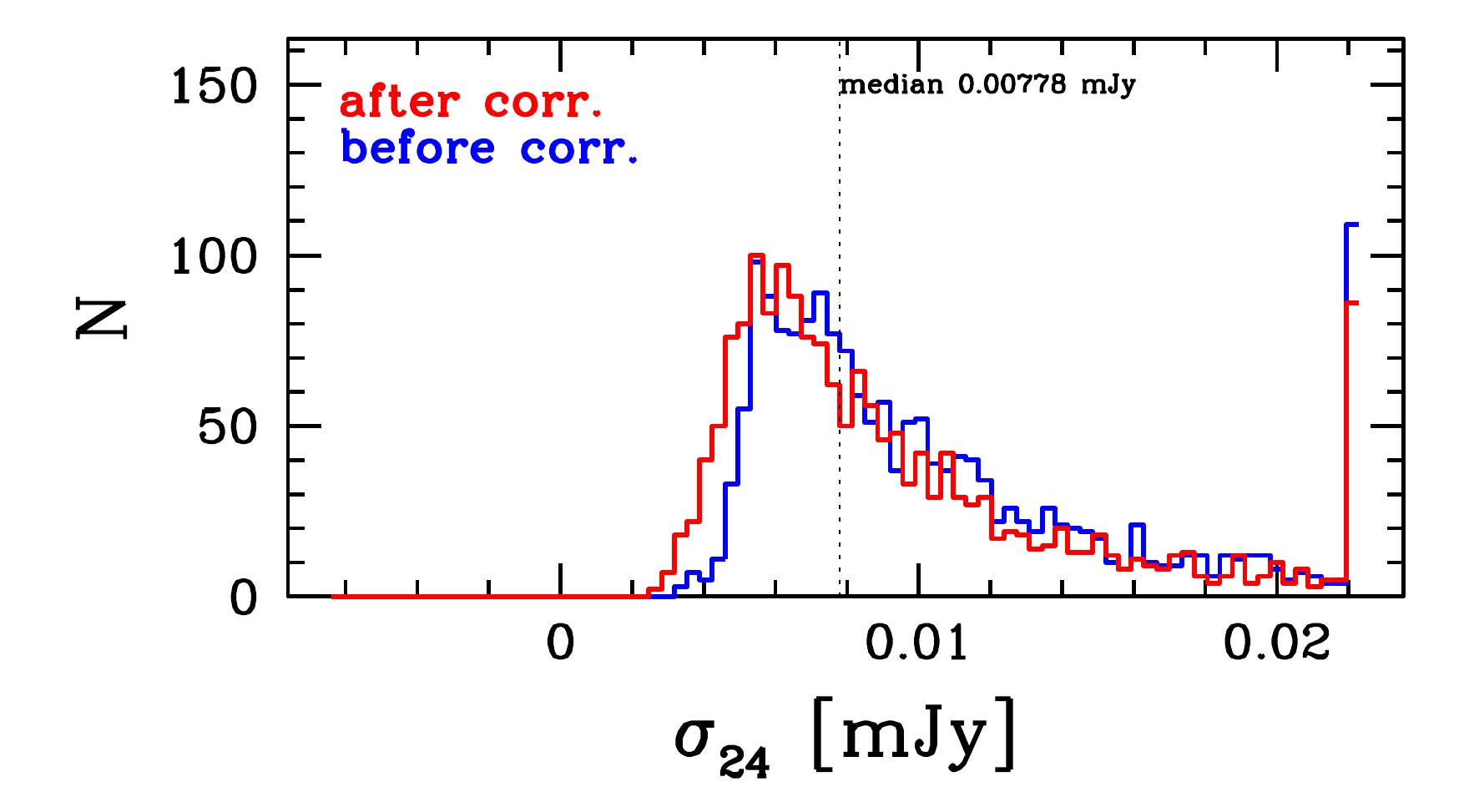}
    \includegraphics[height=2.6cm, trim=0 1cm -1.8cm 0]{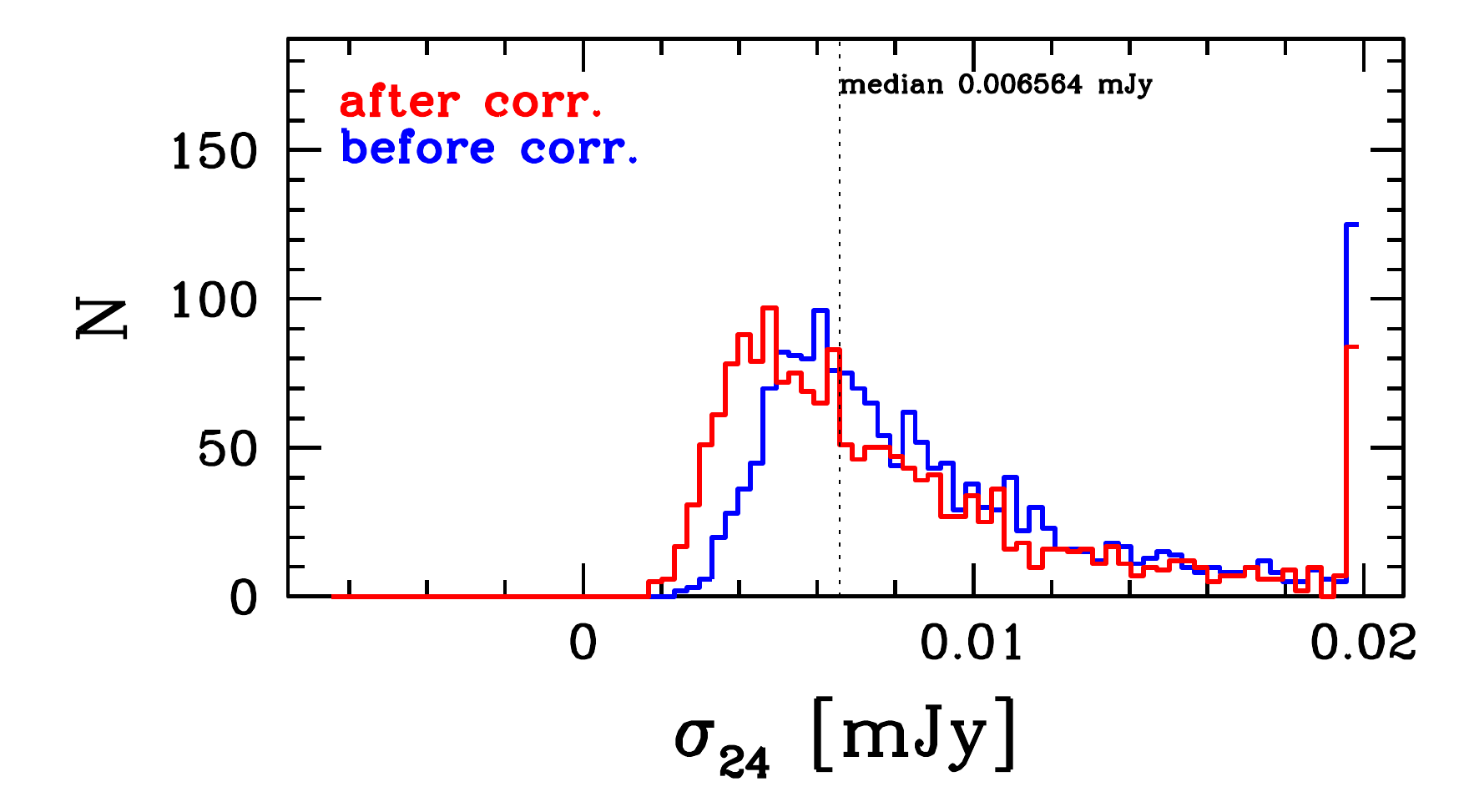}
    \end{subfigure}
    
\caption{%
    Simulation correction analyses at 24~$\mu$m. See descriptions in the text. 
        \label{Figure_galsim_24um_bin}
}
\end{figure}

\clearpage



\clearpage

\section{The cosmic SFR density completeness estimation using SMF-converted SFR histograms}
\label{Section_Appendix_CSFRD_Correction}

\begin{figure}[hb]
\centering
\includegraphics[width=0.32\textwidth, trim=0 1cm 0 0]{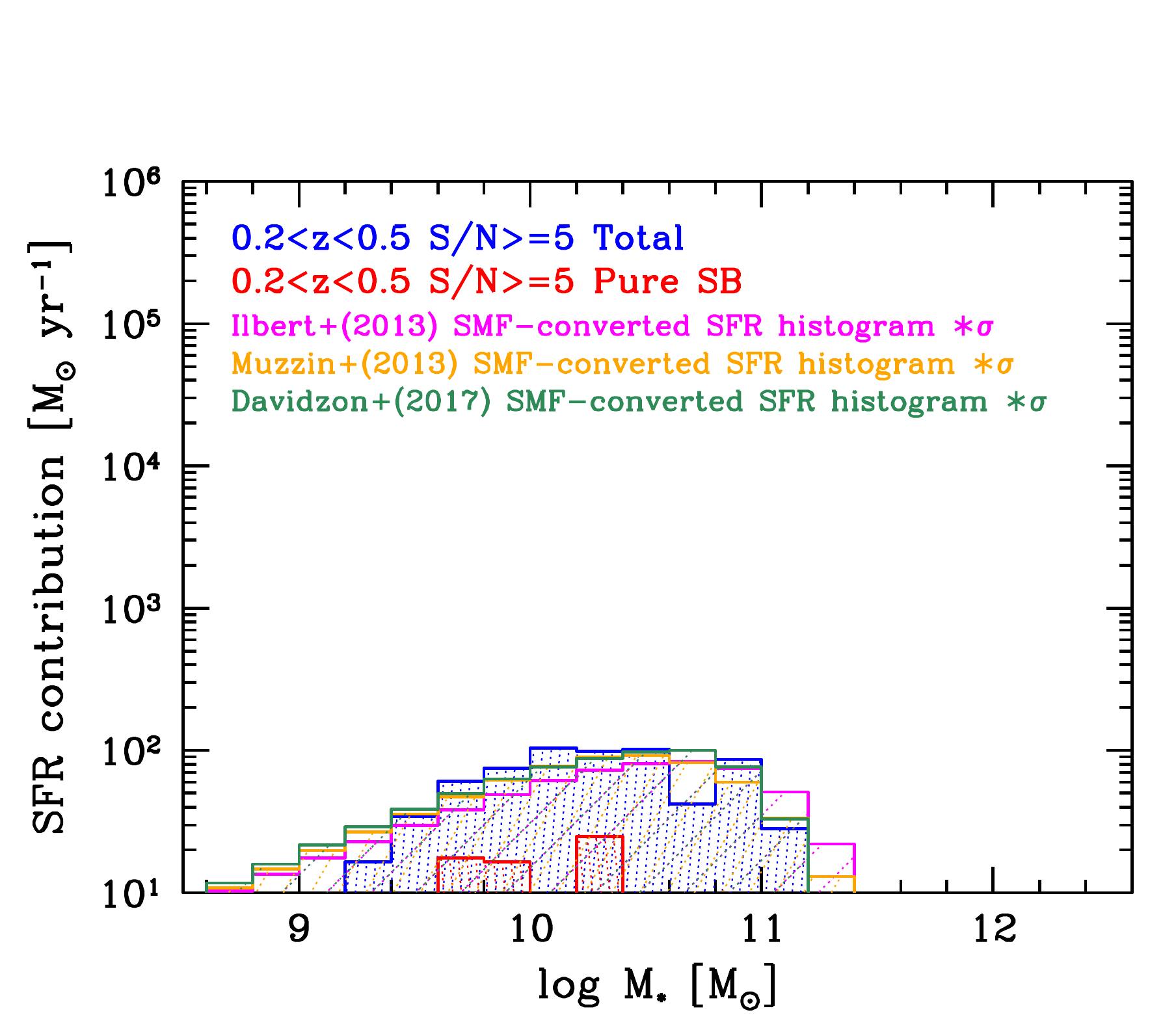}
\includegraphics[width=0.32\textwidth, trim=0 1cm 0 0]{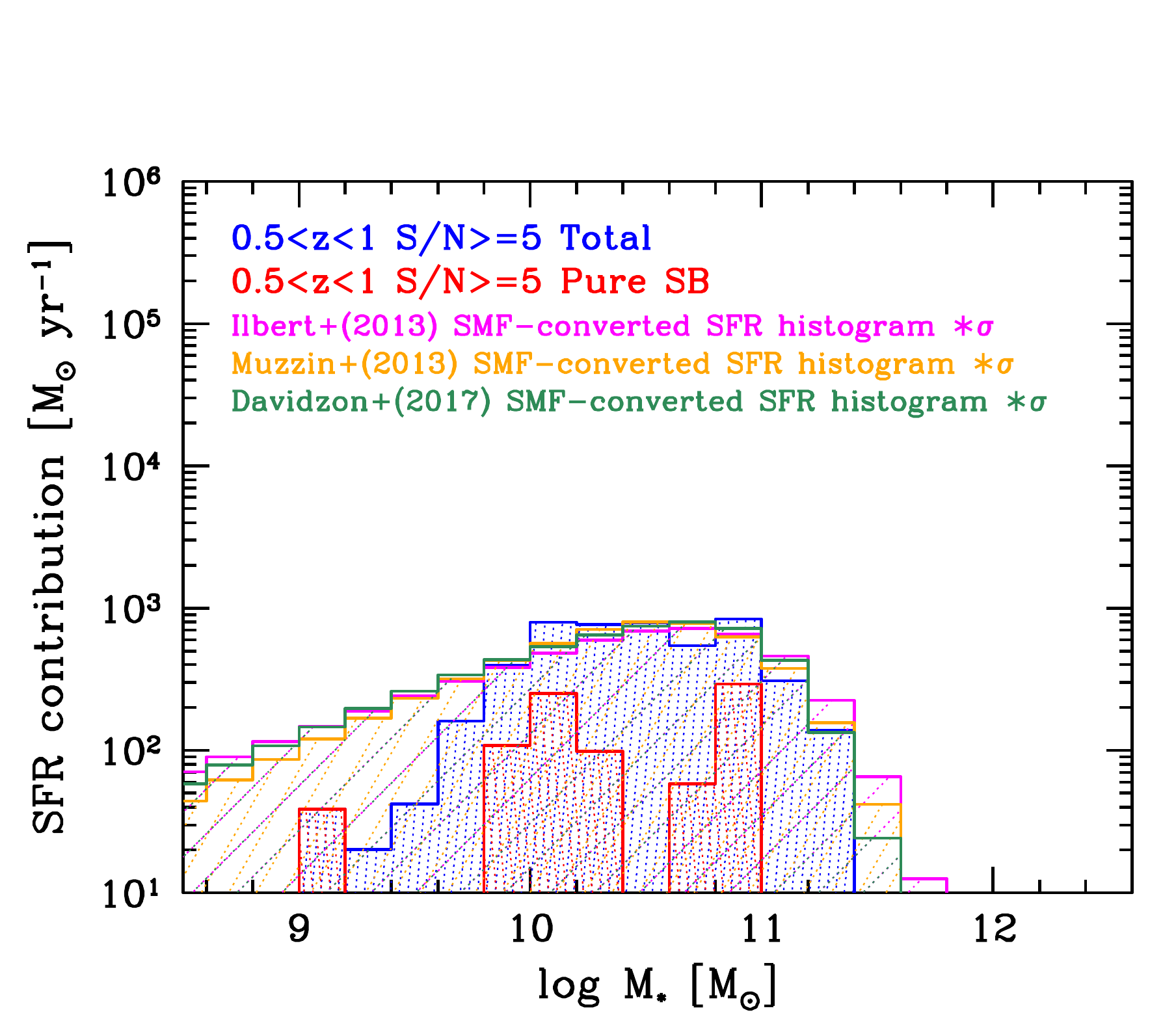}
\includegraphics[width=0.32\textwidth, trim=0 1cm 0 0]{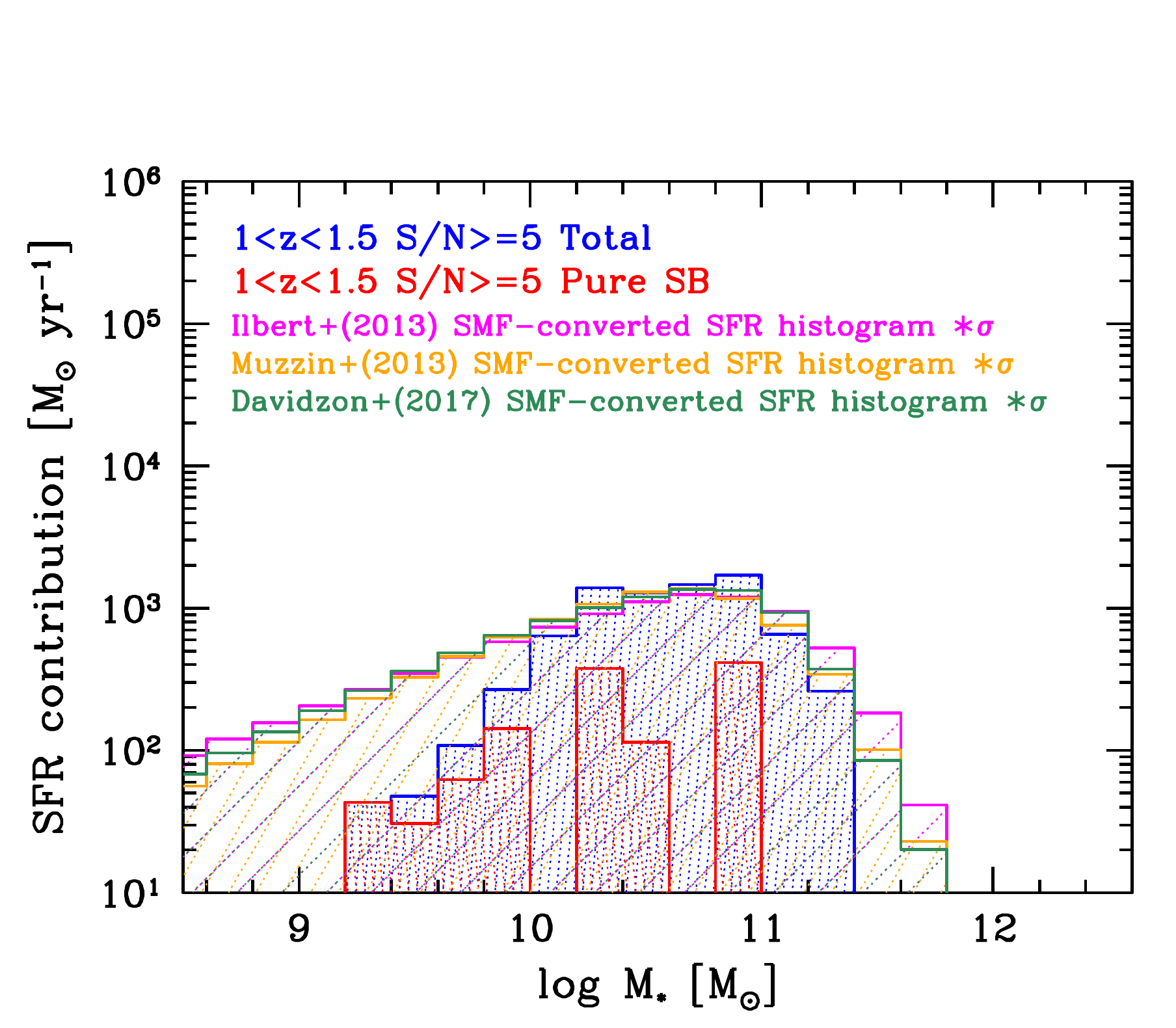}
\includegraphics[width=0.32\textwidth, trim=0 1cm 0 0]{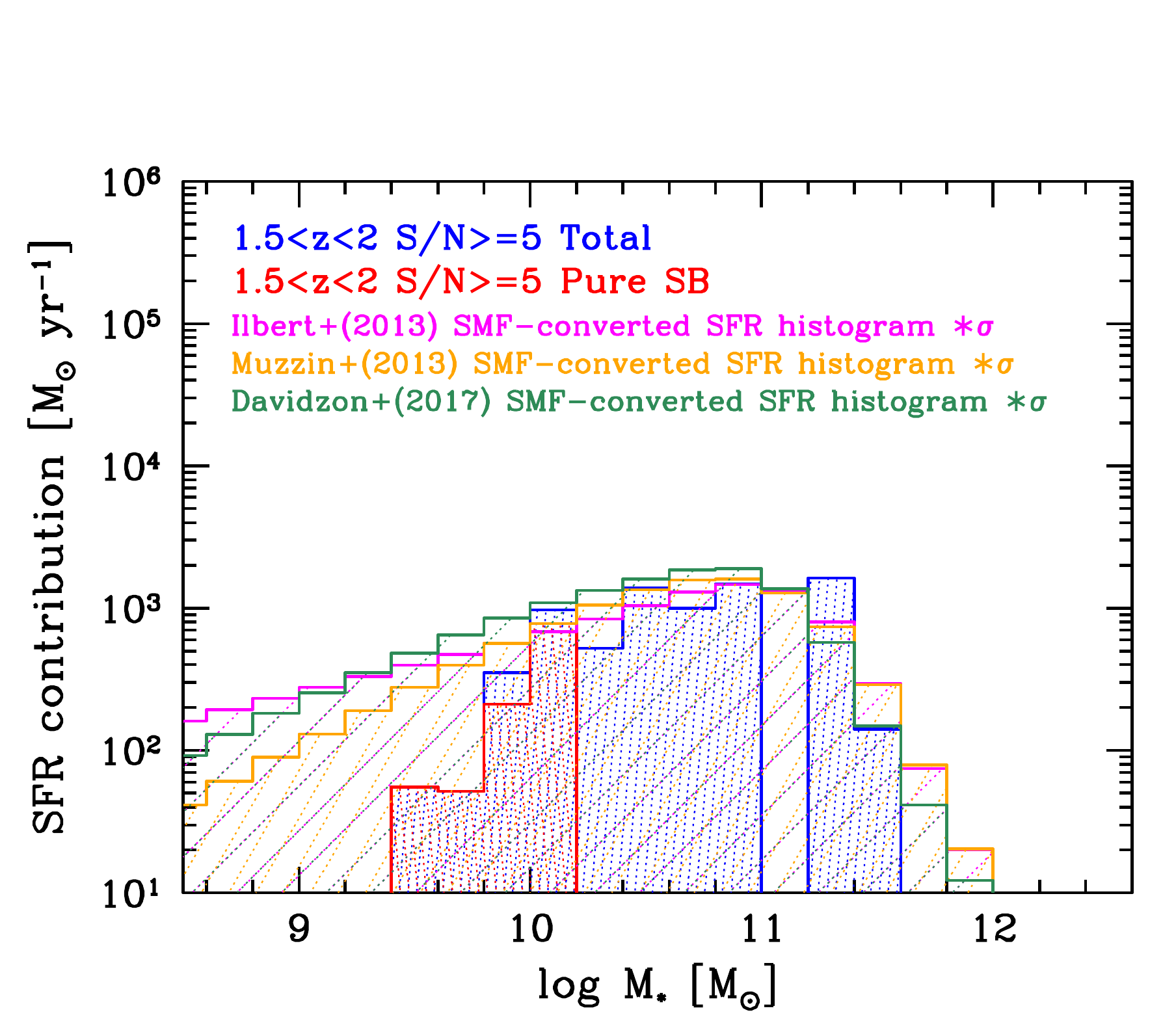}
\includegraphics[width=0.32\textwidth, trim=0 1cm 0 0]{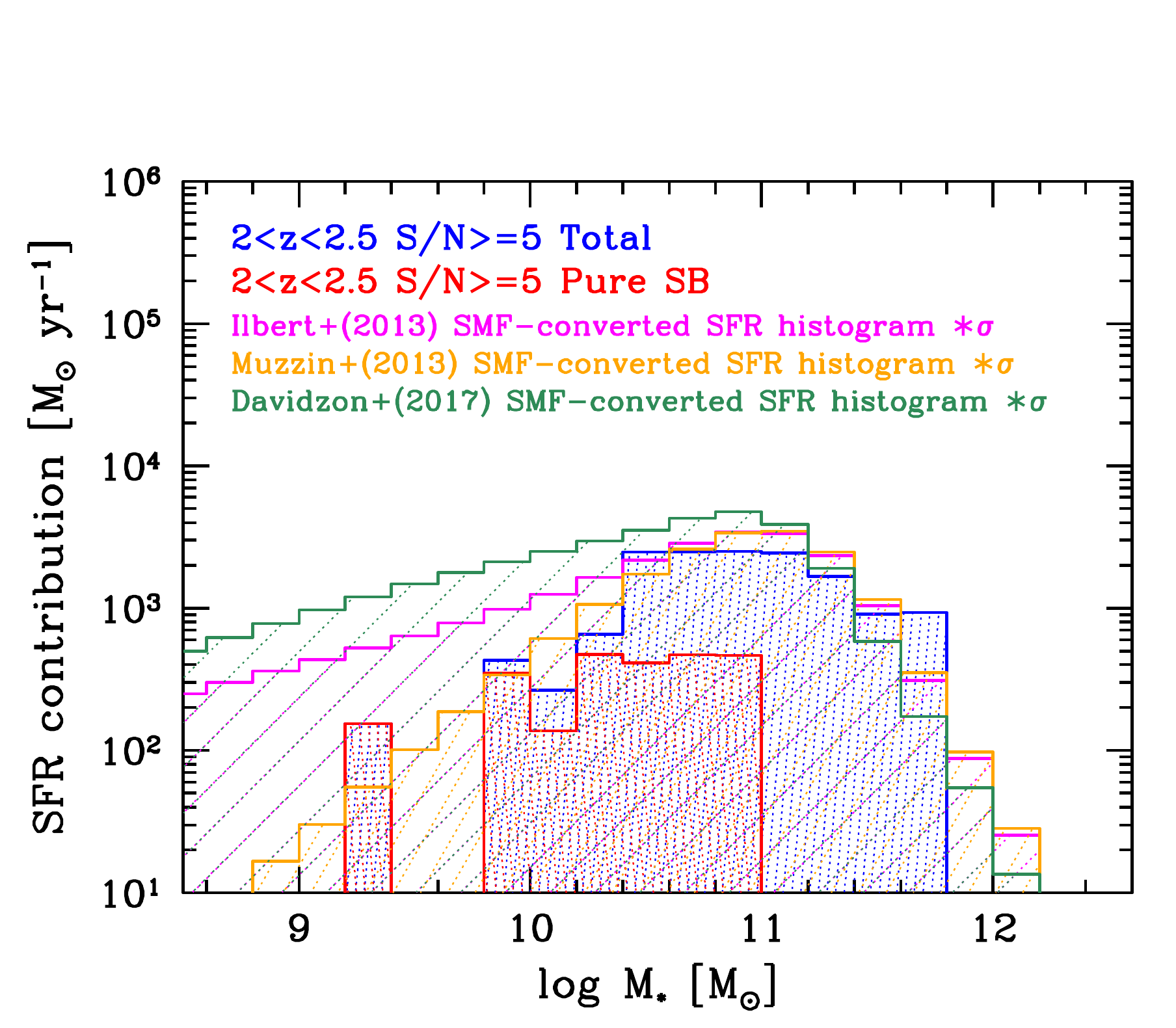}
\includegraphics[width=0.32\textwidth, trim=0 1cm 0 0]{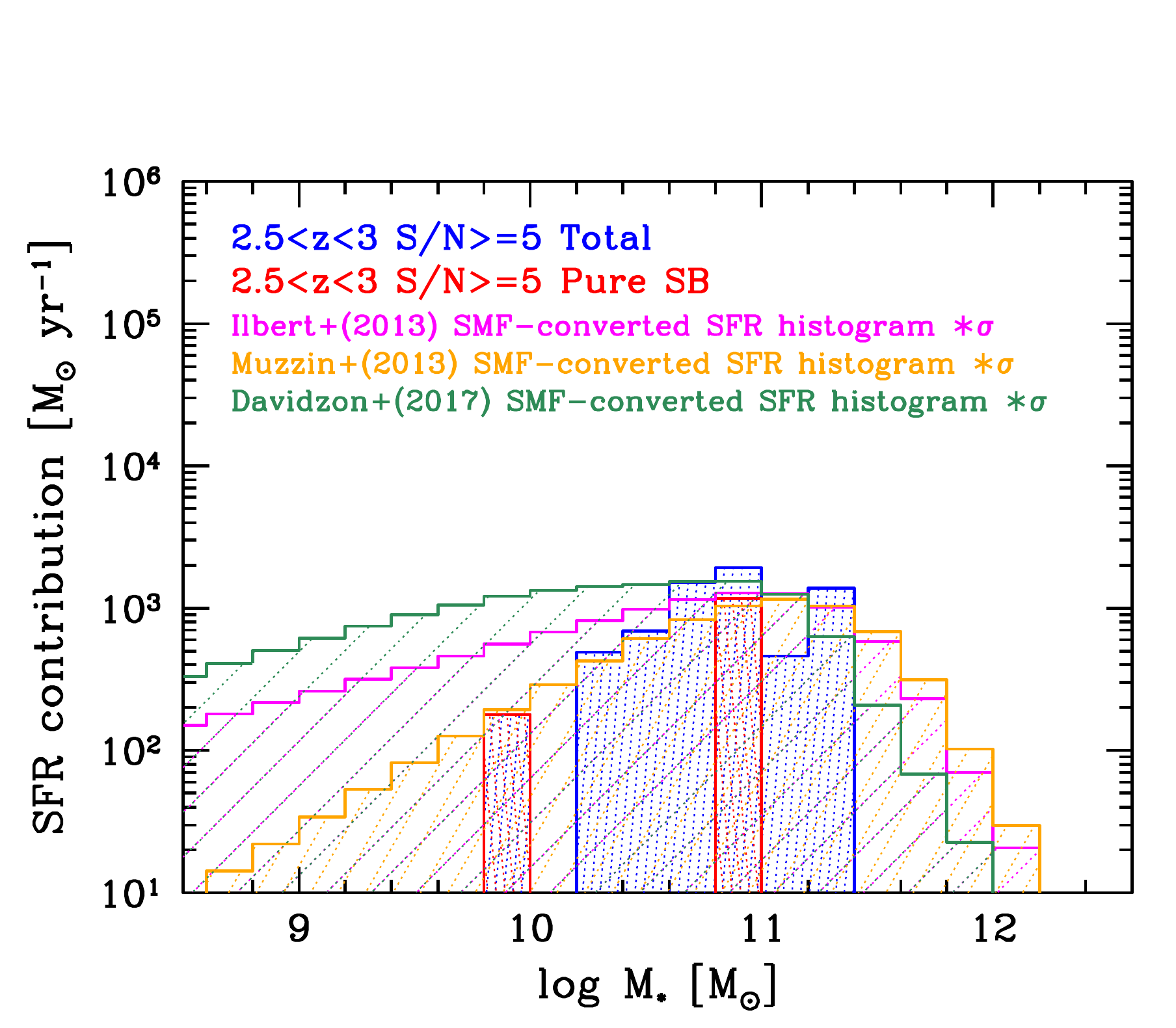}
\includegraphics[width=0.32\textwidth, trim=0 0 0 0]{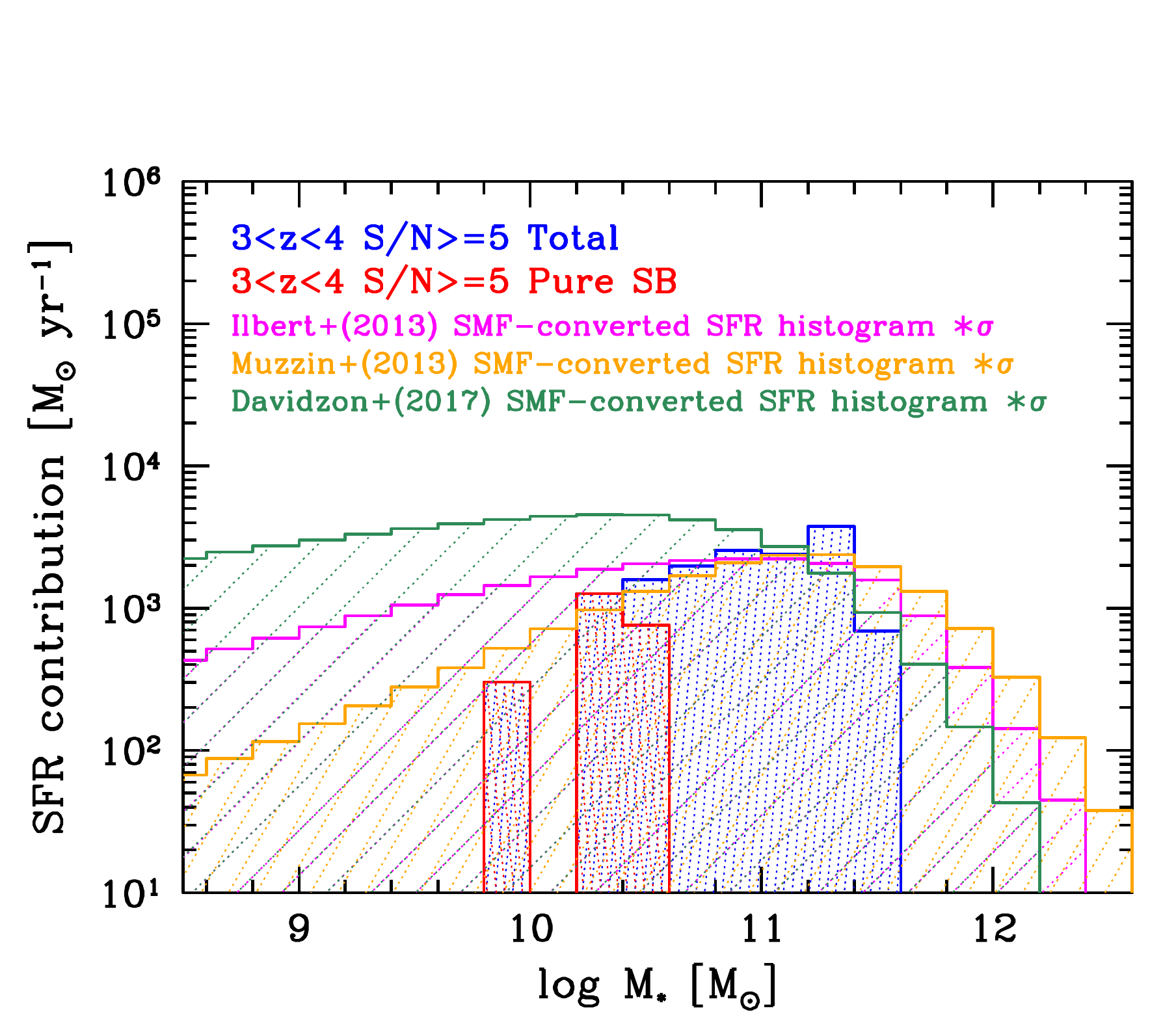}
\includegraphics[width=0.32\textwidth, trim=0 0 0 0]{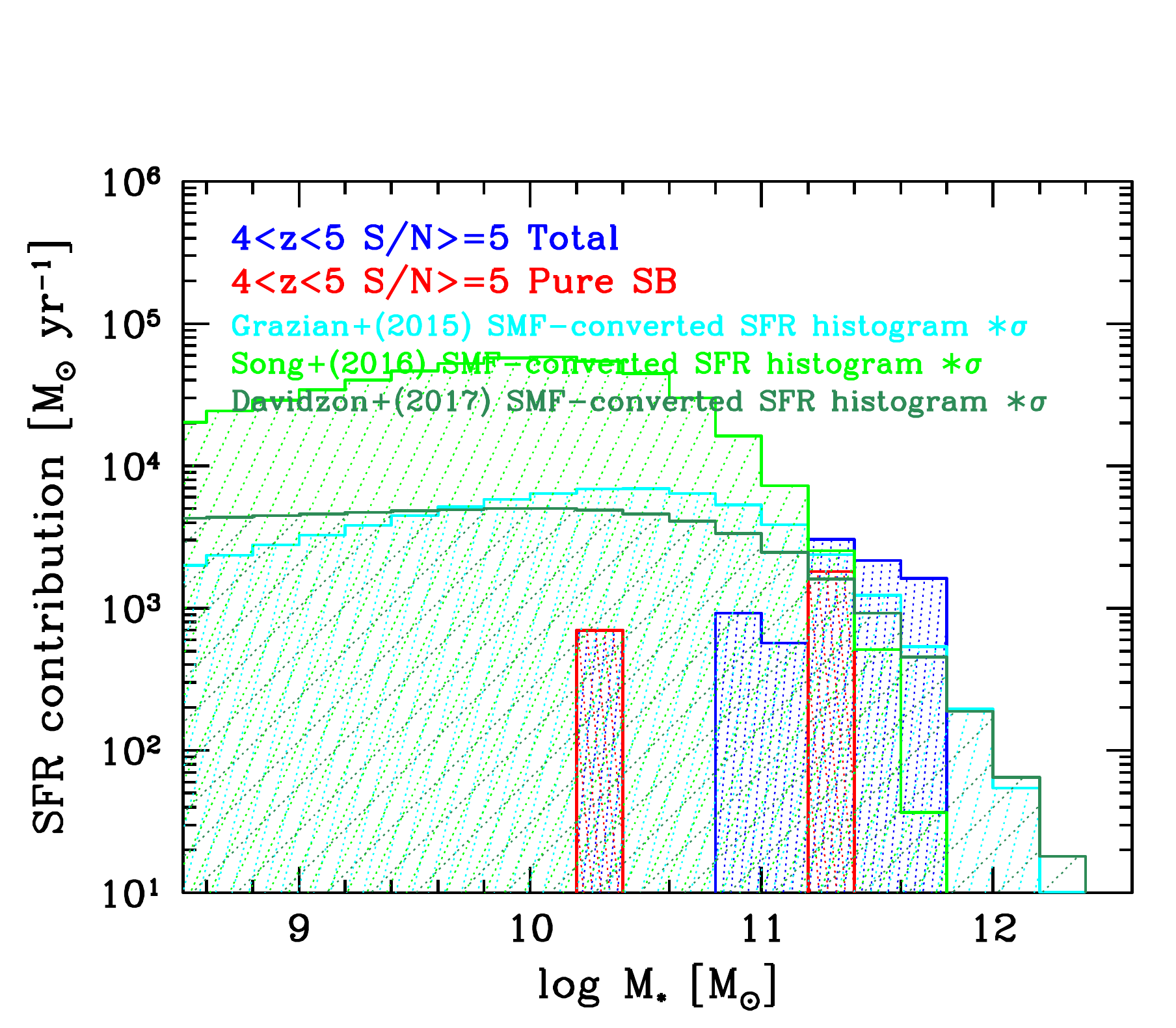}
\includegraphics[width=0.32\textwidth, trim=0 0 0 0]{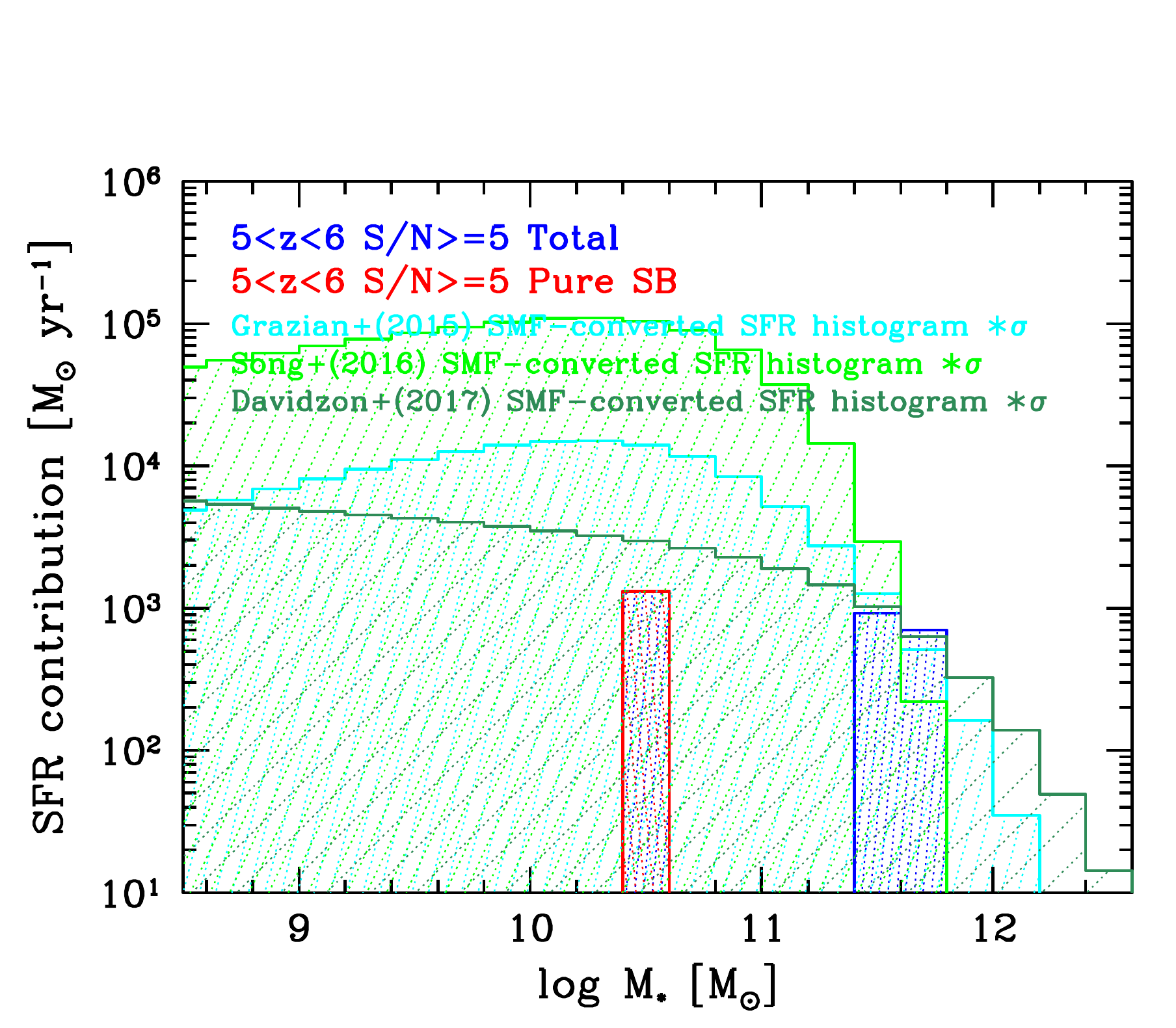}
\caption{%
    The \REVISED[fitted]{renormalized} SMF-converted SFR histograms which are used to estimate the completeness of our sample when deriving the cosmic SFR density (Fig.~\ref{Plot_Cosmic_SFR_Correction}). 
    See text in Section~\ref{Section_CSFRD_Correction}. 
    \label{Plot_Mstar_SFR_Contribution_Renormalized}%
}
\end{figure}

In this work we use a new method to estimate the completeness of our FIR+mm sample by comparing our measurements with \MINORREREVISED[empirical]{literature} SMFs convolved with Main Sequence-like star formation rate distributions. We compute the SMF-converted SFR histograms as described in 
Section~\ref{Section_Mstar_Histograms}, then rescale their normalizations to match the observed SFR histograms. 
We fit only the massive end of the distribution, where the stellar mass histogram has $>50\%$ completeness by number. The re-normalized SMF-converted SFR histograms are shown in Fig.~\ref{Plot_Mstar_SFR_Contribution_Renormalized}. 

\clearpage


\bibliography{biblio}

\end{document}